\documentclass[goeng, submitted, additionalreferee]{thesis}
\pdfoutput=1

\makeatletter
\newif\iffile
\@ifclasswith{thesis}{file}{\filetrue}{\filefalse}
\makeatother

\title{Inverse Problems in Asteroseismology}
\author{Earl Patrick Bellinger}
\town{Albany, New York, USA}

\thesisadvisorycommitteea{\textbf{Dr.~ir.~Saskia Hekker}}
\instthesisadvisorycommitteea{\emph{Max-Planck-Institut f\"ur Sonnensystemforschung, G\"ottingen, Germany\\
Stellar Astrophysics Centre, Aarhus University, Denmark}}
\thesisadvisorycommitteeb{\textbf{Prof.~Dr.~Sarbani Basu}}
\instthesisadvisorycommitteeb{\emph{Department of Astronomy, Yale University, New Haven, CT, USA}}
\thesisadvisorycommitteec{\textbf{Prof.~Dr.~Laurent Gizon}}
\instthesisadvisorycommitteec{\emph{Max-Planck-Institut f\"ur Sonnensystemforschung, G\"ottingen, Germany\\
Institut f\"ur Astrophysik, Georg-August-Universit\"at G\"ottingen, Germany}}
\thesisadvisorycommitteed{\textbf{Prof.~Dr.~Ramin Yahyapour}}
\instthesisadvisorycommitteed{\emph{Institut f\"ur Informatik, Georg-August-Universit\"at G\"ottingen, Germany\\
Gesellschaft f\"ur wissenschaftliche Datenverarbeitung mbH G\"ottingen, Germany}}

\refereea{\textbf{Prof.~Dr.~Ramin Yahyapour}}
\instrefereea{\emph{Institut f\"ur Informatik, Georg-August-Universit\"at G\"ottingen, Germany\\
Gesellschaft f\"ur wissenschaftliche Datenverarbeitung mbH G\"ottingen, Germany}}
\refereeb{\textbf{Prof.~Dr.~Laurent Gizon}}
\instrefereeb{\emph{Max-Planck-Institut f\"ur Sonnensystemforschung, G\"ottingen, Germany\\
Institut f\"ur Astrophysik, Georg-August-Universit\"at G\"ottingen, Germany}}

\refereec{\textbf{Prof.~Dr.~Yvonne Elsworth, FRS}}
\instrefereec{\emph{School of Physics and Astronomy, University of Birmingham, United Kingdom}}

\commissiona{\textbf{Prof.~Dr.~Carsten Damm}}
\instcommissiona{\emph{Institut f\"ur Informatik, Georg-August-Universit\"at G\"ottingen, Germany}}

\commissionb{\textbf{Jun.~Prof.~Dr.~Ing.~Marcus Baum}}
\instcommissionb{\emph{Institut f\"ur Informatik, Georg-August-Universit\"at G\"ottingen, Germany \\
Fakult\"at f\"ur Informatik und Mathematik, Universit\"at Passau, Germany}}

\commissionc{\textbf{Prof.~Dr.~Sarbani Basu}}
\instcommissionc{\emph{Department of Astronomy, Yale University, New Haven, CT, USA}}

\commissiond{\textbf{Dr.~ir.~Saskia Hekker}}
\instcommissiond{\emph{Max-Planck-Institut f\"ur Sonnensystemforschung, G\"ottingen, Germany\\
Stellar Astrophysics Centre, Aarhus University, Denmark}}

\submittedyear{2018}
\publicationyear{2018}
\submitteddate{April~5, 2018}
\examinationdate{May~16, 2018}
\isbn{978-3-944072-61-6}

\usepackage[bookmarksopen=true,%
pdfnewwindow=true,
colorlinks=true,
linkcolor=xlinkcolor,
citecolor=xlinkcolor,
filecolor=xlinkcolor,
urlcolor=xlinkcolor,
final=true]{hyperref}
\hypersetup{pdfstartview=Fit}

\usepackage{bookmark}

\newcommand{\comment}[1]{}
\usepackage[german,english]{babel}

\usepackage[sectionbib]{chapterbib}
\usepackage{natbib}

\usepackage{etoolbox}
\usepackage{calc}
\makeatletter
\patchcmd{\@lbibitem}
  {]}
  {]\makebox[\widthof{999.\ }][r]{\theNAT@ctr.\ }\hspace{2mm}}
  {}{}
\makeatother

\usepackage{printlen}
\usepackage{color}
\usepackage{color}
\usepackage{xcolor}
\usepackage{framed}
\definecolor{shadecolor}{cmyk}{0.02,0.02,0.02,0.02}
\definecolor{xlinkcolor}{cmyk}{1,1,0,0}

\usepackage{longtable}
\usepackage{import}
\usepackage{physics}
\usepackage{ccaption}

\usepackage{subcaption}

\usepackage{afterpage}
\usepackage{array}
\usepackage{rotating}
\usepackage{pdflscape}

\usepackage{graphicx}	
\usepackage{mathtools}
\usepackage{amssymb}	
\usepackage{bm}		    

\usepackage[makeroom]{cancel}

\usepackage{tikz}
\usetikzlibrary{arrows,positioning,shapes,decorations.markings,patterns,fit}
\usepgflibrary{patterns}
\pgfdeclarepatternformonly{fivepointed stars2}
  {\pgfpointorigin}
  {\pgfqpoint{5mm}{5mm}}
  {\pgfqpoint{2mm}{2mm}}
  {
   \pgftransformshift{\pgfqpoint{1mm}{1mm}}
   \pgfpathmoveto{\pgfqpointpolar{18}{1mm}}
   \pgfpathlineto{\pgfqpointpolar{162}{1mm}}
   \pgfpathlineto{\pgfqpointpolar{306}{1mm}}
   \pgfpathlineto{\pgfqpointpolar{90}{1mm}}
   \pgfpathlineto{\pgfqpointpolar{234}{1mm}}
   \pgfpathclose%
   \pgfusepath{fill}
}
\pgfdeclarepatternformonly{fivepointed stars3}
  {\pgfpointorigin}
  {\pgfqpoint{3mm}{3mm}}
  {\pgfqpoint{1.5mm}{1.5mm}}
  {
   \pgftransformshift{\pgfqpoint{1mm}{1mm}}
   \pgfpathmoveto{\pgfqpointpolar{18}{1mm}}
   \pgfpathlineto{\pgfqpointpolar{162}{1mm}}
   \pgfpathlineto{\pgfqpointpolar{306}{1mm}}
   \pgfpathlineto{\pgfqpointpolar{90}{1mm}}
   \pgfpathlineto{\pgfqpointpolar{234}{1mm}}
   \pgfpathclose%
   \pgfusepath{fill}
}

\usepackage{wrapfig}

\usepackage{mathrsfs}

\usepackage{sansmath}

\usepackage[percent]{overpic}
\usepackage[labelsep=period,labelfont={small,bf,sf}]{caption}
\makeatletter
\renewcommand{\fnum@figure}{\normalsize\bfseries\scshape\sffamily{{Figure \thefigure}}}
\renewcommand{\fnum@table}{\normalsize\bfseries\scshape\sffamily{{Table \thetable}}}
\makeatother

\usepackage{needspace}
\usepackage{adjustbox}

\usepackage{savesym}
\savesymbol{tablenum}
\usepackage{siunitx}
\restoresymbol{SIX}{tablenum}

\usepackage{blkarray}

\usepackage{afterpage}
\usepackage{flafter}

\usepackage{eurosym}

\usepackage[sc,osf]{mathpazo} 
\usepackage[euler-digits]{eulervm}
\DeclareSymbolFont{matha}{OML}{txmi}{m}{it}
\DeclareMathSymbol{\varv}{\mathord}{matha}{118}

\usepackage{fancyhdr}
\fancypagestyle{lscape}{%
\fancyhf{} 
\fancyfoot[LE]{}
\fancyfoot[LO] {}
 
}

\def\gridline#1{\vskip6pt\hbox to\hsize{#1}\vskip6pt}

\renewcommand{\mathbf}{\mathbold}
\renewcommand{\boldsymbol}{\mathbold}

\DeclareMathVersion{sans}
\SetSymbolFont{operators}{sans}{OT1}{cmbr}{m}{n}
\SetSymbolFont{letters}{sans}{OML}{cmbrm}{m}{it}
\SetSymbolFont{symbols}{sans}{OMS}{cmbrs}{m}{n}
\SetMathAlphabet{\mathit}{sans}{OT1}{cmbr}{m}{sl}
\SetMathAlphabet{\mathbf}{sans}{OT1}{cmbr}{bx}{n}
\SetMathAlphabet{\mathtt}{sans}{OT1}{cmtl}{m}{n}
\SetSymbolFont{largesymbols}{sans}{OMX}{iwona}{m}{n}

\usepackage{quotchap}
\usepackage{epigraph}
\newcommand{\epigraphrulecolor}{white} 
\makeatletter
\renewcommand{\@epirule}{{\color{\epigraphrulecolor}\rule[.5ex]{\epigraphwidth}{\epigraphrule}}}
\makeatother

\usepackage[raggedright]{titlesec}
\titleformat{\chapter}[display]
  {\sffamily\fontsize{100}{40}\bfseries}
  {\vspace*{-2cm}\hspace*{5in}\textcolor{gray}{\thechapter}}{10pt}{\raggedright\fontsize{30}{1}\normalfont\textit}
\titleformat{\section}
  {\normalfont\sffamily\large\bfseries}
  {\thesection}{1em}{}
\titleformat{\subsection}
  {\normalfont\sffamily\normalsize\bfseries}
  {\thesubsection}{1em}{}
\titleformat{\subsubsection}
  {\normalfont\sffamily\normalsize\bfseries\mathversion{sans}}
  {\thesubsubsection}{1em}{}
\titleformat{\paragraph}
  {\normalfont\sffamily\normalsize\bfseries\mathversion{sans}}
  {\theparagraph}{1em}{}
\titleformat{\subparagraph}
  {\normalfont\sffamily\normalsize\bfseries\mathversion{sans}}
  {\theparagraph}{1em}{}

\newif\ifref
\reffalse
\newcommand{\mb}[1]{\ifref\boldmath\textbf{#1}\unboldmath\else #1\fi}

\usepackage{xspace}
\newif\iflaurent
\laurentfalse
\newcommand{\lr}[1]{\iflaurent\color{red} #1 \color{black}\else #1\fi }

\makeatletter
    \@clubpenalty \clubpenalty
\makeatother

\newcommand{\half}{\frac{1}{2}}
\newcommand{\Ltwo}{\ell(\ell+1)}
\newcommand{\U}{\frac{4\pi\rho r^3}{m}}
\newcommand{\VV}{\frac{G m \rho}{rP}}
\newcommand{\ddr}{\frac{\text{d}}{\text{d}r}}
\newcommand{\ddx}{\frac{\text{d}}{\text{d}\ln r}}

\newcommand{\ddra}[1]{\frac{\text{d} #1}{\text{d}r}}
\newcommand{\ddxa}[1]{\frac{\text{d} #1}{\text{d}\ln r}}
\newcommand{\ddsa}[1]{\frac{\text{d} #1}{\text{d} s}}

\newcommand{\du}{\frac{\delta u}{u}}
\newcommand{\Gr}{\Gamma_{1,\rho}}
\newcommand{\GP}{\Gamma_{1,P}}
\newcommand{\GY}{\Gamma_{1,Y}}
\newcommand{\Kcr}{K_i^{(c, \rho)}}
\newcommand{\Krc}{K_i^{(\rho, c)}}
\newcommand{\Kcsr}{K_i^{(c^2, \rho)}}
\newcommand{\Krcs}{K_i^{(\rho, c^2)}}
\newcommand{\KGr}{K_i^{(\Gamma_1,\rho)}}
\newcommand{\KrG}{K_i^{(\rho, \Gamma_1)}}
\newcommand{\KGcs}{K_i^{(\Gamma_1,c^2)}}
\newcommand{\KcsG}{K_i^{(c^2,\Gamma_1)}}

\newcommand{\dun}{\frac{\delta u'}{u'}}
\newcommand{\dM}{\delta M}
\newcommand{\dR}{\delta R}
\newcommand{\dMM}{\frac{\dM}{M}}
\newcommand{\dRR}{\frac{\dR}{R}}
\newcommand{\dY}{\delta Y}
\newcommand{\KuY}{K_i^{(u',Y)}}
\newcommand{\KYu}{K_i^{(Y,u')}}
\newcommand{\KuYnop}{K_i^{(u,Y)}}
\newcommand{\KYunop}{K_i^{(Y,u)}}

\definecolor{echelle-yellow}{HTML}{F29559}
\definecolor{echelle-blue}{HTML}{0571b0}
\definecolor{echelle-red}{HTML}{DB4D48}
\definecolor{echelle-black}{HTML}{323031}

\definecolor{turn-orange}{HTML}{B16D41}

\definecolor{diff-blue}{HTML}{0571B0}
\definecolor{diff-red}{HTML}{CA0020}
\definecolor{diff-orange}{HTML}{F97100}
\definecolor{diff-purple}{HTML}{551A8B}

\newif\ifhbonecolumn
\hbonecolumntrue
\newcommand{\colwidth}{\ifhbonecolumn 0.5\linewidth\else \linewidth\fi}

\newcommand{\righttrim}{\ifhbonecolumn 0.55 \else 0.45 \fi}

\newcommand{\Mo}{\rm{M}_\odot}

\newcommand{\corr}{\mathrm{Corr}}
  
\defcitealias{2016apj...830...31b}{Bellinger \& Angelou et al. \textcolor{black}{(}2016\textcolor{black}{, BA1 hereinafter)} }

\defcitealias{nelder1965simplex}{Nelder--Mead}
\defcitealias{galileo}{Galileo}
\defcitealias{2017apj...839..116a}{Angelou \& Bellinger et al.\ }
\defcitealias{hadamard}{Jacques Hadamard}
\defcitealias{brunoinfinito}{Giordano Bruno}
\newcommand{\mycitet}[1]{\citetalias{#1} (\citeyear{#1})}
\newcommand{\mycitealt}[1]{\citetalias{#1}\citeyear{#1}}

\makeatletter
\newcommand*{\cleartoleftpage}{%
  \clearpage
    \if@twoside
    \ifodd\c@page
      \hbox{}\newpage
      \if@twocolumn
        \hbox{}\newpage
      \fi
    \fi
  \fi
}
\makeatother

\usepackage{changepage}

\usepackage{pdfpages}

\usepackage[bottom,marginal,splitrule,multiple]{footmisc}

\begin{document}
\label{beginning}
\selectlanguage{english}

\maketitle


\iffile\else
\newpage\thispagestyle{empty}~\newpage
\thispagestyle{empty}
\vspace*{7cm}
\hspace*{0.5cm}\begin{minipage}[c]{0.5\linewidth}
\epigraph{``\emph{Equipped with his five senses, man explores the universe around him and calls \phantom{``}the adventure Science.}''}{--- Edwin Hubble, 1929}
\end{minipage}
\clearpage
\fi

{\hypersetup{linkbordercolor=black,linkcolor=black}
\tableofcontents
\listoffigures
}

\chapter*{Summary\markboth{Summary}{Summary}}
\addcontentsline{toc}{chapter}{Summary}
Asteroseismology allows us to probe the internal structure of stars through their global modes of oscillation. 
Thanks to missions such as the NASA \emph{Kepler} space observatory, we now have high-quality asteroseismic data for nearly $100$ solar-type stars. 
This presents an opportunity to measure the core structures of these stars as well as their ages, masses, radii, and other fundamental parameters. 

This thesis is primarily concerned with two inverse problems in asteroseismology. 
The first is to estimate the fundamental parameters of stars from observations using evolutionary arguments. 
This is inverse to the forward problem of simulating the theoretical evolution of a star, given the initial conditions. 
We solve this problem using supervised machine learning in Chapter~\ref{chap:ML}. 
We find ages, masses, and radii of stars with uncertainties (in the sense of precision) better than $6\%$, $2\%$, and $1\%$, respectively. 
We furthermore use unsupervised machine learning to quantify how each kind of observation of a star is related to its fundamental parameters in Chapter~\ref{chap:statistical}. 

The second problem is to infer the structure of a star from its frequencies of pulsation using asteroseismic arguments. 
This is inverse to the forward problem of calculating the theoretical pulsation frequencies for a known stellar structure. 
Solving this problem presents an opportunity to test the quality of stellar evolution models, as we may then directly compare the asteroseismic structure of a star against theoretical predictions. 
We solve this problem in Chapter~\ref{chap:inversion}. 
Applying this technique to the solar-type stars in 16~Cygni, we find that while the structure of the $1.03$ solar-mass star 16~Cyg~B is in good agreement with theoretical expectations, the more massive 16~Cyg~A differs in its internal structure from best-fitting evolutionary models. 

These inverse problems are both \emph{ill-posed} in the sense that (I) a solution may not exist within the confines of the current theory; (II) if there is a solution, it may not be unique, as many solutions may be consistent with the data; and/or (III) the solutions may be unstable with respect to small fluctuations in the input data. 
Therefore, care must be put into determining possible solutions and applying regularization where necessary. 

Chapter~\ref{chap:intro} introduces this thesis with the history and theory of stellar structure, evolution, and pulsation; and emphasizes the role that variable star astronomy played in shaping our understanding of stellar evolution. 
It also contains the kernels of stellar structure, an introduction to ill-posed inverse problems, and a discussion of some computational issues for the algorithms used to solve these problems. 


\chapter*{Zusammenfassung\markboth{Zusammenfassung}{Zusammenfassung}}
\addcontentsline{toc}{chapter}{Zusammenfassung}
Die Asteroseismologie erlaubt es uns, die innere Struktur der Sterne durch Messungen ihrer globalen Schwingungsmoden zu untersuchen. 
Dank Missionen wie dem Weltraumteleskop \emph{Kepler} der NASA verf\"ugen wir heute \"uber qualitativ hochwertige asteroseismische Daten von fast $100$ sonnen\"ahnlichen Sternen. 
Dies bietet die M\"oglichkeit, das Innere dieser Sterne sowie deren Alter, Masse, Radien und andere fundamentale Parameter zu bestimmen. 

Diese Doktorarbeit besch\"aftigt sich in erster Linie mit zwei inversen Problemen der stellaren Astrophysik. 
Das erste Problem besteht darin, die fundamentalen Parameter eines Sterns aus seinen Beobachtungen mit Hilfe von Argumenten der Sternevolution zu sch\"atzen. 
Dieses Problem ist invers zu dem Vorw\"artsproblem der Simulation der theoretischen Sternentwicklung unter bestimmten Anfangsbedingungen. 
Mit Hilfe von Methoden des \"uberwachten maschinellen Lernens wird dieses Problem in Kapitel~\ref{chap:ML} gel\"ost. 
So ermitteln wir Alter, Masse und Radien mit einer Unsicherheit von weniger als $6\%$, $2\%$ und $1\%$. 
In Kapitel~\ref{chap:statistical} verwenden wir Methoden des un\"uberwachten maschinellen Lernens, um zu quantifizieren wie genau sich die fundamentalen Parametern eines Sterns durch die Kombination verschiedener Arten der Sternbeobachtung bestimmen lassen.

Das zweite Problem besteht darin, die Struktur eines Sterns aus seinen Pulsationsfrequenzen abzuleiten, wobei nur asteroseismische Argumente verwendet werden. 
Dieses Problem ist invers zu dem Vorw\"artsproblem der Berechnung der theoretischen Pulsationsfrequenzen einer bekannten Sternstruktur. 
Die L\"osung dieses Problems bietet die M\"oglichkeit, die Qualit\"at unserer Modelle der Sternentwicklung zu testen, da wir so die asteroseismische Struktur eines Sterns direkt mit theoretischen Vorhersagen vergleichen k\"onnen. 
Dieses Problem wird in Kapitel~\ref{chap:inversion} gel\"ost. 
Wendet man diese Technik auf die beiden sonnen\"ahnlichen Sterne des Systems 16~Cygni an, so stellt man fest, dass die Struktur des $1,03$ Sonnenmassensterns 16~Cyg~B in guter \"Ubereinstimmung mit den theoretischen Vorhersagen ist, w\"ahrend sich der massivere Stern 16~Cyg~A in seiner inneren Struktur von den am besten passenden Evolutionsmodellen unterscheidet. 

Diese inversen Probleme sind im mathematischen Sinne inkorrekt gestellt, sodass (I) eine L\"osung innerhalb der Grenzen der aktuellen Theorie m\"oglicherweise nicht existiert; (II) wenn es eine L\"osung gibt, muss sie nicht eindeutig sein, da viele L\"osungen mit den Daten konsistent sein k\"onnen; und/oder (III) die L\"osungen k\"onnen in Bezug auf kleinere Schwankungen der Ausgangsdaten instabil sein. 
Daher wird viel Sorgfalt darauf verwendet, die Menge der m\"oglichen L\"osungen zu bestimmen und bei Bedarf eine Regularisierung vorzunehmen. 

Kapitel~\ref{chap:intro} leitet diese Arbeit mit der Geschichte und Theorie der Sternstruktur und -evolution ein. Der Schwerpunkt liegt hierbei auf der Theorie der stellaren Pulsationen und wie sie dazu beigetragen hat, unser Verst\"andnis der Sternevolution zu formen. Des Weiteren enth\"alt es Ableitungen der Integralkerne der stellaren Struktur, eine kurze Einf\"uhrung in die mathematisch inkorrekt gestellten inversen Probleme, und eine Diskussion \"uber einige numerische Schwierigkeiten bez\"uglich des maschinellen Lernens und der statistischen Algorithmen die verwendet werden, um diese Probleme zu l\"osen. 


\chapter{Introduction} 
\label{chap:intro}

\section{Variable Stars} 
\label{sec:history}

Points of light in the night sky are not constant but rather they are \emph{variable}: they dim or brighten over time. 
Some of these variations are periodic: they dim and brighten again with a kind of regularity. 
This fact may have been known as early as the time of the ancient Egyptians, who, over $3,200$ years ago, recorded in their calendars the $2.85$-day period of the so-called ``Demon Star,'' Algol \citep[e.g.,][]{jetsu2015shifting}. 
Periodic variables were not known in the Western world however until around the 17th century, after the German pastor David Fabricius and his son observed the reappearance of a faded object that they had previously assumed to be a nova \citep[e.g.,][]{2015pust.book.....C}. 
This object was named \emph{Mira}, Latin for `Wonderful' \citep{hevelius}. 

Regardless of variability, it was still not yet known at this time what these points of light in the sky actually were. 
Extrapolating from \citet{copernicus}, the Italian philosopher \mycitet{brunoinfinito} 
was the among the first in the Western world to suggest that these lights are in fact \emph{stars} not unlike our own Sun. 
Though the 17th century began with Bruno being burned at the stake for this heresy \citep[e.g.,][]{bruno1998giordano}, the recognition of this viewpoint fortunately became commonplace over the following centuries due to the efforts of figures such as \citet{kepler}, \mycitet{galileo}, \citet{newton}, \citet{huygens}, \citet{1838AN.....16...65B}, and \citet{sacchi}. 

The field of research into periodic variable stars arguably began in the year 1638 when the Frisian astronomer Johannes Holwarda measured the period of Mira to be about $11$ months long \citep[e.g.,][]{1997JAVSO..25..115H}. 
Algol itself was not rediscovered in the West as being variable until 1667, although others may have seen it without noting it as such \citep[e.g.,][]{bolt2007biographical}. 
Throughout this and the following century, astronomers such as \citet{10.2307/101080} and \citet{flamsteed} made remarks about a number of stars that seemed to appear, disappear, or otherwise change in brightness; but they did not study them further \citep[e.g.,][]{10.2307/106621}. 

In the 18th century, the English astronomer Edward Pigott and his distant cousin, the short-lived and deaf John Goodricke, calculated the period of Algol as $2.865$ days---a few minutes shorter than the present-day observed value \citep{10.2307/106502,10.2307/106591,2012ApJ...752...20B}. 
They also discovered another variable star, $\beta$~Lyrae, whose symmetric light curve resembled Algol's \citep{1785RSPT...75..127P}. 
Pigott assembled these and ten ``undoubtedly changeable'' others---along with $38$ more candidates---into the first-ever catalog of variable stars \citep{10.2307/106621}. 

To explain the variability, the English polymath John Michell used statistical arguments to reason that stars likely group together and form systems, with ``the odds against the contrary opinion being many million millions to one'' \citep{michell1767inquiry}. 
The light coming from stars could be then eclipsed, with stars or other objects (planets, moons) regularly passing in front of one another in our line of sight to block the light from reaching our eyes. 

At first, Pigott and Goodricke posited that the variability of Algol was caused by eclipses, as Michell had proposed \citep{10.2307/106502}. 
However, within three years they changed their interpretation, then attributing its variability to ``rotation of the star on its axis, having fixed spots that vary only in their size'' \citep{10.2307/106614}. 
This idea may have seemed attractive due to their knowledge of sunspots, which had been known in the Eastern world since at least the time of the Babylonians, though not rediscovered in the West until $150$ years prior when Fabricius and his son turned their telescopes to the Sun following their discovery of Mira \citep{1611mson.book.....F}. 

\begin{figure}[p] 
    \centering
    \includegraphics[width=0.5\textwidth]{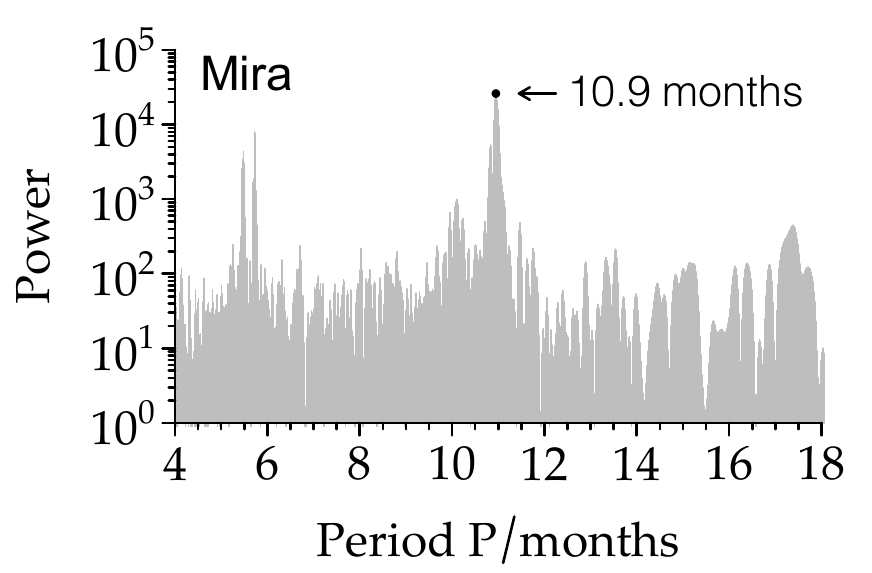}%
    \includegraphics[width=0.5\textwidth]{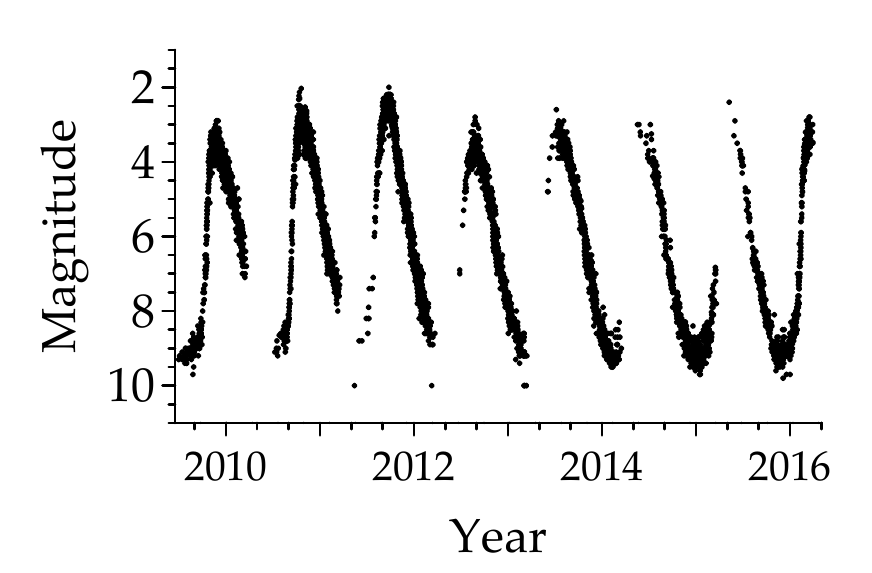}\\%
    \vspace*{0.4cm}
    \includegraphics[width=0.5\textwidth]{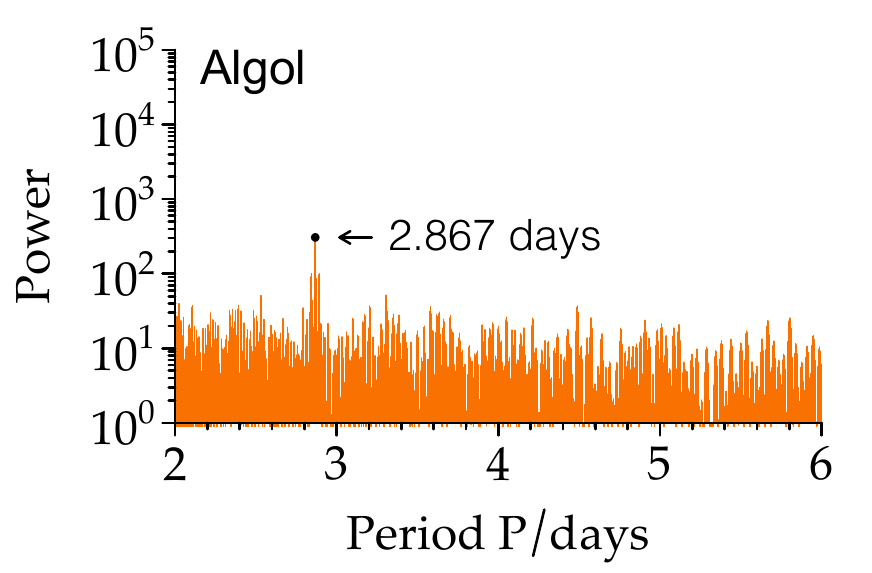}%
    \includegraphics[width=0.5\textwidth]{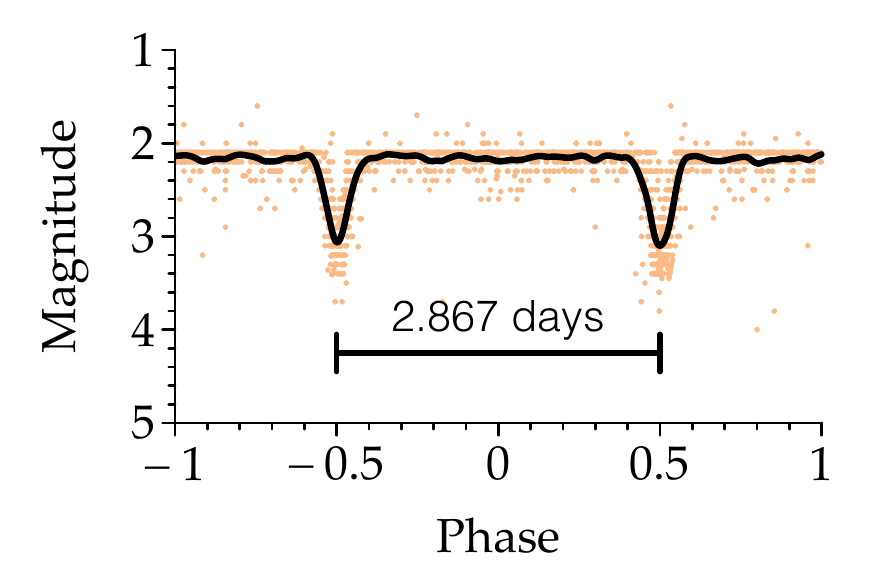}\\%
    \vspace*{0.4cm}
    \includegraphics[width=0.5\textwidth]{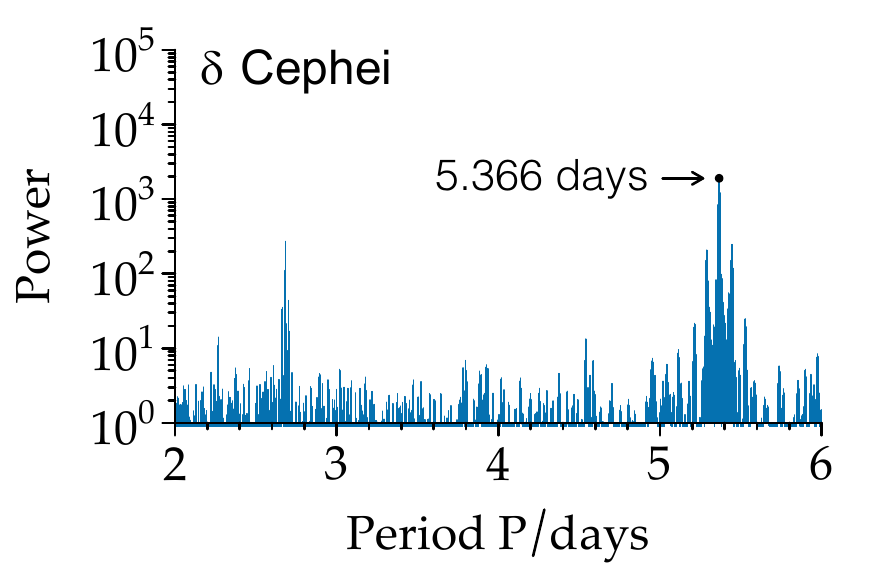}%
    \includegraphics[width=0.5\textwidth]{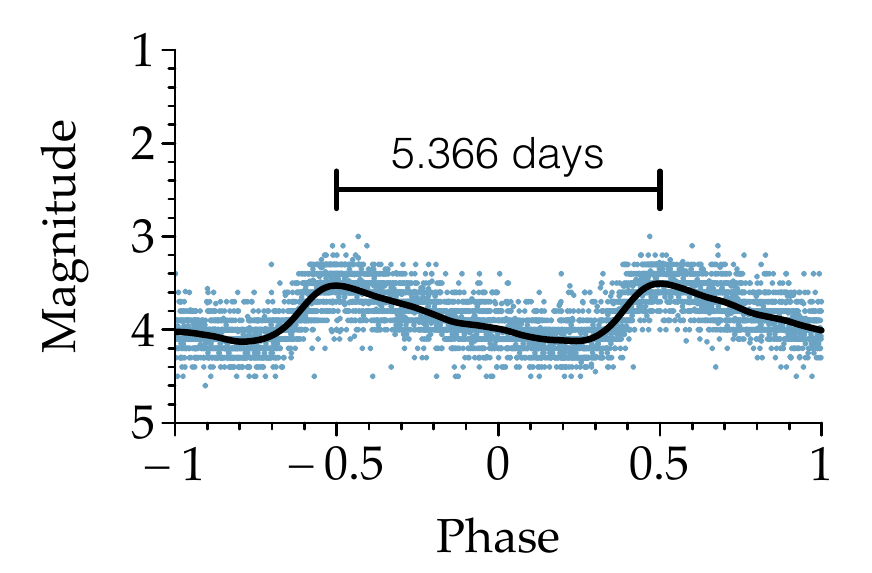}\\%
    \vspace*{0.4cm}
    \caption[Light curves of the first-known periodic variable stars]{Modern-day periodograms and light curves for Mira ($o$~Ceti), Algol ($\beta$~Persei), and $\delta$~Cephei---the ``prototypes'' for the first three discovered classes of periodic variable stars. 
    The light curves for Algol and $\delta$~Cephei are phased by their period. 
    Mira has a long and somewhat irregular period. 
    Unlike the other two, which are constantly changing in brightness, the light from Algol is generally stable with occasional quick dips. 
    \emph{Data acquired from the American Association of Variable Star Observers} \citep[\textsc{AAVSO},][]{AAVSO}. 
    \label{fig:lightcurves}}
\end{figure}

Pigott and Goodricke also discovered two other periodic variable stars, $\eta$~Aquilae and $\delta$~Cephei \citep{1785RSPT...75..127P, 10.2307/106614}. 
These stars earned a new name---\emph{Cepheid} variable stars---as the manner in which their light changed over time was noticeably different from Algol's. 
Rather than quickly dipping and brightening again every so often, these stars appear to change continuously (for a visual comparison, see Figure~\ref{fig:lightcurves}). 
Unlike with Algol, they offered no explanations for Cepheid-type variability. 
Goodricke died that year at the age of $21$, having been elected a Fellow of the Royal Society only days prior, 
but never learning of the honor. 

For a long time thereafter, the discovery of variable stars slowed. 
Less than ten new variables were discovered in the following $60$ or so years. 
These new variables were published by the German astronomer Friedrich Wilhelm Argelander \citep{1844scja.book..122A}, to whom the variable star naming convention\footnote{ Starting with the letter R and the name of the constellation where it is found (e.g., R~Lyrae), then repeating with double letters when the alphabet is exhausted (e.g., RR~Lyrae).} is owed.
The only other major advance in the first half of the 19th century was the development of the least squares method, which improved period estimates \citep[e.g.,][]{1994JHA....25...92Z}. 

In the second half of the 19th century, the fields of astronomical spectroscopy and dry plate astrophotography were born. 
These technologies proved a great aid for the discovery and analysis of variable stars, and even revealed the existence of several new classes of variable stars. 
By 1865, the number of known variable stars had more than doubled, going up to $123$ \citep{1865AN.....63..117C}. 
In the next $30$ years, that number quadrupled with over $300$ new discoveries \citep[e.g.,][]{1997JAVSO..25..115H}. 
Nearly $50$ variable stars were discovered in the year 1896 alone, the majority of which being Mira-type variables, $19$ of which were found by the Harvard ``computer'' Williamina P.~Fleming. 
By the end of the 19th century, the number of known variable stars grew to at least $2000$ \citep[e.g.,][]{Samus2017}. 



The latter half of the 19th century also marked the beginning of a change in attitude toward astronomical research. 
In addition to cataloging the sky, researchers began seeking rigorous physical foundations to understand the nature of the Sun and the stars. 
Applying techniques from the recently-born field of thermodynamics, figures such as William Thomson (a.k.a.\ Lord Kelvin), Julius Robert Mayer, Hermann von Helmholtz and others worked to determine the ages of stars and identify the sources of their energy. 
In particular, they offered the explanation that gravitational energy can be converted into heat via either contraction or the infall of meteoric material. 
For example, Helmholtz demonstrated that the Sun could be powered by contracting merely $380$ feet each year \citep[e.g.,][]{ARNY1990211}. 
Now called the Kelvin-Helmholtz mechanism, this was the only known form of stellar heating at the time. 
Applying it to the study of the Earth and Sun, Kelvin found that the solar system must be at most millions of years old \citep[e.g.,][]{1895Natur..51..438K}, much younger than the currently accepted age of about $4.57$ billion years.\footnote{ A devout Christian, Lord Kelvin used these results to doubt Charles Darwin's recently-published theory of biological evolution, which requires an older Earth \citep{darwin}.}

The calculations that Helmholtz and Kelvin made required details of the structure of the Sun, and so to carry them out, they created the first polytropic models of stellar structure \citep[e.g.,][]{ARNY1990211}. 
These models are characterized by the internal pressure depending only on the density of the stellar material. 
Much as is still done today, they considered a sphere where gravity forces are in balance against pressure forces. 
However, they erroneously assumed that all energy in the Sun is transported by convection. 

\begin{figure}
    \centering
    \begin{overpic}[width=\textwidth,trim={0 0.5cm 0 0}, 
                clip
            ]{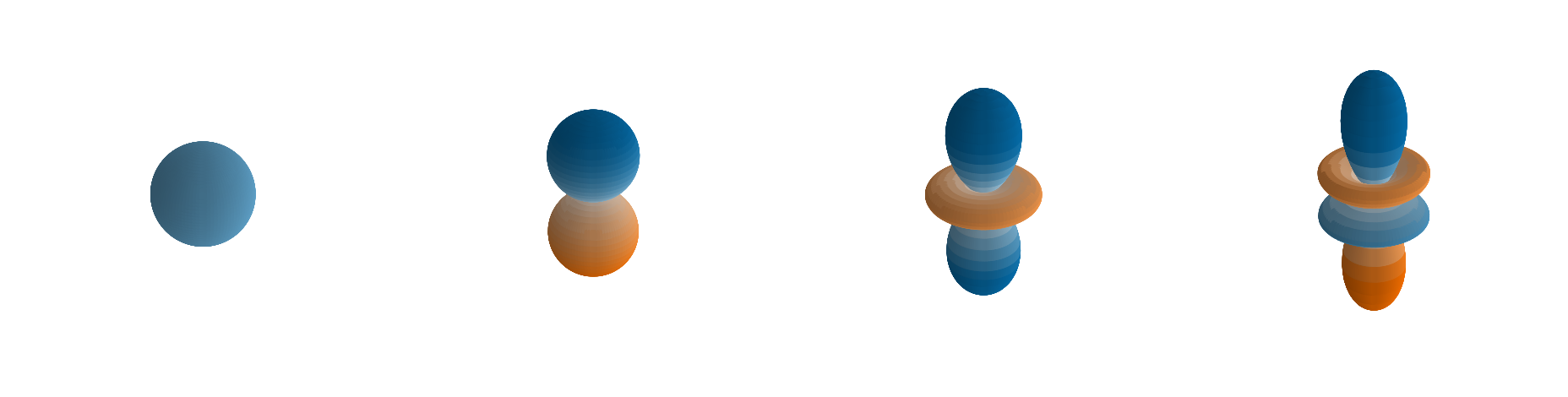}
        \put (9,19)  {\text{radial}}
        \put (33.8,19) {\text{dipole}}
        \put (55.6,19) {\text{quadrupole}}
        \put (82.1,19) {\text{octupole}}
    \end{overpic}
    \caption[Spherical harmonics]{\lr{Radial and non-radial stellar pulsations for a non-rotating star. 
    Mathematically, these show ${r(\theta, \phi) = \Re |Y_{\ell}(\theta, \phi)|}$ in spherical polar coordinates for ${\ell=0}$, $1$, $2$, $3$, where $Y_{\ell}$ is the solution to Laplace's equation on a sphere---special functions known as \emph{spherical harmonics}. The sign of ${\Re(Y_{\ell})}$ is indicated by color. 
    Pulsations with ${\ell=0}$ correspond to the entire star moving toward or away from the center without horizontal motions, i.e., radial pulsations.}
    \label{fig:sph}}
\end{figure}
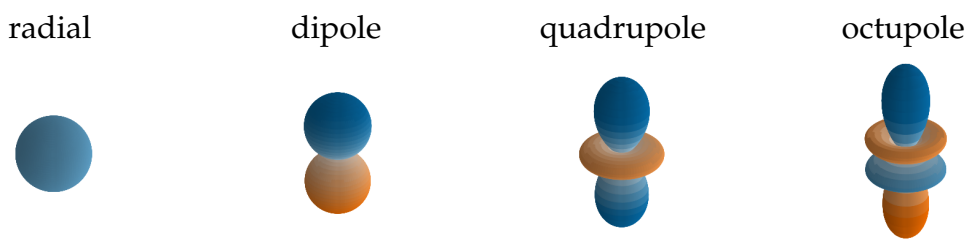

It was also around this time that the idea stars might pulsate was first given serious attention. 
Lord Kelvin was the first to state the equations of non-radial pulsation for chemically homogeneous ``spheroids of incompressible liquid'' \citep{1863RSPT..153..583T}. 
Though this work makes no explicit mention of stars, it was thought at this time that stars might be entirely liquid \citep[e.g.,][]{ARNY1990211}. 
However, it was argued for a long time thereafter that stars could not possibly pulsate non-radially, as these modes of oscillation would be damped out by viscous forces \citep[e.g.,][]{1938ApJ....88..189P}. 
Figure~\ref{fig:sph} shows some of the configurations that a star could take under the pulsation hypothesis.

After completing his Ph.D.\ at the University of G\"ottingen, the German astrophysicist August Ritter wrote a series of $19$ papers over an $11$-year span laying out theory of stellar structure \citep[1878--1889, e.g.,][]{ritter}. 
Ritter had the insight to treat stars as an ideal gas, and derived a relationship between the mass of a star and its luminosity. 
Ritter also developed here the radial theory of stellar pulsations, including the important result connecting the period of stellar pulsation to the mean density of the star. 
Since the source of stellar variability was still an open puzzle, Ritter conjectured that stars might be radial pulsators. 
Unfortunately, this work was largely ignored.\footnote{ In his influential textbook \emph{An Introduction to the Study of Stellar Structure}, Nobel laureate \citet{1939isss.book.....C} characterized this body of work as ``a classic, the value of which has never been adequately recognized,'' and noted that in these works Ritter worked out ``almost the entire foundation for the mathematical theory of stellar structure.''} 

In an attempt to understand the temperature of the Sun, the American theoretical astrophysicist and Yale alumnus J.\ Homer Lane continued work on polytropes \citep[e.g.,][]{1870AmJS...50...57L}. 
Lane discovered the curious fact that stars have a negative heat capacity: i.e., when they lose energy, they contract and heat up. 
Ritter rederived Lane's Law and used it to develop the first physically-motivated (albeit incorrect) theory of stellar evolution: that a star begins its life as a diffuse gaseous mass, which at first contracts and heats; eventually, the star transforms into a liquid, and then undergoes a long period of cooling. 


At the end of that century, the German astronomer Hermann Carl Vogel used spectroscopic measurements to firmly establish that Algol is an eclipsing binary, thereby confirming Goodricke's initial speculation \citep{1889AN....121..241V, 1908ApJ....27....1F}. 
Vogel taught his methods to the Russian astronomer Aristarkh B{\'e}lopolsky, who then took spectra of the Cepheid stars $\delta$~Cephei and $\eta$~Aquilae. 
Though at this time eclipses were widely thought to be the most likely the source of Cepheid variability, B{\'e}lopolsky argued that the radial velocity variations of these stars were inconsistent with the eclipse hypothesis \citep{1897ApJ.....6..393B, 1895ApJ.....1..160B}. 
\epigraph{``\emph{The times of minimum brightness and the times for which the velocity in the line \hphantom{``}of sight is zero do not coincide. For this reason the changes in the brightness of \hphantom{``}the star cannot be explained as the result of eclipses, and some other explanation \hphantom{``}must be sought.}''}{--- Aristarkh Apollonovich B{\'e}lopolsky \\\textit{Researches on the spectrum of the variable star $\eta$~Aquilae} (\citeyear{1897ApJ.....6..393B})}

Several alternative theories for Cepheid variability arose over the years. 
So-called ``veil theories'' suggested that clouds could rapidly form and evaporate, serving to block the source of the light for a short time \citep[e.g.,][]{1889Natur..39..606B}. 
English astronomer Henry Plummer, later President of the Royal Astronomical Society, suggested that Cepheids are radial pulsators \citep{1914MNRAS..74..660P}. 
Others maintained the eclipsing binary hypothesis \citep[e.g.,][]{1909LicOB...5...82D} with some even claiming that B{\'e}lopolsky's measurements had in fact proven it \citep[e.g.,][]{1913Obs....36...59B}. 

\begin{figure}
    \centering
    \includegraphics[width=\textwidth,trim={0 5cm 0 2cm},clip]{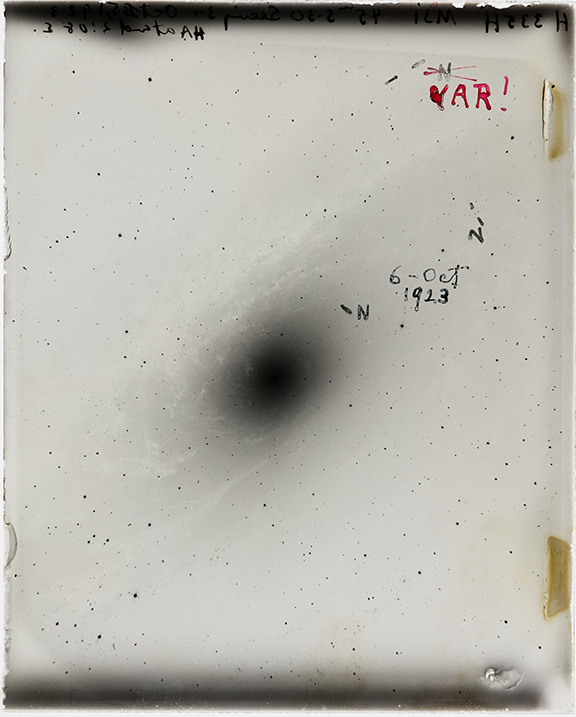}
    \caption[VAR! Day]{Edwin Hubble's photographic plate showing the discovery of a Cepheid variable star in the Andromeda Galaxy (M31). 
    In a series of $17$ papers, Harlow Shapley used the Leavitt Law to estimate the distance to globular clusters and map out the size of the Galaxy, finding that it was substantially larger than previously estimated \citep{1918ApJ....48...89S}. 
    In 1920, Shapley engaged in the ``Great Debate'' of astronomy, in which he argued that the Milky Way comprised the entirety of the Universe \citep{1921BuNRC...2..171S}. 
    Soon thereafter, Edwin Hubble used this same technique to measure the distance to the spiral nebulae M31 and M33 \citep[][see image]{1925Obs....48..139H}. 
    Finding that they were extremely distant, Hubble proved that these nebulae were in fact galaxies external to our Milky Way---instantly expanding the calculated size of the Universe by a factor of $100,000$. 
    Hubble sent these results to Shapley, who, upon viewing them, is said to have remarked: \emph{``Here is the letter that has destroyed my Universe.''} 
    Edwin Hubble subsequently used the Leavitt Law to estimate the distances to several more Cepheid-host galaxies \citep{1929PNAS...15..168H}. 
    Combining these distances with measurements of the speeds at which those galaxies are receding from us, Hubble measured the rate of cosmic expansion, and thus the age of the Universe. 
    Variable star enthusiasts can celebrate October 6 as ``VAR! Day'' (see image). 
    \emph{(Image reprinted with permission from Carnegie Observatories.)}
    \label{fig:var-day}}
\end{figure}

Regardless of the cause of their blinking, Cepheid stars gained near-immediate fame throughout astronomical circles and beyond following the discovery by American astronomer Henrietta Swan Leavitt (another Harvard `computer') that ``the brighter variables have the longer periods'' \citep{1908AnHar..60...87L, 1912HarCi.173....1L}. 
Now known as the Cepheid Period-Luminosity Relation or the \emph{Leavitt Law}, this enabled measurement of vast cosmic distances via comparison of observed brightnesses with those expected from Cepheid periods. 
This discovery thus established Cepheids as standard candles---the first to be discovered---and was quickly put to use in mapping the structure of the Universe (see Figure~\ref{fig:var-day}). 



Around this time, the then-unknown Danish astronomer Ejnar Hertzsprung was working to combine spectroscopy of stars with parallax distance measurements. 
He found that stars form two distinct groups: ``Riesen'' (giants) and ``Zwerge'' (dwarfs). 
Hertzsprung published this work in a photographic journal with little impact \citep{1905WisZP...3..442H, 1907WisZP...5...86H}. 
It did however get the attention of Karl Schwarzschild, director of the G\"ottingen Observatory, who then appointed him to a position there \citep[e.g.,][]{bolt2007biographical}. 
Hertzsprung went on to discover that the pole star Polaris is also a Cepheid-type variable\footnote{ Hence, Caesar is as constant as a variable star \citep{shakespeare}.} \citep{1911AN....189...89H} and furthermore concluded that Cepheids are giant stars \citep{1913AN....196..201H}. 

\begin{figure}
    \centering
    \includegraphics[width=\textwidth]{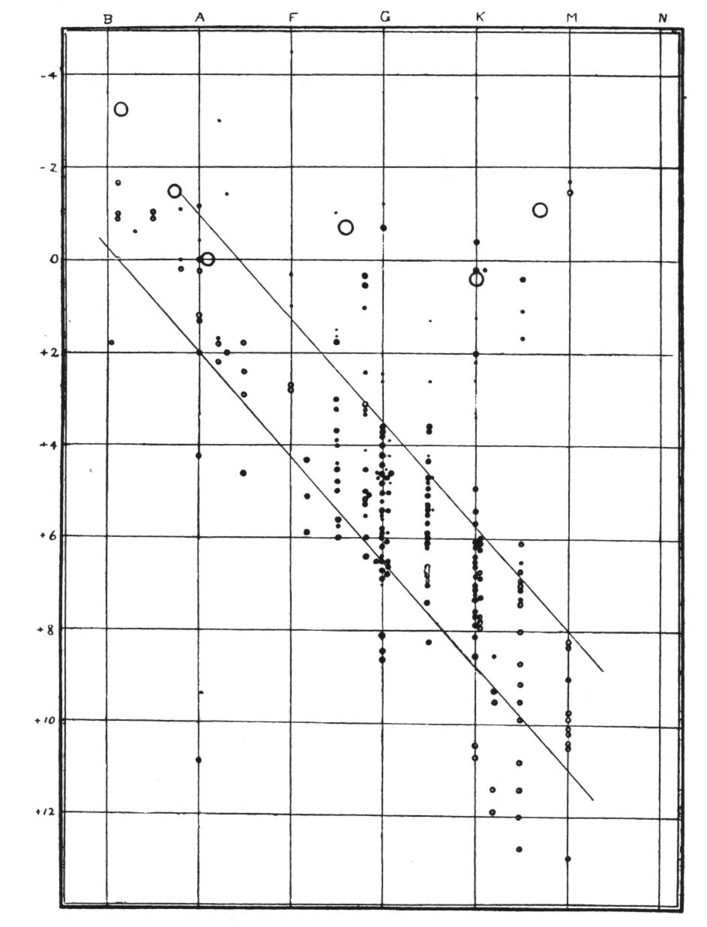}
    \caption[Historical Hertzsprung-Russell Diagram]{One of the first Hertzsprung-Russell diagrams, showing the absolute magnitude of stars against their spectral type. 
        Luminosity increases upward; temperature increases leftward. 
        Dwarf stars reside on the diagonal---the \emph{main sequence}---and giant stars occupy the upper right corner. 
        \emph{(Figure reprinted with permission from \citealt{1914Natur..93..252R}.)} 
        \label{fig:HRD}}
\end{figure}

The director of the Princeton Observatory, Henry Norris Russell, a much more influential astronomer at the time, also came to the same conclusions as Hertzsprung \citep[e.g.,][]{1913Obs....36..324R, 1913Sci....37..651R}. 
Plotting the absolute magnitudes of more stars against their spectral type (see Figure~\ref{fig:HRD}), Russell showed that there was a main diagonal where dwarfs lived, an upper corner where red giants lived, and a lower corner lacking any stars ``except for one star\footnote{ This would later be recognized the first-discovered white dwarf \citep[e.g.,][]{1958whdw.book.....S}.} whose spectrum is very doubtful.'' 
Russell argued that this confirmed Ritter's theory of evolution. 

The following year, Harlow Shapley wrote a seminal paper laying out the collective arguments against the eclipsing binary hypothesis of Cepheid variable stars \citep{1914ApJ....40..448S}. 
First, B{\'e}lopolsky had already shown that the brightness and radial velocity variations did not coincide. 
Second, the periods of some Cepheids are themselves variable. 
Third, the shapes of the light curves for some Cepheids change from cycle to cycle \citep[e.g.,][]{1905ApJ....22..274C}. 
And lastly, ``the best argument,'' since Hertzsprung and Russell had just shown that Cepheids are giant stars, the companion star would need to be inside of the Cepheid in order for eclipses to explain the observed behavior---a ridiculous hypothesis. 
Shapley concluded that Cepheid variability is most likely due to pulsation.\footnote{ It is interesting to note here that John Michell had posed both the theory of earthquakes \citep{Michell01011759} and the explanation of stellar variability in terms of eclipsing stars \citep{michell1767inquiry}, but probably never imagined that stars quake, too.}  

\epigraph{``\emph{Cepheid variables are not binary systems... the explanation of their light-changes \hphantom{``}can much more likely be found in a consideration of internal or surface pulsations \hphantom{``}of isolated stellar bodies.}''}{--- Harlow Shapley\\\textit{On the Nature and Cause of Cepheid Variation} (\citeyear{1914ApJ....40..448S})}

Thus the pulsation hypothesis was born. 
But the theory had its doubters. 
There was no real proof yet---only very strong evidence that the eclipsing binary hypothesis was wrong---and no known mechanism for the pulsation. 
Many, including the eminent star formation theorist James Jeans, rejected the idea of stellar pulsations, Jeans himself arguing that Cepheid variation is rather caused by repeating explosions \citep[e.g.,][]{1919Obs....42...88J}. 
Moreover, many aspects of stellar theory still had major flaws. 
It was still not yet discovered how stars really get their energy, nor how they transport it throughout the interior, nor what they are made of, nor what state of matter they are in, nor how they evolve. 
Jeans himself in fact still held the view that stars are liquid \citep[e.g.,][]{1928Natur.121..173J}. 

The modern view of the stars really began to take hold in the early 20th century with the work of Arthur Eddington. 
Building upon earlier works by \citet{Schwarzschild1906} and \citet{1895MmRAS..51..123S}, Eddington developed the first models of radiative transport in stellar interiors \citep[e.g.,][]{1916MNRAS..77...16E}. 
Combating the view that stellar energy is transported entirely by convection, Eddington worked out the balance between radiative pressure---the outward pressure exerted by the enormous numbers of photons streaming through the star---with the inward pressure exerted by the gaseous stellar material. 
This led to the creation of his ``standard model''---a purely radiative star. 
This treatment complicated stellar models greatly, as the internal structure then depended on the opacity and mean molecular weight of the stellar matter, which were unknown \citep[e.g.,][]{ARNY1990211}. 


The following year, Eddington provided the mechanism for Cepheid variability \citep{1917Obs....40..290E}. 
Applying thermodynamics to the study of the interior, Eddington argued qualitatively that Cepheids pulsate due to an internal heat engine: repeated expansion and collapse due to cyclical ionization and recombination of atoms. 
The following year, he numerically calculated the periods of his stellar models using a linear adiabatic treatment of stellar pulsation, and found good agreement with observations \citep{1918MNRAS..79R...2E}. 
Though further confirmations would come later, this was already strong evidence for the pulsation hypothesis.

Eddington then went on to use observations of Cepheids to dispute the Kelvin-Helmholtz mechanism as being the sole source of stellar longevity \citep{1920SciMo..11..297E}. 
If stars survive on contraction alone, he argued, then their rate of rotation should speed up relatively rapidly due to the conservation of angular momentum. 
This was not what had been observed. 
Similarly, if the pulsation hypothesis is true, then their period of pulsation should change in accordance with changes to their mean density. 
\epigraph{``\emph{Now, on the contraction hypothesis the change of density must amount to at least \hphantom{``}1 per cent.\ in 40 years. The corresponding change of period should be very easily \hphantom{``}detectable. For $\delta$~Cephei the period ought to decrease 40 seconds annually. Now \hphantom{``}$\delta$~Cephei has been under careful observation since 1785, and it is known that \hphantom{``}the change of period, if any, must be very small. S.~Chandler found a decrease of \hphantom{``}period of 1/20 second per annum... I hope the dilemma is plain... Only the inertia \hphantom{``}of tradition keeps the contraction hypothesis alive---or rather, not alive, but an \hphantom{``}unburied corpse.}''}{--- Sir Arthur Stanley Eddington \\\emph{The Internal Constitution of the Stars} (\citeyear{1920SciMo..11..297E})} 

Eddington furthermore rederived Ritter's mass-luminosity relation, and upon applying the relation to stars of spectral types B and A, found that these ``dwarf'' stars are even more massive than the giant stars \citep[e.g.,][]{1924MNRAS..84..308E}. 
This too was difficult to reconcile with the prevailing theory of stellar evolution. 

Eddington therefore sought another explanation. 
During Albert Einstein's ``miracle year,'' Einstein had given his famous equivalence of mass and energy, ${E=mc^2}$ \citep{1905AnP...323..639E}. 
%
%
In 1920, the English chemist and Nobel laureate Francis Aston showed that the mass of one helium atom was approximately $1\%$ less than the sum of four hydrogen atoms \citep{aston1920lix}. 
At this time, it was still assumed that the solar composition was similar to that of the Earth; the amount of hydrogen in the Sun was therefore thought to be relatively small. 
Nevertheless, and despite lacking an exact mechanism, Eddington used these two developments to speculate that the Sun and stars survive via hydrogen fusion \citep{1920SciMo..11..297E}. 
\epigraph{``\emph{A star is drawing on some vast reservoir of energy by means unknown to us. \hphantom{``}This reservoir can scarcely be other than the sub-atomic energy which, it is \hphantom{``}known, exists abundantly in all matter; we sometimes dream that man will one \hphantom{``}day learn how to release it and use it for his service... The atoms of all elements \hphantom{``}are built of hydrogen atoms bound together, and presumably have at one time \hphantom{``}been formed from hydrogen; the interior of a star seems as likely a place as any for \hphantom{``}the evolution to have occurred; whenever it did occur a great amount of energy \hphantom{``}must have been set free; in a star a vast quantity of energy is being set free which \hphantom{``}is hitherto unaccounted for.}''}{--- Sir Arthur Stanley Eddington \\\emph{The Internal Constitution of the Stars} (\citeyear{1920SciMo..11..297E})} 

Within five years, Harlow Shapley's Ph.D.\ student Cecilia Payne showed that hydrogen is about a million times more prevalent in the Sun and stars than on the Earth \citep{1925PhDT.........1P}.
Within two years, the G\"ottinger physicist Friedrich Hund discovered quantum tunnelling, which gives atomic nuclei a probability of penetrating the Coulomb barrier and achieving thermonuclear fusion \citep{hund1927deutung,Nimtz2009}. 
The following year, George Gamow brought this concept to the astrophysical community \citep{1928Natur.122..805G}, and Eddington's speculation was proved. 
Eddington calculated new stellar models that included hydrogen burning, and found that this mechanism could power the Sun for billions of years \citep{1926ics..book.....E}. 

\begin{figure}
    \centering
    \includegraphics[width=0.75\textwidth]{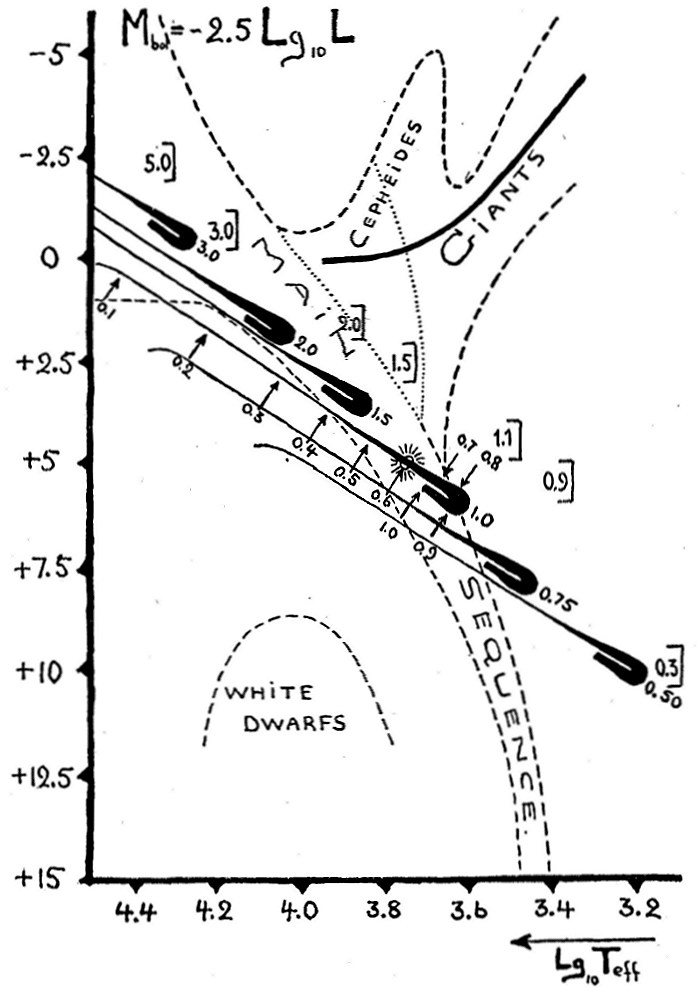}
    \caption[Historical theoretical H-R diagram]{
        Historical theoretical Hertzsprung-Russell diagram showing the evolution of stars with initial masses spanning from $0.5$ to $3$ solar masses. 
        The thickness of each track indicates the time spent at that stage of evolution. 
        The arrowed numbers indicate the amount of hydrogen. 
        The numbers in brackets indicate masses obtained via the mass-luminosity relation. 
        Unlike modern evolutionary tracks, the stars simulated here fail to evolve from dwarfs into giant stars. 
        \emph{(Figure reprinted with permission from \citealt{1938PhRv...53..907G}.)}
    \label{fig:gamow-tracks}}
\end{figure}

This was not the end of the story, however. 
Though hydrogen fusion was now known to fuel the stars, there were still major discrepancies between theory and observation. 
Using the assumption that the stellar interior is chemically homogeneous, George Gamow calculated evolutionary tracks and found that his models failed to become giant stars \citep[][see also Figure~\ref{fig:gamow-tracks}]{1938PhRv...53..907G}.
He furthermore found that he could not reproduce the mass-luminosity relation. 

The solution came that same year, though it would not be widely recognized until long after. 
Discarding Gamow's mixing hypothesis, the Estonian astrophysicist Ernst \"Opik realized that hydrogen fusion could continue burning in a shell after it had been exhausted in the core. 
Applying this insight, \"Opik succeeded in hand-calculating stellar models that evolve from the main sequence up the red giant branch \citep{1938PTarO..30C...1O}. 
Thus, the major features of the H-R diagram were explained. 
Unfortunately, it would be decades before this solution was rediscovered using digital computers \citep[e.g.,][]{ARNY1990211}. 
Although there was still much to do about the evolution beyond the red giant branch---and although debates continue to this day over why stars actually become giants \citep[e.g.,][etc.]{10.1007/978-94-009-8492-9_18,1992ApJ...400..280R,1983A&A...127..411W,1985ApJ...296..554Y,1988ApJ...329..803A,1989MNRAS.236..505W,1991AnPh...16..515W,2000ApJ...538..837S}---this essentially captured the first phases in the modern picture of stellar evolution.





There was still one more major hitch that needed to be reconciled. 
Around the same time that these issues were being resolved, the German-born American astronomer Edward Arthur Fath discovered that $\delta$~Scuti---a star with much resemblance to the Cepheids---has more than one period of pulsation \citep{fath1935photometric}. 
This brought serious challenges to the theory of stellar pulsation, as the second period measured was inconsistent with the mean density of the star \citep{1938ApJ....87..133S,1940PNAS...26..537S}. 
\epigraph{``\emph{One is practically forced to the conclusion that the existence of the pair of periods \hphantom{``}would be inconsistent with the pulsation theory... If the [second period] is correct, \hphantom{``}the pulsation theory is seriously jeopardized.}''}{--- Theodore Eugene Sterne\\\textit{The Secondary Variation of $\delta$~Scuti} (\citeyear{1938ApJ....87..133S})}

Sterne's argument rested on the longstanding assumption that these modes of pulsation needed to be purely radial in nature. 
Challenging this view, \citet{1938ApJ....88..189P} continued Lord Kelvin's work from $75$ years prior to further flesh out the mathematics of non-radial stellar pulsations, only now dealing with heterogeneous chemical compositions---a much more difficult problem. 
\citet{1941MNRAS.101..367C} used this description to calculate the non-radial pulsation frequencies of a stellar model (though his attention was toward binary interactions). 
Such calculations would prove invaluable in the decades to come, as it would be applied to a much more familiar star: the Sun. 






\subsection{Helioseismology} 
The theory that stars pulsate---and that they can pulsate non-radially---was most definitively confirmed with the discovery in the 1960s and 1970s that our own Sun is in fact such a pulsator. 
Obviously, the nature of solar pulsations are of a different character than the ones discovered in other stars to have gone unnoticed for so long. 

\begin{figure}[t!]
    \centering
    \includegraphics[width=0.66\textwidth,
        trim={0 1.5mm 0 1.5mm}, clip]{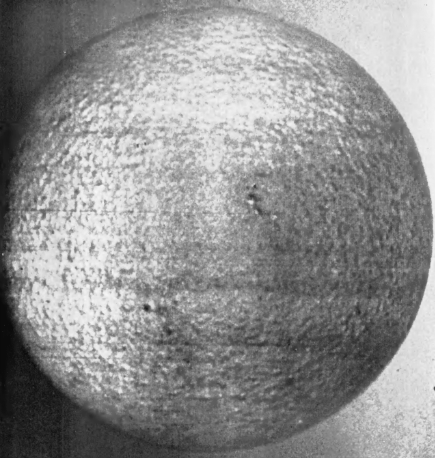}
    \caption[Velocity fields in the solar atmosphere]{Velocity fields in the solar atmosphere revealed by Doppler imaging. 
    \emph{(Figure reprinted with permission from \citealt{1962ApJ...135..474L}.)}
    \label{fig:solar-velocity-fields}}
\end{figure}

Already in 1916, the $23$-year old Canadian solar astronomer Harry Plaskett had found variations in Doppler velocity measurements of the solar surface from a spectroscopic investigation into the solar rotation rate \citep{1916ApJ....43..145P}. 
Whether these variations were intrinsic to the Sun, or, for example, effects from the Earth's atmosphere were unknown until the work by \citet{1954MNRAS.114...17H, 1956MNRAS.116...38H}. 
Many regard the publication of a ``preliminary report'' by Caltech researchers Robert Leighton, Robert Noyes and George Simon as the birth of helioseismology \citep{1962ApJ...135..474L}. 
In this paper, Leighton and colleagues demonstrated that the Sun has multi-periodic variations on the order of about $5$ minutes (see also Figure~\ref{fig:solar-velocity-fields}). 
They were prescient in their speculation that these variations could be used to determine detailed properties of the Sun, or at least its atmosphere. 
\citet{1968ApJ...152..557F} and others furthermore gave evidence that solar oscillations may not merely be confined to the solar atmosphere, but may instead probe deep into the star.

\begin{figure}[t!]
    \centering
    \vspace*{-0.35cm}
    \includegraphics[width=0.73\textwidth
        ]{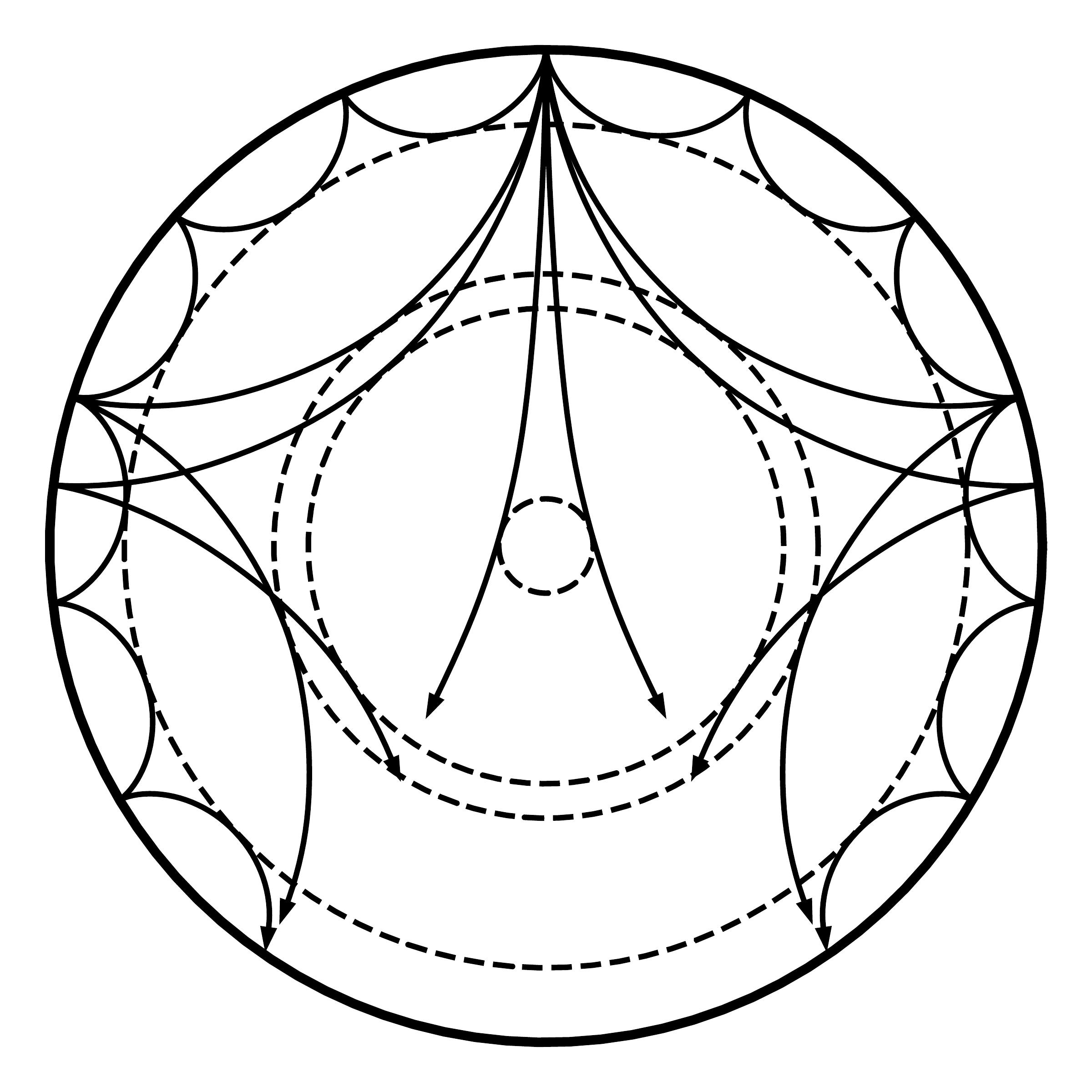}
    \caption[Ray path diagram for solar oscillation modes]{Ray diagram showing the paths of oscillation modes as they propagate through the interior of a solar model. 
    The innermost circle shows the lower turning point of a quadrupole (${\ell=2}$) oscillation mode. 
    Such kinds of modes are observable in the Sun and other stars exhibiting solar-like oscillations. 
    The other modes are ${\ell=20}$, $25$, and $75$, which have so far only ever been observed in the Sun. 
    \emph{(Figure adapted with permission from Warrick Ball [private communication] using the procedure given by \citealt{2000PhDT.........9G}.)}
    \label{fig:rays}}
\end{figure}

\begin{figure}
    \centering
    \includegraphics[width=\textwidth]{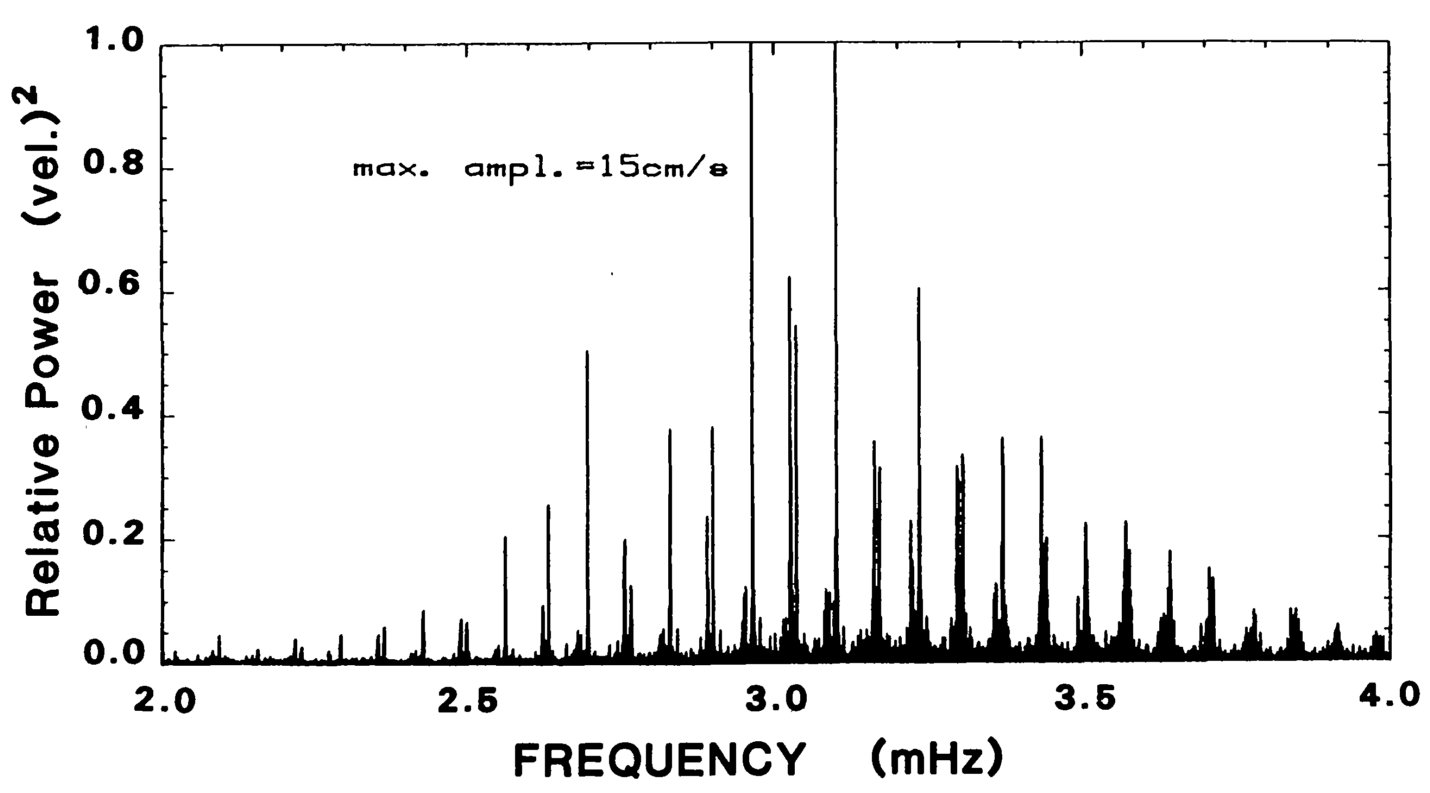}
    \caption[Historical solar power spectrum]{Power spectrum of the Sun from $3$ months of observations showing its $5$-minute ($3$~mHz) oscillations \citep{1981SoPh...74...51C}. 
    Each peak corresponds to an individual mode of oscillation. 
    \emph{(Figure reprinted with permission from the review article of \citealt{1984ARA&A..22..593D}.)}
    \label{fig:solar-power-spectrum}}
\end{figure}

In the early 70s, \citet{1970ApJ...162..993U} and \citet{1971ApL.....7..191L} argued that the oscillations are standing acoustic waves trapped below the solar photosphere, and showed that theoretical periods of this description match the observations. 
\citet{1975A&A....44..371D} and \citet{1977ApJ...218..901R} found that the relationship between the spatial and temporal frequencies of the oscillations are in similar agreement with expectations, giving further credence to the theory. 
\citet{1977ApJ...212..243G} provided a mechanism for the origination of solar oscillations by showing that acoustic waves can be stochastically excited by turbulent convection, which is the dominant source of energy transport in the solar envelope. 
\lr{\citet{1979Natur.282..591C} and \citet{1980Natur.288..541G} made the first identifications of low-degree modes in the Sun, which pass through the entire star, thereby confirming the global nature of the oscillations (see Figures~\ref{fig:rays} and \ref{fig:solar-power-spectrum}).}

The Sun vibrates in a superposition of a great number of low-amplitude modes simultaneously. 
Multiple modes of the same spherical degree $\ell$ (recall Figure~\ref{fig:sph}) can be excited simultaneously. 
These modes are distinguished by their radial order $n$, i.e., the number of nodes (zero crossings) between the center and the surface. 
Additionally, the rotation of the Sun splits each non-radial mode of oscillation into a multiplet of ${2\ell+1}$ modes, which can be distinguished by their azimuthal order $m$, i.e., the number of nodes along the equator. 

Whereas Cepheid and RR~Lyrae stars oscillate in low-order (${n\leq 3}$) radial (${\ell = 0}$) modes, the Sun and other solar-type stars oscillate in high-order (${n\lessapprox 40}$) modes of both radial and non-radial (${\ell\geq 0}$) character, though so far observations of modes with $\ell \geq 4$ have only been confirmed in the Sun, which is made possible by resolving the solar disk. 
Classical pulsators like Mira, Cepheid, RR~Lyrae, and $\delta$~Scuti stars are intrinsically unstable to their oscillations: they are self-excited by their configuration \citep[e.g.,][]{2015EAS....73..111S}. 
Solar-like oscillators, on the other hand, pulsate in stable modes which are both driven and damped by turbulent convection in their outer envelopes. 
Detailed reviews and overviews of global helioseismology have been given by, e.g., \citet{2002RvMP...74.1073C}, \citet{Kosovichev1999,2011lnp...832....3k}, and \citet{2016lrsp...13....2b}. 

\citet{1980ApJS...43..469T} provided asymptotic descriptions for oscillation modes of high radial order (${n\gg\ell}$) as seen in the Sun. 
Mode frequencies of the same spherical degree are equally spaced by a quantity known as the large frequency separation, denoted ${\Delta\nu}$, which is related to the stellar mean density and the inverse sound travel time through the star. 
Modes differing by a spherical degree of two (e.g., ${\ell=0}$ and ${\ell=2}$) and a radial order difference of one (e.g., ${n=21}$ and ${n=20}$) are spaced by the small frequency separation (${\delta\nu}$). 
This quantity is related to the sound-speed gradient, and its measurement provides a good diagnostic of main-sequence age. 
The ratios of these quantities are also useful, because they are insensitive to near-surface layers of the star where several assumptions used to calculate theoretical mode frequencies break down \citep[e.g.,][]{2003A&A...411..215R}. 
To good approximation, these quantities vary little from one radial order to the next, and hence serve as a good summary of the frequency spectrum. 
In the early 1980s, Christensen-Dalsgaard \& Gough applied this asymptotic description to oscillation modes calculated from a solar model and were able to show that the model was in agreement with the observations \citep[e.g.,][]{2002RvMP...74.1073C}. 

\begin{figure}
    \centering
    \includegraphics[width=0.8\textwidth]{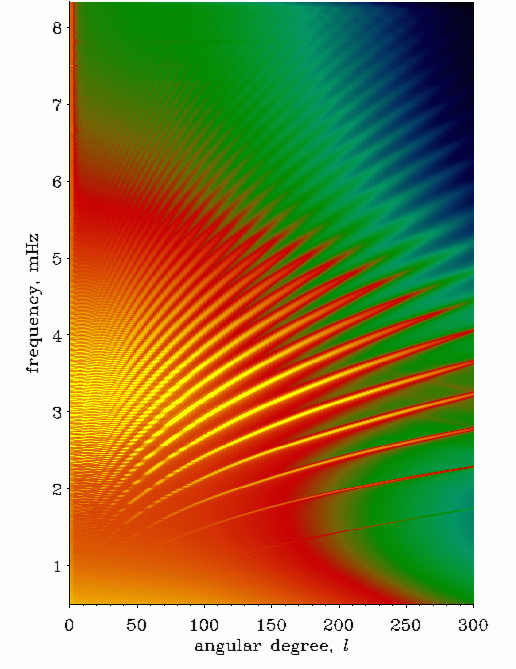}
    \caption[Solar power spectrum from MDI]{Solar power spectrum showing helioseismic oscillation mode frequencies as a function of spherical degree as observed by MDI over a time span of $144$ days. 
    The acoustic oscillation modes of the Sun form the ridges of high power. 
        \emph{(Figure reprinted with permission from \citealt{1997SoPh..175..287R}.)}
        \label{fig:rhodes-mdi}}
\end{figure}

Of course, helioseismic data nowadays are of superb quality. 
Figure~\ref{fig:rhodes-mdi} shows a power spectrum from data obtained by the Michelson Doppler Imager (MDI) instrument onboard the Solar and Heliospheric Observatory (SOHO), a \euro$1$~billion NASA/ESA space mission launched in 1995. 
With such data, thousands of solar oscillation modes have been resolved with high precision \citep[e.g.,][]{1997SoPh..175..287R}. 

\subsubsection*{Helioseismic Inversions}
Many of the confirmations of global helioseismology have come through the comparison of observations to a theoretical models constructed to match the properties of the Sun. 
Such models can be constructed for example via evolutionary modelling; I will discuss the creation of such models in more detail in Section~\ref{sec:evolution}. 
However, even to this day, no solar model matches solar oscillation data exactly \citep[e.g.,][]{1980Natur.288..544C}. 
The question thus arose as to whether these global oscillation modes could be used to make model-independent measurements of the solar interior, in terms of both its structure and its internal rotation rate \citep[e.g.,][]{1976Natur.259...89C, 1981MNRAS.196..731G}. 
This would need to be answered in the context of inverse theory. 

\epigraph{\emph{``The astrophysicists' task is not merely to produce a theoretical model of the Sun \hphantom{``}that is not obviously at variance with observation, but to learn what the internal \hphantom{``}structure actually of the Sun is, and to understand why it is so.''
}}{--- Douglas Owen Gough, FRS \\ 
\textit{Seismic observations of the solar interior} (\citeyear{1991ARA&A..29..627G})}

The forward problem of global helioseismology is to calculate the oscillation mode frequencies for a given model of solar structure (or solar rotation). 
The inverse to this problem is then to calculate the structure (or internal rotation profile) from the mode frequencies. 
The inverse problem is ill-posed because different stellar structures can support the same oscillation pattern, including ones that are clearly nonphysical. 
Furthermore, unless care is taken, small errors to the input data can lead to large errors in the inversion result. 
I will discuss ill-posed problems in more detail in Section~\ref{sec:inverse}. 

In the late 1960s, geophysicists George Backus and James Gilbert developed a stable method for inferring the structure of the Earth from seismic measurements  \citep{1968GeoJ...16..169B, 1970RSPTA.266..123B}. 
This method came to be known as the Gilbert--Backus method or the method of Optimally Localized Averages (OLA) and has been adapted for use and widely applied in helioseismology. 

The idea of OLA is as follows. 
When comparing the model frequencies to the observed frequencies, there are differences, indicating that the structure (or rotation profile) of the model must differ from the structure of the Sun. 
If an oscillation mode were only sensitive to one region of the star, then a difference in frequency for that mode would indicate a difference in structure in that region. 
However, this is not the case: oscillation modes are sensitive to multiple locations in the solar interior, and so it is not possible to disentangle the cause of discrepancy based on only one mode. 

The sensitivities of mode frequencies to perturbations in the structure of the star are called kernels. 
I provide the kernels of stellar structure in Section~\ref{sec:kernels}. 
The OLA method works by combining the modes in such a way that their combination---the averaging kernel---is only sensitive to one region in the star. 
When the combination of frequencies corresponding to that combination of modes differs between the model and the star, then the structure must differ in that region. 
Thus, one can then work out the structure in the locations in the interior where it is possible to construct an averaging kernel. 

By the mid-80s, it became possible to invert frequency splittings and infer the internal rotation rate of the Sun (\citealt{1984Natur.310...22D}, see also e.g.~\citealt{1998ApJ...505..390S, 2009LRSP....6....1H}). 
The following year, the internal solar sound speed profile was deduced via inversion of an asymptotic description known as Duvall's Law, which assumes that the mode frequencies depend exclusively on the speed of sound \citep{1985Natur.315..378C}. 
Soon thereafter, full inversions---which separate the influence on mode frequencies of, e.g., sound speed from density---were used to determine the acoustic structure of the majority of the solar interior (\citealt{1985SoPh..100...65G}, see also e.g.~\citealt{1990MNRAS.244..542D, GoughThompson1991, 1991ARA&A..29..627G, 1994a&as..107..421a, 2009ApJ...699.1403B}). 

Inversions for helioseismic structure have revealed many aspects of the solar interior, such as the depth of the convection zone \citep[e.g.,][]{1991ApJ...378..413C, 1997MNRAS.287..189B}, the helium abundance in the solar envelope \citep[e.g.,][]{1991LNP...388..111D, 1998MNRAS.298..719B}, the equation of state of the solar plasma \citep{1997A&A...322L...5B}, 
and the efficiency of element diffusion \citep{1993ApJ...403L..75C}. 
Rotation inversions have shown that the Sun rotates differentially, having a latitudinally-dependent rotation rate in the convective outer envelope, and rotating as a solid body in the radiative interior \citep[e.g.,][]{2009LRSP....6....1H}. 
These zones are separated by a shear layer that is referred to as the tachocline \citep[][]{1992A&A...265..106S}. 
Finally, investigations based on helioseismic inversions have been instrumental in resolving longstanding issues such as the solar neutrino problem \citep[e.g.,][]{1998PhLB..433....1B}, for which four Nobel prizes have been awarded. 
A detailed review of results that have been obtained via helioseismic inversion has been given by \citet{2016lrsp...13....2b}. 


\subsection{Asteroseismology}
As our Sun is not thought of as being particularly exceptional, it was obviously expected that other stars similar to the Sun should also exhibit solar-like oscillations \citep[e.g.,][]{1984srps.conf...11C}. 
In addition to oscillations in solar-like stars, \citet{1983SoPh...82..469C} further predicted that low-mass giant stars should harbor these kinds of oscillations as well, as these stars also have convective envelopes. 
Moreover, these stars harbor \emph{mixed modes}: modes that behave like acoustic oscillations in the envelope and gravity mode oscillations in the core \citep[e.g.,][]{2001MNRAS.328..601D}. 
However, due to the very small amplitudes of the solar oscillations (on the order of ${10\; \text{cm/s}}$, recall Figure~\ref{fig:solar-power-spectrum}), their discovery in other stars posed a long-standing challenge. 


Already in the late 1980s detections of solar-like oscillations were being claimed \citep{1986A&A...164..383G}. 
These were not however confirmed in follow-up studies \citep[e.g.,][]{1991MNRAS.249..643I}. 
Throughout the 1990s there were more claims of detections in other stars, which mainly served to place upper limits on their amplitudes \citep[e.g.,][]{1990ApJ...350..839B, 1991ApJ...368..599B, 1992A&A...264..138P, 1995MNRAS.276.1295E}. 
Finally, in the 2000s, firm detections of solar-like oscillations in other stars were made, such as in the nearest star, the solar-type star Alpha~Centauri \citep{2001A&A...374L...5B}; 
the subgiant star $\beta$~Hyi \citep{2001ApJ...549L.105B}; 
and the giant stars $\alpha$~Uma \citep{2000ApJ...532L.133B} and $\eta$~Hya \citep{2002A&A...394L...5F}. 
The field of solar-like asteroseismology was born, but in its infancy. 
With the coming space missions, it would soon undergo a revolution.

The first space-based observations came from the NASA \emph{Wide-Field Infrared Explorer} (\textsc{WIRE}), which had failed in its nominal mission, but was fortunately able to be repurposed into an asteroseismology mission \citep{2000ASPC..198..557B}. 
After one month of observation, space photometry yielded solar-like oscillations in the very bright giant star Alpha~Ursae~Majoris \citep{2000ApJ...532L.133B}, and soon thereafter, in Alpha~Centauri as well \citep{2001ESASP.464..391S}. 

The first purposefully dedicated space asteroseismology mission was the Canadian \emph{Microvariability and Oscillations of STars} telescope \citep[\textsc{MOST},][duration 2003--2014]{2003PASP..115.1023W}. 
Though \textsc{MOST} was not sensitive enough to detect oscillations in solar-type stars, \citet{2006ESASP.624E..30B, 2007A&A...468.1033B} did detect radial-mode oscillations in the red giant $\epsilon$~Oph using $28$ days of MOST observations. 
Studying this same star from the ground, \citet{2006A&A...454..943H} was able to detect non-radial pulsations. 
The detection of solar-like oscillations in red giants represents a great confirmation of stellar theory. 
A detailed review on oscillations in red giants has been given by \citet{2017A&ARv..25....1H}. 

\lr{Soon afterwards came the European/French space mission \emph{Convection, Rotation and planetary Transits} \citep[\textsc{CoRoT},][duration 2006--2012]{2006ESASP1306...33B}, which was able to detect solar-like oscillations in solar-type stars \citep[e.g.,][]{2010A&A...515A..87D}. 
Among other successes, CoRoT was particularly valuable for the study of solar-like oscillations in red giant stars, where oscillations in hundreds of these stars were detected (e.g., \citealt{2009Natur.459..398D, 2009A&A...506..465H}).}

\subsubsection*{\emph{Kepler}}

By far the best asteroseismology mission to date has been the \emph{Kepler} space observatory \citep[][duration 2009--2013]{2010ApJ...713L..79K}. 
The data yield from \emph{Kepler} has been enormous; here I will largely restrict discussion to solar-type stars which are relevant for this thesis. 
For detailed reviews and textbooks on asteroseismology, see e.g.\ \citealt{2010aste.book.....a, 2012AN....333..914C, 2013adspr..52.1581h, 2013ARA&A..51..353C}, and \citealt{basuchaplin2017}. 

\emph{Kepler} targeted approximately $150,000$ main sequence stars in a fixed field of view around the constellations of Cygnus, Lyra and Draco. 
Short-cadence and long-cadence targets were observed every $58.89$ seconds and every $29.4$ minutes, respectively. 
Several pipelines were created in preparation of processing the expected asteroseismic yield. 
For example, several groups created pipelines for the automated retrieval of ${\Delta\nu}$ and $\nu_{\max}$ from \emph{Kepler} time series \citep[e.g.,][]{2009CoAst.160...74H, 2009A&A...508..877M, 2010MNRAS.402.2049H, 2010A&A...511A..46M}. 
For detailed stellar modelling, \citet{2009ApJ...699..373M} created the Asteroseismic Modelling Portal (AMP), which fits evolutionary models to the observed asteroseismic data using genetic programming. 
In a hare-and-hound exercise, \citet{2009ApJ...700.1589S} found that the radius determinations from the expected asteroseismic data from \emph{Kepler} are five to ten times better than without.

After launch, the quality of \emph{Kepler} data for asteroseismology was immediately evident, revealing clear signatures of non-radial oscillations in several stars within one month of data collection \citep{2010PASP..122..131G, 2010ApJ...713L.169C}. 
For the majority of stars, only the global properties such as ${\Delta\nu}$ and $\nu_{\max}$ are able to be resolved. 
Even with just these quantities, however, it is possible to infer information about the stars. 
For example, by assumption of homology with the Sun, one can scale oscillation data from solar values to estimate the properties of stars, such as their masses and radii \citep[e.g.,][]{1995A&A...293...87K}. 
This presents the opportunity for ``ensemble asteroseismology.''
\citet{2011Sci...332..213C, 2014ApJS..210....1C} and \citet{2017ApJS..233...23S} used these and other approaches to find the masses, ages, radii, and other fundamental parameters for hundreds of main sequence and subgiant stars observed by \emph{Kepler}. 
In addition, several groups have also worked on improvements to the solar scaling relations \citep[e.g.,][]{2013A&A...550A.126M,2016ApJ...822...15S,2016MNRAS.460.4277G,2017MNRAS.470.2069G,2017ApJ...843...11V}. 

For the best targets, interferometric and spectroscopic measurements have been obtained to complement the asteroseismic data \citep[e.g.,][]{2010MNRAS.405.1907B, 2012MNRAS.423..122B, 2012ApJ...749..152M, 2013MNRAS.433.1262W}. 
These measurements provide the tightest determinations of stellar parameters and the best tests to stellar theory. 
Comparing these data, \citet{2012ApJ...760...32H} found good agreement between radii determined via interferometry and asteroseismology. 
\begin{figure}
    \centering
    \includegraphics[width=\linewidth]{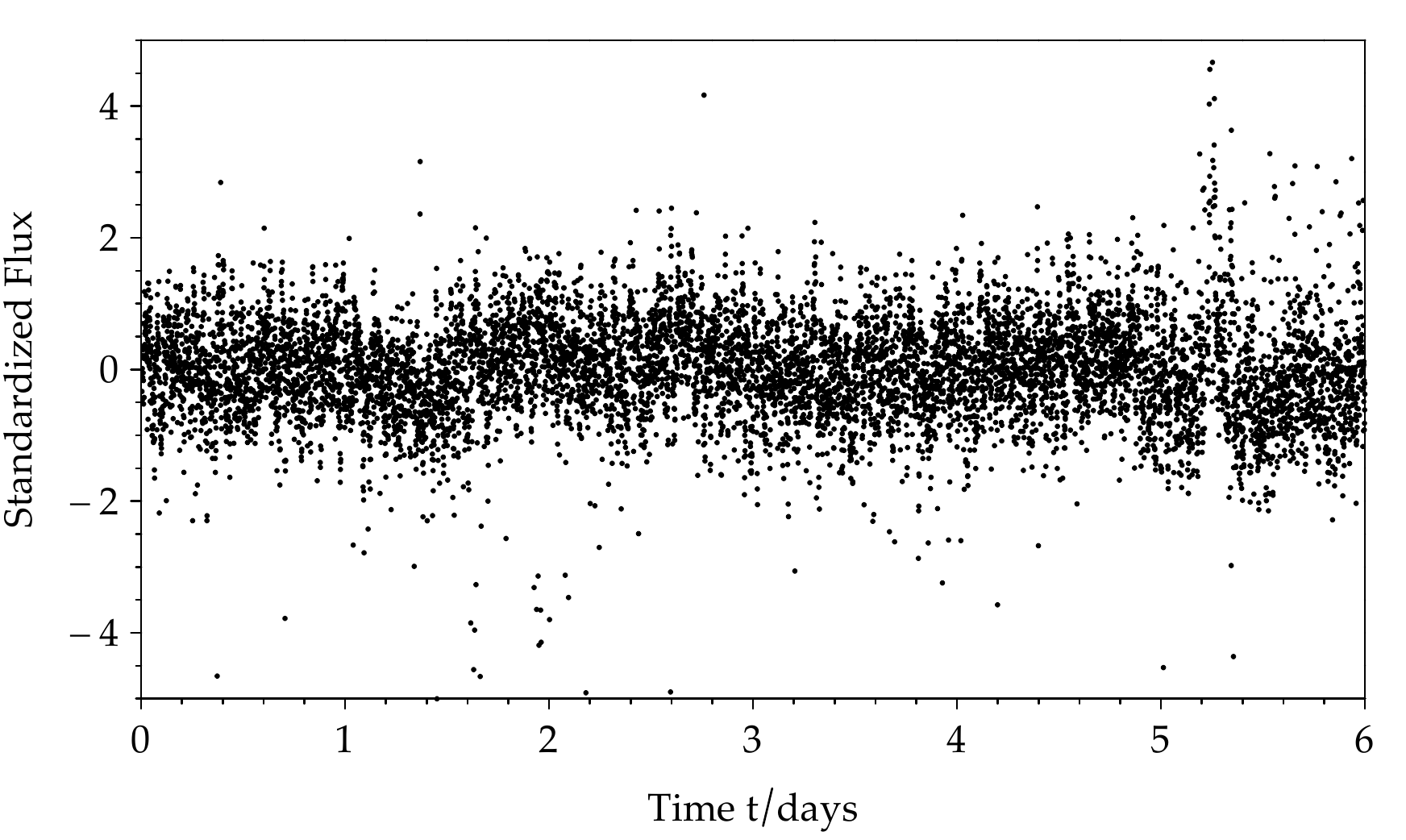}\\
    \includegraphics[width=\linewidth]{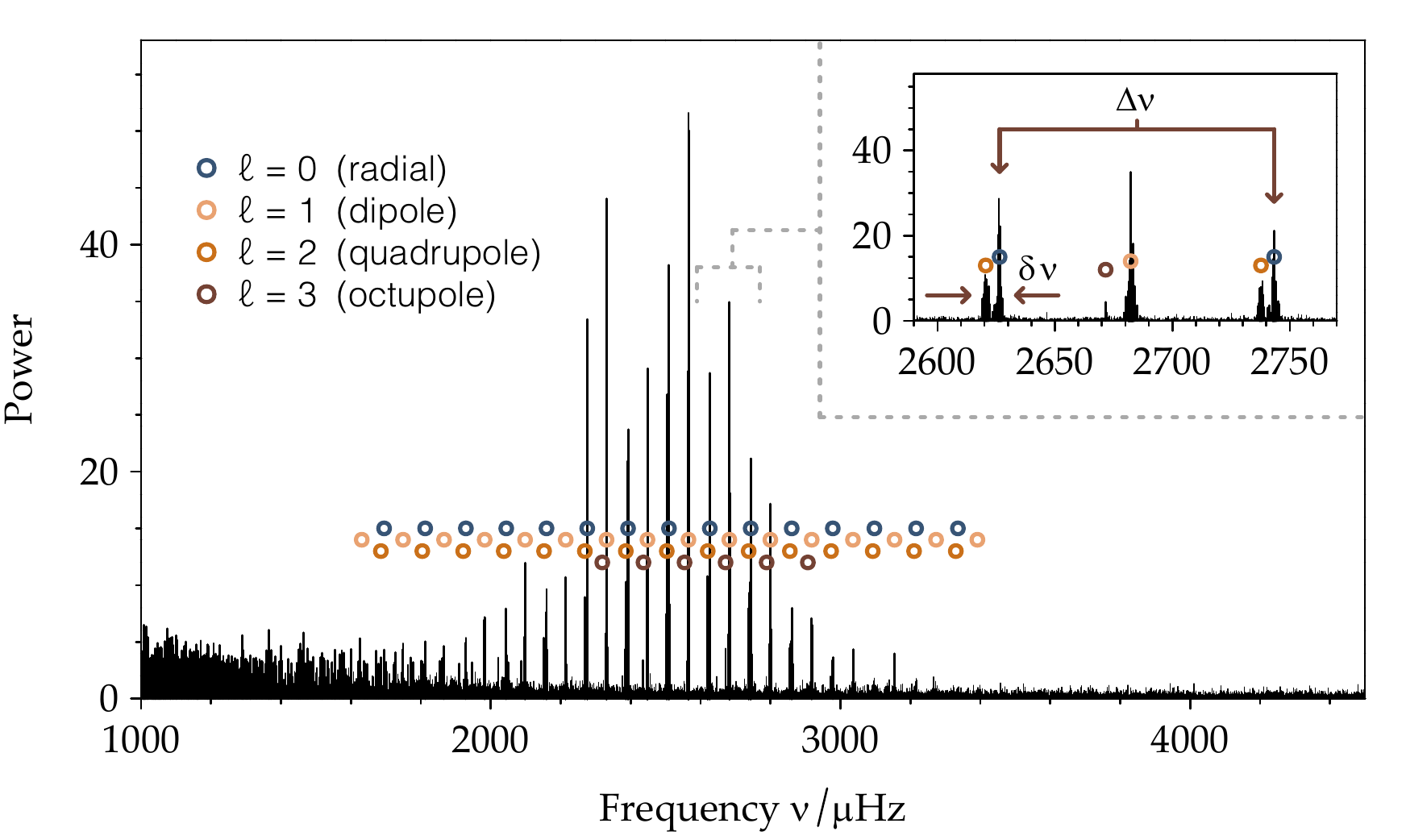}
    \caption[Power spectrum of 16~Cyg~B]{
        Light curve (top) and power spectrum of 16~Cyg~B (bottom) as obtained from the \emph{Kepler} spacecraft. 
        The power spectrum shows $56$ detected oscillation modes, each labelled by their spherical degree (\emph{cf.}~Figure~\ref{fig:sph}). 
        The power excess is roughly Gaussian in shape and centered around a value of ${\nu_{\max}\simeq 2550 \;\mu\text{Hz}}$. 
        The inset figure shows a zoom into the power spectrum with example large (${\Delta\nu \simeq 117\;\mu\text{Hz}}$) and small (${\delta\nu \simeq 6\;\mu\text{Hz}}$) frequency separations. 
        \emph{Data from the Kepler Asteroseismic Science Operations Center \citep{KASOC}.}
    \label{fig:16cygb}}
\end{figure}

The perhaps best solar-like stars observed by \emph{Kepler} are the solar analogs 16~Cygni~A and B. 
These stars form a hierarchical triple system, with 16~Cyg~A being orbited by a red dwarf, and 16~Cyg~B being orbited by a Jovian planet.  
\citet{2012ApJ...748L..10M} ``peak bagged'' these stars (i.e., resolved their frequencies) and found clear detections of ${\ell\leq 3}$ modes (see Figure~\ref{fig:16cygb}). 
They used AMP to determine the evolutionary parameters of these stars, finding a common age of $6.8$~Gyr and common initial chemical compositions, which supports the conatality hypothesis of binary star formation. 
\citet{2015MNRAS.446.2959D} used rotational splittings of the non-radial modes to infer the inclination angles and rotation rates of these stars, in both cases finding a rotation rate of approximately $23$~days.

For approximately $100$ solar-like stars observed by \emph{Kepler}, the data have been good enough for dozens of individual mode frequencies to be resolved. 
These stars form the \emph{Kepler} Ages \citep{2016MNRAS.456.2183D} and \emph{Kepler} LEGACY projects \citep{2017ApJ...835..172L}, the former of which comprises $35$ planet-host candidates. 
\citet{2015MNRAS.452.2127S, 2017ApJ...835..173S} determined the fundamental parameters of these stars using pipelines created by different groups, finding roughly broad agreement. 
\citet{2014ApJ...790..138V, 2017ApJ...837...47V} used seismic glitch analysis to determine the base of the convection zone and helium abundances for the LEGACY sample.
These are the stars analyzed in the coming Sections and Chapters.

\begin{figure}
    \centering
    \begin{minipage}[c]{0.5\textwidth}%
        \includegraphics[width=\textwidth]{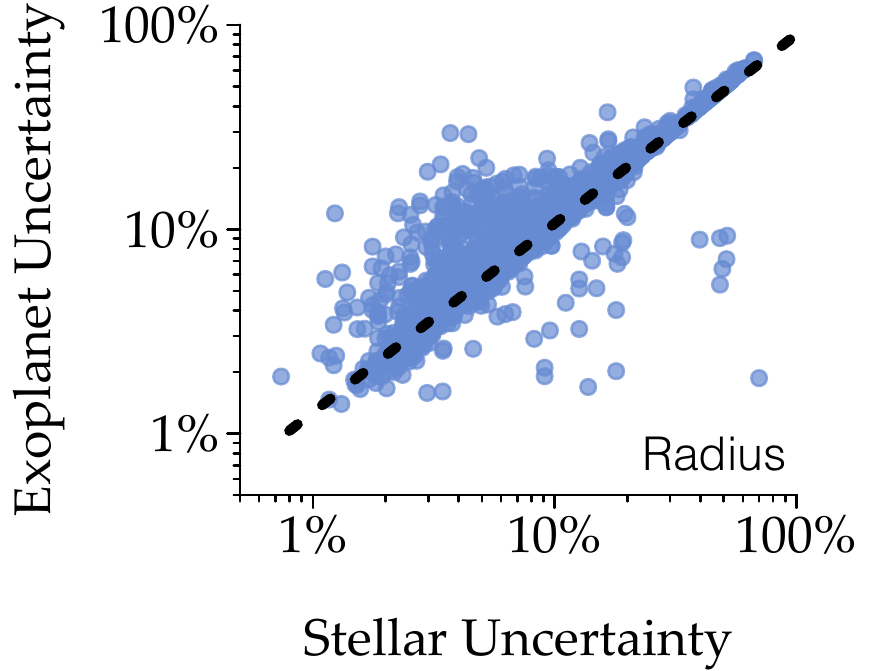}%
    \end{minipage}%
    \hfill
    \begin{minipage}[c]{0.45\textwidth}
        \caption[Exoplanetary uncertainty vs.~host star uncertainty]{Uncertainty in the determination of exoplanetary radii as a function of the uncertainty in the determination of the radius of their host star for nearly $2,400$ exoplanets detected using the transit method, which will also be the method of choice for finding exoplanets in the forthcoming TESS mission. 
        \emph{Data acquired from \href{http://exoplanets.org}{exoplanets.org} \citep{2014PASP..126..827H}.}
        \label{fig:exoplanets}}
    \end{minipage}
\end{figure}

A discussion of the \emph{Kepler} mission would be incomplete without a mention of exoplanets. 
\emph{Kepler} was primarily a plunt-hunting mission, and a very successful one. 
Within \emph{Kepler} data researchers found a plethora of rocky planets, super Earths, and gas giants \citep[e.g.,][]{2008ApJ...680.1450P, 2011ApJ...729...27B, 2012ApJ...745..120B, 2014ApJS..210...20M}. 
Additionally, \emph{Kepler} data were used to find that hot Jupiters are common \citep{2008ApJ...680.1450P}, and that many stellar-planetary systems are misaligned \citep{2013Sci...342..331H}, bringing into question theories of planet formation. 
Of course, asteroseismology is of great aid to the characterization of exoplanets, since the determination of exoplanetary parameters usually depends strongly on the ability to determine the parameters of the host star (see Figure~\ref{fig:exoplanets}). 

Following the failure of its reaction wheels, \emph{Kepler} was repurposed into the wandering K2 mission, which is now in its final stages \citep[][duration 2013--2018]{2014PASP..126..398H}. 
This year, NASA's \emph{Transiting Exoplanet Survey Satellite} mission will launch \citep[TESS,][expected 2018--2020]{2010AAS...21545006R}. 
ESA's \emph{Planetary Transits and Oscillations of stars} mission \citep[PLATO,][expected 2026--2030]{2014ExA....38..249R} is planned for launch in eight years. 
We analyze the anticipated yields of these missions for Sun-like stars in Chapter~\ref{chap:statistical} \citep{2017apj...839..116a}.

\subsubsection*{Asteroseismic Inversions}

Asteroseismic structure inversions are more difficult to perform than in helioseismology for two main reasons. 
\begin{description}
    \setlength{\itemindent}{0pt}
    \item[Mode set.]
    The mode sets available in asteroseismology are much more limited. 
    Due to cancellation effects, only low-degree modes have been observed so far in stars other than the Sun, and so only dozens rather than thousands of mode frequencies are available. 
    \lr{It is only possible to build well-localized averaging kernels in locations where there is a sufficient number of mode lower turning points, as these are the regions where the modes spend most of their time (recall Figure~\ref{fig:rays}). 
    Consequently, asteroseismic inversions using only low-degree modes are generally only capable of making localized probes of the stellar core.} 
    This limitation also rules out the possibility of using techniques such as Regularized Least Squares, which fit the entire internal profile simultaneously \citep[see, e.g.,][]{basuchaplin2017}. 
    
    Furthermore, mode frequencies depend on multiple variables of stellar structure.
    When trying to determine one from asteroseismic information, one must therefore control for other influences. 
    With limited information, this becomes more difficult. 
    In helioseismology, the most common pair of variables is the speed of sound $c$ and the stellar density $\rho$, denoted the ${(c,\rho)}$ kernel pair. 
    
    \item[Mass and radius.]
    The masses and radii of stars are not known to anywhere near the precision for the Sun. 
    This creates difficulties because the kernel functions are derived with respect to a reference model, which is assumed to have the correct mass and radius. 
    Without accounting for this effect, the results of the inversion results will be offset by the differences in mass and volume \citep[][]{2003Ap&SS.284..153B}. 
    Furthermore, the mode frequencies themselves scale with the mass and volume of the star. 
\end{description}

Already in the early 1990s, before the first confirmed asteroseismic detections, \citet{1993ASPC...40..541G} considered the prospect of performing asteroseismic inversions to determine stellar structure. 
In this work, Gough and Kosovichev simulated data sets for a $1.1$ solar mass model that they thought might be likely to be obtained from a future mission. 
They used a solar model as reference. 
Their work was on the one hand pessimistic---assuming only ${\ell\leq 2}$ modes would be available, having mode uncertainties of ${0.1\;\mu\text{Hz}}$---and on the other optimistic, assuming that more than $60$ modes would be observed. 
In comparison, the perhaps best \emph{Kepler} solar-type target, 16~Cyg~B, has approximately $56$ detected modes (though the exact amounts are disputed), $11$ of which being ${\ell=3}$ modes, with uncertainties ranging from ${0.04\;\mu\text{Hz}}$ up to ${5\;\mu\text{Hz}}$. 

Gough and Kosovichev were able to form four well-localized averaging kernels at target radii $0.05$, $0.15$, $0.25$, and $0.35$. 
They simultaneously estimated the difference in mass per volume between the two models while performing the inversion. 
Surface effects were not considered. 

In this work it was already realized that inversions with helium as the second variable could be the most promising route. 
The helium kernels only have amplitude in ionization zones, which are located near to the stellar surface and would require higher-degree modes to resolve anyway. 
\citet{2001ESASP.464..407B} showed that when using the ${(c,\rho)}$ kernel pair with expected asteroseismic data, only one averaging kernel can be formed. 

Some other early attempts with similar setups and results have been reviewed by \citet{2003Ap&SS.284..153B}. 
These works all used mode sets that they thought would be available from future missions: PRISMA, MOST, MONS, and \emph{Eddington}. 
\lr{Unfortunately, PRISMA, MONS, and \emph{Eddington} were not funded, and MOST did not detect any oscillations in solar-like stars. 
It is only now with the CoRoT and \emph{Kepler} missions that the data are good enough to measure internal stellar structure.} 
We invert \emph{Kepler} data to infer the internal structure of 16~Cyg~A and B in Chapter~\ref{chap:inversion} \citep{2017ApJ...851...80B}. 


Several other kinds of inverse problems have been worked on using asteroseismic data. 
Instead of inverting for the full density profile, \citet{2012A&A...539A..63R} introduced an OLA-based technique for estimating stellar mean density. 
They applied the technique to the Sun, $\alpha$~Cen~B, and two stars observed by CoRoT. 
They found that they could estimate mean densities this way to an accuracy of $0.5\%$. 
However, it performed no better than estimating mean densities using the \citet{2008ApJ...683L.175K} surface term corrected solar scaling relation. 

\citet{2015A&A...583A..62B, 2015A&A...574A..42B} extended this work by creating kernels for the acoustic radius and two age indicators: the integral of the sound speed derivative, and a weighted square of the isothermal sound speed derivative. 
They applied these techniques to 16~Cyg~A and B, and, when combining them with interferometric radii, found masses and ages for these stars that were inconsistent with evolutionary modelling \citep{2016A&A...585A.109B, 2016A&A...596A..73B}. 

In addition to the global properties of stars, inversions for stellar rotation rates have also had success. 
\citet{2012ApJ...756...19D, 2014A&A...564A..27D}, \citet{2016ApJ...817...65D}, and \citet{2017A&A...602A..62T} inverted frequency splittings to obtain the core and envelope rotation rates of several sub- and red-giant stars. 
They found, in agreement with theoretical expectations, that the cores of these stars rotate more rapidly than their outer layers. 

\newpage
\subsubsection*{Layout of Thesis}
In this section, we saw that the study of pulsating stars has been a primary driver in the development of the theory of stellar evolution. 
Helioseismic inversions have revealed the structure of the Sun and shown that it is very close (though not identical to) the structure predicted by theoretical models. 
Asteroseismology has confirmed many details predicted by stellar evolution, and asteroseismic inversions show promise for leading to future improvements to evolutionary theory. 

For the interested reader, the following texts contain more details: 
\citet{1958HDP....51..353L} give a thorough overview of variable stars up until the 1950s; 
\citet{ARNY1990211} gives the history of stellar evolution, including later phases of evolution which are not covered here; 
\citet{2016lrsp...13....2b} gives the history of solar oscillations;
\citet{bolt2007biographical} contains an encyclopedia of biographies for astronomers;
and \citet{2015pust.book.....C} give a general overview and history of variable stars.

The remainder of the thesis is organized as follows. 
The following two sections (\ref{sec:evolution}, \ref{sec:pulsation}) give the theoretical background on stellar structure, evolution, and pulsation. 
These enable us to pose and solve the forward problems of simulating the evolution of a star and calculating its frequencies of oscillation. 
In Section~\ref{sec:pulsation}, I furthermore state the kernel functions of stellar structure, which allow us to calculate the differences in mode frequencies between a pair of stellar models of differing structure. 
In the final section of the introduction (Section~\ref{sec:inverse}), I state more formally the inverse problems of asteroseismology that are considered in this thesis, and give some indication of their difficulty. 

In Chapter~\ref{chap:ML}, we perform evolution inversions to determine stellar ages and other fundamental parameters using machine learning \citep{2016apj...830...31b}. 
In Chapter~\ref{chap:statistical}, we use unsupervised machine learning to determine which observations are useful for constraining which properties of stellar models \citep{2017apj...839..116a}. 
In Chapter~\ref{chap:inversion}, we determine the asteroseismic structure of two stars, in one case finding agreement with evolutionary modelling, but in another not \citep{2017ApJ...851...80B}. 
Finally, at the end I give what I assess to be the future prospects for this line of research.

\clearpage\section{Stellar Structure \& Evolution} 
\label{sec:evolution}
\begin{shaded}
\noindent In this section, I will provide a summary of background information on the theory of stellar structure and evolution, with a focus toward the creation of evolutionary models of solar-like stars. 
This will allow us to state the evolution inverse problem: i.e., given observations of a star, to determine its age and evolutionary history. 
Stellar evolution is a well-established field with a rich history and many seminal works on the topic.  Textbooks overviewing the underpinnings of stellar structure and evolution are numerous and include works by \cite{1926ics..book.....E}, \cite{1939isss.book.....C}, \cite{1958ses..book.....S}, \cite{1989fsa..book.....C}, \cite{1990sse..book.....K}, \cite{1994sipp.book.....H}, \cite{2005essp.book.....S}, \cite{pols}, \cite{2012sse..book.....K}, and \cite{brown}. 
The following makes heavy use of these works, along with calculations using the stellar evolution code \emph{Modules for Experiments in Stellar Astrophysics} \citep[\textsc{MESA},][
]{2011apjs..192....3p, 2013apjs..208....4p, 2015apjs..220...15p, 2018ApJS..234...34P}. \end{shaded}

Positing that a star begins as an initially homogeneous cloud of mostly hydrogen that collapses under its own weight until the conditions are ripe for fusion to sustain it, stellar evolution is the collection of physical processes that cause the star to vary over time from this state. 
Reposition in terms of luminosity, radius, density, and color---diagnostics that are visible from the stellar surface---are then predicted from the ensemble of processes that cause the star to transform. 

Many such processes are known. 
Nuclear fusion causes adjustment to the elemental abundances in the core or within shells inside the star. 
Gravitational settling causes heavier elements to sink inward, and radiative levitation selectively resists this sinking. 
Convection induces chemical mixing, which leads to chemical discontinuities when the boundaries of convective zones recede, and dredge-up events when an enveloping convective zone deepens into an area of disparate composition. 
Stars rotate, and this similarly causes material to mix. 
Magnetic fields, binary accretion, thermohaline mixing, and other processes may affect the evolution of stars as well. 


This collection of processes---of which only a subset is ``canonically'' employed in stellar modelling---has been very successful at explaining both the occupations of stars in the Hertzsprung-Russell and Color-Magnitude diagrams, and in predicting the pulsations of stars as well. 
Asteroseismic theory, visited in detail in the section following this one, is capable of determining the character of the stellar oscillations during each stage in a star's life, as well as predicting their corresponding periods. For solar-type stars, these predictions are within seconds of their measured values.

\subsubsection*{Assumptions}
The standard theory describing the evolution of a single star follows from a number of basic assumptions:
\begin{enumerate}
    \item \emph{Stars can be treated as a fluid.} 
    I make this assumption so that we may describe stars using the equations of fluid dynamics rather than considering the motions of individual particles. 
    The fluid approximation is likely a good description for the majority of the stellar interior, but it breaks down above the stellar photosphere. 
    \item \emph{Stars are isolated in space.} 
    I ignore companions and, consequently, the effects of mass transfer and tidal interactions. 
    \item \emph{Stars are spherically symmetric.} 
    I will describe the structure of a star from its core to its surface using only one coordinate (e.g.~radius, but in practice, some quantity that varies monotonically with radius). 
    I ignore rotation, which would distort the star. 
    While all stars rotate, many (such as the Sun) rotate slowly enough that the effects of rotation on their structure can be considered negligible. 
    \lr{\item \emph{Stars are self-gravitating.} 
    I include the effects of a gravitational field, but I ignore electric and magnetic fields. 
    \item \emph{Stars are dynamically stable.} 
    Clearly, stars are pulsating; that is the main subject of this thesis. 
    However, the pulsation timescale is usually much shorter than the evolutionary timescale. 
    These will be treated in detail in the next section. 
    \item \emph{Stars keep their mass.}
    Stars are observed to lose their mass through, for example, stellar winds. 
    However, isolated main-sequence stars lose very little mass. 
    For example, the Sun loses only about one part in $10^{13}$ of its mass each year \citep{2004CeMDA..90..267K}.}
\end{enumerate}
From these assumptions, we may now formulate equations for the structure of a star. 

\subsubsection*{Stellar Structure}
By the structure of a star, I mean the \emph{mechanical} (density $\rho$, pressure $P$), \emph{thermal} (temperature $T$, adiabatic exponents $\boldsymbol\Gamma$), and \emph{chemical} (relative abundances of hydrogen $X$, helium $Y$, and heavy elements $Z$ obeying ${X+Y+Z=1}$) profiles from the core to the `surface.' 
The equations of stellar structure consist of three conservation equations---conservation of mass, momentum, and energy---and the temperature equation. 
These macrophysical equations are supplemented with `microphysics,' numerical inputs for necessary ingredients such as nuclear reaction rates. 
I will present most of the equations of stellar structure essentially without derivation. 
In order to give the reader an idea of the arguments used, however, I will provide derivations based on geometry and basic physics for the conservation of mass and the conservation of (linear) momentum. 

It is natural to consider these quantities spatially (i.e., in one dimension, by the stellar radius). 
However, the radius of a star changes considerably over its lifetime, growing from a dwarf to a giant and then becoming a dwarf again. 
On the other hand, the mass of a star, at least in the main-sequence phase, is very stable. 
The Sun, for example, loses only $10^{-14}$ of its mass per year through fusion and the solar wind. 
Therefore, I will here cast the equations using mass as the independent variable. 
In practice, stellar evolution codes often use a more complex variable which is even more stable than mass. 

\begin{description}
    \setlength{\itemindent}{0pt}
    \item[Conservation of Mass.]
    Geometrically speaking, the mass $m$ contained within a sphere spanning from the centerpoint (${r=0}$) to a radius of $r$ is given by 
    \lr{\begin{equation}
        m(r)
        =
        \int_0^r 4 \pi x^2 \rho(x) \; \text{d}x.
    \end{equation}}
    Differentiating this equation, and dropping arguments, we arrive at 
    \begin{equation} \label{eq:cons-mass} \boxed{
        \frac{\text{d}r}{\text{d}m}
        =
        \frac{1}{4\pi r^2\rho}
    }\end{equation}
    which, as we will see, is the continuity equation in the absence of flows. $\hfill\square\;$
    
    \item[Conservation of Momentum.] 
    The state of balance between gravity and a pressure-gradient force is called hydrostatic support (also known as hydrostatic equilibrium or hydrostatic balance) and is a special case of conservation of momentum. 
    The equation can be derived from either Newton's laws of motion, the Navier--Stokes equations, or from general relativity. 
    Here I show the former. 
    
    Consider a small fluid parcel inside of the star whose base is located at radius $r$ having height ${\text{d}r}$ and a constant area $A$. 
    The parcel has three forces acting upon it: downward and upward forces from pressure, and a downward force from gravity. 
    The upward force on the parcel is 
    \begin{equation}
        F_{\text{upward}} (r) = A \cdot \underbrace{P(r)}_{\makebox[0pt]{\text{\scriptsize pressure below}}}
    \end{equation}
    and the combined downward force is
    \begin{equation}
        F_{\text{downward}} (r)
        = -\left(
            A \cdot \underbrace{P(r+\text{d}r)}_{\makebox[0pt]{\text{\scriptsize pressure above}}} + \underbrace{A \cdot \rho(r) g(r) \cdot \text{d}r}_{\text{gravity}}
        \right).
    \end{equation}
    When these forces are balanced, i.e.~${F_{\text{upward}} = F_{\text{downward}}}$, the parcel is said to be in hydrostatic equilibrium. 
    Thus, we have
    \begin{equation}
        0 = 
        -A\left( 
                \mathrlap{\underbrace{\phantom{\;P(r)}}_{F_{\text{upward}}}}
                \;\mathrlap{\overbrace{\phantom{P(r) - P(r+\text{d}r)}}^{\text{d}P}}
                P(r) - 
                \underbrace{
                    P(r+\text{d}r) - \rho(r) g(r) \cdot \text{d}r
                }_{F_{\text{downward}}}
        \right)
    \end{equation}
    which then gives us
    \begin{align} \label{eq:cons-mom-r}
        \frac{\text{d} P}{\text{d} r} &= -\rho g.
    \end{align}
    We may then apply the equation of conservation of mass (\ref{eq:cons-mass}) and obtain
    \begin{equation} \label{eq:cons-mom} \boxed{
        \frac{\text{d} P}{\text{d} m} = -\frac{Gm}{4\pi r^4}
    }\end{equation}
    where ${G = 6.67408\times 10^{-8}}$~\si{\per\g\cm\cubed\per\square\s} is the gravitational constant. $\hfill\square\;$
    
    These two conservation equations give us the mechanical structure of the star---the pressure and density throughout the stellar interior. 
    Assuming a constant temperature, the ratio of these quantities gives us the speed at which acoustic waves propagate in the star:
    \begin{equation} \label{eq:u}
        u = P/\rho.
    \end{equation}
    This quantity is known as the \emph{\lr{squared} isothermal speed of sound} and will be important in the following investigations. 
    
    \item[Conservation of Energy.]
    The flow of energy $l$ throughout the stellar interior is given by 
    \begin{equation} \label{eq:energy} \boxed{
        \frac{\text{d}l}{\text{d}m}
        =
        \epsilon_{\text{nuc}} - \epsilon_\nu + \epsilon_g 
    }\end{equation}
    where $\epsilon_{\text{nuc}}$ is the energy generated by nuclear reactions, $\epsilon_\nu$ is the energy lost by neutrinos, and $\epsilon_g$ is the gravitational energy from expansion or compression: 
    \begin{equation} \label{eq:eps-g}
        \epsilon_{\text{g}} = -T\, \frac{\partial s}{\partial t}
    \end{equation}
    where $s$ is the specific entropy. 
    The nuclear energy generation rates are supplied externally. 
    Here I use the rates from the \emph{Nuclear Astrophysics Compilation of Reaction Rates} \citep[\textsc{NACRE},][]{1999NuPhA.656....3A}. 
    The neutrino energy loss rates can be calculated using the formulas given by \citet{1996ApJS..102..411I}. 
    
    
    Computing the $\epsilon_g$ term requires an equation of state (EOS). 
    This too is supplied externally. For the low-mass stars considered here, I use the \emph{Opacity Project at Livermore} EOS \citep[\textsc{OPAL},][]{2002apj...576.1064r}. 
    The EOS relates the pressure, density, and temperature of the stellar matter to each other in a thermodynamically-consistent manner. 
    The adiabatic exponents $\boldsymbol\Gamma$, introduced by Chandrasekhar, give these relations as follows: 
    \begin{align} 
        \Gamma_1 
        &= 
        \left(
            \frac{\partial \ln P}{\partial \ln \rho}
        \right)_{\text{ad}} 
        \\
        \frac{\Gamma_2}{\Gamma_2-1}
        &= \label{eq:gamma2}
        \left(
            \frac{\partial \ln P}{\partial \ln T}
        \right)_{\text{ad}} = \frac{1}{\nabla_{\text{ad}}}
        \\
        \Gamma_3 - 1 
        &= 
        \left(
            \frac{\partial \ln T}{\partial \ln \rho}
        \right)_{\text{ad}}
    \end{align}
    which are related to each other as: 
    \begin{equation}
        \frac{\Gamma_1}{\Gamma_3 - 1}
        =
        \frac{\Gamma_2}{\Gamma_2-1}.
    \end{equation}
    The first adiabatic exponent describes how the compression of a layer changes the pressure in that layer, which, as we will see, is important for determining dynamical stability, i.e., stellar pulsations. 
    In particular, \lr{in an anisotropic ideal gas,} the speed at which acoustic waves propagate---the adiabatic speed of sound\footnote{ Not to be confused with the speed of light.}---can be defined as 
    \begin{equation} \label{eq:speed-of-sound}
        c = \sqrt{\Gamma_1 u}. 
    \end{equation}
    The second adiabatic exponent describes how changes in pressure impact upon the temperature, which is important for determining stability against convection. 
    In an ideal monoatomic gas, the adiabatic exponents all equal $5/3$. 

    
    
    \item[Temperature Equation.]
    The temperature throughout the star is given by 
    \begin{equation} \label{eq:temperature} \boxed{
        \frac{\text{d}T}{\text{d}m}
        =
        -\frac{Gm}{4\pi r^4} \frac{T}{P} \nabla_T
    }\end{equation}
    where $\nabla_T$ is a dimensionless temperature gradient:
    \begin{equation}
        \nabla_T 
        =
        \frac{\text{d}\ln T}{\text{d}\ln P}
    \end{equation}
    whose form depends on the mode of energy transport. 
    In the case of pure radiation,
    \begin{equation} \label{eq:radiative-gradiant}
        \nabla_T 
        = 
        \nabla_{\text{rad}} 
        =
        \frac{3}{64\pi \sigma G}
        \frac{\kappa l P}{m T^4}.
    \end{equation}
    where \lr{${\sigma = 5.670367 \cdot 10^{-5} \; \text{erg}\; \text{cm}^{-2}\; \text{s}^{-1}\; \text{K}^{-4}}$}
    is the Stefan-Boltzmann constant and $\kappa$ is the opacity of the stellar matter, which is also supplied externally. 
    Here I use the \textsc{OPAL} opacities \citep{1996ApJ...464..943I}. 
    
    The conductive temperature gradient is negligible for our purposes, though it is relevant e.g.~in white dwarfs. 
    The convective temperature gradient comes from both the adiabatic gradient of the assumed EOS (\emph{cf.}~Equation~\ref{eq:gamma2}) and the specific treatment of convection, which I will discuss later in this section. 
\end{description}

\noindent We thus have four coupled differential equations (\ref{eq:cons-mass}, \ref{eq:cons-mom}, \ref{eq:energy}, \ref{eq:temperature}) that govern stellar structure. 
In order to solve them, we will need four boundary conditions. 

\subsubsection*{Boundary Conditions}

The first boundary is at the central point in the star, where ${m=0}$.
Here we have 
\begin{equation}
    m=0,\qquad r=0,\qquad l=0.
\end{equation} 
The second boundary is at the stellar surface. 
This is where the mass equals the total mass, ${m=M}$; and where the radius equals the total radius, ${r=R}$. 
A simple option is to assume that the temperature and pressure vanish at the surface, i.e.
\begin{equation} \label{eq:zero}
    T(r=R)=0, \qquad P(r=R)=0.
\end{equation}
These are known as \emph{zero-boundary} conditions and we will make use of them later when calculating variational pulsation mode frequencies (see Section~\ref{sec:variational}). 
They are unrealistic, however, as even the interstellar medium has a non-zero temperature. 

A more sophisticated option is to call the surface the region where majority of the radiation escapes from the star, i.e., the photosphere. 
Here I will use a standard Eddington gray atmosphere, which gives the total luminosity and effective temperature 
\begin{equation}
    l(r=R) = L,\qquad T(r=R) = T_{\text{eff}}
\end{equation}
following the Stefan-Boltzmann Law 
for blackbody radiation: 
\begin{align} \label{eq:stefan-boltzmann} 
    L &= 4\pi R^2 \sigma T_{\text{eff}}^4
\end{align}
where $\sigma$ is again the Stefan-Boltzmann constant. 
Finally, the pressure at the surface is given by 
\begin{equation}
    P(r=R)
    =
    \frac{2}{3} 
    \frac{GM}{R^2}
    \frac{1}{\bar \kappa}
\end{equation}
where $\bar \kappa$ is the Rosseland mean opacity. 




\subsubsection*{Stellar Evolution}
For a star to \emph{evolve}, it must change over time. With the exception of one term for the gravitational energy from expansion or compression (Equation~\ref{eq:eps-g}), the equations of stellar structure feature no time derivatives; 
they describe a static star. 
The equations of stellar structure may be supplemented with time-dependent evolution equations describing the internal transport or modification of chemical species. 

\begin{description}
    \setlength{\itemindent}{0pt}
    \item[Nuclear reactions.] Energy generation on the main sequence stems predominately from the conversion of hydrogen atoms (H) into helium atoms (He). 
    The net reaction is 
    \begin{equation}
        4\;^1\text{H}\; \rightarrow\; ^4\text{He} + 2\text{e}^+ + 2\nu_{\text{e}}
    \end{equation}
    where e$^+$ is a positron and $\nu_{\text{e}}$ is a neutrino. 
    Earth-based detections of neutrinos matching the predicted solar output essentially confirm this description. 
    The evolution due to nuclear reactions can be given as: 
    \begin{equation} \boxed{
        \frac{\partial X_i}{\partial t}
        =
        \frac{m_i}{\rho}
        \left( 
            \sum_j r_{ji}
            -
            \sum_k r_{ik}
        \right)
    }\end{equation}
    where $X_i$ is the $i^{\text{th}}$ isotope, $m_i$ is the mass of that isotope, and $r_{i,j}$ is the rate at which $X_i$ is formed from $X_j$. 
    As mentioned, these rates must be supplied externally; here I've chosen to use the \textsc{NACRE} rates. 
    
    \item[Diffusion.] The processes of element diffusion and the gravitational settling of helium and heavy elements can be included via \lr{the diffusion equation: 
    \begin{equation} \label{eq:evol-diffusion} \boxed{
        \frac{\partial X_i}{\partial t}
        =
        D_i\, \frac{\partial^2 X_i}{\partial m^2}
    }\end{equation}
    where $D_i$ is the diffusion coefficient for isotope $X_i$.} 
    Diffusion coefficients must also be externally supplied; a common choice are those of \citealt{1994ApJ...421..828T}.
    
    \item[Convection.] 
    According to the Schwarzschild criterion \citep[e.g.,][]{1958ses..book.....S}, a region is unstable to convection when the radiative gradient exceeds the adiabatic gradient: 
    \begin{equation}
        \nabla_{\text{rad}} 
        > 
        \nabla_{\text{ad}}
    \end{equation}
    (\emph{cf.}~Equations~\ref{eq:gamma2} and~\ref{eq:radiative-gradiant}). 
    Here I will treat convection using the standard \citet{1958ZA.....46..108B} mixing length theory, which approximates the effects of convection by assuming that convective elements travel to some characteristic length $\ell_m$ before mixing the transported material with their newfound surroundings. 
    The mixing length is controlled by a free parameter $\alpha_{\text{MLT}}$, which is scaled by the local \lr{pressure scale height: 
    \begin{align} \label{equation:MLT}
        \ell_m
        &=
        \alpha_{\text{MLT}} \cdot H_p\\
        H_p
        &=
        -\left(\frac{\text{d} \ln P}{\text{d}r}\right)^{-1}.
    \end{align}}
    There is no \emph{a priori} choice for $\alpha_{\text{MLT}}$. 
    Generally, $\alpha_{\text{MLT}}$ is either fixed to a value that has been calibrated to the observed characteristics of the Sun, which we shall address later in this section; or fit on a star-by-star basis (Chapter~\ref{chap:ML}). 
    
    Convection is an efficient mixer. 
    We can model the changes to chemical abundances due to convection as a diffusion process:
    \lr{\begin{equation} \boxed{
        \frac{\partial X_i}{\partial t}
        =
        \frac{\partial}{\partial m} \left( 
            D_{\text{conv}}\, \frac{\partial X_i}{\partial m}
        \right)
    }\end{equation}}
    where ${D_{\text{conv}} \propto v_c \cdot \ell_m}$, with $v_c$ being the convective velocity.
    
    Convective zones can be extended beyond their normal boundaries via convective overshooting. 
    Overshooting is similarly controlled by a free parameter $\alpha_{\text{ov}}$, which extends the boundary by ${\alpha_{\text{ov}} \cdot H_p}$. 
    Like $\alpha_{\text{MLT}}$, the overshooting parameter has no predefined value. 
    While it is not uncommon to exclude the effects of overshooting altogether, $\alpha_{\text{ov}}$ can also be determined from a fit to a stellar population \citep[e.g.,][]{2005ARA&A..43..387G} or on a star-by-star basis (Chapter~\ref{chap:ML}). 
\end{description}

\noindent Calculations generally proceed as follows. 
First, the equations of stellar structure are solved for a given composition. 
Then, time is advanced, and a new composition is computed using the evolution equations. 
The equations of stellar structure are then solved again for the new composition, and the procedure is repeated. 
\citet{1959ApJ...129..628H} introduced an efficient scheme to solve these equations based on iterative application of the Newton-Raphson method. 
We will now solve these equations and model the evolution of the stars. 


\subsubsection*{Solar Calibration}
\label{sec:calibration}

We may begin our calculations by calibrating an evolutionary track to the observed properties of the Sun \citep[e.g.][]{1982MNRAS.199..735C} in accordance with the recommended nominal solar values adopted by the IAU \citep{2015arXiv151007674M}. 
The standard gravitational parameter of the Sun $\mu_\odot$ is known to very high precision from planetary orbits: 
\begin{equation*}
    \mu_\odot = GM_\odot = 1.3271244 \cdot 10^{26} \; \text{cm}^3 \; \text{s}^{-2}.
\end{equation*}
The gravitational constant may be determined experimentally; this then yields the solar mass. 
Next, the Earth-Sun distance as well as direct observations give the solar radius. 
Solar irradiance measurements give the solar luminosity. 
Spectroscopy gives the composition the solar photosphere; I use the mixture as measured by \citet[][hereinafter \textsc{GS98}]{1998SSRv...85..161G} which gives good agreement with helioseismology. 
Finally, radiometric dating of meteorites gives the age of the solar system. 
Putting this all together, the Sun has the following characteristics: 
\lr{\begin{equation} \label{eq:solar-vals}
\begin{aligned}
    \text{mass } M_\odot &= 1.988475 \cdot 10^{33}\; \text{g} \\
    \text{radius } R_\odot &= 6.957 \cdot 10^{10}\; \text{cm} \\
    \text{luminosity } L_\odot &= 3.828 \cdot 10^{33}\; \text{erg}\; \text{s}^{-1} \\
    \makebox[0pt][r]{\text{effective }}\text{temperature } T_{\text{eff},\odot} &= 5772\; \text{K} \\
    \makebox[0pt][r]{\text{heavy mass fraction }} (Z/X)_\odot &= 0.02293 \\
    \text{age } \tau_\odot &= 4.572\cdot 10^{9} \; \text{yr}.
\end{aligned}
\end{equation}}
These are the values that must be reproduced in our solar calibration. 
We will achieve this by altering the initial chemical composition and the efficiency of convective mixing (recall Equation~\ref{equation:MLT}) until these values are reproduced at the solar age. 
Since the Sun is an isolated main-sequence star, its mass has been presumably very stable throughout its lifetime. 
The initial mass of the calibration can therefore remain fixed at the solar value. 
Finally, we only need to check that e.g.~the luminosity and radius are matched, since $R$, $L$, and $T_{\text{eff}}$ are related through the Stefan-Boltzmann Law (Equation~\ref{eq:stefan-boltzmann}).

We therefore have the following optimization problem: 
we wish to tune the initial helium abundance $Y_0$, initial metallicity $Z_0$, and mixing length parameter $\alpha_{\text{MLT}}$ of a solar-mass track such that we minimize ${\log_{10}\left(L/L_\odot\right)}$,  ${\log_{10}\left(R/R_\odot\right)}$, and [Fe/H] at the solar age, where [Fe/H] is defined as
\begin{equation}
    \text{[Fe/H]}
    \equiv
    \log_{10}\left( \frac{Z}{X} \right)_\ast
    -
    \log_{10}\left( \frac{Z}{X} \right)_\odot.
\end{equation}
We may achieve solar calibration by, e.g., iterative application of Newton's rule: 
\begin{equation}
    \mathbf{x}_{t+1}
    =
    \mathbf{x}_t - \mathbf{J}_t^{-1} \mathbf{f}(\mathbf{x}_t)
\end{equation}
where (dropping the $\text{MLT}$ and $0$ subscripts)
\begin{align}
    \mathbf{x}_t
    &=
    \begin{pmatrix} Y_t , & Z_t , & \alpha_t \end{pmatrix}
    \\
    \mathbf{f}(\mathbf{x}_t)
    &=
    \begin{pmatrix} 
        \log_{10} \left\{ L_t/L_\odot \right\}, &
        \log_{10} \left\{ R_t/R_\odot \right\}, &
        \text{[Fe/H]}_t
    \end{pmatrix}
\end{align}
\begin{equation}
    \mathbf{J}_t
    =
    \begin{pmatrix} 
        \dfrac{\strut\partial \log_{10} \left\{ L_t/L_\odot \right\}}{\strut\partial Y} &
        \dfrac{\strut\partial \log_{10} \left\{ L_t/L_\odot \right\}}{\strut\partial Z} &
        \dfrac{\strut\partial \log_{10} \left\{ L_t/L_\odot \right\}}{\strut\partial \alpha}\\
        \dfrac{\strut\partial \log_{10} \left\{ R_t/R_\odot \right\}}{\strut\partial Y} &
        \dfrac{\strut\partial \log_{10} \left\{ R_t/R_\odot \right\}}{\strut\partial Z} &
        \dfrac{\strut\partial \log_{10} \left\{ L_t/L_\odot \right\}}{\strut\partial \alpha}\\
        \dfrac{\strut\partial \text{[Fe/H]}_t}{\strut\partial Y} &
        \dfrac{\strut\partial \text{[Fe/H]}_t}{\strut\partial Z} &
        \dfrac{\strut\partial \text{[Fe/H]}_t}{\strut\partial \alpha}
    \end{pmatrix}.
\end{equation}
Here $t$ refers to the $t^{\text{th}}$ iteration, and the partial derivatives are to be calculated numerically (i.e.\ by running tracks with small changes to those parameters). 
It may also be prudent to enforce some box constraints, for example: ${0.23 \leq Y_0 \leq 0.33}$, ${0 < Z_0 < 0.05}$, ${1 \leq \alpha_{\text{MLT}} \leq 3}$. 
When supplied with a reasonable initial guess, this scheme eventually converges onto a set of parameters that reproduce the observed solar values: 
\begin{alignat}{3} \label{eq:solar-cal-vals} 
    &Y_0 &&\simeq 0.273  &\log_{10}(L/L_\odot) &\simeq 0 \notag\\
    &Z_0 &&\simeq 0.019 \qquad\Rightarrow\qquad &\log_{10}(R/R_\odot) &\simeq 0\\\notag
    &\alpha_{\text{MLT}} &&\simeq 1.84  &\text{[Fe/H]} &\simeq 0.
\end{alignat}
These initial values, as well as the observed values of the Sun (Equations~\ref{eq:solar-vals}) are the ones that will need to be reproduced when we later perform evolutionary inversions on degraded Sun-as-a-star data (see Chapter \ref{chap:ML}), where they are all either unknown or highly uncertain. 

\begin{figure}
    \centering
    \begin{subfigure}[b]{0.5\linewidth}
        \centering
        \includegraphics[width=\textwidth,keepaspectratio]{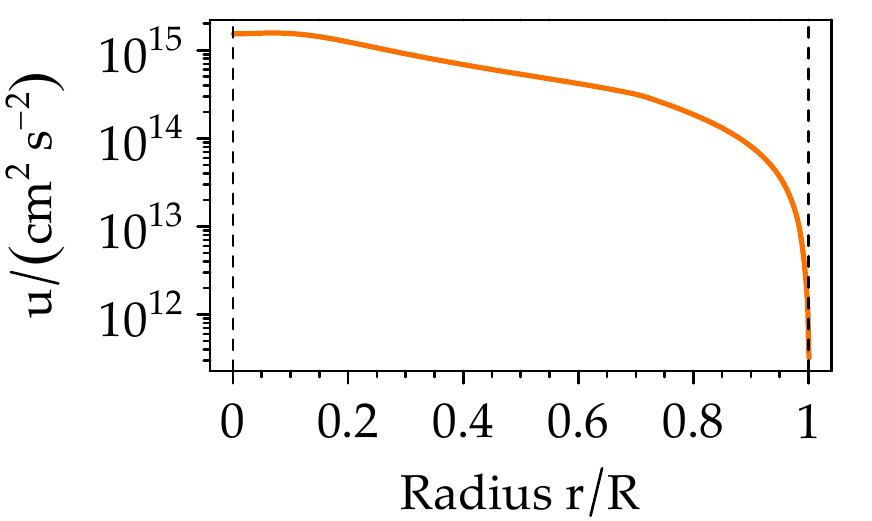}
    \end{subfigure}%
    \begin{subfigure}[b]{0.5\linewidth}
        \centering
        \includegraphics[width=\textwidth,keepaspectratio]{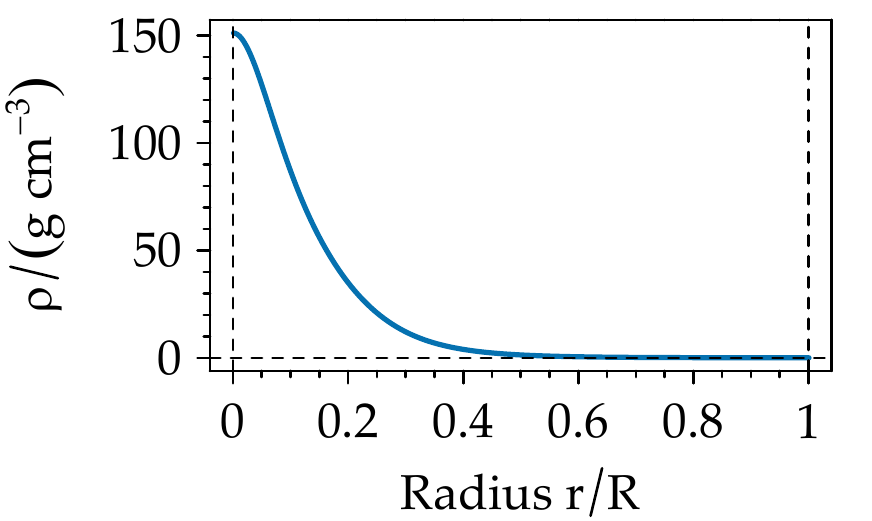}
    \end{subfigure}\\
    \begin{subfigure}[b]{0.5\linewidth}
        \centering
        \includegraphics[width=\textwidth,keepaspectratio]{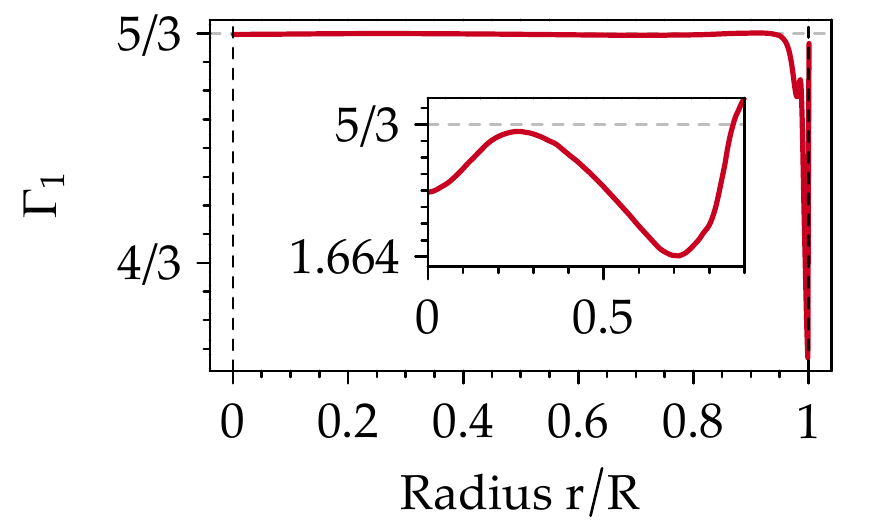}
    \end{subfigure}%
    \begin{subfigure}[b]{0.5\linewidth}
        \centering
        \includegraphics[width=\textwidth,keepaspectratio]{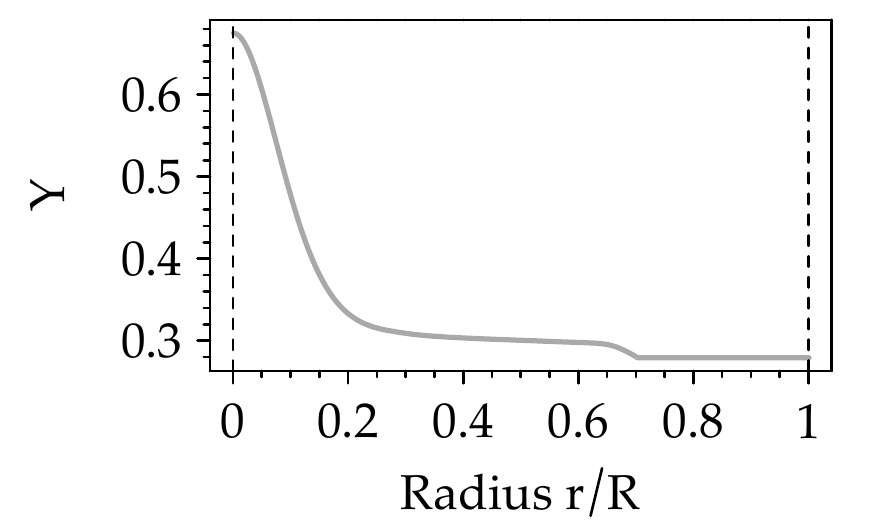}
    \end{subfigure}
    \caption[The Sun's internal mechanical, thermal, and chemical profile]{\lr{Squared isothermal sound speed} (top left), density (top right), first adiabatic exponent (bottom left), and fractional helium abundance (bottom right) profiles for a solar model. 
    \label{fig:profs} } 
    \vspace*{1cm}
    \includegraphics[width=\textwidth]{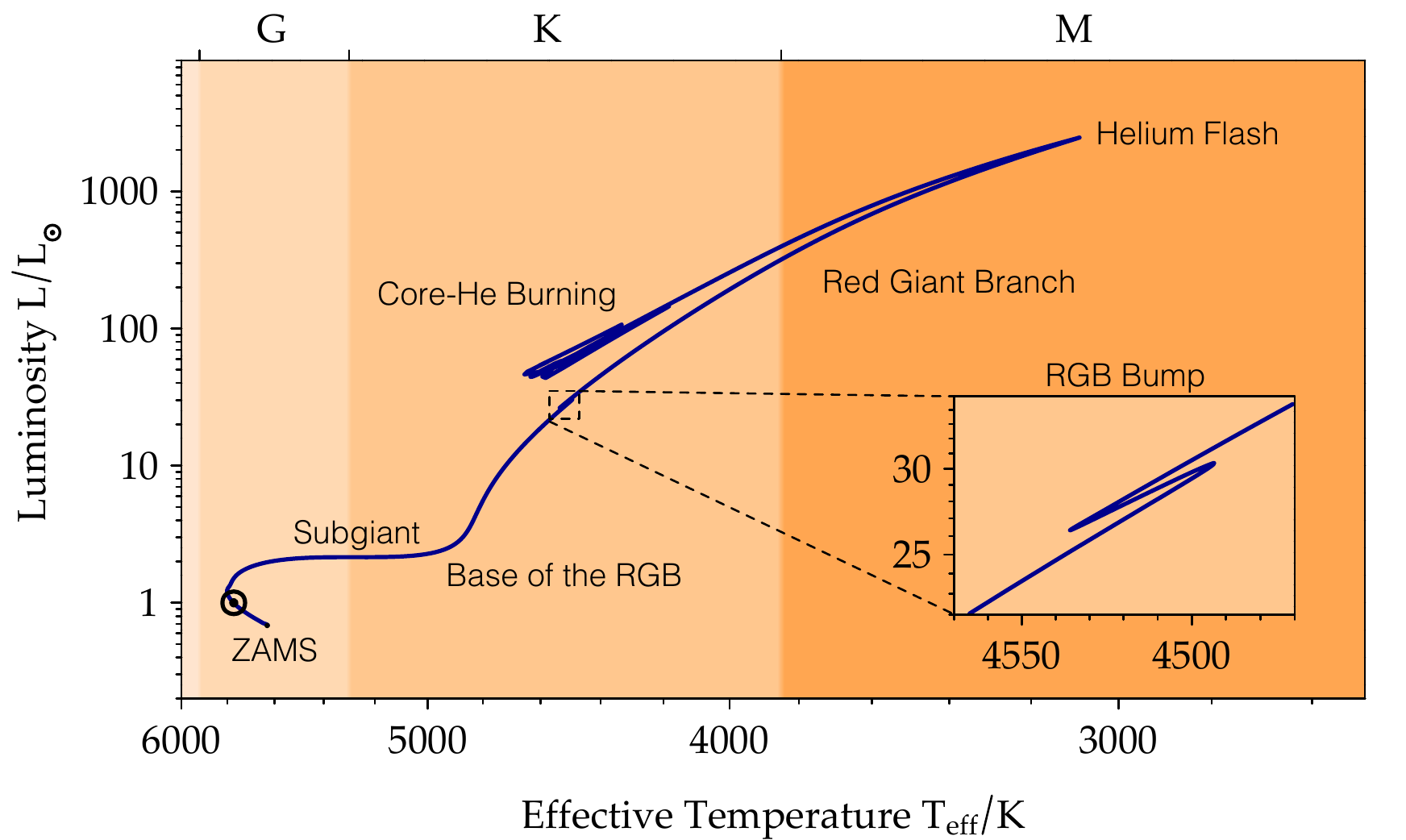}
    \caption[Solar H-R Diagram]{Hertzsprung-Russell diagram showing the evolution of the Sun. 
    The background colors correspond to spectral type (F, G, K, M). 
    The position of the Sun is indicated with the solar symbol ($\odot$). 
    \label{fig:solar-HR}}
\end{figure}

We may now inspect the structure of our solar model. 
Figure~\ref{fig:profs} shows some aspects of the mechanical, thermal, and chemical structure of the model. 
Helioseismology has revealed that these profiles are exceptionally close to the actual interior of the Sun \citep[see, e.g.,][]{2016lrsp...13....2b}. 

A few points are worthy of note here. 
The first adiabatic exponent is close to ${5/3}$ (i.e., nearly the conditions of an ideal gas) for the majority of the solar interior and only deviates from this value close to the solar surface. 
The helium abundance $Y$ in the solar core is maximal due to nearly $5$~Gyr of hydrogen-to-helium fusion. 
Helium is now the dominant element in the core, with the fractional hydrogen abundance being reduced to $0.344$. 
Throughout the convection zone, which extends from ${\sim 0.7\;{r/R}}$ to the solar surface, the helium abundance has a constant value of $0.279$ due to convective mixing. 
This value is somewhat higher than the protosolar value of $0.273$ due to element diffusion. 

The density ranges from around $150$~g/cm$^3$ in the core to less than that of water in the outer half of the star, with the mean density of the Sun being about a hundredth of the core density. 
The pressure in the solar core falls off more rapidly than the density, which causes the speed of sound to rise temporarily when moving away from the centerpoint. 
Furthermore, since $u\propto T/\mu$, where $T$ is the temperature and $\mu$ is the mean molecular weight, the speed of sound in the solar core is related to the age of the Sun via the increased abundance of helium. 

\begin{figure}
    \centering
    \includegraphics[width=\textwidth]{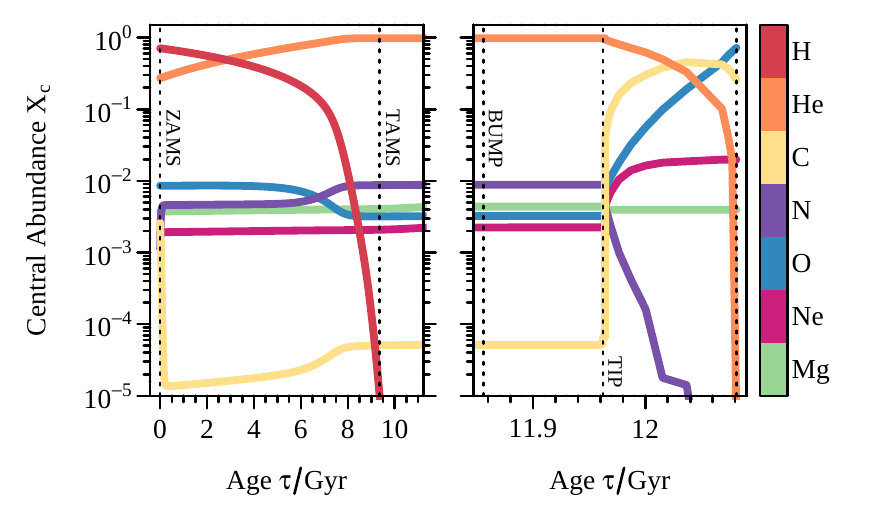}
    \caption[Chemical evolution of the solar core]{The past and future chemical evolution of the core of our Sun. The left panel shows the main sequence evolution, from the zero-age main sequence (ZAMS) to the terminal-age main sequence (TAMS). The right panel shows the evolution from the red giant luminosity bump through to the tip of the red giant branch and eventually to core-helium exhaustion. 
    The core composition does not change throughout the majority of the subgiant and red giant phases. 
    \label{fig:chem_ev}} 
    \vspace*{0.1cm}
    \begin{minipage}[t][1cm][t]{1cm+\widthof{convection}}\vspace{0pt}
\begin{tikzpicture}[scale=3]
  \filldraw[fill=black!5!white]
    (0,0) -- (0.169,0) -- (.169,.1) -- (0,.1) -- (0,0);
  \filldraw[pattern=crosshatch dots, pattern color=black]
    (0,0) -- (0.169,0) -- (.169,.1) -- (0,.1) -- (0,0);
  \node[right] at (0.169, 0.06) {convection}; 
\end{tikzpicture}
\end{minipage}
\begin{minipage}[t][1cm][t]{1cm+\widthof{radiation}}\vspace{0pt}
\begin{tikzpicture}[scale=3]
  \filldraw[fill=orange!30!white]
    (0,0) -- (0.169,0) -- (.169,.1) -- (0,.1) -- (0,0);
  \node[right] at (0.169, 0.06) {radiation}; 
\end{tikzpicture}
\end{minipage}
\begin{minipage}[t][1cm][t]{1cm+\widthof{hydrogen fusion}}\vspace{0pt}
\begin{tikzpicture}[scale=3]
  \filldraw[pattern=fivepointed stars2, pattern color=red!70!black]
    (0,-0.04) -- (0.169,-0.04) -- (.169,.06) -- (0,.06) -- (0,-0.04);
  \node[right] at (0.169, 0) {hydrogen fusion}; 
\end{tikzpicture}
\end{minipage}
\begin{minipage}[t][1cm][t]{1cm+\widthof{helium fusion}}\vspace{0pt}
\begin{tikzpicture}[scale=3]
  \filldraw[pattern=fivepointed stars3, pattern color=blue!70!black]
    (0,0) -- (0.169,0) -- (.169,.1) -- (0,.1) -- (0,0);
  \node[right] at (0.169, 0.06) {helium fusion}; 
\end{tikzpicture}
\end{minipage}

\vspace*{0.25cm}

\begin{tikzpicture}[scale=3] 
  \def\core{{(log10(0.2)-log10(0.001))/3}}
  \def\env{{(log10(0.7)-log10(0.001))/3}}
  \filldraw[fill=orange!30!white]
    (0,0) -- (1,0) arc (0:90:1) -- (0,0);
  \filldraw[fill=orange!15!white]
    (0,0) -- (\core,0) arc (0:90:\core) -- (0,0);
  \filldraw[pattern=fivepointed stars2, pattern color=red!70!black]
    (0,0) -- (\core,0) arc (0:90:\core) -- (0,0);
  \filldraw[fill=black!5!white, draw=black]
    (\env, 0) -- (1, 0) arc (0:90:1) -- (0, \env) arc (90:0:\env);
  \filldraw[pattern=crosshatch dots, pattern color=black]
    (\env, 0) -- (1, 0) arc (0:90:1) -- (0, \env) arc (90:0:\env);
  \draw[color=black]
    (0,0) -- (1,0) arc (0:90:1) -- (0,0);
  \path[inner sep=0] (-0.03, 0) node[left] {$0.001$};
  \path[inner sep=0] (-0.03, {(log10(0.01)-log10(0.001))/3}) node[left] {$0.01$};
  \path[inner sep=0] (-0.03, {(log10(0.1)-log10(0.001))/3}) node[left] {$0.1$};
  \path[inner sep=0] (-0.03, {(log10(1)-log10(0.001))/3}) node[left] {$r/R = 1$};
  \path[inner sep=0] (0.94, 0.9) node {$0.9-1.3~R_\odot$}; 
  \path[inner sep=0] (0.5, -0.15) node {main sequence}; 
\end{tikzpicture}%
\hspace{0.5cm}
\begin{tikzpicture}[scale=3] 
  \def\shellbot{{(log10(0.01)-log10(0.001))/3}}
  \def\shelltop{{(log10(0.02)-log10(0.001))/3}}
  \def\env{{(log10(0.05)-log10(0.001))/3}}
  \filldraw[fill=orange!30!white]
    (0,0) -- (1,0) arc (0:90:1) -- (0,0);
  \filldraw[fill=orange!15!white]
    (\shellbot, 0) -- (\shelltop, 0) arc (0:90:\shelltop) -- (0, \shellbot) arc (90:0:\shellbot);
  \filldraw[pattern=fivepointed stars2, pattern color=red!70!black]
    (\shellbot, 0) -- (\shelltop, 0) arc (0:90:\shelltop) -- (0, \shellbot) arc (90:0:\shellbot);
  \filldraw[fill=black!5!white, draw=black]
    (\env, 0) -- (1, 0) arc (0:90:1) -- (0, \env) arc (90:0:\env);
  \filldraw[pattern=crosshatch dots, pattern color=black]
    (\env, 0) -- (1, 0) arc (0:90:1) -- (0, \env) arc (90:0:\env);
  \draw[color=black]
    (0,0) -- (1,0) arc (0:90:1) -- (0,0);
  \path[inner sep=0] (0.94, 0.9) node {$1.3-175~R_\odot$}; 
  \path[inner sep=0] (0.5, -0.15) node {sub/red giant}; 
\end{tikzpicture} %
\hspace{0.5cm}
\begin{tikzpicture}[scale=3] 
  \def\core{{(log10(0.002)-log10(0.001))/3}}
  \def\shellbot{{(log10(0.0063)-log10(0.001))/3}}
  \def\shelltop{{(log10(0.01)-log10(0.001))/3}}
  \def\env{{(log10(0.2213)-log10(0.001))/3}}
  \def\convcore{{(log10(0.0031)-log10(0.001))/3}}
  \filldraw[fill=orange!30!white]
    (0,0) -- (1,0) arc (0:90:1) -- (0,0);
  \filldraw[fill=black!5!white, draw=black]
    (0,0) -- (\convcore,0) arc (0:90:\convcore) -- (0,0);
  \filldraw[pattern=crosshatch dots, pattern color=black]
    (0,0) -- (\convcore,0) arc (0:90:\convcore) -- (0,0);
  \filldraw[pattern=fivepointed stars3, pattern color=blue!70!black]
    (0,0) -- (\core,0) arc (0:90:\core) -- (0,0);
  \filldraw[fill=orange!15!white]
    (\shellbot, 0) -- (\shelltop, 0) arc (0:90:\shelltop) -- (0, \shellbot) arc (90:0:\shellbot);
  \filldraw[pattern=fivepointed stars2, pattern color=red!70!black]
    (\shellbot, 0) -- (\shelltop, 0) arc (0:90:\shelltop) -- (0, \shellbot) arc (90:0:\shellbot);
  \filldraw[fill=black!5!white, draw=black]
    (\env, 0) -- (1, 0) arc (0:90:1) -- (0, \env) arc (90:0:\env);
  \filldraw[pattern=crosshatch dots, pattern color=black]
    (\env, 0) -- (1, 0) arc (0:90:1) -- (0, \env) arc (90:0:\env);
  \draw[color=black]
    (0,0) -- (1,0) arc (0:90:1) -- (0,0);
  \path[inner sep=0] (0.94, 0.9) node {$10-12~R_\odot$}; 
  \path[inner sep=0] (0.5, -0.15) node {red clump}; 
\end{tikzpicture} 
    \caption[Configurations of the solar interior]{The configuration of the solar interior as the Sun evolves. 
    The present Sun is a main-sequence star with a radiative core where hydrogen fusion is synthesizing helium. 
    The outer ${\sim 30\%}$ of the Sun by radius transports energy by convection. 
    When the Sun depletes its supply of core hydrogen in ${\sim 5}$~Gyr, it will continue burning hydrogen in a shell outside of the core. 
    For the next ${\sim 2.5}$~Gyr, the inert helium core will contract while the convective envelope deepens as the Sun puffs up into a giant star. 
    The Sun will then reach the tip of the red giant branch, where helium in the highly degenerate core will suddenly undergo a flash ignition. 
    The Sun will subsequently become a red clump star, where it will continue fusing hydrogen in a radiative shell while simultaneously fusing helium in its convective core. 
    \label{fig:config}}
\end{figure}

We may additionally inspect the resulting evolutionary path of the solar-calibrated track. 
Figure~\ref{fig:solar-HR} shows the past and future evolution of our Sun, assuming that the theory of stellar evolution is approximately correct; and Figure~\ref{fig:chem_ev} shows the chemical evolution of the solar core. 
The Sun is currently on the main sequence; after several billion years, it will cross the sub-giant branch, climb the red giant branch (RGB), reach the tip of the RGB, and then fall onto the red clump (RC). 
The configurations of the star at these points in its evolution are shown in Figure~\ref{fig:config}. 

Subsequent to these stages is the asymptotic giant branch (AGB, shell-helium \& shell-hydrogen burning) phase, followed by the (misnomered) planetary nebula phase in which the outer layers of the Sun will be shed. 
The Earth and the terrestrial planets of the solar system will almost certainly be consumed or burnt to the point of inhabitability by this point. 
The Sun will then cool nearly indefinitely as a white dwarf---until, after trillions of years, it will finally settle as a black dwarf. 

This is the fairly typical path of a low-mass star and looks roughly the same for stars of solar composition with masses ${0.2 \lessapprox M/M_\odot \lessapprox 1.2}$, with the amount of time taken through this sequence being inversely related to the stellar mass. 
Outside of this range, less massive stars are fully convective and so their evolution can be quite different. 
Even less massive objects (${M \lessapprox 0.1\; M_\odot}$) never achieve hydrogen fusion, and as such, never enter the main sequence. 
More massive stars (${M/M_\odot \gtrapprox 1.2}$) sustain a convective core on the main sequence, and exhibit a feature known as the Henyey hook when leaving it. 
Stars more massive than ${\sim 2.2\;M_{\odot}}$ do not undergo a helium flash on the red giant branch; instead, they gently begin helium burning. 
Finally, stars with a final mass (i.e., after the loss of mass in the later stages of evolution) above about ${1.44\;M_\odot}$ (the Chandrasekhar limit) do not become white and black dwarfs; they rather explode in a supernova, enriching the interstellar medium with heavy mass elements. 
It is to these stars that we owe our astronomical heritage. 

In this thesis, I am mainly focused on the study of stars in their first and longest-lived phase of evolution: the main sequence. 
Currently ongoing work is the application of these techniques developed herein to those later stages of evolution.

\subsubsection*{Evolutionary Paths}
\begin{figure}
    \centering
    \adjustbox{trim=0cm 1.3cm 0cm 0cm, clip}{%
        \includegraphics[width=0.5\textwidth]{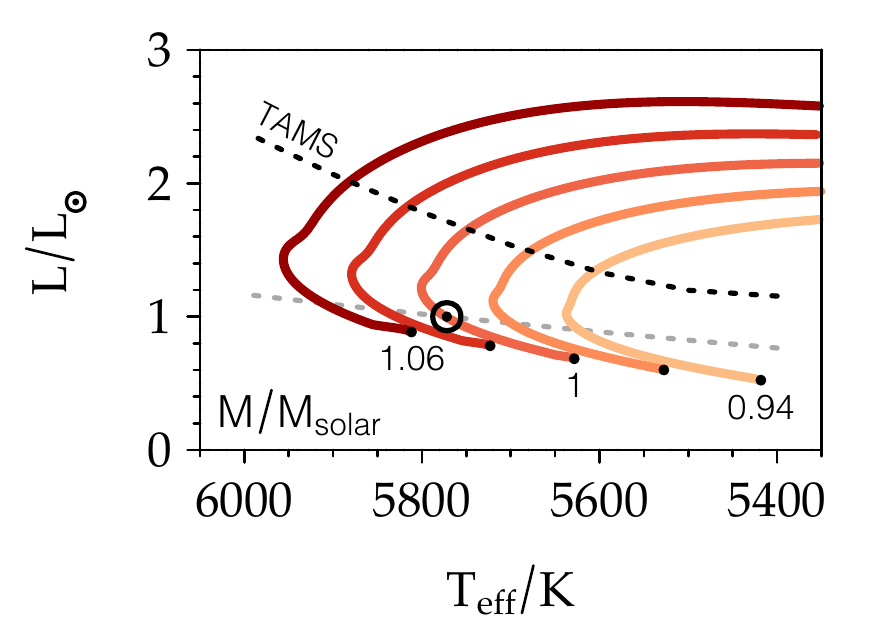}%
    }%
    \adjustbox{trim=1.6cm 1.3cm 0cm 0cm, clip}{%
        \includegraphics[width=0.5\textwidth]{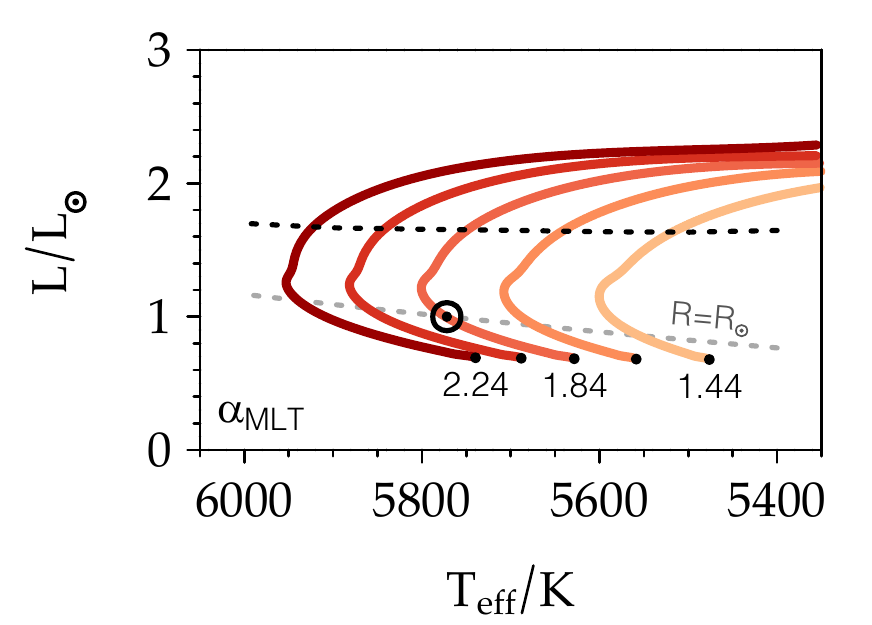}%
    }\\
    \adjustbox{trim=0cm 0cm 0cm 0cm, clip}{%
        \includegraphics[width=0.5\textwidth]{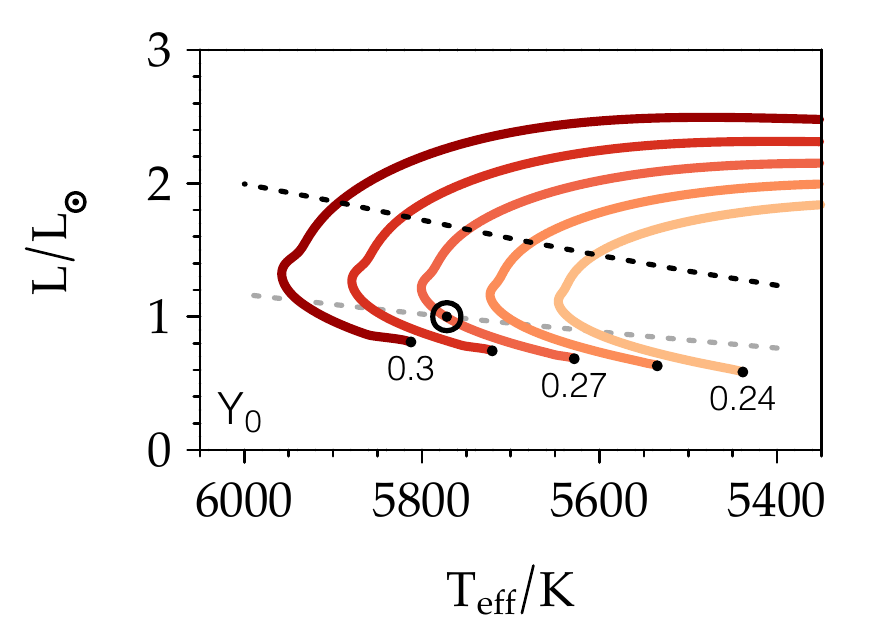}%
    }%
    \adjustbox{trim=1.6cm 0cm 0cm 0cm, clip}{%
        \includegraphics[width=0.5\textwidth]{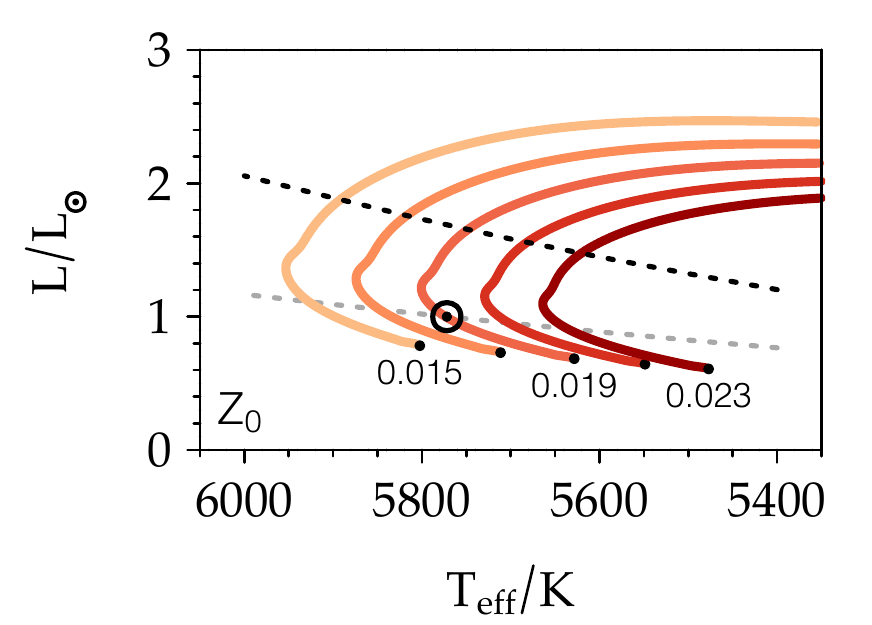}%
    }%
    \caption[Evolutionary tracks]{Theoretical Hertzsprung-Russell diagrams showing the main-sequence and sub-giant phases for evolutionary tracks varied in initial mass (top left), mixing length parameter (top right), initial helium abundance (bottom left) and initial metallicity (bottom right). 
    Aside from the parameter being varied, the remaining parameters are kept fixed at the solar-calibrated values. 
    For each track, ZAMS is marked with a black dot. 
    The solar radius is indicated by the gray dotted line (recall Equation~\ref{eq:stefan-boltzmann}). 
    Core-hydrogen exhaustion (TAMS, ${X_c \sim 10^{-5}}$) is indicated by the black dotted line. 
    The color of the track darkens as the parameter under consideration increases. 
    Notice that unlike the other parameters, an increase to the initial metallicity decreases the effective temperature. 
    The H-R diagram is degenerate in that the sense that the same point can be reached by evolutionary tracks with different input parameters. 
    \label{fig:evolutionary-tracks}}
\end{figure}

The last investigation of this section is focused toward gaining an intuition for what kinds of (in this case: low-mass, main sequence) stars are theoretically possible under the above assumptions. 
This is the forward problem of stellar evolution. 
Figure~\ref{fig:evolutionary-tracks} shows evolutionary tracks for stars under non-solar conditions that I generated by varying the free parameters of stellar evolution from their solar-calibrated values, one at a time. 
Notice that adjustments to different parameters have similar impacts on the resulting evolution of the star. 
Thus it is very difficult, at least on the basis of the position in the H-R diagram, to determine the characteristics of a star. 
As we will see in Section~\ref{sec:inverse}, determining the evolutionary characteristics of a star from observations forms the first of the two inverse problems that are considered in this thesis. 

\clearpage\section{Theory of Stellar Pulsations} 
\label{sec:pulsation}
\begin{shaded}
\noindent The purpose of this section is to give the reader a sufficient background summary on non-radial stellar pulsations in order to be able to understand the remainder of this thesis. 
I draw heavily here from the numerous textbooks that have been written on stellar pulsations, which include works by \cite{1926ics..book.....E}, \cite{1949ptvs.book.....R}, \cite{1979nos..book.....U}, \cite{1980tsp..book.....C}, \cite{2010aste.book.....a}, and \cite{basuchaplin2017}. 
Additionally, the long reviews by \cite{1958HDP....51..353L}, \citet{1993afd..conf..399G}, and \citet{2016lrsp...13....2b} were valuable references. 
I will perform calculations in this section using the \emph{Aarhus adiabatic oscillation package} \citep[\textsc{ADIPLS},][]{2008Ap&SS.316..113C}. 
\end{shaded}

Observations of stellar pulsations grant a new kind of insight into the behavior of stars. 
Whereas classical measurements of stars probe the stellar surface, observations of stellar pulsations, which traverse the stellar interior, bring deeper information to light. 
Measurements of stellar pulsations provide stringent tests on the processes of stellar evolution, as the frequencies of pulsation profoundly depend on the predicted stellar structure. 
Stars exhibiting solar-like oscillations are particularly valuable for this pursuit. 
These stars vibrate in a superposition of a great number of oscillation modes simultaneously, and each mode that can be observed provides additional information that can be used to constrain stellar models. 

The pulsation hypothesis of stellar variability is supported by the fact that the theoretical pulsations of stellar models generally match the observed pulsations of stars. 
Furthermore, theoretically predicted pulsations in stars that were previously not observed to be variable (such as red giants) have been now overwhelmingly confirmed. 
That being said, while the agreement with models is very good, it is not perfect. 
In this section, I will outline the theory of stellar pulsations, thereby allowing us to calculate the time-independent adiabatic pulsation frequencies of our stellar models. 
I will compare the frequencies of my solar-calibrated model to measurements of the Sun. 
I will furthermore present the kernel functions of stellar structure, which quantify how changes to the stellar structure translate into changes in pulsation frequencies. 
This will allow me to state the structure inverse problem: i.e., the problem of determining a star's structure using only asteroseismic arguments. 

\subsubsection*{Assumptions} 

I again begin with my assumptions. In addition to the assumptions for stellar structure, I assume: 

\begin{enumerate} 
    \item \emph{The stellar structure is nearly static.} 
    I ignore all time derivatives (including velocities) in the equilibrium structure of the star. 
    Thus, I am considering only time-independent pulsation frequencies. 
    Clearly, stars evolve over time---the entire preceding section was based on that fact. 
    That said, the evolutionary timescale in the stars considered here (billions of years) is far greater than the pulsation timescale (minutes). 
    
    \item \emph{The pulsations are linear perturbations the static stellar structure.} 
    I ignore non-linear perturbations. 
    This assumption should hold when the pulsation amplitudes are much smaller than the speed of sound. 
    As we've seen, solar oscillations have amplitudes around ${10\;\text{cm/s}}$, whereas the speed of sound at the surface of the solar-calibrated model is on the order of ${10\;\text{km/s}}$. 
    
    \item \emph{The pulsations are adiabatic.} 
    I ignore the transfer of energy between the oscillations and the equilibrium stellar structure. 
    This assumption should hold to good approximation when the pulsation time-scale is much smaller than the thermal timescale. 
    With pulsation periods on the order of minutes, this is true for the majority of the stellar interior.
    However, this assumption too breaks down near to the stellar surface. 
    Furthermore, without consideration of non-adiabatic effects, we will be unable to predict mode amplitudes, and we will not be able to determine whether the modes are excited \citep[e.g.,][]{2015EAS....73..111S}. 
    
    \item \emph{The stellar material is inviscid.} 
    I ignore internal friction. 
    Although the viscosity of the solar core is similar to that of honey (${\sim 100\;\text{cm}^2/\text{s}}$, e.g., \citealt{fox2000geophysical}), the Reynolds numbers throughout the solar interior are large enough to justify this assumption. 
    However, this assumption does break down in convection zones, where turbulent viscosity damps the oscillations. 
\end{enumerate} 
Here and in the previous section I have made several assumptions that are violated in the near-surface layers of stars, or in locations where energy is transported by convection. 
These violations will cause errors in the predicted mode frequencies. 
I will introduce a correction to deal with these errors later in the section. 

\subsubsection*{Fluid Dynamics}
Given a static stellar structure, we consider a small perturbation that displaces all quantities (density, pressure, etc.) from equilibrium. 
For example, the stellar density at position $\vec r$ and time $t$ is 
\begin{align}
    \text{(Eulerian perturbation)} && 
    \rho (\vec r, t)
    &= \label{eq:eulerean}
    \rho_0 (\vec r) 
    + 
    \rho' (\vec r, t)
    \\
    \text{(Lagrangian perturbation)} &&
    \delta\rho(\vec r)
    &= \label{eq:lagrangian}
    \rho'(\vec r_0)
    +
    \vec\xi \cdot \nabla\rho_0 (\vec r)
\end{align}
where $\rho_0$ is the equilibrium density, $\rho'$ is the perturbed density, and ${\vec\xi\equiv\vec r - \vec r_0}$ is the displacement in space.
Here I have made use of the assumption that the equilibrium structure does not depend on time. 
The perturbation induces a velocity field $\vec{\varv}$ given by 
\begin{equation}
    \vec{\varv} (\vec r, t) 
    = 
    \frac{\partial}{\partial t}\vec\xi (\vec r, t).
\end{equation}
This velocity field is then controlled by the following equations: 

\begin{description}
    \setlength{\itemindent}{0pt}
    \item[The continuity equation.]
    As we've seen previously, the equation of continuity is a statement of mass conservation (\emph{cf}.~Equation~\ref{eq:cons-mass}). 
    It states that mass cannot teleport through the star, but rather must travel through it continuously. 
    \lr{The equation can be given as 
    \begin{equation} \label{eq:continuity} 
        \frac{\partial \rho}{\partial t}
        +
        \nabla
        \cdot
        \left(
            \rho \vec{\varv} 
        \right)
        =
        0
    \end{equation}
    where $\nabla\cdot$ is the divergence vector operator.} 
    Substituting the perturbed quantities (Equation~\ref{eq:eulerean}) into Equation~\ref{eq:continuity}, we get 
    \begin{equation}
        \frac{\partial}{\partial t} \left[
            \rho_0(\vec r)
            +
            \rho' (\vec r, t)
        \right]
        +
        \nabla\cdot \left\{
            \left[
                \rho_0(\vec r)
                +
                \rho'(\vec r, t)
            \right]
            \frac{\partial\vec\xi}{\partial t} 
        \right\}
        =
        0.
    \end{equation}
    As we have assumed the equilibrium structure to be static, the corresponding time derivatives vanish. 
    Integrating with respect to time, we then obtain
    \begin{equation} \label{eq:perturbed-continuity} \boxed{
        \rho' 
        + 
        \nabla \cdot \left( 
            \rho_0 \vec\xi 
        \right) 
        = 
        0
    }\end{equation}
    i.e., the perturbed equation of continuity. $\hfill\square\;$
    
    \item[The equation of motion.] \lr{To first order, the general Navier--Stokes momentum equation can be expressed as}
    \begin{equation} 
        \rho
        \left(
            \frac{\partial}{\partial t}
            +
            \vec{\varv}
            \cdot
            \nabla 
        \right)
        \vec{\varv}
        =
        -\nabla P
        +
        \mu \nabla^2 \vec{\varv}
        +
        \frac{1}{3} \mu \nabla\left(
            \nabla \cdot \vec{\varv}
        \right)
        +
        \rho \vec{g}
    \end{equation}
    where $\mu$ is the viscosity of the stellar material and $\vec{g}$ is the gravitational acceleration.
    Since I have assumed that the stellar viscosity is negligible, we can obtain 
    \begin{equation} \label{eq:momentum} 
        \rho
        \left(
            \frac{\partial}{\partial t}
            +
            \vec{\varv}
            \cdot
            \nabla 
        \right)
        \vec{\varv}
        =
        -
        \nabla P
        +
        \rho \vec g
    \end{equation}
    Notice that this equation at equilibrium is the familiar equation of hydrostatic support (\ref{eq:cons-mom-r}):
    \lr{\begin{equation}
        0 = -\nabla P_0 + \rho_0 \vec{g}_0.
    \end{equation}}
    Substituting the perturbations into Equation~(\ref{eq:momentum}) and dropping all higher-order terms, we find the perturbed equation of motion:
    \lr{\begin{equation} \label{eq:perturbed-motion} \boxed{
        \rho_0\,
        \frac{\partial^2 \vec\xi}{\partial t^2}
        =
        -\nabla P'
        -
        \rho_0 \nabla \Phi'
        -
        \rho' \nabla \Phi_0
    }\,.\end{equation}
    Here I have introduced the gravitational potential $\Phi$, the negative gradient of which is the gravitational acceleration: 
    \begin{equation} \label{eq:grav-pot}
        \vec g 
        =
        - \nabla \Phi
        \qquad
        \text{and}
        \qquad
        \Phi(\vec r, t)
        =
        -G
        \int_{V}
            \frac{\rho}{|\vec r - \vec x|}
        \;\text{d}^3 \vec{x}
    \end{equation}
    where $V$ is the volume of the star at equilibrium.} 
    
    \item[Poisson's equation.] 
    Gauss's law for gravity gives that
    \begin{equation}
        \nabla \cdot \vec g
        =
        -4\pi G \rho.
    \end{equation}
    After substituting the gravitational potential and the Eulerian perturbations, we obtain the perturbed Poisson equation to describe the gravitational field: 
    \begin{equation} \label{eq:perturbed-poisson} \boxed{
        \nabla^2 \Phi'
        =
        4\pi G \rho'
    }\,.\end{equation}
    
    \item[The energy equation.]
    The energy equation completes the system by thermodynamically connecting pressure to density. 
    Since I have assumed adiabatic pulsations, the energy equation can be given as
    \begin{equation} 
        \frac{\partial P}{\partial t}
        +
        \vec{\varv}
        \cdot
        \nabla P
        =
        c^2 
        \left(
            \frac{\partial \rho}{\partial t}
            +
            \vec{\varv} \cdot \nabla \rho
        \right)
    \end{equation}
    where $c$ is again the adiabatic speed of sound (\emph{cf.}~Equation~\ref{eq:speed-of-sound}). 
    Substituting the Lagrangian perturbation, we obtain the perturbed energy equation
    \begin{equation} \label{eq:perturbed-energy} \boxed{
        P'
        +
        \vec \xi \cdot \nabla P_0
        =
        c^2_0 \left( 
            \rho' + \vec \xi \cdot \nabla \rho_0
        \right)
    }\,.\end{equation}
\end{description}

\subsubsection*{Symmetry}
\begin{figure}
    \centering
    \includegraphics[width=\textwidth]{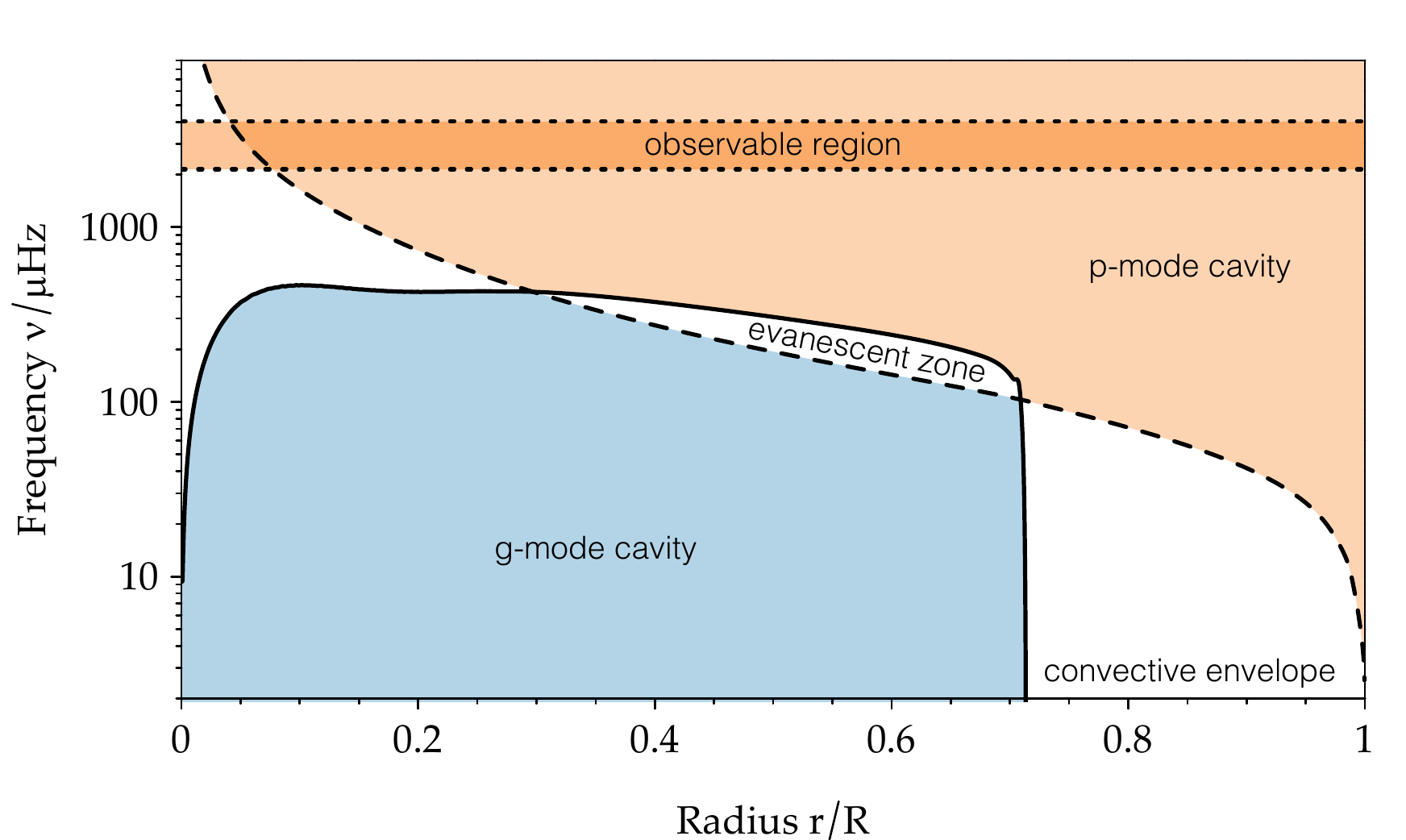}
    \caption[Propagation diagram]{Propagation diagram for a solar model. 
    The blue-shaded area shows the Brunt-V\"ais\"al\"a region where g-modes can propagate (\emph{cf.}~Equation~\ref{eq:brunt-vaisala}). 
    The orange-shaded area shows the ${\ell=1}$ Lamb region where dipolar p-modes can propagate (\emph{cf.}~Equation~\ref{eq:lamb}). 
    Modes are exponentially damped in the evanescent zone; nevertheless, modes of similar frequency can couple in this region, giving rise to mixed modes. 
    The observable region is a few ${\Delta\nu}$ around $\nu_{\max}$; thus, only p-modes are expected to be observed in this range at this stage of evolution. 
    \label{fig:propagation}}
\end{figure}

Now I will apply the assumption of symmetry and consider only oscillatory solutions on a sphere. 
\lr{I separate the displacement vector into radial and horizontal components 
\begin{equation}
    \vec\xi = \xi_r \hat a_r + \vec \xi_h, 
    \qquad 
    \vec\xi_h = \xi_\theta \hat a_\theta + \xi_\phi \hat a_\phi
\end{equation}
where $\hat a$ are unit vectors in indicated directions. 
The radial component of the displacement, for example, can now be expressed as
\begin{align}
    \xi_r(r, \theta, \phi, t)
    &=
    \xi_r(r)
    Y_{\ell}(\theta, \phi)
    \exp \{
        -i\omega t
    \} 
\end{align}
where $\theta$ and $\phi$ are latitude and longitude, $Y_{\ell}$ is Laplace's spherical harmonic for degree $\ell$ (\emph{cf.}~Figure~\ref{fig:sph}), $i$ is the imaginary unit, and ${\omega=2\pi\nu}$ is the cyclic frequency.}
When $\omega^2$ is real, the solution is oscillatory; when it is imaginary, the solution either grows or delays. 
Substituting the spherical, symmetric, harmonic variables into the previous equations (\ref{eq:perturbed-continuity}, \ref{eq:perturbed-motion}, \ref{eq:perturbed-poisson}, \ref{eq:perturbed-energy}) and dropping subscripts for unperturbed quantities, after some manipulations we may find 
\begin{gather} \label{eq:oscillation1}
    \frac{\text{d}\xi_r}{\text{d}r}
    =
    -\left(
        \frac{2}{r}
        +
        \frac{1}{\Gamma_1 P}
        \frac{\text{d}P}{\text{d}r}
    \right)
    \xi_r
    +
    \frac{1}{\rho c^2}
    \left(
        \frac{S_\ell^2}{\omega^2}
        -
        1
    \right)
    P'
    -
    \frac{\ell(\ell+1)}{\omega^2 r^2}
    \Phi' \vphantom{\Bigg(}
    \\
    \frac{\text{d}P'}{\text{d}r}
    =
    \rho \left(
        \omega^2 - N^2
    \right) 
    \xi_r
    +
    \frac{1}{\Gamma_1 P}
    \frac{\text{d}P}{\text{d}r}
    P'
    +
    \rho \frac{\text{d} \Phi'}{\text{d} r}
    \vphantom{\Bigg(}
    \\
    \frac{1}{r^2}
    \frac{\text{d}}{\text{d}r}
    \left(
        r^2\,
        \frac{\text{d}\Phi'}{\text{d}r}
    \right)
    =
    -4\pi G \left(
        \frac{P'}{c^2}
        +
        \frac{\rho}{g}
        \xi_r
        N^2
    \right)
    +
    \frac{\ell(\ell+1)}{r^2}
    \Phi' \vphantom{\Bigg(}
    \label{eq:oscillation3}
\end{gather}%
\lr{as well as 
\begin{gather}
    \vec\xi_h(r, \theta, \phi, t) 
    =
    \sqrt{4\pi}\,
    \xi_h(r)
    \left(
        \frac{\partial Y_\ell}{\partial \theta}\,
        \hat{a}_\theta
        +
        \frac{1}{\sin\theta}
        \frac{\partial Y_\ell}{\partial \phi}\,
        \hat{a}_\phi
    \right)
    \exp \{
        -i\omega t
    \} \vphantom{\Bigg(}
    \\
    \xi_h(r)
    =
    \frac{1}{r\omega^2} \left( 
        \frac{1}{\rho}\, P' - \Phi'
    \right). \vphantom{\Bigg(}
\end{gather}}
Here I have introduced the \emph{Brunt-V\"ais\"al\"a} and \emph{Lamb} \lr{squared} frequencies: 
\begin{gather} \label{eq:brunt-vaisala}
    N^2
    =
    g
    \left(
        \frac{1}{\Gamma_1}
        \frac{\text{d} \ln P}{\text{d}r}
        -
        \frac{\text{d} \ln \rho}{\text{d}r}
    \right) 
    \\ \label{eq:lamb}
    S^2_\ell
    =
    \frac{\ell(\ell+1)c^2}{r^2}
\end{gather}
which give the regions in the star where modes of different character can propagate. 
The former, $N^2$, describes where g-modes can propagate, so called because their restoring force is gravity. 
The latter, $S_\ell^2$, depending on the spherical degree $\ell$, describes where p-modes can propagate, called as such because their restoring force is the pressure gradient. 
These cavities are visualized in Figure~\ref{fig:propagation}. 
Here it can be appreciated that g-modes and convection are two sides of the same coin: when ${N^2<0}$ the fluid is unstable to convection; otherwise, the fluid is unstable to g-mode oscillations. 

\begin{figure}
    \centering
    \includegraphics[width=0.5\textwidth]{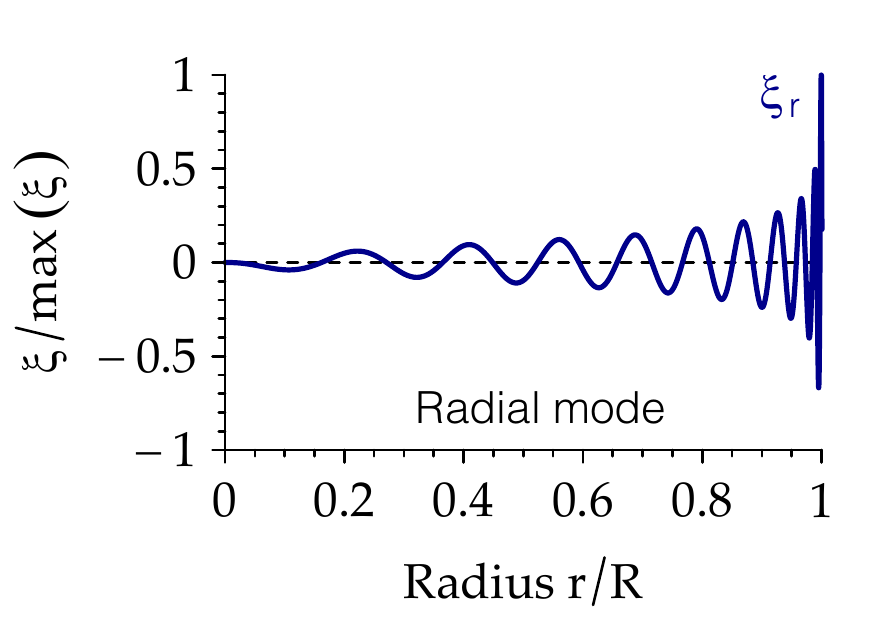}%
    \makebox[0.5\textwidth][c]{%
        \adjustbox{trim={1.7cm 0cm 0cm 0cm},clip}{\includegraphics[width=0.5\textwidth]{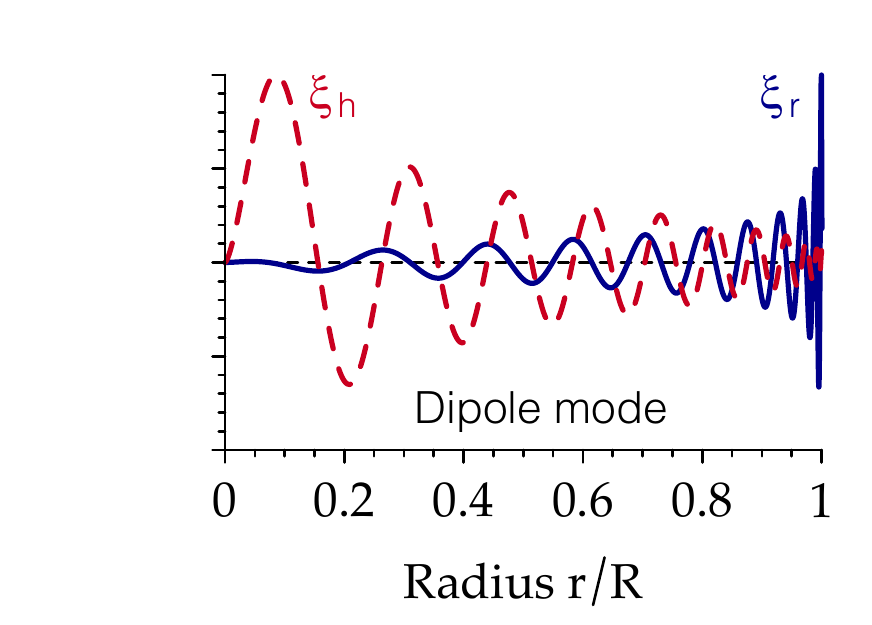}}}
    \caption[Eigenfunctions]{Radial (blue) and horizontal (red) normalized eigenfunctions for radial (${\ell=0}$, left) and dipolar (${\ell=1}$, right) oscillation modes, both having radial order ${n=20}$. 
    The radial displacement of the two modes are quite similar, being only slightly offset in the interior and basically identical in the envelope. 
    The horizontal displacement has zero crossings when the radial displacement is maximal, and vice versa. 
    Radial modes lack horizontal displacement by definition. 
    \label{fig:eigenfunctions}}
\end{figure}

This system of equations (\ref{eq:oscillation1}--\ref{eq:oscillation3}) constitutes a fourth order boundary eigenvalue problem. 
Equipped with suitable boundary conditions, we may numerically calculate the eigenfunctions $\vec\xi$ (see Figure~\ref{fig:eigenfunctions}) and their corresponding eigenfrequencies $\omega$ for a given model of stellar structure. 
This is the forward problem of stellar pulsation.

\subsubsection*{Some Properties of Solar-like Oscillations}

As we have seen in the first section, oscillation modes of the same spherical degree $\ell$ can differ in their radial order $n$ and be excited simultaneously with different frequencies. 
For solar-type stars, it is currently possible to resolve frequencies for modes of low spherical degree (${0~\leq~\ell~\leq~3}$) and `high' radial order (${8~\leq~n~\leq~31}$). 
The frequency range where oscillation power is maximum, called by $\nu_{\max}$, generally corresponds to around ${n=20}$ or so. 
This region of power is proportional to (and obviously lower than) the acoustic cut-off frequency, i.e., the upper frequency bound for oscillations to be reflected back into the star rather than being lost to space: 
\begin{equation} \label{eq:numax}
    \nu_{\max} \propto \nu_{\text{ac}} \propto \frac{g}{\sqrt{T_{\text{eff}}}}.  
\end{equation}
For the Sun, ${\nu_{\max,\odot} \simeq 3090\;\mu\text{Hz}}$ (${\sim 5.4}$~minutes) and ${\nu_{\text{ac},\odot} \simeq 5000\;\mu\text{Hz}}$ (${\sim 3.3}$~minutes). 
Since we lack a proper theoretical treatment of convective transport, which both excites and damps the oscillation modes, we are unable to theoretically predict the amplitudes of the oscillations of our solar model. 
In lieu of this, we may try to predict the general region where oscillations with the greatest amplitudes are to be expected by scaling from the observed solar values \citep[e.g.,][]{1995A&A...293...87K}:
\begin{equation}
    \frac{\nu_{\max,\ast}}{\nu_{\max,\odot}}
    =
    \left(
        \frac{M_\ast}{M_\odot}
    \right)
    \left(
        \frac{R_\ast}{R_\odot}
    \right)^{-2}
    \left(
        \frac{T_{\text{eff},\ast}}{T_{\text{eff},\odot}}
    \right)^{-\frac{1}{2}}
\end{equation}
and likewise for the acoustic cutoff frequency. 

\citet{1980ApJS...43..469T} considered oscillation modes in the asymptotic limit of high radial order (${n\gg\ell}$) and found that theoretical mode frequencies form a pattern. 
In particular, adjacent modes of the same spherical degree are approximately equally spaced, which agrees with the observations that we saw in Figures~\ref{fig:solar-power-spectrum} and \ref{fig:16cygb}. 
The pattern of frequencies can be summarized to first-order approximation as 
\begin{equation} \label{eq:asymptotic}
    \nu_{n,\ell} \simeq \Delta\nu \left( n + \frac{\ell}{2} + \epsilon \right)
\end{equation}
where $\nu_{n,\ell}$ is the frequency of mode (${n,\ell}$) and $\epsilon$ is a phase shift (${\epsilon_\odot\simeq 1.6}$). 
The spacing ${\Delta\nu}$ is called the \emph{large frequency separation} and is related to the inverse sound travel time and proportional to the \lr{root mean density} of the star \citep{1986apj...306l..37u, 1995A&A...293...87K}:
\begin{equation}
    \Delta\nu
    \simeq
    \left( 
        2 \int \frac{\text{d}r}{c}
    \right)^{-1}
    \propto
    \left(
        \frac{M}{R^3}
    \right)^{1/2}. 
\end{equation}
Since the large frequency separation gives the spacing between modes of different orders, it can be calculated empirically with
\begin{equation} \label{eq:Dnu}
    \Delta\nu_{n,\ell} 
    =
    \nu_{n,\ell}
    -
    \nu_{n-1,\ell}.
\end{equation}
Calculating the average large frequency separation of the Sun for radial modes using data from the Birmingham Solar Oscillations Network \citep[\emph{BiSON},][]{2009mnras.396l.100b} we can obtain 
\begin{equation}
    \Delta\nu_\odot = 134.8693 \pm 0.0042\;\mu\text{Hz}. 
\end{equation}
This presents an opportunity to test the quality of our solar model. 
We can calculate the large frequency for our solar-calibrated model either using the inverse sound travel time, or using the frequencies themselves. 
In the former case, we obtain ${\Delta\nu = 136.2970\;\mu\text{Hz}}$. 
In the latter, ${\Delta\nu = 136.2208\;\mu\text{Hz}}$. 

On the one hand, these model values differ by only about one percent from the solar values, which is quite good by astrophysical standards. 
On the other hand, when considering the precision with which ${\Delta\nu_{\odot}}$ can be calculated, this is a highly significant ${\sim 300\sigma}$ difference. 
This difference arises due to our ill treatment of the stellar surface, which we will address later in this section. 

A higher-order expansion of the asymptotic expression additionally gives a term known as the \emph{small frequency separation}, the spacing between modes adjacent in frequency and whose spherical degree differs by two \lr{\citep{1980ApJS...43..469T}}: 
\begin{equation} \label{eq:dnu}
    \delta\nu_{n,\ell}
    =
    \nu_{n,\ell}
    -
    \nu_{n-1,\ell+2}
    \simeq
    -(4\ell + 6)
    \frac{\Delta\nu}{4\pi^2 \nu_{n,\ell}}
    \int
        \frac{\text{d}c}{\text{d}r}
        \frac{\text{d}r}{r}.
\end{equation}
As we can see, the small frequency separation is sensitive to the sound speed gradient, and is therefore a good proxy for the conditions in the stellar core, where the sound speed gradient changes sign (\emph{cf.}~Figure~\ref{fig:profs}). 
This makes ${\delta\nu}$ a diagnostic of main-sequence age. 
We will make use of these relations to infer the properties of stars in Chapter~\ref{chap:ML}, and use computational methods to further understand what properties of stars they reflect in Chapter~\ref{chap:statistical}. 
The average small frequency separation between solar oscillation modes with (${\ell=0},\;{\ell=2}$) is
\begin{equation}
    \delta\nu_\odot \simeq 8.957 \pm 0.059\;\mu\text{Hz}
\end{equation}
and for our solar model, ${\delta\nu=8.939\;\mu\text{Hz}}$, which is good agreement. 

\subsubsection*{A Direct Comparison} 
\begin{figure}[t]
    \centering
    \includegraphics[width=\textwidth,keepaspectratio,trim={0cm 0cm 0cm 0.1cm}, clip]{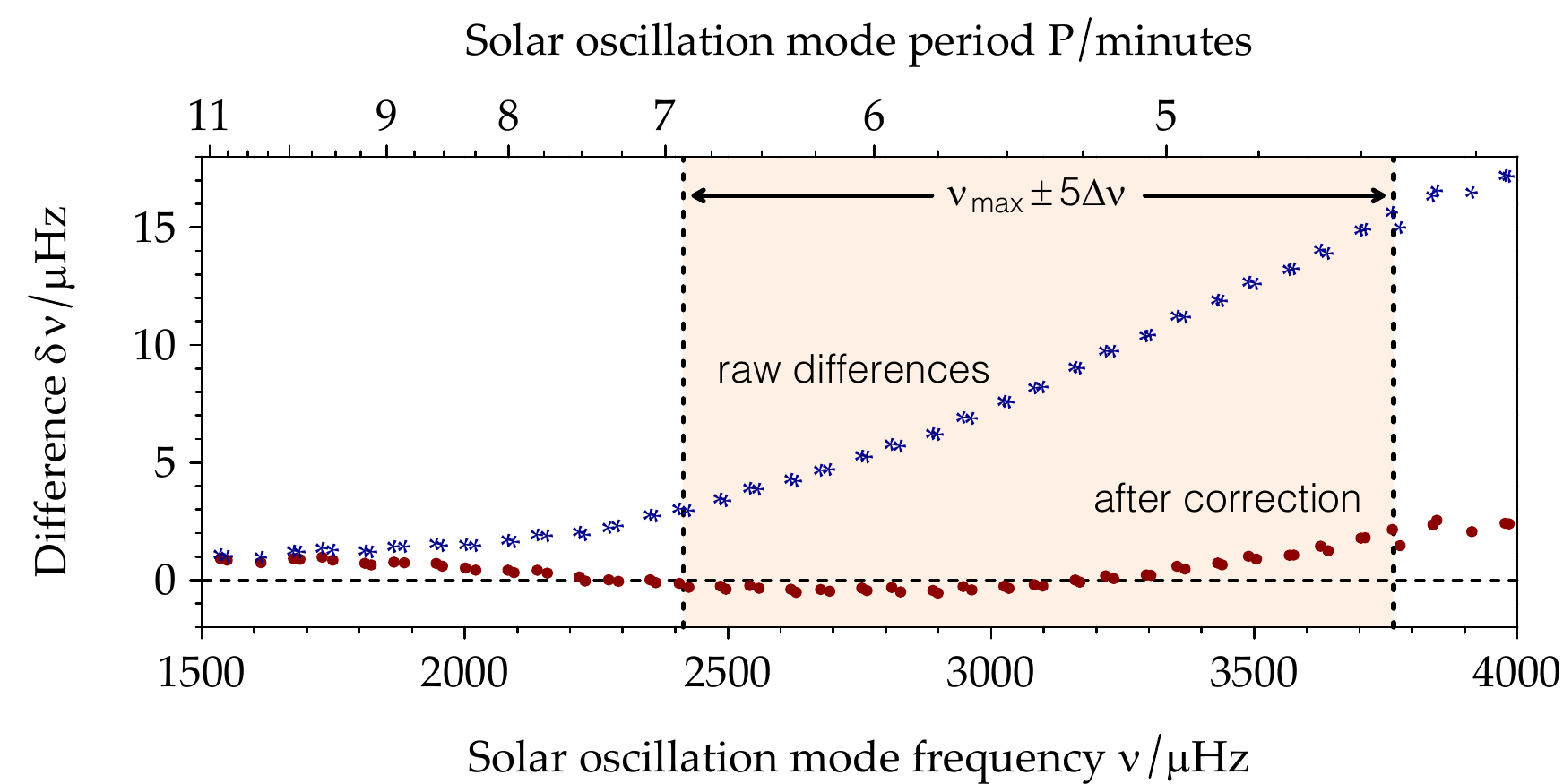}
    \caption[The solar surface effect]{Differences in oscillation frequencies between the Sun and the best-fitting solar model, in the sense of (model $-$ Sun). 
    Even after correcting for the surface term, substantial differences remain. 
    Being that solar frequencies are measured on the order of one part in a thousand, the uncertainties are too small to be visible at this resolution. 
    The offset at zero is likely due to the assumed solar radius differing from the helioseismic radius. 
    The shaded region indicates what the frequency range of the Sun might be if it were a field star observed by \emph{Kepler}. 
    \label{fig:solar_freq_diffs}} 
\end{figure}

We have just compared our solar model against the asymptotic properties of the solar oscillations, finding good agreement with the small frequency separation but less good agreement with the large frequency separation. 
We may now test the quality of our solar model more directly by comparing the individual pulsation mode frequencies themselves to those observed in the Sun. 
This comparison is shown in Figure~\ref{fig:solar_freq_diffs}. 

Immediately it can be seen that there are systematic discrepancies between the model and the actual mode frequencies on the order of ${10\;\mu\text{Hz}}$, i.e., tenths of a percent, which is a difference in period of about $1$ to $2$ seconds. 
In particular, the disagreement gets worse with increasing frequency. 
This phenomenon is called the \emph{surface effect} and has arisen from our improper modelling of the near-surface layers \citep[e.g.,][]{1984srps.conf...11C}. 
The large frequency separation is also sensitive to surface effects, which is why our model ${\Delta\nu}$ differed so significantly from the observed value. 

It is noteworthy that, because all of the waves propagate essentially radially in the near-surface layers (\emph{cf.}~Figures~\ref{fig:rays} and \ref{fig:eigenfunctions}), the surface term is a function of frequency alone and is independent of the spherical degrees of the modes. 
The surface effect is thus often dealt with by introducing a correction that increases with frequency. 
The \citet{2014A&A...568A.123B} treatment of the surface term fits coefficients $\mathbf a$ to the differences between observed and model frequencies according to
\begin{equation} \label{eq:BallGizon-surfterm}
    \delta \nu_{n,\ell} 
    = 
    \frac{1}{I_{n,\ell}} \left[ 
        a_1 \left( 
            \frac{\nu_{n,\ell}}{\nu_{ac}} 
        \right)^{-1} 
        + 
        a_2 \left( 
            \frac{\nu_{n,\ell}}{\nu_{ac}} 
        \right)^{3} 
    \right] 
\end{equation}
where $\nu_{ac}$ is the acoustic cutoff frequency, with ${\nu_{ac,\odot} \approx 5000}$, and $I_{n,\ell}$ is the normalized mode inertia:
\begin{equation} \label{eq:normalized-mode-inertia}
    I_{n,\ell}
    =
    \frac{4\pi}{M}
    \frac{\int \rho
            \left(
                |\xi_r|^2
                +
                \Ltwo |\xi_h|^2
            \right) 
            r^2
        \;\text{d}r}{
            |\xi_r(r=R)|^2
            +
            \Ltwo
            |\xi_h(r=R)|^2
        }.
\end{equation}
However, Figure~\ref{fig:solar_freq_diffs} further shows that even after correcting for the surface term, differences remain. 
This implies that even beyond the near-surface layers, the structure of the Sun differs from the model. 

This motivates the inverse approach. 
We have seen that evolutionary theory can produce a model that agrees with the overall properties of the Sun. 
However, a detailed inspection of the mode frequencies of the model reveals significant disagreement between theory and observation, even after applying corrections. 
We wish to deduce the actual structure of the Sun and the stars using only asteroseismic arguments: i.e., to find the structure that will pulsate identically. 
This problem of deducing the structure of a star from its oscillation frequencies is inverse to the problem of deducing the oscillation frequencies from a given stellar structure. 
In order to pose the inverse problem in a manner that we can solve, however, it is convenient to first make some slight adjustments to our statement of the respective forward problem. 


\subsection{The Relative Forward Problem} \label{sec:variational}
The forward problem of asteroseismology is to calculate the seismic frequencies of a stellar model. 
However, it is not clear how one would go about solving the inverse problem corresponding to this forward problem. 
Instead, we restate the forward problem as the problem of calculating the frequency \emph{differences} with respect to another model---one with a different structure. 
That is: by comparing the differences in structure of two models, what will be the differences in their frequencies? 
I call this the relative forward problem of asteroseismology. 

The benefit of posing the problem in this way is that it facilitates the inverse problem, which is to ask: by comparing the frequencies of the two models, what is the difference in their structure? 
Thus, since we are able to observe frequencies of real stars, we may substitute a star for one of the models, and hence measure the structure of a star. 

\begin{figure}
    \centering
    \begin{subfigure}[b]{0.5\linewidth}
        \centering
        \includegraphics[width=\textwidth,keepaspectratio]{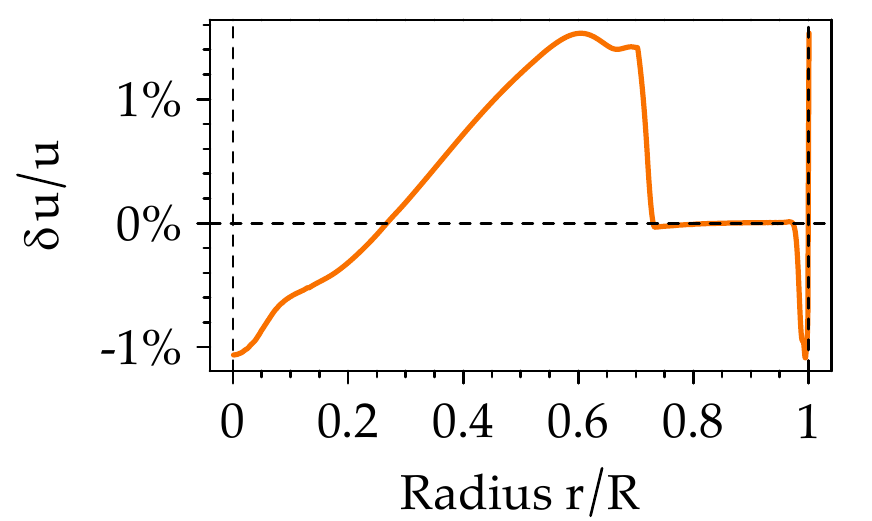}
    \end{subfigure}%
    \begin{subfigure}[b]{0.5\linewidth}
        \centering
        \includegraphics[width=\textwidth,keepaspectratio]{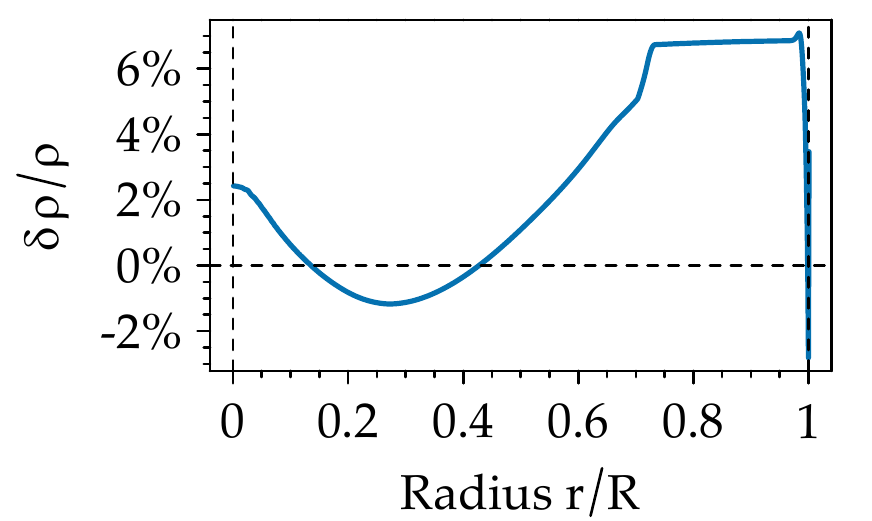}
    \end{subfigure}\\
    \begin{subfigure}[b]{0.5\linewidth}
        \centering
        \includegraphics[width=\textwidth,keepaspectratio]{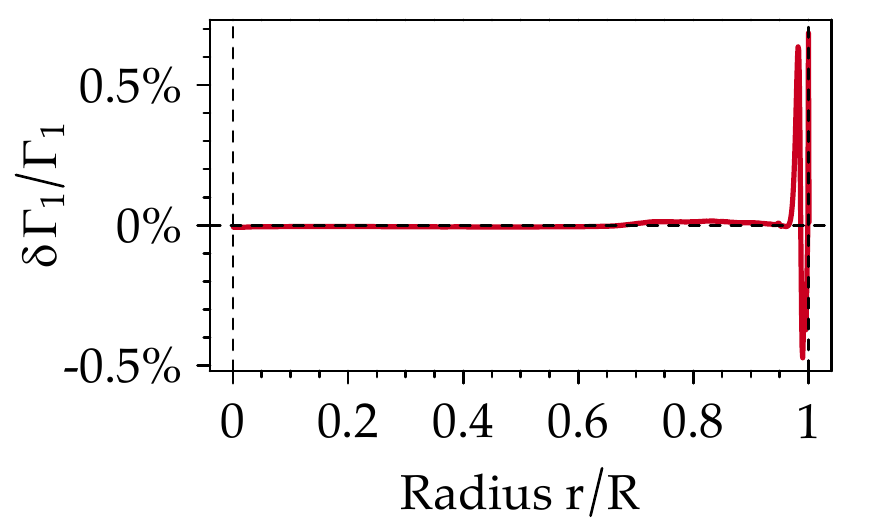}
    \end{subfigure}%
    \begin{subfigure}[b]{0.5\linewidth}
        \centering
        \includegraphics[width=\textwidth,keepaspectratio]{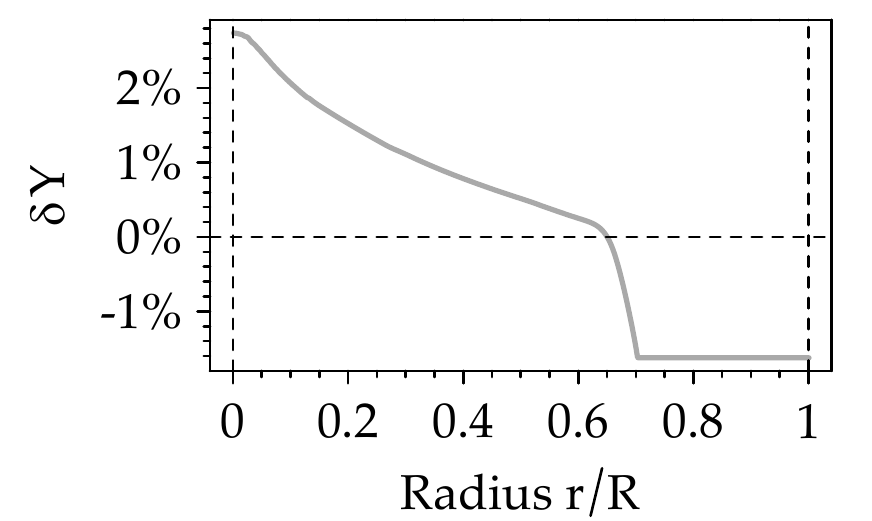}
    \end{subfigure}
    \caption[Structural differences between two solar models]{Relative differences in isothermal sound speed (top left), density (top right), the first adiabatic exponent (bottom left), and helium abundance (bottom right) as a function of radius between two solar-calibrated models with differing input physics (\emph{cf}.~Figure~\ref{fig:profs}). 
    Although the models have the same overall properties (e.g.\ mass \& age); they differ structurally and chemically throughout their interiors. 
    } 
    \label{fig:prof_diffs} 
\end{figure}

To give a concrete example, I have calibrated another solar model using different assumptions on the physics of the stellar interior. 
In particular, this second model differs in that it does not include the effects of elemental diffusion and gravitational settling (i.e., $\mathbf{D}$ is the null matrix in Equation~\ref{eq:evol-diffusion}). 
This model has the same mass, radius, luminosity, metallicity, and age as the diffusion model---yet it differs in internal structure (see Figure~\ref{fig:prof_diffs}). 
The differences in internal structure then give rise to differences in oscillation mode frequencies. 

In order to state the relative forward problem, I will first put the oscillation equations in their so-called \emph{variational formulation}, and then 
linearize the variational frequencies around a reference model. 
The end result will be a Fredholm integral equation relating the relative differences in oscillation mode frequencies to the relative differences in structure, which will then be a suitable starting point for the inverse analysis.

\subsubsection*{Variational Frequencies}
The perturbed hydrodynamical equations  (\ref{eq:oscillation1}--\ref{eq:oscillation3}) feature derivatives of the displacement vector. 
Since we have sought only periodic solutions, we have
\lr{\begin{equation}
    \vec\xi(t)
    =
    \vec\xi\cdot\exp\{i\omega t\}
    \qquad \Rightarrow \qquad
    \frac{\partial\vec\xi}{\partial t}
    =
    -i\omega\vec\xi.
\end{equation}}
Combining the perturbed equations, we can arrive at \citep[e.g.,][]{1979nos..book.....U}
\lr{\begin{equation} \label{eq:first-omega}
    -\omega^2 \rho \vec\xi
    =
    \nabla \left(
        c^2 \rho \nabla \cdot \vec \xi
        +
        \nabla P \cdot \vec \xi
    \right)
    -
    \vec g \,
        \nabla \cdot \left(
            \rho \vec \xi 
        \right)
    +
    \rho \vec g' 
\end{equation}}
where I have dropped the subscripts on the unperturbed quantities. 
This equation relates the cyclic frequency $\omega$ to the properties of the stellar structure. 
Recalling Equation~(\ref{eq:grav-pot}), we can substitute the perturbed gravitational potential with 
\lr{\begin{equation}
    \vec g'
    =
    -\nabla \Phi'
    =
    G \nabla
    \int_V
        \frac{\rho'}{|\vec r - \vec x|}
    \;\text{d}^3 \vec{x}
    =
    -G \nabla \int_V
        \frac{\nabla
            \cdot \left(
                \rho
                \vec\xi
            \right)}{|\vec r - \vec x|}
    \;\text{d}^3 \vec{x}.
\end{equation}}
where the latter substitution makes use of the perturbed equation of continuity (Equation~\ref{eq:perturbed-continuity}). 
%
Thus, all terms in the right hand side of Equation~(\ref{eq:first-omega}) are functions of $\vec\xi$, and so it is an eigenvalue problem of the form
\lr{\begin{equation} \label{eq:operator}
    \mathcal{L}(\vec\xi_i)
    =
    -\omega^2_i
    \vec\xi_i
\end{equation}}
with $\mathcal{L}$ being the linear integro-differential operator satisfying that equation. 
Now ${\vec\xi~\equiv~\vec\xi_i}$ is the displacement eigenfunction for the mode with label ${i\equiv(n,\ell)}$ and ${\omega~\equiv~\omega_i}$ is its corresponding eigenfrequency. 
\citet{1964ApJ...139..664C} 
showed that when $\rho=P=0$ at the outer boundary, this eigenvalue problem is Hermitian, \lr{i.e.,
\begin{equation} \label{eq:hermitian}
    \langle \vec \xi, \mathcal{L}(\vec \eta) \rangle
    =
    \langle \mathcal{L}(\vec \xi), \vec \eta \rangle
\end{equation}
where ${\langle\cdot\rangle}$ denotes the inner product defined by
\lr{\begin{equation} \label{eq:inner-prod}
    \langle
        \vec\xi_i, 
        \vec\eta_i 
    \rangle
    =
    \int_V \rho 
        \vec\xi_i^\ast 
        \cdot 
        \vec \eta_i 
    \; \text{d}^3\vec r
    =
    4\pi
    \int \rho \left(
        \xi_r^\ast \eta_r
        +
        \Ltwo
        \xi_h^\ast \eta_h
    \right)
    r^2
    \;\text{d}r. 
\end{equation}}
Here $^\ast$ is the complex conjugate and $\vec\eta$ is any (suitably regular) vector function of stellar structure.
This is useful because then squared mode frequencies are real and may be calculated via
\lr{\begin{equation} \label{eq:var-freqs}
    -\omega^2_i 
    =
    \frac{
        \langle
            \vec\xi_i, 
            \mathcal{L}(\vec \xi_i) 
        \rangle
    }{
        \langle
            \vec\xi_i, 
            \vec\xi_i 
        \rangle
    }
\end{equation}}}
where $\vec\xi_i$ 
is an eigenvector of the problem and $\omega^2_i$ 
is a real eigenvalue. 
A further property is that the eigenvectors of the problem are orthogonal. 
Finally, we have the variational principle: perturbations to an eigenvector result in only second-order perturbations to the corresponding eigenvalue. 
Frequencies calculated using Equations~(\ref{eq:var-freqs}) are referred to as variational frequencies. 

\subsubsection*{Linearization Around a Reference Model}
We now seek to linearize the problem around a reference model. 
We consider a small perturbation to the eigenfrequency, call it ${\delta\omega^2}$, to the eigenfunction, ${\delta\vec\xi}$, and to the operator, ${\delta\mathcal{L}}$:
\lr{\begin{equation} \label{eq:perturbed-operator}
    \Big(
        \mathcal{L} + \delta\mathcal{L}
    \Big)
    \Big(
        \vec\xi
        +
        \delta\vec\xi
    \Big)
    =
    -\Big(
        \omega
        +
        \delta\omega
    \Big)^2
    \Big(
        \vec\xi
        +
        \delta\vec\xi
    \Big).
\end{equation}
After perturbing all the components from Equation~(\ref{eq:first-omega}), we can find \citep[e.g.,][]{1994a&as..107..421a}
\begin{align} \label{eq:deltaL}
\begin{split}
    \delta\mathcal{L}(\vec\xi)
    ={}
    &\frac{\nabla\rho}{\rho} \delta c^2 \nabla\cdot \vec\xi
    +
    \nabla\left(
        \delta c^2 \nabla \cdot \vec\xi
        +
        \delta\vec g
        \cdot
        \vec\xi
    \right)
    +
    \delta\vec g \,\nabla\cdot\vec\xi
    \\
    &+
    \nabla\left(
        \frac{\delta\rho}{\rho}
    \right)
    c^2\nabla\cdot\vec\xi
    -
    G\nabla\int_V
        \frac{\nabla\cdot \left( \delta\rho\vec\xi \right)}{|\vec r - \vec x|}
    \;\text{d}^3 \vec x.
\end{split}
\end{align} 
Expanding Equation~(\ref{eq:perturbed-operator}), we find at the first order
\begin{equation}
    \mathcal{L}(\delta\vec\xi)
    +
    \delta\mathcal{L}(\vec\xi)
    =
    -\omega^2\delta\vec\xi
    -2\omega\delta\omega\vec\xi.
\end{equation}
Taking the product of both sides with $(\rho\vec\xi^\ast)$ 
and integrating, we obtain
\begin{align} \label{eq:expansion}
\begin{split}
    &\int_V \rho \vec\xi^\ast \cdot  \mathcal{L}(\delta\vec\xi) \; \text{d}^3\vec r
    +
    \int_V \rho \vec\xi^\ast \cdot  \delta\mathcal{L}(\vec\xi) \; \text{d}^3\vec r
    \\= -\omega^2&\int_V \rho \vec\xi^\ast \cdot  \delta\vec\xi \; \text{d}^3\vec r
    -2\omega\delta\omega\int_V \rho \vec\xi^\ast \cdot  \vec\xi \; \text{d}^3\vec r.
\end{split}
\end{align}
Since $\mathcal{L}$ is Hermitian, the first term on both sides cancel to give
\begin{equation} \label{eq:rel-variational}
    \delta\omega
    =
    -\frac{1}{2\omega}\frac{\langle \vec\xi, \delta \mathcal{L}(\vec\xi) \rangle}{\langle \vec\xi, \vec\xi \rangle}.
\end{equation}
Now plugging $\delta\mathcal{L}$ from Equation~(\ref{eq:deltaL}) into Equation~(\ref{eq:rel-variational}) and assuming that $\delta P=0$ at the outer boundary \citep[e.g.,][]{1967MNRAS.136..293L}, one may use integration by parts to obtain, quite generally, a Fredholm integral relation for each mode of oscillation $i$:
\begin{equation} \label{eq:forward} \boxed{
  \frac{\delta\omega_i}{\omega_i} 
  = 
  \int K_i^{(f_1, f_2)} \frac{\delta f_1}{f_1}
                + K_i^{(f_2, f_1)} \frac{\delta f_2}{f_2}
       \;\text{d}r
}\,. \end{equation}}
Here $f_1$ and $f_2$ are two variables of stellar structure (e.g., sound speed and density), and
${\delta f_1}$ and ${\delta f_2}$ are the differences with respect to another model. 
Relative differences in the frequencies ${\delta\omega_i/\omega_i}$ of mode ${i~\equiv~(n,\ell)}$ between two models relate to relative differences in physical quantities of those models via a pair of \emph{kernel functions} $\vec{K}_i$. 

Equation~(\ref{eq:forward}) is the central equation of this thesis, as this is the equation that we will use to infer the internal structures of stars. 
In particular, we will determine the stellar structure profile $f_1$ of a star (for some choice of $\vec{f}$, discussed later) by deducing the relative difference with a best-fitting evolutionary model ${\delta f_1/f_1}$ via inversion of this equation. 
This is the structure inversion problem, which we will revisit in Section~\ref{sec:inverse} and Chapter~\ref{chap:inversion}. 
For now, we will continue by inspecting the kernel functions in detail. 
%

\subsection{Stellar Structure Kernels}
\label{sec:kernels}
We have seen in Equation~(\ref{eq:forward}) that perturbations to the stellar structure translate into perturbations in oscillation mode frequencies, and kernel functions quantify that response. 
The kernels for any given pair of stellar structure variables can be calculated by transforming Equation~(\ref{eq:rel-variational}) into an equation in the form of Equation~(\ref{eq:forward}). 
Because the variables of stellar structure are not independent, kernels must be given with respect to (at least) two variables simultaneously. 
Here I will give the kernels for the following pairs: ${(c,\rho)}$, ${(c^2,\rho)}$, ${(\Gamma_1,\rho)}$, and ${(u,Y)}$. 

\subsubsection*{Kernel Pair \texorpdfstring{$\mathbf{(c, \rho)}$}{(c,rho)}}
\noindent
The kernels for the sound speed and density, i.e.\ ${(f_1, f_2) = (c, \rho)}$ of Equation~(\ref{eq:forward}), can be found as \citep[\emph{cf.}][]{GoughThompson1991}
\lr{\begin{align}
    \omega^2 \mathcal{S} \Kcr ={} & r^2 \rho c^2 \chi^2 
\\  \omega^2 \mathcal{S} \Krc ={} & 
    - \half \left( 
        \xi_r^2 + L^2 \xi_h^2 
    \right) 
    r^2 \rho \omega^2 
    \\& + \frac{1}{2} \rho c^2 \chi^2 r^2 
    - G m \rho \left( 
        \chi 
        + 
        \half \xi_r \ddra{\ln \rho} 
    \right) \xi_r
    \notag\\& - 4\pi G \rho r^2 \int_{r}^R \left( 
        \chi 
        + 
        \half \xi_r \frac{\text{d} \ln \rho}{\text{d} s} 
    \right) \xi_r \rho \; \text{d}s
    \notag\\& + G m \rho \; \xi_r \ddra{\xi_r} 
    + \half G \left(
        m \ddra{\rho} 
        + 
        4\pi r^2 \rho^2 
    \right) \xi_r^2
    \notag\\& - \frac{4 \pi G}{2\ell + 1} \rho \Bigg[ 
        (\ell+1) r^{-\ell} \left(
            \xi_r 
            - 
            \ell \xi_h
        \right) \int_{0}^r \left(
            \rho \chi 
            + 
            \xi_r \ddsa{\rho} 
        \right) s^{\ell+2} \; \text{d}s 
        \notag\\&\hphantom{- \frac{4 \pi G}{2\ell + 1} \rho \Bigg[}
        - \ell r^{\ell+1} \left( 
            \xi_r 
            + 
            \left( 
                \ell+1
            \right) \xi_h 
        \right) \int_r^R \left( 
            \rho \chi 
            + 
            \xi_r \ddsa{\rho} 
        \right) s^{-(\ell-1)} \; \text{d}s 
    \Bigg] \notag
\end{align}}
where 
I have introduced the dilatation
\begin{equation}
    \chi = \ddra{\xi_r} + 2\frac{\xi_r}{r} - \Ltwo \frac{\xi_h}{r}
\end{equation}
and $\mathcal{S}$ is a quantity proportional to the energy of the mode
\begin{equation}
    \mathcal{S} = \int \rho \left( \xi_r^2 + \Ltwo \xi_h^2 \right) r^2 \; \text{d}r. 
\end{equation}

\subsubsection*{Kernel Pair \texorpdfstring{$\mathbf{(c^2, \rho)}$}{(c2,rho)}}
\begin{figure}
    \centering
    \includegraphics[width=0.5\textwidth,trim={0 1.1cm 0 0}, clip]{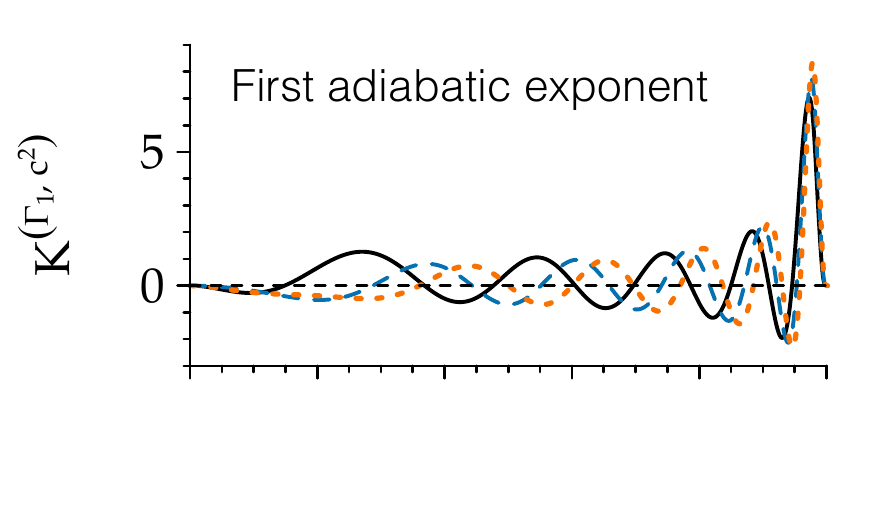}%
    \includegraphics[width=0.5\textwidth,trim={0 1.1cm 0 0}, clip]{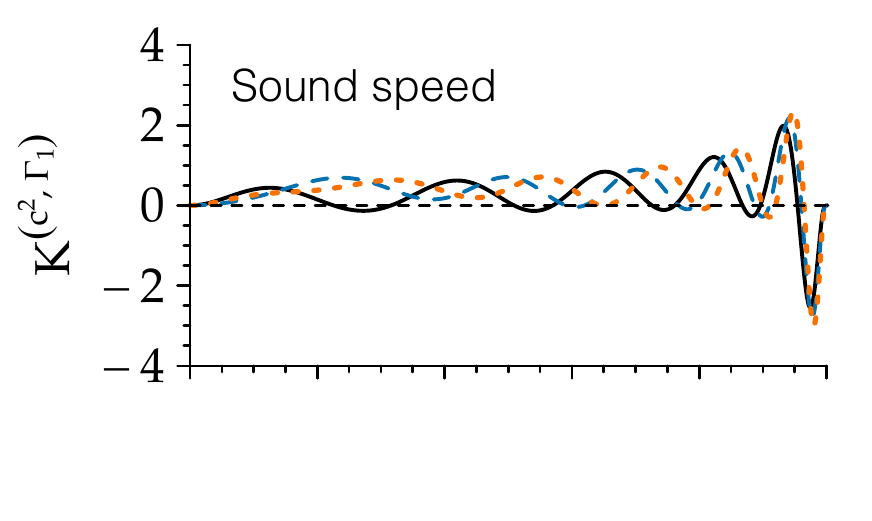}\\
    \includegraphics[width=0.5\textwidth,trim={0 1.1cm 0 0}, clip]{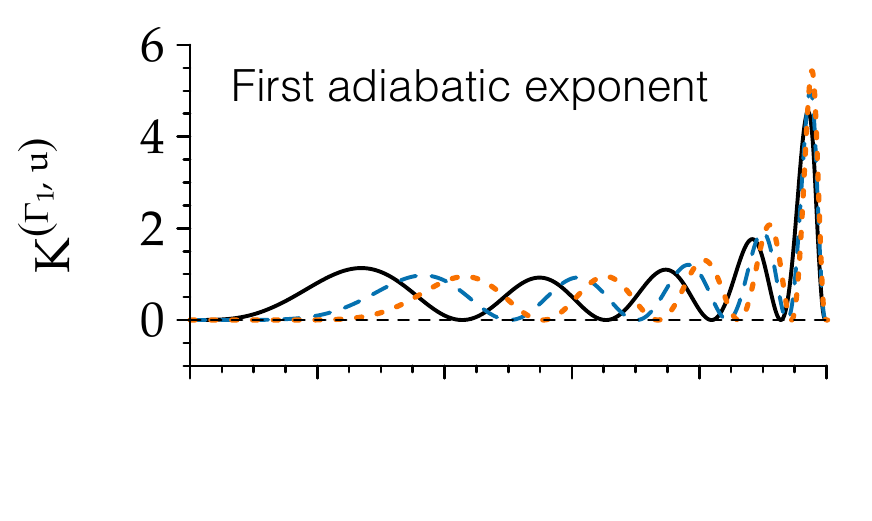}%
    \includegraphics[width=0.5\textwidth,trim={0 1.1cm 0 0}, clip]{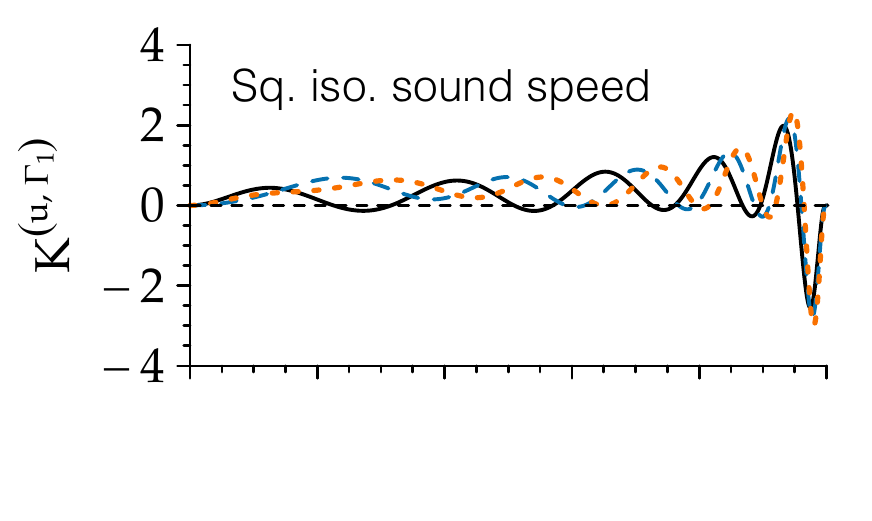}\\
    \includegraphics[width=0.5\textwidth,trim={0 1.1cm 0 0}, clip]{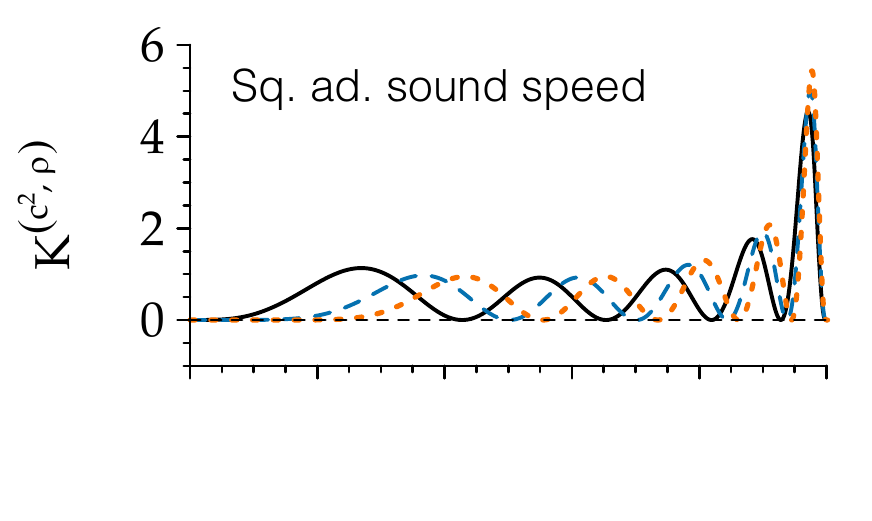}%
    \includegraphics[width=0.5\textwidth,trim={0 1.1cm 0 0}, clip]{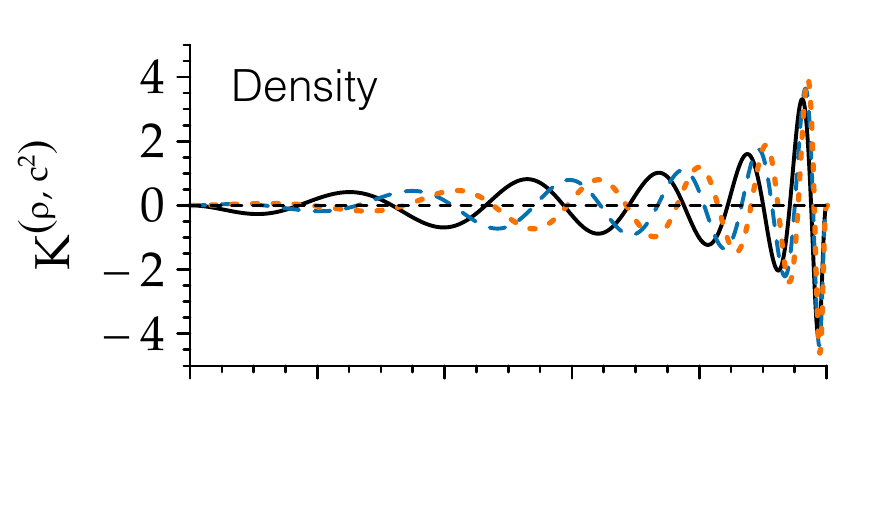}\\
    \includegraphics[width=0.5\textwidth,trim={0 1.1cm 0 0}, clip]{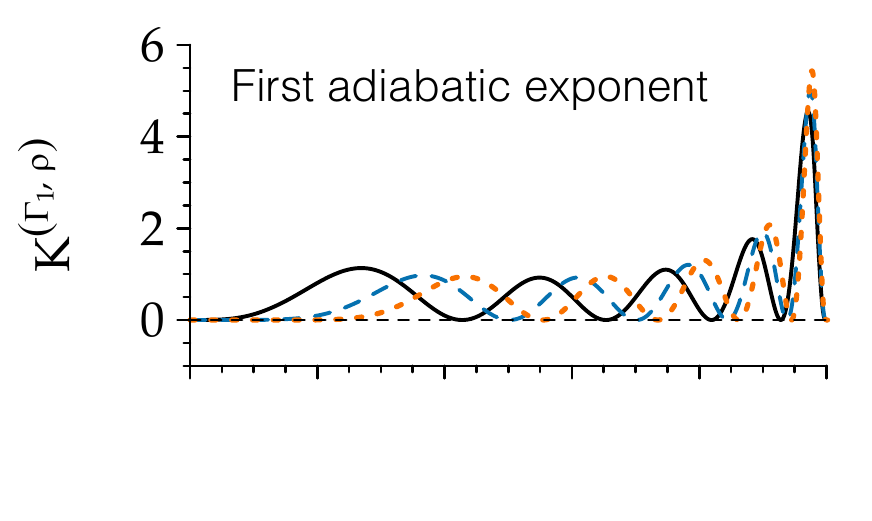}%
    \includegraphics[width=0.5\textwidth,trim={0 1.1cm 0 0}, clip]{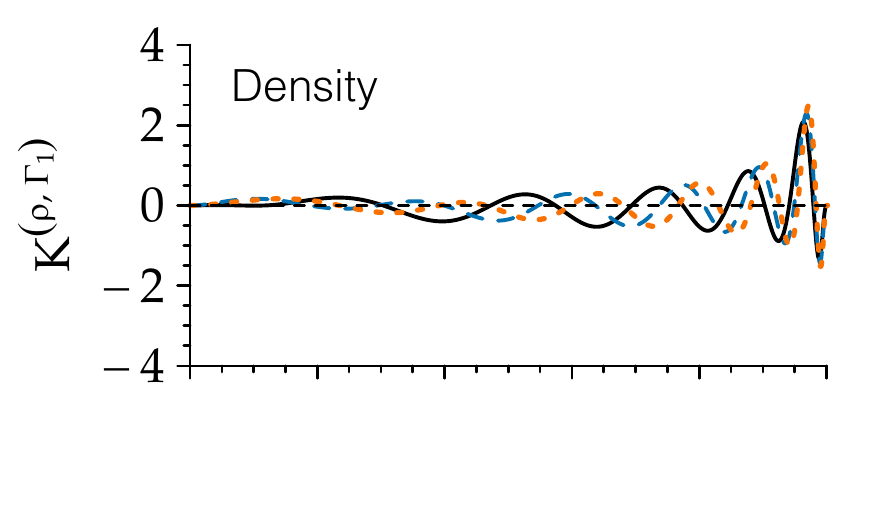}\\
    \includegraphics[width=0.5\textwidth,trim={0 1.1cm 0 0}, clip]{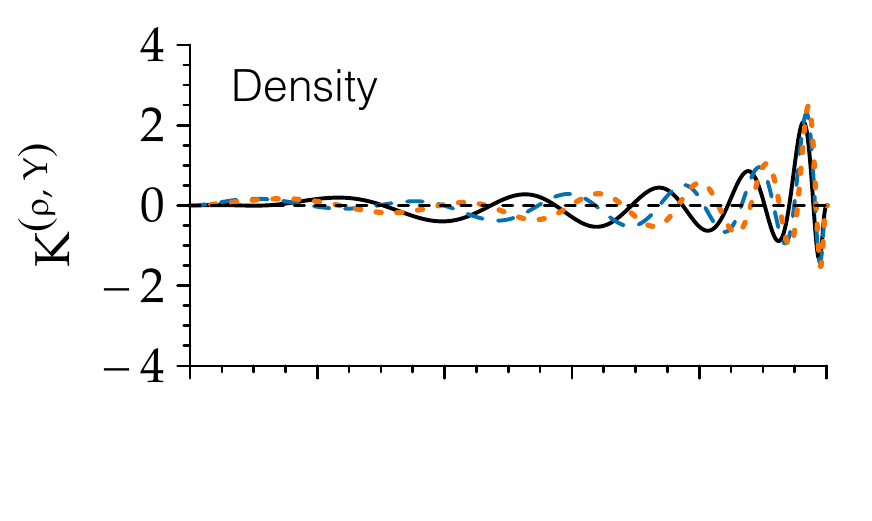}%
    \includegraphics[width=0.5\textwidth,trim={0 1.1cm 0 0}, clip]{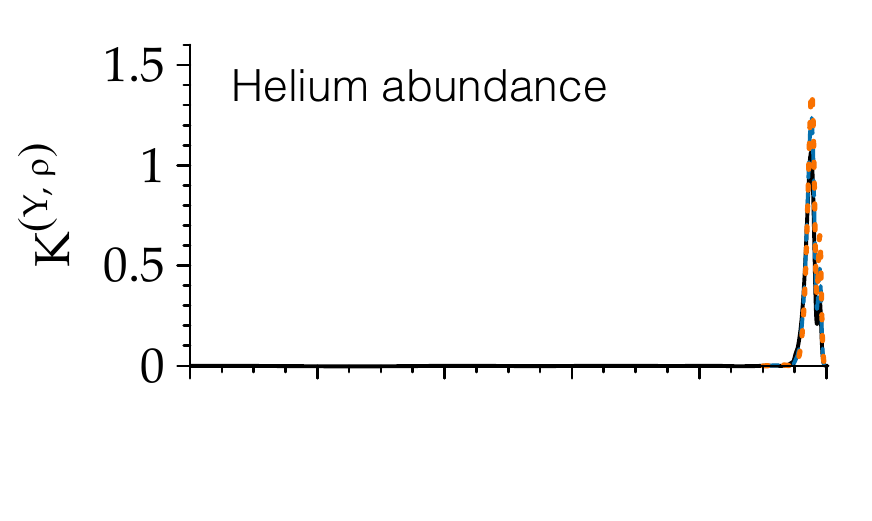}\\
    \includegraphics[width=0.5\textwidth]{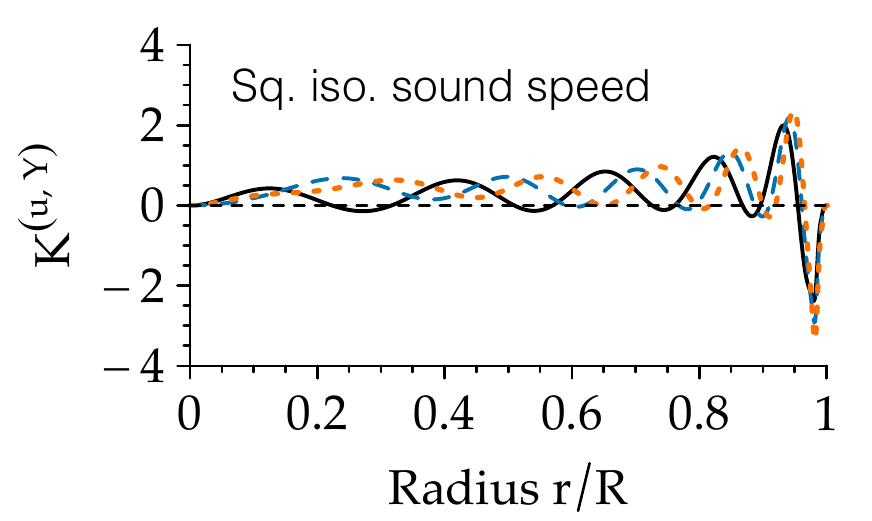}%
    \includegraphics[width=0.5\textwidth]{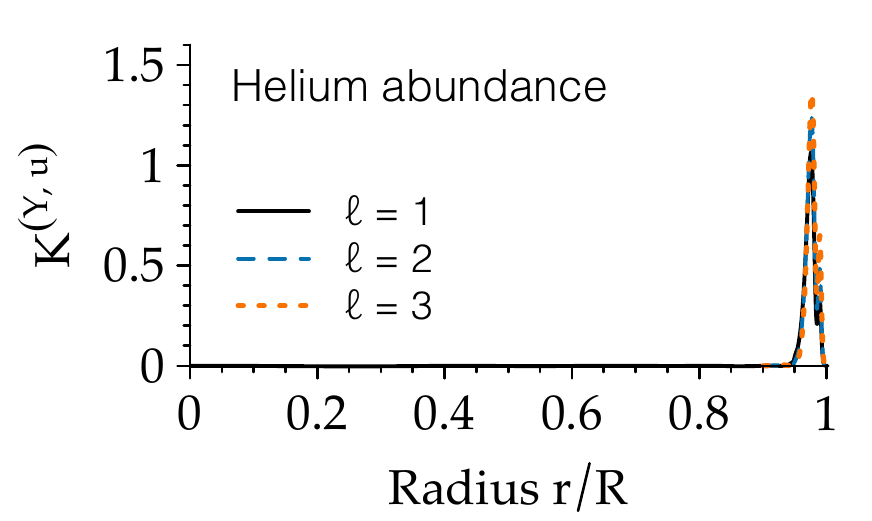}
    \caption[Kernel functions (same $n$, different $\ell$)]{Pairs of kernel functions for modes with the same radial order ${n=5}$ and different spherical degrees ${\ell=1},2,3$. \label{fig:same-n}}
\end{figure}

\begin{figure}
    \centering
    \includegraphics[width=0.5\textwidth,trim={0 1.1cm 0 0}, clip]{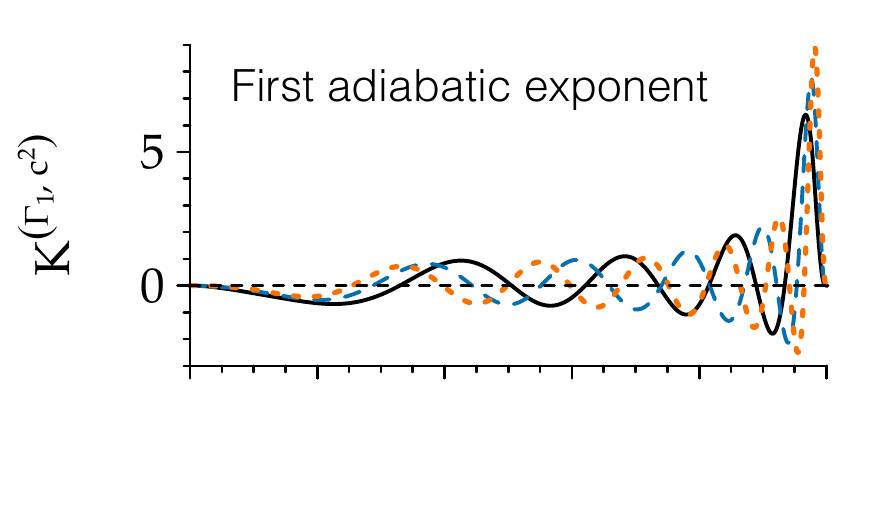}%
    \includegraphics[width=0.5\textwidth,trim={0 1.1cm 0 0}, clip]{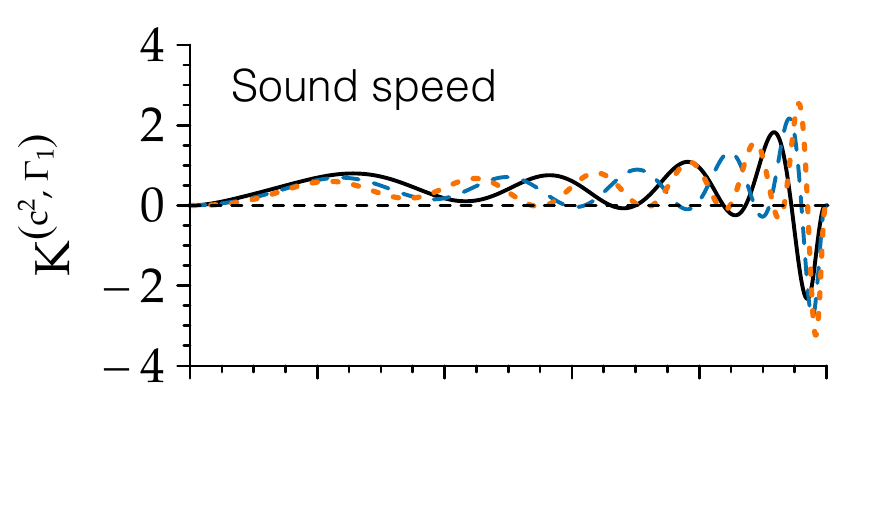}\\
    \includegraphics[width=0.5\textwidth,trim={0 1.1cm 0 0}, clip]{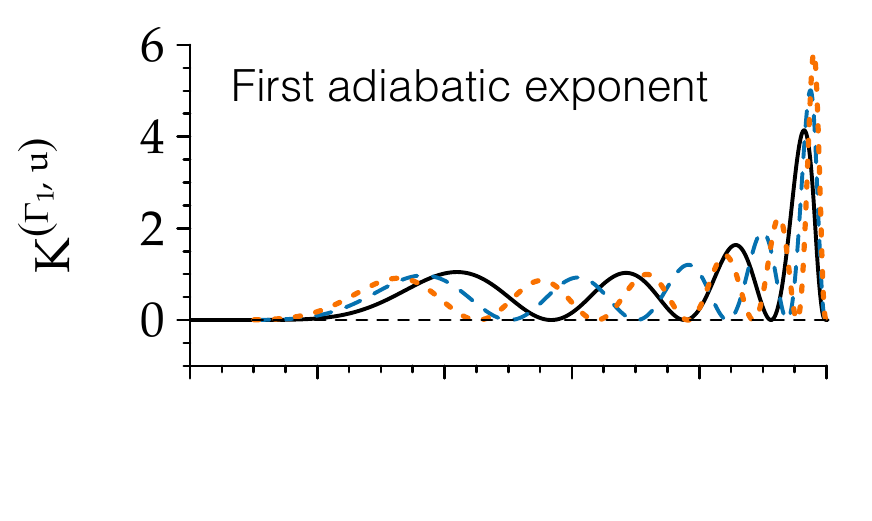}%
    \includegraphics[width=0.5\textwidth,trim={0 1.1cm 0 0}, clip]{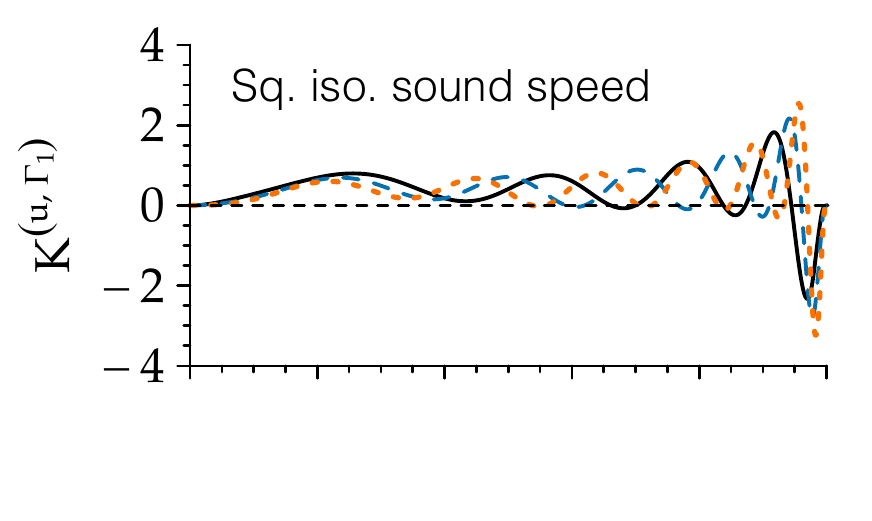}\\
    \includegraphics[width=0.5\textwidth,trim={0 1.1cm 0 0}, clip]{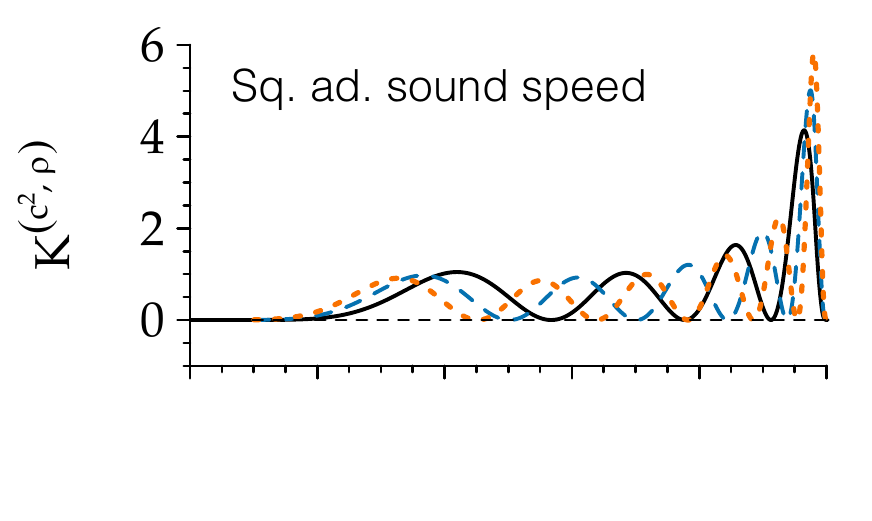}%
    \includegraphics[width=0.5\textwidth,trim={0 1.1cm 0 0}, clip]{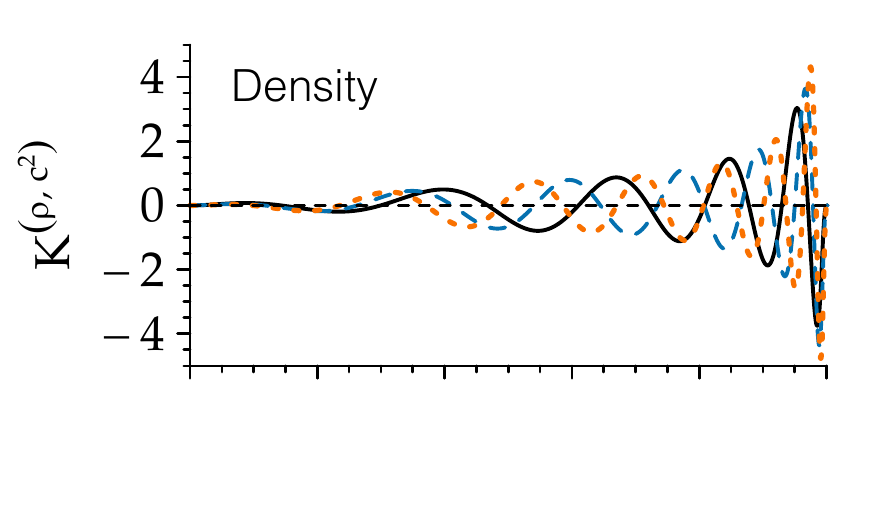}\\
    \includegraphics[width=0.5\textwidth,trim={0 1.1cm 0 0}, clip]{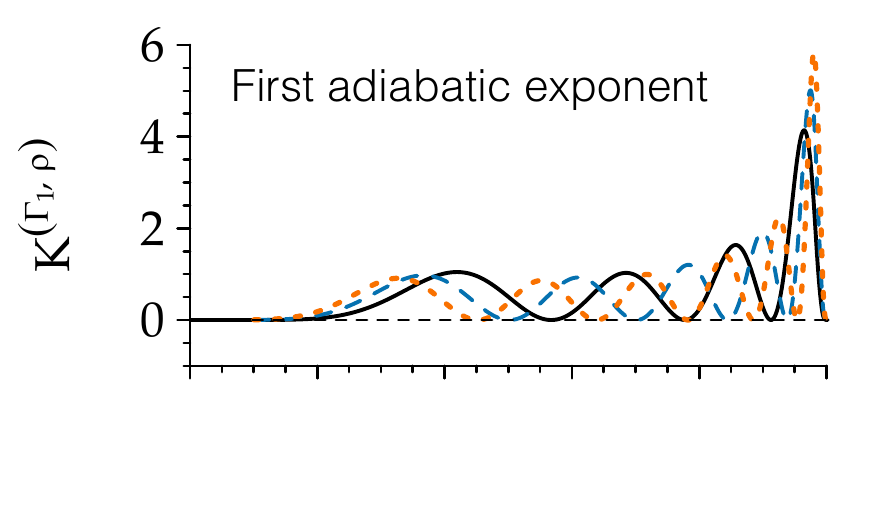}%
    \includegraphics[width=0.5\textwidth,trim={0 1.1cm 0 0}, clip]{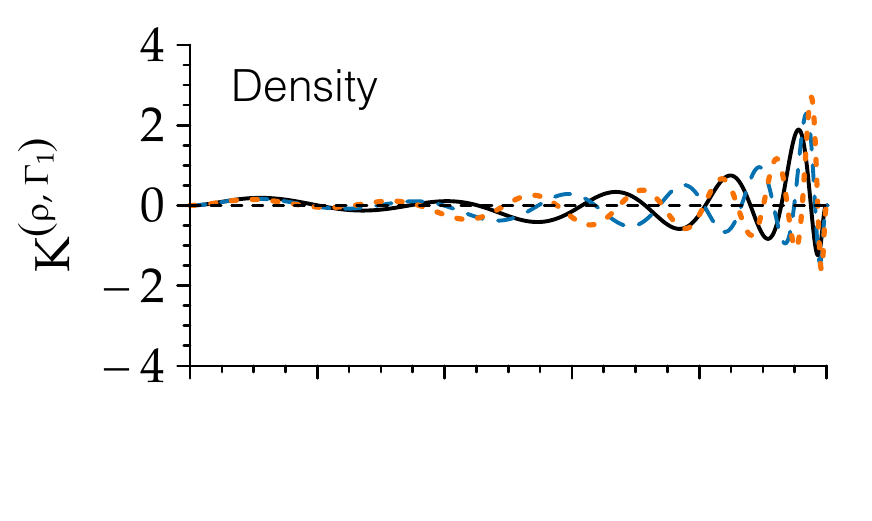}\\
    \includegraphics[width=0.5\textwidth,trim={0 1.1cm 0 0}, clip]{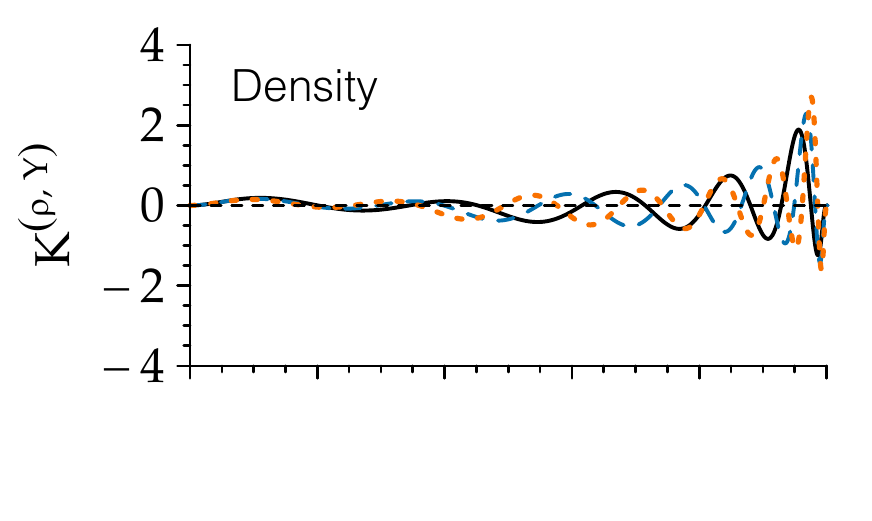}%
    \includegraphics[width=0.5\textwidth,trim={0 1.1cm 0 0}, clip]{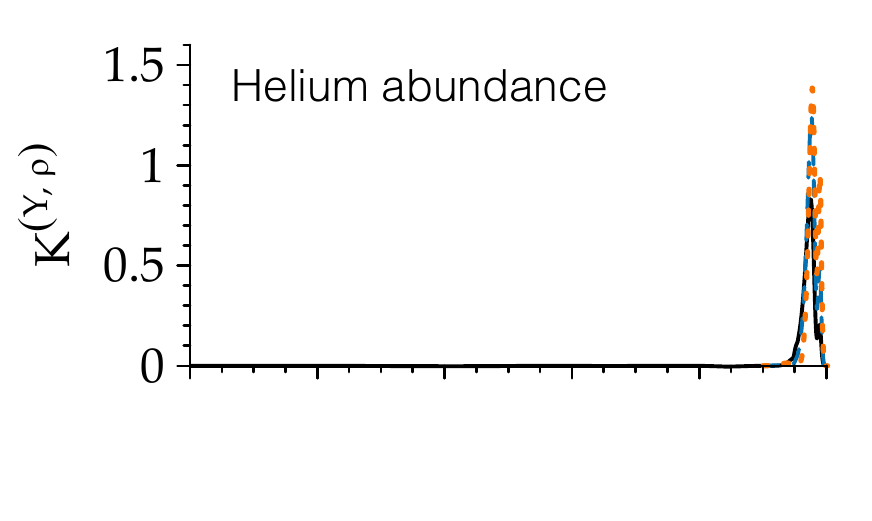}\\
    \includegraphics[width=0.5\textwidth]{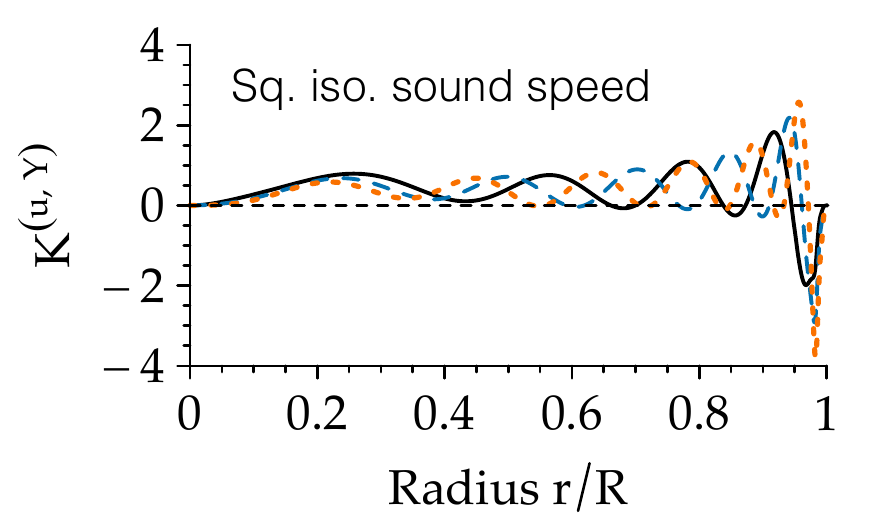}%
    \includegraphics[width=0.5\textwidth]{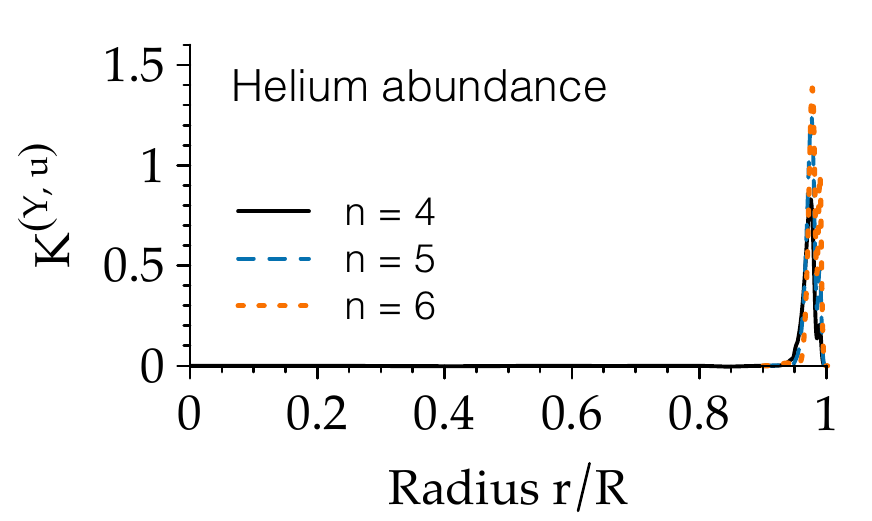}
    \caption[Kernel functions (same $\ell$, different $n$)]{Pairs of kernel functions for modes with the same spherical degree ${\ell=2}$ and different radial order ${n=4},5,6$. \label{fig:same-ell}}
\end{figure}

\noindent
Since all kernel pairs must satisfy Equation~(\ref{eq:forward}), it is straightforward to transform kernel pair ${(c, \rho)}$ to kernel pair ${(c^2, \rho)}$. 
We have that
\begin{equation}
    \int K^{(c,\rho)}_i \frac{\delta c}{c} + K^{(\rho,c)}_i \frac{\delta \rho}{\rho} \; \text{d}x
    =
    \int K^{(c^2,\rho)}_i \frac{\delta c^2}{c^2} + K^{(\rho,c^2)}_i \frac{\delta \rho}{\rho} \; \text{d}x.
\end{equation}
We may expand the sound speed perturbation as
\begin{equation}
    \frac{\delta c^2}{c^2} = \frac{2 c\delta c}{c^2} = 2 \frac{\delta c}{c}
\end{equation}
hence we have
\begin{align}
    \Kcsr &= \frac{1}{2} \Kcr \label{eq:Kcsr}
\\  \Krcs &= \Krc. \label{eq:Krcs}
\end{align}
It is instructive at this point to inspect some kernels and see what they actually look like. Figures \ref{fig:same-n} and \ref{fig:same-ell} show Equations (\ref{eq:Kcsr}) and (\ref{eq:Krcs}) for various different oscillation modes of a solar model. 
These kernels tell us how perturbations to the relevant physical variables would translate into perturbations of the respective oscillation mode frequencies. 
The figure additionally shows more kernel pairs, some of which will be also derived in this section.


\subsubsection*{Kernel Pair \texorpdfstring{$\mathbf{(\Gamma_1, \rho)}$}{(Gamma1,rho)}}
\noindent
Kernel functions for the first adiabatic exponent and density may be transformed from ${(c^2, \rho)}$ kernels via \citep[e.g.][Equations~104-105]{InversionKit}:
\begin{align}
    \KGr ={} & \Kcsr
\\  \KrG ={} & \Krcs - \Kcsr + \frac{G m \rho}{r^2} \int_{s=0}^r \frac{\Gamma_1 \chi^2 s^2}{2 \mathcal{S} \omega^2} \; \text{d}s
\\\notag   &+ \rho r^2 \int_{s=r}^R \frac{4\pi G \rho}{s^2} \left( \int_{t=0}^s \frac{\Gamma_1 \chi^2 t^2}{2 \mathcal{S} \omega^2} \; \text{d}t \right) \; \text{d}s.
\end{align}

\subsubsection*{Kernel Pair $\mathbf{(u,Y)}$}
\noindent
Using additional assumptions, for example under assumption of the EOS, we may formulate kernels for other quantities such as the fractional helium abundance. 
For each mode $i$ we wish to obtain the pair of kernel functions for the isothermal sound speed (recall Equation~\ref{eq:speed-of-sound}) and helium abundance $Y$
\begin{equation}
    \vec K^{(2)}_i = \left[ \KuYnop, \KYunop \right]
\end{equation}
via conversion from the kernel pair of ${(\Gamma_1, \rho)}$
\begin{equation}
    \vec K^{(1)}_i = \left[ \KrG, \KGr \right].
\end{equation}
We can expand the perturbation to the first adiabatic exponent as
\begin{equation}
    \frac{\delta \Gamma_1}{\Gamma_1}
    =
    \Gr \frac{\delta\rho}{\rho}
    +
    \GP \frac{\delta P}{P}
    +
    \GY \delta Y
\end{equation}
where I have introduced the quantities
\begin{equation}
    \Gr \equiv \left( \pdv{\ln \Gamma_1}{\ln \rho} \right)_{P, Y} \qquad 
    \GP \equiv \left( \pdv{\ln \Gamma_1}{\ln P} \right)_{\rho, Y} \qquad
    \GY \equiv \left( \pdv{\ln \Gamma_1}{Y} \right)_{\rho, P}
\end{equation} 
which are calculated from the assumed EOS. 
There are two formulations of these kernels that appear in the literature: the \citet{ThompsonJCD2002} formulation and the \citet{Kosovichev1999} formulation. 
For the sake of completeness, I show both here. 

\begin{description}
\setlength{\itemindent}{0pt}
\item[Thompson--JCD Formulation.]
This kernel pair may be calculated with \citep[][their Equation~A9]{ThompsonJCD2002} 
\begin{align}
    \KYunop &= \GY \cdot \KGr
\\  \KuYnop &= \GP \cdot \KGr - P \cdot \ddr \left( \frac{\psi_i}{P} \right) \label{eq:Gough-JCD-KuY}
\end{align}
where ${\psi(r)}$ is the solution to the system of differential equations 
\begin{equation} \label{eq:bvp}
    \frac{\rho}{r^2 P}
    \psi_i
    =
    \frac{1}{4\pi G} \cdot 
    \ddr \left( 
        \frac{F_i}{r^2 \rho} 
    -
        \frac{1}{r^2 \rho} \cdot \ddra{\psi_i} 
    \right)
\end{equation}
\begin{equation}
    F_i(r) = (\GP + \Gr) \cdot \KGr + \KrG
\end{equation}
with boundary conditions
\begin{equation} \label{eq:bcs1}
    \psi(r=0) = \psi(r=R) = 0.
\end{equation}
In order to calculate these kernels, we must first solve Equation~(\ref{eq:bvp}) for $\psi$ numerically. As it is a system of second-order differential equations, we must first massage it into a first-order system. We may integrate both sides of Equation~(\ref{eq:bvp}) to obtain 
\begin{equation}
    \ddra{\psi_i} = F_i - 4\pi G r^2 \rho \int_{s=r}^R \frac{\rho}{s^2 P} \psi_i \; \text{d}s.
\end{equation}
I use this approach here in this thesis. 

\item[Kosovichev Formulation.]
First let \citep[][his Equations~40; 43-45; 48]{Kosovichev1999}
\begin{equation}
    U = \U \qquad V = \VV
\end{equation}
\begin{align}
A &= \left(
  \begin{bmatrix} 
    V & -V \\
    0 & -U
  \end{bmatrix} 
  +
  \begin{bmatrix} 
    -V & 0 \\
    U & 0
  \end{bmatrix} 
  \begin{bmatrix} 
    1 & 0 \\
    -\Gr & 1
  \end{bmatrix}^{-1}
  \begin{bmatrix} 
    1 & 0 \\
    \GP & 0
  \end{bmatrix} \right)
= \begin{bmatrix}
    0 & -U \\
    V & U
  \end{bmatrix} \\
B &= \left(
  \begin{bmatrix}
    -V & 0 \\
    U & 0
  \end{bmatrix}
  \begin{bmatrix}
    1 & 0 \\
    -\Gr & 1
  \end{bmatrix}^{-1}
  \begin{bmatrix}
    -1 & 0 \\
    0 & \GY
  \end{bmatrix} \right)
= \begin{bmatrix}
    V & 0 \\
    -U & 0
  \end{bmatrix} \\
C &= \left(
  \begin{bmatrix} 
    1 & 0 \\
    -\Gr & 1
  \end{bmatrix}^{-1}
  \begin{bmatrix} 
    1 & 0 \\
    -\GP & 0
  \end{bmatrix} \right) 
= \begin{bmatrix}
    1 & 0 \\
    \Gr + \GP & 0
  \end{bmatrix} \\
D &= \left(
  \begin{bmatrix}
    1 & 0 \\
    -\Gr & 1
  \end{bmatrix}^{-1}
  \begin{bmatrix}
    -1 & 0 \\
    0 & \GY
  \end{bmatrix} \right)
= \begin{bmatrix}
    -1 & 0 \\
    -\Gr & \GY
  \end{bmatrix}.
\end{align}
The kernels can be expressed in matrix form
\begin{equation} \label{eq:vec-k}
    \vec K^{(2)}_i = D^T \vec K^{(1)} - B^T \vec w
\end{equation}
with $\vec w$ being the solution of the differential equation 
\begin{equation} \label{eq:vec-w}
    \ddx \left[ \vec w \right] = -A^T \vec w - C^T \vec K^{(1)}
\end{equation}
having boundary conditions 
\begin{equation} \label{eq:bvs2}
    \frac{\delta \rho}{\rho} w_1 + \frac{\delta m}{m} w_2 = 0 \text{ at } r=0 \text{ and } r=R.
\end{equation}
By substitution of these matrices, we have that $\vec w$ is the solution to 
\begin{align}
    \ddxa{w_1} &= -\U w_2 - \KrG - \left( \Gr + \GP \right) \KGr \\
    \ddxa{w_2} &= \VV w_1 + \U w_2.
\end{align}
Since these derivatives are with respect to a logarithmic quantity, and recalling the identity
\begin{equation}
    \frac{\text{d}x}{\text{d}\ln y} = y\frac{\text{d}x}{\text{d}y} 
\end{equation}
we cast Equation~(\ref{eq:vec-w}) into a useful form as a linear system of first-order differential equations 
\begin{align}
    \ddra{w_1} &= -\frac{4\pi\rho r^2}{m} w_2 - \frac{1}{r} \left[ \KrG + \left( \Gr + \GP \right) \KGr \right] \\
    \ddra{w_2} &= \frac{G m \rho}{r^2P} w_1 + \frac{4\pi\rho r^2}{m} w_2
\end{align}
with the boundary conditions of Equation~(\ref{eq:bvs2}), which without loss of generality may be transformed into
\begin{equation} \label{eq:bvs}
    w_1(r = 0) = w_2(r=R) = 0.
\end{equation}
Finally we may calculate the kernels using this $\vec w$ by substituting the matrices above into Equation~(\ref{eq:vec-k}) to get
\begin{align}
    \KuYnop &= -\KrG - \Gr \cdot \KGr + \VV w_1 - \U w_2 \\
    \KYunop &= \GY \cdot \KGr.
\end{align}
\end{description}
These last kernels---the ${(u,Y)}$ kernel pair---are especially valuable for the following analysis. 
An inspection of their form (Figures~\ref{fig:same-n}~and~\ref{fig:same-ell}) reveals that the $Y$ kernels only have amplitude in ionization zones, which are located near to the stellar surface. 
As we will see later, this implies that it will be possible to isolate the effects of differences in mode frequencies to differences in internal isothermal sound speeds. 

\subsubsection*{Testing the Forward Formulation}

\begin{figure}
        \centering
        \makebox[0.5\textwidth][c]{%
        \adjustbox{trim={0cm 1.1cm 0.3cm 0cm},clip}{%
            \includegraphics[width=0.5\textwidth]{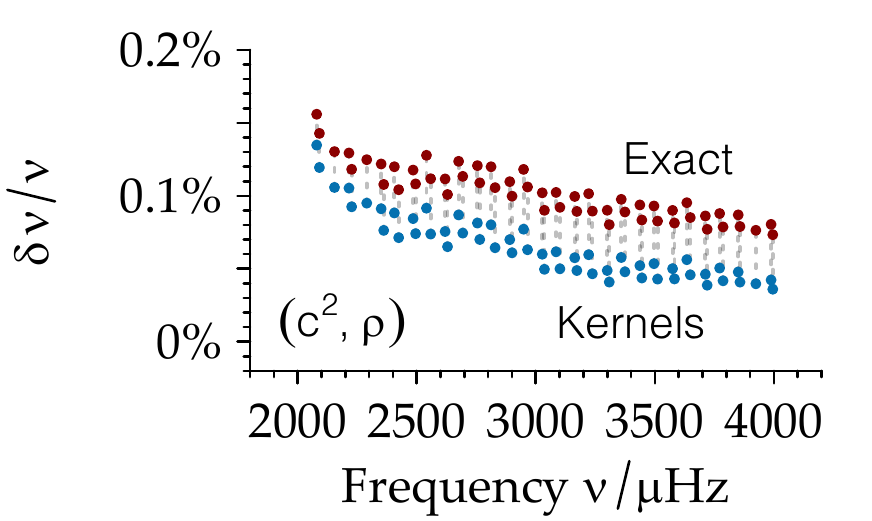}%
        }}\hspace*{-1cm}%
        \makebox[0.5\textwidth][c]{%
            \adjustbox{trim={2cm 1.1cm 0.3cm 0cm},clip}{%
                \includegraphics[width=0.5\textwidth]{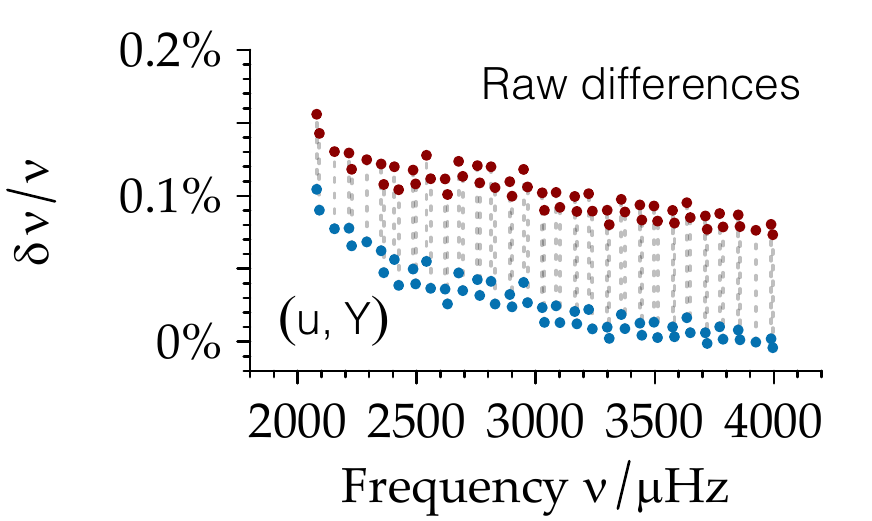}%
            }}\\
        \makebox[0.5\textwidth][c]{%
            \adjustbox{trim={0cm 0cm 0.3cm 0cm},clip}{%
                \includegraphics[width=0.5\textwidth]{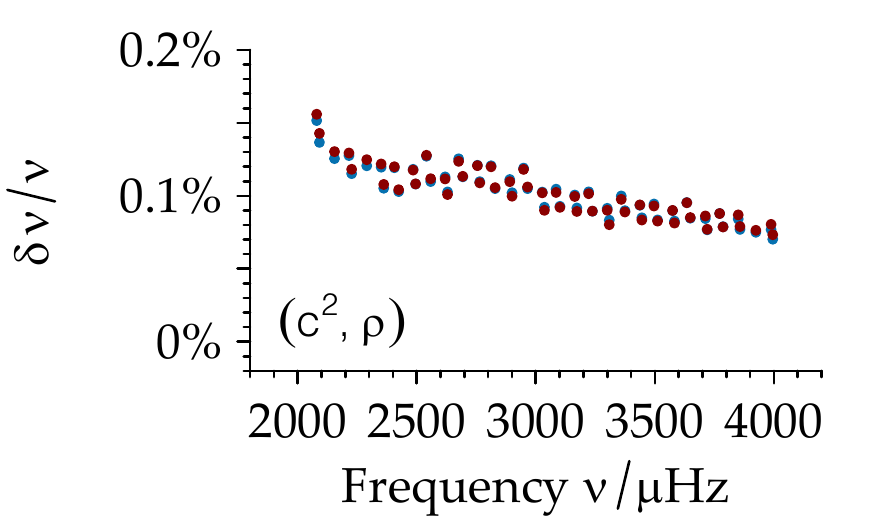}%
            }}\hspace*{-1cm}%
        \makebox[0.5\textwidth][c]{%
            \adjustbox{trim={2cm 0cm 0.3cm 0cm},clip}{%
                \includegraphics[width=0.5\textwidth]{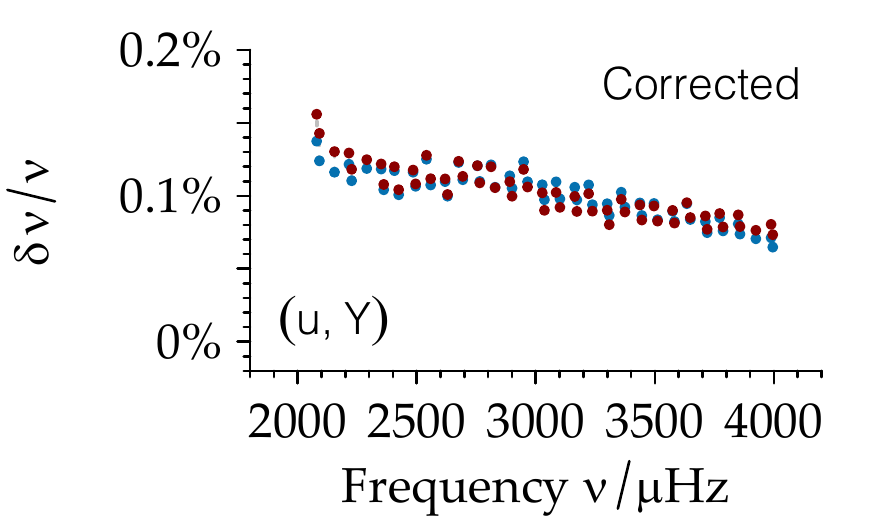}%
        }}
    \caption[Verifying the forward problem]{Top: Relative frequency differences between two solar models using the 16~Cyg~B mode set. 
    The points in red are the exact differences; the points in blue are the differences obtained through Equation~(\ref{eq:forward}) using ${(c^2, \rho)}$ kernels (left) and ${(u, Y)}$ kernels (right). 
    Bottom: the same, but also including the surface-term corrections of Equation~(\ref{eq:forward-surf}). 
    }
    \label{fig:forward}
\end{figure}

We may now compare the actual frequency differences between the two solar models to the differences that we get through the kernel equation (Equation~\ref{eq:forward}). 
The top pair of plots in Figure~\ref{fig:forward} shows this comparison for the ${(c^2, \rho)}$ and ${(u, Y)}$ kernel pairs. 
Here I have shown the comparison using the set of modes (i.e., the ${n,\ell}$ labels) that have been observed in 16~Cyg~B. 
As we have seen previously, the differences again increase as a function of frequency due to surface effects. 
We therefore modify Equation~(\ref{eq:forward}) to take this phenomenon into account by including the \citet{2014A&A...568A.123B} surface term: 
\begin{equation} \label{eq:forward-surf} \boxed{
  \frac{\delta\nu_i}{\nu_i} 
  = 
  \int_0^R \left[ K_i^{(f_1, f_2)} \frac{\delta f_1}{f_1}
                + K_i^{(f_2, f_1)} \frac{\delta f_2}{f_2}
          \right] \; \text{d}r
    + \frac{F(\nu_i)}{I_i}
}\end{equation}
where ${F(\nu_i)}$ is adapted from the surface term of Equation~(\ref{eq:BallGizon-surfterm})
\begin{equation}
    F(\nu_i) 
    = 
    a_1 \left( \frac{\nu_i}{\nu_{ac}} \right)^{-2} + a_2 \left( \frac{\nu_i}{\nu_{ac}} \right)^{2}.
\end{equation}
Figure~\ref{fig:forward} shows that after applying the surface term correction, the agreement between the exact differences and those obtained through the kernels is much better. 
In other words, through the use of the stellar structure kernels, we can translate differences in structure to differences in pulsation frequency. 

\clearpage\section{Inverse Problems} 
\label{sec:inverse}
\begin{shaded}
\noindent In this section, I will provide a general summary of inverse problems, with particular attention toward those that are posed and solved in the subsequent chapters of this thesis. 
Several textbooks discuss inverse problems and their solutions. 
In writing this section, I have made use of the textbooks by \citet{basuchaplin2017}, \citet{kirsch2011introduction} and \citet{neto2012introduction}. 
Additionally, I have found the reviews by \citet{tenorio2001statistical}, \citet{GoughThompson1991}, and \citet{2018ASSP...49...75R} helpful. 
\end{shaded}

\begin{figure}
    \centering
    \begin{tikzpicture}[
        ->, 
        thick, 
        main node/.style={
            circle, 
            draw,
            minimum height=5em,
        },
        node distance=10em, 
        auto,
        pil/.style={
           thick,
           shorten <=2pt,
           shorten >=2pt
        }
    ]
    \node[main node, fill=blue!20!white] (model) {Model};
    \node[main node, right=of model, fill=orange!20!white] (data) {Data}
        edge[pil, <-, bend right=30] node[above, yshift=0.1cm] {\sffamily\small{Forward Problem}} (model)
        edge[pil, ->, bend left=30] node[below, yshift=-0.1cm] {\sffamily\small{Inverse Problem}} (model);
\end{tikzpicture}
    \caption[Forward and Inverse Problems]{A schematic for the relationship between forward and inverse problems. 
            In the forward problem, we use the theory or a model to generate data, such as the types of information that could be observed about a system. 
            In the inverse problem, we seek to reconstruct all the possibilities that are consistent with that observed data. 
        \label{fig:forward-inverse}}
\end{figure}
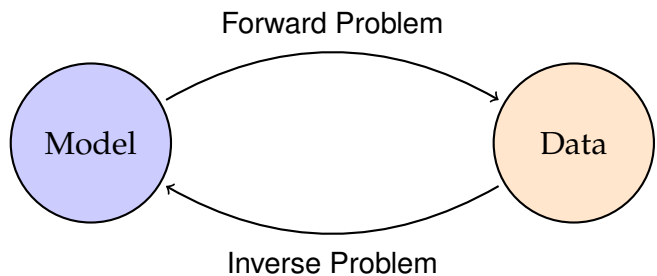

So far we have concerned ourselves with discussions of \emph{forward problems}. 
These can be thought of as problems where we have a theory, we input some initial conditions, and we compute the result deterministically. 
The two topics of the previous chapters have been the theory of stellar evolution and the theory of stellar pulsation. 
In the case of evolution, we supplied the initial conditions (mass, initial composition, mixing length parameter, etc.), and then applied the theory to simulate what such a star would be like at each given time in the future. 
In the case of pulsation, we supplied a static stellar structure, and then applied the theory to calculate the corresponding frequencies of oscillation. 
Now we wish to go in the opposite direction (see Figure~\ref{fig:forward-inverse}). 

\epigraph{``\emph{The cause is hidden, but the result is known.}''}{--- Ovid\\\emph{Metamorphoses} (\citeyear{ovid})}

In the case of evolution, given the observation of a star (e.g., its luminosity, or pulsation data), we wish to determine its overall properties (e.g., mass, radius, age) and evolutionary history (initial composition and so on) using the theory of evolution. 
In the case of pulsation, given the observed oscillation frequencies, we wish to determine the stellar structure that supports those oscillations using the theory of stellar pulsation. 
These are the inverse problems of asteroseismology that form this thesis. 


%

The difficulty in solving these problems comes in part from the fact that they are \emph{ill-posed}. 
At the beginning of the 20th century, the French mathematician \mycitet{hadamard} gave his definition for what constitutes a well-posed problem. 
Hadamard believed that problems worth consideration should have the properties that
\begin{enumerate}
    \item a solution exists (existence), 
    \item the solution is unique (uniqueness), and 
    \item the solution changes continuously with changes to the input (stability). 
\end{enumerate}
A problem that fails to meet one or more of these criteria is then said to be ill-posed. 

\epigraph{``\emph{The respect for Hadamard was so great that incorrectly posed problems were \hphantom{``}considered `taboo' for generations of mathematicians, until comparatively recently \hphantom{``}it became clear that there are a number of quite meaningful problems, the so-called \hphantom{``}`inverse problems,' which are nearly always unstable with respect to fluctuations \hphantom{``}of input data.}''}{--- H.\ Allison\\\emph{Inverse Unstable Problems and Some of Their Applications} (\citeyear{allison1979inverse})}

An example of a well-posed problem is: given the formula for a line and some coordinates, calculate the corresponding points on the line. 
The inverse of this problem---calculating the formula of a line given points belonging to it---also happens to be well-posed. 
Suppose however that we only have one point. 
Then the uniqueness condition is not satisfied, as infinitely many lines pass through that point. 
Suppose instead that we have multiple points, but one of the points does not actually belong to the line. 
Then the existence condition is not satisfied, as no one line passes through all the points. 

One of the most famous inverse problems is the question from mathematician Mark Kac: ``Can One Hear the Shape of a Drum?'' \citep{10.2307/2313748}. 
In a response article entitled ``You Can't Hear the Shape of a Drum,'' \citet{10.2307/29775597} produced two different drums with the same eigenfrequencies. 
The solution to the problem therefore lacks uniqueness, and so it is ill-posed. 

\begin{figure} 
    \centering 
    \begin{tikzpicture}[
        ele/.style={
            minimum width=.8pt,
            inner sep=1pt},
            every fit/.style={ellipse,draw,inner sep=-2pt}
        ]
    \node[ele] (a1) at (0,4) {$1$};
    \node[ele] (a2) at (0,3) {$2$};
    \node[ele] (a3) at (0,2) {$3$};
    \node[ele] (a4) at (0,1) {$4$};
    
    \node[ele] (b1) at (4,4) {$a$};
    \node[ele] (b2) at (4,3) {$b$};
    \node[ele] (b3) at (4,2) {$c$};
    \node[ele] (b4) at (4,1) {$d$};
    
    \node[draw,fit= (a1) (a2) (a3) (a4),minimum width=2cm, thick, fill=orange!20!white] (leftoval) {} ;
    \node[draw,fit= (b1) (b2) (b3) (b4),minimum width=2cm, thick, fill=blue!20!white] (rightoval) {};
    
    \node[above of=leftoval, yshift=1.75cm] {\sffamily Models};
    \node[above of=rightoval, yshift=1.75cm] {\sffamily Data};
    
    \draw[->,thick,shorten <=5pt,shorten >=5pt] (a1) -- (b1);
    \draw[->,thick,shorten <=5pt,shorten >=5pt] (a2) -- (b2);
    \draw[->,thick,shorten <=5pt,shorten >=5pt] (a3) -- (b3);
    \draw[->,thick,shorten <=5pt,shorten >=5pt] (a4) -- (b3);
    \node[ele] (a1) at (0,4) {$1$};
    \node[ele] (a2) at (0,3) {$2$};
    \node[ele] (a3) at (0,2) {$3$};
    \node[ele] (a4) at (0,1) {$4$};
    
    \node[ele] (b1) at (4,4) {$a$};
    \node[ele] (b2) at (4,3) {$b$};
    \node[ele] (b3) at (4,2) {$c$};
    \node[ele] (b4) at (4,1) {$d$};
\end{tikzpicture} 
    \caption[Non-injective and non-surjective functions]{Physical systems are often \emph{non-injective} in the sense that two systems may have different internal conditions but the same external observables. 
    Here models $3$ and $4$ share the same set of observables $c$. 
    This system is also \emph{non-surjective} because the fourth set of observations is not produced by any model. 
        \label{fig:non-injective}}
\end{figure}
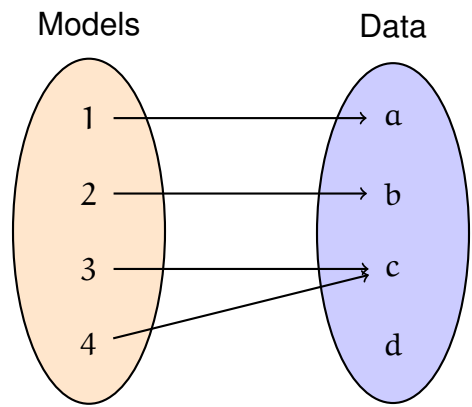

The solutions to physical inverse problems often lack uniqueness. 
At a basic level, measurements are nearly always uncertain, and therefore the solution is uncertain. 
Less obvious however is that two distinct sets of initial conditions can often lead to the same observables (i.e., the forward function is non-injective, see Figure~\ref{fig:non-injective}). 
This is sometimes referred to as degeneracy. 
The evolution inverse problem has the additional issue that there are observations of stars (the Sun is an example) that cannot yet be fully reproduced by any evolutionary model (i.e., the forward function is non-surjective, see again Figure~\ref{fig:non-injective}). 
This is one reason why we separate the two inverse problems, and use the solution from the evolution inversion as the starting point for the structure inversion. 

The word ``inverse'' is especially appropriate because inverse problems can often be stated as finding the inverse of a forward function, operator, or matrix. 
For example, if we have a model $M$ that takes initial conditions $x$ and produces data ${y=M(x)}$, then the inverse problem is to determine ${x=M^{-1}(y)}$ from observations of $y$. 
This is where the condition of stability often runs into problems. 

As an example, consider a simple theory defined by the following linear system of equations:
\begin{alignat}{3}
    &x_1 + x_2& &= y_1 \\
    &x_1 + (1+\epsilon)x_2& &= y_2
\end{alignat}
where $\epsilon$ is an arbitrarily small number. 
The values $y_1$ and $y_2$ are then observed in nature, each coming up to be ${y_1=y_2=2}$. 
We now seek the ``initial conditions'' ${\mathbf{x} \equiv (x_1,x_2)}$ to explain this observation. 
The solution is clearly ${\mathbf{x}=(2,0)}$. 
Now consider that $y_2$ was instead measured to be ${2+\epsilon}$. 
The solution then changes to ${\mathbf{x}=(1,1)}$. 
Recall however that $\epsilon$ was chosen to be arbitrarily small. 
Thus, an arbitrarily small change to the measurement has completely changed the solution. 
To be even more concrete, if we let ${\epsilon=10^{-10}}$ and modify $y_2$ to be, say, ${2+10^{-5}}$, then we obtain ${\mathbf{x}\simeq(-99998, 10000)}$. 
The system is unstable. 

In matrix notation, this system corresponds to
\begin{equation}
    \mathbf{M}\mathbf{x}
    =
    \mathbf{y},
    \qquad
    \mathbf{M}
    =
    \begin{bmatrix}
        1 & 1 \\
        1 & 1+\epsilon 
    \end{bmatrix}. 
\end{equation}
Here our model is the nearly singular matrix $\mathbf{M}$, we have observed the data $\mathbf{y}$, and we've sought the initial conditions ${\mathbf{x}=\mathbf{M}^{-1}\mathbf{y}}$. 
When the condition number ${\kappa(\mathbf{M})=||\mathbf{M}||\;||\mathbf{M}^{-1}||}$ is large, the problem is said to be ill-conditioned. 
When ${\kappa = \infty}$, the problem is ill-posed. 
For this particular system, 
\begin{equation}
    \lim_{\epsilon\to 0} \kappa(\mathbf{M}) = \infty.
\end{equation}

The kernel functions that we derived in the previous section are nearly linearly dependent across the different modes, and so the structure inversion problem is ill-conditioned. 
As we will see later, such problems are generally dealt with by enforcing stability or regularity conditions, i.e., regularization \citep[e.g.,][]{tikhonov1977solutions, tenorio2001statistical}. 


\subsection{Evolution Inversions} 

With the equations of Section~\ref{sec:evolution} and some chosen initial conditions, we can simulate the life of a star, and at each step of the way, determine what observations of that star would yield. 
Thus we have a forward model $M$ which is parameterized by initial conditions $\mathbf{x}$ and time $\tau$, and yields data $\mathbf{y}$:
\begin{align}
    M(\mathbf{x}, \tau)
    &=
    \mathbf{y}
    \\
    \mathbf{x}
    &=
    [M,Y_0,Z_0,\alpha_{\text{MLT}}, \ldots]
    \\
    \mathbf{y}
    &=
    [L, T_{\text{eff}}, \text{[Fe/H]}, \boldsymbol{\nu}, \ldots].
\end{align}
We now seek to interpret observations of a star in the context of the theory of stellar evolution. 
In other words, we seek the inverse function:
\begin{equation}
    M^{-1}(\mathbf{y})
    =
    [\mathbf{x}, \tau].
\end{equation}
Of course, we can also seek a function that outputs additional quantities at the present age, such as the radius if it has not been observed. 
There are several approaches that have been taken to solve this problem, which I will now review. 

\subsubsection*{Scaling Relations}
\label{sec:scaling}
A simple approach to estimate stellar properties is to ``scale'' them from solar values using the equations of stellar structure and pulsation. 
While such an approach does not solve the full evolution inversion problem, it shares a common goal of estimating (a more limited set of) properties such as the stellar mass. 

A simple example comes from the Stefan-Boltzmann law (Equation~\ref{eq:stefan-boltzmann}). 
Replacing this equation with ratios with respect to the solar values, we may obtain 
\begin{equation}
    \frac{R_\ast}{R_\odot}
    =
    \left(
        \frac{L_\ast}{L_\odot}
    \right)^{-2}
    \left(
        \frac{T_{\text{eff},\ast}}{T_{\text{eff},\odot}}
    \right)^4
\end{equation}
from which we can estimate an unknown stellar radius $R_\ast$ from a measured stellar luminosity $L_\ast$ and effective temperature $T_{\text{eff},\ast}$. 
In principle, this relation works; in practice, the luminosities of most stars are unknown, and effective temperatures are measured rather imprecisely (${\gtrapprox 50}$~K uncertainty). 

The same kind of manipulation can be used on the asymptotic equations of stellar pulsation to obtain stellar masses and radii. 
From manipulation of Equations~(\ref{eq:numax}) and (\ref{eq:Dnu}) we find \citep[e.g.,][]{1995A&A...293...87K}:
\begin{align}
    \frac{R_\ast}{R_\odot}
    &=
    \left(
        \frac{\nu_{\max,\ast}}{\nu_{\max,\odot}}
    \right)
    \left(
        \frac{\Delta\nu_\ast}{\Delta\nu_\odot}
    \right)^2
    \left(
        \frac{T_{\text{eff},\ast}}{T_{\text{eff},\odot}}
    \right)^\frac{1}{2}
    \\
    \frac{M_\ast}{M_\odot}
    &=
    \left(
        \frac{\nu_{\max,\ast}}{\nu_{\max,\odot}}
    \right)^3
    \left(
        \frac{\Delta\nu_\ast}{\Delta\nu_\odot}
    \right)^4
    \left(
        \frac{T_{\text{eff},\ast}}{T_{\text{eff},\odot}}
    \right)^\frac{3}{2}
\end{align}
which hold to decent approximation. 
\citet{2017ApJ...843...11V} recently pointed out that the $\nu_{\max}$ scaling relation can be improved by including a term for the mean molecular weight. 

As stars evolve into giants, the assumption of homology breaks down more and more, leading to systematic errors as high as $15\%$ \citep[e.g.,][]{2016ApJ...832..121G}. 
By comparison of theoretical red giant model mode frequencies with those given by the scaling relations, \citet{2016MNRAS.460.4277G, 2017MNRAS.470.2069G} developed metallicity-dependent and mass-dependent corrections to the $\Delta\nu$ scaling relation. 

\begin{figure}
    \centering
    \includegraphics[width=\textwidth]{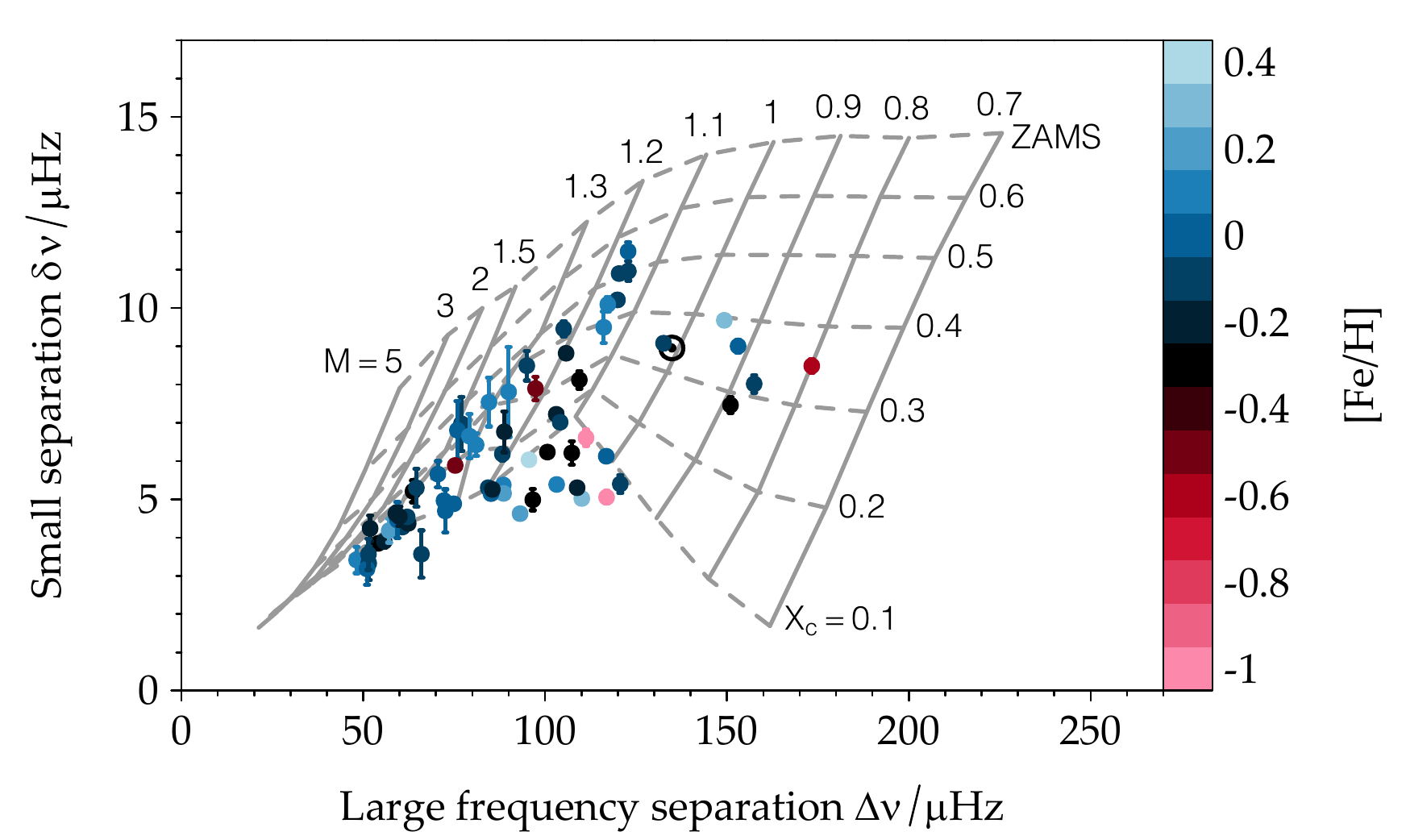}
    \caption[C--D Diagram]{The C--D diagram. 
    The small frequency separation is a proxy for core hydrogen abundance ($X_c$, dashed lines) through the sound speed gradient, and the large frequency separation is a proxy for stellar mass ($M$, solid lines) through the mean density. 
    The gray lines are evolutionary simulations varied in their initial mass and evolved along the main sequence. 
    The frequencies of the models have been calculated using GYRE \citep{2013MNRAS.435.3406T}. 
    Stars with ${M \gtrapprox 1.8\;M_{\odot}}$ do not have convective envelopes on the main sequence and are therefore not theoretically predicted to harbor solar-like oscillations. 
    The points are LEGACY stars observed by \emph{Kepler}, colored by their metallicity \citep{2017ApJ...835..172L}. 
    Many of the stars fall off the diagram, thus illustrating its limitations as a look-up table for stellar properties. 
    \emph{Figure adapted from \citealt{2017EPJWC.16005003B}.} 
    \label{fig:jcd}} 
\end{figure}

These scaling relations do not tell us about the age or evolution of the star. 
We saw previously that the small frequency separation probes the sound speed gradient, which is then an indicator on the main sequence of the conditions in the core, and therefore main-sequence age. 
The so-called C--D diagram shows the core-hydrogen abundance and stellar mass as a function of the frequency separations (\citealt{1984srps.conf...11C}, see also Figure~\ref{fig:jcd}). 
If all stars had the solar abundances and solar mixing length, it would suffice to look up their mass and core-hydrogen abundance in this diagram. 
Since they do not, a more sophisticated approach is required. 

\subsubsection*{Repeated Forward Modelling} 
A more involved approach to determining the properties of stars is through repeated forward modelling. 
Such an approach can also be applied to non-solar-like stars (e.g., evolved stars) where homology relations break down. 
These methods still make no attempt to determine the function $M^{-1}$. 
Though there are variations, they instead try to optimize the result of the forward operator against the observations: 
\begin{equation}
    [\hat{\mathbf x}, \hat\tau]
    =
    \underset{[\mathbf x, \tau]}{\arg\min}\; \left[
        M(\mathbf x, \tau)
        -
        \mathbf y
    \right]^T
    \boldsymbol{\Sigma}_{\mathbf y}^{-1}
    \left[
        M(\mathbf x, \tau)
        -
        \mathbf y
    \right]
\end{equation}
where $\hat\cdot$ means the optimal $\cdot$, and $\boldsymbol\Sigma_{\mathbf y}$ is the covariance matrix for the observations. 
There are several drawbacks with this approach: 
\begin{description}
    \setlength{\itemindent}{0pt}
    \item[Speed.] This approach can be prohibitively slow, especially if new models need to be computed for each input, or if multiple input parameters are being optimized. 
    This is often dealt with by applying additional assumptions to simplify the problem. 
    For example, the mixing length parameter can be kept fixed to the solar-calibrated value \citep[e.g.,][]{2015MNRAS.452.2127S,2017ApJ...835..173S}. 
    Another simplification is to calculate the initial helium abundance from the initial metallicity by assuming a galactic chemical evolution law \citep[e.g.,][]{2015MNRAS.452.2127S, 2017ApJ...835..173S}. 
    This is usually achieved by fitting a line through to two points: the primordial helium abundance from models of Big Bang nucleosynthesis $[Y_p=0.2463,Z_p=0]$ \citep[e.g.,][]{2014JCAP...10..050C} and the calibrated initial solar mixture, e.g., ${[Y_{0,\odot}=0.273,Z_{0,\odot}=0.019]}$, so ${\Delta Y/\Delta Z \simeq 1.4}$. 
    The optimization is then performed over a limited set of input parameters (e.g., ${[M,Z_0]}$) and potentially on a pre-computed grid of models as well. 
    However, the end result then has (typically unpropagated) systematic errors. 
    
    \item[Local Minima.] Commonly, iterative numerical optimization algorithms such as
    \citeauthor{10.2307/43633451}--\citeauthor{10.2307/2098941} (\citeyear{10.2307/43633451,10.2307/2098941})
    and 
    \mycitet{nelder1965simplex}
    are applied for this task \citep[e.g.,][]{2014A&A...569A..21L, 2015A&A...582A..25A}. 
    These approaches can have difficulty finding global minima of the solution. 
    
    There are also no currently known theoretical bounds on the complexity of a Nelder-Mead search \citep{singer1999complexity}. 
    It is however known that this algorithm scales poorly to high dimensions \citep[e.g.,][]{Chen2015}. 
    
    \item[Redundancy.] This approach implicitly assumes that each bit of observable information provides a fully independent constraint to the stellar model, and weights each observation only by its uncertainty. 
    In reality, the observations have some degree of redundancy with respect to the aspects of the model that they constrain (\mycitealt{2017apj...839..116a}, see also Chapter~\ref{chap:statistical}). 
    Matching such an aspect of the model is then arbitrarily upweighted. 
    Some practitioners deal with this problem by applying \emph{ad hoc} weightings \citep[e.g.,][]{2013apjs..208....4p}. 
\end{description}
We therefore seek an approach that naturally avoids these problems. 

\subsubsection*{Random Forest Regression} 
In recent years, machine learning techniques have become increasingly popular for solving inverse problems \citep[e.g.,][]{rosasco2005learning, fai2017inner, 2017InvPr..33l4007A}. 
Some applications include automatic photograph coloration \citep{larsson2016learning}, image reconstruction \citep[e.g.,][]{2017arXiv170300555S}, and medical imaging \citep[e.g.,][]{prato2008inverse, jin2017deep}. 
In fact, supervised learning itself can be viewed as an inverse problem \citep{vito2005learning}. 

In Chapter~\ref{chap:ML} we propose a solution to the evolution inversion problem based on machine learning. 
In particular, we use the variant of random forest regression \citep{breiman2001random} known as extremely randomized trees \citep{geurts2006extremely} to learn the function $M^{-1}$ from a dense grid of evolutionary simulations. 
Ensemble tree-based algorithms are known to be quick to train (especially because the task is `embarrassingly' parallelizable), quick to predict (when the number of trees is not very large), and to have very good predictive performance \citep[e.g.,][]{Caruana:2006:ECS:1143844.1143865}. 
Furthermore, the bootstrap aggregation (``bagging'') that is performed helps with problem degeneracy and dimensionality \citep[e.g.,][]{Skurichina2002}. 
Random forests can suffer from reduced performance if the number of redundant variables is large \citep{louppe2014understanding}, however there are strategies to deal with this drawback \citep{tuv2009feature}. 

\citet[][]{louppe2014understanding} derived the worst-case time complexity of training extremely randomized trees to be ${\mathcal{O}(MKN^2)}$, where $M$ is the number of trees, $N$ is the number of samples, and $K$ is the number of features that is randomly drawn at each node. 
In Chapter~\ref{chap:ML}, we cross-validate $M$ and find satisfactory performance at ${M=256}$. 
The parameter $K$ varies between $2$ and $9$, depending on the types of observations available for a given star. 


To obtain the posterior distribution of solutions for an observed star with measurement uncertainties, we pass random instances of the observations perturbed by their uncertainties through the trained network. 
We have to choose how many random instances that we will use. 
This number should be chosen such that the sample distribution converges to a reasonable degree to the population distribution. 
A useful way to quantify the differences in distributions is the Kullback-Leibler (KL) divergence, also known as relative entropy: 
\begin{equation} \label{eq:KL}
    D_{\text{KL}} (P||Q) = \int_{-\infty}^\infty p(x) \log \frac{p(x)}{q(x)} \;\text{d}x
\end{equation}
where $P$ and $Q$ are two continuous random variables and $p$ and $q$ are their respective densities \citep{10.2307/2236703}. 
A low relative entropy indicates similarity. 

\begin{figure}
    \centering
    \includegraphics[width=0.5\textwidth]{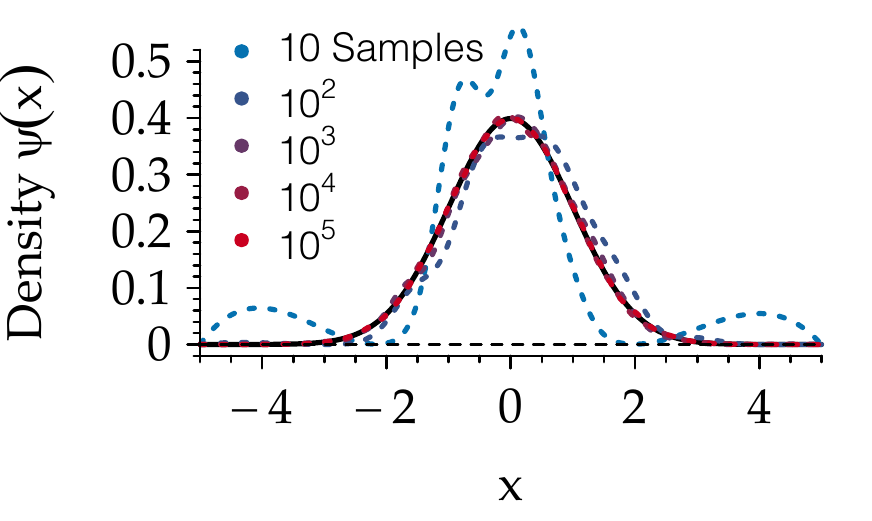}%
    \includegraphics[width=0.5\textwidth]{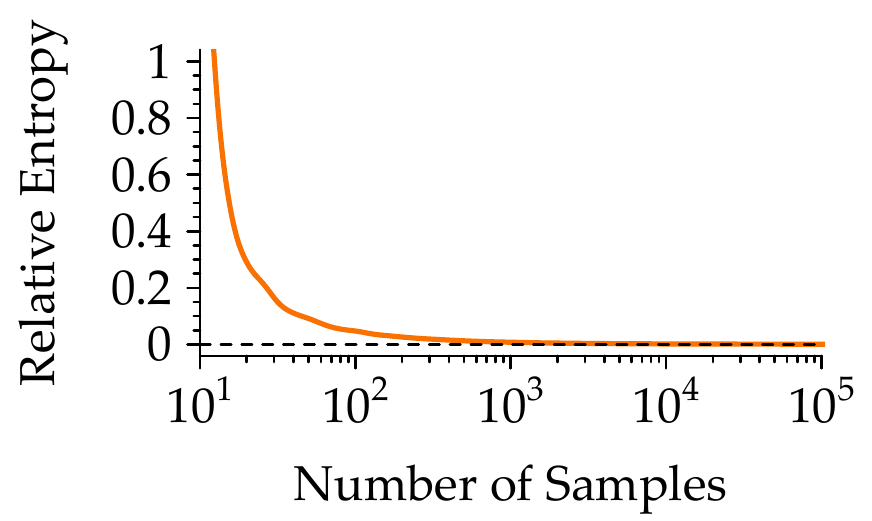}
    \caption[Relative entropy of sample normal distributions]{Left: Normal density distribution (black line) and example sample normal distributions for various sample sizes (dashed lines). 
    Right: Average divergence of sample normal distributions from the standard normal distribution as a function of sample size. 
    \label{fig:kl}}
\end{figure}

\begin{figure}
    \includegraphics[width=\textwidth]{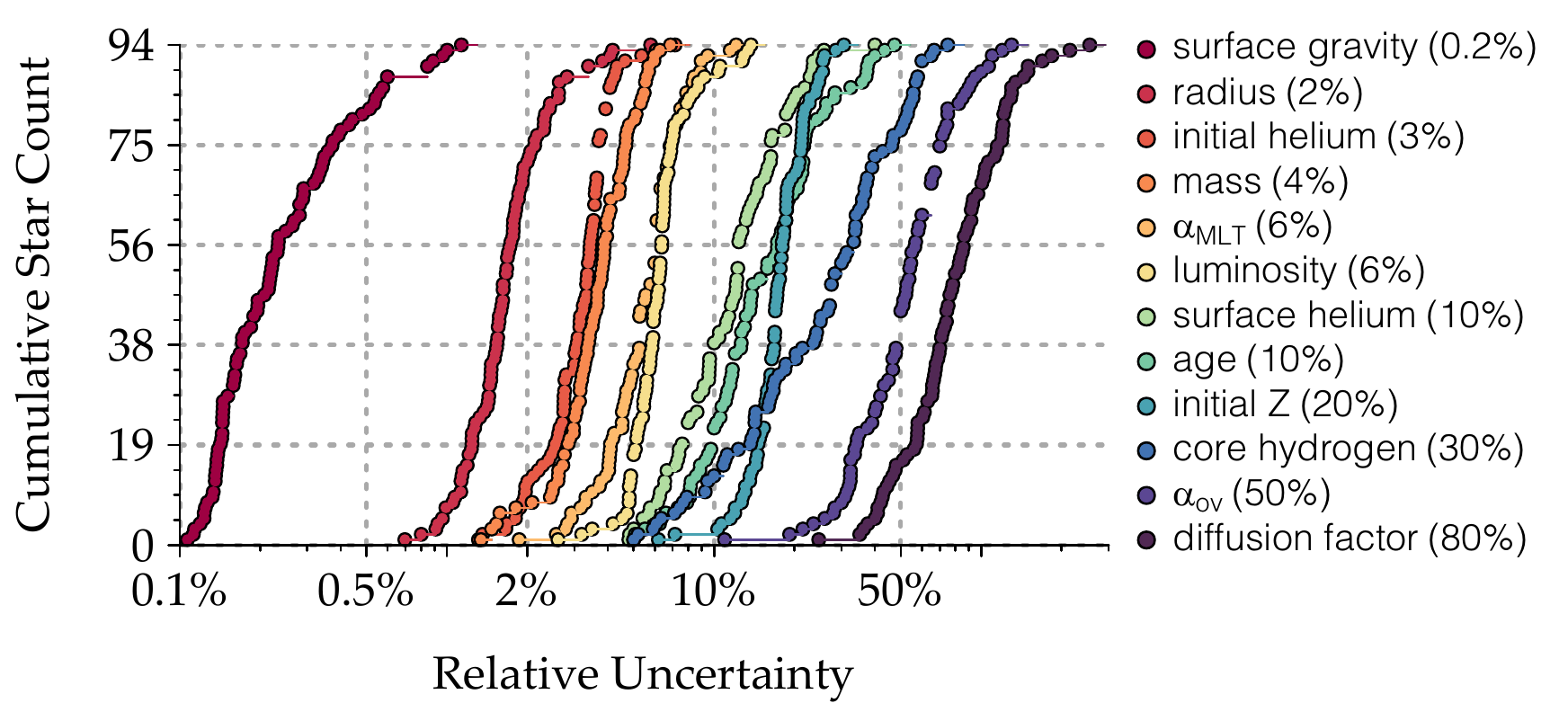}
    \caption[Relative uncertainties in estimated stellar parameters]{Cumulative distribution functions showing the relative uncertainties in estimated stellar parameters for $94$ main-sequence stars. 
    Each type of measurement is sorted by uncertainty. 
    The numbers in parentheses in the legend give the median uncertainty. 
    \emph{Figure adapted from \citealt{2017EPJWC.16005003B}.}
    \label{fig:posterior-cdf}}
\end{figure}

We seek to determine how many random samples we need to generate in order for our posterior distributions to converge to a reasonable degree to their actual distributions. 
A proxy for this would be to determine the KL divergence between the normal distribution and sample normal distributions of varying sizes. 
Figure \ref{fig:kl} shows an example of a standard normal distribution $\psi$ and sample normal densities with different sample sizes. 
The figure furthermore shows the KL divergence of these sample normal distributions as a function of sample size, averaged over $1,000$ random trials. 
The distribution converges around $10,000$ samples. 
Thus, we propagate $10,000$ random instances of the measurement uncertainty through the random forest. 
Applying the technique fleshed out in detail in Chapter~\ref{chap:ML} to $94$ stars observed by \emph{Kepler}, we find the estimates shown in Figure~\ref{fig:posterior-cdf}. 

\subsection{Structure Inversions} 
By solving the evolution inverse problem, we can obtain an evolutionary model for a given observed star. 
However, regardless of the technique used, the mode frequencies of best-fitting models generally fail to match one or more mode frequencies of the star---even after correcting surface effects. 
This implies that the structure of the star differs from the structure of the model. 
This is the starting point for the structure inversion problem. 
We seek to invert Equation~(\ref{eq:forward-surf}) to infer ${f_1(r)}$ from observed mode frequencies, for some choice of $f_1$, by deducing the difference in $f_1$ between the best-fitting model and the star. 
This problem is difficult for multiple reasons: 
\begin{description}
    \setlength{\itemindent}{0px}
    \item[Degeneracy.] As the kernels reveal, a modification to the structure anywhere in a stellar model may cause several or all of its pulsation modes to shift in their frequency of oscillation, and each frequency may shift in a different way. 
    Modifications to different locations in the stellar interior may also cause the same change to the frequency of a mode. 
    
    Furthermore, the mode frequencies are a function of multiple structural quantities. 
    When trying to infer $f_1$, we must ensure that the results are not unduly influenced by $f_2$. 
    With the present quality of asteroseismic data, this restricts us to kernel pairs with ${f_2=Y}$ (recall Section~\ref{sec:kernels}). 
    
    \item[Information Content.] 
    Whereas we are trying to measure a continuous function, which in principle may contain infinite information, we have only a finite set of mode frequencies with which to do it. 
    
    Furthermore, we will only be able to form well-localized averaging kernels in regions where a sufficient number of lower turning points are situated (recall Figure~\ref{fig:rays}). 
    This rules out some inversion methods. 
    
    \item[Stability.]  The kernel functions are nearly linearly dependent, and so the problem is ill-conditioned. 
    Even if the measurements of the mode frequencies were certain, an exact fit to mode frequencies yields highly oscillatory, non-physical solutions \citep[see, e.g.,][]{1990MNRAS.244..542D}. 
    
    \item[Surface Effects.] All of the modes are sensitive to the outermost layers of the star, where our assumptions break down (recall Sections~\ref{sec:evolution} and \ref{sec:pulsation}). 
    Thus, we must take special care to suppress surface effects. 
    However, there may be additional surface effects that the present treatment do not suppress. 
    The treatment of the surface term may furthermore erroneously subtract off more than just surface effects. 
    
    \item[Uniqueness.] The solutions are not unique. 
    From any solution to the inverse problem, a different solution can be generated 
    (see \citealt{GoughThompson1991} for a discussion). 
\end{description}
As discussed in the first section, inversion of asteroseismic data presents some novel challenges over helioseismic inversions \citep[e.g.,][]{2014aste.book...87B}. 
Unlike in helioseismology, in which the solar mass and radius are known to high precision, the masses and radii of solar-type oscillating stars are generally uncertain by at least a percent \citep[see e.g.,][see also Figure~\ref{fig:posterior-cdf}]{2013MNRAS.433.1262W,2015MNRAS.452.2127S,2016apj...830...31b}. 
Although seemingly small, such uncertainties in stellar mass and radius are generally about two orders of magnitude greater than the uncertainties in oscillation mode frequencies. 
The number of observed oscillation modes is also much smaller, and the inner radii at which these modes turn around is much more limited as well. 

The most `obvious' way to invert Equation~(\ref{eq:forward-surf}) would be via a least squares fit to the entire unknown profile. 
That is: replace the functions to be estimated by linear basis functions \citep[e.g., cubic B-splines,][]{de1972calculating}, and then select the coefficients of the basis functions such that the residuals are minimized \citep[e.g.,][]{basuchaplin2017}. 
However, this approach yields oscillatory and nonphysical solutions. 
One can then seek a regularized solution by applying, e.g., the O'Sullivan penalty \citep{o1986automatic}. 
This is a fruitful approach in global helioseismology \citep[e.g.,][]{1990MNRAS.244..542D}, where there is enough information to resolve the majority of the solar interior, to disentangle $f_1$ from $f_2$, and to suppress the surface term. 
For stars, however, there is just not enough information in current observational data for this technique to work. 

The technique of Optimally Localized Averages \citep[OLA,][]{1968GeoJ...16..169B,1970RSPTA.266..123B} provides a path forward. 
As discussed in Section~\ref{sec:history}, the idea of OLA is to linearly combine the modes in such a way that their combination is only sensitive to perturbations in one region in the star. 
Then, if the frequencies of that combination differ between model and star, then the structure of the star differs in that location. 

There are two variants of OLA that appear in the literature: Multiplicative OLA (MOLA), which is based on the original Backus--Gilbert formulation; and Subtractive OLA \citep[SOLA,][]{1992A&A...262L..33P, 1994A&A...281..231P}, which was introduced in helioseismology to reduce computational costs. 
Whereas MOLA requires a matrix inversion at each radius where an averaging kernel is sought (which, as we will see, is computationally intensive), SOLA can use the same matrix inversion for all target radii. 
This speed-up comes at the cost of an additional free parameter. 
We use SOLA to solve the structure inversion problem in Chapter~\ref{chap:inversion}. 

To solve the SOLA problem, we must find the coefficients $c$ that form the linear combination corresponding to (I) a well-localized averaging kernel, (II) a small cross-term kernel, (III) a reasonably suppressed surface term, and (IV) suitably small uncertainties. 
We thus seek to find the coefficients $c$ that minimize
\begin{equation}
    \int\left(
        \sum_i c_i K_i^{(f_1,f_2)} - T(r; r_0, \Delta)
    \right)^2
    \text{d}r
    +
    \beta \int\left(
        \sum_i K_i^{(f_2,f_1)}
    \right)^2
    \text{d}r
    +
    \mu \sum_{i,j} c_i c_j E_{i,j}
\end{equation}
where $\beta$ is a parameter controlling the cross-term kernel, $\mu$ is a parameter controlling the data uncertainties, and $\mathbf{E}$ is the error covariance matrix.
The function we wish the averaging kernel at the target radius $r_0$ to approximate is called the ``target~kernel,'' which I have denoted $T$. 
It may be chosen for example to resemble a localized Gaussian. 

Minimizing this functional amounts to solving the matrix equation ${\mathbf{A}\mathbf{x} = \mathbf{b}}$ that is shown in Equation~(\ref{eq:OLA-mat}), where $\mathbf{A}$ is a symmetric ${(N+3)\times (N+3)}$ matrix with $N$ being the number of observed modes. 
In this matrix I have introduced
\begin{align}
    \mathcal{A}_{i,j} 
    = 
    \int &\; K_i^{(f_1, f_2)} \cdot K_j^{(f_1, f_2)} 
    \; \text{d}r 
    \notag\\\;+\;\beta \int &\; K_i^{(f_2, f_1)} \cdot K_j^{(f_2, f_1)} \; \text{d}r 
    \; + \; \mu E_{i,j}
\end{align}
\begin{equation}
        y_i
        = 
        \int K_i^{(f_1, f_2)}(r) \cdot T(r; r_0, \Delta) \; \text{d}r. 
\end{equation}
Furthermore, I have introduced the Lagrange multipliers $\lambda_1$, $\lambda_2$, and $\lambda_3$ to normalize the averaging kernel and to suppress the surface term. 
Given choices of the parameters $\beta$, $\mu$, and $\Delta$, the matrix $\mathbf{A}$ may be inverted 
to yield ${\mathbf{A}^{-1}\mathbf{b}=\mathbf{x}}$, from which we may deduce ${\mathbf c(r_0)}$ and hence ${f_1(r_0)}$. 
\citet{1998esasp.418..505r,1999MNRAS.309...35R} examined the influence of each of these parameters ($\beta, \mu, \Delta$) on the inversion result. 
In Chapter~\ref{chap:inversion} we introduce a heuristic algorithm to choose these parameters. 
For further details on OLA inversions in helio/asteroseismology, see e.g., \citet{basuchaplin2017}.

\afterpage{
\begin{landscape}
\begin{figure}
    \begin{equation} \label{eq:OLA-mat}
        \begin{blockarray}{rcccccc}
             & {\color{gray} j=1}  & {\color{gray} \ldots} & {\color{gray} N} & {\color{gray} N+1} & {\color{gray} N+2} & {\color{gray} N+3} 
             \\[0.5em]
            \begin{block}{r(cccccc)}
                {\color{gray} i=1} & \mathcal{A}_{1,1}  & \cdots & \mathcal{A}_{1,N}  & \int K_1^{(f_1, f_2)} \; \text{d}r & (\nu_1/\nu_{\text{ac}})^{-2}/I_1  &  (\nu_1/\nu_{\text{ac}})^{2}/I_1  \\[0.75em]
         {\color{gray} \vdots} & \vdots             & \ddots & \vdots             & \vdots &  \vdots  &  \vdots  \\[0.75em]
              {\color{gray} N} & \mathcal{A}_{N,1}  & \cdots & \mathcal{A}_{N,N}  & \int K_N^{(f_1, f_2)} \; \text{d}r &  (\nu_N/\nu_{\text{ac}})^{-2}/I_N  &  (\nu_N/\nu_{\text{ac}})^{2}/I_N  \\[0.75em]
            {\color{gray} 1+N} & \int K_1^{(f_1, f_2)} \; \text{d}r & \cdots & \int K_N^{(f_1, f_2)} \; \text{d}r &  0  & 0  & 0  \\[0.75em]
            {\color{gray} 2+N} & (\nu_1/\nu_{\text{ac}})^{-2}/I_1  &  \cdots   &  (\nu_N/\nu_{\text{ac}})^{-2}/I_N  & 0  & 0  & 0  \\[0.75em]
            {\color{gray} 3+N} & (\nu_1/\nu_{\text{ac}})^{2} /I_1  &  \cdots   &  (\nu_N/\nu_{\text{ac}})^{2}/I_N   & 0  & 0  & 0  \\
            \end{block}\\[-5mm]
            & \BAmulticolumn{6}{c}{\color{gray} \underbrace{\hspace*{14.8cm}}_{\textstyle \mathbf{\vphantom{x}A\vphantom{b}}}} \\%
        \end{blockarray} \hspace*{3mm}
        \begin{blockarray}{c}
             \\[0.5em]
            \begin{block}{(c)}
                c_1 \vphantom{\int K_1^{(f_1)}} \\[0.75em]
                \vdots \\[0.75em]
                c_N \vphantom{\int K_1^{(f_1)}} \\[0.75em]
                \lambda_1 \vphantom{\int K_1^{(f_1)}} \\[0.75em]
                \lambda_2 \vphantom{)^{-1}/I_N} \\[0.75em]
                \lambda_3 \vphantom{)^{-1}/I_N} \\
            \end{block}\\[-5mm]
            {\color{gray} \underbrace{ }_{\textstyle \mathbf{\vphantom{A}x\vphantom{b}}}} \\%
        \end{blockarray} = 
        \begin{blockarray}{c}
            \\[0.5em]
            \begin{block}{(c)}
                y_1 \vphantom{\int K_1^{(f_1)}} \\[0.75em]
                \vdots \\[0.75em]
                y_N \vphantom{\int K_1^{(f_1)}} \\[0.75em]
                1 \vphantom{\int K_1^{(f_1)}} \\[0.75em]
                0 \vphantom{)^{-1}/I_N} \\[0.75em]
                0 \vphantom{)^{-1}/I_N}\\
            \end{block}\\[-5mm]
            {\color{gray} \underbrace{ }_{\textstyle \mathbf{\vphantom{A}b\vphantom{x}}}} \\%
        \end{blockarray} 
    \end{equation}
\end{figure}
\end{landscape}
}

As discussed earlier, the matrix $\mathbf{A}$ is ill-conditioned, and so special care must be taken when trying to obtain the least-squares solution for $\mathbf{x}$ from Equation~(\ref{eq:OLA-mat}). 
Since $\mathbf{A}$ is symmetric, we can use the LDL$^T$ decomposition \citep[e.g.,][]{banerjee2014linear}, which gives 
\begin{equation}
    \mathbf{A} = \mathbf{LDL}^{\text{T}}
\end{equation}
where $\mathbf{D}$ is a square diagonal matrix with entries \lr{${\mathbf{D}=\text{diag}(d_1, d_2, \ldots d_{N+3})}$}; and $\mathbf{L}$ is a lower unitriangular matrix, i.e.\ a matrix of the form 
\begin{equation} \label{eq:lower-diag}
    \begin{pmatrix}
        1       &   0     & 0      & \cdots & 0 \\[3mm]
        L_{2,1} &   1     & 0      & \cdots & 0 \\[3mm]
        L_{3,1} & L_{3,2} & 1      & \ddots & 0 \\[3mm]
        \vdots  & \vdots  & \ddots & \ddots & \vdots \\[3mm]
        L_{n,1} & L_{n,2} & \cdots & L_{n,m-1} & 1
    \end{pmatrix}.
\end{equation}
Substituting the LDL$^T$ decomposition of $\mathbf A$ into our matrix equation, we get
\begin{align}
    \mathbf{LDL}^{\text{T}} \mathbf{x} &= \mathbf{b} \notag\\
    \Rightarrow\; \mathbf{x} &\simeq \mathbf{LD_0}^{-1}\mathbf{L}^{\text{T}}\mathbf{b}.
\end{align}
Since $\mathbf{A}$ is ill-conditioned and hence many of its entries are very nearly zero, I have introduced the pseudo-inverse for the diagonal matrix $\mathbf{D_0}$, which gives 
\lr{\begin{equation}
    \mathbf{D_0}^{-1} = 
    \text{diag}\left( \delta_1, \delta_2, \ldots \delta_{N+3} \right)
    \qquad
    \text{where}
    \qquad
    \delta_i =
    \begin{cases}
        1/d_i & \text{if } |d_i| > t \\
        0 & \text{otherwise}
    \end{cases}
\end{equation}}
where $t$ is a small threshold (e.g., machine precision). 
In this work, I calculate the LDL$^T$ decomposition using {\textsc CHOLMOD} \citep{Chen2008Algorithm8C}. 
The cost to obtain this solution is as follows: 
\begin{itemize}
    \item LDL$^T$ decomposition: ${\mathcal{O}(N^3)}$ \citep{6710599}
    \item conversion and inversion of the diagonal matrix: ${\mathcal{O}(N)}$ 
    \item multiplication of the matrix factors: ${\mathcal{O}(N^6)}$ 
    (although there are more efficient algorithms, e.g., \citealt{COPPERSMITH1990251})
\end{itemize}
where I have here made use of the fact that the matrix is square. 
Hence, the total time complexity is dominated by the final step, yielding $\mathcal{O}(N^6)$.

\newpage
\section{Summary of Thesis}
To conclude the introduction, I will now summarize the ten most important aspects of this thesis: 

\subsection*{\hspace*{0.5cm}\hyperref[chap:ML]{Chapter~\ref{chap:ML}} \citep{2016apj...830...31b}}
\begin{enumerate}
    \item We introduce a new method based on machine learning for precisely determining the ages, masses, radii, and other properties of main sequence stars within seconds. 
    We test this method extensively, including cross-validation, hare-and-hound exercises, on the Sun, and on well-studied stars. 
    \item We apply this method to measure properties of solar-like stars whose frequencies have been resolved using data from \emph{Kepler}. 
    We find age, mass, and radius estimates with uncertainties on the order of $6\%$, $2\%$, and $1\%$, respectively. 
    \item We use this method to recover a diffusion--mass relation, which demonstrates the promise of using this approach to empirically uncover relationships in stellar physics. 
\end{enumerate}

\subsection*{\hspace*{0.5cm}\hyperref[chap:statistical]{Chapter~\ref{chap:statistical}} (\mycitealt{2017apj...839..116a})}
\begin{enumerate}
    \setcounter{enumi}{3}
    \item We systematically investigate the properties of stellar models and determine which kinds of observations of stars are important for constraining unobservable aspects of stars, such as their ages. 
    We find that metallicity measurements are independent and indispensable constraints to stellar models. 
    We furthermore quantify the increase in uncertainty for each stellar parameter that arises from increases in uncertainty of the observational data. 
    \item We analyze the expected asteroseismic yield of the forthcoming space missions TESS and PLATO for solar-like stars. 
    We find that with typical TESS data, we will be able to determine the mass and radius of a Sun-like star to better than $5\%$ uncertainty. 
    This precision will be indispensable in the search for Earth twins. 
\end{enumerate}

\subsection*{\hspace*{0.5cm}\hyperref[chap:inversion]{Chapter~\ref{chap:inversion}} \citep{2017ApJ...851...80B}}
\begin{enumerate}
    \setcounter{enumi}{5}
    \item We introduce an algorithm for inverting asteroseismic data to measure stellar structure, which takes care of imprecise radius and mass estimations and includes the automated determination of inversion parameters. 
    \item We apply our method of asteroseismic structure inversions to measure the internal isothermal speeds of sound in the cores of the solar twins 16~Cyg~A and B. 
    \item We find that in the case of 16~Cyg~B, the asteroseismic structure of the star is in good agreement with the best-fitting evolutionary model. In the case of 16~Cyg~A, however, we find less agreement. 
\end{enumerate}

\subsection*{\hspace*{0.5cm}\hyperref[chap:prospective]{Future Prospects}}
\begin{enumerate}
    \setcounter{enumi}{8}
    \item We solve the structure inverse problem for $18$ more stars, finding even greater disagreements with theoretical models of solar interiors, even when considering a variety of physics inputs. 
    These results seem to indicate that there are improvements needed in our understanding of stellar physics. 
    \item We follow the evolution of the stellar structure kernels past core hydrogen exhaustion and into the sub-giant phase of evolution. 
    We find much greater sensitivity to the deep stellar core, indicating there may soon be the prospect of learning more about the deep interior of another star than we even know about our own Sun. 
\end{enumerate}

\chapter{Fundamental Parameters of Main Sequence Stars in an Instant \\with Machine Learning}
\chaptermark{Fundamental Stellar Parameters with Machine Learning}
\label{chap:ML} 

The contents of this chapter were authored by 
E.~P.~Bellinger, G.~C.~Angelou, S.~Hekker, S.~Basu, W.~H.~Ball, and E.~Guggenberger and published in October of 2016 in \emph{The Astrophysical Journal}, 830 (1), 31.\footnote{Contribution statement: The work of this chapter was carried out by me; the text was mainly written by me, with contributions from G.~C.~Angelou, in collaboration with the other authors.} 
\nocite{2016apj...830...31b}

\section*{Chapter Summary}
Owing to the remarkable photometric precision of space observatories like \emph{Kepler}, stellar and planetary systems beyond our own are now being characterized \emph{en masse} for the first time. These characterizations are pivotal for endeavors such as searching for Earth-like planets and solar twins, understanding the mechanisms that govern stellar evolution, and tracing the dynamics of our Galaxy. The volume of data that is becoming available, however, brings with it the need to process this information accurately and rapidly. While existing methods can constrain \mb{fundamental stellar parameters such as ages, masses, and radii} from these observations, they require substantial computational efforts to do so. 

We develop a method based on machine learning for rapidly estimating fundamental parameters of main-sequence solar-like stars from classical and asteroseismic observations. We first demonstrate this method on a hare-and-hound exercise and then apply it to the Sun, 16~Cyg~A \& B, and $34$ planet-hosting candidates that have been observed by the \emph{Kepler} spacecraft. We find that our estimates and their associated uncertainties are comparable to the results of other methods, but with the additional benefit of being able to explore many more stellar parameters while using much less computation time. We furthermore use this method to present evidence for an empirical diffusion-mass relation. Our method is open source and freely available for the community to use.\footnote{The source code for all analyses and for all figures appearing in this chapter can be found electronically at \url{https://github.com/earlbellinger/asteroseismology} \citep{earl_bellinger_2016_55400}.}

\section{Introduction}

In recent years, dedicated photometric space missions have delivered dramatic improvements to time-series observations of solar-like stars. These improvements have come not only in terms of their precision, but also in their time span and sampling, which has thus enabled direct measurement of dynamical stellar phenomena such as pulsations, binarity, and activity. Detailed measurements like these place strong constraints on models used to determine the ages, masses, and chemical compositions of these stars. This in turn facilitates a wide range of applications in astrophysics, such as testing theories of stellar evolution, characterizing extrasolar planetary systems \citep[e.g.][]{2015ApJ...799..170C, 2015MNRAS.452.2127S}, assessing galactic chemical evolution \citep[e.g.][]{2015ASSP...39..111C}, and performing ensemble studies of the Galaxy \citep[e.g.][]{2011Sci...332..213C, 2013MNRAS.429..423M, 2014ApJS..210....1C}. 

The motivation to increase photometric quality has in part been driven by the goal of measuring oscillation modes in stars that are like our Sun. Asteroseismology, the study of these oscillations, provides the opportunity to constrain the ages of stars through accurate inferences of their interior structures. However, stellar ages cannot be measured directly; instead, they depend on indirect determinations via stellar modelling. 

Traditionally, to determine the age of a star, procedures based on iterative optimization (hereinafter IO) seek the stellar model that best matches the available observations \citep{1994ApJ...427.1013B}. 
Several search strategies have been employed, including exploration through a pre-computed grid of models (i.e.\ grid-based modelling, hereinafter GBM; see \citealt{2011ApJ...730...63G, 2014ApJS..210....1C}); or \emph{in situ} optimization (hereinafter ISO) such as genetic algorithms \citep{2014ApJS..214...27M}, Markov-chain Monte Carlo \citep{2012MNRAS.427.1847B}, or the downhill simplex algorithm (\citealt{2013apjs..208....4p}; see e.g.\ \citealt{2015MNRAS.452.2127S} for an extended discussion on the various methods of dating stars). Utilizing the detailed observations from the \emph{Kepler} and CoRoT space telescopes, these procedures have constrained the ages of several field stars to within $10\%$ of their main-sequence lifetimes \citep{2015MNRAS.452.2127S}. 

IO is computationally intensive in that it demands the calculation of a large number of stellar models (see \citealt{2009ApJ...699..373M} for a discussion). ISO requires that new stellar tracks are calculated for each target, as they do not know \emph{a priori} all of the combinations of stellar parameter values that the optimizer will need for its search. They furthermore converge to local minima and therefore need to be run multiple times from different starting points to attain global coverage. GBM by way of interpolation in a high-dimensional space, on the other hand, is sensitive to the resolution of each parameter and thus requires a very fine grid of models to search through \citep[see e.g.][who use more than five million models that were varied in just four initial parameters]{2010ApJ...725.2176Q}. Additional dimensions such as efficiency parameters (e.g.\ overshooting or mixing length parameters) significantly impact on the number of models needed and hence the search times for these methods. As a consequence, these approaches typically use, for example, a solar-calibrated mixing length parameter or a fixed amount of convective overshooting. Since these values in other stars are unknown, keeping them fixed therefore results in underestimations of uncertainties. This is especially important in the case of atomic diffusion, which is essential when modelling the Sun \citep[see e.g.][]{1994MNRAS.269.1137B}, but is usually disabled for stars with ${M/M_\odot > 1.4}$ because it leads to the unobserved consequence of a hydrogen-only surface \citep{2002A&A...390..611M}. 

These concessions have been made because the relationships connecting \mb{observations} of stars to their internal \mb{properties} are non-linear and difficult to characterize. Here we will show that through the use of machine learning, it is possible to avoid these difficulties by capturing those relations statistically and using them to construct a regression model capable of relating observations of stars to their structural, chemical, and evolutionary properties. The relationships can be learned using many fewer models than IO methods require, and can be used to process entire stellar catalogs with a cost of only seconds per star. 

To date, only about a hundred solar-like oscillators have had their frequencies resolved, allowing each of them be modelled in detail using costly methods based on IO. In the forthcoming era of TESS \citep{2015JATIS...1a4003R} and PLATO \citep{2014ExA....38..249R}, however, seismic data for many more stars will become available, and it will not be possible to dedicate large amounts of supercomputing time to every star. Furthermore, for many stars, it will only be possible to resolve \emph{global} asteroseismic quantities rather than individual frequencies. Therefore, the ability to rapidly constrain stellar parameters for large numbers of stars by means of global oscillation analysis will be paramount. 

In this work, we consider the constrained multiple-regression problem of inferring fundamental stellar \mb{parameters} from observable \mb{quantities}. We construct a random forest of decision tree regressors to learn the relationships connecting observable quantities of main-sequence (MS) stars to their zero-age main-sequence (ZAMS) histories and  current-age structural and chemical attributes. We validate our technique by inferring the parameters of simulated stars in a hare-and-hound exercise, the Sun, and the well-studied stars 16~Cyg~A~and~B. Finally, we conclude by applying our method on a catalog of \emph{Kepler} objects-of-interest (hereinafter KOI; \citealt{2016MNRAS.456.2183D}).  

We explore various model physics by considering stellar evolutionary tracks that are varied not only in their initial mass and chemical composition, but also in their efficiency of convection, extent of convective overshooting, and strength of gravitational settling. We compare our results to the recent findings from GBM \citep{2015MNRAS.452.2127S}, ISO \citep{2015ApJ...811L..37M}, interferometry \citep{2013MNRAS.433.1262W}, and asteroseismic glitch analyses \citep{2014ApJ...790..138V} and find that we obtain similar estimates but with orders-of-magnitude speed-ups.

\section{Method} \label{sec:Method} 
We seek a multiple-regression model capable of characterizing observed stars. To obtain such a model, we build a matrix of evolutionary simulations and use machine learning to discover relationships in the \mb{stellar models} that connect observable quantities of stars to the model quantities that we wish to predict. \mb{The matrix is structured such that each column contains a different stellar quantity and each row contains a different stellar model.} We construct this matrix by extracting models along evolutionary sequences (see Appendix \ref{sec:selection} for details on the model selection process) and summarizing them to yield the same types of information as the stars being observed. Although each star (and each stellar model) may have a different number of \mb{oscillation} modes observed, it is possible to condense this information into only a few numbers by leveraging the fact that the frequencies of these modes follow a regular pattern \citep[for a review of solar-like oscillations, see][]{2013ARA&A..51..353C}. Once the machine has processed this matrix, one can feed the algorithm a catalogue of \mb{stellar observations} and use it to predict the \mb{fundamental} parameters of those stars.

The \mb{observable information obtained from models that can be} used to inform the algorithm may include, but \mb{is} not limited to, combinations of temperatures, metallicities, global oscillation information, surface gravities, luminosities, and/or radii. From these, the machine can learn how to infer stellar parameters such as ages, masses, core hydrogen and surface helium abundances. If luminosities, surface gravities, and/or radii are not supplied, then they may be predicted as well. In addition, the machine can also infer evolutionary parameters such as the initial stellar mass and initial chemical compositions as well as the mixing length parameter, overshoot coefficient, and diffusion multiplication factor needed to reproduce observations, which are explained in detail below. 

\subsection{Model Generation}
\label{sec:models}
We use the open-source 1D stellar evolution code \emph{Modules for Experiments in Stellar Astrophysics} \citep[MESA;][]{2011apjs..192....3p} to generate main-sequence stellar models from solar-like evolutionary tracks varied in initial mass $M$, helium $Y_0$, metallicity $Z_0$, mixing length parameter $\alpha_{\text{MLT}}$, overshoot coefficient $\alpha_{\text{ov}}$, and \mb{diffusion multiplication factor} $D$. \mb{The diffusion multiplication factor} serves to amplify or diminish the effects of diffusion, where a value of zero turn\mb{s} it off and a value of two double\mb{s} all velocities. The initial conditions are varied in the ranges ${M\in [0.7, 1.6]\;M_\odot}$, ${Y_0\in [0.22, 0.34]}$, ${Z_0\in [10^{-5}, 10^{-1}]}$ (varied logarithmically), ${\alpha_{\text{MLT}}\in [1.5, 2.5]}$, ${\alpha_{\text{ov}}\in [10^{-4}, 1]}$ (varied logarithmically), and ${D\in [10^{-6}, 10^2]}$ (varied logarithmically). We put a cut-off of $10^{-3}$ and $10^{-5}$ on $\alpha_{\text{ov}}$ and $D$, respectively, below which we consider them to be zero and \mb{disable them}. The initial parameters of each track are chosen in a quasi-random fashion so as to populate the initial-condition hyperspace as homogeneously and rapidly as possible (shown in Figure~\ref{fig:inputs}; see Appendix \ref{sec:grid} for more details). 

\begin{landscape}
\begin{figure}
    \centering
    \includegraphics[width=\linewidth, keepaspectratio]{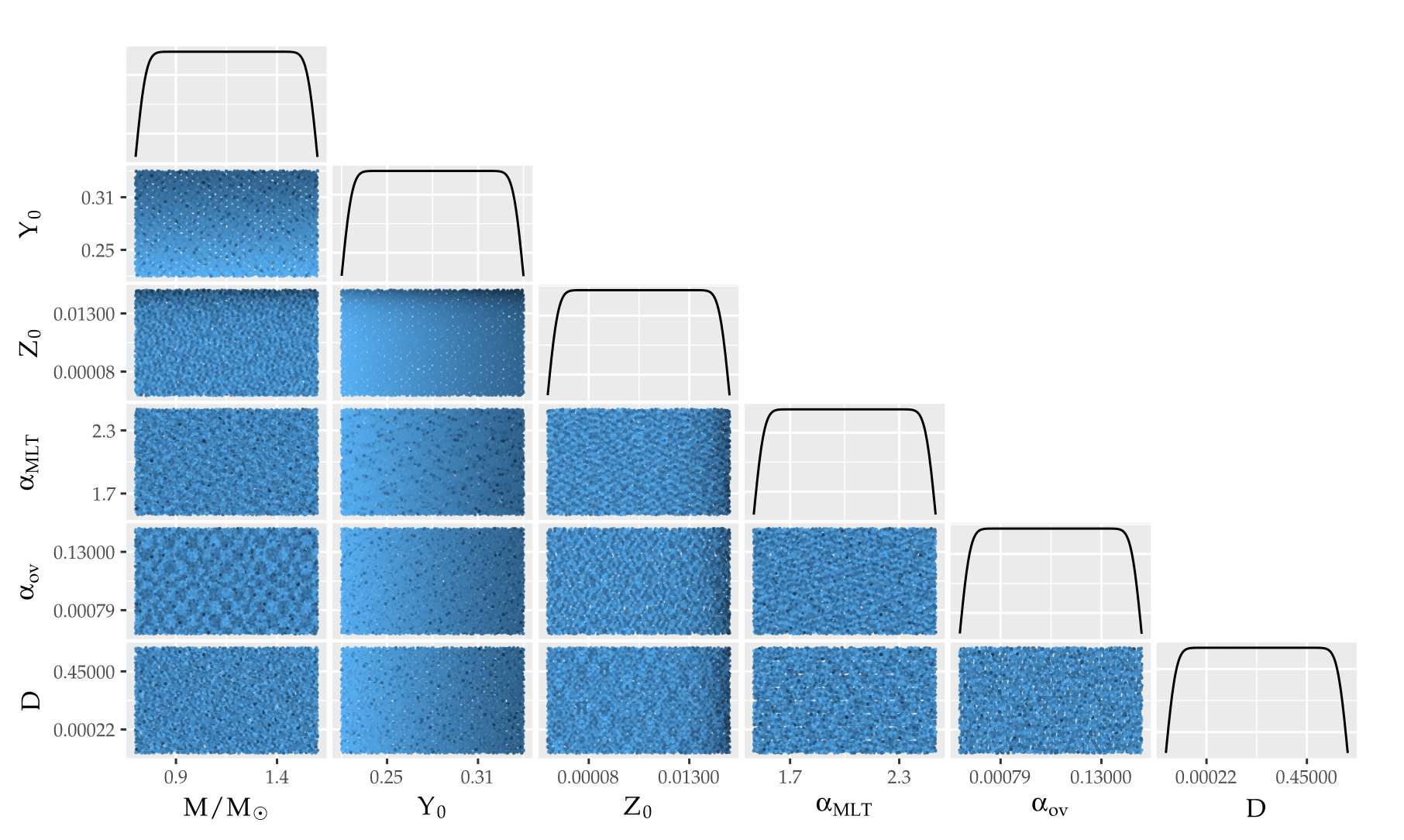}
    \caption[Initial conditions for evolutionary model grid]{(Caption on other page.)\label{fig:inputs}}
\end{figure}
\end{landscape}
\begin{figure}
    \contcaption{Scatterplot matrix (lower panels) and density plots (diagonal) of evolutionary track initial conditions considered. Mass ($M$), initial helium ($Y_0$), initial metallicity ($Z_0$), mixing length parameter ($\alpha_{\text{MLT}}$), overshoot ($\alpha_{\text{ov}}$), and diffusion multiplication factor ($D$) were varied in a quasi-random fashion to obtain a low-discrepancy grid of model tracks. Points are colored by their initial hydrogen ${X_0=1-Y_0-Z_0}$, with blue being \mb{high} $X_0$ (${\approx 78\%}$) and black being \mb{low} $X_0$ (${\approx 56\%}$). The parameter space is densely populated with evolutionary tracks of maximally different initial conditions.}
\end{figure}

We use MESA version r8118 with the Helmholtz-formulated equation of state that allows for radiation pressure and interpolates within the 2005 update of the OPAL EOS tables \citep{2002apj...576.1064r}. We assume a \citet{1998SSRv...85..161G} solar composition for our initial abundances and opacity tables. Since we restrict our study to the main sequence, we use an eight-isotope nuclear network consisting of $^1$H, $^3$He, $^4$He, $^{12}$C, $^{14}$N, $^{16}$O, $^{20}$Ne, and $^{24}$Mg. We use a step function for overshooting and set a scaling factor ${f_0 = \alpha_{\text{ov}}/5}$ to determine the radius ${r_0 = H_p \cdot f_0}$ inside the convective zone at which convection switches to overshooting, where $H_p$ is the pressure scale height. \mb{The overshooting parameter applies to all convective boundaries and is kept fixed throughout the course of a track's evolution, so a non-zero value does not imply that the model has a convective core at any specific age.} 
All pre-main-sequence (PMS) models are calculated with a simple photospheric approximation, after which an Eddington $T-\tau$ atmosphere is appended on at ZAMS. We call ZAMS the point at which the nuclear luminosity of the models make up $99.9\%$ of the total luminosity. We calculate atomic diffusion with gravitation settling and without radiative levitation on the main sequence using five diffusion class representatives: $^1$H, $^3$He, $^4$He, $^{16}$O, and $^{56}$Fe \citep{burgers1969flow}.\footnote{The atomic number of each representative isotope is used to calculate the diffusion rate of the other isotopes allocated to that group; see \citet{2011apjs..192....3p}.} 
Following their most recent measurements, we correct the defaults in MESA of the gravitational constant (${G=6.67408\times 10^{-8}}$~\si{\per\g\cm\cubed\per\square\s}; \citealt{2015arXiv150707956M}), the gravitational mass of the Sun (${M_\odot = 1.988475\times 10^{33}}$~\si{\g}~${= \mu G^{-1} = 1.32712440042\times 10^{11}}$~\si{\km\per\s}~$G^{-1}$, where $\mu$ is the standard gravitational parameter; \citealt{pitjeva2015determination}), and the solar radius (${R_\odot = 6.95568\times 10^{10}}$~\si{\cm}; \citealt{2008ApJ...675L..53H}). 

Each track is evolved from ZAMS to either an age of ${\tau=16}$ Gyr or until terminal-age main sequence (TAMS), which we define as having a fractional core hydrogen abundance ($X_{\text{c}}$) below $10^{-3}$. Evolutionary tracks with efficient heavy-element settling can develop discontinuities in their surface abundances if they lack sufficient model resolution. We implement adaptive remeshing by recomputing any track with abundance discontinuities in its surface layers using finer spatial and temporal resolutions (see Appendix \mb{\ref{sec:remeshing}} for details). Running stellar physics codes in a batch mode like this requires care, so we manually inspect multiple evolutionary diagnostics to ensure that proper convergence has been achieved. 


\subsection{Calculation of Seismic Parameters}
\label{sec:seis}
We use the ADIPLS pulsation package \citep{2008Ap&SS.316..113C} to compute p-mode oscillations up to spherical degree ${\ell=3}$ below the acoustic cut-off frequency. We use on average of around $4,000$ points per stellar model and therefore have adequate resolution to calculate frequencies without remeshing. We denote any frequency separation $S$ as the difference between a frequency $\nu$ of spherical degree $\ell$ and radial order $n$ and another frequency, that is: 
\begin{equation} 
  S_{(\ell_1, \ell_2)}(n_1, n_2) \equiv \nu_{\ell_1}(n_1) - \nu_{\ell_2}(n_2).
\end{equation}
The large frequency separation is then
\begin{equation} 
  \Delta\nu_\ell(n) \equiv S_{(\ell, \ell)}(n, n-1)
\end{equation}
and the small frequency separation is
\begin{equation}
  \delta\nu_{(\ell, \ell+2)}(n) \equiv S_{(\ell, \ell+2)}(n, n-1).
\end{equation}
Near-surface layers of stars are poorly-modeled, which induces systematic frequency offsets \citep[see e.g.][]{1999A&A...351..689R}. The ratios between the large and small frequency separations (Equation~\ref{eqn:LSratio}), and also between the large frequency separation and five-point-averaged frequencies (Equation~\ref{eqn:rnl}) have been shown to be less sensitive to the surface term than the aforementioned separations and are therefore valuable asteroseismic diagnostics of stellar interiors \citep{2003A&A...411..215R}. They are defined as
\begin{equation} 
  \mathrm{r}_{(\ell,\ell+2)}(n) \equiv \frac{\delta\nu_{(\ell, \ell+2)}(n)}{\Delta\nu_{(1-\ell)}(n+\ell)} \label{eqn:LSratio}
\end{equation}
\begin{equation} 
  \mathrm{r}_{(\ell, 1-\ell)}(n) \equiv \frac{\mathrm{dd}_{(\ell,1-\ell)}(n)}{\Delta\nu_{(1-\ell)}(n+\ell)} \label{eqn:rnl}
\end{equation}
where
\begin{align} 
  \mathrm{dd}_{0,1}(n) \equiv \frac{1}{8} \big[\nu_0(n-1) &- 4\nu_1(n-1) 
                                 +6\nu_0(n) \notag\\&- 4\nu_1(n) +  \nu_0(n+1)\big]\\ 
  \mathrm{dd}_{1,0}(n) \equiv -\frac{1}{8} \big[\nu_1(n-1) &- 4\nu_0(n) 
                                 +6\nu_1(n) \notag\\&- 4\nu_0(n+1) + \nu_1(n+1)\big].
\end{align}
Since the set of radial orders that are observable differs from star to star, we collect global statistics on $\Delta\nu_0$, $\delta\nu_{0,2}$, $\delta\nu_{1,3}$, $r_{0,2}$, $r_{1,3}$, $r_{0,1}$, and $r_{1,0}$. We mimic the range of observable frequencies in our models by weighting all frequencies by their position in a Gaussian envelope centered at the predicted frequency of maximum oscillation power $\nu_{\max}$ and having full-width at half-maximum of ${0.66\cdot\nu_{\max}{}^{0.88}}$ as per the prescription given by \citet{2012A&A...537A..30M}. We then calculate the weighted median of each variable, which we denote with angled parentheses (e.g.\ $\langle r_{0,2}\rangle$). We choose the median rather than the mean because it is a robust statistic with a high breakdown point, meaning that it is much less sensitive to the presence of outliers (for a discussion of breakdown points, see \citealt{hampel1971general}, who attributed them to Gauss). This approach allows us to predict the fundamental \mb{stellar} parameters of any solar-like oscillator with multiple observed modes irrespective of which exact radial orders have been detected. Illustrations of the methods used to derive the frequency separations and ratios of a stellar model are shown in Figure~\ref{fig:ratios}. 

\afterpage{
\begin{landscape}
\begin{figure}
    \centering
    \includegraphics[width=0.45\linewidth,keepaspectratio]{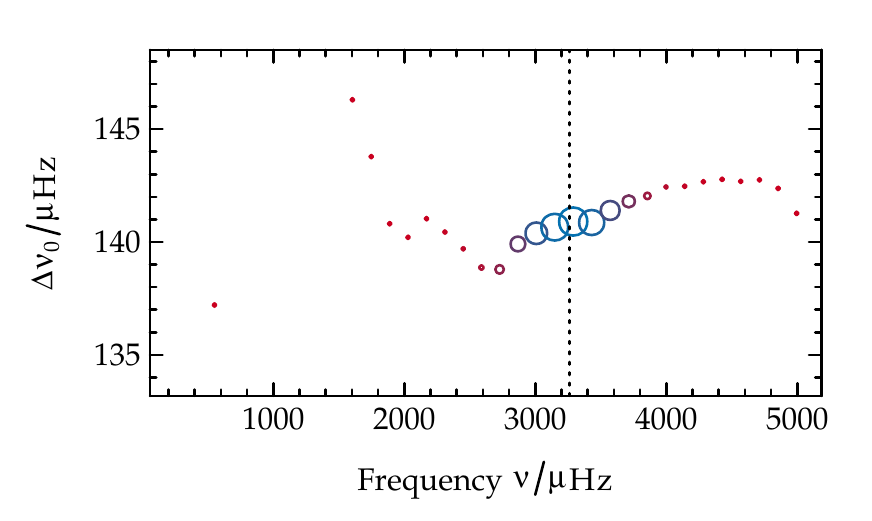}%
    \includegraphics[width=0.45\linewidth,keepaspectratio]{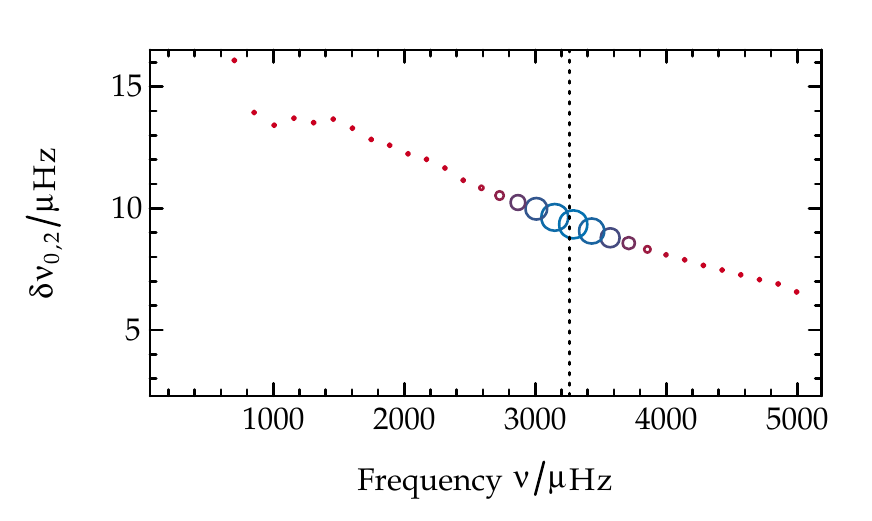}\\
    \includegraphics[width=0.45\linewidth,keepaspectratio]{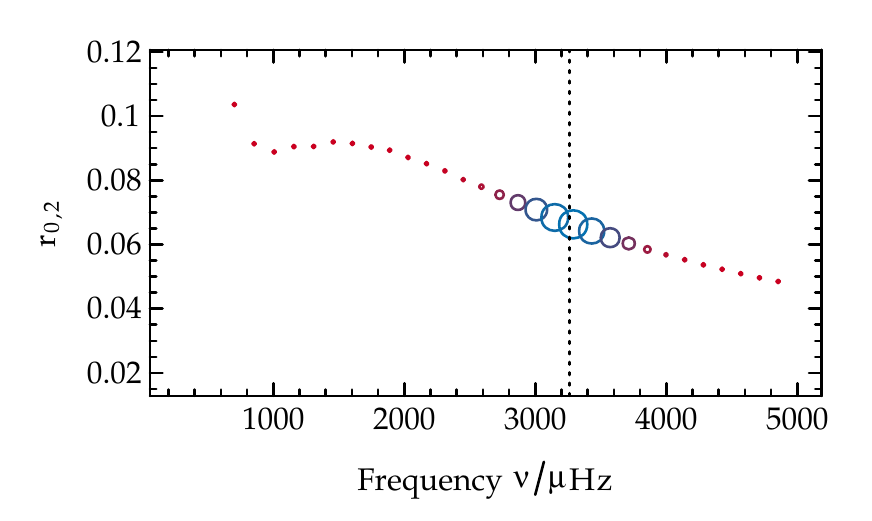}%
    \includegraphics[width=0.45\linewidth,keepaspectratio]{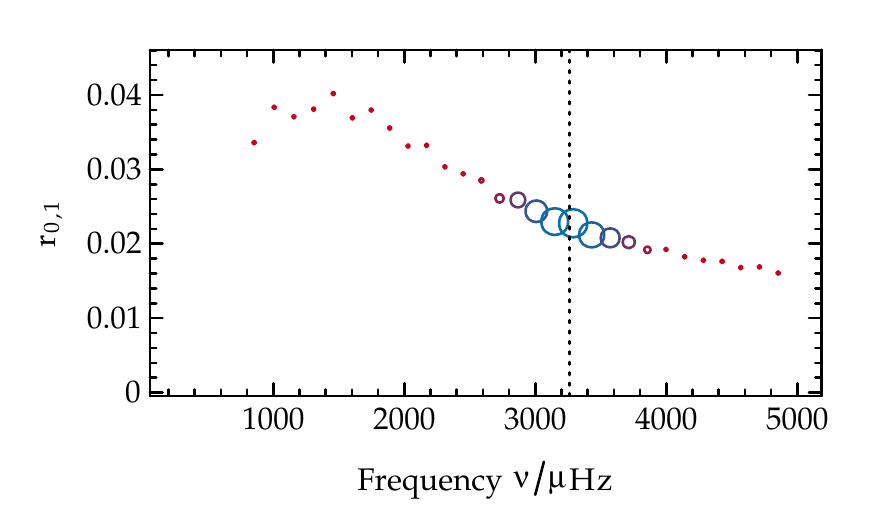}\\
    \caption[Seismic parameters of a stellar model]{Calculation of seismic parameters for a stellar model. 
    The large and small frequency separations $\Delta\nu_0$ (top left) and $\delta\nu_{0,2}$ (top right) and frequency ratios $r_{0,2}$ (bottom left) and $r_{0,1}$ (bottom right) are shown as a function of frequency. The vertical dotted line in these bottom four plots indicates $\nu_{\max}$. Points are sized and colored proportionally to the applied weighting\mb{, with large blue symbols indicating high weight and small red symbols indicating low weight.} }%
    \label{fig:ratios}
\end{figure}
\end{landscape}
}

\subsection{Training the Random Forest} \label{sec:forest}
We train a random forest regressor on our matrix of evolutionary models to discover the relations that facilitate inference of stellar parameters from \mb{observed} quantities. A schematic representation of the topology of our random forest regressor can be seen in Figure~\ref{fig:rf}. \mb{Random forests arise in machine learning through the family of algorithms known as CART, i.e. Classification and Regression Trees.} There are several good textbooks that discuss random forests \citep[see e.g.][Chapter 15]{hastie2005elements}. \mb{A random forest is an ensemble regressor, meaning that it is composed of many individual components that each perform statistical regression, and the forest subsequently averages over the results from each component \citep{breiman2001random}. The components of the ensemble are decision trees, each of which learns a set of decision rules for relating \mb{observable quantities} to \mb{stellar parameters}. An ensemble approach is preferred because using only a single decision tree that is able to see all of the training data may result in a regressor that has memorized the training data and is therefore unable to generalize to as yet unseen values. This undesirable phenomenon is known in machine learning as over-fitting, and is analogous to fitting $n$ data points using a degree $n$ polynomial: the fit will work perfectly on the data that was used for fitting, but fail badly on any unseen data. To avoid this, each decision tree in the forest is given a random subset of the evolutionary models and a random subset of the observable quantities from which to build a set of rules relating observed quantities to stellar parameters. This process, known as statistical bagging \citep[][Section~8.7]{hastie2005elements}, prevents the collection of trees from becoming over-fit to the training data, and thus results in a regression model that is capable of generalizing the information it has learned and predicting values for data on which it has not been trained. } 

\afterpage{
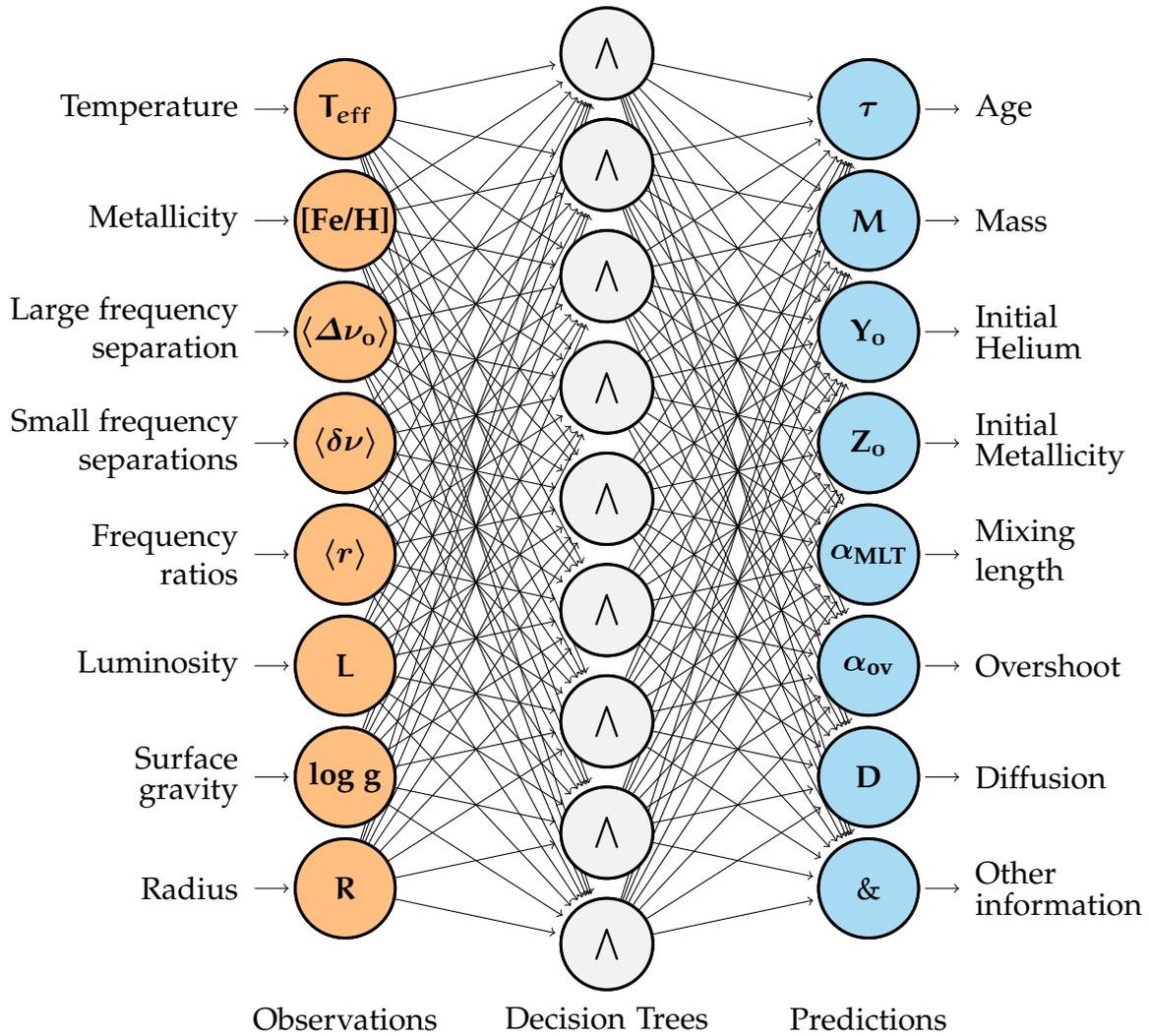
\begin{figure*}[ht]
    \centering
    \begin{adjustwidth*}{}{-0.35cm}
    \centering


\definecolor{mDarkBrown}{HTML}{604c38}
\definecolor{mDarkTeal}{HTML}{23373b}
\definecolor{DodgerBlue}{HTML}{1E90FF}
\definecolor{DeepPurple}{HTML}{800080}
\definecolor{mLightBrown}{HTML}{EB811B}
\definecolor{mMediumBrown}{HTML}{C87A2F}
\definecolor{bubi}{HTML}{FF8000}
\definecolor{awesome}{HTML}{0395DE}

\begin{tikzpicture}
    [   shorten >=2pt,
        ->,
        draw=black,
        node distance=7cm, 
        every node/.style={font=\normalsize}
    ]
    \tikzstyle{every pin edge}=[<-,shorten <=2pt];
    \tikzstyle{neuron}=[circle,fill=black!25,minimum size=38pt,inner sep=0pt, line width=0.4mm, draw=black];
    \tikzstyle{input neuron}=[neuron, fill=bubi!50!white,
    text=black];
    \tikzstyle{output neuron}=[neuron, fill=awesome!35!white, text=black];
    \tikzstyle{hidden neuron}=[neuron, fill=black!5, text=black, minimum size=35pt];
    \tikzstyle{annot}=[text width=1.5in, text centered];
    
    \tikzstyle{path}=[inner sep=0pt];

    \node[input neuron, pin=left:Temperature] (I-1) at (0,-1) {};
    \node[input neuron, pin=left:Metallicity] (I-2) at (0,-2.5) {};
    \node[input neuron, pin=left:{\shortstack[r]{Large frequency\\separation}}] (I-3) at (0,-4) {};
    \node[input neuron, pin=left:{\shortstack[r]{Small frequency\\separations}}] (I-4) at (0,-5.5) {};
    \node[input neuron, pin=left:{\shortstack[r]{Frequency\\ratios}}] (I-5) at (0,-7) {};
    \node[input neuron, pin=left:Luminosity] (I-6) at (0,-8.5) {};
    \node[input neuron, pin=left:{\shortstack[r]{Surface\\gravity}}] (I-7) at (0,-10) {};
    \node[input neuron, pin=left:Radius] (I-8) at (0,-11.5) {};

    \foreach \name / \y in {1,...,9}
        \path[yshift=1.25cm]
            node[hidden neuron] (H-\name) at (3.5cm,{-\y*1.5 cm}) {$\bigwedge$};
    
    \node[output neuron,pin={[pin edge={->}]right:Age}] (O-1) at (7cm, -1 cm) {};
    \node[output neuron,pin={[pin edge={->}]right:Mass}] (O-2) at (7cm, -2.5 cm) {};
    \node[output neuron,pin={[pin edge={->}]right:\shortstack[l]{Initial\\Helium}}] (O-3) at (7cm, -4 cm) {};
    \node[output neuron,pin={[pin edge={->}]right:\shortstack[l]{Initial\\Metallicity}}] (O-4) at (7cm, -5.5 cm) {};
    \node[output neuron,pin={[pin edge={->}]right:{\shortstack[l]{Mixing\\length}}}] (O-5) at (7cm, -7 cm) {};
    \node[output neuron,pin={[pin edge={->}]right:Overshoot}] (O-6) at (7cm, -8.5 cm) {};
    \node[output neuron,pin={[pin edge={->}]right:Diffusion}] (O-7) at (7cm, -10 cm) {};
    \node[output neuron,pin={[pin edge={->}]right:{\shortstack[l]{Other\\information}}}] (O-8) at (7cm, -11.5 cm) {};
    
    \foreach \source in {1,...,8}
        \foreach \dest in {1,...,9}
            \path (I-\source) edge (H-\dest);
            
    \foreach \source in {1,...,9}
        \foreach \dest in {1,...,8}
            \path (H-\source) edge (O-\dest);
    
    \node[input neuron] (I-1) at (0,-1) {$\mathbf{T_{\mathbf{eff}}}$};
    \node[input neuron] (I-2) at (0,-2.5) {\textbf{[Fe/H]}};
    \node[input neuron] (I-3) at (0,-4) {$\bm{\langle\Delta\nu_0\rangle}$};
    \node[input neuron] (I-4) at (0,-5.5) {$\bm{\langle\delta\nu\rangle}$};
    \node[input neuron] (I-5) at (0,-7) {$\bm{\langle r\mathbf\rangle}$};
    \node[input neuron] (I-6) at (0,-8.5) {$\mathbf{L}$};
    \node[input neuron] (I-7) at (0,-10) {\textbf{log~g}};
    \node[input neuron] (I-8) at (0,-11.5) {$\mathbf{R}$};
    
    \foreach \name / \y in {1,...,9}
        \path[yshift=1.25cm]
            node[hidden neuron] (H-\name) at (3.5cm,{-\y*1.5 cm}) {$\bigwedge$};
    
    \node[output neuron] (O-1) at (7cm, -1 cm) {$\bm\tau$};
    \node[output neuron] (O-2) at (7cm, -2.5 cm) {$\mathbf{M}$};
    \node[output neuron] (O-3) at (7cm, -4 cm) {$\textbf{Y}_{\bm{0}}$};
    \node[output neuron] (O-4) at (7cm, -5.5 cm) {$\textbf{Z}_{\bm{0}}$};
    \node[output neuron] (O-5) at (7cm, -7 cm) {$\bm{\alpha_{\textbf{\text{MLT}}}}$};
    \node[output neuron] (O-6) at (7cm, -8.5 cm) {$\bm{\alpha_{\textbf{\text{ov}}}}$};
    \node[output neuron] (O-7) at (7cm, -10 cm) {\textbf{D}};
    \node[output neuron] (O-8) at (7cm, -11.5 cm) {$\bm{\&}$};
    
    \node[annot, below of=H-9, node distance=1cm] (hl) {Decision Trees};
    \node[annot, left of=hl, node distance=3.5cm] {Observations};
    \node[annot, right of=hl, node distance=3.5cm] {Predictions};
    
\end{tikzpicture}

    \end{adjustwidth*}
    \caption[Random Forest]{A schematic representation of a random forest regressor for inferring fundamental stellar parameters. \mb{Observable quantities} such as \mb{$T_{\text{eff}}$ and [Fe/H]} and global asteroseismic quantities like \mb{$\langle\Delta\nu\rangle$ and} $\langle\delta\nu_{0,2}\rangle$ are input on the left side. These quantities are then fed through to some number of hidden decision trees, which each independently predict \mb{parameters} like age and mass. The predictions are then averaged and output on the right side. All inputs and outputs are optional. For example, surface gravities, luminosities, and radii are not always available \mb{from observations} (e.g.\ with the KOI stars\mb{, see Section~\ref{sec:koi} below}). In their absence, these quantities can be predicted instead of being supplied. In this case, those nodes can be moved over to the ``prediction'' side instead of being on the ``observations'' side. Also, in addition to potentially unobserved inputs like stellar radii, other interesting model parameters can be predicted as well, such as core hydrogen mass fraction or surface helium abundance. \label{fig:rf} }
\end{figure*}
}

\subsubsection*{Feature Importance} \label{sec:importances}

\mb{The CART algorithm uses} information theory to decide which rule is the best choice for inferring \mb{stellar parameters} like age and mass from the supplied information \citep[][Chapter 9]{hastie2005elements}. At every stage, the rule that creates the largest decrease in mean squared error (MSE) is crafted. A rule may be, for example, ``all models with ${L <0.4\;L_\odot}$ have ${M< 1\;M_\odot}$.'' Rules are created until every \mb{stellar model} that was supplied to that particular tree is fully explained by a sequence of decisions. We moreover use a variant on random forests known as \emph{extremely} randomized trees \citep{geurts2006extremely}, which further randomize attribute splittings (e.g.\ split on L) and the location of the cut-point (e.g.\ split on ${0.4 \; L/L_\odot}$) used when creating decision rules. 

\begin{figure}
    \centering
    \includegraphics[width=300.887892pt, keepaspectratio]{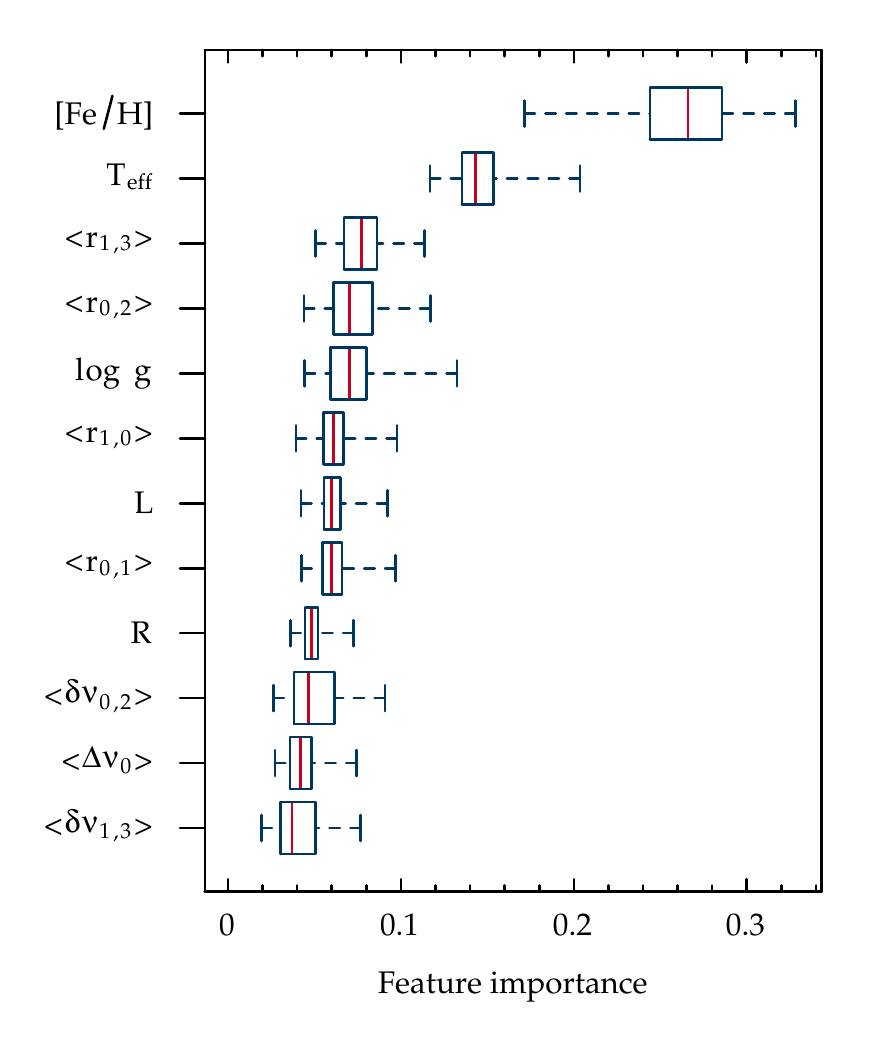}
    \caption[Feature Importances]{Box-and-whisker plots of relative importance for each observable feature in inferring fundamental stellar parameters as measured by a random forest regressor grown from a grid of evolutionary models. The boxes display the first ($16\%$) and third ($84\%$) quartile of feature importance over all trees, the center line indicates the median, and the whiskers extend to the most extreme values.}
    \label{fig:importances}
\end{figure}

\mb{The process of constructing a random forest} presents an opportunity for not only inferring stellar parameters from observations, but also for understanding the relationships that exist in the \mb{stellar models}. Each decision tree explicitly ranks the relative ``importance'' of each observable quantity \mb{for inferring stellar parameters}, where importance is defined in terms of both the reduction in MSE after defining a decision rule based on that quantity and the number of models that use that rule. \mb{In machine learning, the variables that have been measured and are supplied as inputs to the algorithm are known as ``features.'' Figure~\ref{fig:importances} shows a feature importance plot, i.e.~distributions of relative importance over all of the trees in the forest for each feature used to infer stellar parameters. The features that are used most often to construct decision rules are metallicity and temperature, which are each significantly more important features than the rest.} The importance of [Fe/H] is due to the fact that the determinations of quantities like the $Z_0$ and $D$ depend nearly entirely on it \citep[see also][]{2017apj...839..116a}. Note that importance does not indicate indispensability: an appreciable fraction of decision rules being made based off of \mb{one feature} does not mean that another forest without that \mb{feature} would not perform just as well. That being said, these results indicate that the best area to improve measurements would be in metallicity determinations, because for stars being predicted using this random forest, less precise values here means exploring many more paths and hence arriving at less certain predictions. 

For many stars, \mb{stellar quantities} such as radii, luminosities, surface gravities, and/or oscillation modes with spherical degree ${\ell=3}$ are not available from observations. For example, the KOI data set \mb{(see Section~\ref{sec:koi} below)} lacks all of this information, and the hare-and-hound exercise data \mb{(see Section~\ref{sec:hnh} below)} lack all of these except luminosities. We therefore must train random forests that predict those quantities instead of using them as \mb{features}. We show the relative importance for \mb{the remaining features that were} used to train these forests in Figure~\ref{fig:importances2}. When ${\ell=3}$ modes and luminosities are omitted, effective temperature jumps in importance and ties with [Fe/H] as the most \mb{important feature}. 

\afterpage{
\begin{landscape}
\begin{figure}
    \centering
    \includegraphics[width=0.45\linewidth,keepaspectratio]{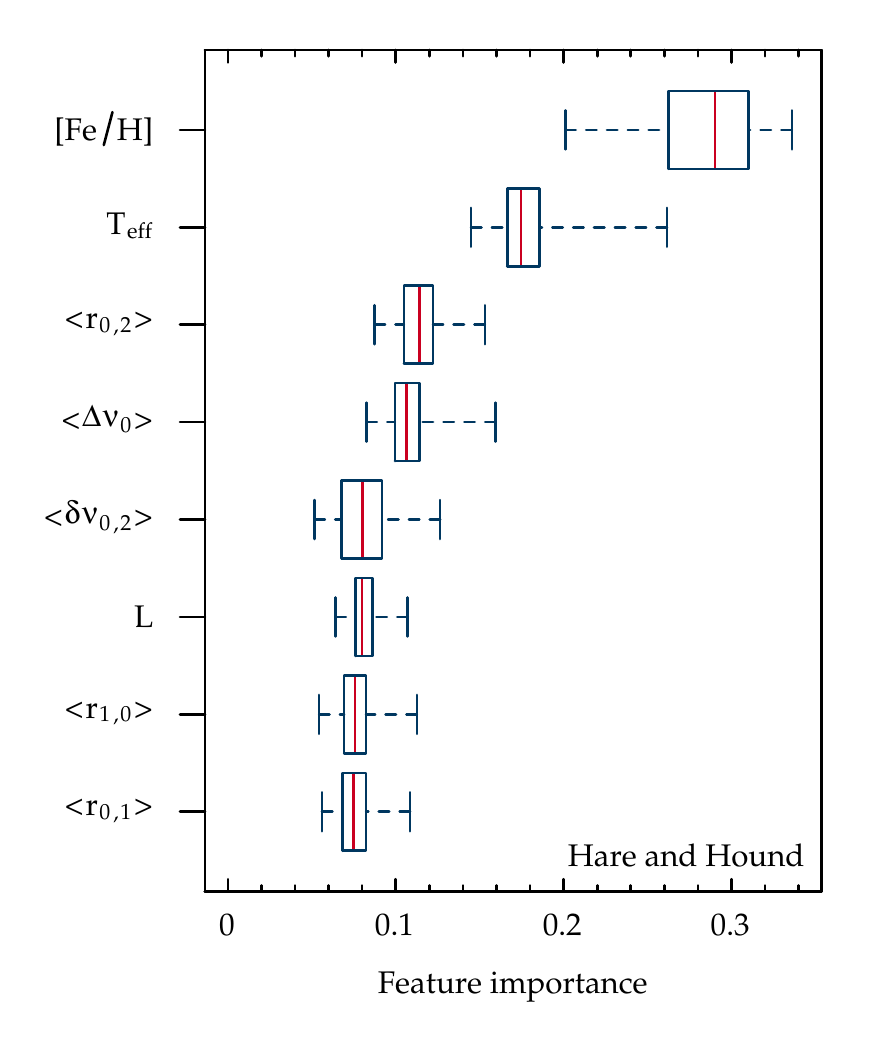}
    \includegraphics[width=0.45\linewidth, keepaspectratio]{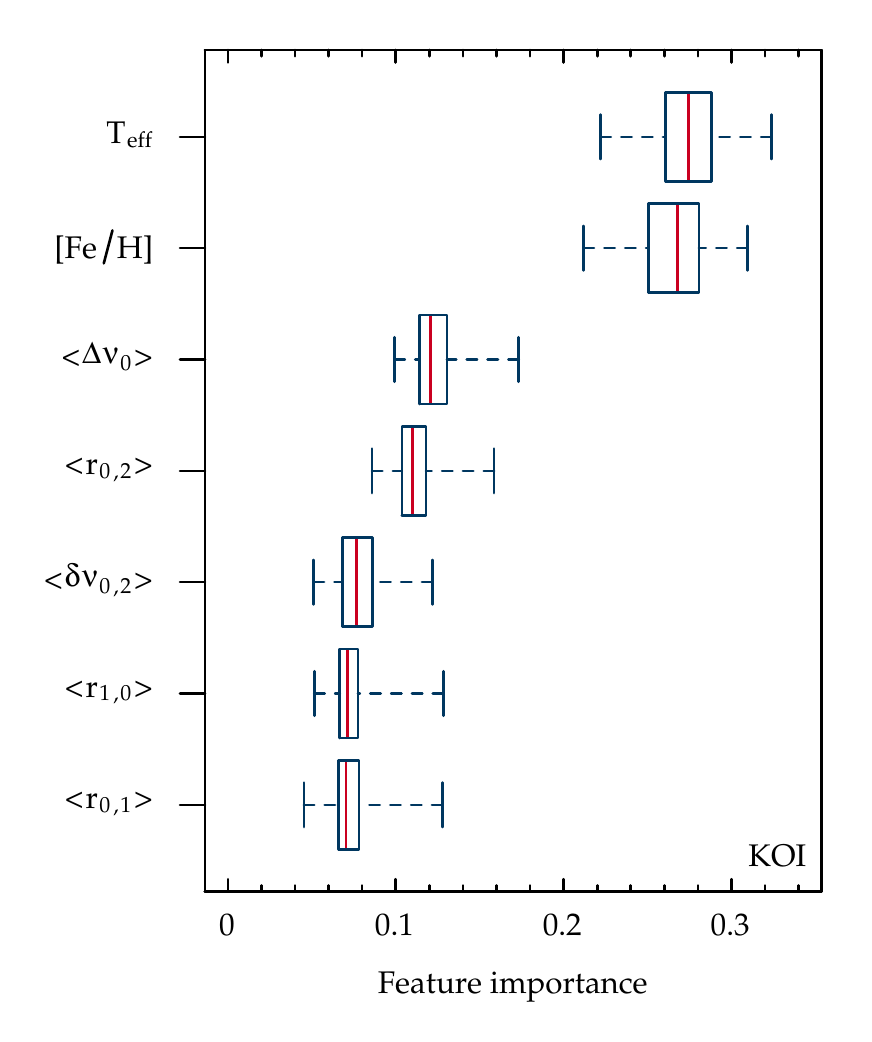}
    \caption[Feature Importances (Hare-and-Hound, KAGES)]{Box-and-whisker plots of relative importance for each feature in measuring fundamental stellar parameters for the hare-and-hound exercise data (left), where luminosities are available; and the \emph{Kepler} objects-of-interest (right), where they are not. Octupole (${\ell=3}$) modes have not been measured in any of these stars, so ${\langle\delta\nu_{1,3}\rangle}$ and ${\langle r_{1,3}\rangle}$ from evolutionary modelling are not supplied to these random forests. The boxes are sorted by median importance.
    \label{fig:importances2} }
\end{figure}
\end{landscape}
}

\subsubsection*{Advantages of CART}
We choose random forests over any of the many other non-linear regression routines (e.g.\ neural networks, support vector regression, etc.) for several reasons. 
First, random forests perform \emph{constrained} regression; that is, they only make predictions within the boundaries of the supplied training data \citep[see e.g.][Section~9.2.1]{hastie2005elements}. This is in contrast to other methods like neural networks, which ordinarily perform unconstrained regression and are therefore not prevented from predicting non-physical quantities such as negative masses or from violating conservation requirements. 

Secondly, due to the decision rule process that is explained below, random forests are insensitive to the scale of the data. Unless care is taken, other regression methods will artificially weight some observable \mb{quantities} like temperature as being more important than, say, luminosity, solely because temperatures are written using larger numbers (e.g., $5777$ vs.\ $1$, see for example section 11.5.3 of \citealt{hastie2005elements} for a discussion). 
Consequently, solutions obtained by other methods will change if they are \mb{run using features that are} expressed using different units of measure. 
For example, other methods will produce different regressors if trained on luminosity values expressed in solar units verses values expressed in erg\mb{s}, whereas random forests will not. \mb{Commonly, this problem is mitigated in other methods by means of variable standardization and through the use of Mahalabonis distances \citep{mahalanobis1936generalized}. 
However, these transformations are arbitrary, and handling variables naturally without rescaling is thus preferred. } 

Thirdly, random forests take only seconds to train, which can be a large benefit if different stars have different \mb{features} available. For example, some stars have luminosity information available whereas others do not, so a different regressor must be trained for each. In the extreme case, if one wanted to make predictions for stars using all of their respectively observed frequencies, one would need to train a new regressor for each star using the subset of simulated frequencies that correspond to the ones observed for that star. Ignoring the difficulties of surface-term corrections and mode identifications, such an approach would be well-handled by random forest, suffering only a small hit to performance from its relatively small training cost. On the other hand, it would be infeasible to do this on a star-by-star basis with most other routines such as deep neural networks, because \mb{those} methods can take days or even weeks to train. 

And finally\mb{, as we saw in the previous section,} random forests provide the opportunity to extract insight about the actual regression being performed by examining the importance of each \mb{feature} in making predictions. 

\subsubsection*{Uncertainty}
\label{sec:uncertainties}
There are three separate sources of uncertainty in predicting stellar parameters. The first is the systematic uncertainty in the physics used to model stars. These uncertainties are unknown, however, and hence cannot be propagated. The second is the uncertainty belonging to the observations of the star. We propagate measurement uncertainties $\sigma$ into the predictions by perturbing all measured quantities ${n=10,000}$ times with normal noise having zero mean and standard deviation $\sigma$. We account for the covariance between asteroseismic separations and ratios by recalculating them upon each perturbation. 

The final source is regression uncertainty. Fundamentally, each parameter can only be constrained to the extent that observations are able to bear information pertaining to that parameter. Even if observations were error-free, there still may exist a limit to which information gleaned from the surface may tell us about the physical \mb{qualities} and evolutionary history of a star. We quantify those limits via cross-validation: we train the random forest on only a subset of the simulated evolutionary tracks and make predictions on a held-out validation set. We randomly hold out a different subset of the tracks $25$ times to serve as different validation sets and obtain averaged accuracy scores.

We calculate accuracies using several scores. The first is the explained variance score V$_{\text{e}}$:
\begin{equation}
  \text{V}_{\text{e}} = 1 - \frac{\text{Var}\{ y - \hat y \}}{\text{Var}\{ y \}}
\end{equation}
where $y$ is the \mb{true} value we want to predict from the validation set (e.g.\ stellar mass), $\hat y$ is the predicted value from the random forest, and Var is the variance, i.e.\ the square of the standard deviation. This score tells us the extent to which the regressor has reduced the variance in the parameter it is predicting. The value ranges from negative infinity, which would be obtained by a pathologically bad predictor; to one for a perfect predictor, which occurs if all of the values are predicted with zero error. 

The next score we consider is the residuals of each prediction, i.e.\ the absolute difference between the true value $y$ and the predicted value $\hat y$. Naturally, we want this value to be as low as possible. We also consider the precision of the regression $\hat \sigma$ by taking the standard deviation of predictions across all of the decision trees in the forest. Finally, we consider these scores together by calculating the distance of the residuals in units of precision, i.e.\ ${\abs{\hat y - y} / \hat{\sigma}}$. 

Figure~\ref{fig:evaluation-tracks} shows these accuracies as a function of the number of evolutionary tracks used in the training of the random forest. Since the residuals and standard deviations of each parameter are incomparable, we normalize them by dividing by the maximum value. We also consider the number of trees in the forest and the number of models per evolutionary track\mb{. In this work, we use $256$ trees in each forest, which we have selected via cross-validation by choosing a number of trees that is greater than the point at which we saw that the explained variance was no longer increasing greatly;} see Appendix \ref{sec:evaluation} for an extended discussion. 

\afterpage{
\begin{landscape}
\begin{figure}
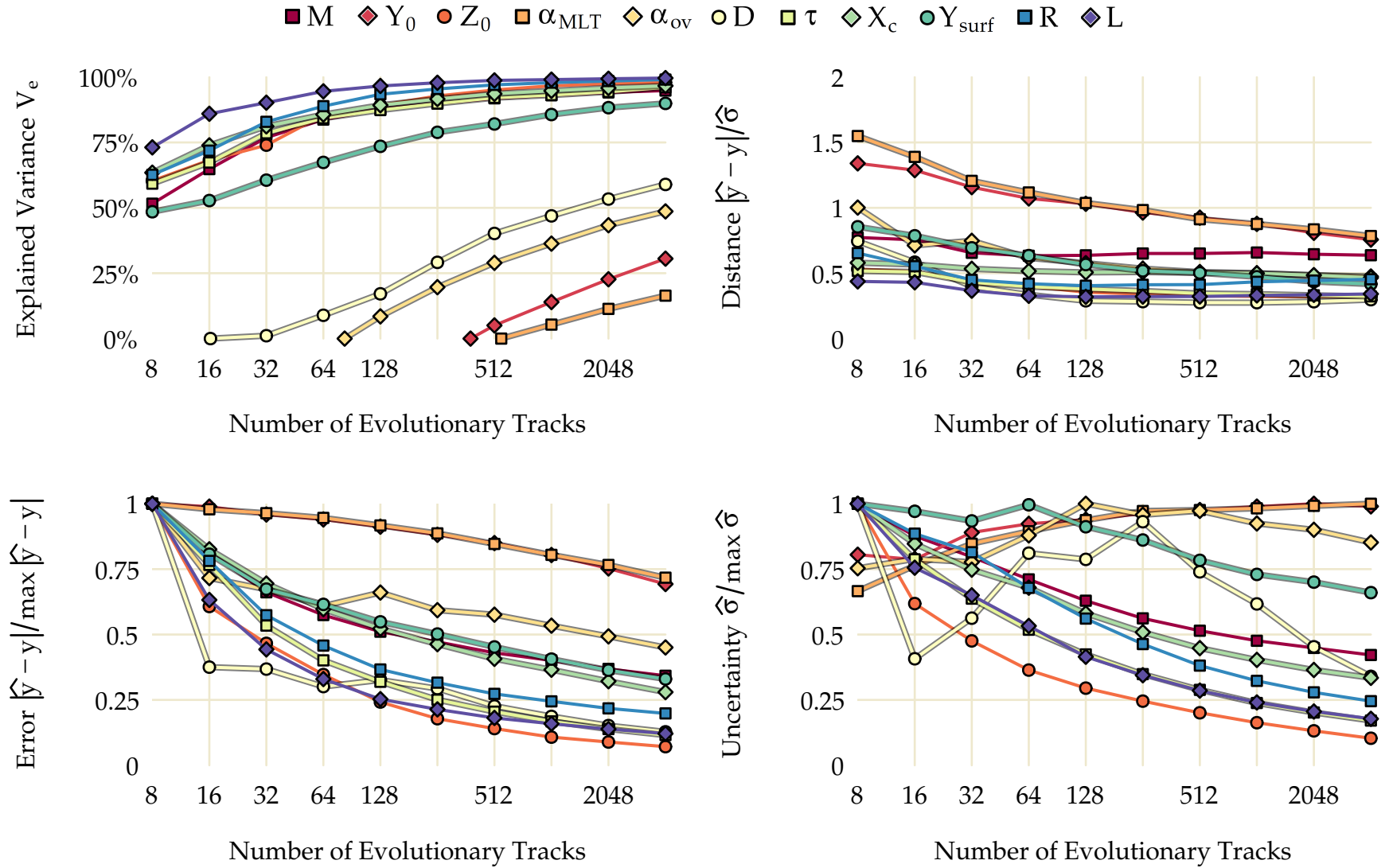

    \vspace*{-1cm}
    \centering
    \includegraphics[width=0.6\linewidth,keepaspectratio]{legend.png}\\
    \includegraphics[width=0.5\linewidth,keepaspectratio]{num_tracks-ev.pdf}
    \includegraphics[width=0.5\linewidth,keepaspectratio]{num_tracks-dist.pdf}\\
    \includegraphics[width=0.5\linewidth,keepaspectratio]{num_tracks-diff.pdf}
    \includegraphics[width=0.5\linewidth,keepaspectratio]{num_tracks-sigma.pdf}
    \caption[Evaluations of regression accuracy]{Evaluations of regression accuracy. Explained variance (top left), accuracy per precision distance (top right), normalized absolute error (bottom left), and normalized uncertainty (bottom right) for each stellar parameter as a function of the number of evolutionary tracks used in training the random forest. These results use $64$ models per track and $256$ trees in the random forest. \label{fig:evaluation-tracks}} 
\end{figure}
\end{landscape}
}

When supplied with enough \mb{stellar models}, the random forest reduces the variance in each parameter and is able to make precise inferences. The forest has very high predictive power for most \mb{parameters}, and as a result, essentially all of the uncertainty when predicting quantities such as stellar radii and luminosities will stem from observational uncertainty. However, for some model \mb{parameters}---most notably the mixing length parameter---there is still a great deal of variance in the residuals. Prior to \mb{the point where the regressor has been trained on} about $500$ evolutionary tracks, the differences between the true and predicted mixing lengths actually have a greater variance than just the true mixing lengths themselves. Likewise, the diffusion multiplication factor is difficult to constrain because a star can achieve the same present-day [Fe/H] by either having a large initial non-hydrogen abundance and a large diffusion \mb{multiplication} factor, or by having the same initial [Fe/H] as present [Fe/H] but with diffusion disabled. These difficult-to-constrain \mb{parameters} will therefore be predicted with substantial \mb{uncertainties} regardless of the precision of the observations. 


\section{Results}
We perform three tests of our method. We begin with a hare-and-hound simulation exercise to show that we can reliably recover parameters. We then move to the Sun and the solar-like stars 16~Cyg~A \& B, which have been the subjects of many investigations; and we conclude by applying our method to $34$ \emph{Kepler} objects-of-interest. In each case, we train our random forest regressor on the subset of observational data that is available for the stars being processed. In the case of the Sun and 16~Cygni, we know very accurately their radii, luminosities, and surface gravities. For other stars, we will predict this information instead of supplying it.

\subsection{Hare and Hound} 
\label{sec:hnh}
We performed a blind hare-and-hound exercise to evaluate the performance of our predictor. Author S.B.\ prepared twelve models varied in mass, initial chemical composition, and mixing length parameter with only some models having overshooting and only some models having atomic diffusion included. The models were evolved without rotation using the Yale rotating stellar evolution code \citep[YREC;][]{2008ApSS.316...31D}, which is a different evolution code than the one that was used to train the random forest. Effective temperatures, luminosities, [Fe/H] and $\nu_{\max}$ values as well as ${\ell=0},1,2$ frequencies were obtained from each model. Author G.C.A.\  perturbed the ``observations'' of these models according to the scheme devised by \citet{spaceinn}. \mb{Appendix \ref{sec:hare-and-hound} lists the true values and the perturbed observations of the hare-and-hound models}. The perturbed observations and their uncertainties were given to author E.P.B.\@, who used the described method to recover the stellar parameters of \mb{these} models without being given access to the true values. Relative differences between the true and predicted ages, masses, and radii for these models are plotted against their true values in Figure~\ref{fig:hare-comparison}. The method is able to recover the true model values within uncertainties even when they have been perturbed by noise. We do not compare the predicted mixing length parameter, overshooting parameter or diffusion \mb{multiplication} factor \mb{the interpretation of these parameters depends on how they have been defined and their precise implementation.}

\afterpage{
\begin{landscape}
\begin{figure}
    \centering
    \includegraphics[width=0.487\linewidth,keepaspectratio]{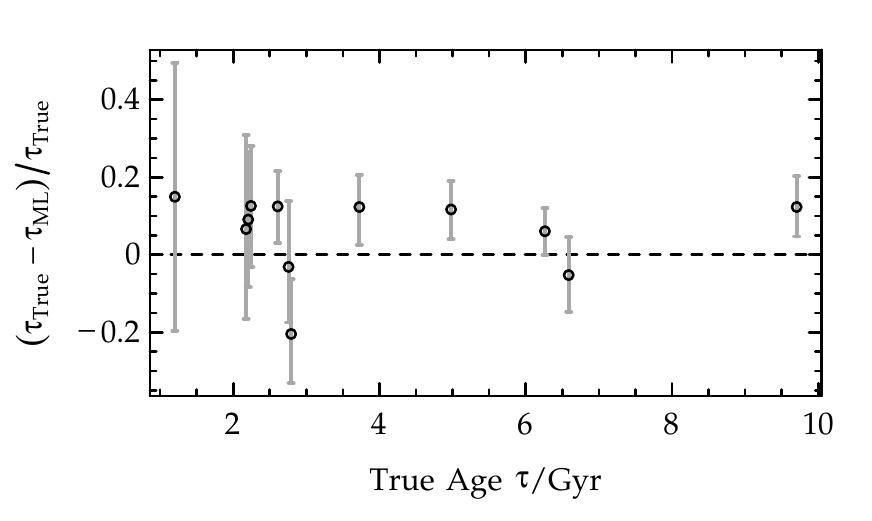}%
    \includegraphics[width=0.487\linewidth,keepaspectratio]{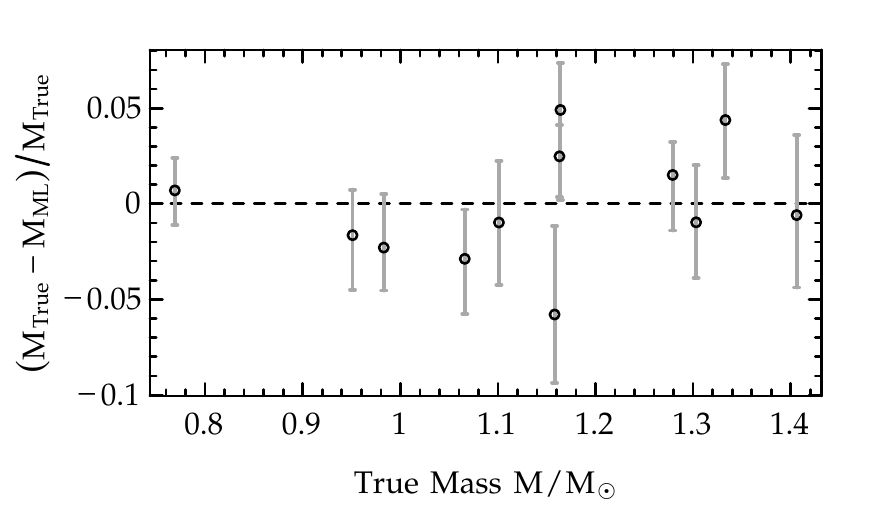}\\
    \includegraphics[width=0.487\linewidth,keepaspectratio]{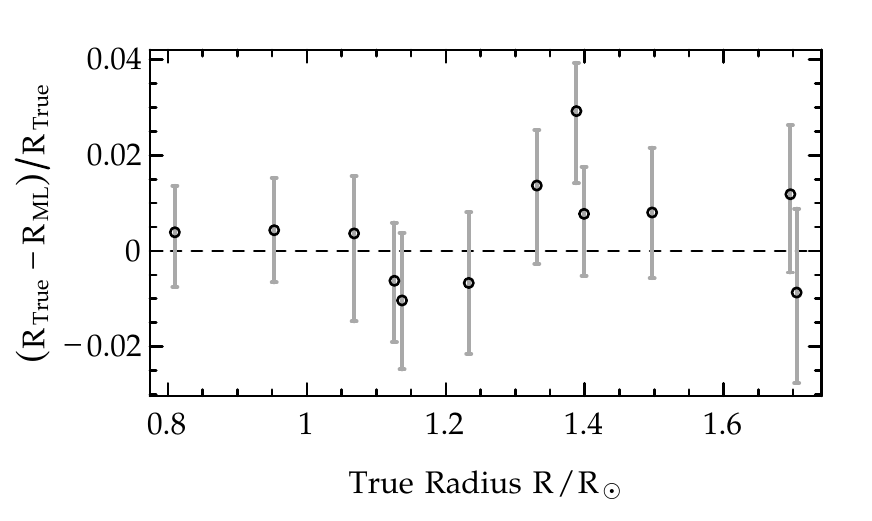}
    \caption[Hare-and-hound results]{Relative differences between the predicted and true values for age (top left), mass (top right), and radius (bottom) as a function of the true values in a hare-and-hound simulation exercise. \vspace*{5mm} 
    \label{fig:hare-comparison}}
\end{figure}
\end{landscape}
}

\subsection{The Sun and the 16~Cygni System}
To ensure confidence in our predictions on \emph{Kepler} data, we first degrade the frequencies of the Sun at solar minimum that were obtained by the Birmingham Solar-Oscillations Network \citep[BiSON;][]{2014MNRAS.439.2025D} to the level of information that is achievable by the spacecraft. We also degrade the Sun's uncertainties of \mb{other} observations by applying 16~Cyg~B's uncertainties of effective temperature, luminosity, surface gravity, metallicity, $\nu_{\max}$, radius, and radial velocity. Finally, we perturb each value with random Gaussian noise according to its uncertainty to reflect the fact that the measured value of an uncertain observation is not \emph{per se} the true value. We use the random forest whose feature importances were shown in Figure~\ref{fig:importances} to predict the values of the Sun; i.e.\ the random forest trained on effective temperatures, metallicities, luminosities, surface gravities, radii, and global asteroseismic \mb{quantities} $\langle \Delta\nu_0 \rangle$, $\langle \delta\nu_{0,2} \rangle$, $\langle \delta\nu_{1,3} \rangle$, $\langle r_{0,2} \rangle$, $\langle r_{1,3} \rangle$, $\langle r_{0,1} \rangle$, and $\langle r_{1,0} \rangle$. We show in Figure~\ref{fig:corner} the densities for the predicted mass, initial composition, mixing length parameter, overshoot coefficient, and diffusion multiplication factor needed for fitting an evolutionary model to degraded data of the Sun as well as the predicted solar age, core hydrogen abundance, and surface helium abundance. \mb{As discussed in Section~\ref{sec:uncertainties}, these densities show the distributions resulting from running $10,000$ different noise perturbations fed through the random forest.} Relative uncertainties ${\epsilon=100\cdot\sigma/\mu}$ are also indicated, where $\mu$ is the mean and $\sigma$ is the standard deviation of the quantity being predicted. Our predictions are in good agreement with the known values (see also Table~\ref{tab:results} and Table~\ref{tab:results-ca}, and \emph{cf}.~Equation~\ref{eq:solar-cal-vals}). 

Several parameters show multimodality due to model degeneracies. For example, two solutions for the initial helium are present. This is because it covaries with the mixing length parameter: the peak of \mb{higher} $Y_0$ corresponds to the peak of \mb{lower} $\alpha_{\text{MLT}}$ and vice versa. Likewise, high values of surface helium correspond to low values of the diffusion \mb{multiplication} factor. 

Effective temperatures, surface gravities, and metallicities of 16~Cyg~A~and~B were obtained from \citet{2009A&A...508L..17R}; radii and luminosities from \citet{2013MNRAS.433.1262W}; and frequencies from \citet{2015MNRAS.446.2959D}. We obtained the radial velocity measurements of 16~Cyg~A~and~B from \citet{2002ApJS..141..503N} and corrected frequencies for Doppler shifting as per the prescription in \citet{2014MNRAS.445L..94D}. We tried with and without line-of-sight corrections and found that it did not affect the predicted quantities or their uncertainties. We use the same random forest as we used for the degraded solar data to predict the \mb{parameters} of these stars. The initial parameters---masses, chemical compositions, mixing lengths, diffusion \mb{multiplication} factors, and overshoot coefficients---for 16~Cygni as predicted by machine learning \mb{are shown} in Table~\ref{tab:results}, and the predicted current parameters---age, surface helium and core hydrogen abundances---\mb{are shown} in Table~\ref{tab:results-ca}. For reference we also show the predicted solar values from these inputs there as well. These results support the hypothesis that 16~Cyg~A~and~B were co-natal; i.e.\ they formed at the same time with the same initial composition. 

We additionally predict the radii and luminosities of 16~Cyg~A~and~B instead of using them as \mb{features}. Figure~\ref{fig:interferometry} shows our inferred radii, luminosities and surface helium abundances of 16~Cyg~A~and~B plotted \mb{along with} the values determined by interferometry \citep{2013MNRAS.433.1262W} and an asteroseismic estimate \citep{2014ApJ...790..138V}. Here again we find excellent agreement between our method and the measured values. 

\citet{2015ApJ...811L..37M} performed detailed modelling of 16~Cyg~A~and~B using the Asteroseismic Modeling Portal (AMP), a genetic algorithm for matching individual frequencies of stars to stellar models. They calculated their results without heavy-element diffusion (i.e.\ with helium-only diffusion) and without overshooting. In order to account for systematic uncertainties, they multiplied the spectroscopic uncertainties of 16~Cyg~A~and~B by an arbitrary constant ${C=3}$. Therefore, in order to make a fair comparison between the results of our method and theirs, we generate a new matrix of evolutionary models with those same conditions and also increase the uncertainties on [Fe/H] by a factor of $C$. In Figure~\ref{fig:16Cyg-hist}, we show probability densities of the predicted parameters of 16~Cyg~A~and~B that we obtain using machine learning in comparison with the results obtained by AMP. We find the values and uncertainties agree well. To perform their analysis, AMP required more than $15,000$ hours of CPU time to model 16~Cyg~A~and~B using the world's 10th fastest supercomputer, the Texas Advanced Computing Center Stampede \citep{TOP500}. Here we have obtained comparable results in roughly one minute \mb{on a computing cluster with $64$ $2.5$~GHz cores} using only global asteroseismic \mb{quantities} and no individual frequencies. Although more computationally expensive than our method, detailed optimization codes like AMP do have advantages in that they are additionally able to obtain detailed structural models of stars. 

\afterpage{
\begin{landscape}
\begin{figure}
    \vspace*{-1cm}
    \centering
    \includegraphics[width=\linewidth,keepaspectratio]{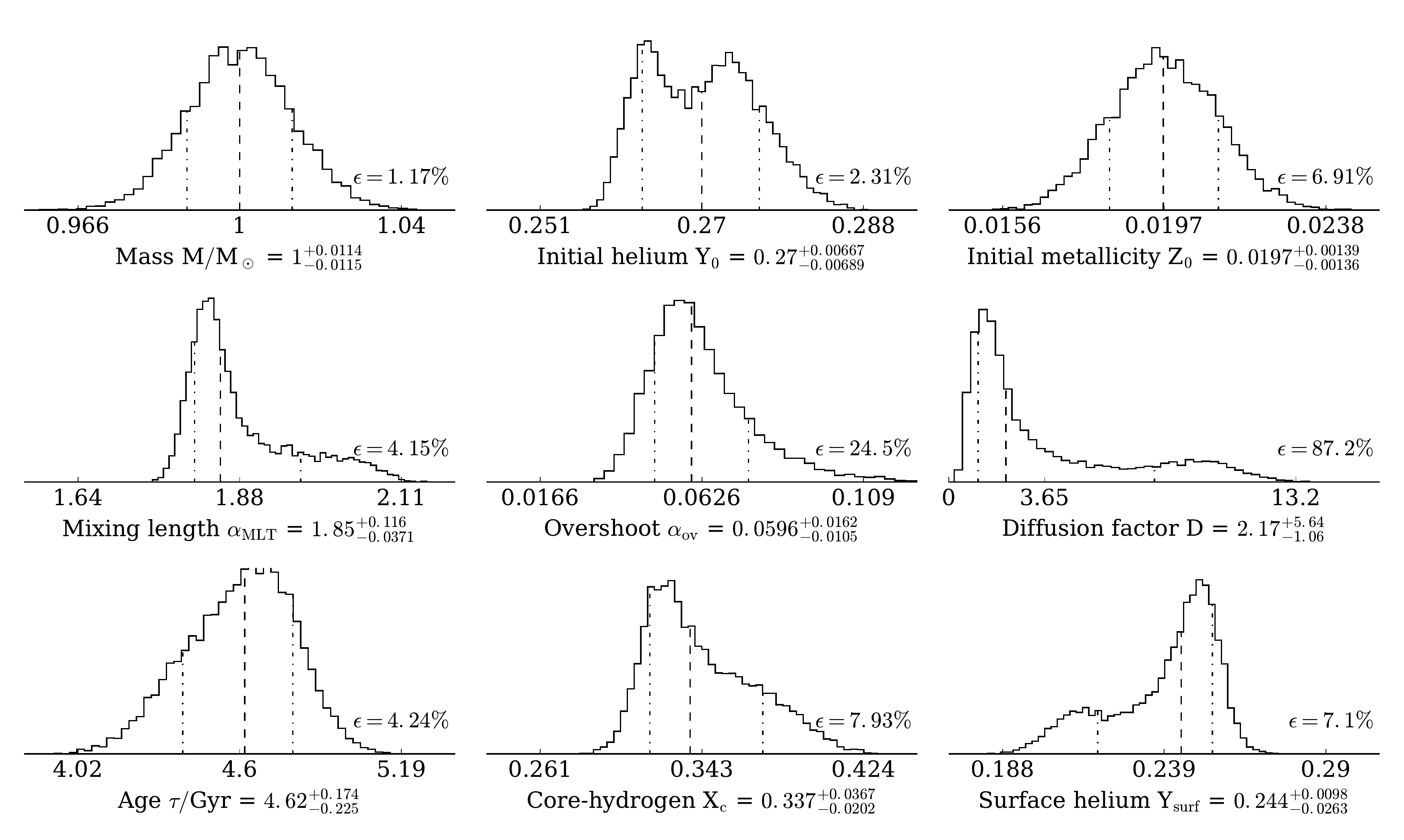}
    \caption[Posterior distributions for degraded solar data]{Predictions from machine learning of initial (top six) and current (bottom three) stellar parameters for degraded solar data. Labels are placed at the mean and $3\sigma$ levels. \mb{Dashed and dot-dashed} lines indicate the median and quartiles\mb{, respectively}. Relative uncertainties $\epsilon$ are shown beside each plot. Note that the overshoot parameter applies to all convective boundaries and is not modified over the course of evolution, so a non-zero value does not imply a convective core. 
    \label{fig:corner} } 
\end{figure}
\end{landscape}
}

\afterpage{
\begin{table} \centering 
\caption{Means and standard deviations for predicted initial stellar parameters of the Sun (degraded data) and 16~Cyg~A~and~B.
\label{tab:results}}
\hspace*{-1.7cm}\begin{tabular}{cccccccc}
Name & $M/M_\odot$ & $Y_0$ & $Z_0$ & $\alpha_{\mathrm{MLT}}$ & $\alpha_{\mathrm{ov}}$ & D \\ \hline \hline
Sun      & 1.00 $\pm$ 0.012 & 0.270 $\pm$ 0.0062 & 0.020 $\pm$ 0.0014 & 1.88 $\pm$ 0.078 & 0.06 $\pm$ 0.015 & 3.7 $\pm$ 3.18 \\
16~Cyg~A & 1.08 $\pm$ 0.016 & 0.262 $\pm$ 0.0073 & 0.022 $\pm$ 0.0014 & 1.86 $\pm$ 0.077 & 0.07 $\pm$ 0.028 & 0.9 $\pm$ 0.76 \\
16~Cyg~B & 1.03 $\pm$ 0.015 & 0.268 $\pm$ 0.0065 & 0.021 $\pm$ 0.0015 & 1.83 $\pm$ 0.069 & 0.11 $\pm$ 0.029 & 1.9 $\pm$ 1.57
\\ \hline \end{tabular}
\end{table}

\begin{table} \centering 
\caption{Means and standard deviations for predicted current-age stellar \mb{parameters} of the Sun (degraded data) and 16~Cyg~A~and~B. \label{tab:results-ca}}
\begin{tabular}{cccc}
Name & $\tau/$Gyr & X$_{\mathrm{c}}$ & $Y_{\mathrm{surf}}$ \\ \hline \hline
Sun      & 4.6 $\pm$ 0.20 & 0.34 $\pm$ 0.027 & 0.24  $\pm$ 0.017 \\
16~Cyg~A & 6.9 $\pm$ 0.40 & 0.06 $\pm$ 0.024 & 0.246 $\pm$ 0.0085 \\
16~Cyg~B & 6.8 $\pm$ 0.28 & 0.15 $\pm$ 0.023 & 0.24  $\pm$ 0.017
\\ \hline \end{tabular}
\end{table}
}

\afterpage{
\begin{landscape}
\begin{figure}
    \centering
    \includegraphics[width=0.45\linewidth, keepaspectratio]{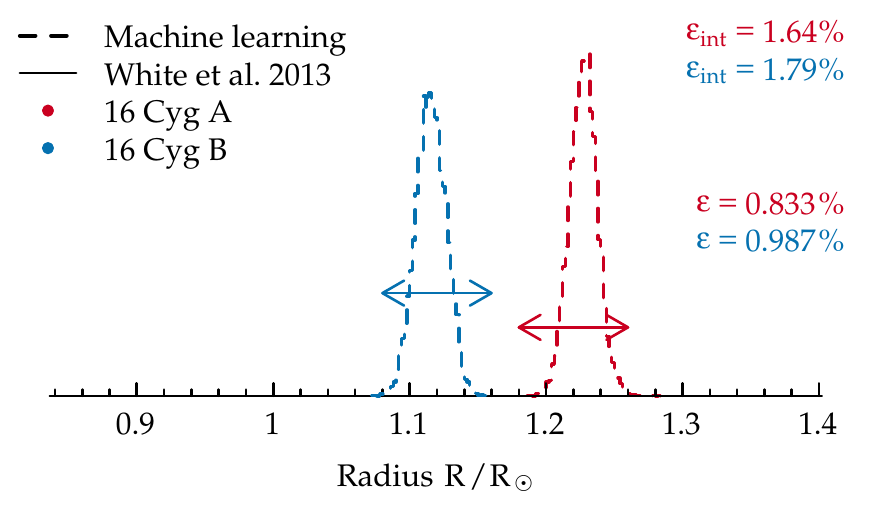}%
    \includegraphics[width=0.45\linewidth, keepaspectratio]{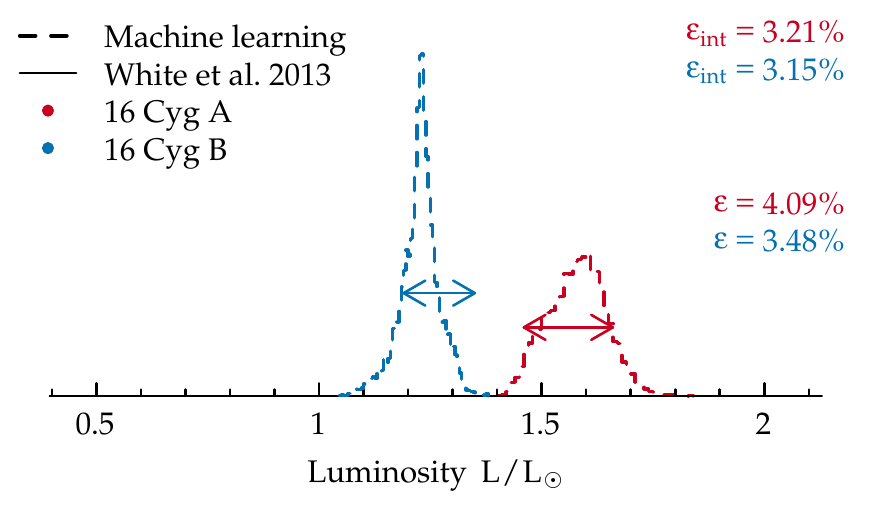}\\
    \includegraphics[width=0.45\linewidth, keepaspectratio]{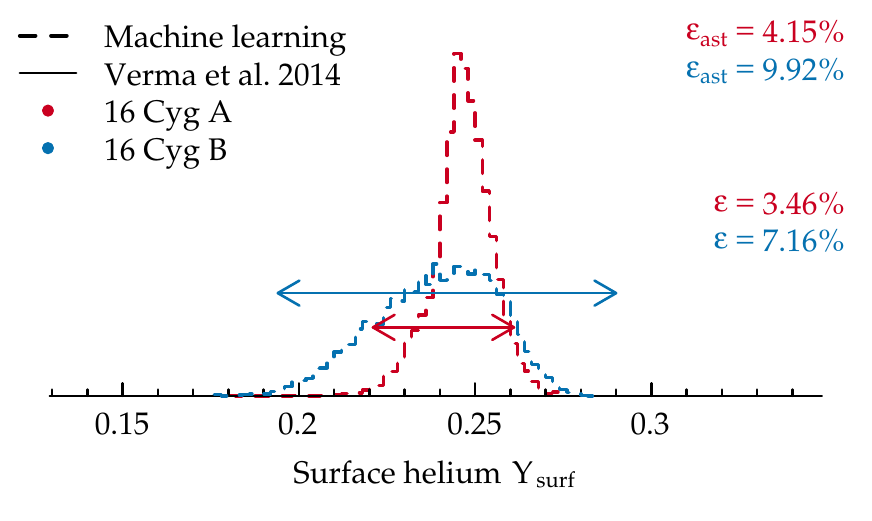}
    \caption[Posterior distributions for 16~Cygni (radius, luminosity, surface helium abundance)]{Probability densities for predictions of 16~Cyg~A (red) and B (blue) from machine learning of radii (top left), luminosities (top right), and surface helium abundances (bottom). Relative uncertainties $\epsilon$ are shown beside each plot. Predictions and $2\sigma$ uncertainties from interferometric (``int'') measurements and asteroseismic (``ast'') estimates are shown with arrows.}
    \label{fig:interferometry}
\end{figure}
\end{landscape}
}

\afterpage{
\begin{landscape}
\begin{figure}
    \centering
    \includegraphics[width=0.47\linewidth, keepaspectratio]{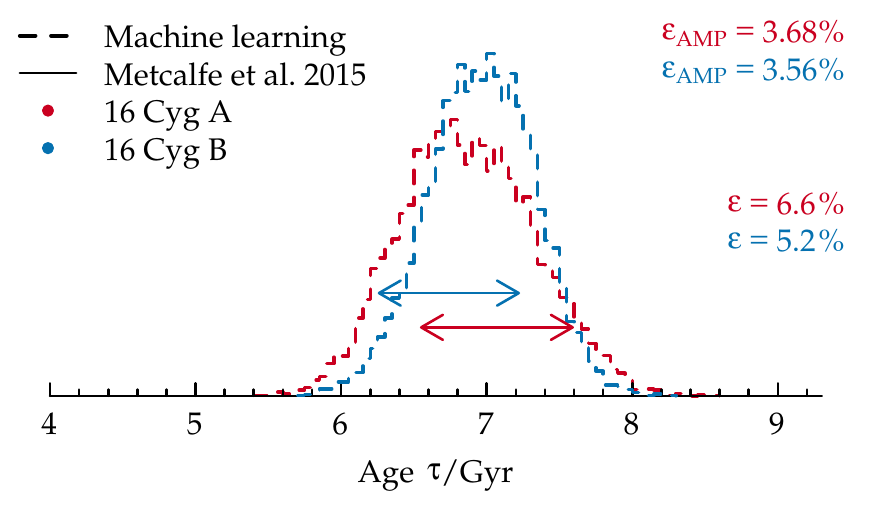}\hfill
    \includegraphics[width=0.47\linewidth, keepaspectratio]{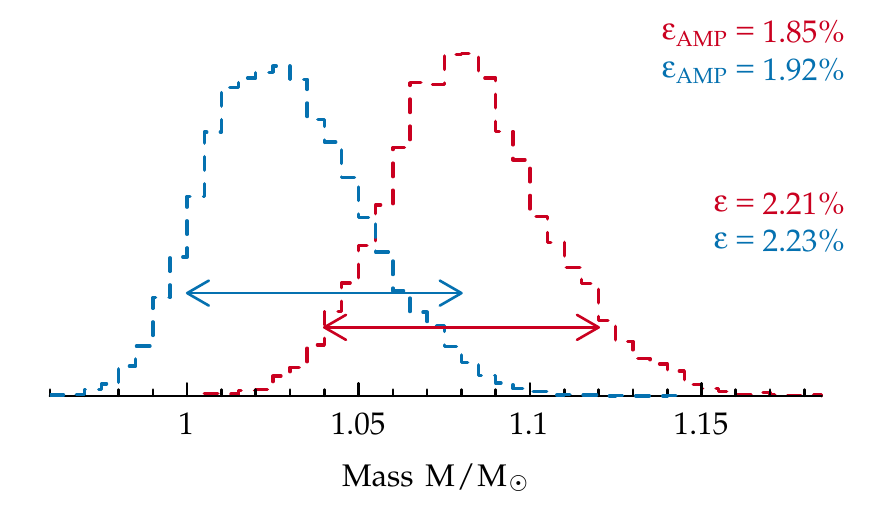}\\
    \includegraphics[width=0.47\linewidth, keepaspectratio]{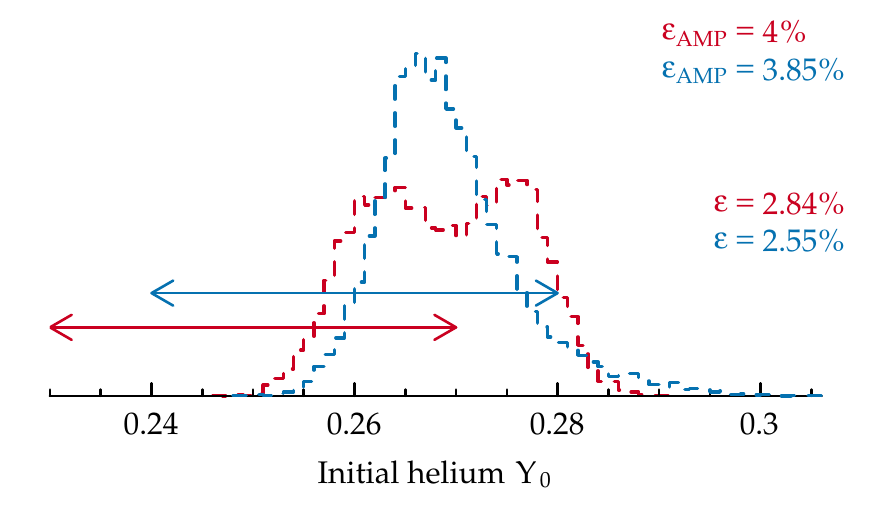}\hfill
    \includegraphics[width=0.47\linewidth, keepaspectratio]{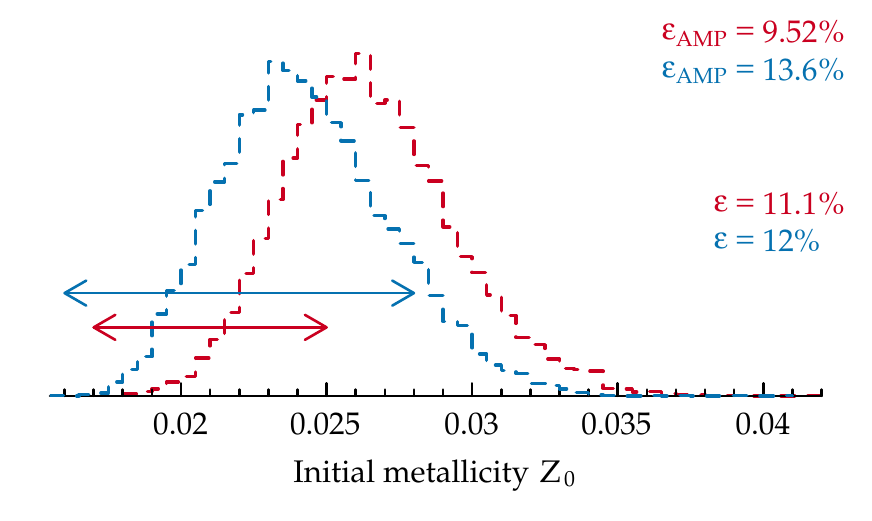}
    \caption[Posterior distributions for 16~Cygni (age, mass, initial helium abundance, initial metallicity)]{Probability densities showing predictions from machine learning of fundamental stellar parameters for 16~Cyg~A (red) and B (blue) \mb{along with} predictions from AMP modelling. Relative uncertainties are shown beside each plot. Predictions and $2\sigma$ uncertainties from AMP modelling are shown with arrows. \vspace*{5mm}
    \label{fig:16Cyg-hist}}
\end{figure}
\end{landscape}
}

\subsection{\emph{Kepler} Objects of Interest}
\label{sec:koi}
We obtain observations and frequencies of the KOI targets from \citet{2016MNRAS.456.2183D}. We use line-of-sight radial velocity corrections when available, which was only the case for KIC 6278762 \citep{2002AJ....124.1144L}, KIC 10666592 \citep{2013A&A...554A..84M}, and KIC 3632418 \citep{2006AstL...32..759G}. We use the random forest whose feature importances were shown in Figure~\ref{fig:importances2} to predict the fundamental \mb{parameters} of these stars; that is, the random forest that is trained on effective temperatures, metallicities, and asteroseismic quantities ${\langle \Delta\nu_0 \rangle}$, ${\langle \delta\nu_{0,2} \rangle}$, ${\langle r_{0,2} \rangle}$, ${\langle r_{0,1} \rangle}$, and ${\langle r_{1,0} \rangle}$. The predicted initial conditions---masses, chemical compositions, mixing lengths, overshoot coefficients, and diffusion \mb{multiplication} factors---are shown in Table~\ref{tab:results-kages}; and the predicted current conditions---ages, core hydrogen abundances, surface gravities, luminosities, radii, and surface helium abundances---are shown in Table~\ref{tab:results-kages-curr}. Figure~\ref{fig:us-vs-them} shows the fundamental parameters obtained from our method plotted against those obtained by \citet[hereinafter KAGES]{2015MNRAS.452.2127S}. We find good agreement across all stars. 

Although still in statistical agreement, the median values of our predicted ages are systematically lower and the median values of our predicted masses are systematically higher than those predicted by KAGES. We conjecture that these discrepancies arise from differences in input physics. We vary the efficiency of diffusion, the extent of convective overshooting, and the value of the mixing length parameter to arrive at these estimates, whereas \mb{the KAGES} models are calculated using fixed \mb{amounts} of diffusion, without overshoot, and with a solar-calibrated mixing length. Models with overshooting, for example, will be more evolved at the same age due to having larger core masses. Without direct access to their models, however, the exact reason is difficult to pinpoint. 

We find a significant linear trend in the \emph{Kepler} objects-of-interest between the diffusion multiplication factor and stellar mass needed to reproduce observations (${P = 0.0001}$ from a two-sided t-test with ${N-2=32}$ degrees of freedom). Since the values of mass and diffusion \mb{multiplication} factor are uncertain, we use Deming regression to estimate the coefficients of this relation without regression dilution \citep{deming1943statistical}. We show the diffusion multiplication factors as a function of stellar mass for all of these stars in Figure~\ref{fig:diffusion}. We find that the diffusion \mb{multiplication} factor linearly decreases with mass, i.e.\ 
\begin{equation} \label{eq:diffusion}
    \text{D} = ( 8.6 \pm 1.94 ) - ( 5.6 \pm 1.37 ) \cdot \text{M}/\text{M}_\odot
\end{equation}
and that this relation explains observations better than any constant factor (e.g., ${D=1}$ or ${D=0}$). 

\afterpage{
\begin{table}
\caption{Means and standard deviations for initial conditions of the KOI data set inferred via machine learning. The values obtained from degraded solar data predicted on these quantities are shown for reference. \label{tab:results-kages}}
\hspace*{-1.9cm}\begin{tabular}{ccccccc}
KIC & $M/M_\odot$ & $Y_0$ & $Z_0$ & $\alpha_{\mathrm{MLT}}$ & $\alpha_{\mathrm{ov}}$ & $D$ \\ \hline\hline
 3425851 & 1.15 $\pm$ 0.053  & 0.28  $\pm$ 0.020  & 0.015 $\pm$ 0.0028 & 1.9  $\pm$ 0.23  & 0.06 $\pm$ 0.057 &  0.5 $\pm$  0.92 \\
 3544595 & 0.91 $\pm$ 0.032  & 0.270 $\pm$ 0.0090 & 0.015 $\pm$ 0.0028 & 1.9  $\pm$ 0.10  & 0.2  $\pm$ 0.11  &  4.9 $\pm$  4.38 \\
 3632418 & 1.39 $\pm$ 0.057  & 0.267 $\pm$ 0.0089 & 0.019 $\pm$ 0.0032 & 2.0  $\pm$ 0.12  & 0.2  $\pm$ 0.14  &  1.1 $\pm$  1.01 \\
 4141376 & 1.03 $\pm$ 0.036  & 0.267 $\pm$ 0.0097 & 0.012 $\pm$ 0.0025 & 1.9  $\pm$ 0.12  & 0.1  $\pm$ 0.11  &  4.0 $\pm$  4.09 \\
 4143755 & 0.99 $\pm$ 0.037  & 0.277 $\pm$ 0.0050 & 0.014 $\pm$ 0.0026 & 1.77 $\pm$ 0.033 & 0.37 $\pm$ 0.071 & 13.4 $\pm$  5.37 \\
 4349452 & 1.22 $\pm$ 0.056  & 0.28  $\pm$ 0.012  & 0.020 $\pm$ 0.0043 & 1.9  $\pm$ 0.17  & 0.10 $\pm$ 0.090 &  7.3 $\pm$  8.82 \\
 4914423 & 1.19 $\pm$ 0.048  & 0.274 $\pm$ 0.0097 & 0.026 $\pm$ 0.0046 & 1.8  $\pm$ 0.11  & 0.08 $\pm$ 0.043 &  2.3 $\pm$  1.6 \\
 5094751 & 1.11 $\pm$ 0.038  & 0.274 $\pm$ 0.0082 & 0.018 $\pm$ 0.0030 & 1.8  $\pm$ 0.11  & 0.07 $\pm$ 0.041 &  2.3 $\pm$  1.39 \\
 5866724 & 1.29 $\pm$ 0.065  & 0.28  $\pm$ 0.011  & 0.027 $\pm$ 0.0058 & 1.8  $\pm$ 0.13  & 0.12 $\pm$ 0.086 &  7.0 $\pm$  8.38 \\
 6196457 & 1.31 $\pm$ 0.058  & 0.276 $\pm$ 0.005  & 0.032 $\pm$ 0.0050 & 1.71 $\pm$ 0.050 & 0.16 $\pm$ 0.055 &  5.7 $\pm$  2.34 \\
 6278762 & 0.76 $\pm$ 0.012  & 0.254 $\pm$ 0.0058 & 0.013 $\pm$ 0.0017 & 2.09 $\pm$ 0.069 & 0.06 $\pm$ 0.028 &  5.3 $\pm$  2.23 \\
 6521045 & 1.19 $\pm$ 0.046  & 0.273 $\pm$ 0.0071 & 0.027 $\pm$ 0.0044 & 1.82 $\pm$ 0.074 & 0.12 $\pm$ 0.036 &  3.2 $\pm$  1.31 \\
 7670943 & 1.30 $\pm$ 0.061  & 0.28  $\pm$ 0.017  & 0.021 $\pm$ 0.0045 & 2.0  $\pm$ 0.23  & 0.06 $\pm$ 0.064 &  1.0 $\pm$  2.55 \\
 8077137 & 1.23 $\pm$ 0.070  & 0.270 $\pm$ 0.0093 & 0.018 $\pm$ 0.0028 & 1.8  $\pm$ 0.14  & 0.2  $\pm$ 0.11  &  2.9 $\pm$  2.08 \\
 8292840 & 1.15 $\pm$ 0.079  & 0.28  $\pm$ 0.010  & 0.016 $\pm$ 0.0049 & 1.8  $\pm$ 0.15  & 0.1  $\pm$ 0.12  & 11.  $\pm$ 10.7  \\
 8349582 & 1.23 $\pm$ 0.040  & 0.271 $\pm$ 0.0069 & 0.043 $\pm$ 0.0074 & 1.9  $\pm$ 0.12  & 0.11 $\pm$ 0.060 &  2.5 $\pm$  1.11 \\
 8478994 & 0.81 $\pm$ 0.022  & 0.272 $\pm$ 0.0082 & 0.010 $\pm$ 0.0012 & 1.91 $\pm$ 0.054 & 0.21 $\pm$ 0.068 & 17.  $\pm$  9.74 \\
 8494142 & 1.42 $\pm$ 0.058  & 0.27  $\pm$ 0.010  & 0.028 $\pm$ 0.0046 & 1.70 $\pm$ 0.064 & 0.10 $\pm$ 0.051 &  1.6 $\pm$  1.65 \\
 8554498 & 1.39 $\pm$ 0.067  & 0.272 $\pm$ 0.0082 & 0.031 $\pm$ 0.0032 & 1.70 $\pm$ 0.077 & 0.14 $\pm$ 0.079 &  1.7 $\pm$  1.17 \\
 8684730 & 1.44 $\pm$ 0.030  & 0.277 $\pm$ 0.0075 & 0.041 $\pm$ 0.0049 & 1.9  $\pm$ 0.14  & 0.29 $\pm$ 0.094 & 15.2 $\pm$  8.81 \\
 8866102 & 1.26 $\pm$ 0.069  & 0.28  $\pm$ 0.013  & 0.021 $\pm$ 0.0048 & 1.8  $\pm$ 0.15  & 0.08 $\pm$ 0.070 &  5.  $\pm$  7.48 \\
 9414417 & 1.36 $\pm$ 0.054  & 0.264 $\pm$ 0.0073 & 0.018 $\pm$ 0.0028 & 1.9  $\pm$ 0.13  & 0.2  $\pm$ 0.1   &  2.2 $\pm$  1.68 \\
 9592705 & 1.45 $\pm$ 0.038  & 0.27  $\pm$ 0.010  & 0.029 $\pm$ 0.0038 & 1.72 $\pm$ 0.064 & 0.12 $\pm$ 0.056 &  0.6 $\pm$  0.47 \\
 9955598 & 0.93 $\pm$ 0.028  & 0.27  $\pm$ 0.011  & 0.023 $\pm$ 0.0039 & 1.9  $\pm$ 0.10  & 0.2  $\pm$ 0.13  &  2.2 $\pm$  1.76 \\
10514430 & 1.13 $\pm$ 0.053  & 0.277 $\pm$ 0.0046 & 0.021 $\pm$ 0.0039 & 1.78 $\pm$ 0.059 & 0.30 $\pm$ 0.097 &  4.7 $\pm$  1.77 \\
10586004 & 1.31 $\pm$ 0.078  & 0.274 $\pm$ 0.0055 & 0.038 $\pm$ 0.0071 & 1.8  $\pm$ 0.13  & 0.2  $\pm$ 0.13  &  4.3 $\pm$  3.99 \\
10666592 & 1.50 $\pm$ 0.023  & 0.30  $\pm$ 0.013  & 0.030 $\pm$ 0.0032 & 1.8  $\pm$ 0.11  & 0.06 $\pm$ 0.043 &  0.2 $\pm$  0.14 \\
10963065 & 1.09 $\pm$ 0.031  & 0.264 $\pm$ 0.0083 & 0.014 $\pm$ 0.0025 & 1.8  $\pm$ 0.11  & 0.05 $\pm$ 0.027 &  3.1 $\pm$  2.68 \\
11133306 & 1.11 $\pm$ 0.044  & 0.272 $\pm$ 0.0099 & 0.021 $\pm$ 0.0040 & 1.8  $\pm$ 0.16  & 0.04 $\pm$ 0.033 &  5.  $\pm$  5.75 \\
11295426 & 1.11 $\pm$ 0.033  & 0.27  $\pm$ 0.010  & 0.025 $\pm$ 0.0036 & 1.81 $\pm$ 0.084 & 0.05 $\pm$ 0.035 &  1.3 $\pm$  0.87 \\
11401755 & 1.15 $\pm$ 0.039  & 0.271 $\pm$ 0.0057 & 0.015 $\pm$ 0.0023 & 1.88 $\pm$ 0.055 & 0.33 $\pm$ 0.071 &  3.8 $\pm$  1.81 \\
11807274 & 1.32 $\pm$ 0.079  & 0.276 $\pm$ 0.0097 & 0.024 $\pm$ 0.0051 & 1.77 $\pm$ 0.083 & 0.11 $\pm$ 0.066 &  5.4 $\pm$  5.61 \\
11853905 & 1.22 $\pm$ 0.055  & 0.272 $\pm$ 0.0072 & 0.029 $\pm$ 0.0050 & 1.8  $\pm$ 0.12  & 0.18 $\pm$ 0.086 &  3.3 $\pm$  1.85 \\
11904151 & 0.93 $\pm$ 0.033  & 0.265 $\pm$ 0.0091 & 0.016 $\pm$ 0.0030 & 1.8  $\pm$ 0.13  & 0.05 $\pm$ 0.029 &  3.1 $\pm$  2.09 \\
Sun & 1.00 $\pm$ 0.0093 & 0.266 $\pm$ 0.0035 & 0.018 $\pm$ 0.0011 & 1.81 $\pm$ 0.032 & 0.07 $\pm$ 0.021 &  2.1 $\pm$  0.83 \\ \hline \end{tabular}
\end{table}
\clearpage
\begin{table}
\caption{Means and standard deviations for current-age conditions of the KOI data set inferred via machine learning. The values obtained from degraded solar data predicted on these quantities are shown for reference. \label{tab:results-kages-curr}}
\hspace*{-2.2cm}\begin{tabular}{ccccccc}
KIC & $\tau/$Gyr & X$_{\mathrm{c}}$ & log g & L$/$L$_\odot$ & R$/$R$_\odot$ & $Y_{\mathrm{surf}}$ \\ \hline\hline
 3425851 &  3.7 $\pm$ 0.76  & 0.14  $\pm$ 0.081  & 4.234 $\pm$ 0.0098 & 2.7  $\pm$ 0.16  & 1.36  $\pm$ 0.022  & 0.27  $\pm$ 0.026 \\
 3544595 &  6.7 $\pm$ 1.47  & 0.31  $\pm$ 0.078  & 4.46  $\pm$ 0.016  & 0.84 $\pm$ 0.068 & 0.94  $\pm$ 0.020  & 0.23  $\pm$ 0.023 \\
 3632418 &  3.0 $\pm$ 0.36  & 0.10  $\pm$ 0.039  & 4.020 $\pm$ 0.0076 & 5.2  $\pm$ 0.25  & 1.91  $\pm$ 0.031  & 0.24  $\pm$ 0.021 \\
 4141376 &  3.4 $\pm$ 0.67  & 0.38  $\pm$ 0.070  & 4.41  $\pm$ 0.011  & 1.42 $\pm$ 0.097 & 1.05  $\pm$ 0.019  & 0.24  $\pm$ 0.022 \\
 4143755 &  8.0 $\pm$ 0.80  & 0.07  $\pm$ 0.022  & 4.09  $\pm$ 0.013  & 2.3  $\pm$ 0.12  & 1.50  $\pm$ 0.029  & 0.17  $\pm$ 0.023 \\
 4349452 &  2.4 $\pm$ 0.78  & 0.4   $\pm$ 0.10   & 4.28  $\pm$ 0.012  & 2.5  $\pm$ 0.14  & 1.32  $\pm$ 0.022  & 0.22  $\pm$ 0.043 \\
 4914423 &  5.2 $\pm$ 0.58  & 0.06  $\pm$ 0.032  & 4.162 $\pm$ 0.0097 & 2.5  $\pm$ 0.16  & 1.50  $\pm$ 0.022  & 0.24  $\pm$ 0.023 \\
 5094751 &  5.3 $\pm$ 0.67  & 0.07  $\pm$ 0.039  & 4.209 $\pm$ 0.0082 & 2.2  $\pm$ 0.13  & 1.37  $\pm$ 0.017  & 0.23  $\pm$ 0.024 \\
 5866724 &  2.4 $\pm$ 0.96  & 0.4   $\pm$ 0.12   & 4.24  $\pm$ 0.017  & 2.7  $\pm$ 0.13  & 1.42  $\pm$ 0.022  & 0.23  $\pm$ 0.038 \\
 6196457 &  4.0 $\pm$ 0.73  & 0.18  $\pm$ 0.061  & 4.11  $\pm$ 0.022  & 3.3  $\pm$ 0.21  & 1.68  $\pm$ 0.041  & 0.24  $\pm$ 0.016 \\
 6278762 & 10.3 $\pm$ 0.96  & 0.35  $\pm$ 0.026  & 4.557 $\pm$ 0.0084 & 0.34 $\pm$ 0.022 & 0.761 $\pm$ 0.0061 & 0.19  $\pm$ 0.023 \\
 6521045 &  5.6 $\pm$ 0.370 & 0.027 $\pm$ 0.0097 & 4.122 $\pm$ 0.0055 & 2.7  $\pm$ 0.15  & 1.57  $\pm$ 0.025  & 0.22  $\pm$ 0.019 \\
 7670943 &  2.3 $\pm$ 0.59  & 0.32  $\pm$ 0.088  & 4.234 $\pm$ 0.0099 & 3.3  $\pm$ 0.23  & 1.44  $\pm$ 0.025  & 0.26  $\pm$ 0.029 \\
 8077137 &  4.4 $\pm$ 0.96  & 0.08  $\pm$ 0.052  & 4.08  $\pm$ 0.016  & 3.7  $\pm$ 0.24  & 1.68  $\pm$ 0.044  & 0.22  $\pm$ 0.031 \\
 8292840 &  3.4 $\pm$ 1.48  & 0.3   $\pm$ 0.14   & 4.25  $\pm$ 0.023  & 2.6  $\pm$ 0.20  & 1.34  $\pm$ 0.026  & 0.19  $\pm$ 0.049 \\
 8349582 &  6.7 $\pm$ 0.53  & 0.02  $\pm$ 0.012  & 4.16  $\pm$ 0.012  & 2.2  $\pm$ 0.12  & 1.52  $\pm$ 0.016  & 0.23  $\pm$ 0.015 \\
 8478994 &  4.6 $\pm$ 1.75  & 0.50  $\pm$ 0.055  & 4.55  $\pm$ 0.012  & 0.51 $\pm$ 0.036 & 0.79  $\pm$ 0.014  & 0.21  $\pm$ 0.022 \\
 8494142 &  2.8 $\pm$ 0.52  & 0.18  $\pm$ 0.067  & 4.06  $\pm$ 0.018  & 4.5  $\pm$ 0.32  & 1.84  $\pm$ 0.043  & 0.24  $\pm$ 0.029 \\
 8554498 &  3.7 $\pm$ 0.79  & 0.09  $\pm$ 0.060  & 4.04  $\pm$ 0.015  & 4.1  $\pm$ 0.20  & 1.86  $\pm$ 0.043  & 0.25  $\pm$ 0.018 \\
 8684730 &  3.0 $\pm$ 0.38  & 0.24  $\pm$ 0.065  & 4.06  $\pm$ 0.046  & 4.1  $\pm$ 0.53  & 1.9   $\pm$ 0.11   & 0.17  $\pm$ 0.040 \\
 8866102 &  1.9 $\pm$ 0.71  & 0.4   $\pm$ 0.11   & 4.27  $\pm$ 0.014  & 2.8  $\pm$ 0.16  & 1.36  $\pm$ 0.024  & 0.24  $\pm$ 0.039 \\
 9414417 &  3.1 $\pm$ 0.31  & 0.09  $\pm$ 0.030  & 4.016 $\pm$ 0.0058 & 5.0  $\pm$ 0.32  & 1.90  $\pm$ 0.032  & 0.21  $\pm$ 0.026 \\
 9592705 &  3.0 $\pm$ 0.38  & 0.05  $\pm$ 0.026  & 3.973 $\pm$ 0.0087 & 5.7  $\pm$ 0.37  & 2.06  $\pm$ 0.035  & 0.26  $\pm$ 0.015 \\
 9955598 &  7.0 $\pm$ 0.98  & 0.37  $\pm$ 0.035  & 4.494 $\pm$ 0.0061 & 0.66 $\pm$ 0.041 & 0.90  $\pm$ 0.013  & 0.25  $\pm$ 0.020 \\
10514430 &  6.5 $\pm$ 0.89  & 0.06  $\pm$ 0.022  & 4.08  $\pm$ 0.014  & 2.9  $\pm$ 0.17  & 1.62  $\pm$ 0.026  & 0.22  $\pm$ 0.021 \\
10586004 &  4.9 $\pm$ 1.39  & 0.12  $\pm$ 0.090  & 4.09  $\pm$ 0.041  & 3.1  $\pm$ 0.27  & 1.71  $\pm$ 0.070  & 0.24  $\pm$ 0.021 \\
10666592 &  2.0 $\pm$ 0.24  & 0.15  $\pm$ 0.036  & 4.020 $\pm$ 0.0066 & 5.7  $\pm$ 0.33  & 1.98  $\pm$ 0.018  & 0.29  $\pm$ 0.014 \\
10963065 &  4.4 $\pm$ 0.58  & 0.16  $\pm$ 0.054  & 4.292 $\pm$ 0.0070 & 2.0  $\pm$ 0.1   & 1.24  $\pm$ 0.015  & 0.22  $\pm$ 0.029 \\
11133306 &  4.1 $\pm$ 0.84  & 0.22  $\pm$ 0.079  & 4.319 $\pm$ 0.0096 & 1.7  $\pm$ 0.11  & 1.21  $\pm$ 0.019  & 0.22  $\pm$ 0.036 \\
11295426 &  6.2 $\pm$ 0.78  & 0.09  $\pm$ 0.036  & 4.283 $\pm$ 0.0059 & 1.65 $\pm$ 0.095 & 1.26  $\pm$ 0.016  & 0.24  $\pm$ 0.012 \\
11401755 &  5.6 $\pm$ 0.630 & 0.037 $\pm$ 0.0053 & 4.043 $\pm$ 0.0071 & 3.4  $\pm$ 0.19  & 1.69  $\pm$ 0.026  & 0.21  $\pm$ 0.026 \\
11807274 &  2.8 $\pm$ 1.05  & 0.3   $\pm$ 0.11   & 4.17  $\pm$ 0.024  & 3.5  $\pm$ 0.22  & 1.57  $\pm$ 0.038  & 0.22  $\pm$ 0.035 \\
11853905 &  5.7 $\pm$ 0.78  & 0.04  $\pm$ 0.020  & 4.11  $\pm$ 0.011  & 2.7  $\pm$ 0.16  & 1.62  $\pm$ 0.030  & 0.23  $\pm$ 0.022 \\
11904151 &  9.6 $\pm$ 1.43  & 0.08  $\pm$ 0.037  & 4.348 $\pm$ 0.0097 & 1.09 $\pm$ 0.06  & 1.07  $\pm$ 0.019  & 0.21  $\pm$ 0.026 \\
Sun &  4.6 $\pm$ 0.16  & 0.36  $\pm$ 0.012  & 4.439 $\pm$ 0.0038 & 1.01 $\pm$ 0.041 & 1.000 $\pm$ 0.0066 & 0.245 $\pm$ 0.0076 \\ \hline \end{tabular}
\end{table}
}

\afterpage{
\begin{landscape}
\begin{figure*}
    \centering
    \includegraphics[width=0.28\linewidth,keepaspectratio]{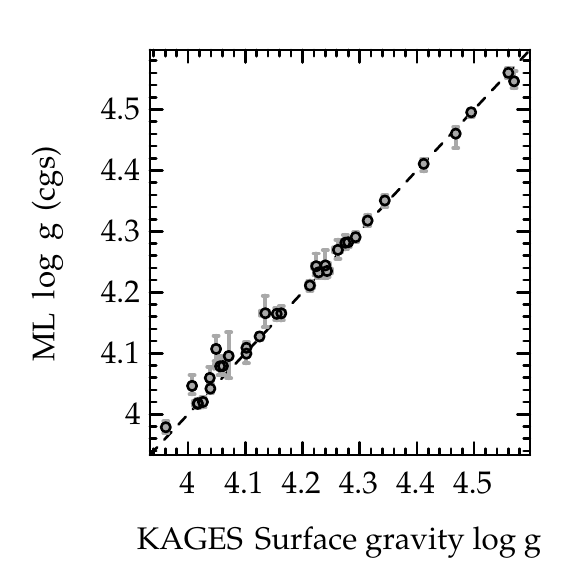}%
    \includegraphics[width=0.28\linewidth,keepaspectratio]{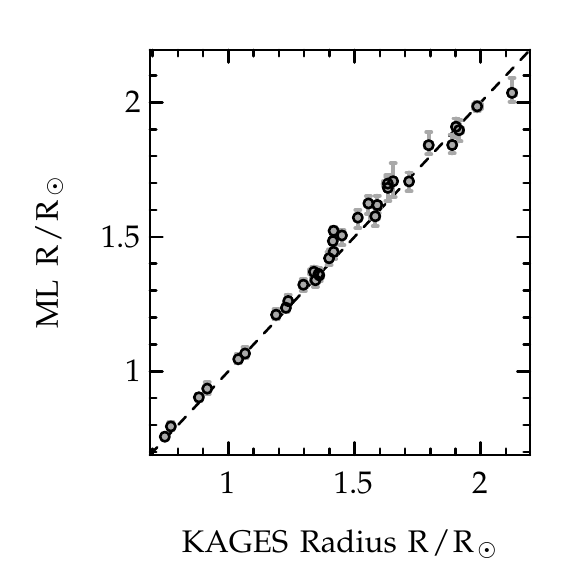}%
    \includegraphics[width=0.28\linewidth,keepaspectratio]{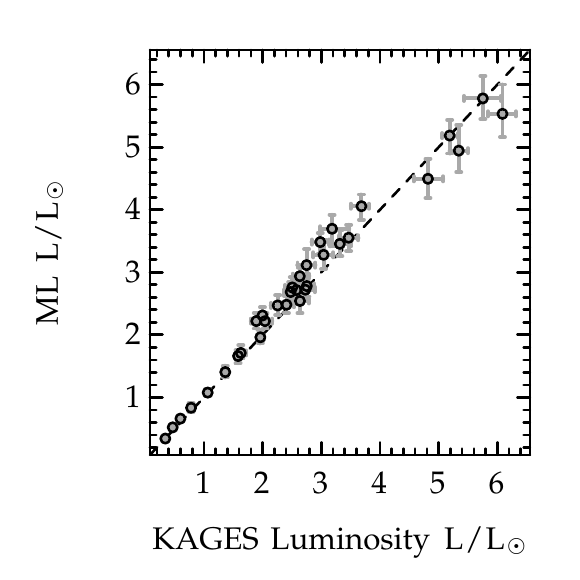}\\
    \includegraphics[width=0.28\linewidth,keepaspectratio]{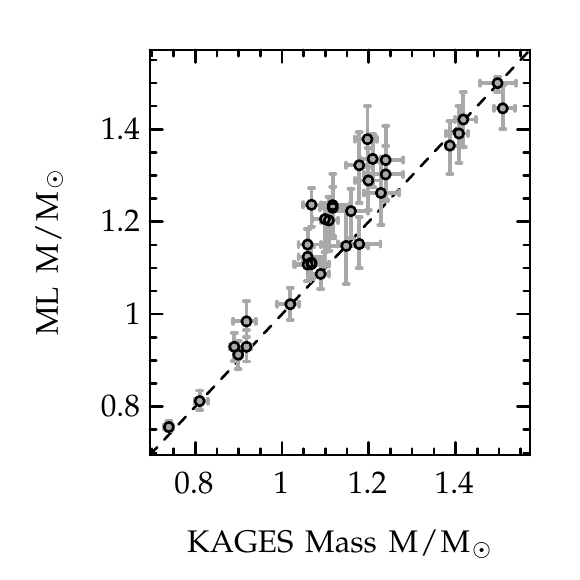}%
    \includegraphics[width=0.28\linewidth,keepaspectratio]{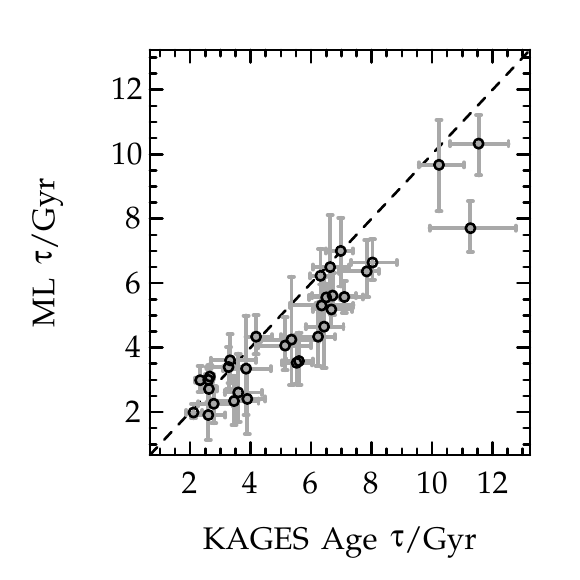}
    \caption[Surface gravities, radii, luminosities, masses, and ages for $34$ \emph{Kepler} objects-of-interest]{Predicted surface gravities, radii, luminosities, masses, and ages of $34$ \emph{Kepler} objects-of-interest plotted against the suggested KAGES values. Medians, $16\%$ quantiles, and $84\%$ quantiles are shown for each point. A dashed line of agreement is shown in all panels to guide the eye. }
    \label{fig:us-vs-them}
\end{figure*}
\end{landscape}
}

\afterpage{
\begin{landscape}
\begin{figure*}
    \centering
    \includegraphics[width=0.98\linewidth,keepaspectratio]{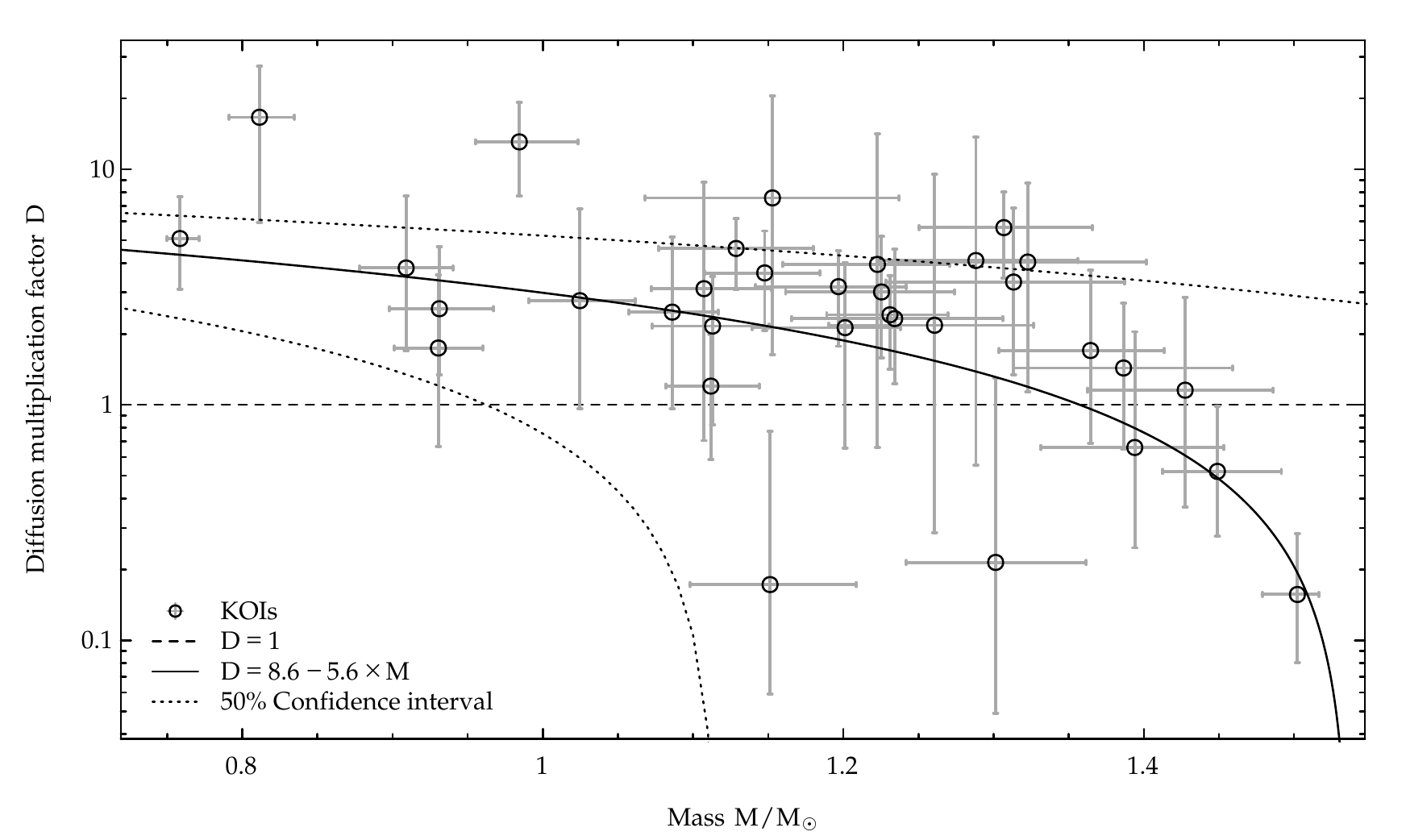}
    \caption[Empirical diffusion-mass relation]{Logarithmic diffusion multiplication factor as a function of stellar mass for $34$ \emph{Kepler} objects-of-interest. The solid line is the line of best fit from Equation~(\ref{eq:diffusion}) and the dashed lines are the $50\%$ confidence interval around this fit. \label{fig:diffusion} } 
\end{figure*}
\end{landscape}
}

\section{Discussion}
The amount of time it takes to make predictions for a star using a trained random forest can be decomposed into two parts: the amount of time it takes to calculate perturbations to the observations of the star \mb{(see Section~\ref{sec:uncertainties})}, and the amount of time it takes to make a prediction on each perturbed set of observations. Hence we have
\begin{equation}
    t = n(t_p + t_r)
\end{equation}
where $t$ is the total time, $n$ is the number of perturbations, $t_p$ is the time it takes to perform a single perturbation, and $t_r$ is the random forest regression time. We typically see times of ${t_p = (7.9 \pm 0.7) \cdot 10^{-3} \; (\si{\s})}$ and ${t_r = (1.8 \pm 0.4) \cdot 10^{-5} \; (\si{\s})}$. We chose a conservative ${n=10,000}$ for the results presented here, which results in a time of around a minute per star. Since each star can be processed independently and in parallel, a computing cluster could feasibly process a catalog containing millions of objects in less than a day. Since \mb{${t_r \ll t_p}$}, the calculation depends almost entirely on the time it takes to perturb the observations.\footnote{Our perturbation code uses an interpreted language (R), so if needed, there is still room for speed-up.} There is also the one-time cost of training the random forest, which takes less than a minute and can be reused without retraining on every star with the same information. It does need to be retrained if one wants to consider a different combination of input or output parameters. 

There is a one-time cost of generating the matrix of training data. We ran our simulation generation scheme for a week on our computing cluster and obtained $5,325$ evolutionary tracks with $64$ models per track, \mb{which resulted in a} $123$~MB \mb{matrix of stellar models}. This is at least an order of magnitude fewer models than the amount that other methods use. Furthermore, this is in general more tracks than is needed by our method: we showed in Figure~\ref{fig:evaluation-tracks} that for most parameters---most notably age, mass, luminosity, radius, initial metallicity, and core hydrogen abundance---one needs only a fraction of the models that we generated in order to obtain good predictive accuracies. Finally, unless one wants to consider a different range of parameters or different input physics, this matrix would not need to be calculated again; a random forest trained on this matrix can be re-used for all future stars that are observed. Of course, our method would still work if trained using a different matrix of models, and our grid should work with other grid-based modelling methods. 

Previously, \citet{pulone1997age} developed a neural network for predicting stellar age based on the star's position in the Hertzsprung-Russell diagram. More recently, \citet{2016arXiv160200902V} have worked on incorporating seismic information into that analysis as we have done here. Our method provides several advantages over these approaches. Firstly, the random forests that we use perform constrained regression, meaning that the values we predict for quantities like age and mass will always be non-negative and within the bounds of the training data, which is not true of the neural networks-based approach that they take. Secondly, using \emph{averaged} frequency separations allows us to make predictions without need for concern over which radial orders were observed. Thirdly, we have shown that our random forests are very fast to train, and can be retrained in only seconds for stars that are missing observational constraints such as luminosities. In contrast, deep neural networks are computationally intensive to train, potentially taking days or weeks to converge depending on the breadth of network topologies considered in the cross-validation. Finally, our grid is varied in six initial parameters---$M$, $Y_0$, $Z_0$, $\alpha_{\text{MLT}}$, $\alpha_{\text{ov}}$, and $D$, which allows our method to explore a wide range of stellar model parameters.

\section{Conclusions}
Here we have considered the constrained multiple-regression problem of inferring fundamental stellar parameters from observations. We created a grid of evolutionary tracks varied in mass, chemical composition, mixing length parameter, overshooting coefficient, and diffusion \mb{multiplication} factor. We evolved each track in time along the main sequence and collected \mb{observable quantities} such as effective temperatures and metallicities as well as global statistics on the modes of oscillations from models along each evolutionary path. We used this matrix of \mb{stellar models} to train a machine learning algorithm to be able to discern the patterns that relate observations to \mb{fundamental stellar parameters}. We then applied this method to hare-and-hound exercise data, the Sun, 16~Cyg~A~and~B, and $34$ planet-hosting candidates that have been observed by \emph{Kepler} and rapidly obtained precise initial conditions and current-age values of these stars. 
Remarkably, we were able to empirically determine the value of the diffusion \mb{multiplication} factor and hence the efficiency of diffusion required to reproduce the observations instead of inhibiting it \emph{ad hoc}. A larger sample size \mb{will} better constrain the diffusion \mb{multiplication} factor and determine what other variables are relevant in its parameterization. \mb{This is work in progress.}

The method presented here has many advantages over existing approaches. First, random forests can be trained and used in only seconds and hence provide substantial speed-ups over other methods. Observations of a star simply need to be fed through the forest---akin to plugging numbers into an equation---and do not need to be subjected to expensive iterative optimization procedures. 
Secondly, random forests perform non-linear and non-parametric regression, which means that the method can use orders-of-magnitude fewer models for the same level of precision, while additionally attaining a more rigorous appraisal of uncertainties for the predicted quantities. 
Thirdly, our method allows us to investigate wide ranges and combinations of stellar parameters. 
And finally, the method presented here provides the opportunity to extract insights from the statistical regression that is being performed, which is achieved by examining the relationships in stellar physics that the machine learns by analyzing simulation data. This contrasts the blind optimization processes of other methods that provide an answer but do not indicate the elements that were important in doing so. 

We note that the predicted quantities reflect a set of choices in stellar physics. Although such biases are impossible to propagate, varying model parameters that are usually kept fixed---such as the mixing length parameter, diffusion \mb{multiplication} factor, and overshooting coefficient---takes us a step in the right direction. Furthermore, the fact that quantities such as stellar radii and luminosities---quantities that have been measured accurately, not just precisely---can be reproduced both precisely and accurately by this method, gives a degree of confidence in its efficacy. 

The method we have presented here is currently only applicable to main-sequence stars. We intend to extend this study to later stages of evolution.

\paragraph*{Acknowledgements} 
\noindent The research leading to the presented results has received funding from the European Research Council under the European Community's Seventh Framework Programme (FP7/2007-2013) / ERC grant agreement no 338251 (StellarAges). This research was undertaken in the context of the International Max Planck Research School \mb{for Solar System Research}. S.B.\ acknowledges partial support from NSF grant AST-1514676 and NASA grant NNX13AE70G. W.H.B. acknowledges research funding by Deutsche Forschungsgemeinschaft (DFG) under grant SFB 963/1 ``Astrophysical flow instabilities and turbulence'' (Project A18).

\paragraph*{Software}
\noindent Analysis in this chapter was performed with \mb{python 3.5.1} libraries scikit-learn \mb{0.17.1} \citep{scikit-learn}, NumPy \mb{1.10.4} \citep{van2011numpy}, and pandas \mb{0.17.1} \citep{mckinney2010data} as well as \mb{R 3.2.3} \citep{R} and the R libraries magicaxis \mb{1.9.4} \citep{magicaxis}, RColorBrewer \mb{1.1-2} \citep{RColorBrewer}, parallelMap \mb{1.3} \citep{parallelMap}, data.table \mb{1.9.6} \citep{data.table}, lpSolve \mb{5.6.13} \citep{lpSolve}, ggplot2 \mb{2.1.0} \citep{ggplot2}, GGally \mb{1.0.1} \citep{GGally}, scales \mb{0.3.0} \citep{scales}, deming \mb{1.0-1} \citep{deming}, and matrixStats \mb{0.50.1} \citep{matrixStats}.

\section{Appendix}
\subsection{Model Selection}
\label{sec:selection}
To prevent statistical bias towards the evolutionary tracks that generate the most models, i.e.\ the ones that require the most careful calculations and therefore use smaller time-steps, or those that live on the main sequence for a longer amount of time; we select ${n=64}$ models from each evolutionary track such that the models are as evenly-spaced in core hydrogen abundance as possible. \mb{We chose $64$ because it is a power of two, which thus allows us to successively omit every other model when testing our regression routine and still maintain regular spacings.}

Starting from the original vector of length $n$ of core hydrogen abundances $\mathbf x$, we find the subset of length $m$ that is closest to the optimal spacing $\mathbf b$, where\lr{\footnote{\lr{This equation has been corrected from the original publication.}}
\begin{equation}
    b_i
    =
    X_T + (i-1) \cdot \frac{X_Z - X_T}{m-1},
    \qquad
    i=1,\ldots,m
\end{equation}}
with $X_Z$ being the core hydrogen abundance at ZAMS and $X_T$ being that at TAMS. To obtain the closest possible vector to $\mathbf b$ from our data $\mathbf x$, we solve a transportation problem using integer optimization \citep{23145595}. First we set up a cost matrix $\boldsymbol{C}$ consisting of absolute differences between the original abundances $\mathbf x$ and the ideal abundances $\mathbf b$:
\begin{equation} 
  \boldsymbol{C} = \left[
  \begin{array}{cccc}
    \abs{b_1-x_1} & \abs{b_1-x_2} & \dots & \abs{b_1-x_n} \\
    \abs{b_2-x_1} & \abs{b_2-x_2} & \dots & \abs{b_2-x_n} \\
    \vdots & \vdots & \ddots & \vdots \\
    \abs{b_m-x_1} & \abs{b_m-x_2} & \dots & \abs{b_m-x_n}
  \end{array} \right].
\end{equation}
We then require that exactly $m$ values are selected from $\mathbf x$, and that each value is selected no more than one time. Simply selecting the closest data point to each ideally-separated point will not work because this could result in the same point being selected twice; and selecting the second closest point in that situation does not remedy it because a different result could be obtained if the points were processed in a different order. 

We denote the optimal solution matrix by $\hat{\boldsymbol{S}}$, and find it by minimizing the cost matrix subject to the following constraints:
\begin{align}
  \hat{\boldsymbol{S}} = \underset{\boldsymbol S}{\arg\min} \; & \sum_{ij} S_{ij} C_{ij} \notag\\
  \text{subject to } & \sum_j S_{ij} \leq 1 \; \text{ for all } i=1\ldots n \notag\\
  \text{and } & \sum_i S_{ij} = 1 \; \text{ for all } j=1\ldots m.
  \label{eq:optimal-spacing}
\end{align}
The indices of $\mathbf x$ that are most near to being equidistantly-spaced are then found by looking at which columns of $\hat{\boldsymbol S}$ contain ones, and we are done. The solution is visualized in Figure~\ref{fig:nearly-even}.

\afterpage{
\clearpage
\begin{landscape}
\begin{figure}
    \centering
    \includegraphics[width=0.95\linewidth, keepaspectratio]{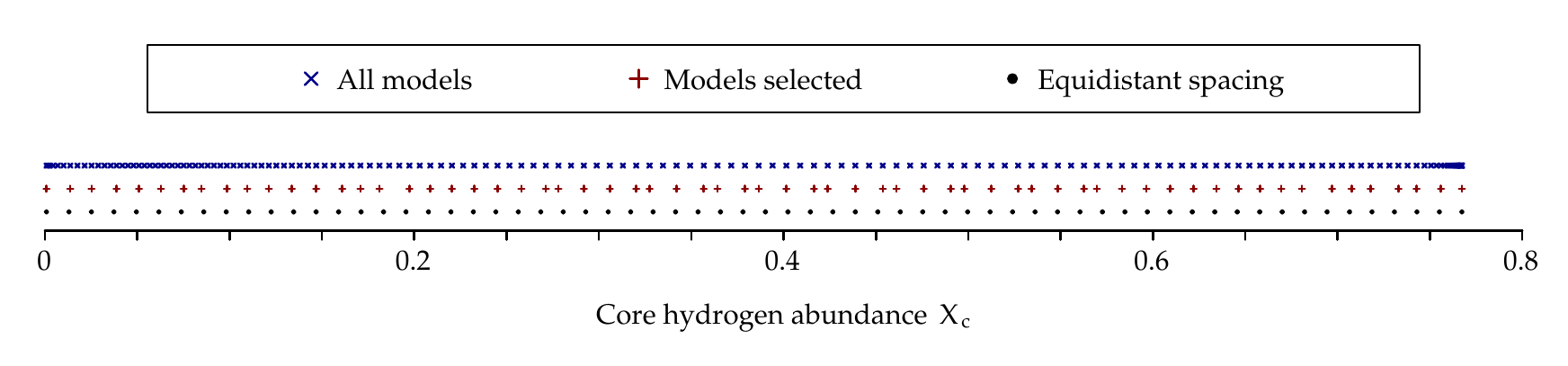}
    \caption[Model selection]{A visualization of the model selection process performed on each evolutionary track in order to obtain the same number of models from each track. The blue crosses show all of the models along the evolutionary track as they vary from ZAMS to TAMS in core hydrogen abundance and the red crosses show the models selected from this track. The models were chosen via linear transport such that they satisfy Equation~(\ref{eq:optimal-spacing}). For reference, an equidistant spacing is shown with black points. \vspace*{5mm}
    \label{fig:nearly-even} }%
\end{figure}
\end{landscape}
}

\subsection{Initial Grid Strategy}
\label{sec:grid}
The initial conditions of a stellar model can be viewed as a \mb{six-dimensional hyperrectangle} with dimensions $M$, $Y_0$, $Z_0$, $\alpha_{\text{MLT}}$, $\alpha_{\text{ov}}$, and $D$. In order to vary all of these parameters simultaneously and fill the \mb{hyperrectangle} as quickly as possible, we construct a grid of initial conditions following a quasi-random point generation scheme. This is in contrast to linear or random point generation schemes, over which it has several advantages. 

A linear grid subdivides all dimensions in which initial quantities can vary into equal parts and creates a track of models for every combination of these subdivisions. Although in the limit such a strategy will fill the \mb{hyperrectangle} of initial conditions, it does so very slowly. It is furthermore suboptimal in the sense that linear grids maximize redundant information, as each varied quantity is tried with the exact same values of all other parameters that have been considered already. In a high-dimensional setting, if any of the parameters are irrelevant to the task of the computation, then the majority of the tracks in a linear grid will not contribute any new information.

A refinement on this approach is to create a grid of models with randomly varied initial conditions. Such a strategy fills the space more rapidly, and furthermore solves the problem of redundant information. However, this approach suffers from a different problem: since the points are generated at random, they tend to ``clump up'' at random as well. This results in random gaps in the parameter space, which are obviously undesirable. 

\begin{figure*}
    \centering
    \includegraphics[width=0.32\textwidth,keepaspectratio,
        trim={7.75cm 9.8cm 7.75cm 9.8cm}, clip]{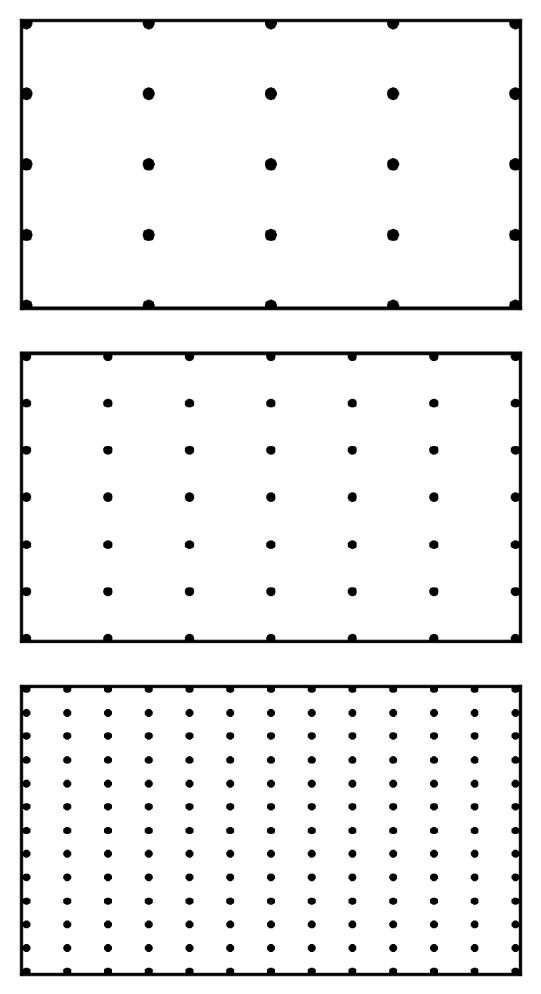}\hfill
    \includegraphics[width=0.32\textwidth,keepaspectratio,
        trim={7.75cm 9.8cm 7.75cm 9.8cm}, clip]{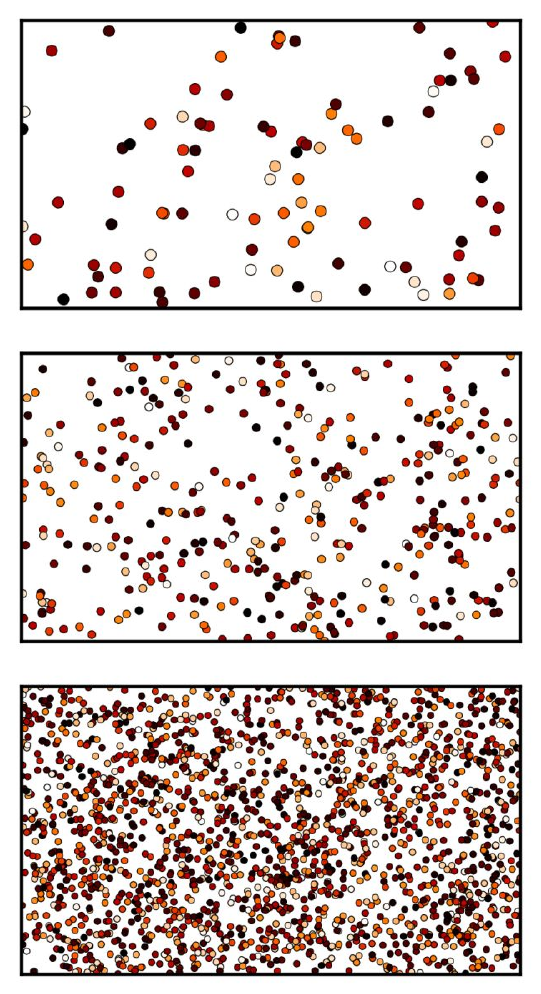}\hfill
    \includegraphics[width=0.32\textwidth,keepaspectratio,
        trim={7.75cm 9.8cm 7.75cm 9.8cm}, clip]{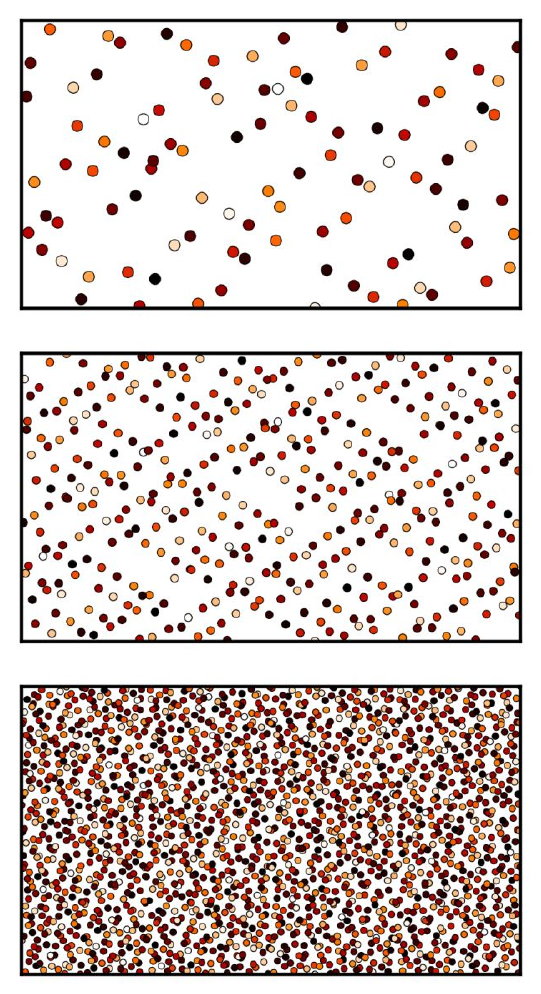}\\
    \parbox{0.32\textwidth}{\centering Linear}\hfill
    \parbox{0.32\textwidth}{\centering Random}\hfill
    \parbox{0.32\textwidth}{\centering Quasi-random}
    \caption[Comparison of point generation schemes (linear, random, quasi-random)]{Results of different methods for generating multidimensional grids portrayed via a unit cube projected onto a unit square. Linear (left), random (middle), and quasi-random (right) grids are generated in three dimensions, with color depicting the third dimension, i.e., the distance between the reader and the screen. From top to bottom, all three methods are shown with 100, 400, and 2000 points generated, respectively. }%
    \label{fig:grids}
\end{figure*}

Therefore, in order to select points that do not stack, do not clump, and also fill the space as rapidly as possible, we generate Sobol numbers \citep{sobol1967distribution} in the unit 6-cube and map them to the parameter ranges of each quantity that we want to vary. Sobol numbers are a sequence of $m$-dimensional vectors ${x_1 \ldots x_n}$ in the unit hypercube $I^m$ constructed such that the integral of a real function $f$ in that space is equivalent in the limit to that function evaluated on those numbers, that is,
\begin{equation}
    \int_{I^m} f = \lim_{n \to \infty} \frac{1}{n}\sum_{i=1}^n f(x_i)
\end{equation}
with the sequence being chosen such that the convergence is achieved as quickly as possible. By doing this, we both minimize redundant information and furthermore sample the hyperspace of possible stars as uniformly as possible. Figure~\ref{fig:grids} visualizes the different methods of generating multidimensional grids: linear, random, and the quasi-random strategy that we took. This method applied to initial model conditions was shown in Figure~\ref{fig:inputs} with 1- and 2D projection plots of the evolutionary tracks generated for our grid. 

\subsection{Adaptive Remeshing}
\label{sec:remeshing}

\begin{figure}[t!]
    \centering
    \includegraphics[width=0.7\linewidth,keepaspectratio,trim={0 1.4cm 0 0.25cm},clip]{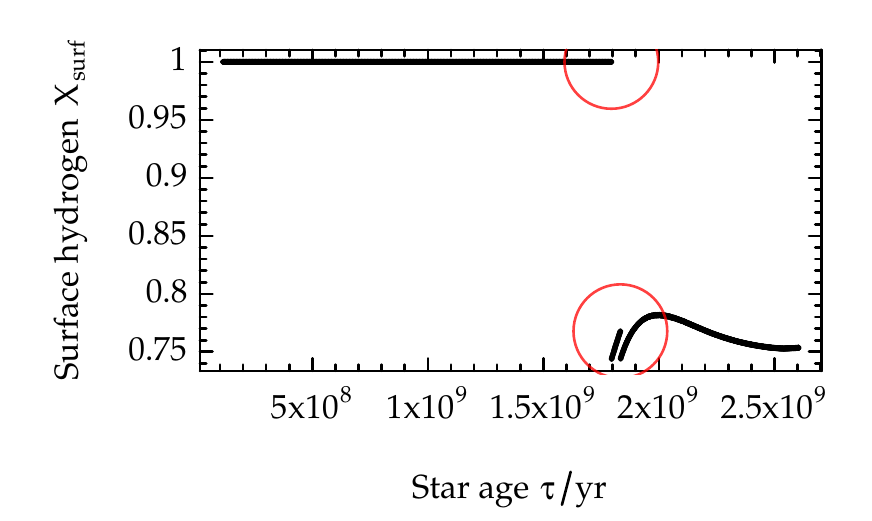}\\
    \includegraphics[width=0.7\linewidth,keepaspectratio,trim={0 1.4cm 0 0.25cm},clip]{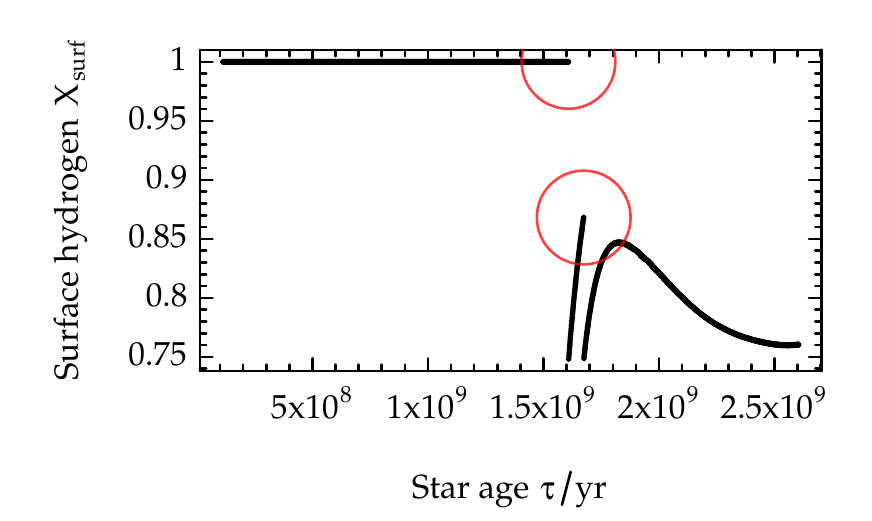}\\
    \includegraphics[width=0.7\linewidth,keepaspectratio,trim={0 0 0 0.25cm},clip]{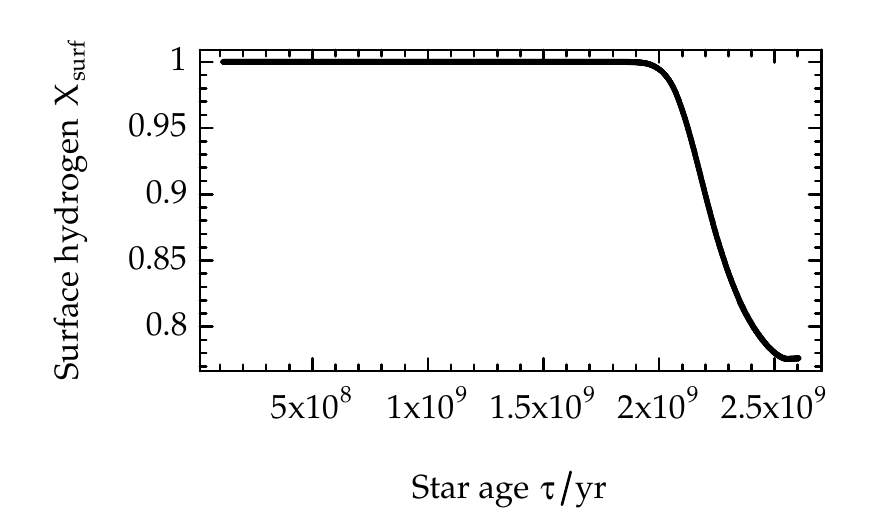}\\
    \caption[Surface abundance discontinuity detection]{Three iterations of surface abundance discontinuity detection and iterative remeshing for an evolutionary track. The detected discontinuities are encircled in red. The third iteration has no discontinuities and so this track is considered to have converged. \vspace*{5mm} \label{fig:discontinuity} }
\end{figure}

When performing element diffusion calculations in MESA, the surface abundance of each isotope is considered as an average over the outermost cells of the model. The number of outer cells ${N}$ is chosen such that the mass of the surface is more than ten times the mass of the ${(N+1)^{\text{th}}}$ cell. Occasionally, this approach can lead to a situation where surface abundances change dramatically and discontinuously in a single time-step. These abundance discontinuities then propagate as discontinuities in effective temperatures, surface gravities, and radii. An example of such a difficulty can be seen in Figure~\ref{fig:discontinuity}. 

Instead of being a physical reality, these effects arise only when there is insufficient mesh resolution in the outermost layers of the model. We therefore seek to detect these cases and re-run any such evolutionary track using a finer mesh resolution. We consider a track an outlier if its surface hydrogen abundance changes by more than $1\%$ in a single time-step. We iteratively re-run any track with outliers detected using a finer mesh resolution, and, if necessary, smaller time-steps, until convergence is reached. The process and a resolved track can also be seen in Figure~\ref{fig:discontinuity}. 

Some tracks still do not converge without surface abundance discontinuities despite the fineness of the mesh or the brevity of the time-steps, and are therefore not included in our study. These troublesome evolutionary tracks seem to be located only located in a thin ridge of models having sufficiently high stellar mass (${M > M_\odot}$), a deficit of initial metals (${Z_0 < 0.001}$) and a specific inefficiency of diffusion (${D \simeq 0.01}$). A visualization of this can be seen in Figure~\ref{fig:diffusion-gap}.

\afterpage{
\clearpage
\begin{landscape}
\begin{figure*}
    \centering
    \includegraphics[width=0.452\linewidth,keepaspectratio]{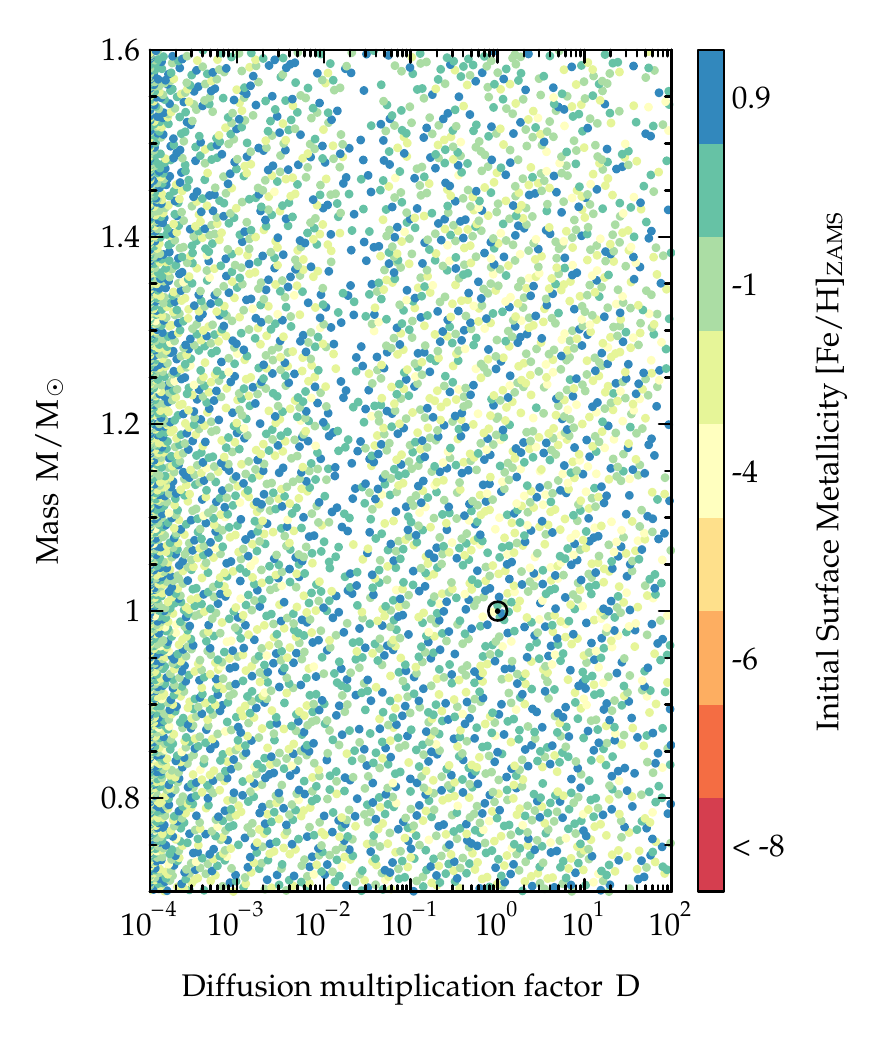}
    \includegraphics[width=0.452\linewidth,keepaspectratio]{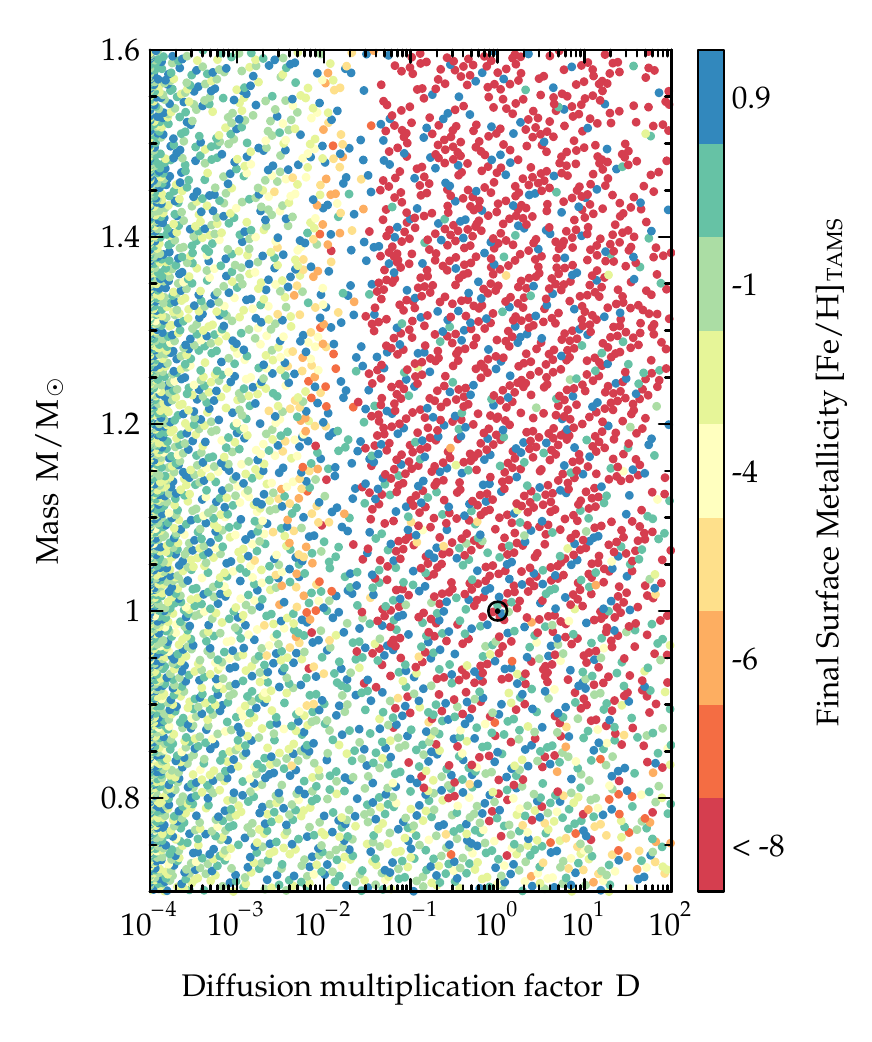}
    \caption[Model convergence as a function of mass and diffusion]{Stellar mass as a function of diffusion \mb{multiplication} factor colored by initial surface metallicity (left) and final surface metallicity (right). A ridge of \mb{missing points indicating} unconverged evolutionary tracks can be seen around a diffusion \mb{multiplication} factor of $0.01$. Beyond this ridge, tracks that were initially metal-poor end their main-sequence lives with all of their metals drained from their surfaces. \label{fig:diffusion-gap} }
\end{figure*}
\end{landscape}
}

\subsection{Evaluating the Regressor}
\label{sec:evaluation}
In training the random forest regressor, we must determine how many evolutionary tracks $N$ to include, how many models $M$ to extract from each evolutionary track, and how many trees $T$ to use when growing the forest. As such it is useful to define measures of gauging the accuracy of the random forest so that we may evaluate it with different combinations of these parameters. 

By far the most common way of measuring the quality of a random forest regressor is its so-called ``out-of-bag'' (OOB) score \citep[see e.g.\ Section~3.1 of][]{breiman2001random}. While each tree is trained on only a subset (or ``bag'') of the stellar models, all trees are tested on all of the models that they did not see. This provides an accuracy score representing how well the forest will perform when predicting on observations that it has not seen yet. We can then use the scores defined in Section~\ref{sec:uncertainties} to calculate OOB scores. 

However, such an approach to scoring is too optimistic in this scenario. Since a tree can get models from every simulation, predicting the \mb{parameters} of a model when the tree has been trained on one of that model's neighbors leads to an artificially inflated OOB score. This is especially the case for quantities like stellar mass, which do not change along the main sequence. A tree that has witnessed neighbors on either side of the model being predicted will have no error when predicting that model's mass, and hence the score will seem artificially better than it should be. 

Therefore, we opt instead to build validation sets containing entire tracks that are left out from the training of the random forest. We omit models and tracks in powers of two so that we may roughly maintain the regular spacing that we have established in our grid of models (refer back to Appendices \ref{sec:grid} and \ref{sec:selection} for details). 

We have already shown in Figure~\ref{fig:evaluation-tracks} these cross-validated scores as a function of the number of evolutionary tracks. Figure~\ref{fig:app-evaluation-models} now shows these scores as a function of the number of models obtained from each evolutionary track, and Figure~\ref{fig:app-evaluation-trees} shows them as a function of the number of trees in the forest. Naturally, $\hat\sigma$ increases with the number of trees, but this is not a mark against having more trees: this score is trivially minimal when there is only one tree, as that tree must agree with itself! We find that although more is better for all quantities,  there is not much improvement after about ${T=32}$ and ${M=16}$. It is also interesting to note that the predictions do not suffer very much from using only four models per track, which results in a random forest trained on only a few thousand models. 

\afterpage{
\clearpage
\begin{landscape}
\begin{figure*}
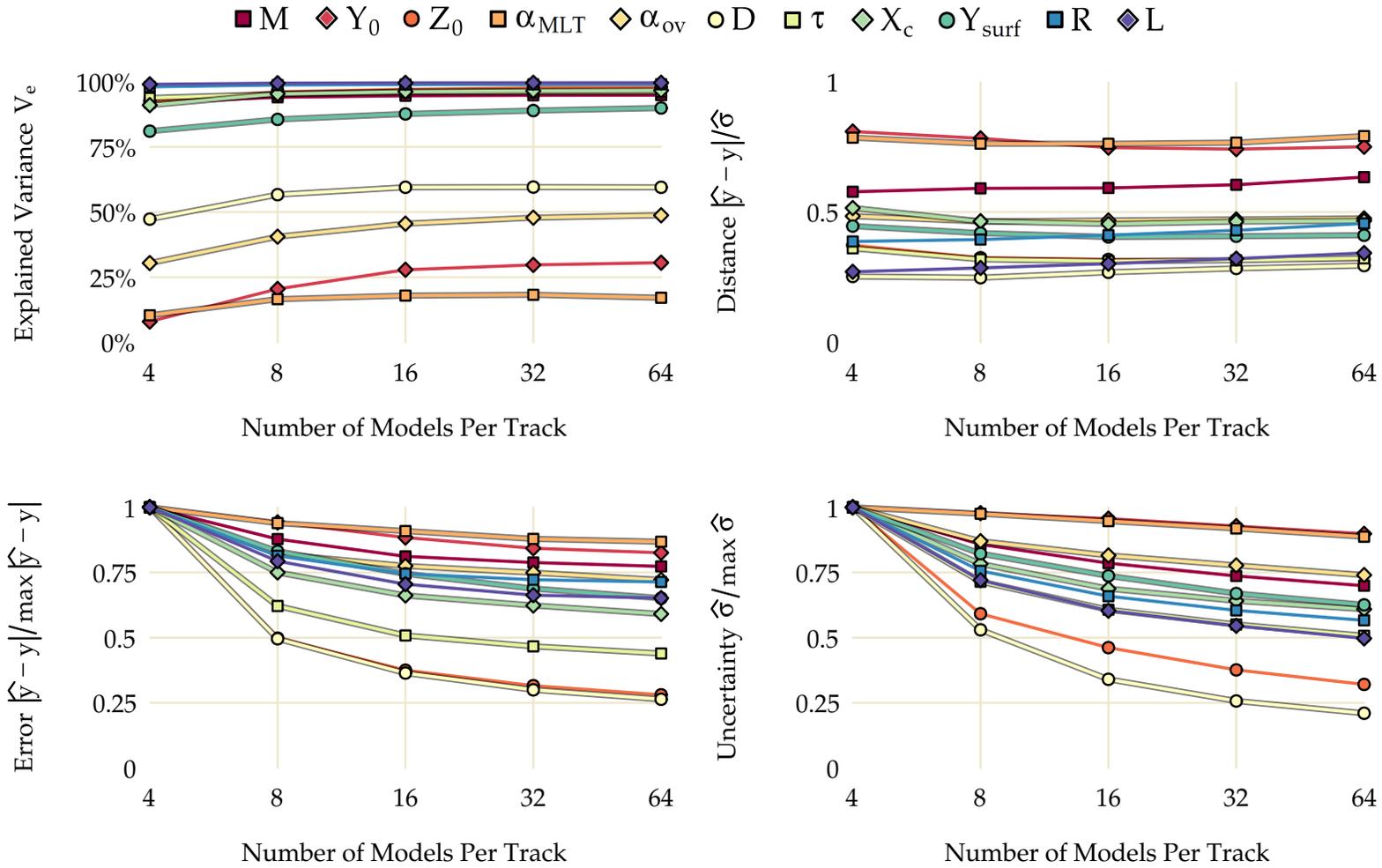

    \centering
    \includegraphics[width=0.6\linewidth,keepaspectratio]{legend.png}\\
    \includegraphics[width=0.45\linewidth,keepaspectratio]{num_points-ev.pdf}%
    \includegraphics[width=0.45\linewidth,keepaspectratio]{num_points-dist.pdf}\\
    \includegraphics[width=0.45\linewidth,keepaspectratio]{num_points-diff.pdf}%
    \includegraphics[width=0.45\linewidth,keepaspectratio]{num_points-sigma.pdf}\\
    \caption[Evaluations of regression accuracy against the number of models per evolutionary track]{
    Explained variance (top left), accuracy per precision distance (top right), normalized absolute error (bottom left), and normalized standard deviation of predictions (bottom right) for each stellar parameter as a function of the number of models per evolutionary track. \label{fig:app-evaluation-models}} 
\end{figure*}
\end{landscape}
\clearpage
}

\afterpage{
\clearpage
\begin{landscape}
\begin{figure}
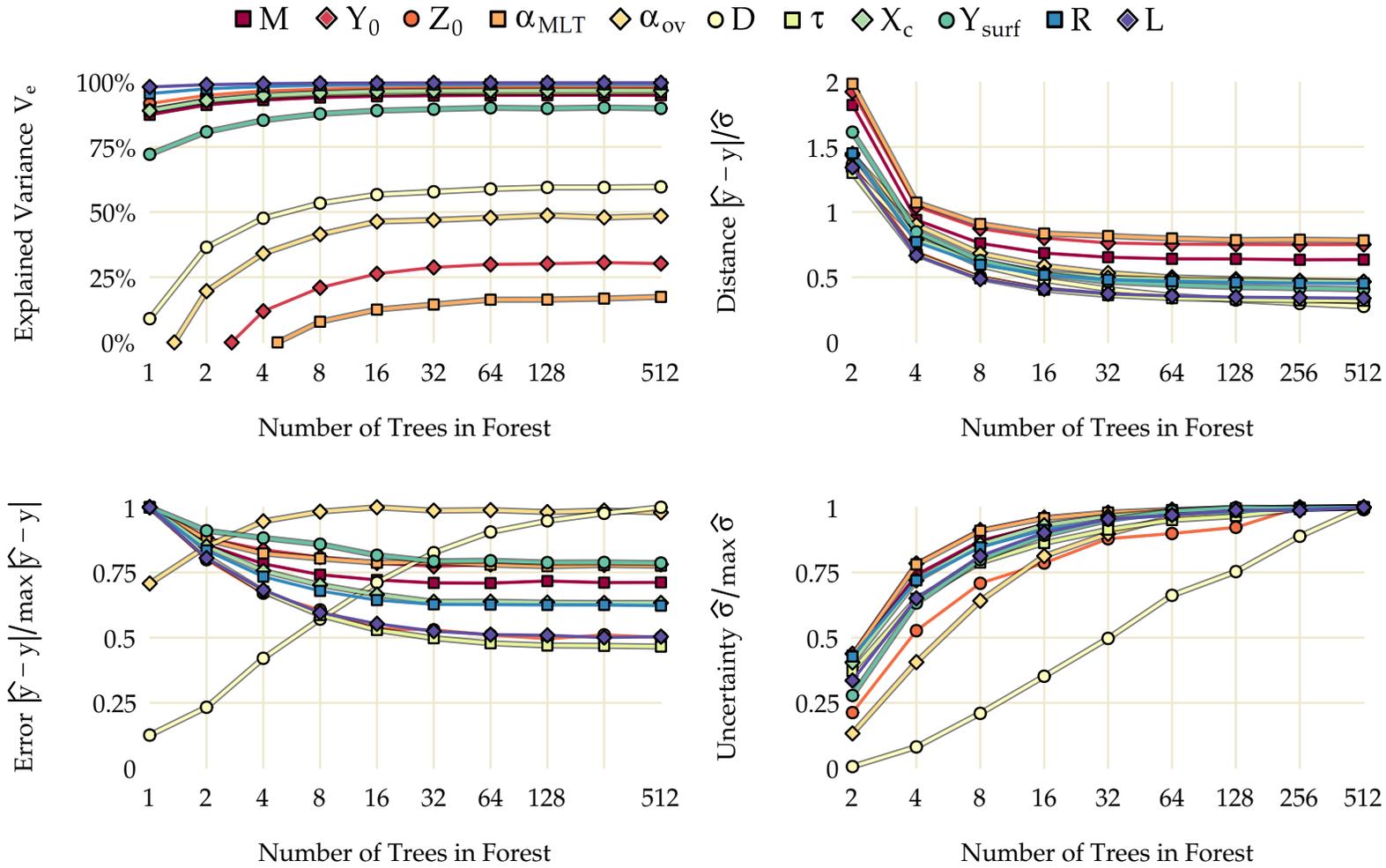

    \centering
    \includegraphics[width=0.6\linewidth,keepaspectratio]{legend.png}\\
    \includegraphics[width=0.45\linewidth,keepaspectratio]{num_trees-ev.pdf}%
    \includegraphics[width=0.45\linewidth,keepaspectratio]{num_trees-dist.pdf}\\
    \includegraphics[width=0.45\linewidth,keepaspectratio]{num_trees-diff.pdf}%
    \includegraphics[width=0.45\linewidth,keepaspectratio]{num_trees-sigma.pdf}\\
    \caption[Evaluations of regression accuracy against the number of trees]{Explained variance (top left), accuracy per precision distance (top right), normalized absolute error (bottom left), and normalized model uncertainty (bottom right) for each stellar parameter as a function of the number of trees used in training the random forest. \label{fig:app-evaluation-trees}} 
\end{figure}
\end{landscape}
}

\subsection{Hare and Hound}
\label{sec:hare-and-hound}
Table~\ref{tab:hnh-true} lists the true values of the hare-and-hound exercise performed here, and Table~\ref{tab:hnh-perturb} lists the perturbed inputs that were supplied to the machine learning algorithm.

\afterpage{
\clearpage
\begin{table} \centering 
\caption{True values for the hare-and-hound exercise. \label{tab:hnh-true}}
\hspace*{-0.66cm}\begin{tabular}{ccccccccccc}
Model & $R/R_\odot$ & $M/M_\odot$ & $\tau$ & $T_{\text{eff}}$ & $L/L_\odot$ & [Fe/H] & $Y_0$ & $\nu_{\max}$ & $\alpha_{\text{ov}}$ & $D$ \\ \hline\hline
0 & 1.705 & 1.303 & 3.725 & 6297.96 & 4.11 & 0.03 & 0.2520 & 1313.67 & No & No \\
1 & 1.388 & 1.279 & 2.608 & 5861.38 & 2.04 & 0.26 & 0.2577 & 2020.34 & No & No \\
2 & 1.068 & 0.951 & 6.587 & 5876.25 & 1.22 & 0.04 & 0.3057 & 2534.29 & No & No \\
3 & 1.126 & 1.066 & 2.242 & 6453.57 & 1.98 & -0.36 & 0.2678 & 2429.83 & No & No \\
4 & 1.497 & 1.406 & 1.202 & 6506.26 & 3.61 & 0.14 & 0.2629 & 1808.52 & No & No \\
5 & 1.331 & 1.163 & 4.979 & 6081.35 & 2.18 & 0.03 & 0.2499 & 1955.72 & No & No \\
6 & 0.953 & 0.983 & 2.757 & 5721.37 & 0.87 & -0.06 & 0.2683 & 3345.56 & No & No \\
7 & 1.137 & 1.101 & 2.205 & 6378.23 & 1.92 & -0.31 & 0.2504 & 2483.83 & No & No \\
8 & 1.696 & 1.333 & 2.792 & 6382.22 & 4.29 & -0.07 & 0.2555 & 1348.83 & No & No \\
9 & 0.810 & 0.769 & 9.705 & 5919.70 & 0.72 & -0.83 & 0.2493 & 3563.09 & No & No \\
10 & 1.399 & 1.164 & 6.263 & 5916.71 & 2.15 & 0.00 & 0.2480 & 1799.10 & Yes & Yes \\
11 & 1.233 & 1.158 & 2.176 & 6228.02 & 2.05 & 0.11 & 0.2796 & 2247.53 & Yes & Yes
\\ \hline \end{tabular}
\end{table}

\begin{table} \centering 
\caption{Supplied (perturbed) inputs for the hare-and-hound exercise. \label{tab:hnh-perturb}}
\begin{tabular}{ccccc}
Model & $T_{\text{eff}}$ & $L/L_\odot$ & [Fe/H] & $\nu_{\max}$ \\ \hline\hline
 0 & 6237 $\pm$ 85 & 4.2 $\pm$ 0.12 & -0.03 $\pm$ 0.09 & 1398 $\pm$  66 \\
 1 & 5806 $\pm$ 85 & 2.1 $\pm$ 0.06 &  0.16 $\pm$ 0.09 & 2030 $\pm$ 100 \\
 2 & 5885 $\pm$ 85 & 1.2 $\pm$ 0.04 & -0.05 $\pm$ 0.09 & 2630 $\pm$ 127 \\
 3 & 6422 $\pm$ 85 & 2.0 $\pm$ 0.06 & -0.36 $\pm$ 0.09 & 2480 $\pm$ 124 \\
 4 & 6526 $\pm$ 85 & 3.7 $\pm$ 0.11 &  0.14 $\pm$ 0.09 & 1752 $\pm$  89 \\
 5 & 6118 $\pm$ 85 & 2.2 $\pm$ 0.06 &  0.04 $\pm$ 0.09 & 1890 $\pm$ 101 \\
 6 & 5741 $\pm$ 85 & 0.8 $\pm$ 0.03 &  0.06 $\pm$ 0.09 & 3490 $\pm$ 165 \\
 7 & 6289 $\pm$ 85 & 2.0 $\pm$ 0.06 & -0.28 $\pm$ 0.09 & 2440 $\pm$ 124 \\
 8 & 6351 $\pm$ 85 & 4.3 $\pm$ 0.13 & -0.12 $\pm$ 0.09 & 1294 $\pm$  67 \\
 9 & 5998 $\pm$ 85 & 0.7 $\pm$ 0.02 & -0.85 $\pm$ 0.09 & 3290 $\pm$ 179 \\
10 & 5899 $\pm$ 85 & 2.2 $\pm$ 0.06 & -0.03 $\pm$ 0.09 & 1930 $\pm$ 101 \\
11 & 6251 $\pm$ 85 & 2.0 $\pm$ 0.06 &  0.13 $\pm$ 0.09 & 2360 $\pm$ 101
\\ \hline \end{tabular}
\end{table}
}
\clearpage












\chapter{On the Statistical Properties of the Lower Main Sequence}
\label{chap:statistical}

The contents of this chapter were authored 
by G.~C.~Angelou, E.~P.~Bellinger, S.~Hekker, and S.~Basu and published in April of 2017 in \emph{The Astrophysical Journal}, 839 (2), 116.\footnote{Contribution statement: The work and writing of this chapter were done in equal parts between G.~C.~Angelou and myself, under the supervision of S.~Hekker and S.~Basu.} 
\nocite{2017apj...839..116a} 




\section*{Chapter Summary}
Astronomy is in an era where all-sky surveys are mapping the Galaxy. 
The plethora of photometric, spectroscopic, asteroseismic and astrometric data allows us to characterize the comprising stars in detail.  
Here we quantify to what extent precise stellar observations reveal information about the properties of a star, including properties that are unobserved, or even unobservable. 
We analyze the diagnostic potential of classical and asteroseismic observations for inferring stellar parameters such as age, mass and radius from evolutionary tracks of solar-like oscillators on the lower main sequence. 
We perform rank correlation tests in order to determine the capacity of each observable quantity to probe structural components of stars and infer their evolutionary histories. We also analyze the principal components of classic and asteroseismic observables to highlight the degree of redundancy present in the measured quantities and demonstrate the extent to which information of the model parameters can be extracted.
We perform multiple regression using combinations of observable quantities in a grid of evolutionary simulations and appraise the predictive utility of each combination in determining the properties of stars.
We identify the combinations that are useful and provide limits to where each type of observable quantity can reveal information about a star. We investigate the accuracy with which targets in the upcoming TESS and PLATO missions can be characterized.  We demonstrate that the combination of observations from GAIA and PLATO will allow us to tightly constrain stellar masses, ages and radii with machine learning for the purposes of galactic and planetary studies.

\section{Introduction} 

The main sequence is generally considered the most well-understood phase of stellar evolution. 
Our Sun is a main-sequence star, and its proximity provides a wealth of constraints to the physics that may occur in low-mass counterparts during this phase \citep[e.g.,][]{2015SSRv..196...49B,2016lrsp...13....2b}. 
Core-hydrogen burning stars are long-lived and hence numerous: indeed, the majority of the stars for which we can resolve parallaxes reside on the main sequence \citep{2016arXiv160904172G}. 
Additionally, many stars of this type display stochastic or ``solar-like'' oscillations that serve to reveal the stellar interior (see, for example, \citealt{2013ARA&A..51..353C} for a review on solar-like oscillators).  
Main-sequence stars are important astrophysical laboratories for testing theories of stellar physics, structure, and evolution; and are a testbed for general physical theories such as nuclear fusion, diffusion, and convection \citep[e.g.,][]{1994MNRAS.269.1137B,1990ARAA..28..263S}. 

Despite all of this, however, the ages of main-sequence stars remain uncertain to at least  $10\%$. This uncertainty stems not only from observational imprecision, but also from the inability of observations to fully constrain stellar parameters.  
Recently, \citetalias{2016apj...830...31b} showed that even for stellar models without observational uncertainties, some model attributes of stars---such as their initial helium abundance or efficiency of convection---could not be fully resolved via global information that can be gleaned from their surfaces.

It is well-known that different observable quantities of stars constrain different model properties. For example, in the now-famous Christensen-Dalsgaard diagram (C--D diagram, the so-called ``asteroseismic HR diagram''), in which the large frequency separation is plotted against the small frequency separation (Appendix \ref{sec:sdefs}), the large frequency separation covaries with the mass of the star and the small frequency separation covaries with its core-hydrogen abundance. Hence, observing one of these quantities sheds light on its unobservable counterpart.
However, to date, a systematic investigation of the extent to which each observable quantity constrains each model property has not been performed.

The equations dictating stellar structure and evolution, and the corresponding microphysics that these equations respond to, give rise to emergent behaviors that are difficult to characterize through examination of the constituting ingredients themselves. To elucidate these opaque relationships, we seek to determine the extent to which observable stellar properties are capable of constraining the internal structures, chemical mixtures, and evolutionary histories of stars. Here we employ the methodology of exploratory data science, a statistical philosophy by which underlying structure in data---simulated or otherwise---can be unearthed.

BA1 used machine learning to build a statistical description of main-sequence stellar evolution. They trained a random forest (RF) of decision trees to learn the relationships that exist between model input parameters and their resultant observable quantities. The technique was developed with particular focus on the determination of stellar ages.  Ages are essential for understanding stellar evolution, characterizing extrasolar planetary systems and advancing models of galactic chemical evolution. 
Notably, the RF developed by BA1 was able to accurately predict stellar properties such as radii and luminosities using other information collected from the stars in their sample. This illustrates that there is redundant information in the stellar quantities, and that there exist model covariances between these quantities that can be characterized and exploited.

The philosophy employed in BA1 is a departure from the standard practice of stellar model fitting. Ordinarily, stellar parameters of observed stars are sought via $\chi^2$-minimization.
The difference in approaches give rise to two points that motivate this paper:
\begin{enumerate}

    \item Methods based on $\chi^2$-minimization assume that each bit of observed information contributes to the objective of constraining the model properties of a star in an exact proportion to how precisely it has been measured. However, two quantities may be measured independently with no measured covariance, and yet still provide redundant information about the star. The result of such a minimization procedure will therefore be a model that is biased towards that redundant information.  The RF developed in BA1, on the other hand, uses the process of statistical bagging to mitigate over-fitting of the data (see also \citealt{hastie2005elements}).  Here we demonstrate the degree to which the observables  carry redundant information about the star. 

    \item The 
    optimization searches of iterative model finding procedures provide solutions but do not indicate the elements that were important in doing so. The use of regression requires that the observables correlate with those model parameters that we wish to infer.  We therefore identify to what extent each observable constrains each model property, and how well the observables must be measured to achieve a desired precision from the regression.  
    
\end{enumerate}

The method developed in BA1 makes use of an artificial intelligence strategy known as supervised learning. The RF that they train seeks relations in evolutionary simulations that enable model properties to be inferred as precisely as possible. Although the RF performs the analysis quickly, precisely, and automatically; supervised machine learning strategies do not provide much insight into how the end result is obtained. The algorithm essentially produces a formula for inferring stellar properties from observations, but one that is too complex for people to use analytically by hand. 

Here we incorporate a complementary strategy. We use the counterpart of supervised learning---\emph{unsupervised learning}---to explicitly uncover the relations between observable properties of stars and their model parameters. Hence, BA1 is of a strictly practical nature: stellar parameters can be inferred rapidly without regard for the how or why; and this paper is aimed to further an understanding of the processes actually involved in such a deduction. 

In this study we draw heavily from the work presented in BA1.
Our analysis initially focuses on elucidating the inherent statistical properties of the grid of stellar models used to train the BA1 RF. 
We determine the relationships and covariances between a chosen subset of stellar parameters and asteroseismic quantities (see Table~\ref{tab:parmdefs}). 
We carry out simultaneous rank correlation tests on the chosen parameters and identify the necessary, dispensable, and irrelevant information for determining each stellar property.
Then, using principal component analysis 
we reduce the dimensionality of the observable quantities and identify to what extent they reveal information of the model parameters.
We subsequently shift the focus of our analysis to how the grid properties are used by the RF and how the choices in the parameters impact on the precision of the regression. 
We train RFs using all combinations of observable quantities in our dataset. The purpose of this is two-fold: first, it is often the case that we wish to quickly characterize a star from a few easily observed quantities---the Hertzsprung-Russell (HR) diagram serves as the classic example. Training and scoring all possible RF combinations provides a means to \emph{quantify} the utility and predictive power of classical and asteroseismic parameters for inferring stellar properties. Secondly, it provides insight into the relationships determined by machine learning algorithms. 
Finally, we identify the observational accuracy required to satisfactorily constrain key stellar parameters. We investigate the observable quantities independently as well as consider the measurements expected from the upcoming TESS and PLATO missions.

\section{Stellar Models and Parameters}
\begin{table}
\centering
{
\renewcommand{\arraystretch}{1.2}
\centering
\begin{tabular}{lll} 
\hline \hline  
\textbf{Qty} & \textbf{Definition} & \textbf{Unit} \\ 
\hline
\multicolumn{3}{l}{Model Input Parameters} \\ 
$M$ & Initial mass & M$_{\odot}$ \\
$Y_0$ & Initial helium mass fraction & \\
$Z_0$ & Initial metal mass fraction & \\
$\alpha_{\text{MLT}}$ & Mixing length parameter & \\
$\alpha_{\text{ov}}$ & Overshoot parameter & \\ 
$D$ & Diffusion efficiency factor & \\[8pt] 
\multicolumn{3}{l}{Stellar Attributes} \\
$\tau$ & Age & yr \\
$\tau_{\text{MS}}$ & Normalized main-sequence lifetime &  \\
M$_{\text{cc}}$ & Convective core mass & M$_{\odot}$ \\
X$_{\text{surf}}$ & Surface hydrogen mass fraction & \\
$Y_{\text{surf}}$ & Surface helium mass fraction & \\
X$_c$ & Central hydrogen mass fraction & \\
$L$ & Luminosity & L$_{\odot}$ \\ 
$R$ & Radius & R$_{\odot}$ \\[8pt]
\multicolumn{3}{l}{Classical Observables} \\
$[\text{Fe/H}]$ & Surface metallicity & \\
$\log{} g$ & Logarithmic surface gravity &  \\ 
$T_{\text{eff}}$ & Effective temperature & K \\[8pt]

\multicolumn{3}{l}{Asteroseismic Observables} \\
$\nu_{\max}$ & Frequency of maximum oscillation power & $\mu$Hz \\
$\langle\Delta\nu_0\rangle$ & Large frequency separation ($\ell=0$) & $\mu$Hz \\
$\langle\delta\nu_{02}\rangle$ & Small frequency separation ($\ell=0,2$) & $\mu$Hz \\
$\langle\delta\nu_{13}\rangle$ & Small frequency separation ($\ell=1,3$) & $\mu$Hz \\
$\langle r_{02}\rangle$ & Frequency separation ratio ($\ell=0,2$) &  \\
$\langle r_{13}\rangle$ & Frequency separation ratio ($\ell=1,3$) &  \\
$\langle r_{01}\rangle$ & Frequency average ratio ($\ell=0,1$) & \\
$\langle r_{10}\rangle$ & Frequency average ratio ($\ell=1,0$) & \\[8pt]\hline 
\end{tabular}
}
\caption{Definitions of the quantities analyzed in this study separated into four parts: model input parameters, stellar attributes, classical observables, and asteroseismic observables. 
Asteroseismic definitions are in Appendix \ref{sec:sdefs}. 
Angled parenthesis indicate the quantity is a calculated weighted median.} 
\label{tab:parmdefs} 
\end{table}

We used  \emph{Modules for Experiments in Stellar Astrophysics} \citep[MESA,][]{2011apjs..192....3p} to generate a grid of stellar evolutionary sequences initially for the purpose of training a random forest. The tracks are varied in initial mass $M$, helium $Y_0$, metallicity $Z_0$, mixing length parameter $\alpha_{\text{MLT}}$, overshoot coefficient $\alpha_{\text{ov}}$, and atomic diffusion multiplication factor $D$ (see BA1 Section~2.1 for details).
Initial model parameters were chosen in a quasi-random fashion from the parameter ranges listed in Table~\ref{tab:prange}.
In total $5325$ evolutionary tracks were evolved from ZAMS to either an age of ${\tau=15}$~Gyr or until terminal-age main sequence (TAMS), which we define as having a fractional core-hydrogen abundance $X_{\text{c}}$ below $10^{-3}$. We conduct our analysis on a subset of stellar models chosen from each sequence so not to bias our statistics towards longer lived stars or numerically challenging evolutionary tracks. 
Details of the choice of input physics, grid generation strategy, and model selection procedure are further outlined in BA1. 
In addition to computing the stellar structure we post process  each model with  the ADIPLS pulsation package \citep{2008Ap&SS.316..113C}.  P-mode oscillations up to spherical degree ${\ell=3}$ below the acoustic cut-off frequency are computed, and from these, frequency separations and separation ratios calculated (see Appendix \ref{sec:sdefs} for mathematical definitions).

There are many quantities that could be included in the current analysis. 
The $25$ parameters we have selected to investigate are listed in Table~\ref{tab:parmdefs}. 
They  comprise key asteroseismic and structural quantities and reflect our focus on characterizing the relationships between observable quantities (observables hereinafter) and those variables that allow us to generate detailed stellar models.

\begin{table}
\centering
\begin{tabular}{llll}
\hline \hline
Parameter	&	Min Value	&	Max Value	&	Variation 	\\ 
\hline 
Mass	&	0.7	&	1.6	&	linear\\
$Y_0$	&	0.22	&	0.34	&	linear\\
$Z_0$	&	$10^{-5}$	&	$10^{-1}$	&	logarithmic\\
$\alpha_{\text{MLT}}$	&	1.5	&	2.5	&	linear\\
$\alpha_{\text{ov}}$	&	$10^{-4}$	&	1	&	logarithmic\\
D	&	$10^{-6}$	&	$10^{2}$	&	logarithmic	\\
\hline
\end{tabular}
\caption{Ranges and sampling strategy for the initial model parameters in the BA1 grid.}
\label{tab:prange}
\end{table}

We consider two parameters not included in the RF training data. 
BA1 elected to omit the frequency of maximum oscillation power, $\nu_{\max}$ (Equation~\ref{equ:nmax}), in their regression model\footnote{$\nu_{\max}$ does have some role in the algorithm developed by BA1, as it is responsible for the location of the Gaussian envelope used to weight and derive averaged/median frequency separations.}. This quantity displays a strong correlation with ${\langle\Delta\nu_0\rangle}$  (see Figure~\ref{fig:filt_corr}  or \citealt{2009A&A...506..465H,2009MNRAS.400L..80S}) and thus offers very little additional information when frequencies are known. 
We include it in the current analysis because $\nu_{\max}$ is the simplest global asteroseismic parameter to extract from time-series observations, and because recent work by Theme{\ss}l~et~al.~(private communication) indicates that the $\nu_{\max}$ scaling relation more accurately reproduces stellar parameters in well-constrained binary systems than the ${\langle\Delta\nu_0\rangle}$ relation (Equation~\ref{equ:dnu}). This is despite the fact that ${\langle\Delta\nu_0\rangle}$ can be measured more precisely and that the relation can be corrected for temperature and metallicity dependencies (Equation~\ref{eq:corrfunc2}) to yield greater accuracy \citep{2016MNRAS.460.4277G, 2016ApJ...822...15S}. 

To complement $\tau$, we have also added normalized main-sequence age, $\tau_{\text{MS}}$, which describes how parameters change as a function of stellar evolution. 
Many low-mass stars in the grid do not reach the terminal-age main sequence (TAMS) before their evolution is stopped. 
Their main-sequence lifetime is estimated by linearly extrapolating the rate at which the central hydrogen is depleted,
\begin{equation}
\tau_{\text{TAMS}} =  \frac{\tau_{\text{last}}}{1-(X_{\text{c, last}}/X_{\text{c, init}})}
\end{equation}
where $\tau_{\text{TAMS}}$ is the TAMS age,  $\tau_{\text{last}}$ is the age of the last model in  the track,  $X_{\text{c, last}}$ is the corresponding core-hydrogen abundance for that model and $X_{\text{c, init}}$ is the core-hydrogen abundance of the initial model in that track. 
For the longest-lived stars we find such an extrapolation is within about  $25\%$ of the true TAMS age. The uncertainty in the extrapolation for these stars stems from the fact we only capture the hydrogen depletion in the early part of the main sequence i.e., when  ${X_{\text{c, last}} > 0.3}$.  Estimating the TAMS age in this manner, however,  will not impact our conclusions.
Large discrepancies are limited to a small number of tracks (192) and differences between the true and extrapolated ages are reduced as ${X_{\text{c, last}} \to 0}$. 
Main sequence lifetime  provides insight into the general correlations that develop as a function of main-sequence stellar evolution. Thus it is the monotonicity of $\tau_{\text{MS}}$ within a given track that is key.  The stellar age parameter, on the other hand, is useful for exploring correlations across the whole parameter space.

\needspace{5\baselineskip}
\section{Rank Correlation Test}
\label{sec:RCT}

\begin{figure}
    \centering
    \includegraphics[width=0.6\linewidth]{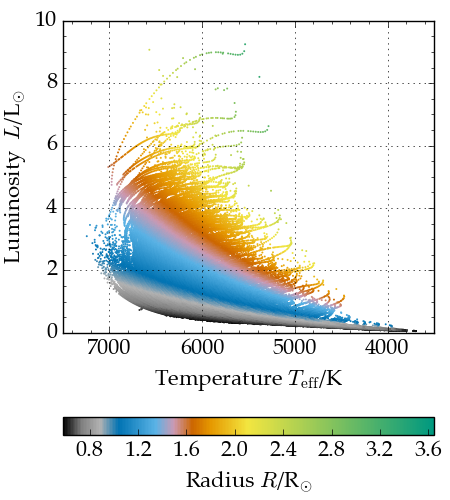}
    \caption[Hertzsprung-Russell diagram for the grid of models]{Hertzsprung-Russell diagram for those tracks in the truncated grid (see text for details). Here each model is coloured by stellar radius.}
    \label{fig:HRDRad}
\end{figure}

\begin{figure*}
    \centering
    \includegraphics[trim={1.5cm 0 2cm 1cm},clip, width=\textwidth]{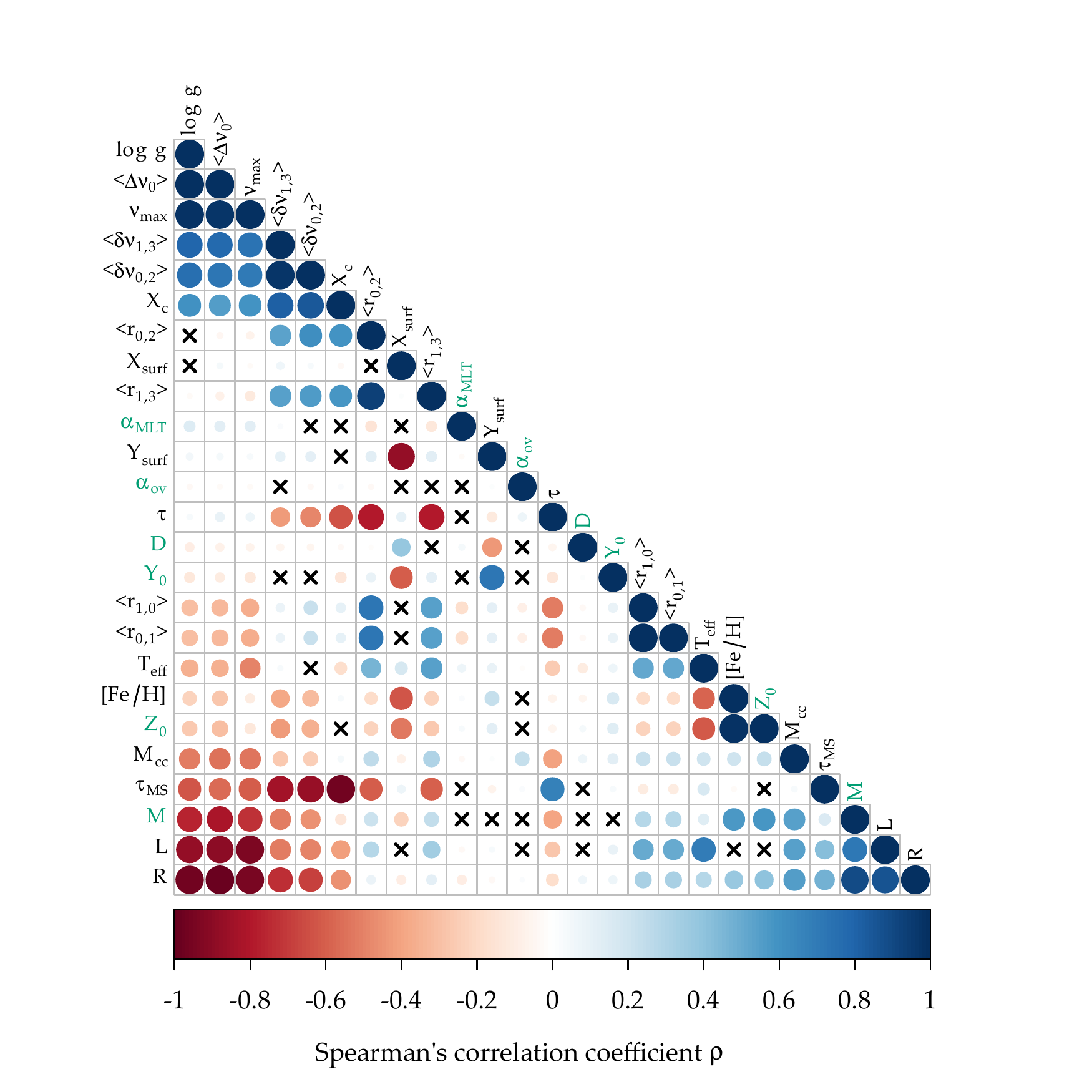}
    \caption[Rank correlation diagram]{Spearman rank correlation matrix comprising various stellar and asteroseismic parameters. The quantities are as described in Table~\ref{tab:parmdefs}
    with model input parameters marked in green.
    The size and the color of each circle both indicate the magnitude of the Spearman coefficient with red and blue denoting negative and positive correlations respectively.  The presence of a cross indicates that 
the two parameters have failed our significance test; i.e., the correlation is indistinguishable from nil. 
The variables are ordered according to their correlation with the first eigensolution of the correlation matrix$^2$.}
    \label{fig:filt_corr}
\end{figure*}

\footnotetext{As principal component analysis is the eigensolution of the correlation (or covariance) matrix, the first eigenvalue indicates the maximum variance in the variables that can be accounted for by a linear model with a single underlying `factor.'
Ordering the parameters in this way  demonstrates the direction of the first principal component (PC$_1$) vector. Figure~\ref{fig:filt_corr} thus  offers a visual representation of principal component analysis which we employ in Section~\ref{sec:PCA}.}

We begin our analysis with a rank correlation test, the purpose of which being to understand the 
statistical properties of the collective lower main sequence.
This is distinct from typical analyses that focus on the evolutionary properties within 
individual stellar tracks or chemically homogeneous isochrones. By identifying correlations present across the entire parameter space we reveal exploitable relationships available to model fitting and regression methods.

Since many quantities (see Table~\ref{tab:parmdefs}) are known to vary in a highly non-linear fashion, we opt to study \emph{rank} statistics. In particular, we replace each quantity by its rank, i.e., an integer representing how big or small a particular quantity is compared to the other models; and calculate Spearman's correlation coefficient $\rho$ between all variables. We further calculate the significance of these correlations (p-values) using the Spearman $\rho$ test. We adopt a conservative significance cut-off of ${\alpha = 10^{-5}}$ and use the Bonferroni correction to account for the fact that we are making multiple ($625$) comparisons \citep[e.g.,][]{doi:10.1080/01621459.1955.10501294}. 

This analysis allows us to determine whether quantities vary monotonically in the same direction (${\rho \approx 1}$), i.e.~both increasing or both decreasing; monotonically apart (${\rho \approx -1}$), i.e.~one increases while the other decreases; or neither (${\rho \approx 0}$)\footnote{Spearman's~$\rho$ is equivalent to Pearson's~$r$ on ranked quantities. We note also that ${\rho = 0}$ does not necessarily indicate a  relationship does not exist; simply that the relationship is not monotonic. A parabolic function for example would result in ${\rho = 0}$.}. When $\abs{\rho}$ is nearly one, the information from one parameter can be used to determine information about the other. Therefore, this is a valuable tool for probing the relationships that exist in and across evolutionary tracks and determining which model properties can be inferred from which observable quantities. 

In the current analysis, we are strictly interested in the relationships expected from the observational data. We apply cuts to the grid computed by BA1 as it spans a wide parameter range\footnote{When training a RF for the purposes of characterizing stellar systems,  sampling the parameter space well beyond the expected ranges of each quantity is prudent. RFs do not extrapolate---doing so would be undesirable anyway---so characterizing a star requires that all of its observations are firmly within the boundaries of the grid used to train the RF. Doing this furthermore avoids pre-conceived biases in the analysis: it allows the observations to dictate the interesting regions of the parameter space rather than limiting the ranges to the values we \emph{expect} the parameters to take.}. The full set of tracks in the BA1 grid includes models with temperatures exceeding the limit in which solar-like oscillations are thought to develop (${ T_{\text{eff}} \approx 6700}$~K, i.e., the approximate surface temperature beyond which the stellar envelopes are radiative rather than convective). 
Evolutionary tracks in the training grid with more than half of the constituent models having $T{_{\text{eff}} > 6700}$~K are excluded from the rank correlation analysis.
Note that the grid will still contain models with $T{_{\text{eff}} > 6700}$~K if more than half the models in a track display temperatures below this cutoff; there is some chance we may observe such stars. 
Likewise, we omit tracks where high atomic-diffusion rates significantly drain metals from the surface, i.e., tracks where more than half the models display surface-hydrogen mass fractions ${> 0.95}$. The dearth of stars observed at zero metallicity indicates that there are some physical processes not included in our models (e.g., radiative levitation or turbulent diffusion) which inhibit the unabated flow of metals from the stellar surface.
This is a common result in models of high-mass stars that include gravitational settling and therefore the process is \emph{ordinarily} suppressed once  ${M \gtrsim 1.1\;M_{\odot}}$. Metal depletion may also arise in cases when settling is made to operate extremely efficiently. 
The removal of these sequences reduces the BA1 training set from $5325$ to $2010$ evolutionary tracks (truncated grid hereinafter) for the current analysis. 
In Figure~\ref{fig:HRDRad} we plot the truncated grid in the HR diagram and color the models according to radius.

Figure~\ref{fig:filt_corr} shows the results of the correlation analysis for the truncated grid.
We defer correlation analysis on the full grid of models to Appendix \ref{sec:fullcorr}. 
Care is needed when interpreting Figure~\ref{fig:filt_corr}. 
First, it is important to remember that correlation is not transitive\footnote{This is irrespective of whether one is using Pearson's~$r$, Spearman's~$\rho$ or Kendall's~$\tau$.} \citep{lang}, i.e.,
\begin{equation}
\corr(X,Y) \wedge \corr(Y,Z) \not \Rightarrow \corr(X,Z)
\end{equation}
even when the correlations are due to causative relationships \citep{stav}.
In fact one can only draw inference on the direction of ${\corr(X,Z)}$ in cases when 
\begin{equation}
\rho_{X,Y}^2 + \rho_{Y,Z}^2 > 1
\end{equation}
(transitive criterion hereinafter).

Second, recall that these correlations hold only for the main sequence. During the main sequence there is generally a positive correlation between, say, $L$ and $T_{\text{eff}}$. 
This relationship will change as the stars evolve further beyond the main-sequence turnoff. 

Third, save for correlations with $\tau_{\text{MS}}$, the relationships presented here do not describe how parameters correlate internally throughout an evolutionary track. Rather, they describe how they correlate across \emph{all} tracks. For example, as a star ascends the main sequence, luminosity increases and therefore one may expect a strong positive correlation between $\tau$ and $L$. The fact that we report a negative correlation is because higher-mass stars are shorter lived -- thus high $L$ corresponds to a lower $\tau$  when the whole parameter space is considered. This correlation is in fact stronger in the analysis of the complete grid used in BA1 which we report in Appendix \ref{sec:fullcorr}, as our grid truncation preferentially selects against higher-mass stars. 
Furthermore we note that some initial model variables ($M$, $Y_0$, $Z_0$, $\alpha_{\text{MLT}}$, $\alpha_{\text{ov}}$ and $D$; all indicated in green) correlate with other parameters.  This would not be the case if we reported correlations within tracks, as these parameters do not change within a given track. 

It should be noted that there is some bias present in the grid as the low-mass stars are not computed to the end of their main-sequence lifetime. The strengths of some correlations would change had we considered evolution beyond the age of the Universe. 
\subsection{Interpreting the Correlations}
Having set the general context in which to interpret Figure~\ref{fig:filt_corr}, we highlight some statistical features of the lower main sequence that can be extracted: 

\begin{itemize}
\item Most pairs of parameters with ${| \rho | \approx 1}$ correspond to well known main-sequence and/or asteroseismic relations. Pairs displaying strong correlations include:

\begin{tabular}{lll}
  $\langle\Delta\nu_0\rangle  -  \log{} g$; 
& $\langle\Delta\nu_0\rangle - \nu_{\max}$; 
& $\log{} g - \nu_{\max}$;  \\
  $\langle\Delta\nu_0\rangle - R$; 
& $\log{} g - R$; 
& $M  -  R$;  \\
  $L - R$; 
& $\langle\delta\nu_{02}\rangle - X_c$. 
&
\end{tabular}

\item Figure~\ref{fig:HRDRad} illustrates why $T_{\text{eff}}$ and its correlations with $R$ and $L$ are weaker than those listed above. 
Many of the tracks evolve past the main-sequence turn off before exhausting their core-hydrogen abundance. The change in morphology of the HR diagram and resultant increase in radius impacts on the monotonicity of the respective correlations.

\item The mass of the convective core (M$_{\text{cc}}$) displays a moderate negative correlation with age whereas it barely registers a relationship with $\tau_{\text{MS}}$. 
It is the higher-mass and hence shorter-lived stars that preferentially develop convective cores. A negative correlation with age is therefore according to expectations.  
In stars that burn hydrogen radiatively no correlation will develop between M$_{\text{cc}}$ and $\tau_{\text{MS}}$. 
In those stars that burn convectively, the size of the convective core will grow but then recede as the CNO-burning region becomes more centrally condensed. 
These two factors lead to an (essentially) null result between M$_{\text{cc}}$ and $\tau_{\text{MS}}$.

\item The correlations between $\tau$ and the ratios ${\langle r_{02}\rangle}$ and ${\langle r_{13}\rangle}$ are stronger than the correlation between $\tau$ and  X$_c$. 
The grid comprises large ranges in mass and metallicity and hence stars at different ages can possess the same X$_c$, thereby weakening the strength of that correlation.
Conversely, as one might expect,  $\tau_{\text{MS}}$ exhibits a stronger relationship with X$_c$ than the ratios. 

\item The small frequency separations and the asteroseismic frequency ratios strongly correlate with both $\tau$ and X$_c$. The large frequency separation, however, demonstrates a much stronger correlation with X$_c$ than it does with $\tau$. The rate at which stars burn their central fuel will largely depend on their mass, thus the 
models can attain the same density (which is proportional to the large frequency separation) at a range of ages. Both $\tau_{\text{MS}}$ and X$_c$ are evolutionary variables and display the expected correlations with ${\langle\Delta\nu_0\rangle}$. 

\item We lack the necessary information to constrain some of the initial model variables. Indeed ${[\text{Fe/H}]}$ provides some constraints on the diffusion efficiency factor $D$, but there is much degeneracy: a model can attain the same surface Y starting with a low ${[\text{Fe/H}]}$ and low diffusion rate as a track with a high ${[\text{Fe/H}]}$ and high diffusion rate. It is possible that fitting for the base of the convective envelope through seismic analysis of the acoustic glitch signal \citep{2014ApJ...782...18M, 2014ApJ...794..114V} could help further constrain these parameters.

\end{itemize}

Figure~\ref{fig:filt_corr} immediately reveals information about the relationships utilized in the machine learning algorithms. 
For those parameter pairs that failed the significance test, neither is likely to feature in the regression model that predicts the other, except in a circumstance where a subset of models exhibit a trend that is absent from the general case of all the models being considered together.
Conversely, where possible, the regressor will attempt to draw on information from pairs that display the strongest correlations.
Quantities such as radius illustrate that there is indeed redundant information in independently measured parameters.
This is useful as the observables measured, and their corresponding accuracy, will vary from survey to survey. 
If a key piece of datum is missing or unreliable, a new regression model can be trained using an appropriate substituted quantity in its place.   
This requires that the redundant information in the observables are treated correctly, if however they are not, then they will lead to biases in model finding procedures. We explore this point further in the next section.

\section{Principal Component Analysis}
\label{sec:PCA}
\begin{figure}
    \centering
    \includegraphics[width=0.8\linewidth]{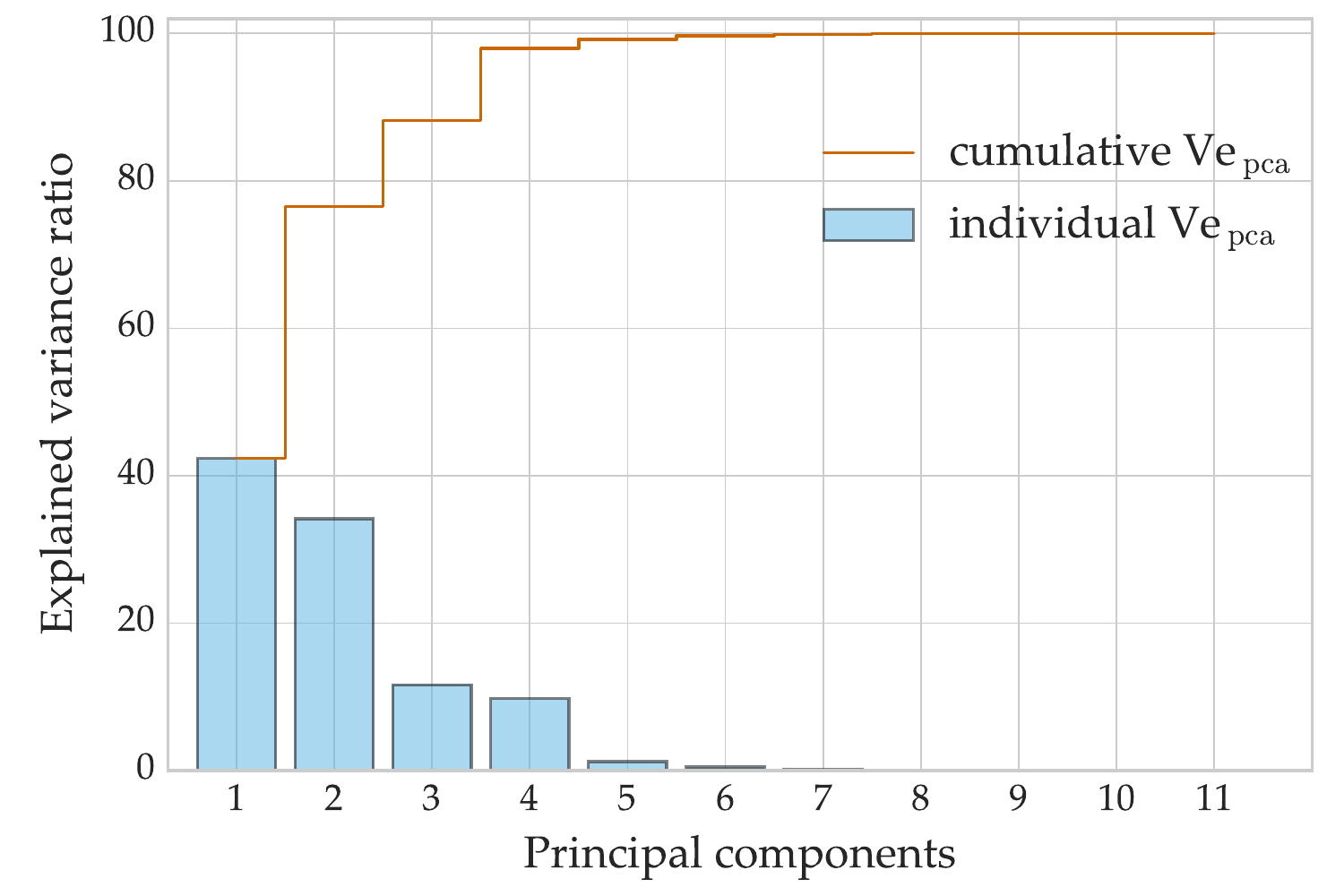}
    \caption[Explained variance of principal components]{{The explained variance (Ve$_{\,\text{pca}}$) and cumulative Ve$_{\,\text{pca}}$ of the principal components comprising the \emph{observable} quantities listed in Table~\ref{tab:parmdefs}. The figure demonstrates that  $98\%$ of the variance in the $11$ observational parameters can be explained by four independent components and $99.2\%$ of the variance explained when a fifth component is considered. The Ve$_{\,\text{pca}}$ of each component is  also presented in the second column of Table~\ref{tab:PCAEV}. }}
    \label{fig:GCA-pca}
\end{figure}

\begin{figure} \centering
\includegraphics[width=0.7\textwidth]{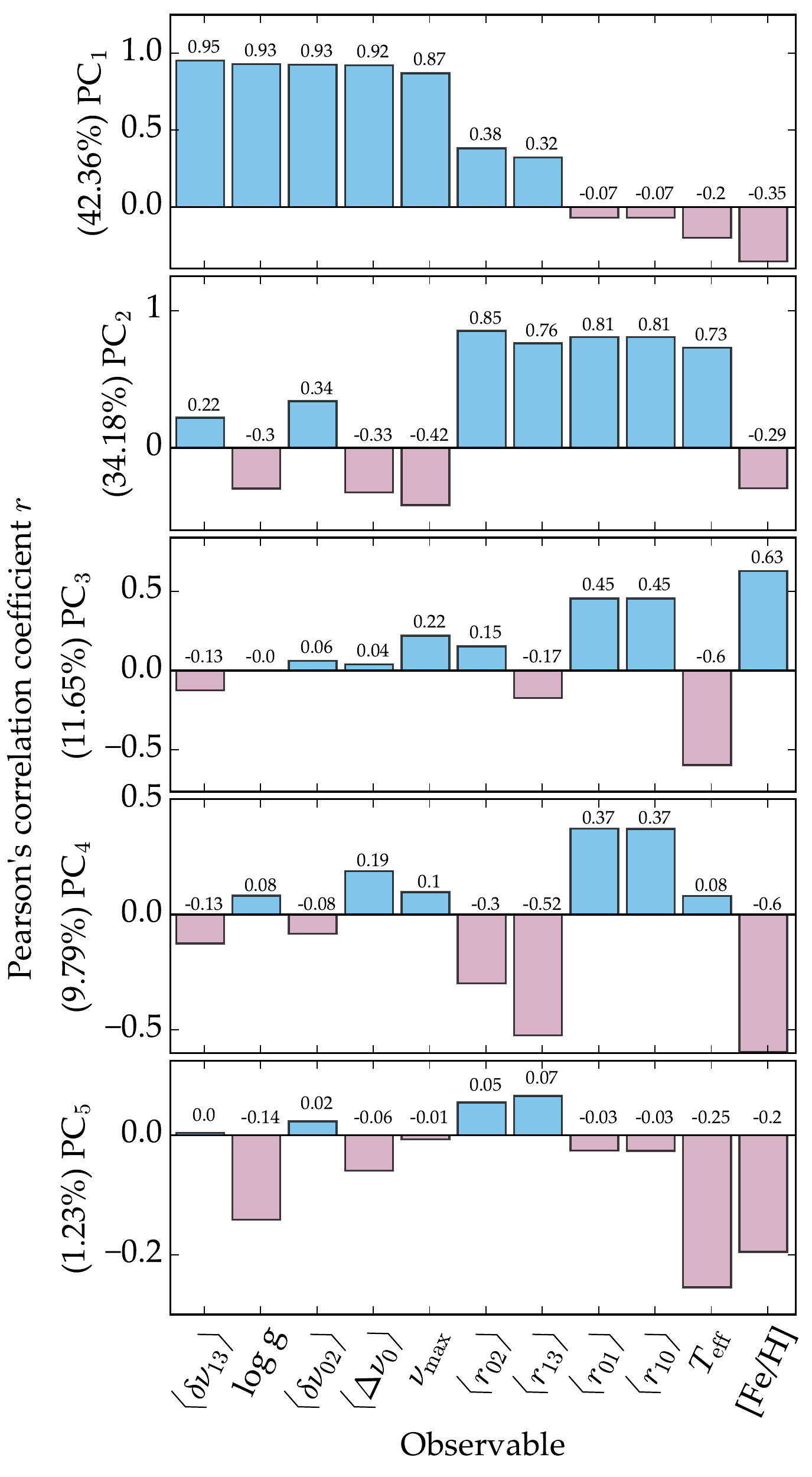}%
\caption[Correlation between principal components and stellar observables]{ Pearson correlation strength between the first five principal components and the stellar observables.  Quantities are ordered according to their correlation strength with the first principal component.  Strong correlations indicate that much of the variance of the quantity is captured by the given PC. Note that the ordinate axes in this figure are on different scales. }
\label{fig:GCA-pcabar}
\end{figure}

\begin{figure} \centering
\includegraphics[width=0.7\textwidth]{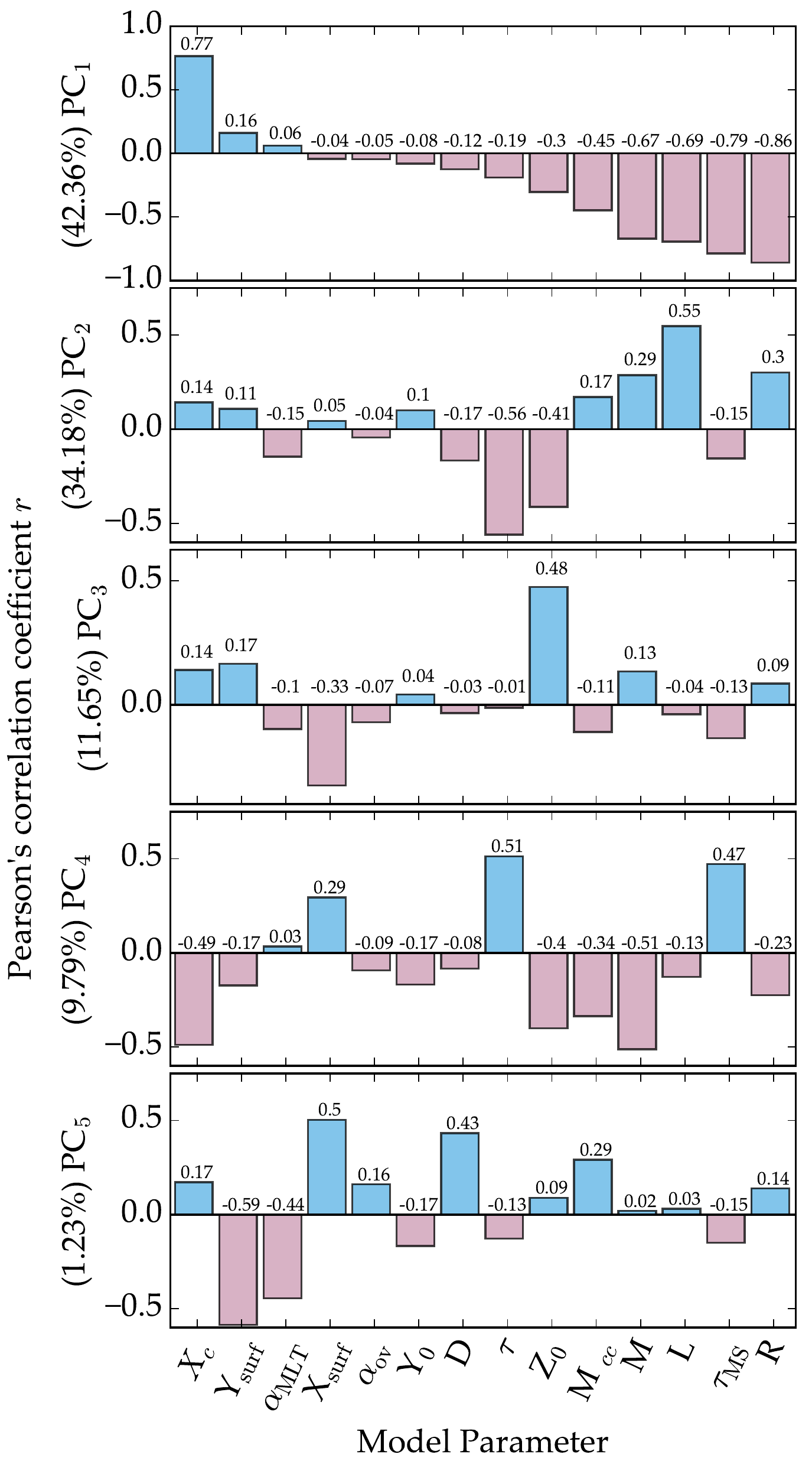}%
\caption[Correlation between principal components and model parameters]{ Pearson correlation strength between the first five principal components and the model parameters (\emph{cf.} Figure~\ref{fig:GCA-pcabar}). }
\label{fig:GCA-pcabarb}
\end{figure}

Past studies, particularly \citet{1994ApJ...427.1013B}, have argued that redundancies and covariances in the stellar observables should be taken into account during any model fitting procedure. 
They demonstrated one particular method (singular value decomposition, SVD hereinafter) of avoiding such biases. In the previous section we identified correlations present in the lower main sequence. Here we demonstrate the degree of redundant information contained in the observables by applying dimensionality reduction. 
We perform principal component analysis (PCA) in order to discover latent structure in observable stellar quantities such that they may be related more directly---and without redundancy---to parameters of stellar modelling. 
Through the principal components (PCs) we quantify the extent to which the observables capture information of the model parameters.

A natural strategy for dealing with high-dimensional data is to reduce the dimensionality in search of \emph{latent variables}; i.e., hidden variables that are more useful than the original quantities under consideration. 
Principal component analysis (PCA) is a technique to transform data into a sequence of orthogonal, and hence independent,  linear combinations of the variables.
Each successive component is constructed to maximize the residual variance from the original data whilst remaining orthogonal to the previous components.
It is a linear transformation in which the change of basis captures the variance contained in original data. 
If parameters in the data are highly correlated, then PCA can potentially produce a lower-dimensional representation without significant loss of the information. 
The method can therefore introduce a new set of variables capable of revealing the underlying structure of an originally high-dimensional space.

PCA  belongs to a family of matrix decomposition techniques that also include methods such as non-negative matrix factorization and independent components analysis as well as variations such as sparse PCA and kernel PCA. 
It has previously been employed in an astrophysical context \citep{2008ApJ...686.1349B, 1987ASSL..131.....M} along with SVD \citep{1994ApJ...427.1013B, 2009ApJ...699..373M} to handle correlated errors in observational data. 
The PCs in this work are calculated from the eigensolution of the correlation matrix, the results of which are not scale invariant. 
We note that PCA can be interpreted as the singular value decomposition of a data matrix in cases where the columns have first been centered by their means. 
Thus SVD analysis\footnote{This method is in fact more numerically stable but more computationally expensive for extracting PCs.} is an alternative method for extracting the PCs (see also Appendix \ref{sec:lambdaa}). 
We indeed compare both methods as a check on our methodology and find that the magnitude of PC scores are identical although the direction (sign) of the vector may differ on occasion.

\subsection{Explained Variance of the Principal Components}
\label{sec:ev}
We perform PCA on $11$ classical and asteroseismic observables listed in Table~\ref{tab:parmdefs}.
The chosen parameters reflect the quantities typically  extracted\footnote{Radius and luminosity are in some cases observable, but not ubiquitously available in the pre-GAIA era. We concede that the inclusion of ${\ell=3}$ modes is an optimistic assumption.} from stars in the \emph{Kepler} \citep{2004SPIE.5487.1491K,2010Sci...327..977B} field.
Our analysis focuses on the  truncated grid of models\footnote{To extract a robust interpretation of the PCs we consider different subsets of the BA1 grid (see Appendix \ref{sec:fullPCA}).} (see Section~\ref{sec:RCT}).
The truncated grid reduces our matrix to size ${128640 \times 11}$ on which we perform the PCA (there are $340,800$ models in the full BA1 grid). 

The PCs throughout this analysis are calculated from the eigendecomposition of observables in the correlation matrix. Here we wish to explain the variance in the data values rather than their rankings. 
 We employ Pearson's~$r$ in the computation of the correlation matrix for the PCA analysis rather than Spearman's~$\rho$. This allows us to transform freely back and forth between the original data space and the space of Pearson PCs.  
 
A given data matrix $\mathbf{X}$ (grid) is of size ${n \times p}$ where $n$ is the number of models  and $p$ is the number of observable parameters. 
Each entry $x_{np}$ in $\mathbf{X}$ is centered and scaled such that 
\begin{equation}
\bar{x}_{np} = (x_{np} -\hat{x_n})/\sigma_{x_n}
\end{equation}
where $\bar{x}_{np}$ is the centered and scaled value,  $x_{np}$ is the original entry, $\hat{x}_n$ is the mean of the particular parameter and $\sigma_{x_n}$ is its standard deviation. 
With all variables having zero mean and unit variance ($\mathbf{\bar{X}}$), our analysis is equivalent to performing eigendecomposition on the covariance matrix\footnote{We are essentially performing the eigendecompostion of the normalized covariance matrix.}. 
We compute $\boldsymbol\Sigma$, the matrix of Pearson's~$r$  coefficients, between all entries in $\mathbf{\bar{X}}$; and compute the eigenvalues and eigenvectors of $\boldsymbol\Sigma$ to determine the PCs.
The eigenvalues, $\lambda_i$, of $\boldsymbol\Sigma$ indicate the absolute variance explained by the eigenvectors. We use this to compute the fraction of variance explained by the eigenvector in the dataset, Ve$_{\, \text{pca}}$, such that:
\begin{equation}
\text{Ve}_{\, \text{pca}} \ (\text{PC}_i)=\frac{\lambda_i}{\sum_{i=1}^p \lambda_i}, 
\label{eqn:pcaev}
\end{equation}
where the number of observables in the data matrix, $p$, is equivalent to the number of principal components we extract.

The Ve$_{\,\text{pca}}$ and the cumulative explained variance of the PCs are reported in Figure~\ref{fig:GCA-pca} (see also the second column in Table~\ref{tab:PCAEV}).
Remarkably, we find that $99.2\%$ of the variance in our 11-dimensional observable space can be explained by a space of five components.
Hence, observable stellar quantities are clearly highly redundant in what they reveal, as only five dimensions contain original information about the star.

Further insight into the PCs can be gained through correlation analysis between the transformed data (i.e., data matrix projected onto the new PC features) and the original data matrix of observables.
Any observable that correlates with a PC contributes to the linear combination of parameters that comprise that PC -- the PC is capturing part of the variance in that observable/dimension. 
Multiple parameters that simultaneously have a large fraction of their variance explained by the same PC, must therefore carry redundant information about the star\footnote{The correlation analysis is in general similar to reporting the PC loadings. 
In PCA loadings are the elements of the eigenvector scaled by the square roots of the respective eigenvalues.
The elements of the eigenvector are coefficients that indicate the weighting of the original data parameters that combine to form that PC. 
As we have centred and scaled the data before performing the PCA, the correlation coefficients are equivalent to the loadings.}.
In Figure~\ref{fig:GCA-pcabar} we quantify, through Pearson's~$r$ coefficient, the extent to which each observable correlates with the first five PCs.
The parameters in the top panel of Figure~\ref{fig:GCA-pcabar} are ordered by their correlation with the first principal component. 
PC$_1$ accounts for a significant fraction of the variance in the observables (Ve${_{\, \text{pca}}=42.36\%}$). 
The top panel of  Figure~ \ref{fig:GCA-pcabar} reveals that this component correlates very strongly 
(${r > 0.85}$)  with   
$\nu_{\max}$, 
${\langle\Delta\nu_0\rangle}$,
${\langle\delta\nu_{02}\rangle}$, 
${\langle\delta\nu_{13}\rangle}$,
and ${\log{} g}$.  
The strong correlations imply that the basis vector captures most of the variance across the five parameters simultaneously and points to a common latent variable.

\subsection{Interpreting the Principal Components}
\label{sec:intPC}

In Figure~ \ref{fig:GCA-pcabar} and Figure~\ref{fig:GCA-pcabarb} we plot the results of correlation analysis between all parameters in the grid and the transformed observables (PCs). 
The figures offer a quantitative overview of the PCs allowing  us to identify what interpretable features the PCs have captured.
We have seen that Figure~\ref{fig:GCA-pcabar} demonstrates the extent to which each observable correlates with the first five PCs, similarly  Figure~\ref{fig:GCA-pcabarb} demonstrates how the principal components correlate with the model parameters. The corresponding correlation coefficients between the parameters and \emph{all} PCs are listed in Tables \ref{tab:ocoefs} \& \ref{tab:mcoefs}.

Any interpretation of the PCs based on Figures \ref{fig:GCA-pcabar} and \ref{fig:GCA-pcabarb} are only valid for the truncated grid of models to which this PCA has been applied. For results on other sub grids we refer the reader to Appendices \ref{sec:fullPCA} and \ref{sec:PCAg}. 
We draw upon the figures for generality in the discussion section (Section~\ref{sec:disc}).

Information about \emph{direct} correlations between parameters can be extracted from PCA which further helps with interpreting the underlying features.
Any two parameters that correlate with a given principal component and meet the transitive criterion will be positively correlated if they both have the same sign with respect to the PC, and negatively correlated if their signs differ. 

As is often the case with PCA, the first few principal components can be interpreted as describing the large-scale physical behavior of the system.
We interpret that the underlying feature that PC$_1$ captures is straightforwardly the stellar radius.  
This is the physical property that has the greatest impact on the observables.
From PC$_1$ in Figures \ref{fig:GCA-pcabar} and \ref{fig:GCA-pcabarb} we can infer (from the transitive criterion) that as a star evolves along the main sequence, i.e., $\tau_{\text{MS}}$ increases or $X_c$ decreases, radius (and for the most part L) will increase.  
The consequence of increasing radius being $\nu_{\max}$, 
${\langle\Delta\nu_0\rangle}$,
${\langle\delta\nu_{02}\rangle}$, 
${\langle\delta\nu_{13}\rangle}$,
 ${\log{} g}$ all decrease and thus their variance is explained by PC$_1$. 
We note that this PC also correlates with $M$ as stars with larger $M$ will have larger radii.

PC$_2$ can be interpreted as a `core-surface' feature.
PC$_2$ correlates strongly with different combinations of seismic ratios and small frequency separations. 
With strong weightings from the core it is no surprise that PC$_2$ features a moderate-to-strong correlation with $\tau$.  
This direction of maximal variance comprises information from all the observables and correlates with (mostly) all the dependent model variables further suggesting some form of time evolution. 
The information from the surface is provided by $T_{\text{eff}}$.  
There is a degree to which the variance in $T_{\text{eff}}$ is captured by the time-evolutionary aspect of this component. 
However PC$_2$ also displays a moderate correlation with the time-independent $Z_0$ and thus there is a second aspect to  PC$_2$.
$Z_0$ dictates the temperature at the surface through opacities and nuclear burning in the core.

PC$_3$ appears to have the role of capturing the more extreme models in the grid. 
In the truncated grid the correlations with $[\text{Fe/H}]$ and $T_{\text{eff}}$ suggest that the focus of this PC to account for the variance in the observations imparted by low-metallicity models. 

PC$_4$ appears to be a secondary `core-surface' feature much like PC$_2$.  
It uses surface information, in this case ${[\text{Fe/H}]}$, in conjunction with some information from the core in the form of the ${\langle r_{02}\rangle}$ and ${\langle r_{13}\rangle}$ ratios.  

PC$_5$ encapsulates the mixing processes that impact upon the surface abundances of the star but it is only required to explain a small fraction of the total variance in the data.

\subsection{Inferring Stellar Parameters} 
\label{sec:ISP}

The dimensionality reduction achieved by the PCA quantifies the degree of redundancy in the stellar observables alluded to by Figure~\ref{fig:filt_corr}. 
However, we also wish to quantify the extent to which the observed stellar properties constrain the internal structures and chemical mixtures of the star, i.e., the model properties.

In our application of RF regression the machine tries to fit for each model parameter, the success of which we can appraise (see Section~\ref{sec:qu}). 
Here we conduct a more fundamental evaluation: how well can we capture the variance in the model parameters simply by explaining the variance in the observed data? 
In other words: having removed the redundancies, to what extent is information of the model parameters encoded in the observables? We hence devise a score, $\Lambda$, such that:
\begin{equation}
   \Lambda(X) = \  \sum^{p}_{i=1}  r(X, PC_i)^2
\end{equation}
where $X$ is the parameter of interest, $p$ is the number of PCs ($11$ in our case)  and  ${r(X, PC_i)}$ is the Pearson coefficient between the parameter and the PC.
As we centred and scaled our data before computing the correlation matrix and extracting the PCs, the $\Lambda(X)$ score is equivalent to summing the square of the PC loadings. The square of each loading indicates the variation in an observable that is explained by the component. A useful property of having scaled our data is that ${\Lambda(X) =1}$  for each of our observables. We demonstrate these properties further in Appendix \ref{sec:lambdaa}.

In Figure~\ref{fig:GCA-pcabarb} we projected the parameter space of our model quantities onto the PC space.  Whilst these are not the optimum vectors to explain our model parameters, that is not their purpose; we instead wish to determine what we can learn about the model quantities by understanding the observables. 
As the square of the correlation coefficients (loadings) will indicate the fraction of explained variance for the parameter by a given PC, determining the  
$\Lambda(X)$ score for the model parameters gives an indication of the extent the model data are retrievable from the observables.  

In Table~\ref{tab:corrL} we list the $\Lambda$ score for each of the model parameters in Table~\ref{tab:parmdefs}. Parameters with larger $\Lambda$ scores have much of their variance captured by the linear models used to explain the observables.  We expect  to be able to infer parameters such as $R$, $L$ and $\tau_{\text{MS}}$ with a great deal of confidence through regression. 
Parameters with intermediate values of $\Lambda$ ($\tau$, M$_{cc}$) we can expect to recover with some success by employing more sophisticated modelling, however, it is not clear that there is enough information contained in the observables to always do so. In cases with the lowest values of $\Lambda$, such as the initial model parameters $\alpha_{\text{MLT}}$, $Y_0$ and $\alpha_{\text{ov}}$, explaining the variance in the observables does not explain the variance in the model parameters. New observables that provide independent information about the star are required to recover these parameters with higher confidence. Fitting the acoustic glitch for example may (eventually) provide constraints on the degree of convective envelope overshoot or atomic diffusion \citep{2017ApJ...837...47V}.

\begin{table}
    \centering
    \begin{tabular}{ccc}
    \hline \hline
   
Parameter &  & $\Lambda_{\text{param}}$ \\ \hline     
R	&	&	0.97	\\
L	&	&	0.96	\\
$X_c$	&	&	0.94	\\
$\tau_{\text{MS}}$	&	&	0.93	\\
M	&	&	0.91	\\
$\tau$	&	&	0.79	\\
$Z_0$	&	&	0.73	\\
M$_{cc}$	&	&	0.61	\\
$Y_{\text{surf}}$	&	&	0.50	\\
X$_{\text{surf}}$	&	&	0.48	\\
$\alpha_{\text{MLT}}$	&	&	0.38	\\
$Y_0$	&	&	0.31	\\
D	&	&	0.29	\\
$\alpha_{\text{ov}}$	&	&	0.08	\\
 \hline
    \end{tabular}
    \caption{The $\Lambda$ score is a sum of the squares of  $r_{\text{PC, param}}$. Any parameter with high $\Lambda$ is explained well by a linear model and can be confidently inferred. We have insufficient information to constrain those parameters with the lowest $\Lambda$.}
    \label{tab:corrL}
\end{table}

\section{Quantifying the Utility of Stellar Observables} 
\label{sec:qu}

There is certainly value and a degree of intuition in dimensionality reduction.
PCA has demonstrated the significant information redundancy in our data. 
It has also allowed us to identify information from the model parameters manifested in the observables, and indicated to what extent those parameters can be extracted.  
We now turn to another strategy of exploratory data science, which is to let machine learning algorithms fit complicated models to the data. As we shift our focus from \emph{what} information is present to \emph{how} it can be exploited, we transition from unsupervised to supervised learning methods.

In the PCA we determined orthogonal vectors that are the best fit to the observables. 
Here we utilize a RF to perform non-parametric, multiple regression in order to create the best functions capable of inferring each stellar parameter. With this particular form of supervised learning 
the relationships between observables and model parameters remain hidden. Though some insight into the regression function can be gained through examination of the feature importances, the tree topology makes further interpretation difficult. 
We thus seek to elucidate the RF's decision making processes by appraising
how well different combinations of parameters can predict the quantities in Table~\ref{tab:parmdefs}. 
 
 This approach not only illustrates the RF's ability to recover missing observational data, say for a rapid stellar evolution calculation, but also systematically \emph{quantifies} the usefulness of each parameter in predicting all other quantities in the limit of perfect information. It is analogous to a seismic inversion in that it demonstrates the inherent uncertainty with which information can be reconstructed from the available observations. 
Whereas PCA serves to remove the redundant stellar information in the parameters, the analysis here is designed to highlight them. 

Using the full grid of BA1 models, we perform multiple regression on every unique combination of observables in Table~\ref{tab:parmdefs}. We omit those combinations that contain the quantity we are training for and include models with $R$ and $L$ as observables, resulting in the calculation of $49,153$ RFs.

We divide the full grid into a testing (${\approx 15,000}$ models) and training set as per the method ascribed in Appendix D of BA1 so not to over-estimate the performance of the regression. 
We perform two-fold cross-validation on each RF and, as in BA1, measure their success on the test data with an explained variance score, V$_{\text{e}}$:
\begin{equation}
\label{eqn:ve}
  \text{V}_{\text{e}} = 1 - \frac{\text{Var}\{ y - \hat y \}}{\text{Var}\{ y \}}.
\end{equation}
Here $y$ is the \emph{true} value we want to predict, $\hat{y}$ is the predicted value from the random forest, and Var is the variance. This score tells us the extent to which the regressor has reduced the variance in the parameter it is predicting with a score of one implying that the model predicts all values with zero error. This is a different but equivalent definition by which to measure the same quantity in Equation~(\ref{eqn:pcaev}). We have adopted the same notation as BA1 for evaluating the RF which we use to distinguish from the definition used in the PCA (Section~\ref{sec:PCA}). We also provide a measure of the `typical' error in the predictions, ${\mu (\epsilon)}$,  which is calculated by averaging the absolute difference ($\epsilon$) between the predicted and true values for each parameter. More formally: 
\begin{equation}
\label{eqn:mu}
    \mu (\epsilon) = \frac{1}{n} \sum^n_{i=1} | \hat y_i - y_i |,
\end{equation}
where $n$ is the number of models in the test data. Through ${\mu (\epsilon)}$  we provide an indicative error associated with the regression model, over the whole parameter space, and  in units of the quantity of interest.  

The best combinations of parameters for inferring each quantity of interest are listed in Table~\ref{tab:regparms}. 
We present combinations of up to five parameters after which there is negligible improvement to the predictions. 
We mark with a dash the occasions where the regressor is unable to produce a positive $V_e$ score.  
It is important to remember that while a score of one implies a perfect predictor, any ${V_e < 1}$ implies there is still \emph{some} error in the model.  We thus opt for truncation rather than rounding when listing the scores. 
Predictions of the seismic quantities are omitted here.  They strongly co-vary and are easily recovered when other seismic parameters are known; they are discussed separately in  Section~\ref{sec:seispr}. 
Their strong covariances also mean that many of the ratios and separations used in the regression models are interchangeable (e.g., ${\langle r_{02}\rangle}$ for 
${\langle r_{13}\rangle}$ or ${\langle r_{01}\rangle}$ for ${\langle r_{10}\rangle}$) resulting in negligible differences to our two scores. 

\afterpage{
\clearpage
\begin{landscape}
\thispagestyle{lscape}
\pagestyle{lscape}
\begin{table} \scriptsize
\caption{The best combinations of observables for constraining the non-seismic parameters in Table~\ref{tab:parmdefs}. For each combination we provide the ${V_e}$ score (Equation~\ref{eqn:ve}) and ${\mu (\epsilon)}$ score (Equation~\ref{eqn:mu}, given in the units indicated by the predicted quantity column).\label{tab:regparms}}
\hspace*{-1.25cm}
\begin{tabular}{  l | l  l  l | l  l  l  l | l  l  l  l | l  l  l  l | l  l  l  l   }
Predicted   &
\multicolumn{3}{c|}{One Parameter}    &
\multicolumn{4}{c|}{Two Parameters}    &
\multicolumn{4}{c|}{Three Parameters}    &
\multicolumn{4}{c|}{Four Parameters}    &
\multicolumn{4}{c}{Five Parameters}        \\ \cline{2-20}
Quantity         &        
Observable    &  
$V_e$    & 
$\mu (\epsilon)$    & 
\multicolumn{2}{c}{Observables} &
$V_e$    & 
$\mu (\epsilon)$    & 
 \multicolumn{2}{c}{Observables} &
$V_e$    & 
$\mu (\epsilon)$    & 
 \multicolumn{2}{c}{Observables} &
$V_e$    & 
$\mu (\epsilon)$    & 
 \multicolumn{2}{c}{Observables} &
$V_e$    & 
$\mu (\epsilon)$ \\ \hline\hline
  $R/R_\odot$ & $\langle\Delta\nu_0\rangle$ & 0.955 & 0.046 & $\langle\Delta\nu_0\rangle$, $\nu_{\max}$ && 0.985 & 0.027 & $\langle\Delta\nu_0\rangle$, $\nu_{\max}$, && 0.999 & 0.009 & $\langle\Delta\nu_0\rangle$, $\nu_{\max}$, && 0.999 & 0.008 & $\langle\Delta\nu_0\rangle$, $\nu_{\max}$, $T_{\text{eff}}$, && 0.999 & 0.008\\
  &  &  &  &  &  &  &  &  $T_{\text{eff}}$ &  &  &  & $T_{\text{eff}}$, $\log{} g$ &&  & & $\log{} g$, $\langle r_{10}\rangle$ &  &  &\\[3pt]
  $\log{} g$ & $\langle\Delta\nu_0\rangle$ & 0.86 & 0.046 & $T_{\text{eff}}$, $\nu_{\max}$ && 0.999 & 0.004 & $T_{\text{eff}}$, $\nu_{\max}$, && 0.999 & 0.003 & $T_{\text{eff}}$, $\nu_{\max}$, && 0.999 & 0.002 & $T_{\text{eff}}$, $\nu_{\max}$, $[\text{Fe/H}]$, && 0.999 & 0.002\\
  &  &  &  &  &  &  &  &  $[\text{Fe/H}]$ &  &  &  & $[\text{Fe/H}]$, $\langle r_{13}\rangle$ & & & & $\langle r_{02}\rangle$, $\langle r_{13}\rangle$ &  &  &\\[3pt]
  $L/L_\odot$ & $T_{\text{eff}}$ & 0.739 & 1.583 & $T_{\text{eff}}$, $\langle\Delta\nu_0\rangle$ && 0.993 & 0.254 & $T_{\text{eff}}$, $\langle\Delta\nu_0\rangle$, &&0.999 & 0.13 & $T_{\text{eff}}$, $\langle\Delta\nu_0\rangle$, && 0.999 & 0.136 & $T_{\text{eff}}$, $\langle\Delta\nu_0\rangle$, $\nu_{\max}$, && 0.999 & 0.135\\
  &  &  &  &  &  &  &  &  $\nu_{\max}$ &  &  &  & $\nu_{\max}$, $\langle r_{10}\rangle$ &&  & & $\log{} g$, $\langle r_{10}\rangle$ &  &  &\\[3pt]   
 $T_{\text{eff}}$/K & $[\text{Fe/H}]$ & 0.298 & 1216 & $\log{} g$, $\nu_{\max}$ && 0.989 & 104 & $\log{} g$, $\nu_{\max}$, && 0.991 & 95 & $\log{} g$, $\nu_{\max}$, && 0.992 & 96 & $\log{} g$, $\nu_{\max}$, $\langle r_{01}\rangle$, && 0.992 & 96\\
  &  &  &  &  &  &  &  &  $\langle r_{01}\rangle$ &  &  &  & $\langle r_{01}\rangle$, $\langle\delta\nu_{13}\rangle$ &&  & & $\langle\Delta\nu_0\rangle$, $\langle\delta\nu_{13}\rangle$ &  &  &\\[3pt]
    $Z_0$ & $[\text{Fe/H}]$ & 0.927 & 0.003 & $[\text{Fe/H}]$, $\langle\delta\nu_{02}\rangle$ && 0.96 & 0.002 & $[\text{Fe/H}]$, $T_{\text{eff}}$, && 0.982 & 0.001 & $[\text{Fe/H}]$, $T_{\text{eff}}$, && 0.986 & 0.001 & $[\text{Fe/H}]$, $T_{\text{eff}}$, $\langle\Delta\nu_0\rangle$, && 0.987 & 0.001\\
  &  &  &  &  &  &  &  &  $\langle\Delta\nu_0\rangle$ &  &  &  & $\langle\Delta\nu_0\rangle$, $\langle r_{13}\rangle$ &&  & & $\langle r_{01}\rangle$, $\langle r_{13}\rangle$ &  &  &\\[3pt]
  $M/M_\odot$ & $\langle\Delta\nu_0\rangle$ & 0.348 & 0.157 & $\langle\Delta\nu_0\rangle$, $\log{} g$ && 0.857 & 0.072 & $\langle\Delta\nu_0\rangle$, $T_{\text{eff}}$, && 0.982 & 0.022 & $\langle\Delta\nu_0\rangle$, $\log{} g$, && 0.986 & 0.02 & $\langle\Delta\nu_0\rangle$, $\log{} g$, $\nu_{\max}$, && 0.982 & 0.024\\
  &  &  &  &  &  &  &  &  $\nu_{\max}$ &  &  &  & $\nu_{\max}$, $T_{\text{eff}}$ &&  & & $T_{\text{eff}}$, $\langle r_{10}\rangle$ &  &  &\\[3pt] 
    $\tau_{\text{MS}}$ & $\langle\delta\nu_{02}\rangle$ & 0.543 & 0.147 & $\langle r_{02}\rangle$, $\langle r_{01}\rangle$ && 0.846 & 0.077 & $\langle\Delta\nu_0\rangle$, $\nu_{\max}$, && 0.957 & 0.038 & $\langle r_{02}\rangle$, $\nu_{\max}$, && 0.977 & 0.025 & $\langle r_{02}\rangle$, $\nu_{\max}$, $\langle r_{10}\rangle$, && 0.981 & 0.021\\
  &  &  &  &  &  &  &  &  $\langle r_{13}\rangle$ &  &  &  & $\langle r_{10}\rangle$, $T_{\text{eff}}$ &&  & & $T_{\text{eff}}$, $[\text{Fe/H}]$ &  &  &\\[3pt] 
   X$_c$ & $\langle\delta\nu_{02}\rangle$ & 0.508 & 0.113 & $\nu_{\max}$, $\langle r_{13}\rangle$ && 0.842 & 0.062 & $\nu_{\max}$, $\langle r_{13}\rangle$, && 0.958 & 0.031 & $\nu_{\max}$,  $\langle r_{13}\rangle$,& & 0.978 & 0.023 & $\nu_{\max}$, $\langle r_{13}\rangle$, $\langle\Delta\nu_0\rangle$, && 0.979 & 0.022\\
  &  &  &  &  &  &  &  &  $\langle\Delta\nu_0\rangle$ &  &  &  & $\langle\Delta\nu_0\rangle$, $\langle r_{10}\rangle$ &&  & & $\log{} g$, $\langle r_{10}\rangle$ &  &  &\\[3pt]
  $\tau$ (Gyr)& $\langle r_{02}\rangle$ & 0.645 & 0.995 & $\langle r_{02}\rangle$, $\nu_{\max}$ && 0.844 & 0.642 & $\langle r_{13}\rangle$, $\nu_{\max}$, && 0.907 & 0.468 & $\langle r_{02}\rangle$,  $T_{\text{eff}}$, && 0.931 & 0.332 & $\langle r_{02}\rangle$, $\nu_{\max}$, $\langle r_{01}\rangle$, && 0.943 & 0.282\\
  &  &  &  &  &  &  &  &  $\langle r_{10}\rangle$ &  &  &  & $\langle r_{01}\rangle$, $\langle\Delta\nu_0\rangle$ &&  & & $T_{\text{eff}}$, $[\text{Fe/H}]$ &  &  &\\[3pt] 
   X$_{\text{surf}}$ & $[\text{Fe/H}]$ & 0.655 & 0.051 & $[\text{Fe/H}]$, $\log{} g$ && 0.772 & 0.041 & $[\text{Fe/H}]$, $\langle\Delta\nu_0\rangle$, && 0.895 & 0.027 & $[\text{Fe/H}]$, $\langle\Delta\nu_0\rangle$, && 0.928 & 0.024 & $[\text{Fe/H}]$, $\langle\Delta\nu_0\rangle$, $T_{\text{eff}}$, && 0.936 & 0.022\\
  &  &  &  &  &  &  &  &  $\nu_{\max}$ &  &  &  & $T_{\text{eff}}$, $\langle r_{02}\rangle$ &&  & & $\langle r_{02}\rangle$, $\nu_{\max}$ &  &  &\\[3pt] 
  $M_{\text{cc}}/M_\odot$ & --- & --- & --- & $\langle r_{13}\rangle$, $\langle\delta\nu_{02}\rangle$ && 0.679 & 0.015 & $\langle r_{13}\rangle$, $\nu_{\max}$, & &0.862 & 0.009 & $\langle r_{13}\rangle$, $\nu_{\max}$, && 0.908 & 0.007 & $\langle r_{13}\rangle$, $\nu_{\max}$, $\langle r_{10}\rangle$, && 0.928 & 0.006\\
  &  &  &  &  &  &  &  &  $\langle r_{10}\rangle$ &  &  &  & $\langle r_{10}\rangle$, $T_{\text{eff}}$ &&  & & $T_{\text{eff}}$, $[\text{Fe/H}]$ &  &  &\\[3pt]
   $Y_{\text{surf}}$ & $[\text{Fe/H}]$ & 0.597 & 0.052 & $[\text{Fe/H}]$, $\log{} g$ && 0.736 & 0.041 & $[\text{Fe/H}]$, $\langle\Delta\nu_0\rangle$, && 0.887 & 0.025 & $[\text{Fe/H}]$, $\langle\Delta\nu_0\rangle$, && 0.916 & 0.024 & $[\text{Fe/H}]$, $\langle\Delta\nu_0\rangle$, $\langle r_{02}\rangle$, && 0.927 & 0.022\\
  &  &  &  &  &  &  &  &  $\nu_{\max}$ &  &  &  & $\langle r_{02}\rangle$, $T_{\text{eff}}$ &&  & & $T_{\text{eff}}$, $\nu_{\max}$ &  &  &\\[3pt] 
   $Y_0$ & --- & --- & --- & $\langle\Delta\nu_0\rangle$, $\nu_{\max}$ && 0.077 & 0.027 & $\langle\Delta\nu_0\rangle$, $\nu_{\max}$, && 0.471 & 0.02 & $\langle\Delta\nu_0\rangle$, $\nu_{\max}$, && 0.536 & 0.019 & $\langle\Delta\nu_0\rangle$, $\nu_{\max}$, $[\text{Fe/H}]$, && 0.625 & 0.017\\
  &  &  &  &  &  &  &  &  $[\text{Fe/H}]$ &  &  &  & $[\text{Fe/H}]$, $\log{} g$ &&  & & $\log{} g$, $\langle\delta\nu_{13}\rangle$ &  &  &\\[3pt] 
    $\alpha_{\text{ov}}$ & --- & --- & --- & $\langle r_{13}\rangle$, $\langle r_{02}\rangle$ && 0.231 & 0.089 & $\langle r_{13}\rangle$, $\langle r_{10}\rangle$, && 0.44 & 0.075 & $\langle r_{13}\rangle$,  $\langle r_{10}\rangle$, && 0.524 & 0.068 & $\langle r_{13}\rangle$, $\langle r_{10}\rangle$, $\nu_{\max}$, && 0.55 & 0.067\\
  &  &  &  &  &  &  &  &  $\nu_{\max}$ &  &  &  & $\nu_{\max}$, $T_{\text{eff}}$ &&  & & $T_{\text{eff}}$, $[\text{Fe/H}]$ &  &  &\\[3pt] 
   D & --- & --- & --- & $[\text{Fe/H}]$, $\langle\delta\nu_{02}\rangle$ && 0.022 & 5.393 & $[\text{Fe/H}]$, $T_{\text{eff}}$, && 0.295 & 4.483 & $[\text{Fe/H}]$, $T_{\text{eff}}$, && 0.446 & 3.706 & $[\text{Fe/H}]$,  $T_{\text{eff}}$, $\langle r_{02}\rangle$, && 0.519 & 3.333\\
  &  &  &  &  &  &  &  &  $\langle\Delta\nu_0\rangle$ &  &  &  & $\langle r_{02}\rangle$, $\langle\Delta\nu_0\rangle$ &&  & & $\log{} g$, $\langle r_{10}\rangle$ &  &  &\\[3pt] 
  $[\text{Fe/H}]$ &--- & --- & --- & $\nu_{\max}$, $\log{} g$ && 0.179 & 2.777 & $\nu_{\max}$, $\log{} g$, && 0.273 & 2.439 & $\nu_{\max}$, $\log{} g$, && 0.309 & 2.312 & $\nu_{\max}$, $\log{} g$, $\langle r_{02}\rangle$, && 0.312 & 2.277\\
  &  &  &  &  &  &  &  &  $\langle r_{02}\rangle$ &  &  &  & $\langle r_{02}\rangle$, $\langle r_{10}\rangle$ &&  & & $\langle r_{01}\rangle$, $\langle r_{13}\rangle$ &  &  &\\[3pt] 
  $\alpha_{\text{MLT}}$ & --- & --- & --- & --- --- && --- & --- & $T_{\text{eff}}$, $\nu_{\max}$, && 0.069 & 0.234 & $T_{\text{eff}}$, $\nu_{\max}$, && 0.201 & 0.211 & $T_{\text{eff}}$, $\nu_{\max}$, $\langle r_{01}\rangle$, && 0.229 & 0.207\\
  &  &  &  &  &  &  &  &  $\langle\delta\nu_{13}\rangle$ &  &  &  & $[\text{Fe/H}]$, $\langle r_{02}\rangle$ &&  & & $[\text{Fe/H}]$, $\langle r_{13}\rangle$ &  &  &\\[3pt] 
 \end{tabular} 
 \end{table}
\end{landscape}
}

Many of the RFs we trained do not provide a satisfactory regression model for the quantity we are training for. Below we provide a deeper analysis for some of the more interesting results, focusing primarily on the predictions of ages and surface abundances. 

\subsection{Ages}
\label{sec:sages}
The current exercise allows us to evaluate the theoretical limit in which parameter pairs, such as those used in the C--D diagram, can constrain stellar ages. 
Recall that there are six initial model parameters varied simultaneously in the BA1 grid. 
Describing a six dimensional parameter space with two quantities invariably leads to degenerate solutions for age and necessarily high uncertainties.
The parameter pairs that offer similarly the best constraints on $\tau$ are listed in Table~\ref{tab:cd}. The combination of ${\langle r_{02}\rangle}$ and $\nu_{\max}$  marginally provide the best probe, explaining the largest fraction of the variance and inferring ages with uncertainty ${\mu (\epsilon) = \pm 642}$~Myr.  

\begin{table}
\centering
\caption{The best two-parameter combinations of observables for constraining stellar age. Below the dividing horizontal line we include the best spectroscopic pair for comparison as well as ${\log{} g}$ -- ${\langle\Delta\nu_0\rangle}$ to highlight the necessity of the small frequency separation in determining stellar ages. The BA1 grid is varied in six dimensions and with such a high-dimensional parameter space the quantities in the C--D diagram (fifth row) constrain age with `typical' uncertainty of $701$~Myr.}
    \begin{tabular}{cccc}
    \hline \hline
\multicolumn{2}{c}{Parameters} & $V_e$ & $\mu (\epsilon)$ [Gyr] \\ \hline 
$\langle r_{02}\rangle$     & $\nu_{\max}$               &0.844  & 0.642\\
$\langle r_{02}\rangle$     & $\log{} g$                  &0.833  & 0.683\\
$\langle r_{13}\rangle$     &$\nu_{\max}$                &0.827  & 0.711\\
$\langle r_{02}\rangle$     & $\langle\Delta\nu_0\rangle$    &0.825  & 0.694\\
$\langle\Delta\nu_0\rangle$ &$\langle\delta\nu_{02}\rangle$  &0.824  & 0.701\\
$\langle r_{02}\rangle$     & $\langle\delta\nu_{02}\rangle$ &0.821  & 0.701\\
PC$_2$ & PC$_8$ & 0.788 & 0.767 \\
PC$_2$ & PC$_4$ & 0.776 & 0.762 \\
\hline
 $\log{} g$ & $\langle\Delta\nu_0\rangle$ & 0.481 & 1.29 \\
 $\log{} g$ & $T_{\text{eff}}$ & 0.321 & 1.53 \\
 \hline
    \end{tabular}
    \label{tab:cd}
\end{table}

This is in comparison to ${\mu (\epsilon) = \pm 701}$~Myr for ${\langle\Delta\nu_0\rangle}$ and ${\langle\delta\nu_{02}\rangle}$  as per the C--D diagram. 
In Table~\ref{tab:cd} we also include results from regression calculated with the PCs and find they perform comparably well. The results here omit any uncertainty stemming from the surface effect suggesting that the ${\langle r_{02}\rangle}$ and $\nu_{\max}$ pair are indeed the preferable choice.

It is clear from Tables~\ref{tab:regparms} and \ref{tab:cd}  how important the small frequency separation and frequency ratios are for the determination of stellar ages on the MS. 
If we limit the combinations to the classical observables, we find that  ${\log{} g}$ and $T_{\text{eff}}$ can explain just $32.1\%$ of the variance in $\tau$ with uncertainty ${\mu (\epsilon) = \pm 1.5}$~Gyr across the whole grid. The introduction of the large separation offers little improvement. The parameter pair ${\log{} g}$ and ${\langle\Delta\nu_0\rangle}$  explain $48.1\%$ of the variance with  ${\mu (\epsilon) = \pm 1.29}$~Gyr.
If we permit the RF to draw upon five observables for its regression model, some of the degeneracy in $\tau$ is lifted.  The last column in Table~\ref{tab:regparms}
indicates that the RF can reduce the average uncertainty in predicting  $\tau$ such that  ${\mu (\epsilon) =\pm 282}$~Myr.

\subsection{Abundances}
\begin{figure}
    \centering
    \includegraphics[width=0.9\linewidth]{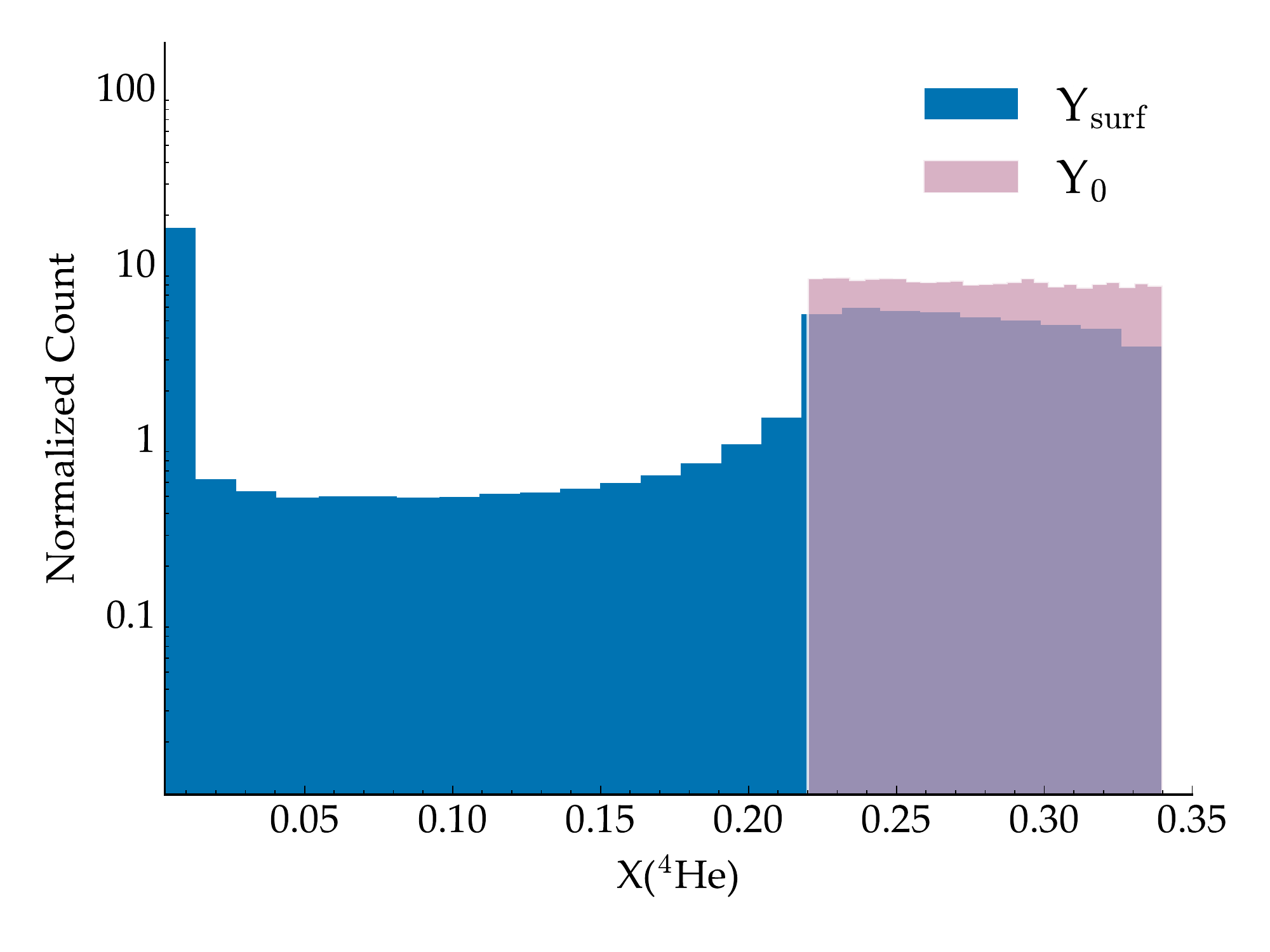}
    \caption[Distributions of initial and surface helium abundances in the generated stellar models]{Distributions of $Y_{\text{surf}}$ and $Y_0$ in the BA1 grid.}
    \label{fig:Hehist}
\end{figure}
The small frequency separations and separation ratios are integral for the determination of ages. 
However, the feature importances in BA1 (their Figure~5) indicate that the RF relies predominately on  $T_{\text{eff}}$ and $[\text{Fe/H}]$ to infer other model parameters. 
Table~\ref{tab:regparms} confirms how important measuring $[\text{Fe/H}]$ is for characterizing stars. This quantity is preferentially selected in the many RFs and their regression models, whilst $[\text{Fe/H}]$ itself cannot be determined from the other observables with any degree of confidence.   
$[\text{Fe/H}]$ is an indispensable piece of independent information.

Accurate determination of $[\text{Fe/H}]$ is paramount for inferring many of the current-age stellar attributes. $[\text{Fe/H}]$ also features prominently in the retrodiction of the initial model parameters but these quantities are characterized by large uncertainties. Foremost, we have no observable that satisfactorily constrains diffusion; $D$ demonstrates an average uncertainty spanning three orders of magnitude. This in turn introduces uncertainty in retrodicting the initial metal content. 

Predictions for $Z_0$ at first glace appear to be robust; we report $V_e$ and  ${\mu (\epsilon) = \pm 0.001}$. However we contend that a reported error of  ${\mu (\epsilon) = \pm 0.001}$ is not all that insightful given that the grid is sampled down to ${Z_0 = 10^{-5}}$. $Z_0$ is sampled logarithmically and takes a small (linear) range in values. In such cases   a relative error is a more useful measure of performance than an absolute difference.

In Table~\ref{tab:relabunds} we devise a series of measures that better appraise the performance of the RF in predicting abundances. We report the average absolute difference as per Table~\ref{tab:regparms} [${\mu (\epsilon)}$], the maximum absolute difference [${\max(\epsilon)}$] and the median absolute difference [${\tilde{\epsilon}}$]. We also consider the average relative error [${\mu (\eta)}$], the maximum relative error [${\max(\eta)}$] and median relative error [$\tilde{\eta}$], where the relative error is a percentage defined as
\begin{equation}
  \eta= \frac{| \hat y_i - y_i |}{|y_i|} \cdot 100.
\end{equation}

\begin{table}
\centering
\caption{Different measures of uncertainty in predicting stellar abundances with the RF. See text for definitions and motivations.}
    \begin{tabular}{lccc}
 \hline \hline
 Error Measure& $Y_{\text{surf}}$ & $Y_0$ & $Z_0$ \\ \hline
 $\mu (\epsilon)$                 & 0.02 &0.017 & 0.001 \\
 Max($\epsilon$)             & 0.25 &0.09   & 0.037\\    
$\tilde {\epsilon}$          & 0.016  &0.02     &0.00019\\
$\mu (\eta)$ [\%]       & $10^{13}$ &8.92& 124.5 \\ 
  Max($\eta$) [\%]   & $10^{14}$ &40.34  & 9052 \\
$\tilde {\eta}$  [\%] & 10.88 &7.68  & 13.5 \\
\hline
\end{tabular}
\label{tab:relabunds}
\end{table}

We find  ${\mu (\eta) = 125\%}$ in the retrodiction of metallicity. We attribute the seemingly large uncertainty to the bias imparted by extreme models that have undergone significant diffusion -- we report a maximum relative error of $9000\%$. With less sensitivity to the outlying metal-depleted models, the median relative uncertainty, ${\tilde{\eta} = 13.5\%}$, offers the most appropriate measure of error in the regression. 
Likewise, the extreme ${\mu (\eta)}$ and $\max(\eta)$ scores for $Y_{\text{surf}}$ also stem from models with high diffusion leading to very small non-zero abundances by which we normalize. 

It is interesting to compare the regressor's ability to infer $Y_{\text{surf}}$ and $Y_0$ abundances. We find that  $Y_{\text{surf}}$ can be well fit (${V_e = 0.927}$) with ${\mu (\epsilon) = \pm 0.022}$. In contrast, the initial abundance, $Y_0$, cannot be confidently retrodicted  (${V_e=0.625}$) yet results in a smaller average error [${\mu (\epsilon) = \pm 0.017}$]. 
This initially surprising result can be understood through examination of the respective parameter distributions in the BA1 grid (Figure~\ref{fig:Hehist}).
The grid is uniformly sampled in initial helium with ${Y_0 \in [0.22, 0.34]}$. 
Atomic diffusion acts to drain helium from the surface layers and in fact, in some models, completely depletes this species from the envelope. 
The surface helium abundance of a stellar model can thus attain values in the larger range
${Y_{\text{surf}} \in [0.0, 0.34]}$.  
In a  uniform distribution, such as we have for $Y_0$, the largest theoretical uncertainty  is 
\begin{equation}
    \max \left( \frac{\sigma^2(Y_0)}{Y_0} \right) = \frac{|b-a|}{|a|} \cdot 100 = 54.51\%,
\end{equation}
where $a$ and $b$ are the respective minimum and maximum values in our parameter range.  
This means that if the regressor was unable to explain \emph{any} of the variance in this quantity and was randomly choosing $Y_0$ values from the initial distribution, the worst relative uncertainty we would expect is $54.51\%$. The fact that we do go someway to predicting this quantity results in ${\mu (\eta) \approx 8\% }$ and more accurate inferences than for $Y_{\text{surf}}$.

\subsection{Other Results}
We mention briefly other interesting results from the approximately $50,000$ RFs not necessarily reported in Table~\ref{tab:regparms}.
Stellar masses can be accurately inferred from spectroscopic measurements. The combination of  ${\log{} g}$, $T_{\text{eff}}$ and $[\text{Fe/H}]$ constrains mass equally well as the pair ${\langle\Delta\nu_0\rangle}$ -- ${\log{} g}$. Both combinations explain 
 $86\%$ of the variance in mass with ${\mu (\epsilon) = \pm 0.07 \; M_{\odot}}$. With six degrees of freedom in the BA1 grid, we cannot determine mass to an accuracy better than  ${\mu (\epsilon) = \pm 0.02\;M_{\odot}}$.  
 Whilst all observables correlate with $M$, they do not contain sufficient information to separate out the redundant structures that are possible by tweaking the other initial model parameters. We in fact find no improvement in our regression for $M$ beyond three parameters\footnote{Numerics accounts for the differences in the third decimal place for scores in Table~\ref{tab:regparms}.}. 

If required, the RF can determine $T_{\text{eff}}$ with high accuracy. 
Although this is almost certainly always an input for the RF, with two or more observables $T_{\text{eff}}$ can be determined with  ${\mu (\epsilon) \approx 100}$~K -- an uncertainty comparable to typical spectroscopic errors. If one of $L$ or $R$ are provided as an input to the RF, a factor of two reduction in the uncertainty is achieved with ${\mu (\epsilon) \lesssim 50}$~K. Furthermore, our testing of the RF (not included here) indicates that if both $L$ and $R$ are provided as observables the Stefan-Boltzmann law is recovered with ${\mu (\epsilon) = 4}$~K.

\subsection{Seismic Quantities} \label{sec:seispr}
We did not include the predictions for the seismic parameters in Table~\ref{tab:regparms} as they often carry redundant information. Indeed we accomplish little 
by reporting how the different combinations of ratios and separations can be used to recover each other. 
We thus opt to analyze the seismic parameters separately, where we can employ discretion to present useful comparisons and highlight noteworthy results. 

\subsubsection*{The large frequency separation -- $\langle\Delta\nu_0\rangle$}

In lieu of a direct measurement, ${\langle\Delta\nu_0\rangle}$ can be estimated from stellar models via an asteroseismic scaling relation (Equation~\ref{equ:dnu}). Alternatively, it may be inferred from the observables through an empirical power law that relates ${\langle\Delta\nu_0\rangle}$ to $\nu_{\max}$ \citep{2009A&A...506..465H,2009MNRAS.400L..80S}. 
The power law estimates ${\langle\Delta\nu_0\rangle}$ within  $15\%$ of its measured value \citep{2009MNRAS.400L..80S}.
We compare the RF's ability to likewise predict ${\langle\Delta\nu_0\rangle}$ from $\nu_{\max}$ in Table~\ref{tab:dnu}. We also consider two and three parameter combinations for inferring ${\langle\Delta\nu_0\rangle}$ with the requirement that they do not comprise the remaining seismic observables. 

\begin{table}
\centering
\caption{Combinations of observables that best constrain $\langle\Delta\nu_0\rangle$.}
    \label{tab:dnu}
    \begin{tabular}{ccccccc}
    \hline \hline
\multicolumn{3}{c}{Parameters} && $V_e$ & $\mu (\epsilon)$ & $\mu (\eta)$  \\  
&&&&&[$\mu$Hz] & [\%] \\ \hline
$\nu_{\max}$ &  &   &&0.930 & 7.815 & 6.11 \\ 
$T_{\text{eff}}$     & $\nu_{\max}$&     &&0.990  & 3.09 & 2.46 \\
$\log{} g$     &$\nu_{\max}$ &               &&0.990  & 2.95 & 2.34 \\
$\log{} g$     &  $T_{\text{eff}}$&                 & &0.990  & 2.92 & 2.31\\
$[\text{Fe/H}]$  & $\nu_{\max}$ &                 &&0.991  & 2.81 & 2.24\\
$T_{\text{eff}}$ & $[\text{Fe/H}]$ & $\nu_{\max}$ && 0.995 & 1.67 & 2.13 \\
$\log{} g$ & $[\text{Fe/H}]$ & $\nu_{\max}$ && 0.995 & 1.65 & 2.11\\
\hline
    \end{tabular}
\end{table}

We find that the RF predicts ${\langle\Delta\nu_0\rangle}$  from $\nu_{\max}$  with ${\mu (\eta) \approx 6\%}$. These results are based on error free information (cross-validation hence no measurement noise) and the inclusion of $\nu_{\max}$ from a scaling law. 
In order to conduct a more faithful comparison with \citet{2009MNRAS.400L..80S}, we analyze the same data used in the derivation of their power law.  
Their Table~1 is a compilation of  $\nu_{\max}$ and ${\langle\Delta\nu_0\rangle}$ values from the literature. The data are predominately from radial velocity studies and measured with less precision than we have come to expect from \emph{Kepler} timeseries; they provide a robust test of the RF. 
We feed the RF the quoted $\nu_{\max}$ measurements and predict associated ${\langle\Delta\nu_0\rangle}$ values. We compare our predictions to the ${\langle\Delta\nu_0\rangle}$ values from the literature which are used to calculate corresponding 
$\epsilon$ and $\eta$ scores. Our results are presented in Table~\ref{tab:stello}. 
We omit entries from  the \citet{2009MNRAS.400L..80S} dataset that are outside the parameter ranges of our training grid.
For the remaining $17$ stars we find ${\mu (\eta) \approx 8\%}$ which is comparable to  ${\mu (\eta) \approx 6\%}$ accuracy achieved from cross-validation test (approximately $15,000$ stars).  

\begin{table}
\centering
\caption{Predictions of ${\langle\Delta\nu_0\rangle}$ for stars listed in \citet{2009MNRAS.400L..80S}. Results pertain to a random forest trained with $\nu_{\max}$ as the only input. Predictions are compared to literature values from the sources listed in Table~1 of  \citet{2009MNRAS.400L..80S}. The RF performs as well as the power-law relation ($10$-$15\%$) even on data measured with less precision than stars observed by \emph{Kepler}.}
\label{tab:stello}
\begin{tabular}{llllll}
\hline  \hline
Star & $\nu_{\max}$ & $\langle\Delta\nu_0\rangle_{\text{lit}}$ & $\langle\Delta\nu_0\rangle_{\text{pred}}$ & $\epsilon$ & $\eta$  \\ 
& ($\mu$Hz) & ($\mu$Hz)  &($\mu$Hz) &($\mu$Hz) & (\%) \\ \hline
$\tau\;$Cet &4500 & 170 & 171 & 1  & 1  \\
$\alpha\;$Cen B &4100 & 161 & 184 & 22& 14  \\
Sun &3100 & 135 & 138 & 3 & 2    \\
$\iota\;$Hor &2700 & 120   & 136 &  16&14  \\
$\gamma\;$Pav  &2600 & 120 & 122 & 1  & 1  \\
$\alpha\;$Cen A &2400 & 106 & 124 & 18 & 17  \\
HD 175726 &2000 & 97    & 100 & 3  & 3  \\
$\mu\;$Ara &2000 & 90    & 100 & 10 & 11  \\
HD 181906 &1900 & 88  & 97  & 10 & 11 \\
HD 49933 &1760 & 86  & 101 & 15 & 18  \\
HD 181420 &1500 & 75    & 76  & 1  & 1  \\
$\beta\;$Vir &1400 & 72    & 77   & 5 & 8  \\
$\mu\;$Her  &1200 & 57  & 63   & 7 & 12  \\
$\beta\;$Hyi &1000 & 57  & 57  & 0  & 0  \\
Procyon &1000 & 55    & 57 & 2 & 4    \\
$\eta\;$Boo &750  & 40  & 45   & 5 & 13  \\
$\nu\;$Ind &320  & 25  & 23  & 3 & 10  \\ \hline
\end{tabular}
\end{table}

The last column in Table~\ref{tab:stello} indicates that the accuracy from the RF is similar to that of the power law. In addition, we find that parameterizing the RF regression as a function of two observables reduces the uncertainty by a factor of $2$--$3$   (Table~\ref{tab:dnu}).
This hints that the inclusion of a temperature or metallicity dependence may also improve the fit offered by the power law\footnote{Symbolic regression will help determine whether, in this case, the fitting by the RF has a sensible functional form that can be straightforwardly expressed by two independent variables. This result seems reasonable as the additional information is likely providing a better handle on the stellar mass.}. 
 
\begin{figure}
\centering
\includegraphics[width=0.9\linewidth]{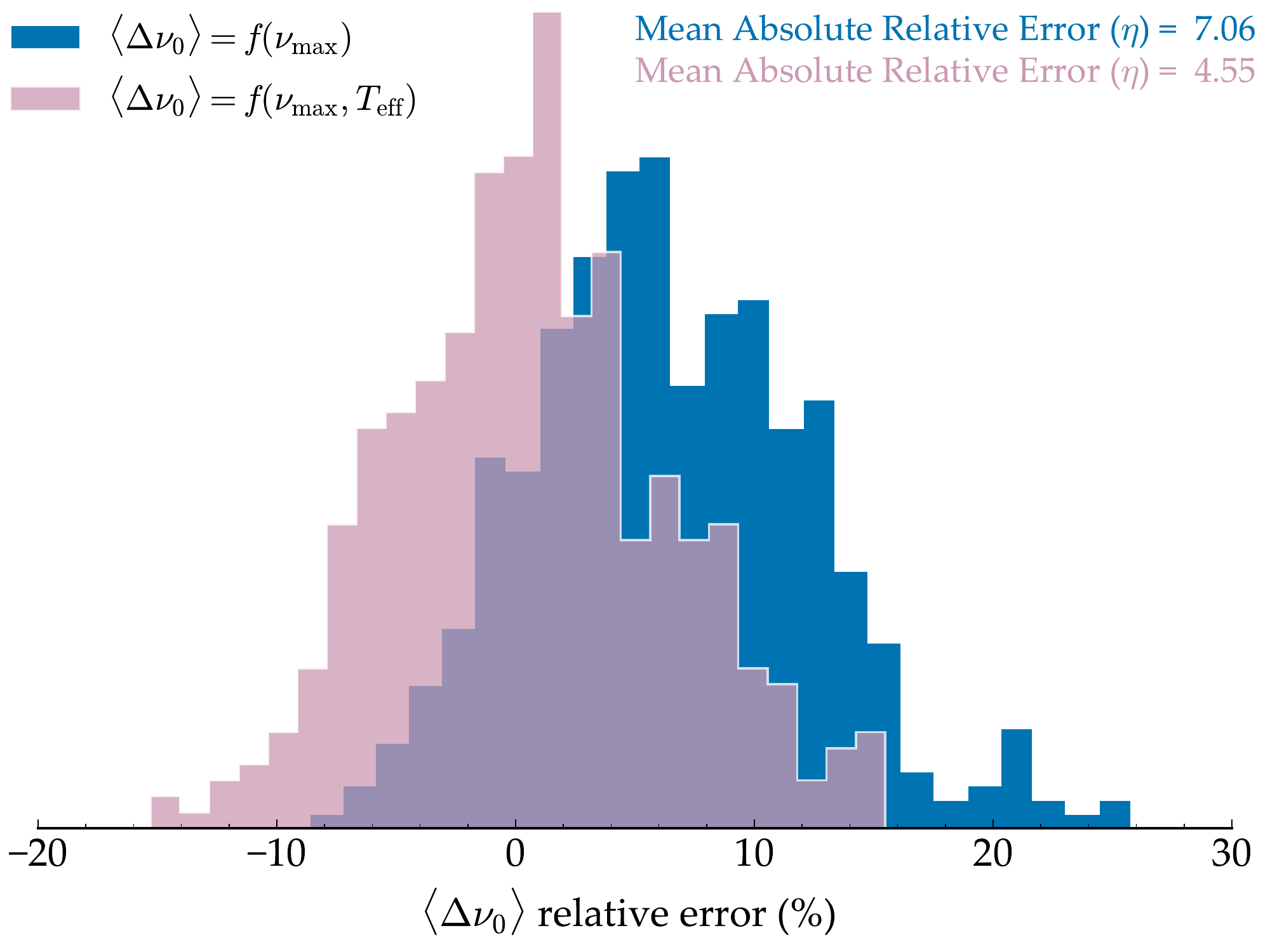}
\caption[Relative error in predictions for $\Delta\nu$]{Relative error (\%) in the predictions for ${\langle\Delta\nu_0\rangle}$ for $467$ stars reported in \citet{2014ApJS..210....1C}. The blue colored distribution indicates the error in the predictions from the random forest using $\nu_{\max}$ as the only input observation whilst the distribution marked in lavender are the results from providing $\nu_{\max}$ and $T_{\text{eff}}$. In the calculations we employ the effective temperatures determined from \citet{2012ApJS..199...30P} based on  SDSS photometry.} 
\label{fig:chap}
\end{figure}

Analysis of recent \emph{Kepler} data yields a similar result. In Figure~\ref{fig:chap} we present the percentage error in our predictions of $467$ stars measured by \emph{Kepler} as reported in Table~1 of \citet{2014ApJS..210....1C}. We analyze stars for which $\nu_{\max}$, ${\langle\Delta\nu_0\rangle}$ have been measured from the oscillation spectra along with  $T_{\text{eff}}$ as determined by \citet{2012ApJS..199...30P} based on Sloan Digital Sky Survey (SDSS) photometry. Results from the \emph{Kepler} sample confirm that predictions for ${\langle\Delta\nu_0\rangle}$ are improved with the inclusion of $T_{\text{eff}}$ (lavender distribution). The blue distribution indicates that ${\langle\Delta\nu_0\rangle}$ is systematically overestimated when the RF only has access to information from $\nu_{\max}$ -- a bias that may very well be present in the power-law fit. With the inclusion of $T_{\text{eff}}$ our predictions become more accurate and precise with the bias from the single parameter function mitigated. We do not quite reproduce the accuracy achieved in the cross validation (Table~\ref{tab:dnu}) using error free information.  Unsurprisingly,  measurement uncertainty, which we do not consider here, does not permit the accuracy attained in the ideal case.

\subsubsection*{The frequency of maximum oscillation power -- $\nu_{\max}$}

Currently we are unable to predict the frequency of maximum oscillation power from first principles. \citet{1991ApJ...368..599B} and \citet{1995A&A...293...87K} showed that this quantity does scale with the acoustic cut-off frequency and can thus be estimated via the Equation~(\ref{equ:nmax}) scaling relation. It is therefore expected that Table~\ref{tab:nmx} indicates that  $\nu_{\max}$ is best inferred from ${\log{} g}$  and $T_{\text{eff}}$. These are the two observables that correlate strongest those parameters used to calculate $\nu_{\max}$ in the training grid. 

\begin{table}
\centering
\caption{Combinations of observables that  best constrain $\nu_{\max}$.}
    \begin{tabular}{ccccc}
\hline    \hline
\multicolumn{2}{c}{Parameters} && $V_e$ &$\mu (\epsilon)$ [$\mu$Hz] \\ \hline 
$\langle\Delta\nu_0\rangle$ &&&0.923  & 7.88 \\
$\log{} g$ & $[\text{Fe/H}]$ && 0.888 & 9.99 \\
$\log{} g$     & $\langle\Delta\nu_0\rangle$     &&0.954  & 5.38\\
$T_{\text{eff}}$     & $\langle r_{10}\rangle$ &   &0.960  & 5.11\\
$[\text{Fe/H}]$    &$\langle\Delta\nu_0\rangle$        &       &0.992  & 2.90\\
$T_{\text{eff}}$     & $\langle\Delta\nu_0\rangle$    &  &0.992  & 2.84\\
$\log{} g$  & $T_{\text{eff}}$                 &&0.999  & 0.83\\
\hline
    \end{tabular}
    \label{tab:nmx}
\end{table}

\subsubsection*{The small frequency separation -- $\langle\delta\nu_{02}\rangle$}
The small frequency separation is an indispensable piece of independent information for determining stellar age. In the asymptotic limit \citep{1980ApJS...43..469T}
\begin{equation}
\langle\delta\nu_{13}\rangle = \frac{5}{3} \langle\delta\nu_{02}\rangle 
\end{equation}
and as Table~\ref{tab:d02} demonstrates, the RF recovers  ${\langle\delta\nu_{02}\rangle}$ in the unlikely case that it is not extracted but ${\langle\delta\nu_{13}\rangle}$ is.
If we disregard combinations that include the seismic ratios, which also contain information of the local small frequency separation, we lack sufficient information to satisfactorily constrain ${\langle\delta\nu_{02}\rangle}$. 
Clearly much of the evolutionary aspect of this quantity can be explained though parameters 
that correlate with main-sequence lifetime e.g.,  ${\log{} g}$, ${\langle\Delta\nu_0\rangle}$,   $\nu_{\max}$ and  $T_{\text{eff}}$. However the associated errors of ${\mu (\epsilon) > 1.0 \; \mu}$Hz can correspond to large age uncertainties for main sequence stars (${\eta > 10\%}$).

\begin{table}
\centering
\caption{Combinations of observables, without the asteroseismic ratios, that best constrain $\langle\delta\nu_{02}\rangle$.}
    \begin{tabular}{cccccc}
    \hline
\multicolumn{3}{c}{Parameters} && $V_e$ & $\mu (\epsilon)$ [$\mu$Hz] \\ \hline \hline
$\langle\delta\nu_{13}\rangle$ &  &   &&0.944 & 0.66 \\ 
$\log{} g$ &  &   &&0.542 & 2.08 \\ 
$\langle\delta\nu_{13}\rangle$ &  $\langle r_{10}\rangle$  &   &&0.987 & 0.320 \\ 
$T_{\text{eff}}$     & $\nu_{\max}$&     &&0.776  & 1.40\\
$\log{} g$     &  $T_{\text{eff}}$&                 & &0.775  & 1.40\\
$\log{} g$     &$\nu_{\max}$ &               &&0.772  & 1.41\\
$\log{} g$     &$\langle\Delta\nu_0\rangle$ &               &&0.723  & 1.54\\
$T_{\text{eff}}$     & $\langle\Delta\nu_0\rangle$&     &&0.720  & 1.58\\
$\log{} g$ & $[\text{Fe/H}]$ &  && 0.720 & 1.59 \\
$\log{} g$ & $[\text{Fe/H}]$ & $\langle\Delta\nu_0\rangle$ && 0.861 & 1.06 \\
$\log{} g$ & $\nu_{\max}$ & $\langle\Delta\nu_0\rangle$ && 0.860 & 1.09 \\
\hline
    \end{tabular}
    \label{tab:d02}
\end{table}

\section[Quantifying the Required Measurement Accuracy]{Quantifying the Required Measurement Accuracy of Stellar Observables}
\label{sec:accu}
In the previous section we used RF regression to appraise how well combinations of observables constrain other stellar parameters. The ${\approx 50,000}$ RFs were   
evaluated using cross-validation. The tests are a pure measure of the regressor's performance as we have error-free information that we attempt to reproduce (withheld models).
 As we have already alluded to, like all procedures that seek to infer stellar parameters, we must also consider the consequences of measurement uncertainty in our method.

Measurement uncertainty will impact the RF results in a manner that is different to model finding algorithms. Consider an iterative model finding procedure in which we seek an  optimum model for a set of observations. We can typically expect $T_{\text{eff}}$ as a constraint with an associated  uncertainty of ${\sigma = 100}$~K.  The minimization algorithm will identify a set of candidate models, many with quite different structures. Hence the uncertainty in $T_{\text{eff}}$  will impact 
all stellar quantities simultaneously.  The  RF, on the other hand,
 builds a statistical description of stellar evolution by calculating a regression model for each individual parameter from the training data. 
 The BA1 method requires that each input observable is perturbed with random Gaussian noise according to its measurement uncertainty. Monte Carlo perturbations are performed $10,000$ times and each instantiation evaluated by the RF to yield individual density distributions for each stellar parameter. Thus the uncertainty in $T_{\text{eff}}$, or any observable for that matter, will only impact on the predictions 
 of each parameter in proportion to the degree to which it features in that parameter's regression model.

 The methodology, combined with the speed of the RF, provides a tractable means to 
 asses how the individual measurement uncertainty of an observable will impact upon each predicted stellar quantity. We hence determine  how accurately the observables must be measured in order to achieve a desired precision from the RF.

We train a RF on the observables listed in Table~\ref{tab:sun}.
We take the (approximate) solar value of each observable as our measurement and consider
`observational uncertainties' ($\sigma$) within the ranges specified in Table~\ref{tab:sun}. 
We first perturb the measurement values with Gaussian noise assuming the minimum $\sigma$ values listed. We produce $10,000$ instantiations for that set of $\sigma$ values, ensuring each perturbed observable remains within the limits of our training grid.
We evaluate stellar parameters and determine detailed distributions for that set of uncertainties. We repeat the process increasing the $\sigma$ for a single observable  always keeping the  $\sigma$ values of the other observables at their minimum. 
We draw $50$ $\sigma$ values for each observable sampling their specified ranges evenly.  We produce  probability density distributions for $250$ sets of $\sigma$ values, the results of which are summarized in Figure~\ref{fig:uncert1}.

 \begin{table}
 \centering
    \caption{Central values and uncertainty ranges used for predicting the Sun in Figure~\ref{fig:uncert1}.}
    \begin{tabular}{lccc}
    \hline \hline
Quantity & Value & Min($\sigma$) & Max($\sigma$) \\ \hline 
$T_{\text{eff}}$ (K)  & 5777 & 10  & 500\\
$\log{} g$ &  4.43812 & 0.00013 & 1.0\\
$[\text{Fe/H}]$ & 0.0 & 0.05 & 0.2 \\
$\langle\Delta\nu_0\rangle$ ($\mu$Hz) & 136.0 & 0.5 & 10\\
$\langle\delta\nu_{02}\rangle$ ($\mu$Hz)& 9.0 & 0.5 & 5 \\
\hline
    \end{tabular}
    \label{tab:sun}
\end{table}

In Figure~\ref{fig:uncert1} we plot the median value (solid line) and the $68\%$  confidence interval (shaded region)
 for $M$, $\tau$, $L$ and $R$ as a function of the uncertainty applied to each observable. The figure is organised such that each row (and color) corresponds to the observable that has had its uncertainty increased and each column corresponds to the model parameter of interest. In this Figure, the left axis indicates the predicted value from the RF and the right axis indicates the relative error with reference to the true values of the Sun.  The horizontal dotted grey lines mark the reference value in each case whilst the dotted vertical lines indicate a typical uncertainty for the perturbed observable. 

The particular RF we have trained does not significantly rely on $T_{\text{eff}}$ in its regression model for $M$, $\tau$ or $R$. As the radius is supplemented by the seismic quantities, any uncertainty in $T_{\text{eff}}$ is propagated as uncertainty in the luminosity. We find a typical uncertainty of $100$~K corresponds to an error of ${\pm 0.2}$ $L$/L$_{\odot}$ at the $68\%$ confidence level.

The inference on solar mass is affected once ${\delta \, \log{} g > 0.03}$. 
However, even at unreasonably large values of  ${\delta \, \log{} g = 1}$, the uncertainties for mass and age remained relatively constrained by additional seismic information. We find 
that $L$ and $R$ are far more reliant on ${\log{} g}$ in their regression function with uncertainties in these quantities growing significantly once  ${\delta \, \log{} g > 0.1}$.

The feature importances in BA1 indicate that ${[\text{Fe/H}]}$ is used most often by the RF in  crafting its decision rules.  The four stellar parameters we investigate here indeed all rely on information from ${[\text{Fe/H}]}$, however, they are supplemented by seismic information which helps to constrain the uncertainty in their predictions. It is the model parameters such as the mixing length, degree of overshoot and initial metallicty  that become much less certain as we increase ${\sigma([\text{Fe/H}])}$ (not shown here).

The seismic diagnostics are very sensitive to the stellar structure, and hence also those  parameters we use to characterize a star ($M$, $\tau$, $L$ and $R$). We have seen how reliant the RF is on the seismic diagnostics in the  regression models, allowing us to still predict the structural properties with relatively good precision in the face of large spectroscopic uncertainties. Without accurate measurement of ${\langle\Delta\nu_0\rangle}$ the uncertainty in structure parameters increase significantly. Whilst the uncertainty in ${\langle\delta\nu_{02}\rangle}$ does introduce some small uncertainty in $M$, $L$ and $R$, as expected, its accuracy significantly impacts upon our ability constrain stellar age.

\begin{landscape}
\begin{figure}
    \centering\vspace*{-0.75cm}
    \includegraphics[width=0.9\linewidth,keepaspectratio]{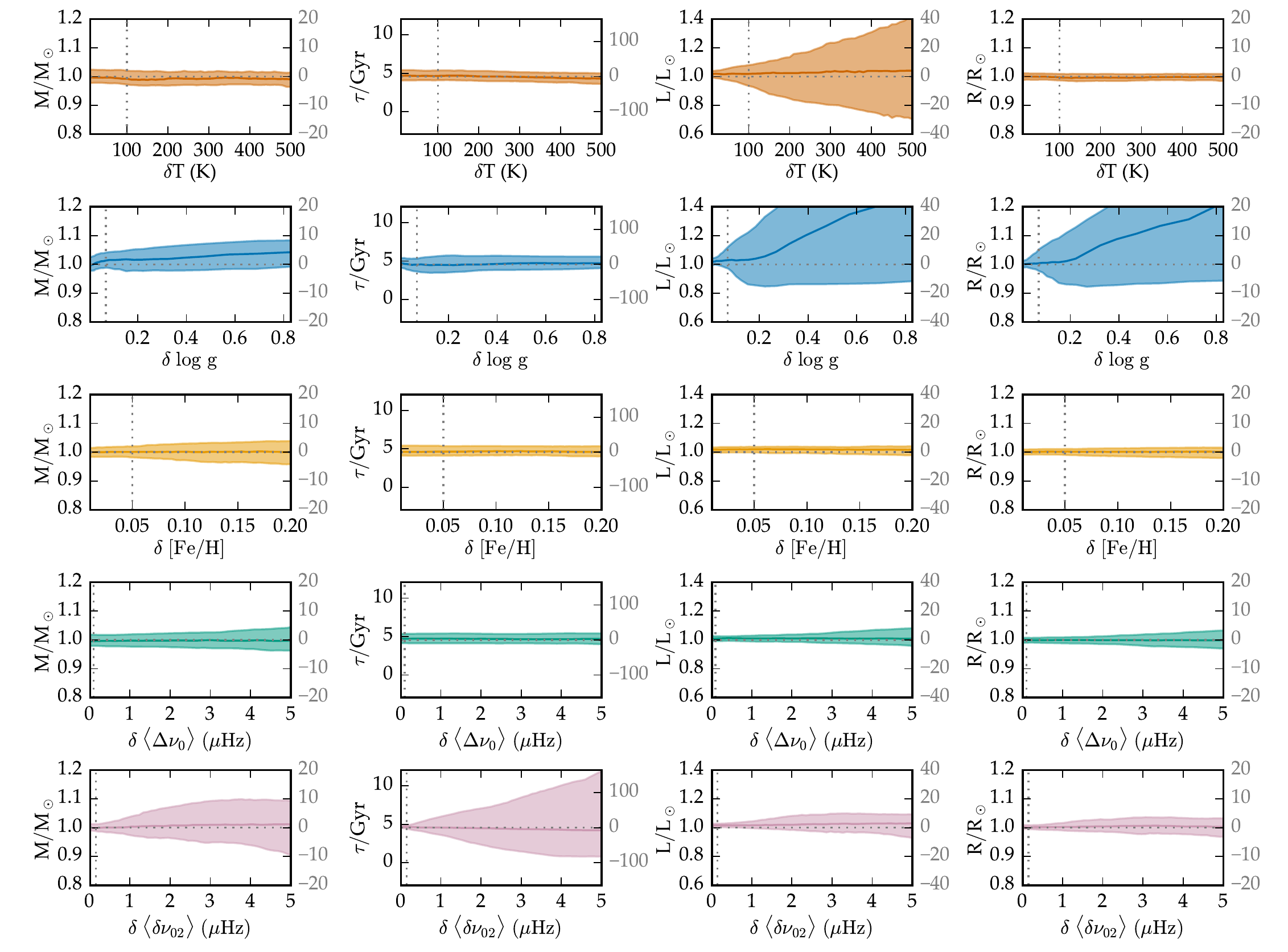}
    \caption[Impact of uncertainties on predictions of mass, age, luminosity and radius]{(Caption on other page.) \label{fig:uncert1}}
\end{figure}
\end{landscape}
\begin{figure}
  \contcaption{Predictions for the solar mass, age, luminosity and radius as a function of the uncertainties applied to key observables. In each panel we have perturbed the quantity on the abscissa in isolation, centred around the measured value listed in  Table~\ref{tab:sun} and with the uncertainties in the ranges specified therein. We indicate the median predicted value (solid line) and the  $68\%$  confidence interval (shaded region). The dotted horizontal lines mark the zero point or true value in each panel and the vertical line indicates a typical observational uncertainty for the perturbed quantity.}
\end{figure}

\section{Discussion}
\label{sec:disc}

Advances in stellar evolution theory are usually sought through refinement of the standard canonical model.  In this classical approach, observations reveal  behaviour that cannot be explained by the current stellar theory, a model is constructed, analysis of the resultant predictions are carried out and conclusions on the efficacy of that model drawn. In this study we adopted a complementary approach: an exploratory based method whereby we performed statistical analysis of models covering a large range of known physics. Rather than first develop a new model to evaluate, we explored the current paradigm to quantify existing relationships and draw new conclusions.  

Some of the techniques employed in this analysis are over $100$ years old and in many areas of research are powerful standalone tools. They have rarely featured in the field of stellar modelling. Here we comment briefly on the timing of our manuscript which we attribute to two main factors: the advent of supervised machine learning techniques and modern computing resources.  

Random forests are an integral part of the present analysis and are a modern technology.
They help place the use of statistical methods in stellar evolution in a wider practical context.  Elucidating both the relationships found by RF and the exploitable information inherent in the model data provided motivation for the use of techniques such as PCA and correlation analysis.  The RF further facilitated the application of these methods due to the requirement that the models be cast  into a comprehensive evolutionary matrix; something that is not strictly necessary for grid based searches.     

Our approach shares similarities to that taken by \citet{1994ApJ...427.1013B} although we differ in methodology. Since their work, we have seen the necessary increase in computing power and the success of the \emph{Kepler} and CoRoT space missions.  The statistical analysis here requires a well sampled grid of stellar models both with structure and oscillations computed. It cost a week of modern supercomputing time to generate the matrix upon which these operations are performed. Evaluating and training approximately $50,000$  RFs itself is also a computationally expensive endeavour.

\subsection{Features of the Dataset}
It is not clear \emph{a priori} through inspection of the equations of stellar structure, if and how any two emergent quantities of the models co-vary. 
There are, of course, combinations of parameters whose covariances are well-founded in stellar theory, but there exist quantities whose diagnostic power remain underutilized and could in fact offer additional insight into the underlying models.
Bringing such relationships to light over the collective lower main sequence is a key aim of our statistical investigation. 
The correlations in the truncated grid (Figure~\ref{fig:filt_corr}) and full BA1 grid (\ref{fig:corr}) reveal the relationships that can be utilized to constrain each of the quantities listed in Table~\ref{tab:parmdefs}.
Many of the model properties that we wish to infer correlate with several observables simultaneously.  This indicates that the observables carry redundant information about the star. 
In addition, observables co-vary amongst themselves. During  iterative model searches some of the covariances, such as between the seismic ratios, are taken into account.  However, for example, it is possible to obtain independent measurements of $\nu_{\max}$, ${\langle\Delta\nu_0\rangle}$, and ${\log{} g}$. Treating these as independent degrees of freedom without considering model covariances then biases the fit towards the parameters to which these quantities pertain and can result in a solution that is overfit. 

We determined the degree of degeneracy in the observables through PCA dimensionality reduction. 
As mentioned previously, RF regression falls under the umbrella of \emph{supervised} learning, whereas PCA is a form of \emph{unsupervised} learning. 
The difference is that in supervised learning, there is a correct answer that the algorithm is trying to understand how to reproduce.
In the case of unsupervised learning, the machine attempts to directly infer properties of data without any help from the supervisor. Hence, regression and classification analyses are forms of supervised learning, whereas cluster and factor analyses are examples of unsupervised learning.
In the case of supervised learning there is a clear measure of success in the resultant model. 
There is a desired output that the inputs try to match. 
The efficacy can be quantified and evaluated via, say, cross-validation or information-theoretic metrics. 
Unsupervised learning methods simply try to identify features and in the case of PCA these features are not necessarily interpretable. 

The PCA in Section~\ref{sec:PCA} focused on the truncated grid. It comprises $11$ stellar observables of all which carry information on the model properties to varying degrees. 
 We found that $99.2\%$ of the variance in the observables could be explained by five components with
nearly $98\%$ of the data are explained by four components. It could be argued that PC$_5$ explains noise rather than features, however, we found that PC$_5$ displays distinct enough correlations (i.e., with near surface physics) that it warrants inclusion in our analysis. 
The clear dimensionality reduction, from $11$ observables to five PCs,  highlights the value in performing PCA: had we found comparable contributions from each component, we would have instead confirmed a clear dominance from higher order relations and an inadequacy of an approach based on linear analysis.

Our primary goal in Section~\ref{sec:PCA} was to reduce the dimensionality of the observables.
We initially considered regions of the parameter space where observations have shown stars to occupy. 
Following on from the rank correlation tests in Section~\ref{sec:RCT} we applied PCA to a truncated version of the BA1 grid.  
However, the results of the PCA depend on the properties of the data and will change depending on features such as the parameter ranges and number of models in the grid. 
For example performing PCA on the full set of evolutionary tracks ($340,800$ models) demands that components are dedicated to explaining variance in (wider) unobserved regions of the parameter space. 
In order to demonstrate that our interpretations of the PCs are robust, we repeated the PCA on four different subsets of the BA1 grid. 
 We made cuts to the mass and metallicity ranges on the training data the results of which are included in Appendix \ref{sec:PCAg} by means of qualitative correlation plots. 
 
The PCs of the respective grids explain a similar percentage of the variance in each grid: PC$_1$ accounts for approximately  $40\%$ of the variance,  PC$_2$  approximately  $35\%$  etc.,  with more than  $75\%$ of the variance in the observables explained by the first two PCs. 
We interpret this result as the PCA capturing essentially the same five inherent `features' in the observables. 
It follows that the choices in grid size and parameter range have only a small effect on the explained variances.
Analysis of all four grids helps further illustrate that there is redundant information carried in some observables, particularly the seismic separations and ratios. 
Varying the parameter ranges changes the correlations between the PCs and observables (loadings) yet the PCs still explain a similar percentage of the variance in each case.  
Due to the information redundancies the PCs can be constructed such that same features are captured with different linear combinations of the observables.
How exactly a PC is constructed in a particular grid will depend on the amount of variance in the observables imparted by the chosen parameter ranges.

With respect to the independent model parameters, it is no surprise that in general PC$_1$ is strongly correlated with the stellar mass ($M$) and  and PC$_2$ with initial metallicity ($Z_0$). 
These are the principal determinants of stellar evolution in that order and both impact upon the stellar structure independently.
In the two grids where we have cut the mass and metallicity ranges we find that the loading of $T_{\text{eff}}$ is larger in PC$_1$.
This is because in the more solar-like tracks $T_{\text{eff}}$ is a strongly monotonic function of evolution.
The surface aspect of PC$_2$ is then supplemented with some information from  ${\log{} g}$ and ${[\text{Fe/H}]}$.

Reducing the dimensionality of the observables and relating them back to the model parameters without redundancy aided with the interpretation of the PCs. 
Whilst it is useful to have the observables so succinctly described, it does not provide insight into the model parameters we wish to infer. 
We thus condensed the information from the correlation plots into a $\Lambda$ score which 
is the sum of the square of the correlation coefficients  between the model parameters and the PCs (determined for the observables). Squaring the correlation coefficients is equivalent to the squaring the PC loadings of the centered and scaled observables. 
The score is a means to quantify the extent to which information from the model parameters, dependent and independent, are encoded in the observables.
We calculated $\Lambda$ scores for all four grids upon which PCA was performed (Appendix \ref{sec:lambdaa}) and indeed found mostly consistent results. We note some differences arise in the initial model parameters such as $\alpha_{\text{MLT}}$ and $\alpha_{\text{OV}}$ which reflect their underlying 
distributions from the choices in grid truncation. The above analyses can be applied to any combination of observables and model parameters to gauge their utility.

\subsection{Exploiting the Inherent Relationships}

Understanding the inherent properties of the collective lower main sequence is the first step in elucidating the BA1 RF regression. The statistical analysis quantified what information was present in the training data for the RF to exploit. We illustrated why the available data permit BA1 to predict parameters such as $M$, $R$ and $L$ with such high precision and why initial model parameters such as $D$ and $\alpha_{\text{MLT}}$ remain uncertain in comparison. Whilst  Section~\ref{sec:RCT} and Section~\ref{sec:PCA} demonstrated the breadth of information available to the RF, in Section~\ref{sec:qu} we determined how the information could best be used. 

RFs are amongst the most powerful tools in mathematics for non-linear regression. 
The BA1 RF uses the observables, creating a set of decision rules that reduce the variance in the parameter it is fitting.
Whilst feature importances provide some insight into this process as a whole it does not provide specific details for the individual parameters. 
By performing non-parametric multiple regression with every combination of observable in our grid,  
we demonstrated how the correlations in Figure~\ref{fig:filt_corr} could best be exploited and best combined to reveal the most information about each stellar quantity.
Two of the observables, ${[\text{Fe/H}]}$ and ${\langle\delta\nu_{02}\rangle}$ (or as a ratio),  are of vital importance in model fitting procedures as they provide indispensable pieces of independent information that cannot be inferred from other quantities.  

We in effect invert the observations for the model parameters based on functions learnt from the training data. 
Thus we can determine the relative importance of each observable for inferring the model parameters. We, in addition, provide a precision with which we can determine each model parameter \emph{directly} from the information contained in the observables. 
The attainable precision is a function of the number of initial model parameters that are varied and the model degeneracy in the data. 
For example, with perfect information from the observables, the six dimensions in the BA1 grid limits our inference on mass to 
${\mu (\epsilon) =0.02 \ \Mo}$.

Many of the Tables in Section~\ref{sec:qu} demonstrated an important property of the RF. 
In the case of missing or unreliable measurements of an observable, the RF can draw upon information redundancies in the data to determine new regression rules for the model parameters.
In principle, such redundancies can lead to biases and overfitting in iterative model finding methods.
During such search procedures the best fitting stellar model is the one that best matches all of the observations but each observation only bares on some parts of the model, and observations can contain redundant information.

Through statistical bagging and multiple regression the RF is less likely to overfit. 
These underlying methodologies are the reason why in Section~\ref{sec:accu} many of the parameters we inferred remained well constrained despite large uncertainties in some of the observables.   
In statistical bagging different subsets of the training grid are sent to different nodes. Each node will use information theory to create a set of decision trees to explain the parameter of interest. The nodes will differ in their rules and choice of parameters. Thus the uncertainty in an observable will only impact on the parameter we infer to the extent to which the observable is used in the rules.  
Take the example from Figure~\ref{fig:uncert1} where with a ${5\;\mu}$Hz uncertainty in ${\langle\Delta\nu_0\rangle}$ the RF still predicts the solar properties albeit with slightly less confidence.
The other observables help constrain the predictions. 

Part of the analysis in Section~\ref{sec:qu} demonstrated the best possible (average) precision in which we can hope to infer stellar parameters. Our error analysis in Section~\ref{sec:accu} is an extension of this. Rather than assume perfect information we determined  
 the measurement accuracy required of the observables to attain a desired precision from the RF.
Our analysis focused on the Sun and is indicative of solar-like analogues.
In Table~\ref{tab:relabunds} we saw some of the large uncertainties associated with retrodicting abundances in low-metallicity stars. We have greater degeneracy with the efficiency of diffusion and the initial abundances. These large error scores by no means indicate that the RF is incapable of characterizing low-metallicity stars. Rather it is an honest appraisal of stellar uncertainties when we do not make assumptions of the initial abundance say through a dY/dZ  chemical evolution ``law"  or a fixed diffusion efficiency. Our error analysis here does not take into account covariances and was designed to investigate the impact on an observable-by-observable basis.
A more detailed error analysis and the associated issues at low metallicity form the focus of a forthcoming paper.

\subsection{Implications for the TESS and PLATO missions}

The NASA TESS mission \citep{2015JATIS...1a4003R} and ESA's PLATO \citep{2014ExA....38..249R} herald a new age for the space-based photometry and the detection of planetary transits. Due to launch in 2018 and 2025 respectively, their common primary science mission is to identify terrestrial planets around bright stars. The pre-selection of bright targets will ensure that the stellar hosts can be further analyzed with spectroscopy and it is expected that many of the planet candidates will be suitable for atmospheric follow-up (ideally) with the James Webb Space Telescope. As was the case with the \emph{Kepler} and \emph{CoRoT} missions, the photometric time-series observations will prove useful to asteroseismology. In the case of PLATO the study of the stellar structure through asteroseismology is a key science goal in the mission design \citep{2014ExA....38..249R}.  

TESS will monitor photometric variations of ${> 10^5}$ low-mass main-sequence stars. Under its `step and stare' pointing strategy, fields will be monitored for periods ranging from one month to one year depending primarily on their ecliptic latitude. With its two minute and $30$ minute cadences, TESS will be able to detect small rocky planets around solar like stars at $\le$ 7th magnitude. It is expected to detect of the order $1,700$ planets with sub-Neptune masses \citep{2016ApJ...830..138C} and will identify many more larger planets around dimmer targets.  The asteroseismic potential of TESS has been rigorously investigated by \citet{2016ApJ...830..138C}. Their analysis of the expected TESS photometry indicates the presence of an oscillation power excess 
in low-mass main-sequence stars when there is no systematic noise present in the data. With an expected systematic noise level of  $60$ ppm hr$^{1/2}$ from the mission, their analysis indicates a detectable power-excess in F-dwarfs as well as sub giants and red giants -- this owing to the higher luminosity and hence larger mode amplitudes in these stars. For a majority of stars the $27$ day pointing is insufficient to extract detailed asteroseismic diagnostics such as mode frequencies or separations. Rather, the seismic information will be limited to the determination of $\nu_{\max}$ in stars where the power-excess is detected. As a consequence, masses and radii for the TESS targets are to be determined using a combination of  GAIA data, the  $\nu_{\max}$ -- ${\langle\Delta\nu_0\rangle}$ power law \citep{2009A&A...506..465H,2009MNRAS.400L..80S} , asteroseismic scaling relations and grid-based searches.

The number of small planet detections from the PLATO mission is expected to eclipse the number found by \emph{Kepler}
and TESS by up to three orders of magnitude. In addition, the PLATO pointing strategy will allow for the measurement of oscillation frequencies in $> 80,000$ dwarf and subgiant stars with magnitudes less than $11$.  In total the mission will provide approximately one million light curves for stars with brightness $\le$ 13th magnitude \citep{2014ExA....38..249R}. In many stars modes up to spherical degree ${\ell =3}$ will be detected with typical frequency uncertainties in the range  $0.1$ -- $0.3\;\mu$Hz. The second major science goal of PLATO is to 
probe stellar structure and evolution by asteroseismology and provide support to exoplanet science through determining
\begin{itemize}
\item stellar masses with an accuracy of better than  $10\%$,
\item stellar radii accurate to $1$--$2\%$, and
\item ages of solar-like stars accurate to  $10\%$.
\end{itemize}

Here we treat the `Sun as a star' in order to quantify how well we can characterize target systems observed by the 
upcoming space missions and to determine the prospect of meeting the accuracy requirements.  
In Table~\ref{tab:tpl} we indicate the observables the missions are likely to provide. 
We degrade the corresponding solar data according to the expected uncertainty from the respective measurements.   
As GAIA is complete down to 20th magnitude we have assumed that distances and hence luminosities will be available for all targets in these missions. We consider data for TESS targets assuming both $60$~ppm~hr$^{1/2}$ and no systematic noise in the photometry. Thus in the case of the latter we anticipate that an oscillation power excess can be extracted for a solar-like star and $\nu_{\max}$ determined. The large and small frequency separations for the PLATO data are determined by degrading a subset of solar frequencies using the method described in BA1. We take a conservative approach in this calculation and assume that the ${\ell=3}$ modes are not extracted.

\afterpage{
\cleardoublepage
\begin{figure}
    \centering
    \includegraphics[width=\textwidth,height=1.02\textheight, keepaspectratio ]{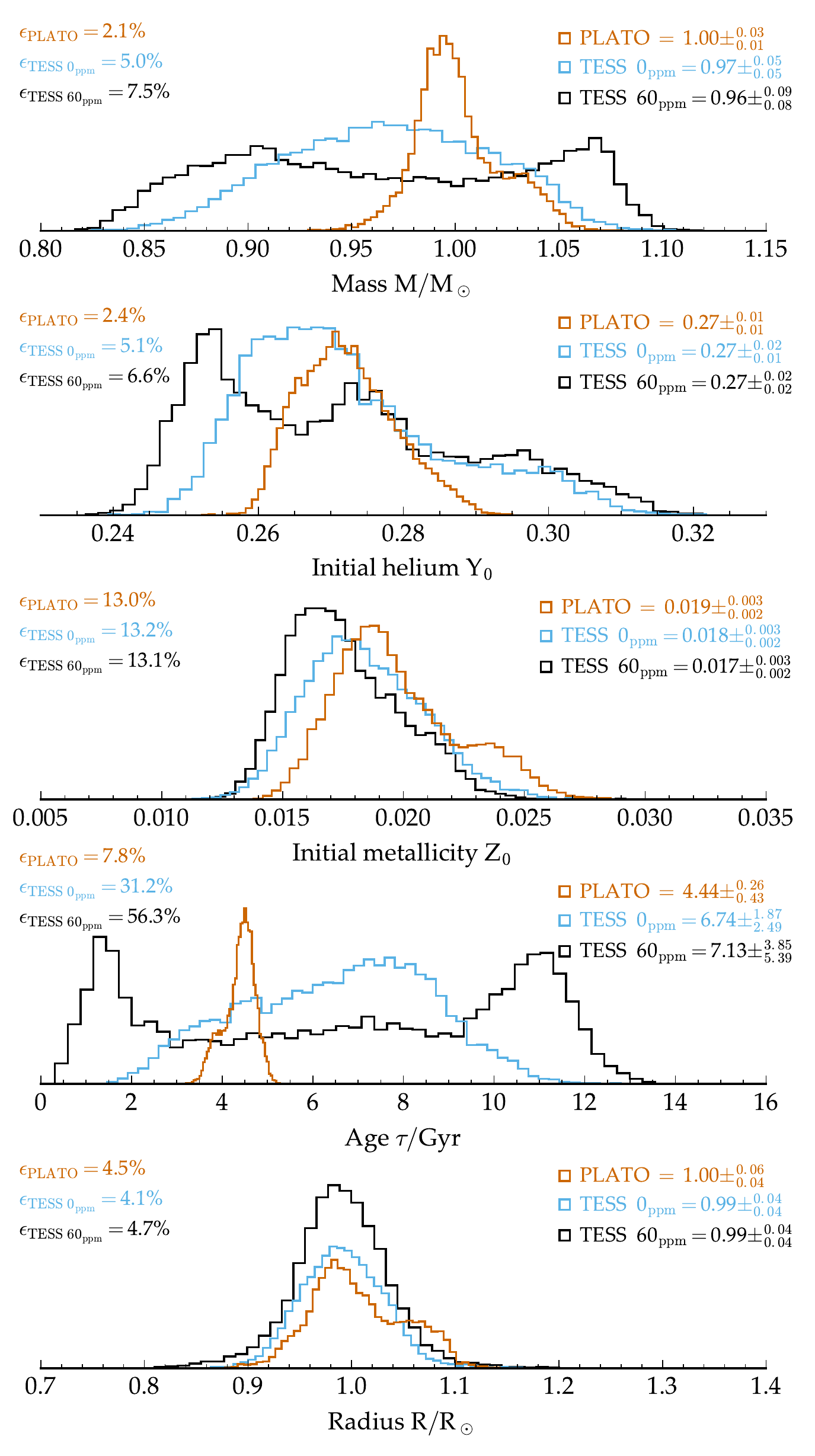}
    \caption[Recovering solar parameters using observations expected for targets from TESS and PLATO]{(Caption on other page.)  \label{fig:TessPlato}}
\end{figure}
\begin{landscape}
\begin{figure}
    \contcaption{Predictions for the `Sun as a star' using observations expected for targets from TESS (assuming two different systematic noise levels) and PLATO space missions. In each panel we list the median with uncertainties ($84\%$--$50\%$ confidence intervals and $50\%$--$16\%$ confidence intervals)   for the quantities as well as the relative error in our prediction.} 
\end{figure}
\begin{table} 
\centering
\caption{Solar data degraded to the level expected for sun-like stars in: the TESS catalogue assuming systematic  noise of $60$~ppm~hr$^{1/2}$ from the mission, TESS assuming no systematic noise and from PLATO. For each set of observables we include the feature importances from the random forest used in characterizing the `Sun as a star.' Note that in the case of the expected PLATO data we have perturbed a subset of frequencies according to their distance from $\nu_{\max}$. The numbers reported for the separations and ratios are thus the respective means and standard deviations of $10,000$ perturbations to the data which we evaluate to determine our parameter distributions. \label{tab:tpl}} 
\begin{tabular}{l|ccc|ccc|ccc}
\multicolumn{1}{c}{} &
\multicolumn{3}{c}{TESS ($60$~ppm~hr$^{1/2}$)}   &
\multicolumn{3}{|c|}{TESS ($0$~ppm~hr$^{1/2}$)}   &
\multicolumn{3}{c}{PLATO} \\\hline
Parameter  &
Value &
Uncertainty &  
Importance &
Value &
Uncertainty &
Importance &
Value &
Uncertainty &
Importance \\\hline
$T_{\text{eff}}$ (K) & 5778 &100 &29.3\% & 5778 &100 &26.7\%& 5778 &100 &16.2\%\\
$[\text{Fe/H}]$ &-0.014 &0.021 &34.3\% &-0.014 &0.021 &33.4\% &-0.014 &0.021 &27.9\%\\
$\log{} g$ &4.43 &0.07 &18.5\% &4.43 &0.07 &12.4\% &4.43 &0.07 &8.8\\
$L \ (L/\text{L}_{\odot})$   &0.98  &0.04 &18.0\% &0.98  &0.04  &16.7\% &0.98  &0.04  &7.8\%\\
$\nu_{\max} \ (\mu \text{Hz})$  & -- & -- & -- &3093 &100 &10.8\% & -- &-- & --\\
$\langle\Delta\nu_0\rangle  \ (\mu \text{Hz})$  & -- & -- & -- & -- & -- & -- &134.81 &0.05 &6.4\%\\
$\langle\delta\nu_{02}\rangle  \ (\mu \text{Hz})$  & -- & -- & -- & -- & -- & -- &9.02 &0.15 &7.1\% \\
$\langle r_{01}\rangle$ & -- & -- & -- & -- & -- & -- &0.0226 &0.0005 &7.4\%\\
$\langle r_{10}\rangle$ & -- & -- & -- & -- & -- & -- &0.0227 &0.0005 &7.3\% \\
$\langle r_{02}\rangle$& -- & -- & -- & -- & -- & -- &0.0668 &0.0011 & 11.1\%
\end{tabular}
\end{table}
\end{landscape}
}

Figure~\ref{fig:TessPlato} shows our predictions for masses, radii, ages, initial helium and metallicity for a `Sun-as-a-star' exercise. In each panel we indicate the median of the probability density distribution and the corresponding  uncertainty from the $16\%$ and $84\%$ confidence intervals for the parameter we are predicting. In addition we determine the relative error which we define as ${\epsilon = 100 \cdot \sigma/\mu}$  where $\mu$ is the mean and $\sigma$ is the standard deviation of the distributions. In Appendix \ref{sec:PSM} we further demonstrate the impact of the measurement uncertainty on the prediction of each quantity as per Figure~\ref{fig:uncert1}.

Although we can expect accurate mass determinations for targets in both missions, the supplementary seismic data from PLATO allows us to improve the precision with which we determine mass by approximately a factor of two. This is despite the fact the RF has identified a less-likely but not impossible (slightly) younger, higher-mass, higher-metallicity solution from the  PLATO  data (we find bimodalities for most quantities predicted with the PLATO observables). In the case of TESS, the absence of the large frequency separation leads to greater uncertainty. One of the methods discussed  by \citet{2016ApJ...830..138C} for the mass determination of TESS targets is to use the power law linking $\nu_{\max}$ to ${\langle\Delta\nu_0\rangle}$ (which has been shown to be accurate to $10$--$15\%$) and apply the asteroseismic scaling laws (Equations \ref{equ:nmax} and \ref{equ:dnu}).  In Section~\ref{sec:seispr} we demonstrated that the random forest exploits further information from  temperature or metallicity measurements to improve the accuracy of the $\nu_{\max}$ -- ${\langle\Delta\nu_0\rangle}$ relation.  Thus we expect the accuracy with which we predict mass from TESS data to represent an upper limit to that attainable by  applying the power-law and scaling relations. 

The assumption of GAIA distances and hence stellar luminosities ensure that radii can be determined for targets in both missions; the seismology is essentially redundant for the inference of the stellar radius.  We note that the relative error for  PLATO in our `Sun-as-a-star' test  is a factor of two higher than the $1$--$2\%$ expected by the consortium. This is a consequence of having identified bimodal solutions. Their target accuracy can likely be met if the uncertainties in the measurements are further reduced and a unimodal solution found. 

The analysis in Section~\ref{sec:seispr} has highlighted the necessity of the small frequency separation  in order to 
tightly constrain the ages of field stars. The predictions for age in Figure~\ref{fig:TessPlato} are therefore as expected. 
The inclusion of oscillation frequencies and determination of the small frequency separation (and ratios) from PLATO data result in age uncertainties for solar-like stars to within the  $10\%$ level.  Without information from the core, ages for TESS targets remain largely unconstrained and consistent with the accuracy typically expected when dating field stars spectroscopically.

\section{Conclusions}
In this work we examined the processes that allow random forest regression to rapidly and accurately infer stellar parameters \citep{2016apj...830...31b}. We shed light on the inherent properties of the model training data that the algorithm can exploit. 

\begin{itemize}
   
             \item We demonstrated that there is a large amount of information redundancy in the stellar parameters which is integral to the efficacy of the random forest algorithm. Through statistical bagging, the random forest  creates sets of decision rules using different combinations of observables to infer a given quantity. The methodology results in  robust predictions and includes the ability to compensate for data that are missing or unreliable. 
             
          \item We illustrated the behaviour of parameters across the collective lower main sequence with the relationships that arise (e.g., age -- luminosity) different to those that develop internally along an evolutionary track. This is the inherent information the random forest draws upon in its regression.
     
         \item  We found the parameter pairs that exhibit the strongest correlations correspond to well known asteroseismic and main-sequence relations.

   \item The random forest works well in cases when there is sufficient information and sufficient redundancy.
   Through principal component analysis we quantified the degree of degeneracy in the observables. 
   Our analysis demonstrated that $99.2\%$ of the variance in the $11$ stellar observables could be explained by five principal components.

 \item The observables we have considered only carry five pieces of independent information.
             During  iterative model searches it is common that independently determined parameters such as $\nu_{\max}$, ${\langle\Delta\nu_0\rangle}$, and ${\log{} g}$ are treated as independent degrees of freedom. The composition of the principal components indicate that by not considering their model covariances, any fit is biased towards the common stellar information to which these parameters pertain.

  \item We devised a score  which  allows us to rank the degree to which model parameters can be inferred from the observables.  Radius, luminosity, and main-sequence lifetime can be extracted  with confidence, however, the initial model parameters such as $\alpha_{\text{MLT}}$, $Y_0$ and $\alpha_{\text{ov}}$ are not sufficiently constrained by the observables and cannot be inferred directly from the data.  Our analysis can be extended in a straightforward manner to model parameters and observables not considered here.

  \item Having elucidated the statistical properties of the training data, we sought to better understand how the random forest uses the data in its decision making rules.
By performing non-parametric multiple regression with every combination of observable in our grid we determined:
\begin{enumerate}
    \item  which observables are the most important/useful for each model parameter,
    \item  the minimum set of observables that satisfactorily constrain each model parameter, and
    \item  the precision with which we can determine each model parameter \emph{directly} from the information contained in the observables. 
\end{enumerate}

\item We examined the quantities on a parameter by parameter basis and here highlight the results for mass and age.  In a grid of stellar evolution models varied in six initial parameters we find that the average error in predicting mass across the grid is ${\pm 0.02 \ \Mo}$ and  ${\pm 282}$ Myr for age. The average error in age increases by a factor of three when we are limited to information from only two observables such as in the Christensen-Dalsgaard diagram. Three parameters are sufficient for constraining mass whereas we require five observables to determine age.

\item We determined whether the random forest could reproduce the well-known power law that relates ${\langle\Delta\nu_0\rangle}$ to $\nu_{\max}$ and found that additional information from  $T_{\text{eff}}$ or  $[\text{Fe/H}]$ reduces the average error in the relation by a factor of two. 

\item We investigated the measurement accuracy required of the observables to attain a desired precision from the random forest. 
The processes of statistical bagging and multiple regression help mitigate the impact of large spectroscopic errors as the random draws upon complementary seismic information when devising its decision rules. 
The results confirm that $[\text{Fe/H}]$  and  ${\langle\delta\nu_{02}\rangle}$ are indispensable independent pieces of information for model fitting algorithms.

\item Finally, we determined the accuracy and precision with which we can expect to characterize solar-like stars observed by the upcoming TESS and PLATO space missions. In both cases masses can be accurately inferred and measurements from GAIA will ensure that radii are well constrained. Oscillation frequencies will not be detectable in most low-mass main sequence stars observed by TESS. In contrast, the availability of the small frequency separation for PLATO targets will permit accurately determined stellar ages. 
\end{itemize}

\paragraph*{Acknowledgements}
\noindent The research leading to the presented results has received funding from the European Research Council under the European Community's Seventh Framework Programme (FP7/2007-2013) / ERC grant agreement no 338251 (StellarAges). E.B.\ undertook this research in the context of the International Max Planck Research School for Solar System Research. S.B.\ acknowledges partial support from NSF grant AST-1514676 and NASA grant NNX13AE70G. We thank Alexey Mints and the anonymous referee for their useful comments and discussions which helped improve this manuscript.

\paragraph*{Software} 
\noindent Stellar models were calculated with  \emph{Modules for Experiments in Stellar Astrophysics} r8118 \citep[MESA,][]{2011apjs..192....3p} and stellar oscillations with the ADIPLS pulsation package 0.2 \citep{2008Ap&SS.316..113C}. 
Analysis in this manuscript was performed with python 3.5.1 libraries scikit-learn 0.17.1 \citep{scikit-learn}, NumPy 1.11.0 \citep{van2011numpy}, matplotlib 1.5.1  \citep{Hunter:2007}, biokit 0.3.2 \citep{biokit} and pandas 0.19.0 \citep{mckinney2010data} as well as R 3.3.2 \citep{R} and the R libraries magicaxis 2.0.0 \citep{magicaxis}, RColorBrewer 1.1-2 \citep{RColorBrewer}, parallelMap 1.3 \citep{parallelMap}, data.table 1.9.6 \citep{data.table}, ggplot2 2.1.0 \citep{ggplot2}, GGally 1.2.0 \citep{GGally}, scales 0.4.0 \citep{scales} and  Corrplot 0.77. 


\section{Appendix}

\subsection{Seismic Definitions} 
\label{sec:sdefs}
We denote any frequency separation $S$ as the difference between a frequency $\nu$ of spherical degree $\ell$ and radial order $n$ and another frequency: 
\begin{equation} 
  S_{(\ell_1, \ell_2)}(n_1, n_2) \equiv \nu_{\ell_1}(n_1) - \nu_{\ell_2}(n_2).
\end{equation}
The large-frequency separation is defined as
\begin{equation} 
  \Delta\nu_\ell(n) \equiv S_{(\ell, \ell)}(n, n-1)
\end{equation}
and the small-frequency separation is
\begin{equation}
  \delta\nu_{(\ell, \ell+2)}(n) \equiv S_{(\ell, \ell+2)}(n, n-1).
\end{equation}
\citet{2003A&A...411..215R} have demonstrated that taking the ratio of the \emph{local} large and small-frequency separations reduces the systematic offset introduced from improper modelling of the near-surface super-adiabatic region. This ratio is defined as: 
\begin{equation}    
  \mathrm{r}_{(\ell,\ell +2)}(n) \equiv \frac{\delta\nu_{(\ell, \ell+2)}(n)}{\Delta\nu_{(1-\ell)}(n+\ell)}.
\end{equation}
In addition, it was shown that the frequency-dependent offset can be somewhat mitigated by constructing ratios from five-point frequency separations and the \emph{local} large separation:
\begin{equation} 
  \mathrm{r}_{(\ell, 1-\ell)}(n) \equiv \frac{\mathrm{dd}_{(\ell,1-\ell)}(n)}{\Delta\nu_{(1-\ell)}(n+\ell)} 
\end{equation}
where the five point separations are defined as:
\begin{align} 
  \mathrm{dd}_{0,1} \equiv \frac{1}{8} \big[&\nu_0(n-1) - 4\nu_1(n-1) \notag\\
                                 &+6\nu_0(n) - 4\nu_1(n) + \nu_0(n+1)\big]\\ 
  \mathrm{dd}_{1,0} \equiv -\frac{1}{8} \big[&\nu_1(n-1) - 4\nu_0(n) \notag\\
                                &+6\nu_1(n) - 4\nu_0(n+1) + \nu_1(n+1)\big]. \label{eqn:dlast}
\end{align}
We calculate dozens of oscillation frequencies per star with the mode sets available dependent on the internal structure of an individual model. We thus determine a single representative value 
by following the prescription of \citet{2012A&A...537A..30M}. In order to mimic how the oscillation spectra would appear in an observational data, we weight all frequencies by their position in a Gaussian envelope with  full-width at half-maximum of ${0.66\cdot\nu_{\max}{}^{0.88}}$ and centered at the predicted frequency of maximum oscillation power $\nu_{\max}$.  We then calculate the weighted median of each variable, which we denote with angled parentheses (e.g.\ ${\langle r_{1,0}\rangle}$).

\subsection{Asteroseismic Scaling Relations}

\begin{equation}
\nu_{\max} \approx \frac{ M/M_{\odot}(T_{\text{eff}}/T_{\text{eff},\odot})^{3.5}}{L/L_{\odot}} \nu_{\max,\odot} \: 
\label{equ:nmax}
\end{equation}

\begin{equation}
\Delta\nu \approx \frac{(M/M_{\odot})^{0.5}(T_{\text{eff}}/T_{\text{eff},\odot})^{3}}{(L/L_{\odot})^{0.75}}
\Delta\nu_{\odot} \: 
\label{equ:dnu}
\end{equation}

\citet{2016MNRAS.460.4277G} have shown that a metallicity-dependent correction is required for the Equation~(\ref{equ:dnu}) scaling relation.
The ${\Delta\nu_{\odot}}$ term can be replaced with a more appropriate reference value which can be calcuated according to:

\begin{equation}
\Delta\nu_{\text{ref}}= A \cdot e^{\lambda T_{\text{eff}}/10^4K} \cdot (\cos(\omega \cdot T_{\text{eff}}/10^4K+\phi))+B,
\label{eq:corrfunc2}
\end{equation}
and where the unkown terms are listed in Table~\ref{tab:pars2}.

\begin{table}
	\centering
	\caption{Parameters of the correction function.}
	\label{tab:pars2}
	\begin{tabular}{lc} 
		\hline
		A & 0.64$\cdot$[Fe/H] + 1.78  $\mu Hz$ \\
		$\lambda$ & $-$0.55$\cdot$[Fe/H] + 1.23  \\
		$\omega$ & 22.21 rad/K \\
		$\phi$ & 0.48$\cdot$[Fe/H] + 0.12 \\
		B & 0.66$\cdot$[Fe/H] + 134.92 $\mu Hz$ \\
		\hline
	\end{tabular}
\end{table}

\subsection{Correlation Plot} 
\label{sec:fullcorr}

\begin{figure*}
    \centering
    \includegraphics[trim={1cm 0 2cm 1cm},clip,
    width=\textwidth]{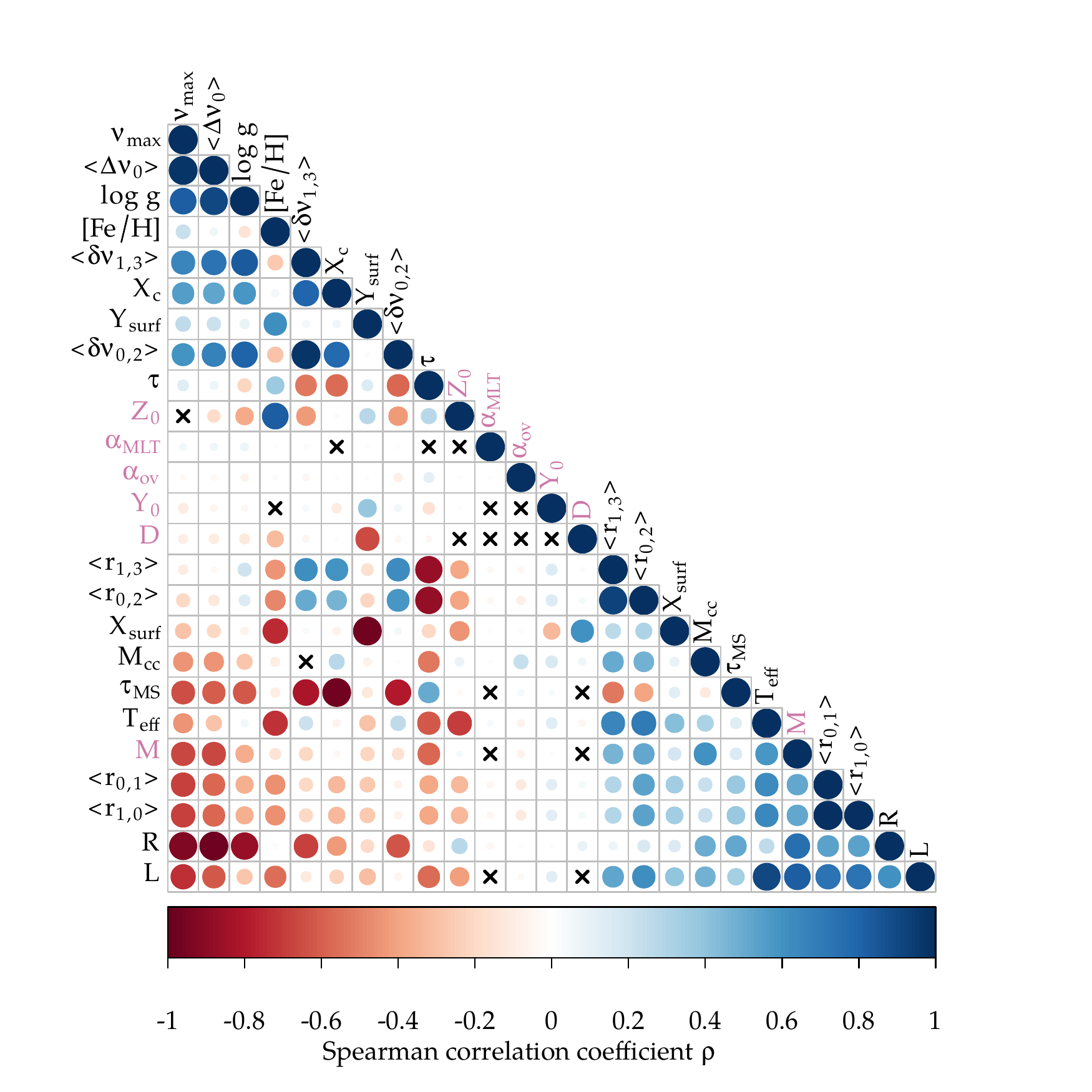}
    \caption[]{Spearman rank correlation matrix comprising various stellar and asteroseismic parameters. The quantities are as described in Table~\ref{tab:parmdefs}
    with model input parameters marked in purple above. The complete grid of models are considered here. The size and color of each circle indicates the sign and magnitude of the Spearman rank coefficient, $\rho$, between two variables.  All correlations are significant excepting the entries indicated with a cross. The variables are ordered by the first principal component of the correlation matrix. }
    \label{fig:corr}
\end{figure*}

The full BA1 grid introduces some biases in our correlation analysis, particularly from tracks with calculated with  high-mass and/or high-diffusion. Correlation analysis with all models included are presented in Figure~\ref{fig:corr}. 

A major difference that arises between Figure~\ref{fig:filt_corr} and Figure~\ref{fig:corr} is in the ordering of variables. Recall that we report the quantities according to the first principal component of the correlation matrix. Different combinations of variables are required to maximise the variance of each principal component in the new parameter space. Although the PCA analysis in Figures \ref{fig:corr-pcaobs} and Figures \ref{fig:corr-pcamods} rely on Pearson rather than Spearman correlations, they do demonstrate the difference in the composition of the PCs in each grid. 

We also find differences in the correlations that pertain to current surface abundance parameters. Consider the pair M -- $Y_{\text{surf}}$. In Figure~\ref{fig:corr} we find a small but non-negligible negative correlation. The reason being that higher mass tracks diffuse the helium from their surface more efficiently than low-mass stars.  Without the influence of these stars in our sample, our significance test yields a null correlation in Figure~\ref{fig:filt_corr}; the expected result from a quasi-random distribution of initial abundances. 

Two interesting features emanating from our grid selection relates to the parameter pairs ${\langle\delta\nu_{02}\rangle}$ -- $T_{\text{eff}}$ and ${\langle r_{02}\rangle}$ -- ${\log{} g}$. 
We find a null correlation  between ${\langle\delta\nu_{02}\rangle}$ -- $T_{\text{eff}}$ in truncated grid however this emerges as a small positive correlation when the full grid is considered. 
In Section~\ref{sec:sages} we discussed the redundancy in the C--D diagram when projecting stellar models varied in six dimensions into a two-dimensional parameter space.  Thus the  null correlation 
arising from the truncated grid reflects the fact there many combinations of (primarily) mass and metallicity and hence temperature at a given age.  
The full grid, however, consists of a large number of hot short-lived stars that impart a noticeable trend. 

A similar argument applies to ${\langle r_{02}\rangle}$ -- ${\log{} g}$. There are a great number of combinations of ${\langle\Delta\nu_0\rangle}$ and
${\langle\delta\nu_{02}\rangle}$ for a given ${\langle r_{02}\rangle}$ thus in the truncated grid no correlation with ${\log{} g}$ is registered. 
Once again the number of massive short-lived stars bias this previous null correlation.

Finally we note two minor results. Some pairs of parameters in the truncated grid which report null correlations in  Figure~\ref{fig:filt_corr},  show very weak correlations in Figure~\ref{fig:corr}. 
We refer to  L -- $\alpha_{\text{ov}}$ and $\alpha_{\text{MLT}}$ -- ${\langle\delta\nu_{02}\rangle}$ as cases in point. The correlations remain very weak in the current analysis and the larger sample size has introduced a minor trend that in this case passes our conservative significance criterion. We note also that most variables display a much stronger correlation with age in the full grid.

\subsection{Principal Component Analysis Explained Variance} 
\label{sec:fullPCA}

The PCs and their correlations will change depending on the number of dimensions included in the grid and the range of values each parameter takes; the PCs identify vectors of maximal variance. 
Our aim is to determine whether the PCs capture fundamental features ubiquitously encoded in the observables. 
Thus, we wish to investigate the information inherent to the dimensions and mitigate the impact of parameter ranges on our PCs. 
In order to provide a more robust interpretation we have calculated the PCs and their correlations with four different considerations given to the BA1 grid:
\begin{description}
    \item[\textbf{Grid A}] The full BA1 training grid;
    \item[\textbf{Grid B}] The truncated grid;
    \item[\textbf{Grid C}] A grid where more than half the models in each track possess metallicities of [Fe/H] $> -2$; and
     \item[\textbf{Grid D}] A grid with masses limited to ${M < 1.2}$~M$_{\odot}$.
\end{description}
Qualitative correlations between the stellar parameters and the PCs in each grid are presented in Figures \ref{fig:corr-pcaobs} and Figures \ref{fig:corr-pcamods}.

\begin{table}
\centering
\caption{Percentage of the variance explained by each principal component. We report the explained variance percentages for the complete grid of training models (Grid A) and for the truncated set (Grid B, see Section~\ref{sec:RCT}) that better encompasses the observational parameter space. In each case we consider the grid with and without the inclusion of $\nu_{\max}$ which is estimated using the \citet{1995A&A...293...87K} scaling relations rather than calculated from first principle equations. We also consider the explained variances when limits are placed on the metallicity (Grid C) and mass (Grid D) ranges of the models.  These grids are used in Section~\ref{sec:disc} to help interpret the PCs. \label{tab:PCAEV}} 
\begin{tabular}{c|cccc|cc}
        \multicolumn{1}{c}{}  &
        \multicolumn{4}{c|}{$\nu_{\max}$ Included}               &
        \multicolumn{2}{c}{$\nu_{\max}$ Excluded}\\\hline
        Component  &
        Grid A &
        Grid B &
        Grid C &
        Grid D &
        Grid A &
        Grid B \\\hline
PC$_1$ 	&	41.79	&	42.36	& 42.49 & 42.74 &	40.89	&	41.47	\\
PC$_2$ 	&	36.12	&	34.18	& 37.49 & 35.89 &	36.52	&	33.65	\\
PC$_3$	&	9.17	&	11.65	& 9.39  & 10.25 &	8.99	&	12.21	\\
PC$_4$	&	7.69	&	9.79	& 7.69  &  6.89 &	8.27	&	10.58	\\
PC$_5$ 	&	4.23	&	1.23	& 2.14  &  3.36 &	4.55	&	1.36	\\
PC$_6$	&	0.54	&	0.48	&0.41    & 0.53   &	0.48	&	0.51	\\
PC$_7$	&	0.25	&	0.18	&0.24    &0.18    &	0.16	&	0.12	\\
PC$_8$ 	&	0.12	&	0.08	&0.09    &0.10    &	0.10	&	0.09	\\
PC$_9$	&	0.05	&	0.03	&0.04    &0.04    &	0.03	&	0.02	\\
PC$_{10}$ 	&	0.02	&	0.01 &0.01   &0.01	&	0.01	&	0.00	\\
PC$_{11}$	&	0.01	&	0.00 &0.00   &0.00     &	--	    &	--	
\end{tabular}
\end{table}

\subsection{PCA Correlation Analysis} \label{sec:ccoefs}
Figures \ref{fig:GCA-pcabar} and \ref{fig:GCA-pcabarb} demonstrate the correlation strengths between our stellar parameters and the first five PCs. In Tables \ref{tab:ocoefs} and \ref{tab:ocoefs} we list the coefficients between all parameters and all PCs. The table is useful for  determining whether the transitive criterion applies to parameters within a given PC. It also aids in the calculation of the $\Lambda$ scores in Section~\ref{sec:ISP}.

\afterpage{
\clearpage
\begin{landscape}
\begin{table}
\centering
\caption{Pearson's~$r$ coefficients between the principal components and observables in the truncated grid.}
\label{tab:ocoefs}
\begin{tabular}{c|ccccccccccc}
& $\log{} g$ &$T_{\text{eff}}$  & $[\text{Fe/H}]$ & $\langle\Delta\nu_0\rangle$  & $\langle\delta\nu_{02}\rangle$ & $\langle r_{02}\rangle$    & $\langle r_{01}\rangle$    & $\langle\delta\nu_{13}\rangle$ &  $\langle r_{13}\rangle$   &  $\langle r_{10}\rangle$   & $\nu_{\max}$        \\ \hline \hline
PC$_1$    & 0.93  & -0.20 & -0.35 & 0.92  & 0.93  & 0.38  & -0.07 & 0.95  & 0.32  & -0.07   & 0.87  \\
PC$_2$    & -0.30 & 0.73  & -0.29 & -0.33 & 0.34  & 0.85  & 0.81  & 0.22  & 0.76  & 0.81    & -0.42 \\
PC$_3$    & 0.00  & -0.60 & 0.63  & 0.04  & 0.06  & 0.15  & 0.45  & -0.13 & -0.17 & 0.45    & 0.22  \\
PC$_4$    & 0.08  & 0.08  & -0.60 & 0.19  & -0.08 & -0.30 & 0.37  & -0.13 & -0.52 & 0.37    & 0.10  \\
PC$_5$    & 0.14  & 0.25  & 0.20  & 0.06  & -0.02 & -0.05 & 0.03  & 0.00  & -0.07 & 0.03    & 0.01  \\
PC$_6$    & 0.11  & -0.01 & -0.04 & 0.01  & -0.11 & 0.09  & 0.00  & -0.12 & 0.06  & 0.00    & 0.04  \\
PC$_7$    & -0.09 & 0.03  & 0.01  & 0.05  & -0.03 & -0.01 & 0.00  & -0.01 & 0.03  & 0.00    & 0.09  \\
PC$_8$    & 0.02  & -0.02 & 0.00  & 0.00  & -0.04 & -0.05 & 0.02  & 0.03  & 0.05  & 0.02    & -0.01 \\
PC$_9$    & 0.01  & 0.01  & 0.00  & -0.04 & 0.02  & -0.02 & 0.00  & -0.02 & 0.01  & 0.00    & 0.03  \\
PC$_{10}$   & 0.00  & 0.00  & 0.00  & 0.02  & 0.02  & -0.01 & 0.00  & -0.02 & 0.01  & 0.00    & -0.01 \\
PC$_{11}$   & 0.00  & 0.00  & 0.00  & 0.00  & 0.00  & 0.00  & 0.01  & 0.00  & 0.00  & -0.01   & 0.00  \\ \hline
\end{tabular}
\end{table}

\begin{table} 
\centering
\caption{Pearson's~$r$ coefficients between the principal components and model parameters in the truncated grid.}
\label{tab:mcoefs}
\hspace*{-0.5cm}
\begin{tabular}{c|cccccccccccccc}
    &$M$   & $Y$         & $Z$         & $\alpha_{\text{MLT}}$     & $\alpha_{\text{ov}}$ & $D$ & $\tau$      & $\tau_{\text{MS}}$     & $X_c$      & $M_{\text{cc}}$  & $X_{\text{surf}}$   & $Y_{\text{surf}}$   & $R$    & $L$                  \\ \hline \hline
PC$_1$  & -0.67 & -0.08 & -0.30 & 0.06      & -0.05     & -0.12 & -0.19 & -0.72 & 0.77     & -0.45   & -0.04   & 0.16   & -0.86 & -0.69 \\
PC$_2$  & 0.29  & 0.10  & -0.41 & -0.15     & -0.04     & -0.17 & -0.56 & -0.26 & 0.14     & 0.17    & 0.05    & 0.11   & 0.30  & 0.55  \\
PC$_3$  & 0.13  & 0.04  & 0.48  & -0.10     & -0.07     & -0.03 & -0.01 & -0.04 & 0.14     & -0.11   & -0.33   & 0.17   & 0.09  & -0.04 \\
PC$_4$  & -0.51 & -0.17 & -0.40 & 0.03      & -0.09     & -0.08 & 0.51  & 0.48  & -0.49    & -0.34   & 0.29    & -0.17  & -0.23 & -0.13 \\
PC$_5$  & -0.02 & 0.17  & -0.09 & 0.44      & -0.16     & -0.43 & 0.13  & 0.13  & -0.17    & -0.29   & -0.50   & 0.59   & -0.14 & -0.03 \\
PC$_6$  & -0.02 & -0.20 & 0.08  & -0.01     & -0.15     & -0.10 & 0.28  & 0.07  & -0.02    & -0.25   & 0.05    & -0.10  & -0.14 & -0.03 \\
PC$_7$  & 0.09  & 0.05  & -0.05 & 0.14      & 0.11      & -0.15 & -0.10 & 0.03  & 0.10     & 0.30    & -0.14   & 0.17   & 0.18  & 0.38  \\
PC$_8$  & -0.11 & 0.08  & -0.06 & -0.25     & 0.05      & -0.07 & 0.03  & -0.03 & -0.02    & -0.12   & -0.05   & 0.08   & -0.07 & -0.09 \\
PC$_9$  & 0.21  & -0.35 & 0.22  & 0.20      & -0.06     & -0.08 & -0.22 & -0.12 & 0.15     & 0.04    & -0.01   & -0.10  & 0.06  & 0.07  \\
PC$_{10}$ & -0.17 & 0.26  & -0.14 & -0.15     & 0.02      & 0.06  & 0.17  & 0.12  & -0.10    & -0.01   & 0.01    & 0.06   & -0.09 & -0.04 \\
PC$_{11}$ & -0.01 & -0.01 & -0.01 & 0.01      & 0.01      & 0.00  & 0.00  & 0.00  & 0.00     & -0.02   & 0.01    & -0.01  & 0.00  & -0.01 \\ \hline
\end{tabular} 
\end{table}
\end{landscape}
}

\subsection{PC correlations with different grids}
\label{sec:PCAg}
In Section~\ref{sec:intPC}  we presented the correlation strengths between the PCs and observables (Figure~\ref{fig:GCA-pcabar}) and the PCs and the model parameters (\ref{fig:GCA-pcabarb}). 
 Here we perform the same analysis with the different subsets of the BA1 grid described in Appendix \ref{sec:fullPCA}. In order to compare the results for each grid, in Figures \ref{fig:corr-pcaobs} and \ref{fig:corr-pcamods} we employ a correlation plot rather than the quantitative bar chart used in Section~\ref{sec:intPC}. This allows an inspection of the qualitative behaviour of the PCs in each case. We find a similar explained variance from the corresponding PCs in each grid. This suggests that the PCs capture essentially the same inherent features in model data and that the PCs are not due to the number of models in our analysis or the chosen parameter ranges.

\afterpage{
\clearpage
\begin{landscape}
\begin{figure}
    \centering
    \includegraphics[width=\linewidth]{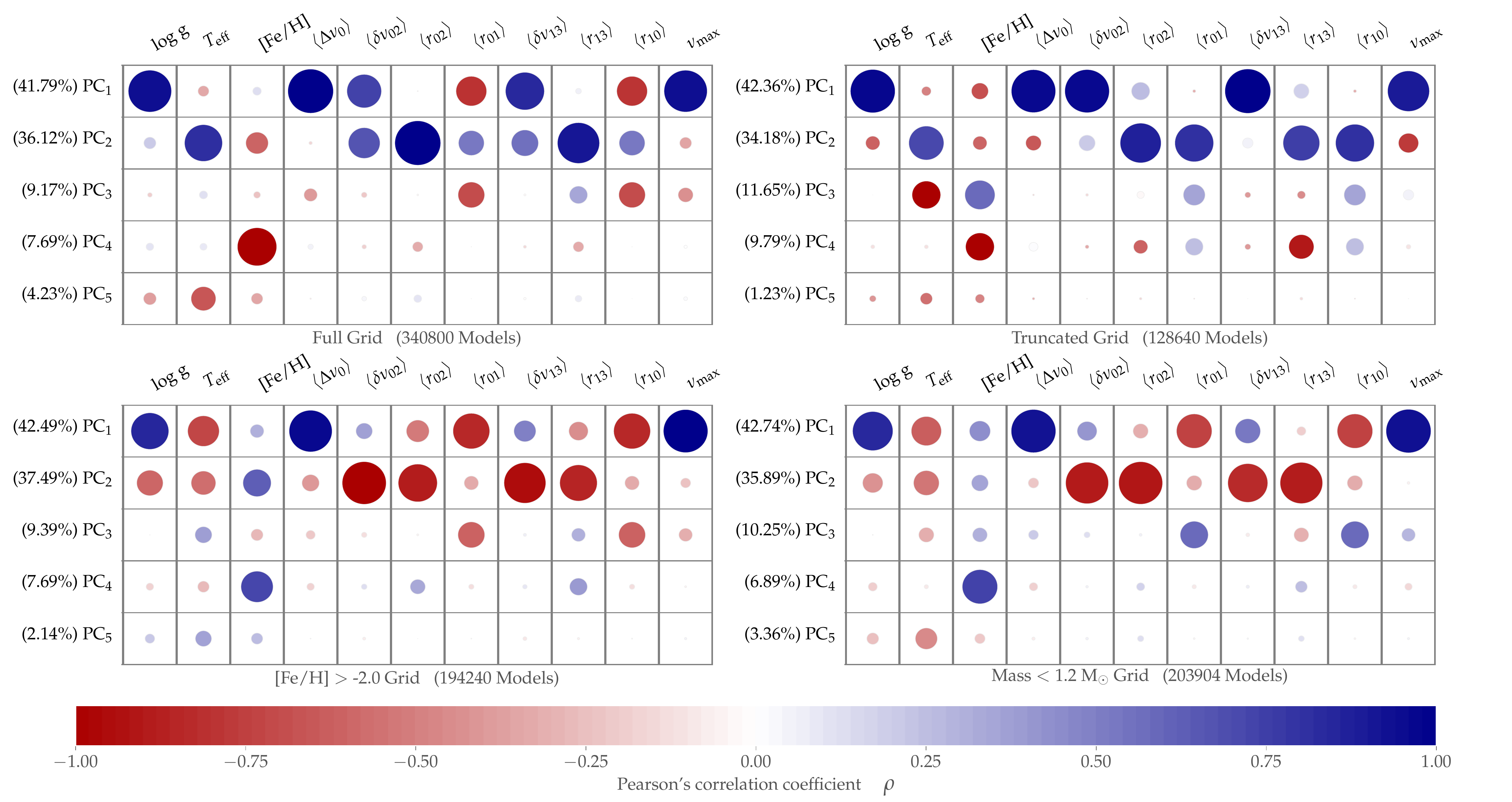}
    \caption[]{Pearson correlation matrices relating the principal components back to the stellar observables in each of the four grids described in Appendix \ref{sec:fullPCA}. \label{fig:corr-pcaobs}}
\end{figure}

\begin{figure}
    \centering
    \includegraphics[width=\linewidth]{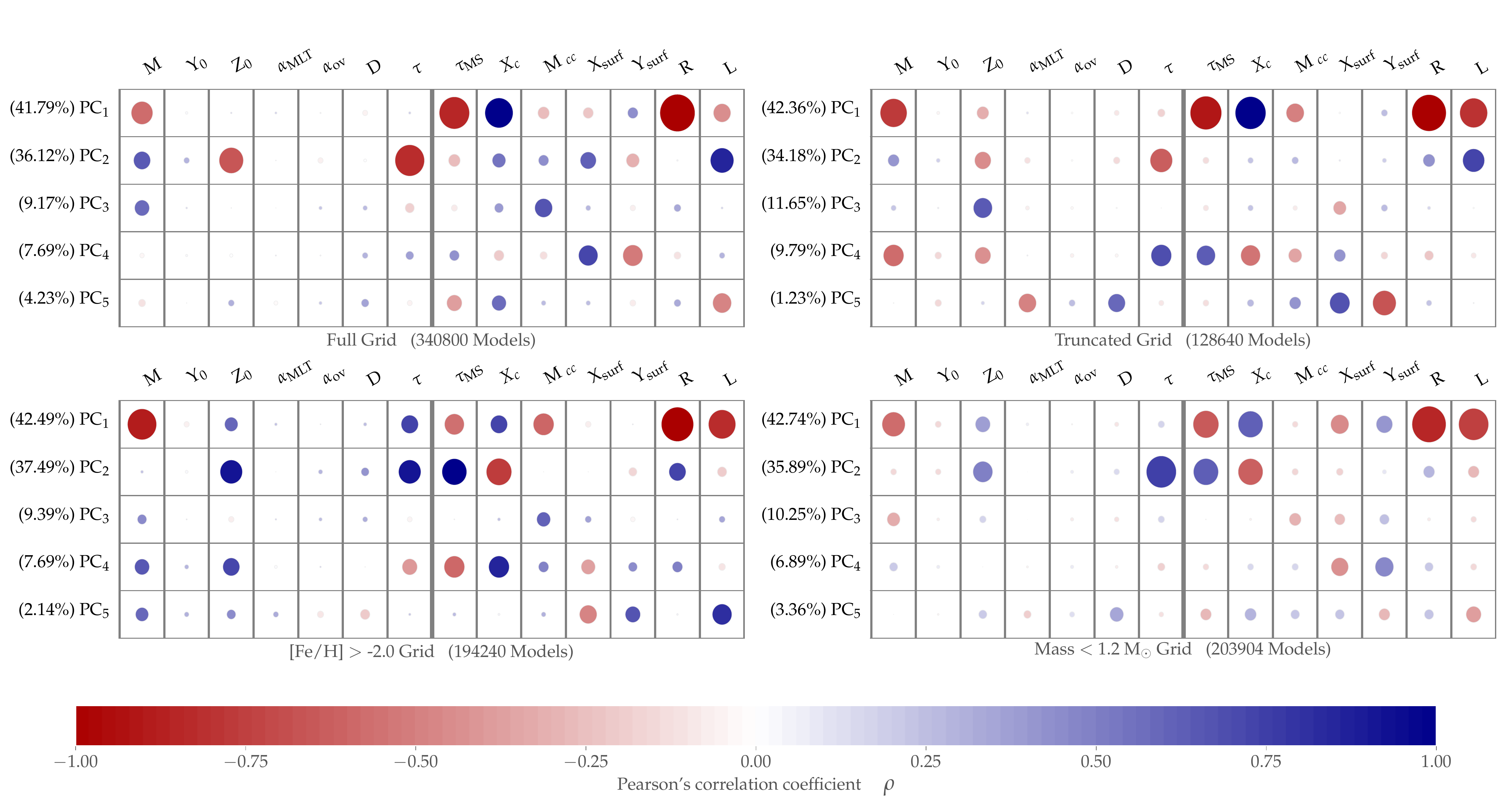}
    \caption[]{Pearson correlation matrices relating the principal components back to the model quantities in each of the four grids described in Appendix \ref{sec:fullPCA}. \label{fig:corr-pcamods}} 
\end{figure}
\end{landscape}
}

\subsection{$\Lambda$ Analysis}
\label{sec:lambdaa}

The data matrix of observables $\mathbf{X}$ is size ${n \times p}$ where n is the number of training models and p the number of parameters.
We centre and scale the entries according to the mean and standard deviation of each parameter.
The resultant matrix,  $\mathbf{\bar{X}}$, therefore has the property that for each parameter, $p$, ${\mu(p) =0}$ and ${\sigma(p) =1}$. 
We compute the correlation matrix, $R$, for the matrix  $\mathbf{\bar{X}}$ :
\begin{eqnarray}
    \mathbf{R}  &= \text{Corr}(\mathbf{\bar{X}}) \\[3pt] \nonumber
                &= \mathbf{\bar{X}}\mathbf{\bar{X}^\top}.  
\end{eqnarray}
As the correlation and covariance matrices are symmetric we calculate the eigendecomposition of R such that:
\begin{equation}
\mathbf{R}=\mathbf{VLV^\top},
\end{equation}  
where \textbf{V} a matrix of eigenvector columns and \textbf{L} a diagonal matrix of eigenvalues.
The eigenvectors specify the principal axes of the data and the eigenvalues indicate the amount of variance there is in the data in the direction of the corresponding eigenvector.
We can define the projection matrix \textbf{P} such that we project/transform our data into the new space    
\begin{equation}
\mathbf{P} = \mathbf{\bar{X}}  \mathbf{V}.
\end{equation}

The correlation matrix is a special case of the covariance matrix in that the former is normalised.
For generality let us consider the covariance matrix, such that the original data matrix was centred but not scaled ($\mathbf{\hat{X}}$), then
\begin{eqnarray}
    \mathbf{C}  &= \text{Cov}(\mathbf{\hat{X}}) \\[3pt] \nonumber
                &= \frac{1}{n-1}\mathbf{\hat{X}}\mathbf{\hat{X}^\top}\\[3pt] \nonumber
                &= \mathbf{V} \mathbf{L} \mathbf{V^\top},
\end{eqnarray}
where we divide by (n-1) to unbias to covariance (the covariance entries will have different scales).

Alternatively and equivalently, we may extract our PCs through SVD of  $\mathbf{\hat{X}}$ such that: 
\begin{equation}
    \mathbf{\hat{X}} = \mathbf{U}  \mathbf{\Sigma} \mathbf{ V^\top}
\end{equation}
where \textbf{U} is the left matrix of singular orthogonal vectors with dimensions ${n \times n}$,
$\mathbf{\Sigma}$ is a diagonal matrix of singular values with dimensions $n \times p$, 
and $\mathbf{V^{\top}}$ is the right matrix of singular  orthogonal vectors with diemsnions ${p \times p}$. 
The diagonal elements of $\mathbf{\Sigma}$ assign a  relative  importance  to  each  vector whereas the vectors of \textbf{V} are the principal directions/axes.
As the matricies \textbf{U} and \textbf{V} comprise orthogonal components they have the property
\begin{eqnarray}
\label{eqn:ident}
\mathbf{U^{\top}U}=\mathbf I_{n \times n} \\ \nonumber
\mathbf{V^{\top}V}=\mathbf I_{p \times p}.
\end{eqnarray}
We note also that 
\begin{align}
    \left(\mathbf {A \cdot B \cdot C }\right)^\top &= \mathbf{C^\top \cdot B^\top \cdot A^\top} \\
    \implies  (\mathbf U\mathbf \Sigma\mathbf V^\top)^\top &= (\mathbf V\mathbf \Sigma\mathbf U^\top)
\end{align}
as $\mathbf{\Sigma}$ is a diagonal matrix.

We can reconstruct the eigendecomposition of the covariance matrix from the SVD: 
\begin{eqnarray}
 \frac{1}{n-1}\mathbf{\hat{X}}\mathbf{\hat{X}}^\top &= \frac{1}{n-1} (\mathbf U\mathbf \Sigma\mathbf V^\top)(\mathbf U\mathbf \Sigma\mathbf V^\top)^\top\\[3pt] \nonumber
&= \frac{1}{n-1}(\mathbf U\mathbf \Sigma\mathbf V^\top)(\mathbf V\mathbf \Sigma\mathbf U^\top)
\end{eqnarray}
and from our identities in Equation~(\ref{eqn:ident})
\begin{equation}
 \frac{1}{n-1}\mathbf{\hat{X}}\mathbf{\hat{X}}^\top=\mathbf U \frac{\mathbf \Sigma^2}{n-1} \mathbf U^\top.
\end{equation}
We therefore find that the square roots of the eigenvalues of $\mathbf{C}$ are the singular values of  $\mathbf{\bar{X}}$ and that the vectors in the right singular matrix, \textbf{V}, are the principal directions/axes. The projection matrix can be calculated from the SVD such that 
\begin{align}
\mathbf{P} &= \mathbf{\hat{X}}  \mathbf{V} \\[3pt] \nonumber
           &= \mathbf U \mathbf \Sigma \mathbf V^\top \mathbf{V}  \\[3pt] \nonumber
           &= \mathbf U \mathbf \Sigma.
\end{align}

The PCA loadings are the columns of \textbf{L} which implies that 
\begin{equation}
\mathbf{L}=\mathbf{V}\frac{\mathbf \Sigma}{\sqrt{n-1}}.
\end{equation}
We can see that the loadings are the eigenvectors scaled by the square roots of the respective eigenvalues.
With these definitions we can compute the cross-covariance matrix between original variables and the standardized projection matrix. 
To calculate the standardized PC scores for \textbf{P} we require each column of \textbf{U} to have unit variance. As $\mathbf{\Sigma}$ is diagonal it is simply a scaling matrix and can be dropped here yielding: 
\begin{align}
\frac{1}{n-1}\mathbf{X}^\top(\sqrt{n-1}\mathbf{U}) &=\frac{1}{\sqrt{n-1}}\mathbf{V}\mathbf{\Sigma}\mathbf{U}^\top\mathbf{U} \\&=\frac{1}{\sqrt{n-1}}\mathbf{V}\mathbf{\Sigma}
\\&=\mathbf{L}.
\end{align}
We find that the covariance matrix between the standardized PCs and original variables is in fact given by the loadings. 
In  Section~\ref{sec:ev} we computed the \emph{correlations} between the observables and their PCs rather than the covariances, requiring that the observables are normalized by their standard deviation. As we centred and scaled our data prior to performing the PCA, their values are unity and our correlation analysis is therefore equivalent to reporting the loadings. 

The correlation analysis allowed us to project the model data onto the PC space and determine the `equivalent' loadings for each parameter. Through the $\lambda$ score we can therefore determine to what extent the variance in the model data is captured by the PCs.  In Table~\ref{tab:corrLFull} we compare the results of the analysis for each grid. We find similar results for most parameters 
with differences in some of the initial model parameters due to their underlying distributions as a result of the grid truncations.

\subsection{Impact of Uncertainties for Upcoming Photometric Space Missions} 
Below we demonstrate the impact of measurement uncertainty on the prediction of parameters from the upcoming TESS (Figure~\ref{fig:uncerttess}) and PLATO (Figure~\ref{fig:uncertplato}) space missions.  We produce  probability density distributions for $250$ sets of $\sigma$ values for each parameter we predict. The ranges for each parameter from which we draw our $\sigma$ values are listed in Table~\ref{tab:sunstar}. We restrict out observables to those we are likely to possess from the respective missions. 
In each figure we plot the median value (solid line) and the  $68\%$  confidence interval (shaded region).

\afterpage{
\begin{table}
    \centering
    \caption{The $\Lambda$ score is a sum of the squares of  ${r(X, PC_i)}$ indicating the variance explained for a given parameter. These scores are by definition unity for our observables.}
    \label{tab:corrLFull}
    \begin{tabular}{ccccc}
    \hline \hline
    & \multicolumn{4}{c}{$\Lambda_{\text{param}}$} \\
Parameter & Grid A & Grid B & Grid C & Grid D \\ \hline     
R	&	0.97	&	0.97	&	0.98	&	0.97	\\
L	&	0.93	&	0.96	&	0.93	&	0.95	\\
$X_c$	&	0.93	&	0.94	&	0.93	&	0.94	\\
$\tau_{\text{MS}}$	&	0.93	&	0.93	&	0.93	&	0.94	\\
M	&	0.91	&	0.91	&	0.92	&	0.88	\\
$\tau$	&	0.74	&	0.79	&	0.78	&	0.76	\\
$Z_0$	&	0.76	&	0.73	&	0.78	&	0.80	\\
M$_{cc}$	&	0.58	&	0.61	&	0.68	&	0.41	\\
$Y_{\text{surf}}$	&	0.48	&	0.50	&	0.55	&	0.54	\\
X$_{\text{surf}}$	&	0.50	&	0.48	&	0.53	&	0.55	\\
$\alpha_{\text{MLT}}$	&	0.02	&	0.38	&	0.04	&	0.06	\\
$Y_0$	&	0.10	&	0.31	&	0.27	&	0.09	\\
D	&	0.13	&	0.29	&	0.22	&	0.21	\\
$\alpha_{\text{ov}}$	&	0.10	&	0.08	&	0.11	&	0.12	\\

 \hline
    \end{tabular}
\end{table}

 \begin{table}
    \centering
    \caption{Central solar values and uncertainty ranges used for predictions in Figures \ref{fig:uncerttess} and \ref{fig:uncertplato}.}
    \begin{tabular}{lcccccc}
    \hline \hline
 \multicolumn{1}{c}{} & \multicolumn{3}{c}{TESS}& \multicolumn{3}{c}{PLATO}  \\
Quantity & Value & Min($\sigma$) & Max($\sigma$) & Value & Min($\sigma$) & Max($\sigma$)\\ \hline 
$T_{\text{eff}}$ (K)  & 5777 & 10  & 500& 5777 & 10  & 500\\
$\log{} g$ &  4.44 & 0.0001 & 1.0&  4.44 & 0.0001 & 1.0\\
$[\text{Fe/H}]$ & 0.0 & 0.05 & 0.5 & 0.0 & 0.05 & 0.5\\
$L$ & 1.0 & 0.001 & 10 & 1.0 & 0.001 & 10 \\
$\nu_{\max}$ & 3050 & 10 & 500 & -- & -- & --\\
$\langle\Delta\nu_0\rangle$ ($\mu$Hz) & -- & -- & --& 136.0 & 0.5 & 50\\
$\langle\delta\nu_{02}\rangle$ ($\mu$Hz)& -- & -- & -- & 9.0 & 0.5 & 5 \\
\hline
    \end{tabular}
    \label{tab:sunstar}
\end{table}
}

\clearpage

\label{sec:PSM}

\afterpage{
\begin{landscape}
\begin{figure}\vspace*{-0.75cm}
    \centering
    \includegraphics[width=0.9\linewidth,keepaspectratio]{tess2.pdf}
    \caption[]{(Caption on other page.)\label{fig:uncerttess}}
\end{figure}
\end{landscape}

\begin{figure}
  \contcaption{Predictions for the solar mass, age, luminosity and radius as a function of the uncertainties applied to key observables. In each panel we have perturbed the quantity on the abscissa in isolation, centred around the measured value listed in  Table~\ref{tab:sunstar} and with the uncertainties in the ranges specified therein. We indicate the median predicted value (solid line) and the  $68\%$  confidence interval (shaded region). Here the observables comprise those expected from the TESS space mission assuming that the p-mode power excess can be extracted.} 
\end{figure}


\afterpage{
\begin{landscape}
\begin{figure}\vspace*{-0.75cm}
    \centering
    \includegraphics[width=0.9\linewidth,keepaspectratio]{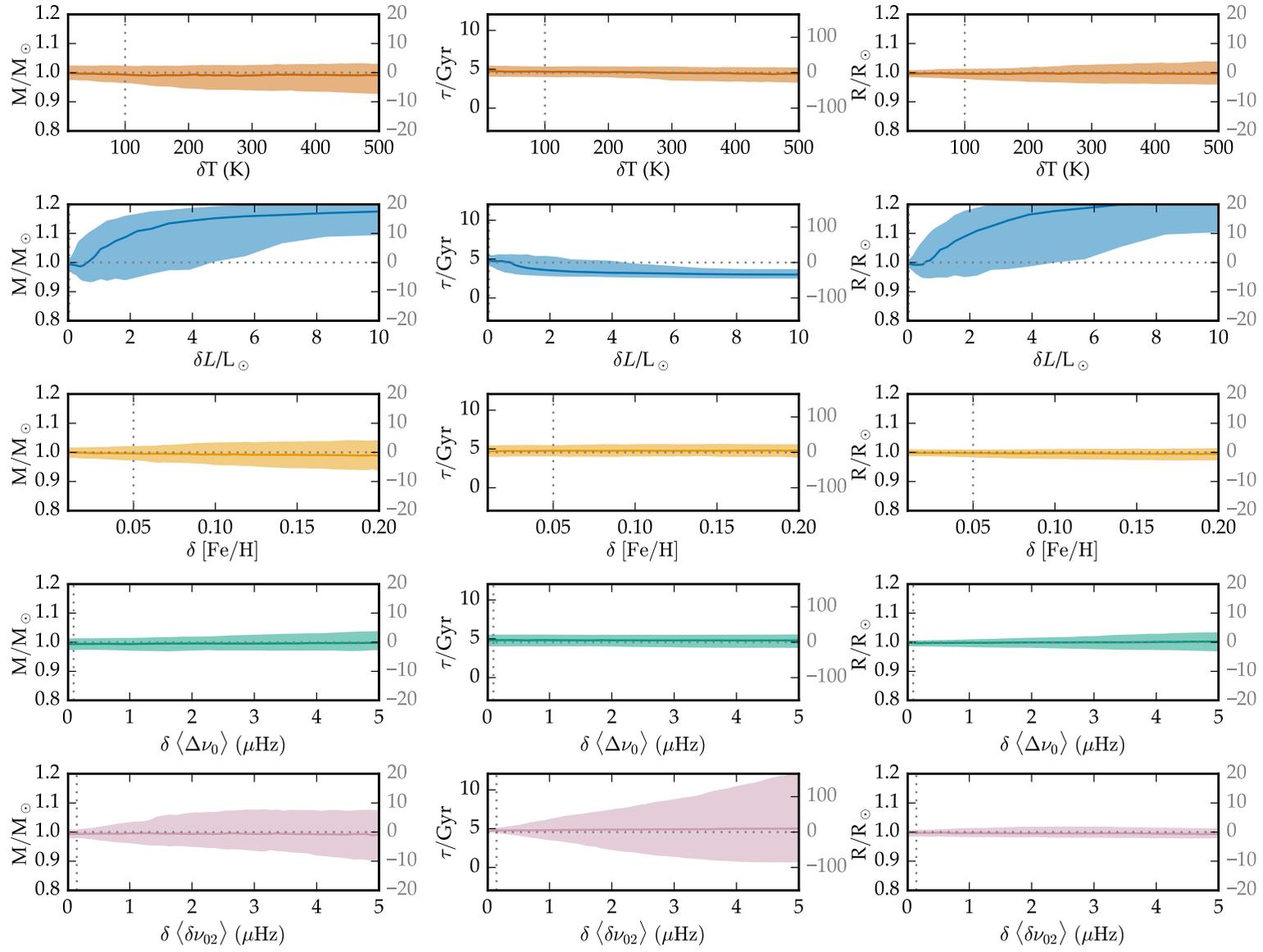}
    \caption[]{The same as Figure~\ref{fig:uncerttess}, but for PLATO. \label{fig:uncertplato}} 
\end{figure}
\end{landscape}
}
}
\clearpage

\chapter{Model-Independent Measurement of Internal Stellar Structure \\in 16~Cygni A \& B}
\chaptermark{Model-Independent Measurement of Internal Stellar Structure}
\label{chap:inversion}

\vspace{2cm}
The contents of this chapter were authored by E.~P.~Bellinger, S.~Basu, S.~Hekker, and W.~H.~Ball and published in December of 2017 in \emph{The Astrophysical Journal}, 851 (2), 80.\footnote{Contribution statement: The work of this chapter was carried out and written by me, under the supervision of S.~Basu and S.~Hekker and in collaboration with W.~H.~Ball. } 
\nocite{2017ApJ...851...80B}

\vspace*{1cm}

\section*{Chapter Summary}
We present a method for measuring internal stellar structure based on asteroseismology that we call ``inversions for agreement.'' 
The method accounts for imprecise estimates of stellar mass and radius as well as the relatively limited oscillation mode sets that are available for distant stars. 
\mb{By construction, the results of the method are independent of stellar models.} 
We apply this method to measure the isothermal sound speeds in the cores of the solar-type stars 16~Cyg~A and B using asteroseismic data obtained from \emph{Kepler} observations. 
We compare the asteroseismic structure that we deduce against best-fitting evolutionary models and find that the sound speeds in the cores of these stars exceed those of the models. 
\newpage
\section{Introduction} 

The detection and study of internal waves in stars---asteroseismology---provides a unique view into stellar interiors. 
As the structure of a star dictates the varieties and frequencies of its normal modes of oscillation, asteroseismic data can be used to set limits on the conditions inside a star. 
This is usually achieved by evolving stellar models, and the structure of the best-fitting model is then assumed to be a proxy for the structure of the star. 
However, theoretical pulsation frequencies of even the best stellar models have significant discrepancies with observations, implying that the structure of the star differs from the structure of the model. 
This is true for the Sun and other stars alike. 
A way to proceed from this point would be to quantify what internal conditions do support the oscillations that have been observed. 
This problem is the inverse of determining the mode frequencies of a known stellar structure, and is thus known as a \emph{structure inversion}. 
Structure inversions are of value because their results are independent of models. 
However, the structure inversion problem is ill-posed in the sense described by \citet{hadamard} and therefore difficult to solve, especially given the relatively limited data that are available for other stars. 
Consequently, structure inversions for internal properties such as the sound-speed profile have thus far been restricted to the Sun and other bodies within the solar system. 
In this paper, we present results of structure inversions performed to probe core structure in other stars. 
More specifically, we invert measured p-mode frequencies to deduce the squared isothermal sound speed (${u\equiv P/\rho}$, where $P$ is pressure and $\rho$ is density), in the cores of the two solar analogs 16~Cyg~A and 16~Cyg~B. 
We achieve this by introducing an algorithm that we call ``inversions for agreement'' that works with the available data.

Helioseismic inversions, i.e.~inversions for the Sun, have revealed that sound-speed profiles of solar-calibrated evolutionary models differ by only fractions of a percent from the actual structure of the Sun---a rare triumph of accuracy by astrophysical standards. 
Furthermore, even before all flavors of solar neutrinos could be detected, helioseismic inversions were instrumental in showing that the solar neutrino problem was external to solar modeling \citep[e.g.][]{1997MNRAS.289L...1A,1998PhLB..433....1B}.
Additionally, the importance of some physical processes in stellar physics have been revealed by helioseismic inversions as well. 
For example, by comparing solar models with and without diffusion and gravitational settling of helium and heavy elements, \citet{1993ApJ...403L..75C} showed that it is important to take these effects into account (see also Figure~20 of \citealt{2016lrsp...13....2b}), and it has now become common practice to include these processes when modeling other solar-like stars. 
Hence, structure inversions are useful for verifying and improving models both within stellar physics and beyond.

The stars we wish to study with structure inversions are pulsating solar-type stars observed by \emph{Kepler}. 
They are cool dwarf stars on the main sequence that pulsate in pure p-modes and show no signs of mode mixing \citep[for a review of solar-like oscillations, see, e.g.,][]{2013ARA&A..51..353C}. 
The precise measurement of pulsation frequencies in these and other similar stars has enabled estimates of their ages, masses, and radii to better than $15\%$, $4\%$, and $2\%$, respectively \citep{2015MNRAS.452.2127S, 2017ApJ...835..173S, 2016apj...830...31b, 2017EPJWC.16005003B, 2017apj...839..116a}. 
The solar-type stars belonging to the triple system of 16~Cygni are two of the most well-studied stars in this field. 
Though stellar models of these stars match the overall characteristics of the stars, such as their radii, luminosities, temperatures, and metallicities; an inspection of their mode frequencies reveals significant disagreements. 
Figure~\ref{fig:echelle} shows a comparison of mode frequencies between models \citep[][models \emph{GOE}]{2017ApJ...835..173S} and observations \citep{2015MNRAS.446.2959D} of 16~Cyg~A and B, with Sun-as-a-star data shown for reference. 
Clear differences can be seen between the mode frequencies of the evolutionary models and the measured mode frequencies of the stars.

\afterpage{
\begin{landscape}
\begin{figure}
    \centering
    \makebox[\linewidth][c]{
        \includegraphics[width=0.94\linewidth]{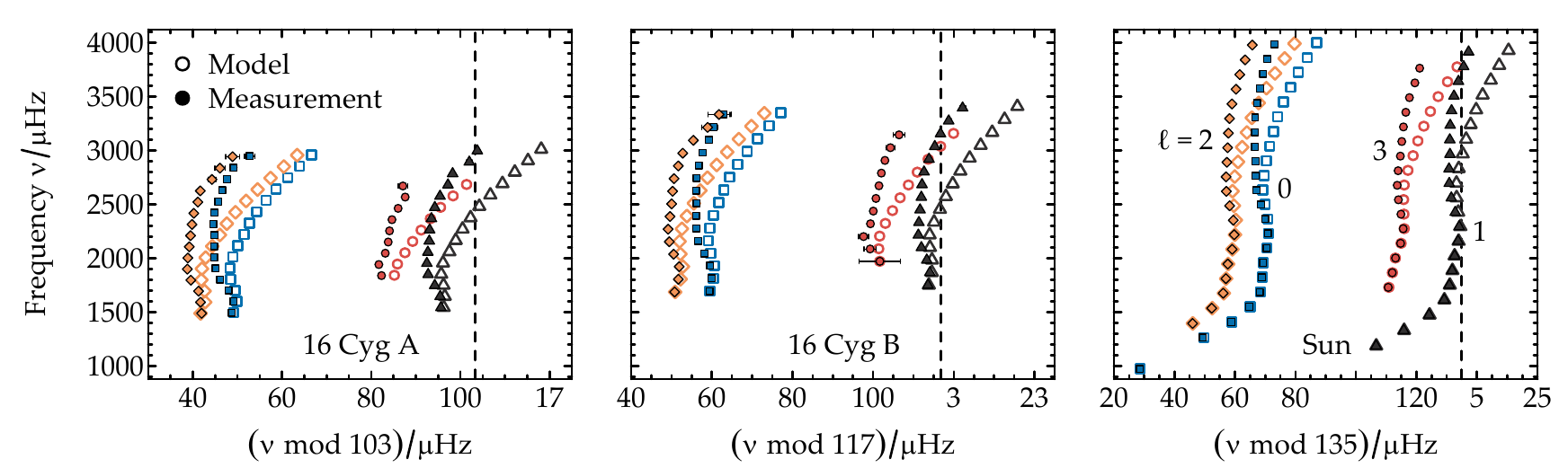}
    }
    \caption[\'Echelle diagrams for 16~Cygni and the Sun]{\'Echelle diagrams comparing \emph{GOE} evolutionary models of 16~Cyg~A (left) and B (center) to frequencies extracted from \emph{Kepler} data. 
    For reference, the right panel shows the solar model Model S \citep{1996Sci...272.1286C} in comparison with low-degree frequencies of the quiet Sun from BiSON data \citep[][]{2014MNRAS.439.2025D}. 
    The dashed line indicates the large frequency separation ($\Delta\nu$). 
    Open symbols are model frequencies and filled symbols are observed frequencies. 
    Spherical degrees $\ell$ are indicated with color and shape: 
    \textcolor{echelle-blue}{0} (blue squares), 
    \textcolor{echelle-black}{1} (black triangles), 
    \textcolor{echelle-yellow}{2} (yellow diamonds), and
    \textcolor{echelle-red}{3} (red circles). 
    Error bars show $1\sigma$ uncertainties, which in most cases are not visible. Model frequencies significantly differ from observed frequencies in nearly all cases. } 
    \label{fig:echelle} 
\end{figure} 
\end{landscape}
}

The most conspicuous difference between the oscillations of stars and stellar models is an offset that increases with frequency. This offset arises due to inadequacies in modeling the effects of convection in the near-surface layers \citep[see, e.g.,][]{1984srps.conf...11C} as well as neglected treatment of pulsation-convection interaction \citep{2017MNRAS.464L.124H}. 
These are collectively known as ``surface effects,'' and the offset they produce is usually called the ``surface term.'' 
For modes of low spherical degree $\ell$, the surface term is a function of frequency alone. 
There are a number of methods for correcting the disparities imposed by surface effects, such as those given by \citet{2008ApJ...683L.175K}, \citet[][hereinafter BG14]{2014A&A...568A.123B}, and \citet{2015A&A...583A.112S}. 
Each of these methods work by assuming that the frequency offset due to the surface term has a particular form that can be fitted to the frequency differences and subtracted off. 
Even after correction for the surface term, however, differences remain. 
Figure~\ref{fig:bg-corr} shows the remaining discrepancies between mode frequencies of models and observations of 16~Cygni after subtracting off the two-term ``BG14-2'' surface effect. 
More than half of the surface-term corrected mode frequencies still have significant differences with the observed values. 
Moreover, the disparities are most significant in the radial and dipole modes, which probe the deep interior of the star. 

\afterpage{
\begin{landscape}
\begin{figure}
    \centering
    \makebox[\linewidth][c]{%
        \includegraphics[width=0.94\linewidth]{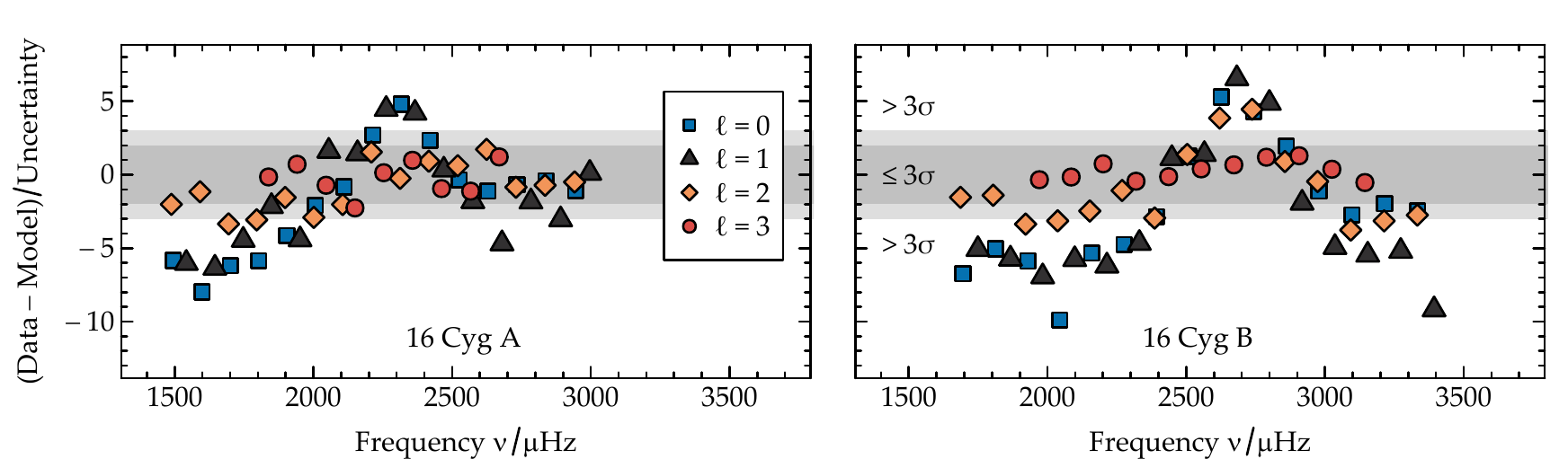}
    }
    \caption[Differences in oscillation mode frequencies between models and observations after correcting for surface effects]{Differences in oscillation mode frequencies between models and observations after correcting for surface effects. 
    Mode frequencies that lie outside of the shaded regions, demarcating the $2\sigma$ and $3\sigma$ boundaries, have significant differences that are caused by differences in internal structure. 
    \label{fig:bg-corr}} 
\end{figure} 
\end{landscape}
}

Since mode frequencies of models produced by stellar evolution codes have significant differences with respect to observations even after correction for the surface term, we pursue the use of inversion techniques to make more direct determinations of stellar structure.

\Needspace{3\baselineskip}
\subsection{The Inversion Problem}
Structure inversions can be posed as the problem of deducing small differences in structure between a star and a sufficiently close reference model by comparison of their mode frequencies. 
The basic problem is the same as the structure inversion problem for the Sun \citep[for reviews of solar structure inversions, see for example][]{Kosovichev1999, 2016lrsp...13....2b}. 
The dependence of mode frequencies on the radial structure of a star is nonlinear and involves unobservable displacement eigenfunctions. 
However, the oscillation equations are, to first order, a set of Hermitian eigenvalue equations \citep{1964ApJ...139..664C}, and hence they can be linearized around a known model using the variational principle. 
The linearization links the differences in frequencies between the reference model and the star to the differences in their internal structure. 
A byproduct of the linearization is the fact that the differences must be considered with respect to at least two stellar structure functions simultaneously, as variables such as the sound~speed~$c$ and density~$\rho$ are not independent but rather related through the equations of stellar structure. 
The equations resulting from the linearization can be written as 
\begin{equation} \label{eq:inversion}
    \mathscr{P} [\nu_i] = \int \mathbf{K}_i(r) \cdot \mathscr{P} [\boldsymbol{f}(r)] \; \text{d}r + \epsilon_i, \quad i \in \mathscr{M}
\end{equation}
where $\mathscr{M}$~is the set of observed modes, 
$\boldsymbol{\nu}$~are the oscillation frequencies of those modes, 
$\boldsymbol{f}$~contains two stellar structure functions (i.e., ${f_1(r)}$ and ${f_2(r)}$; e.g.~${c(r)}$ and ${\rho(r)}$), 
$r$~is the fractional stellar radius, and 
${\mathscr{P}}$~is a perturbation operator (in this case, the relative difference operator). 
Since measurements are uncertain, we include a term $\boldsymbol \epsilon$ for the differences between the true and the measured values. 
Each mode of oscillation $i$ has its own pair of kernels $\mathbf {K}_i$ that relate changes in $\mathbf f$ to changes in $\nu_i$. 
The kernels are derived from the perturbation analysis (see, e.g., \citealt{GoughThompson1991} or Sec.~6.2.~of \citealt{2016lrsp...13....2b} for details) and can be computed for a given reference model. 
Since the eigenproblem is Hermitian, perturbations to the oscillation mode eigenfrequencies do not depend to the first order on perturbations to the mode eigenfunctions. 
The inverse problem is thus to deduce $\mathbf f$ from the data $\boldsymbol \nu$, given that the kernels are known. 
There is no analytic solution to this problem and numerical methods must be employed. 
In practice, another term must also be added in order to account for the aforementioned surface effects. 
Although the technique makes use of a reference model, the results are independent; all stellar models within the linear regime produce essentially the same inference about the star \citep{2000ApJ...529.1084B}. 
We expand Equation~(\ref{eq:inversion}) explicitly in the next section.

Like many inverse problems, the structure inversion problem is ill-posed: the solutions are not unique, and they are also unstable with respect to small fluctuations in the oscillation data (see \citealt{GoughThompson1991} for a discussion). 
Solutions must therefore be regularized \citep[for a review of statistical regularization, see, e.g.,][]{tenorio2001statistical}. 
There are two popular ways of inverting Equation~(\ref{eq:inversion}): the Regularized Least Squares \citep[RLS;][]{tikhonov1977solutions} fitting method, which attempts to determine the stellar structure functions $\mathbf f$ that best fit to the observed data; and (2) the method of Optimally Localized Averages \citep[OLA;][]{1968GeoJ...16..169B}, which attempts to make linear combinations of the data that correspond to localized averages of one of the two components of $\mathbf f$. 
Both methods have been used extensively in the case of the Sun.
Details of how the inversions are implemented can be found in \citealt{2016lrsp...13....2b} and references therein.

\begin{figure}
    \centering%
        \adjustbox{trim={0.02\width} 0.005cm {0.04\width} 0.01cm, clip}{%
            \includegraphics[width=0.75\linewidth]{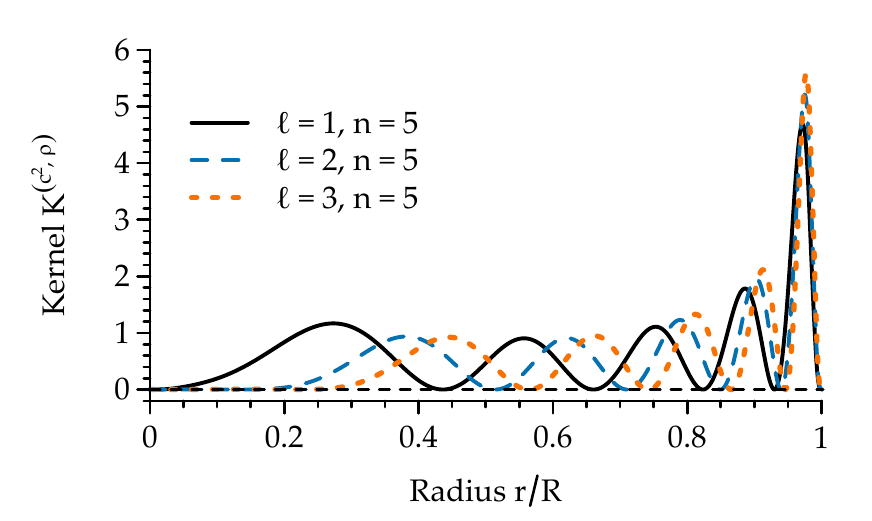}%
        }%
    \\%
        \adjustbox{trim={0.02\width} 0.005cm {0.04\width} 0.01cm, clip}{%
            \includegraphics[width=0.75\linewidth]{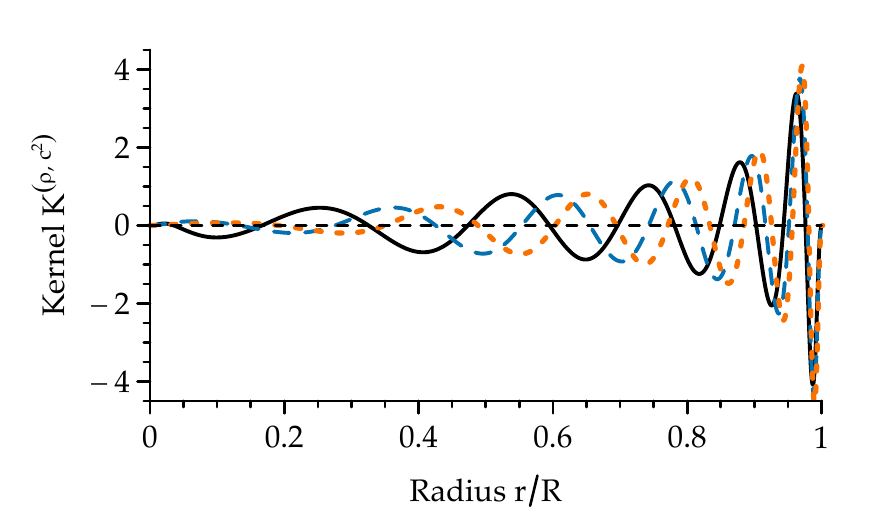}}%
    \caption[$(c^2,\rho)$ kernels for 16~Cyg~A]{Kernels for the squared adiabatic sound speed and density, $K^{(c^2, \rho)}$ (top), and the reverse, $K^{(\rho, c^2)}$ (bottom), as a function of fractional radius for oscillation modes of model \emph{GOE} of 16~Cyg~A. Kernels are shown for modes with the same radial order $n$ but different spherical degree $\ell$ (see the legend). 
    \label{fig:c2-rho} }
\end{figure}

In helioseismic investigations, the most common choice of $\mathbf f$ is the combination of squared adiabatic sound speed $c^2$ and density $\rho$.
The kernels for this pair are shown in Figure~\ref{fig:c2-rho}. 
The basic ingredients of helioseismic inversion are the thousands of precisely measured solar mode frequencies whose spherical degrees range up to ${\ell \simeq 200}$ or higher. 
Reference models have the same mass, radius, and age as the Sun. 
Inversion of helioseismic data yields inferences of solar structure throughout most of the solar interior \citep[see, e.g.,][]{2009ApJ...699.1403B}.

There are two major difficulties in trying to invert for the structure of other stars.
The first difficulty is the lack of data. 
Even for the best solar-type targets, only about $55$ mode frequencies have been able to be measured. 
Furthermore, due to cancellation effects, we only get data for low-degree modes, usually of degree ${\ell=0--2}$ and sometimes $3$. 
This limits the regions in the star that we are able to probe, the inversion techniques that we are able to employ, and the pair of stellar structure functions that we are able to use. 
Second, when compared with the Sun, masses and radii of stars are not known with the same precision. 
This is problematic because differences in mass and radius between the reference model and the proxy star cause systematic errors in the inversion results \citep[see][]{2003Ap&SS.284..153B}. 
Most of the time, these quantities are not known independently and need to be determined from the same set of data. 
Even where independent estimates are available, such as radii from interferometric measurements, the uncertainties are non-negligible. 
Both the amount of data and the precision to which the stellar mass and radius are known cause difficulties in inversion of asteroseismic data, and therefore the inversion methods need to be modified.

\Needspace{3\baselineskip}
\subsection{Asteroseismic Inversions} 
\label{sec:asteroseismic-inversions}

Even before {\it CoRoT} and {\it Kepler} detected oscillations in a large number of stars, there were a number of studies that investigated the possibility of inverting asteroseismic p-mode oscillations to determine the core structures of solar-like stars \citep{1993ASPC...40..541G,
1998mons.proc...33G, 2001ESASP.464..411B, 2001ESASP.464..407B,
2002ESASP.485..249B, 2003Ap&SS.284..153B}. 
Additionally, there was at least one inconclusive study that tried to perform an inversion of seismic data from Procyon~A \citep{2004ESASP.559..186D}. 
The theoretical investigations of structure inversions all used mode sets and data uncertainties that were expected to be available from future missions to determine how well the structure differences between the cores of pairs of models could be determined. 
Unfortunately, the assumptions about the available mode sets and uncertainties were rather optimistic when compared with data available today.

\Needspace{3\baselineskip}
\subsubsection*{Mode Set}
\begin{figure}
    \centering
    \makebox[\linewidth][c]{%
    \includegraphics[width=0.75\linewidth]{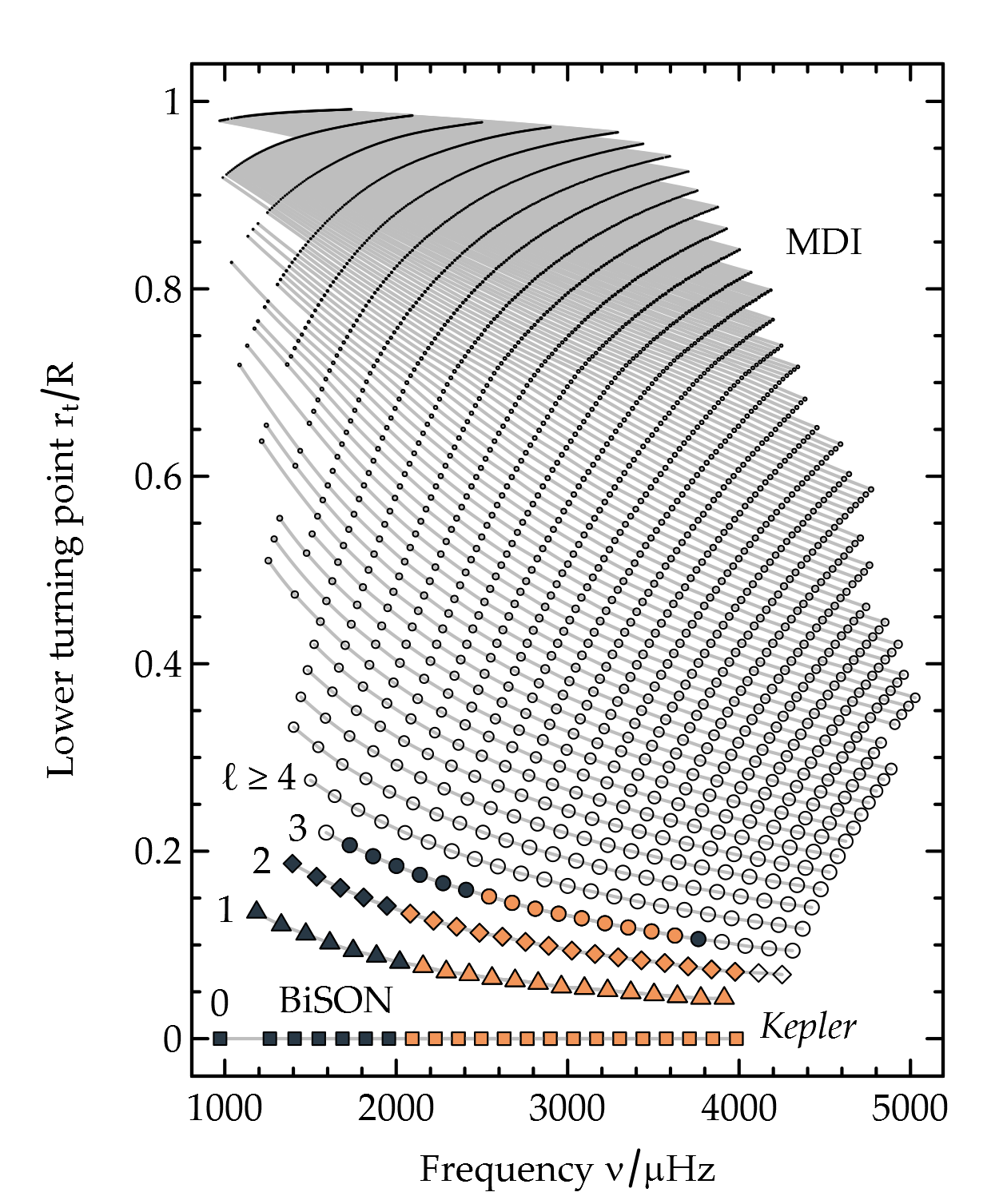}
    }
    \caption[Lower turning points of a solar model]{Lower turning points as a function of frequency for oscillation modes of a solar model with the MDI mode set (all points), BiSON mode set (all filled points) and the 16~Cyg~A mode set \mb{from \emph{Kepler}} (\textcolor{turn-orange}{orange} filled points). 
    Modes of the same spherical degree are connected by lines, with modes of spherical degree ${\ell = 0}$, $1$, $2$, and $3$ shown with squares, triangles, diamonds, and circles, respectively. Compared to the Sun, asteroseismology of solar-like oscillators is restricted to low-degree, high-frequency modes. 
    \label{fig:turning-points}} 
\end{figure}

The limited mode set available for stars other than the Sun makes the inversion problem more difficult. 
The fact that we cannot make resolved-disk observations of other stars generally restricts the detection of modes to ${\ell \le 3}$. 
The lower turning points of these modes are within the stellar core; consequently, lacking more shallowly trapped modes, we will be unable to resolve the details of the stellar envelope. 
Figure~\ref{fig:turning-points} illustrates this difficulty by comparing the propagation cavities of oscillation modes with different degrees from a solar model. 
The figure shows lower turning points for low-degree Sun-as-a-star modes obtained by the Birmingham Solar Oscillation Network \citep[BiSON;][]{2014MNRAS.439.2025D} and the ${\ell > 3}$ modes obtained by the Michaelson Doppler Imager (MDI) mission on board the Solar and Heliospheric Observatory \citep[SOHO,][]{1997SoPh..175..287R}. 
The figure further shows the mode set that would be available if the Sun were a star in the {\it Kepler} field. 
Such a restricted mode set eliminates the possibility of using an inversion technique, such as RLS, that requires simultaneous determination of $f_1$ and $f_2$ over as large a part of the star as possible. 
Instead, we are confined to investigations of the stellar core.

\begin{figure}
    \centering%
        \adjustbox{trim={0.02\width} 0.005cm {0.04\width} 0.01cm, clip}{%
            \includegraphics[width=0.75\linewidth]{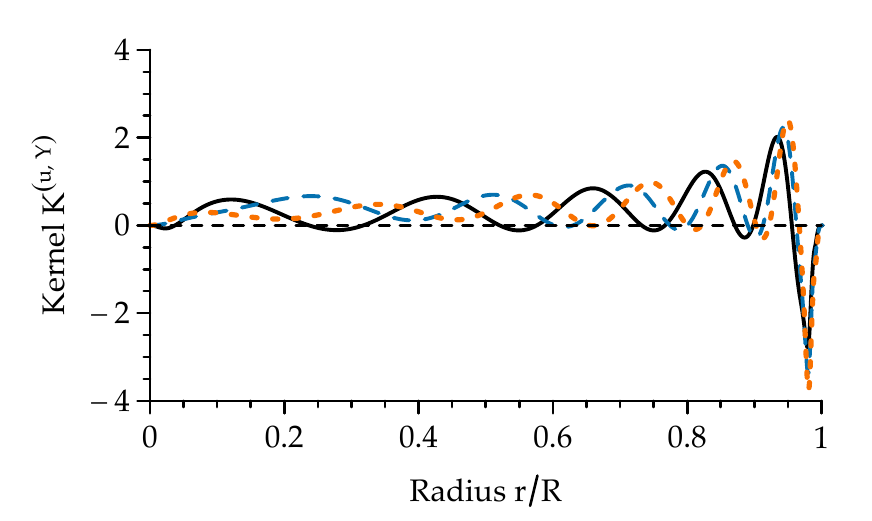}%
        }%
    \\%
        \adjustbox{trim={0.02\width} 0.005cm {0.04\width} 0.01cm, clip}{%
            \includegraphics[width=0.75\linewidth]{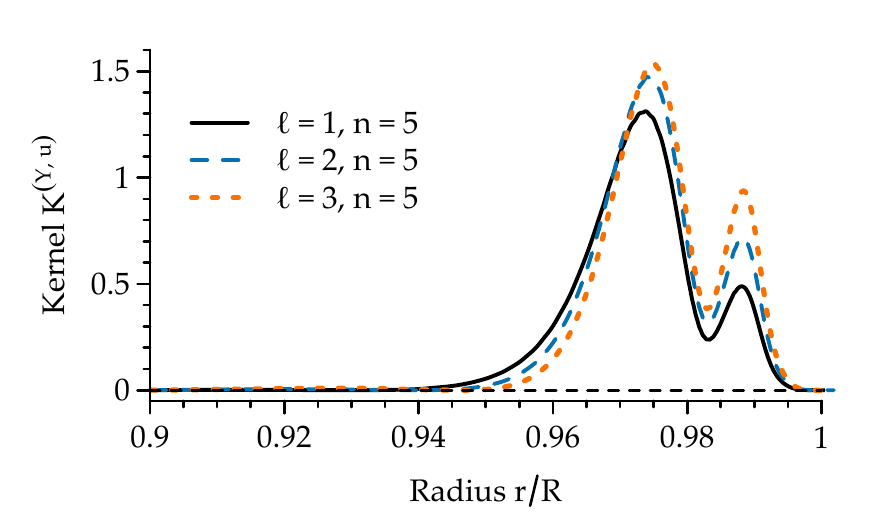}}%
    \caption[${(u,Y)}$ kernels for 16~Cyg~A]{Kernels for the squared isothermal sound speed and helium abundance, $K^{(u, Y)}$ (top), and the reverse, $K^{(Y, u)}$ (bottom), as a function of fractional radius for oscillation modes of model \emph{GOE} of 16~Cyg~A. Notice that in contrast to the $K^{(\rho, c^2)}$ kernels shown in Figure~\ref{fig:c2-rho}, the $K^{(Y, u)}$ kernels have very small values (${0 < K(r) < 0.01}$) in the interior ${r < 0.9\; R}$. \label{fig:same-n-uY} }
\end{figure}

Inversions using the OLA method or its variants are most suited for asteroseismic inversions, since OLA allows inversions over a small part of the star. 
\citet{2003Ap&SS.284..153B} showed that instead of the ${(c^2,\rho)}$ pair of variables used in solar inversions, the ${(u, Y)}$ pair is better suited for asteroseismic structure inversions\mb{, where $Y$ is the fractional helium abundance}. 
This is because the kernels for $Y$ are nonzero only in the helium ionization zone, as shown in Figure~\ref{fig:same-n-uY}. 
Thus from the point of view of Equation~(\ref{eq:inversion}) the data, i.e., the frequency differences, are almost completely determined by differences in $u$, thereby making $u$ easier to determine. 
However, in order to derive the kernels for the ${(u, Y)}$ pair, we have to assume that the EOS of the star is the same as that of the reference model \citep{1990MNRAS.244..542D, Kosovichev1999, ThompsonJCD2002}. 
In other words, we are artificially adding information to the system. 
\citet{1997A&A...322L...5B} have shown that in the case of the Sun, this results in systematic errors in the inversion result; 
however, for other stars, we expect the errors caused by data uncertainties to be much larger than the systematic errors caused by an incorrect EOS. 
Thus, we proceed with this pair of variables. 

\Needspace{3\baselineskip}
\subsubsection*{Mass and Radius}
The reduced precision of mass $M$ and radius $R$ estimates for stars other than the Sun also makes the problem more difficult. 
Frequencies scale as the square root of mean density, i.e., ${\nu^2 \propto M/R^3}$, so an unaccounted for difference in $M$ and $R$ between the star and the reference model gives rise to additional systematic errors in the inversion result. 
As these errors are proportional to the uncertainties in $M$ and $R$, they are much larger than those expected from an incorrect EOS. 
Solar inversions as well as trial inversions for stellar models have hitherto been performed under the assumption that the mass and radius of the star are known. 
Having imprecise estimates of the stellar mass and radius means that the mass and radius of the reference model are likely to differ from those of the star. 
\citet{2001ESASP.464..411B} accounted for this effect in their tests of asteroseismic inversions with pairs of models by adding terms for $\delta M$ and $\delta R$ to the inversion procedure. 
However, they assumed $\delta M$ and $\delta R$ to be known exactly, and the impact of uncertainties was not explored in that work.

Another difficulty arises from the fact that the inversion equation and the kernels are usually derived using dimensionless units, with the relative differences in $f_1$ and $f_2$ being calculated at constant fractional radii.
This raises complications alluded to earlier: the $u$ inversion result itself is also systematically offset by the differences in mass and radius \citep{2003Ap&SS.284..153B}.
In short, since kernels are derived using dimensionless variables, instead of a dimensional $u$, we actually have ${u' \equiv P'/\rho'}$, where $'$ denotes a dimensionless variable. 
It is straightforward to see from the equation governing conservation of mass that ${\rho \propto M/R^3}$. 
Likewise, from the equation of hydrostatic support one finds that ${P \propto M^2/R^4}$. 
Hence ${u'=uR/M}$, and so an inversion whose reference model has a different $M$ or $R$ will result in a $u$ profile that differs by
\begin{equation} \label{eq:MR}
    \dun - \du = \dRR - \dMM.
\end{equation}
Thus the inversion procedure must be modified in order to accommodate the reduced precision of mass and radius estimates. 

\vspace{1cm}

These difficulties---limited mode sets and the uncertainties in stellar mass and radius estimates---have so far prevented structure inversions from widespread application in other stars. 
In this paper, we propose a way to circumvent the systematic error that results from the reference model having an incorrect mass and radius by extending the inversion procedure to use multiple reference models spanning the uncertainties in mass and radius. 
Furthermore, we introduce a new algorithm for the automated determination of inversion parameters. 
To put it concisely, this algorithm works by selecting the inversion parameters that maximize the agreement in the inversion result from different reference models. 
We apply this technique to the areas where the limited set of observed asteroseismic modes have resolving power, i.e., in the interior $30\%$ of the star. 
We first demonstrate the efficacy of the algorithm by inverting the frequency differences between known models to determine that we are capable of producing the correct result. 
We then apply the method to the solar-type components of the 16~Cyg system with data obtained from the \emph{Kepler} mission.

\Needspace{3\baselineskip}
\section{Methods}
We seek to measure the difference in internal structure between stars and their best-fitting evolutionary models, which we assume to be sufficiently close in structure such that linear perturbation theory applies. 
We begin by explicitly expanding Equation~(\ref{eq:inversion}) using the ${(u', Y)}$ kernel pair. 
Given a set of $\mathscr{M}$ pulsation modes whose frequencies $\boldsymbol\nu$ have been measured, e.g.\ $$\mathscr{M}=\left\{(\ell=0, n=10), (\ell=1, n=12), \ldots\right\}$$ for each mode of oscillation ${i\in\mathscr{M}}$ we have an equation relating a frequency perturbation to perturbations in stellar structure: 
\ifhbonecolumn
\begin{equation} \label{eq:forward2} 
         \frac{\delta\nu_i'}{\nu_i'} 
         = 
          \int \KuY(r) \cdot \dun(r) \; \text{d}r 
         +\int \KYu(r) \cdot \dY(r)  \; \text{d}r 
         +\frac{F_{\text{surf}}(\nu_i')}{\nu_i'\cdot I_i} 
         +\epsilon_i. 
\end{equation} 
\else
\begin{align} \label{eq:forward2} 
         \frac{\delta\nu_i'}{\nu_i'} 
         = 
         &\int \KuY(r) \cdot \dun(r) \; \text{d}r 
\notag\\+&\int \KYu(r) \cdot \dY(r)  \; \text{d}r 
         +\frac{F_{\text{surf}}(\nu_i')}{\nu_i'\cdot I_i} 
         +\epsilon_i. 
\end{align} 
\fi
Here $\delta \nu'$ is the difference in dimensionless oscillation mode frequency in the sense of (model - star), 
$\delta u'(r)$ is the difference in the dimensionless squared isothermal sound speed between a given stellar model and the star at fractional radius $r$, 
and $\dY(r)$ is the difference in the helium abundance. 
We assume the unknown differences between the true and the measured frequencies $\boldsymbol\epsilon$ to be independent and normally distributed with zero mean and known standard deviations $\boldsymbol{\sigma}$. 
The kernel functions $\boldsymbol{K}^{(u', Y)}$ and $\boldsymbol{K}^{(Y, u')}$ are known functions of the reference model and serve to relate changes in $u'$ and $Y$ to changes in oscillation mode frequencies. 
Finally, $F_{\text{surf}}$ is a surface term that depends on frequency and is normalized by mode inertiae $\boldsymbol{I}$. Here we use the BG14-2 surface term, which \citet{2015ApJ...808..123S} showed to be a good choice. This relation has 
\begin{equation}
    F_{\text{surf}}(\nu'; \nu'_{\text{ac}}, \mathbf a) = 
           a_1 \left( \frac{\nu'}{\nu_{\text{ac}}'} \right)^{-1} 
         + a_2 \left( \frac{\nu'}{\nu_{\text{ac}}'} \right)^3 
\end{equation}
where $\mathbf a$ are coefficients that must be estimated during the inversion procedure and $\nu_{\text{ac}}'$ is the dimensionless acoustic frequency cut-off, which, under assumption of ideal gas, can be approximated \mb{by scaling from solar values with \citep{1991ApJ...368..599B}} 
\begin{equation}
    \nu_{\text{ac}}' 
    = 
    \nu_{\text{ac},\odot} \cdot
    \frac{g}{g_{\odot}} 
    \left( 
        \frac{T_{\text{eff}}}{T_{\text{eff},\odot}} 
    \right)^{-1/2} 
    \left( 
        \frac{R^3}{G M} 
    \right)^{1/2}
\end{equation}
with $g$ being the surface gravity of the reference model, $T_{\text{eff}}$ its effective temperature, $G$ the gravitational constant, \mb{and quantities subscripted with $\odot$ indicating the solar value.} 
The next step is to invert Equation~(\ref{eq:forward2}) to infer ${\delta u'/u'(r)}$, for which we will use the OLA technique. 

\Needspace{5\baselineskip}
\subsection{Optimally Localized Averages}
We invert Equation~(\ref{eq:forward2}) using the OLA method.
If, for the sake of argument, the ${(u',Y)}$ kernel function of an oscillation mode were a $\delta$ function located at $r_0$ and zero elsewhere, and also if the ${(Y,u')}$ kernel were zero everywhere, then a departure in frequency of this mode from the observed value would demand that ${u'(r_0)}$ differs between model and star. 
According to Equation~(\ref{eq:forward2}), the relative difference in ${u'(r_0)}$ between the model and the star would be proportional to the relative difference in that mode's frequency. The OLA inversion technique works based on this concept. 

OLA combines the kernels of the observed modes into an \emph{averaging~kernel}~$\mathscr{K}$ resembling a localized function that is peaked at a chosen target radius inside the star. 
This is done via a linear combination of Equation~(\ref{eq:forward2}) over the observed modes, where each mode ${i \in \mathscr{M}}$ is weighted by a coefficient $c_i$. 
If a vector of coefficients $\mathbf c$ exists such that an averaging kernel with the desired properties can be formed, the inversion result, i.e., the relative difference in $u'$ between the model and the star, is then given by that same combination of the data. 
The process that creates the averaging kernel for $u'$ also combines the kernels of $Y$ to create a \emph{cross-term~kernel},~$\mathscr{C}$, and a reliable inversion result depends on $\mathscr{C}$ being as small as possible. 
Under these conditions, and assuming the surface term has been removed, the inversion result corresponds to an average of the underlying true difference weighted by the averaging kernel, i.e.,
\begin{equation}
    \left\langle\dun\right\rangle(r_0)
    =
    \int{\mathscr{K}}(r,r_0)\cdot \dun(r) \; \text{d}r
\end{equation}
assuming that ${\int{\mathscr{K}}\;\text{d}r=1}$.
Of course, the influence of data uncertainties must  be controlled as well.

More formally, for a given target radius $r_0$, the OLA procedure aims to construct an averaging kernel $\mathscr{K}(r)$ that is well-localized around ${r=r_0}$. 
Recalling Equation~(\ref{eq:forward2}), OLA proceeds by constructing a linear combination over all the observed modes: 
\begin{align} \label{eq:OLA}
    \sum_{i \in \mathscr{M}} c_i(r_0) \frac{\delta\nu_i'}{\nu_i'}
    =
    &\int \mathscr{K}(r; r_0, \mathbf c) \cdot \dun(r) \; \text{d}r \notag
\\ +&\int \mathscr{C}(r; r_0, \mathbf c) \cdot \dY(r) \; \text{d}r \notag
\\ +&\sum_{i \in \mathscr{M}} c_i(r_0) \cdot F_{\text{surf}}(\nu_i'; \nu'_{\text{ac}}, \mathbf a)/\left(\nu_i' \cdot I_i \right) \notag
\\ +&\sum_{i \in \mathscr{M}} c_i(r_0) \cdot \epsilon_i
\end{align}
where the vector $\mathbf c$ are inversion coefficients that will need to be determined for each given $r_0$ and
\begin{align}
    \mathscr{K}(r; r_0, \mathbf c) &= \sum_{i \in \mathscr{M}} c_i(r_0) \cdot K_i^{(u, Y)}(r)
\\  \mathscr{C}(r; r_0, \mathbf c) &= \sum_{i \in \mathscr{M}} c_i(r_0) \cdot K_i^{(Y, u)}(r)
\end{align}
subject to the constraint that 
\begin{equation} \label{eq:sum-to-one}
    \int \mathscr{K}(r; r_0) \; \text{d}r = 1.
\end{equation}
Provided that the averaging kernel is well-localized at the target radius and the cross-term kernel, the surface-term contributions, and the combined data uncertainties are all small; this combination of relative frequency differences gives a localized average of ${\delta u'/u'}$ at the target radius $r_0$:
\begin{equation} \label{eq:local-avg}
    \left\langle \dun \right\rangle (r_0) = \sum_{i \in \mathscr{M}} \left( c_i(r_0) \cdot \frac{\delta\nu_i'}{\nu_i'} \right). 
\end{equation}
Here we have chosen to express relative differences in the sense
\begin{equation} \label{eq:rel-diff}
    \frac{\delta q}{q} = \frac{(\text{model} - \text{star})}{\text{model}} = \frac{ (q_{\text{ref}} - q_{\text{star}}) }{q_{\text{ref}}}
\end{equation} 
where $q$ can refer to any quantity. 
Thus, Equation~(\ref{eq:local-avg}) can be redimensionalized using Equation~(\ref{eq:MR}) to infer $u_{\text{star}}$ with
\begin{equation} \label{eq:dimensional}
     u_{\text{star}}(r) = \left( 1 - \dun(r) + \dRR - \dMM \right) \cdot u_{\text{ref}}(r).
\end{equation}
We now turn our attention to determining the coefficients $\mathbf c$ that make this estimate possible.

\Needspace{3\baselineskip}
\subsection{Inversion Coefficients Using Subtractive OLA}
The optimal inversion coefficients $\boldsymbol {\hat c}$ must strike a balance between forming a well-localized averaging kernel and forming a small cross-term kernel, while still having small uncertainty. 
In Subtractive OLA \citep[SOLA,][]{1992A&A...262L..33P, 1994A&A...281..231P}, the averaging kernel is formed according to a specified well-localized form (the ``target~kernel''), and the coefficients $\mathbf c$ are determined by minimizing the difference between the averaging kernel obtained and the target kernel. 
This is a fast implementation of the OLA method. 
It comes at the price of a free parameter in the form of the properties of the target kernel. 
SOLA determines optimal coefficients $\mathbf{\hat{c}}$ for a given target radius $r_0$ by solving the optimization problem 
\begin{align} \label{eq:opt}
        \boldsymbol {\hat c}(r_0; \beta, \mu, \Delta)
        =
        {}&\underset{\mathbf{c}}{\arg\min} \; \Bigg\{
            \mathscr{F}(\mathbf c; r_0, \Delta) 
          + \beta \int \mathscr{C}(r; r_0, \mathbf c)^2 \; \text{d}r 
          + \mu \sum_{i\in\mathscr{M}} \left( c_i^2 \cdot \sigma_i^2 \right)
         \Bigg\} 
         \notag\\\text{subject~to} & \; 
        \int{\mathscr{K}(r; r_0, \mathbf{c})} \; \text{d}r = 1
\quad\text{ and }\quad  \sum_{i\in\mathscr{M}} c_i \cdot \frac{F_{\text{surf}}(\nu'_i; \nu_{\text{ac}})}{\nu'_i \cdot I_i} = 0.
\end{align}
Here $\beta$ and $\mu$ are parameters that must be chosen to penalize the amplitude of the cross-term kernel and the effect of data uncertainties, respectively. 
A third parameter, $\Delta$, gives the width of the target kernel(s). 
The function $\mathscr{F}$ penalizes deviations of the averaging kernel from the target~kernel~$T$ and can be calculated as
\begin{equation}
        \mathscr{F}(\mathbf c; r_0, \Delta)
        =  
        \int \left[ \mathscr{K}(r; r_0, \mathbf c) - T(r; r_0, \Delta) \right]^2 \; \text{d}r. 
\end{equation}
The functional form of $T$ can be chosen, e.g.~as a modified Gaussian that decays to zero at ${r=0}$ but remains peaked at ${r=r_0}$ \citep[e.g.][]{1999MNRAS.309...35R} with 
\begin{align}
    T(r; r_0, \Delta) &= A\cdot r\cdot \exp\left\{-\mathcal{G}(r; r_0, \Delta)^2\right\} \\
    \mathcal{G}(r; r_0, \Delta) &= \frac{r-r_0}{D(r_0, \Delta)} + \frac{D(r_0, \Delta)}{2 r_0}.
\end{align}
The normalization factor $A$ is chosen to ensure ${\int T \; \text{d}r = 1}$. 
Since the resolution ultimately depends on the internal sound speed $c_s$ \citep[][]{1993ASPC...42..141T}, the function $D$ gives the width of the kernels according to variations in $c_s$ and a free parameter $\Delta$ that describes a fiducial width as 
\begin{align}
    D(r_0, \Delta) &= \Delta \cdot \frac{c_s(r_0)}{c_s(r_f)} 
\end{align}
with $r_f$ being an arbitrary reference point (e.g. we choose ${r_f=0.2}$, although the result is rather insensitive to the choice). 
We note that other choices of $\mathscr{F}$, $T$, $\mathcal{G}$, and $D$ are possible \citep[see, e.g.,][]{1985SoPh..100...65G, 1989ApJ...343..526B}, but they will not be explored here. 

The SOLA inversion problem can be cast into a system of linear equations with the constraints enforced using Lagrange multipliers. 
Given choices of $\beta$, $\mu$, and $\Delta$, Equation~(\ref{eq:opt}) can be solved via matrix inversion, the details of which can be found, for example, in Chapter~10 of \citealt{basuchaplin2017}. 
See \citet{1999MNRAS.309...35R} for a description of how inversion parameters are usually selected in helioseismology. 
Depending on the data that are available, it may be possible to form zero, one, or more well-localized averaging kernels with correspondingly small cross-term kernels and well-controlled uncertainties at different locations in the stellar interior.

\Needspace{3\baselineskip}
\subsection{Selecting Inversion Parameters with Multiple Reference Models (``Inversions for Agreement'')} 
\label{sec:inversion-for-agreement}
It is not clear \emph{a priori} which inversion parameters should be chosen, nor is there a reliable algorithm for their selection. 
Here we propose an algorithm for selecting inversion parameters based on the following information. 
First, besides the effects that stem from differences in $M$ and $R$, inversion results do not otherwise depend on the choice of reference model: with proper selection of inversion parameters, a wide range of reference models are capable of producing the correct inference \citep{2000ApJ...529.1084B}. 
\mb{Furthermore, for a given mode set, and setting aside the surface term, the values of the mode frequencies themselves do not play a role in determining the averaging and cross-term kernels.} 
Thus, provided the differences in the kernels between models are small, the same inversion parameters can be used for different models. 
Instead of performing single-model inversions, we invert using an array of reference models that span the uncertainties in $M$ and $R$. 
We simultaneously estimate the inversion parameters and the stellar $M$ and $R$ such that the inferred stellar $u$ profile from the different models are in agreement. 
We achieve this via repeated iterative optimization with random noise realizations. 
We constrain $M$ and $R$ with normal priors based on past studies, and set uniform priors on the inversion parameters. 
\mb{We have also tried this procedure with each reference model having its own individual set of inversion parameters ($\beta, \mu, \Delta$) to optimize, and we found that it did not have a substantial impact on the results. }

We generate an array of nine reference models that are calibrated to span the $1\sigma$ uncertainties in mass and radius for each star whose interior structure we seek to infer. 
We optimize a vector of five inversion parameters 
${\boldsymbol\alpha = (\beta, \mu, \Delta, M_{\text{star}}, R_{\text{star}})}$ 
which are shared among the nine models. 
We take an average among their inferred values of $u_{\text{star}}$, and finally we choose the $\boldsymbol \alpha$ that minimizes the variance of this average, weighted by the priors on $M_{\text{star}}$ and $R_{\text{star}}$. 
Formally, we postulate that the optimal inversion parameters $\boldsymbol{\hat{\alpha}}$ across all of the reference models is 
\begin{align} \label{eq:invert-for-agree}
    \boldsymbol{\hat{\alpha}}
    =
    \underset{\boldsymbol{\alpha}}{\arg\min} \Bigg\{ 
    &\sum_{r_j\in \boldsymbol{r_0}} 
            \log \text{Var} \left[ 
                \tilde u\left(r_j; \boldsymbol {\alpha} \right) 
            \right] - \log \Psi(\boldsymbol\alpha)
    \Bigg\}
\end{align}
where 
$\text{Var}$ is the variance operator, 
$\mathbf{r_0}$ are the target radii, 
and $\tilde u$ is a vector whose $k$th element 
${u_k(r_0; \boldsymbol \alpha)}$ gives the inferred value of $u_{\text{star}}$ at target radius $r_0$ via the $k$th reference model using the inversion parameters $\boldsymbol \alpha$
(\emph{cf}.~Equations.~\ref{eq:local-avg}-\ref{eq:opt}). 
Finally, $\Psi$ is the prior distribution, which in this case has
\begin{equation}
    \Psi(\boldsymbol \alpha)
    =
    \psi\left(M_{\text{star}}; \mu_M, \sigma^2_M\right) \cdot
    \psi\left(R_{\text{star}}; \mu_R, \sigma^2_R\right)
\end{equation}
with $\psi$ being the normal density function and 
$\mu_x$ and $\sigma_x$ being the mean and standard deviation of $x$. 
In each iteration of the algorithm, each of the non- and redimensionalizations are performed with the current estimate of $M_{\text{star}}$ and $R_{\text{star}}$. 
For example,
\ifhbonecolumn
\begin{equation}
    \frac{\delta \nu'}{\nu'}
    =
    \left[
        \left(
            \frac{R_{\text{ref}}^{3/2}}{M_{\text{ref}}^{1/2}}
        \right) 
        \nu_{\text{ref}}
        -
        \left(
            \frac{R_{\text{star}}^{3/2}}{M_{\text{star}}^{1/2}}
        \right)
        \nu_{\text{star}}
    \right]
    / 
    \left[
        \left(
            \frac{R_{\text{ref}}^{3/2}}{M_{\text{ref}}^{1/2}}
        \right)
        \nu_{\text{ref}}
    \right]. 
\end{equation} 
\else
\begin{gather}
    \frac{\delta \nu'}{\nu'}
    =
    \left[
        \left(
            \frac{R_{\text{ref}}^{3/2}}{M_{\text{ref}}^{1/2}}
        \right) 
        \nu_{\text{ref}}
        -
        \left(
            \frac{R_{\text{star}}^{3/2}}{M_{\text{star}}^{1/2}}
        \right)
        \nu_{\text{star}}
    \right]
    / \notag \\
    \left[
        \left(
            \frac{R_{\text{ref}}^{3/2}}{M_{\text{ref}}^{1/2}}
        \right)
        \nu_{\text{ref}}
    \right]. 
\end{gather} 
\fi
In summary, Equation~(\ref{eq:invert-for-agree}) says that the optimal inversion parameters are the ones that give the same inference of $u_{\text{star}}$ across all the reference models.

Since the inversion results depend on uncertain measurements, we perform repeated trials with random realizations of noise. 
Specifically, in each trial, we perturb each frequency $\nu$ with normal noise according its uncertainty $\sigma_{\nu}$, and the mass and radius estimates $\mu_M$ and $\mu_R$ via their uncertainties $\sigma_M$ and $\sigma_R$. 
We then use the 
\mycitet{nelder1965simplex}
downhill simplex method to numerically search for the parameters that satisfy Equation~(\ref{eq:invert-for-agree}) for that realization of noise. 
Because each inversion parameter is strictly non-negative and can potentially take on a large range of values, we optimize $\log \boldsymbol \alpha$. 
We stop each trial after either the relative change in the objective function is reduced by less than the square root of the machine precision for double precision floating point numbers (${\sim 10^{-8}}$), or a maximum number of $512$ iterations is reached. 
In the majority of cases, the former condition is met. 
We perform $128$ trials and report the averaged results. 
Finally, we visually inspect the resulting averaging kernels and cross-term kernels to ensure that the averaging kernels are well-localized at the target radii and that the cross-term kernels have small amplitude everywhere.

\Needspace{3\baselineskip}
\section{Results} 
\subsection{Tests on Models} 
In order to validate our technique, we first apply the method to known models; this allows us to check that the procedure does indeed produce the correct result.
Specifically, we determine whether or not we can accurately recover the internal $u$ profiles of the \emph{GOE} models of 16~Cyg~A and B using an array of different reference models as reference.

For the test, we generate an array of reference models for each star by calibrating models to their estimated masses (${\pm 1 \sigma}$, \citealt{2016apj...830...31b}), radii (${\pm 1 \sigma}$, \citealt{2013MNRAS.433.1262W}), ages \citep{2016apj...830...31b}, luminosities \citep{2013MNRAS.433.1262W}, and metallicities \citep{2009A&A...508L..17R}. 
The estimates we use for these stars are given in Table~\ref{tab:stellar-parameters}. 
\mb{We calculate the models using the given mean values of their ages, luminosities, and metallicities.} 
We construct the models using the MESA stellar evolution code \citep[\emph{Modules for Experiments in Stellar Astrophysics},][]{2011apjs..192....3p}. 
For each model, we use ADIPLS \citep[\emph{the Aarhus adiabatic oscillation package},][]{2008Ap&SS.316..113C} to calculate the adiabatic oscillation mode frequencies corresponding to the $54$ and $56$ oscillation modes that have been identified in 16~Cyg~A and B, respectively. 
We use the same treatments of evolution and pulsation that are described in Section~2.1 of \citealt{2016apj...830...31b}. 
None of the reference models have exactly the same mass or radius as the two \emph{GOE} models that we are treating as our proxy stars. 
We perturb the proxy star frequencies with noise prior to beginning the procedure.

\begin{table}
\caption{Fundamental parameters of 16~Cyg~A and B. \label{tab:stellar-parameters}}
\hspace*{-0.4cm}
\begin{tabular}{l|ccccccccc} 
    Name &
    Mass &
    Radius &
    Age &
    Luminosity &
    Metallicity &
    \\ \hline
    &
    $\mu_M \pm \sigma_M$ &
    $\mu_R \pm \sigma_R$ &
    $\tau$ &
    $L$ &
    $[$Fe$/$H$]$ &
    \\
    &
    $[\text{M}_\odot]$ &
    $[\text{R}_\odot]$ &
    $[\text{Gyr}]$ &
    $[\text{L}_\odot]$ &
    (dex) \\ \hline
    16~Cyg~A & 1.080 $\pm$ 0.016 & 1.22 $\pm$ 0.02 & 6.90 $\pm$ 0.40 & 1.56 $\pm$ 0.05 & 0.096 $\pm$ 0.026 \\ 
    16~Cyg~B & 1.030 $\pm$ 0.015 & 1.12 $\pm$ 0.02 & 6.80 $\pm$ 0.28 & 1.27 $\pm$ 0.04 & 0.052 $\pm$ 0.021 \\ \hline
\end{tabular}
\end{table}

We apply the inversion-for-agreement procedure described in Section~\ref{sec:inversion-for-agreement}. 
The results are shown in Figure~\ref{fig:model-test}. 
The procedure gets the correct result. 
The uncertainties in ${\delta u/u}$ are given by the average over the $128$ trials. 
The ``uncertainties'' in fractional radius ${r/R}$ are a measure of the resolution of the inversion and are given by the width at half maximum of an average over the averaging kernels of the different trials. 
The averaging kernels are reasonably well-localized and the cross-term kernels are small everywhere. 
The averaging kernels placed at ${r_0=0.3}$ begin to develop some amplitude outside of the target region; this is why we do not attempt to probe shallower layers.

\afterpage{
\begin{landscape}
\begin{figure}\vspace*{-0.75cm}
    \centering
    \makebox[\linewidth][c]{%
        \adjustbox{trim={0cm 1.4cm 0.5cm 0.5cm},clip}{%
            \includegraphics[width=0.47\linewidth]{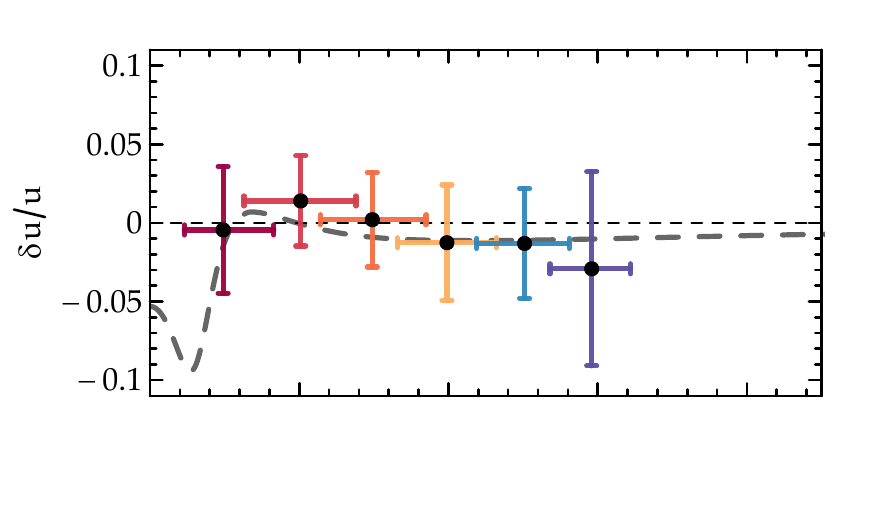}%
        }%
        \adjustbox{trim={1.4cm 1.4cm 0.5cm 0.5cm},clip}{%
            \includegraphics[width=0.47\linewidth]{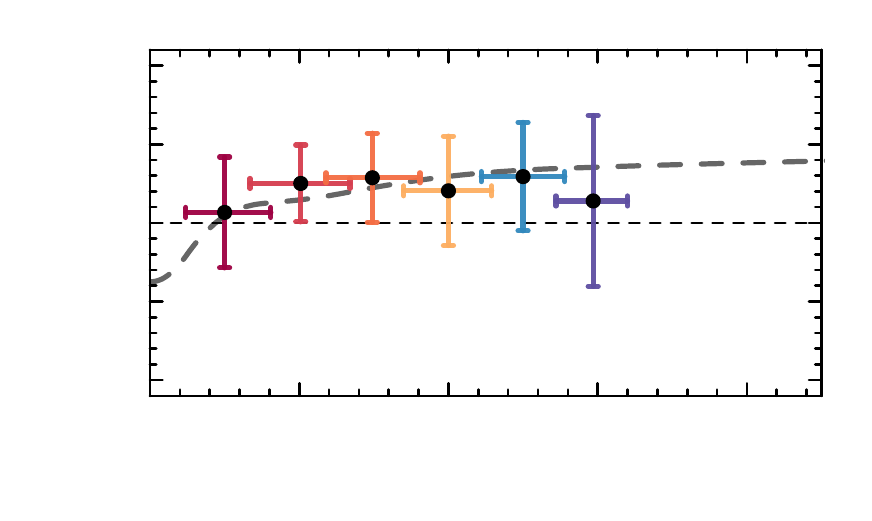}%
        }%
    }\\%
    \makebox[\linewidth][c]{%
        \adjustbox{trim={0cm 1.4cm 0.5cm 0.5cm},clip}{%
            \includegraphics[width=0.47\linewidth]{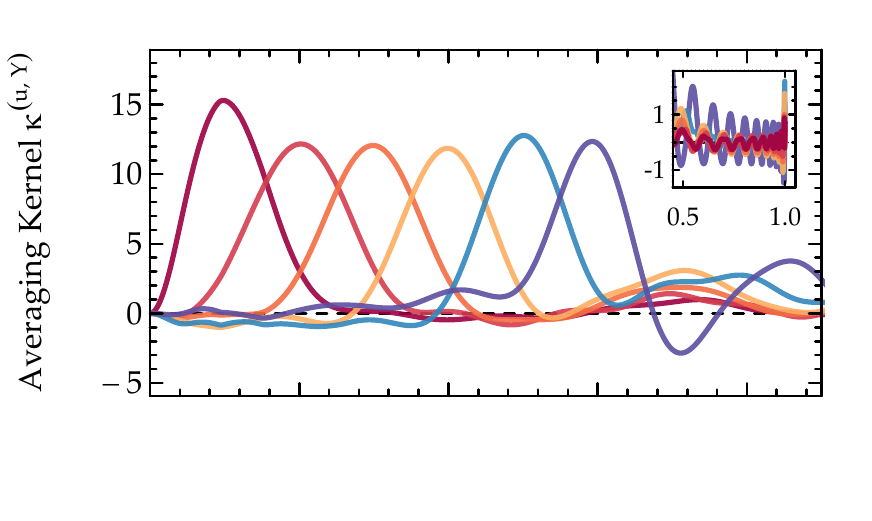}%
        }%
        \adjustbox{trim={1.4cm 1.4cm 0.5cm 0.5cm},clip}{%
            \includegraphics[width=0.47\linewidth]{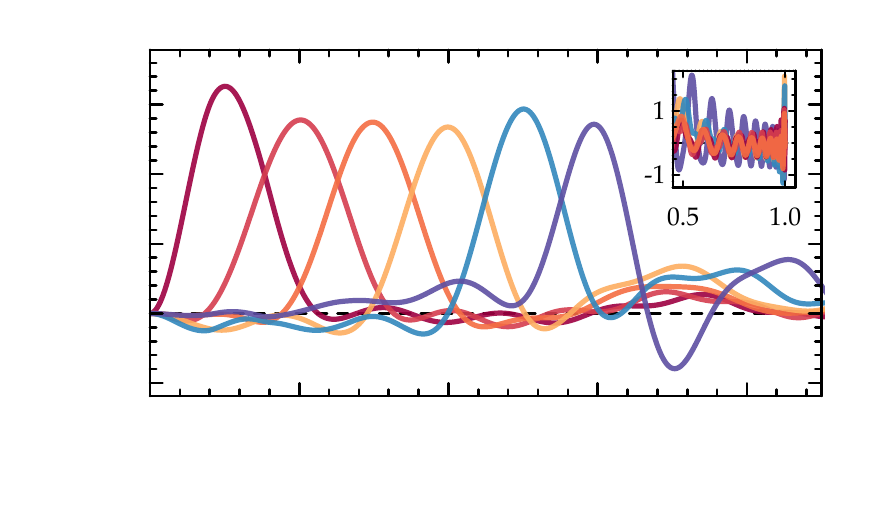}%
        }%
    }\\%
    \makebox[\linewidth][c]{%
        \adjustbox{trim={0cm 0cm 0.5cm 0.5cm},clip}{%
            \includegraphics[width=0.47\linewidth]{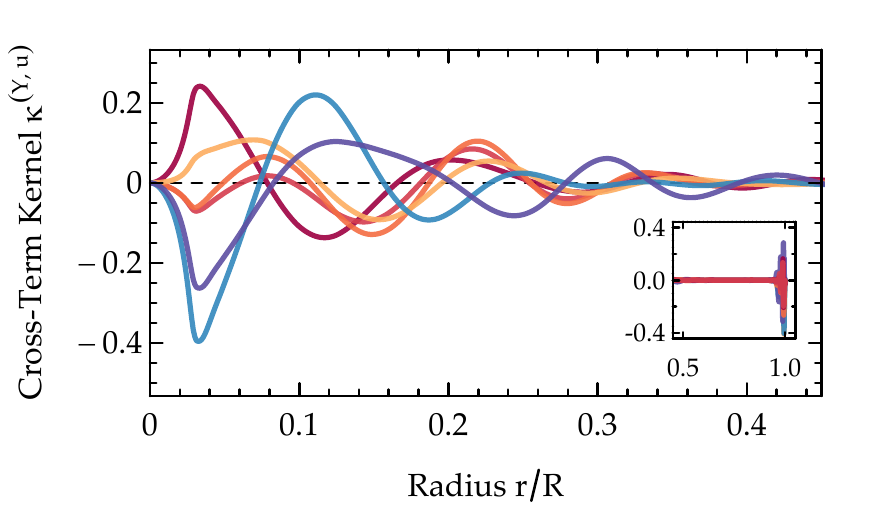}%
        }%
        \adjustbox{trim={1.4cm 0cm 0.5cm 0.5cm},clip}{%
            \includegraphics[width=0.47\linewidth]{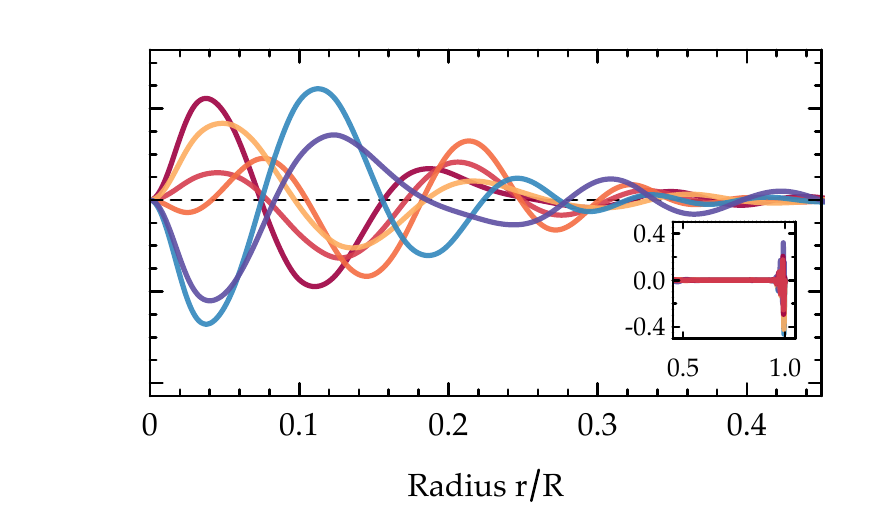}%
        }%
    }%
    \caption[Hare-and-hound structure inversions]{(Caption on other page.) 
    \label{fig:model-test} }
\end{figure}
\end{landscape}
}
\begin{figure}
    \contcaption{Structural inversions for the internal squared isothermal sound-speed profile $u$ of evolutionary models of 16~Cyg~A (left) and 16~Cyg~B (right). 
    \textbf{Top}: actual relative difference ${\delta u/u}$ between the evolutionary model and \mb{a reference model from the corresponding array of reference models for that star} (dashed gray line), and the result of the inversion-for-agreement procedure presented here (colored points). 
    The colors serve to associate the inversion results with their respective averaging and cross-term kernels. 
    \textbf{Middle}: averaged averaging kernels, sensitive to changes in $u'$, which have been placed at target radii $\mathbf{r_0} = [0.05, 0.1, 0.15, 0.2, 0.25, 0.3]$. 
    \textbf{Bottom}: averaged cross-term kernels that are sensitive to changes in helium abundance, whose amplitudes should be small everywhere relative to the averaging kernels. 
    \textbf{Insets}: the behavior of the averaging and cross-term kernels closer to the surface, where their amplitudes are small as desired (note the change in axes). }
\end{figure}

\subsection{Inversions for Stellar Structure} 
We now apply our structure inversion-for-agreement procedure on asteroseismic data of 16~Cyg~A and B. 
The relative differences with respect to the \emph{GOE} evolutionary models of these stars are shown in Figure~\ref{fig:Cyg-inversions}. 
As the mode sets are the same as in our tests with models, the averaging kernels and cross-term kernels are nearly identical to those shown in Figure~\ref{fig:model-test}.  
The results are also tabulated in Tables~\ref{tab:CygA}~and~\ref{tab:CygB}. 
We find that the sound speeds throughout the cores of 16~Cyg~A and B exceed those of these evolutionary models.

\afterpage{
\begin{landscape}
\begin{figure}
    \centering
    \makebox[\colwidth][c]{%
        \adjustbox{trim={0 0 \righttrim cm 0.1cm},clip}{
            \includegraphics[width=0.47\linewidth]{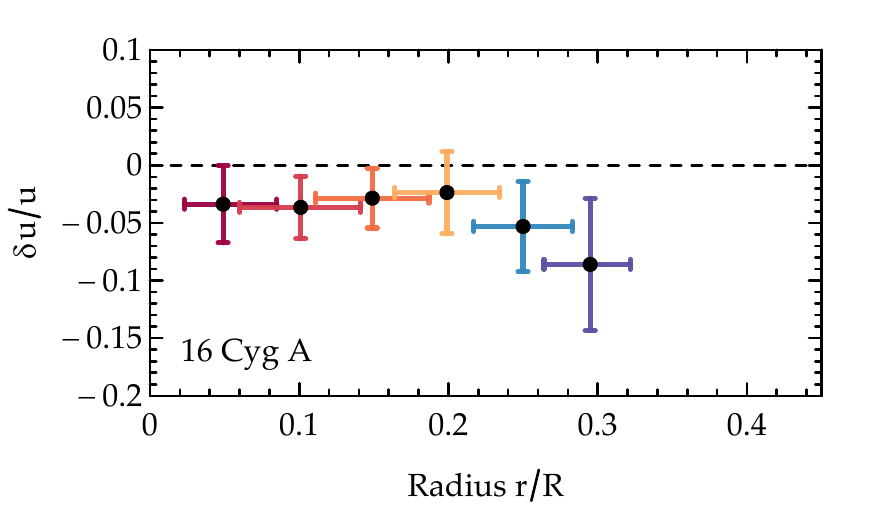}
        }%
        \adjustbox{trim={1.45cm 0 0 0.1cm},clip}{
            \includegraphics[width=0.47\linewidth]{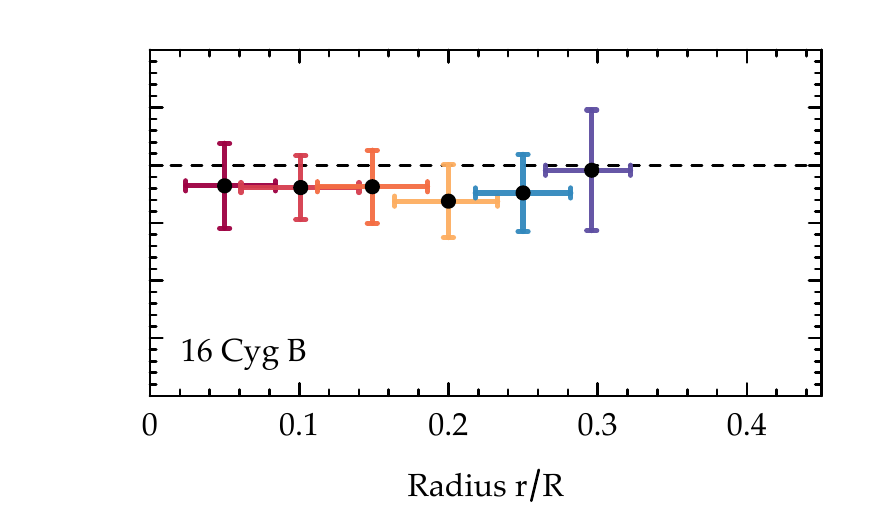}
        }
    }
    \caption[Structure inversions of 16~Cyg~A and B]{ Structural inversions for the internal squared isothermal sound-speed profile $u$ of 16~Cyg~A (left) and 16~Cyg~B (right) using the inversion-for-agreement technique introduced in this paper. 
    Results are shown in terms of relative differences with respect to the \emph{GOE} evolutionary models of these stars (\emph{cf}.~Equation~\ref{eq:rel-diff}). 
    The sound speeds in the cores of 16~Cyg~A and B are greater than those of the evolutionary models. 
    \label{fig:Cyg-inversions} 
    }
\end{figure}
\end{landscape}
}

\afterpage{
\begin{table}
\centering
\caption{Results of Inversions for the Squared Isothermal Sound Speed $u$ inside of 16~Cyg~A at Different Target Radii $r_0$ in the Stellar Core \label{tab:CygA}}
\begin{tabular}{ccccccc} \hline\hline
    Target radius & Peak of $\mathscr{K}$    & Relative $u$ difference          & Sq.~iso.~sound~speed \\ \hline
    $r_{0}$       & $r_{\max} \pm \text{FWHM}$ & $\delta u/u \pm \sigma_{\delta}$ & $u \pm \sigma_u$ \\
    $[$R$]$           & $[$R$]$                      & $[\text{w.r.t.~model~\emph{GOE}}]$                 & [$10^{15}$ cm$^2$ s$^{-2}$] \\ \hline
    0.05 & 0.049 $\pm$ 0.031 & -0.033 $\pm$ 0.033 & 1.515 $\pm$ 0.049 \\ 
    0.10 & 0.101 $\pm$ 0.041 & -0.036 $\pm$ 0.027 & 1.580 $\pm$ 0.041 \\ 
    0.15 & 0.149 $\pm$ 0.038 & -0.028 $\pm$ 0.025 & 1.404 $\pm$ 0.035 \\ 
    0.20 & 0.199 $\pm$ 0.035 & -0.023 $\pm$ 0.035 & 1.181 $\pm$ 0.041 \\ 
    0.25 & 0.250 $\pm$ 0.033 & -0.053 $\pm$ 0.039 & 1.019 $\pm$ 0.037 \\ 
    0.30 & 0.295 $\pm$ 0.029 & -0.086 $\pm$ 0.057 & 0.910 $\pm$ 0.048 \\ \hline
\end{tabular}
\vspace{2cm}
\caption{Results of inversions for the squared isothermal sound speed $u$ inside of 16~Cyg~B. \label{tab:CygB}}
\begin{tabular}{ccccccc} \hline\hline
    Target radius & Peak of $\mathscr{K}$    & Relative $u$ difference          & Sq.~iso.~sound~speed \\ \hline
    $r_{0}$       & $r_{\max} \pm \text{FWHM}$ & $\delta u/u \pm \sigma_{\delta}$ & $u \pm \sigma_u$ \\
    $[$R$]$           & $[$R$]$                      & $[\text{w.r.t.~model~\emph{GOE}}]$                 & [$10^{15}$ cm$^2$ s$^{-2}$] \\ \hline
    0.05 & 0.050 $\pm$ 0.030 & -0.017 $\pm$ 0.036 & 1.485 $\pm$ 0.053 \\ 
    0.10 & 0.101 $\pm$ 0.039 & -0.019 $\pm$ 0.027 & 1.533 $\pm$ 0.041 \\ 
    0.15 & 0.149 $\pm$ 0.037 & -0.018 $\pm$ 0.031 & 1.402 $\pm$ 0.043 \\ 
    0.20 & 0.200 $\pm$ 0.034 & -0.031 $\pm$ 0.031 & 1.216 $\pm$ 0.037 \\ 
    0.25 & 0.250 $\pm$ 0.032 & -0.024 $\pm$ 0.033 & 1.025 $\pm$ 0.033 \\ 
    0.30 & 0.296 $\pm$ 0.028 & -0.004 $\pm$ 0.052 & 0.870 $\pm$ 0.045 \\ \hline
\end{tabular}
\end{table}
}

In the case of 16~Cyg~A, each of the individual measurements hovers around a ${1\sigma}$ difference. 
On the one hand, all of the model sound speeds are found to be lower than in the star, indicating that there are systematic differences between the model and the star.
Viewed this way, the overall result is more significant than each of the measurements taken separately. 
On the other hand, there is covariance between the different measurements, because the different averaging kernels overlap to some degree. 
Thus, assigning an overall level of statistical significance to these results is challenging.

To assess whether the differences may stem from the \emph{GOE} models having wrong masses or radii, we compare the inversion results against other models of different mass and radius. 
Following Equation~(\ref{eq:MR}), the spread in sound speeds caused by mass and radius estimates are largest for the models with either a high radius and a low mass, or models with a low radius and high mass. 
Thus, we show in Figure~\ref{fig:ref-mods} these inversion results against models with masses and radii that differ by ${1\sigma}$ in opposite directions from the mean estimated masses and radii of these stars. 
In both cases, the models with higher masses and lower radii are preferred. 
However, while the 16~Cyg~B models show roughly broad agreement, the 16~Cyg~A models do not agree quite as well. 

\afterpage{
\begin{landscape}
\begin{figure}
    \centering
    \makebox[\colwidth][c]{%
        \adjustbox{trim={0 0 \righttrim cm 0.1cm},clip}{
            \includegraphics[width=0.47\linewidth]{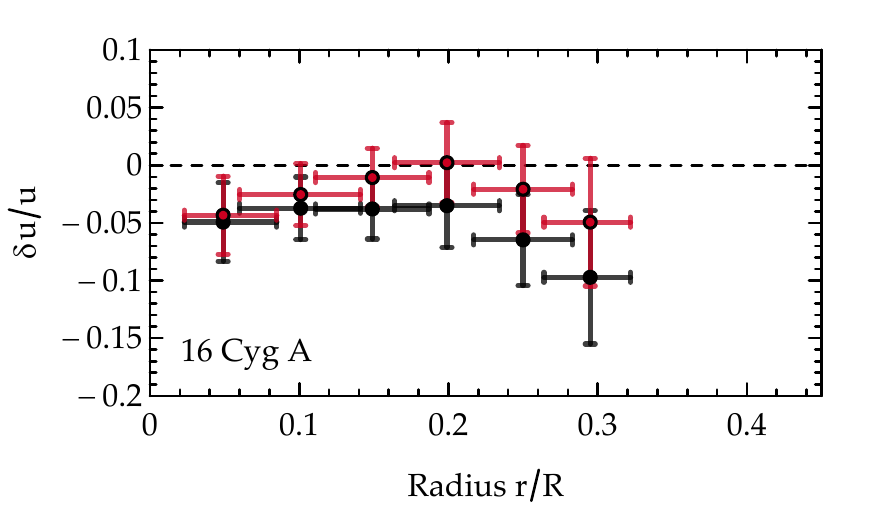}
        }%
        \adjustbox{trim={1.45cm 0 0 0.1cm},clip}{
            \includegraphics[width=0.47\linewidth]{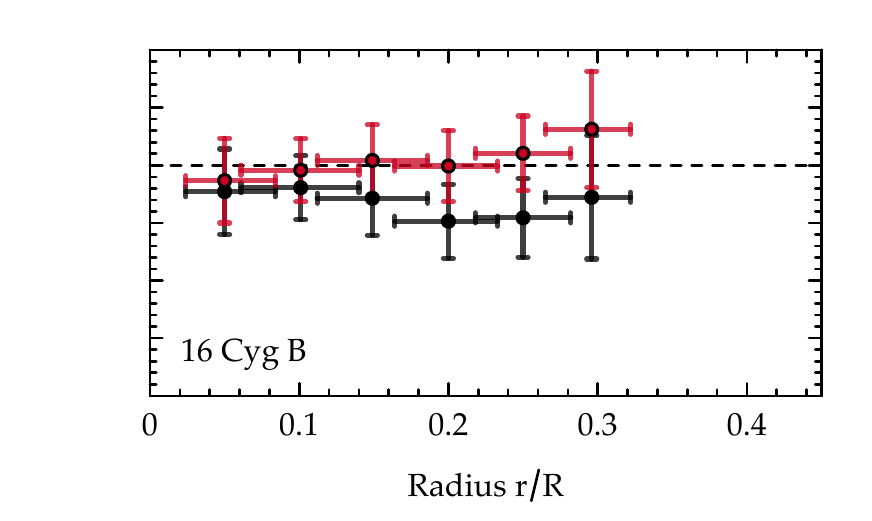}
        }
    }
    \caption[Impact of mass and radius on inversion results]{Relative differences between the inferred sound speeds $u$ of 16~Cyg~A (left) and 16~Cyg~B (right) shown against a model with a high radius and low mass (black points) and a model with a low radius and high mass (\textcolor{diff-red}{red} points). 
    \label{fig:ref-mods} }
\end{figure}
\end{landscape}
}

The isothermal speed of sound depends principally on the inverse of the mean molecular weight $\mu$ of the fluid. 
Fusion alters the core composition and increases $\mu$; thus, with all else equal, older stars will have a lower $u$ in the core. 
To assess the effect of stellar age in the context of these results, we evolve two models to match the characteristics of 16~Cyg~A (\emph{cf}.~Table~\ref{tab:stellar-parameters}) with ages of ${\tau=6}$~Gyr and ${\tau=5}$~Gyr, which are significantly lower than the estimated age of ${\tau=6.90\pm 0.40}$~Gyr. 
The relative differences between the core ${u}$ of 16~Cyg~A and these models are shown in Figure~\ref{fig:low-age}. 
In the deep core (${r=0.05}$), the young age models have smaller differences when compared with the \emph{GOE} model. 
However, the differences farther out are not explained with smaller ages. 
Furthermore, although it seems the inner core is better with the low-age models, frequency combinations such as ${r_{0,2}}$ \citep{2003A&A...411..215R} indicate that the low-age models are not appropriate. 
A comparison of ${r_{0,2}}$ values for these models is shown in Figure~\ref{fig:r02}.
This may explain why the differences in ${u}$ worsen just outside the core.

\afterpage{
\begin{figure}
    \centering
    \makebox[\colwidth][c]{%
        \includegraphics[width=0.72\linewidth]{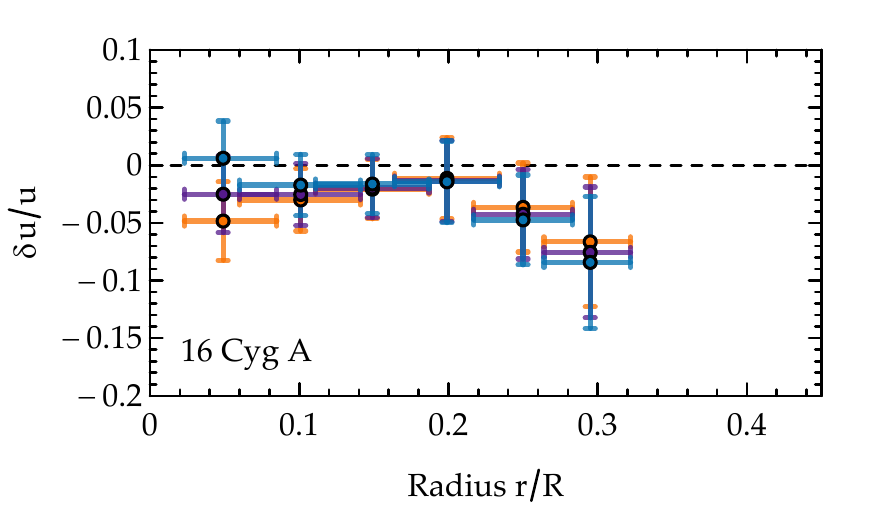}
    }
    \caption[Impact of stellar ages on inversion results]{Relative differences between the isothermal sound speed $u$ in the core of 16~Cyg~A and models with lower ages (6~Gyr in \textcolor{diff-purple}{purple}, $5$~Gyr in \textcolor{diff-blue}{blue}). 
    A model of 16~Cyg~A at the mean estimated present age (6.9~Gyr in \textcolor{diff-orange}{orange}) is shown for reference. 
    \label{fig:low-age} }
\vspace{2cm}
    \centering
    \makebox[\colwidth][c]{%
        \includegraphics[width=0.72\linewidth]{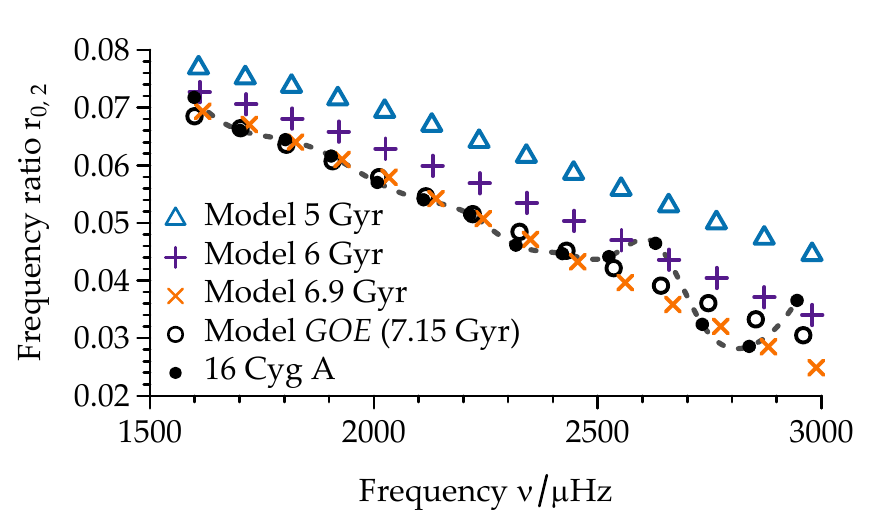}
    }
    \caption[Frequency ratios for models of different ages]{Ratio of the small frequency separation to the large frequency separation---a core-conditions indicator that is insensitive to surface effects---against mode frequency for asteroseismic data of 16~Cyg~A in comparison with models of various ages. 
    \label{fig:r02} }
\end{figure}
}

\Needspace{3\baselineskip}
\section{Discussion and Conclusions} 
In this paper, we examined the problem of deducing the core structures of solar-like stars based on the frequencies of their normal modes of oscillation. 
We applied the SOLA inversion technique to infer the radial dependence of the squared isothermal sound speed throughout the interiors of two solar-type main-sequence stars. 
We inverted using the $(u',Y)$ kernel pair because the influence of the second variable (${Y}$) is very low in the regions of our interest. 
We presented a new algorithm for the automated determination of inversion parameters that also accounts for imprecise/inaccurate stellar mass and radius estimates. 
We validated this technique on models, and then applied it to the well-studied stars 16~Cyg~A and B. 
We measured $u$ at several different radii within these stars and compared these values to best-fitting evolutionary models of these stars. 
We found that the sound speeds in the cores of these stars are greater than in the \emph{GOE} models. 
This is to our knowledge the first time the radial variation in sound speed has been measured in a star other than the Sun.

In the case of 16~Cyg~B, it seems plausible that adjustments to the mass and radius of the \emph{GOE} model may serve to fix the differences that we find. 
In the case of 16~Cyg~A, however, the source of the disparities is more difficult to pinpoint. 
Lower age models help with the differences in the deeper parts of the core, but do not aid with the differences farther out. 
Furthermore, the lower age models fail to reproduce the asteroseismic frequency ratios of 16~Cyg~A, which effectively rules age out as the culprit. 
Missing physical processes, incorrect application of known processes, or inadequate inputs in the calculations of the models may therefore be at fault. 
\mb{For example, while the GOE model of 16~Cyg~A does not have a convective core at the present age, it did have one during the first $1.75$~Gyr of its evolution. 
As core convection modifies the mean molecular weight, the duration of its existence may leave a footprint in the sound speed. 
It may then be the case that an incorrect prescription of convection in stellar cores is the cause of these discrepancies. }

16~Cyg~A and B are stars either on the main sequence or nearly at the main-sequence turnoff. 
The main sequence is a well-studied phase of evolution, and the different types of observations that are possible for main-sequence stars lead to estimates of their ages, masses, and radii in a well-known way. 
Being the first and also the longest-lived stage of evolution, getting the details of the main-sequence evolution right is necessary for also getting the later stages of stellar evolution right as well. 
Any neglected processes that cause substantial errors on the core structure of main-sequence stars will subsequently propagate into the later stages of evolution.

As is always the case with ill-posed inverse problems, there is no guarantee that the end result will be the true profile of the star. 
That being said, the procedure has worked well in blind tests on models with known structure. 
Therefore, some confidence can be put in the results.

\paragraph*{Acknowledgements} 
\noindent The research leading to the presented results has received funding from the European Research Council under the European Community's Seventh Framework Programme (FP7/2007-2013) / ERC grant agreement no 338251 (StellarAges). This research was undertaken in the context of the International Max Planck Research School for Solar System Research. E.P.B.\ acknowledges support from the National Physical Science Consortium Fellowship. S.B.\ acknowledges partial support from NSF grant AST-1514676 and NASA grant NNX13AE70G. \mb{We thank the anonymous referee for their very helpful report.} 

\paragraph*{Software} 
\noindent R 3.2.3 \citep{R}, magicaxis 2.0.0 \citep{magicaxis2}, 
kernel calculations by \citet{kerexact}, \mb{MESA \citep{2011apjs..192....3p, 2013apjs..208....4p, 2015apjs..220...15p}, and ADIPLS \citep{2008Ap&SS.316..113C}.} 

\stepcounter{chapter}
\chapter*{Future Prospects\markboth{Future Prospects}{Future Prospects}}
\addcontentsline{toc}{chapter}{Future Prospects}
\label{chap:prospective}

Though stars are, overall, generally considered to be well-understood, a number of open problems remain in asteroseismology and, more widely, the field of stellar astrophysics as a whole. 
At a basic level, we currently are unable to predict stellar radii from first principles. 
This is due to the fact that we use time-independent one-dimensional theories of convection in evolutionary models---approximations which are controlled by free parameters. 
Properly modelling convection in stellar interiors seems to be among the biggest goals in modern theoretical stellar astrophysics. 
Furthermore, for similar reasons, we generally fail to predict pulsation frequencies of stars, even after making post-hoc corrections for near-surface effects. 

Along similar lines, one of the most basic facts about stars (and astronomical bodies in general) is that they rotate. 
Yet canonical stellar modelling often neglects the effects of rotation, and other similarly `obvious' phenomena such as magnetic fields. 
The very long-term future of research into stars may feature fully 3D magnetohydrodynamical stellar modelling, or even a full treatment of every individual particle that make up the star; however, it is clear that we are far away from that point. 

In terms of the continuation of the research presented in this thesis, there are a few avenues in particular that I intend to explore in the coming months and years: 
\begin{description}
    \setlength{\itemindent}{0pt}
    \item[Structure inversions of more stars.]
    The next step is to apply the technique developed in Chapter~\ref{chap:inversion} to as many stars as possible. 
    This will allow us to determine whether the theory of stellar evolution produces models with the correct interior structures. 
    
    Figures~\ref{fig:phy} and \ref{fig:phy2} show structure inversions for $20$~stars from the \emph{Kepler} LEGACY sample \citep{2017ApJ...835..172L}. 
    The reference models have been constructed under four different assumptions of input physics: with/without diffusion, and with/without convective core overshooting. 
    While some stars show broad agreement throughout their interior with evolutionary models (e.g., KIC~$5184732$), most of the models disagree substantially with the interior structure of the stars. 
    Furthermore, there seems to be no set of input physics considered here that repairs the differences. 
    This indicates that important ingredients may be missing from canonical models of stellar interiors, such as mixing induced by internal rotation. 

\begin{figure}
    \centering
    \adjustbox{trim=0cm 1.4cm 0cm 0cm, clip}{%
        \includegraphics[width=0.55\textwidth]{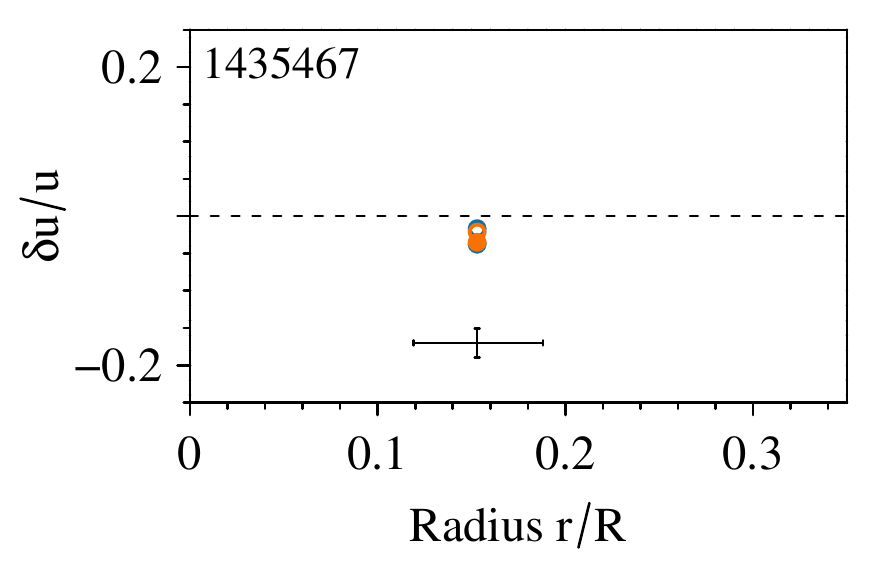}%
    }%
    \adjustbox{trim=1.6cm 1.4cm 0cm 0cm, clip}{%
        \includegraphics[width=0.55\textwidth]{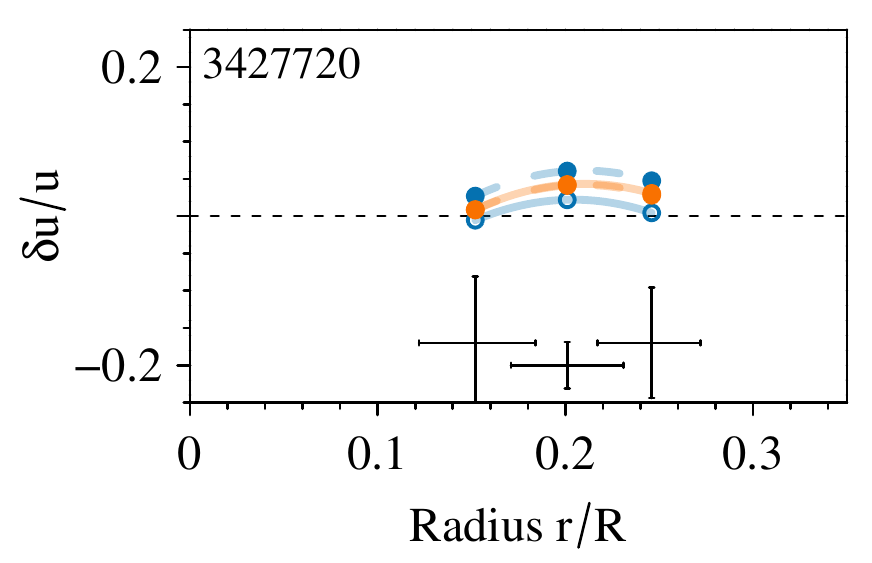}%
    }\\%
    \adjustbox{trim=0cm 1.4cm 0cm 0cm, clip}{%
        \includegraphics[width=0.55\textwidth]{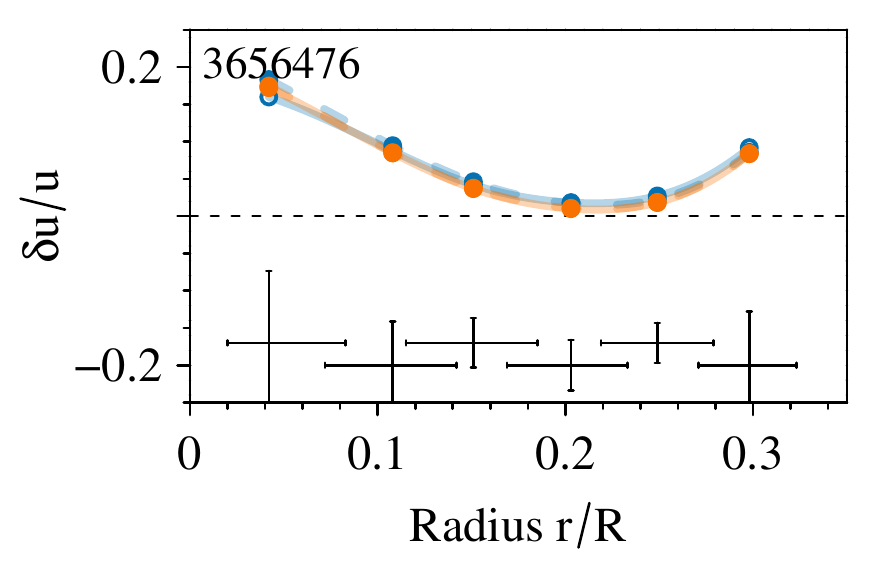}%
    }%
    \adjustbox{trim=1.6cm 1.4cm 0cm 0cm, clip}{%
        \includegraphics[width=0.55\textwidth]{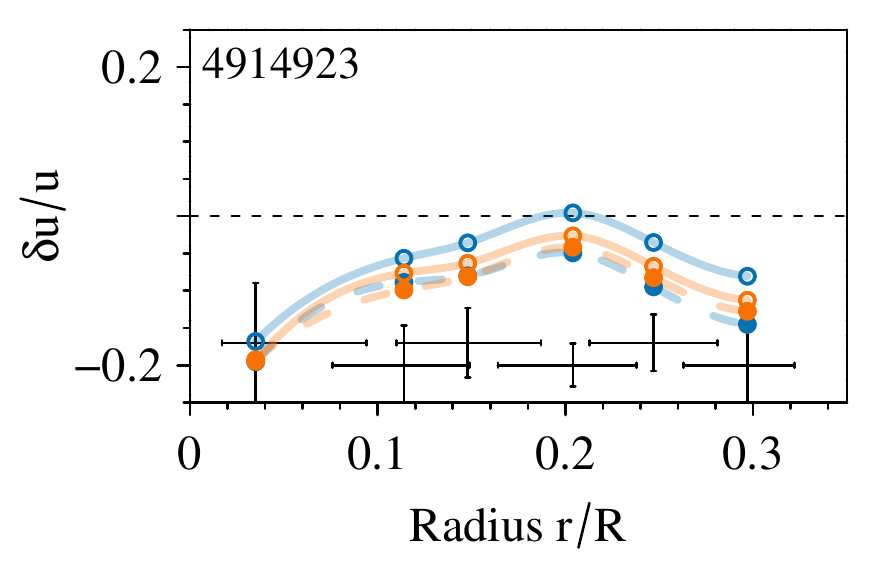}%
    }\\%
    \adjustbox{trim=0cm 1.4cm 0cm 0cm, clip}{%
        \includegraphics[width=0.55\textwidth]{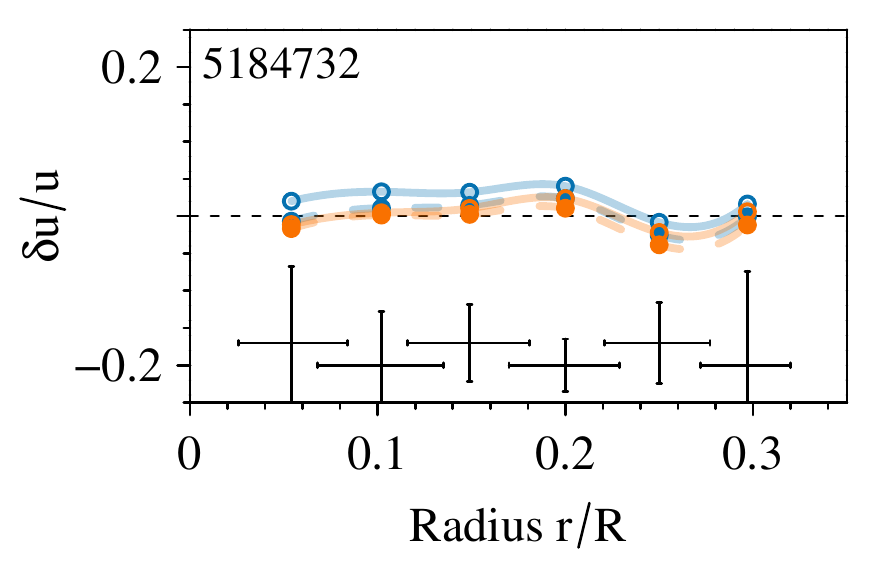}%
    }%
    \adjustbox{trim=1.6cm 1.4cm 0cm 0cm, clip}{%
        \includegraphics[width=0.55\textwidth]{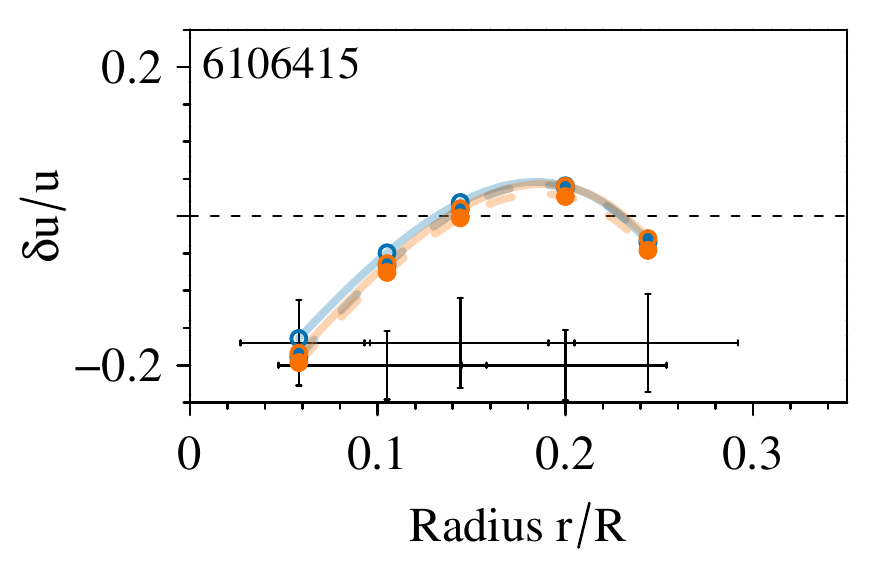}%
    }\\%
    \adjustbox{trim=0cm 1.4cm 0cm 0cm, clip}{%
        \includegraphics[width=0.55\textwidth]{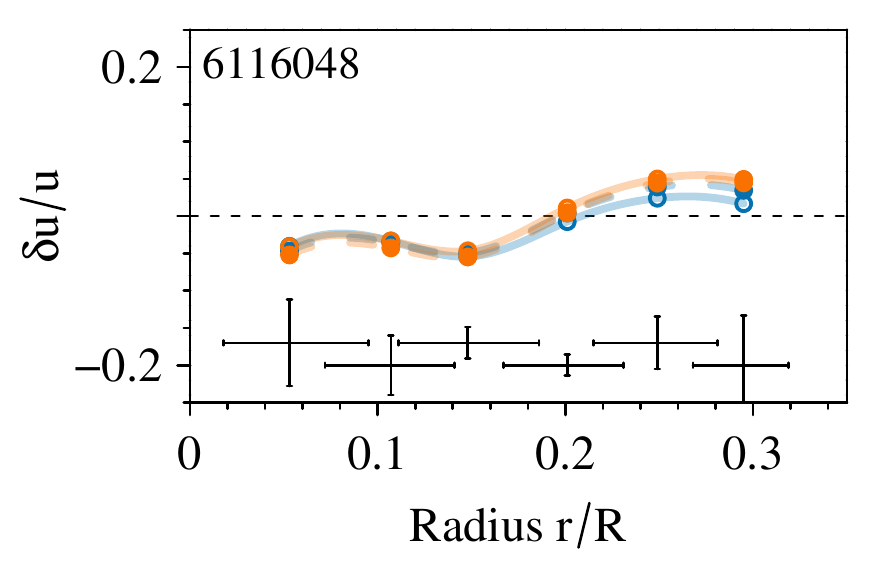}%
    }%
    \adjustbox{trim=1.6cm 1.4cm 0cm 0cm, clip}{%
        \includegraphics[width=0.55\textwidth]{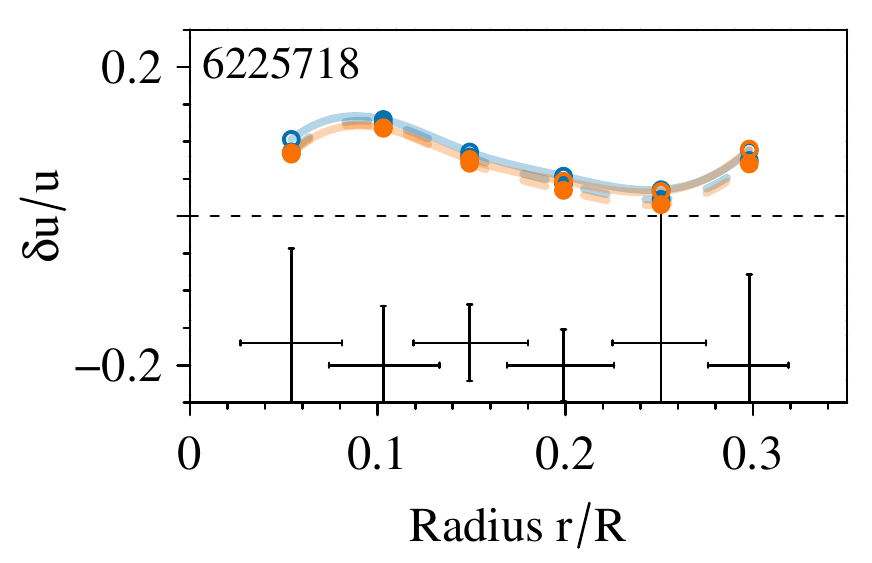}%
    }\\%
    \adjustbox{trim=0cm 0cm 0cm 0cm, clip}{%
        \includegraphics[width=0.55\textwidth]{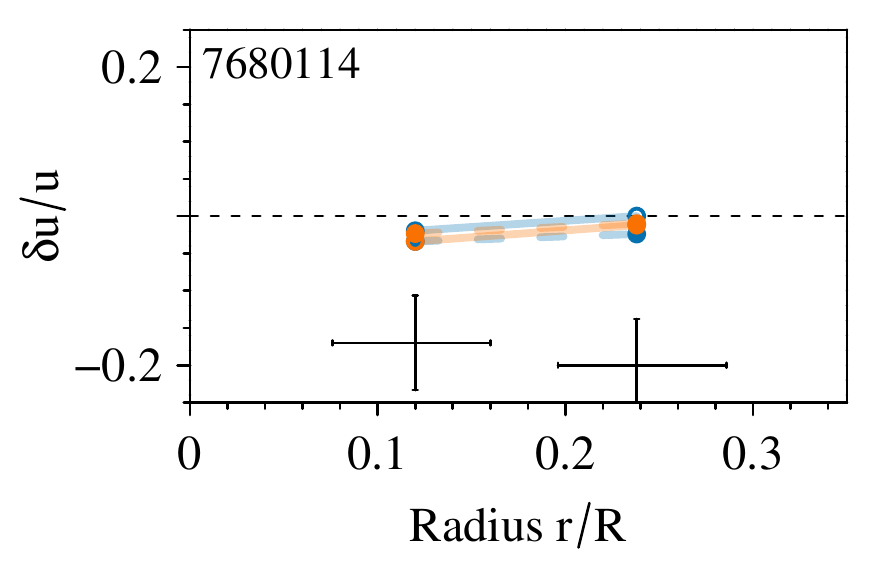}%
    }%
    \adjustbox{trim=1.6cm 0cm 0cm 0cm, clip}{%
        \includegraphics[width=0.55\textwidth]{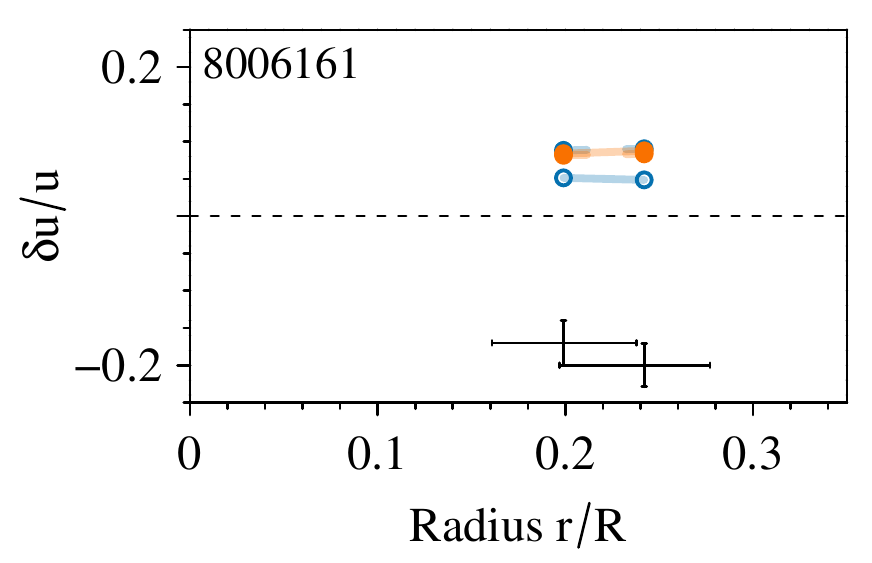}%
    }%
    \caption[Inversion Zoo]{(Continued in Figure~\ref{fig:phy2}.) \label{fig:phy}}
\end{figure}
\begin{figure}
    \centering
    \adjustbox{trim=0cm 1.4cm 0cm 0cm, clip}{%
        \includegraphics[width=0.55\textwidth]{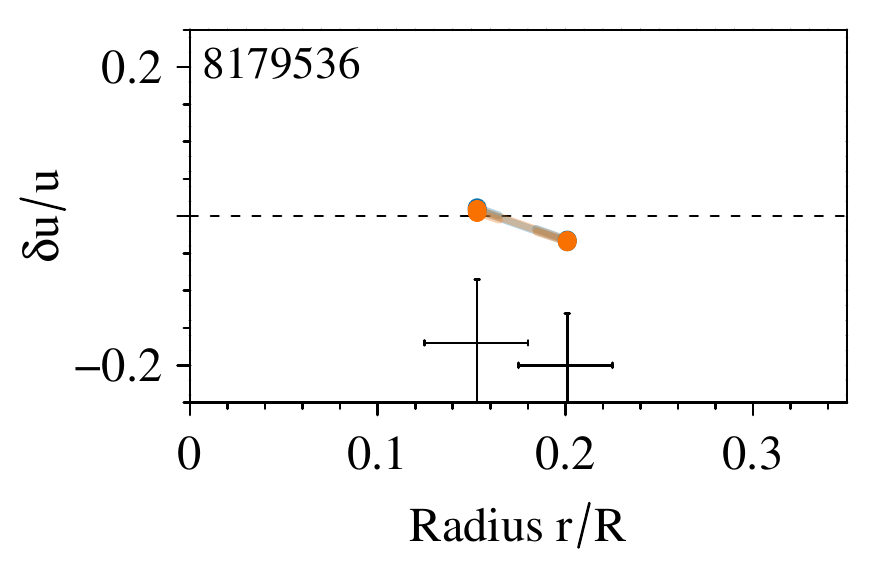}%
    }%
    \adjustbox{trim=1.6cm 1.4cm 0cm 0cm, clip}{%
        \includegraphics[width=0.55\textwidth]{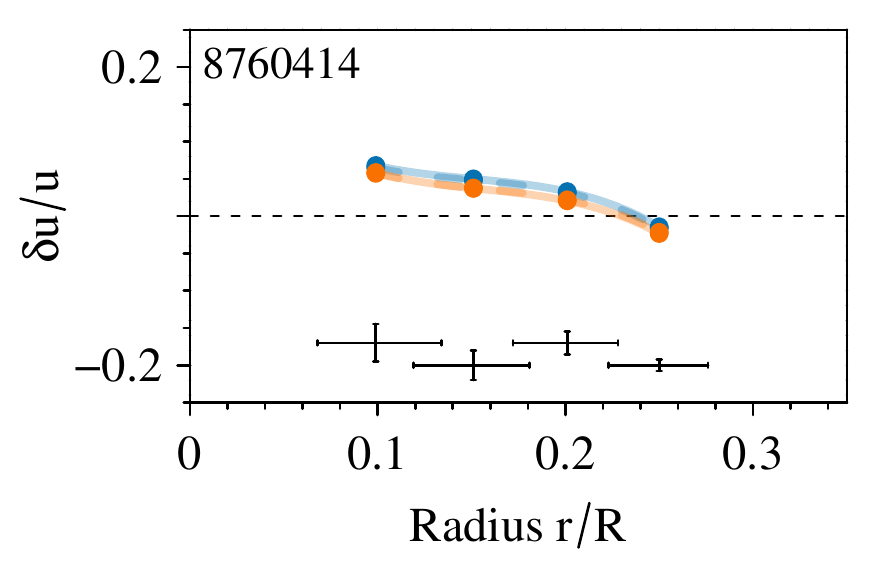}%
    }\\%
    \adjustbox{trim=0cm 1.4cm 0cm 0cm, clip}{%
        \includegraphics[width=0.55\textwidth]{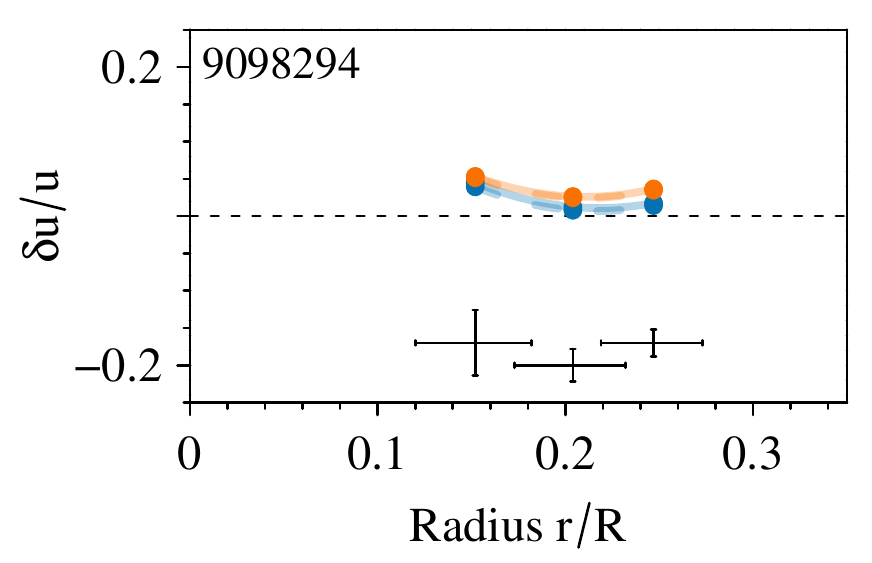}%
    }%
    \adjustbox{trim=1.6cm 1.4cm 0cm 0cm, clip}{%
        \includegraphics[width=0.55\textwidth]{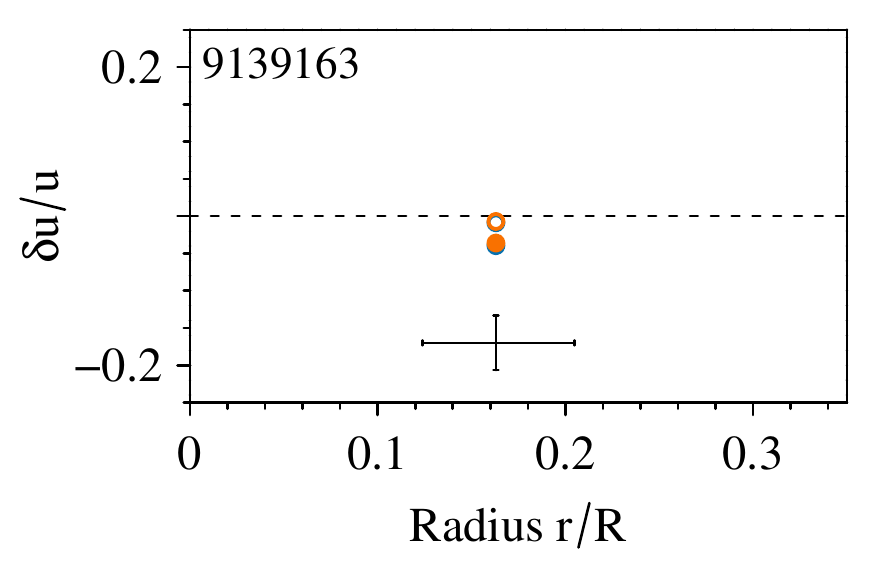}%
    }\\%
    \adjustbox{trim=0cm 1.4cm 0cm 0cm, clip}{%
        \includegraphics[width=0.55\textwidth]{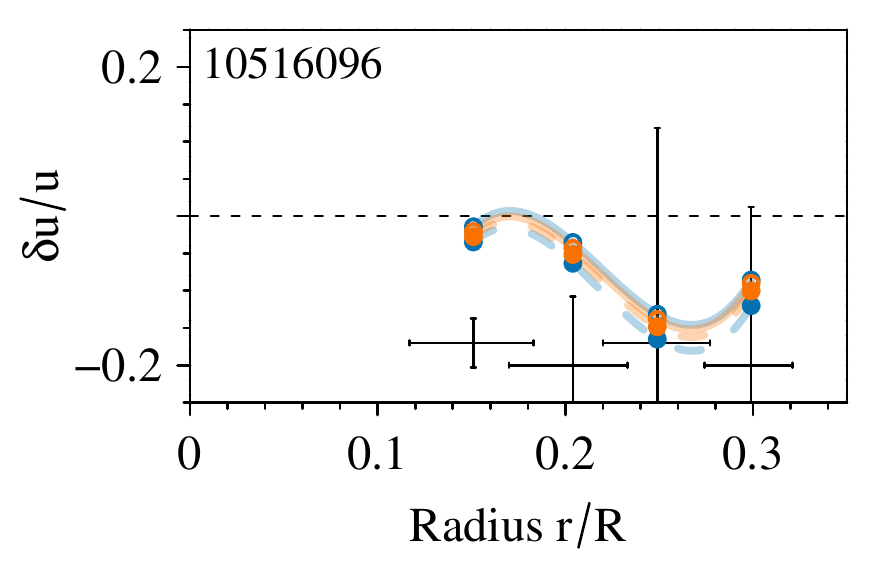}%
    }%
    \adjustbox{trim=1.6cm 1.4cm 0cm 0cm, clip}{%
        \includegraphics[width=0.55\textwidth]{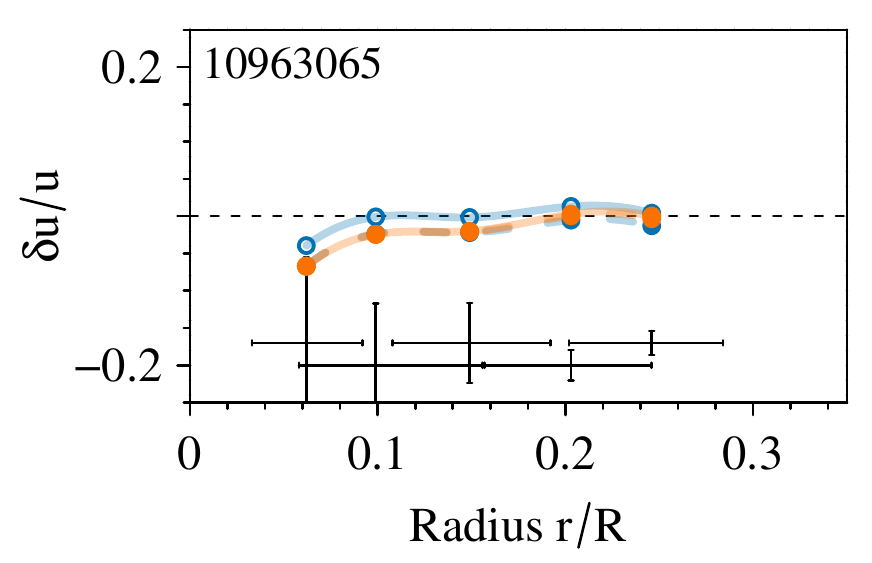}%
    }\\%
    \adjustbox{trim=0cm 1.4cm 0cm 0cm, clip}{%
        \includegraphics[width=0.55\textwidth]{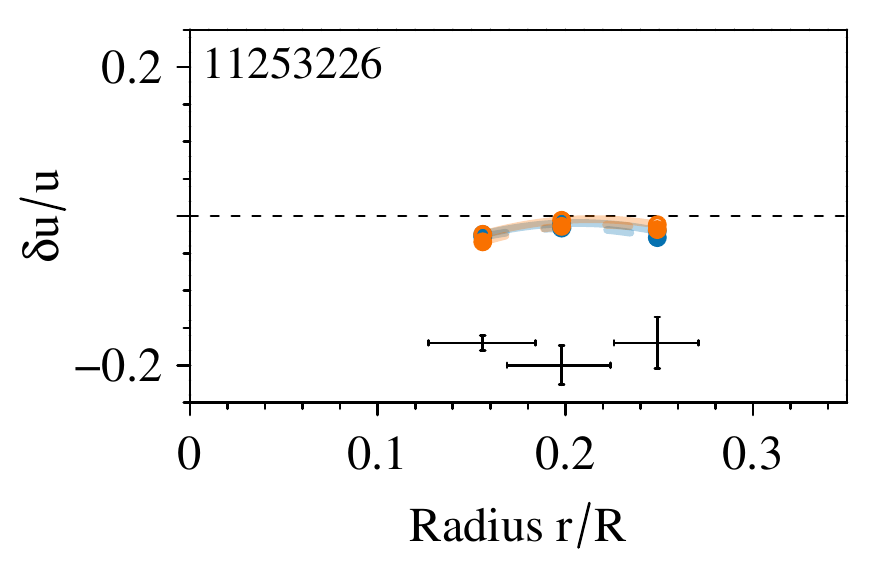}%
    }%
    \adjustbox{trim=1.6cm 1.4cm 0cm 0cm, clip}{%
        \includegraphics[width=0.55\textwidth]{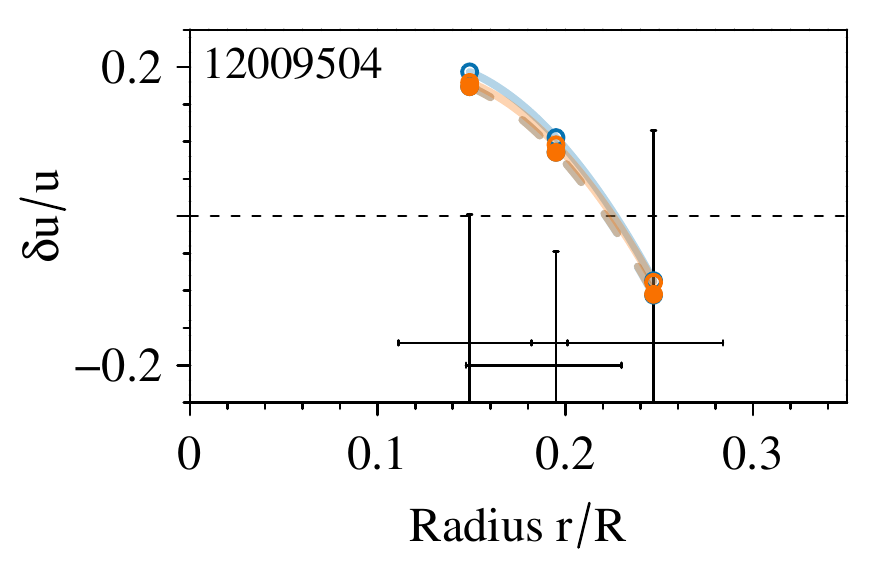}%
    }\\%
    \adjustbox{trim=0cm 0cm 0cm 0cm, clip}{%
        \includegraphics[width=0.55\textwidth]{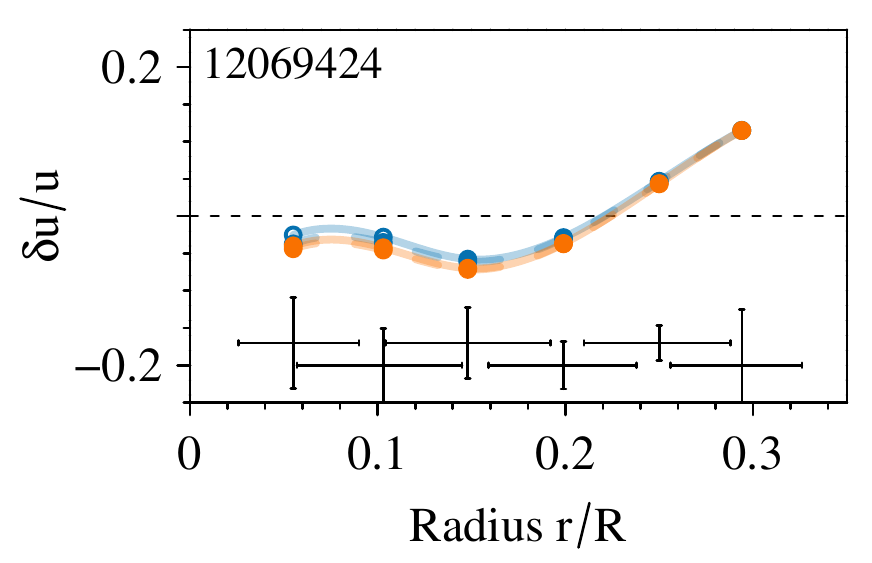}%
    }%
    \adjustbox{trim=1.6cm 0cm 0cm 0cm, clip}{%
        \includegraphics[width=0.55\textwidth]{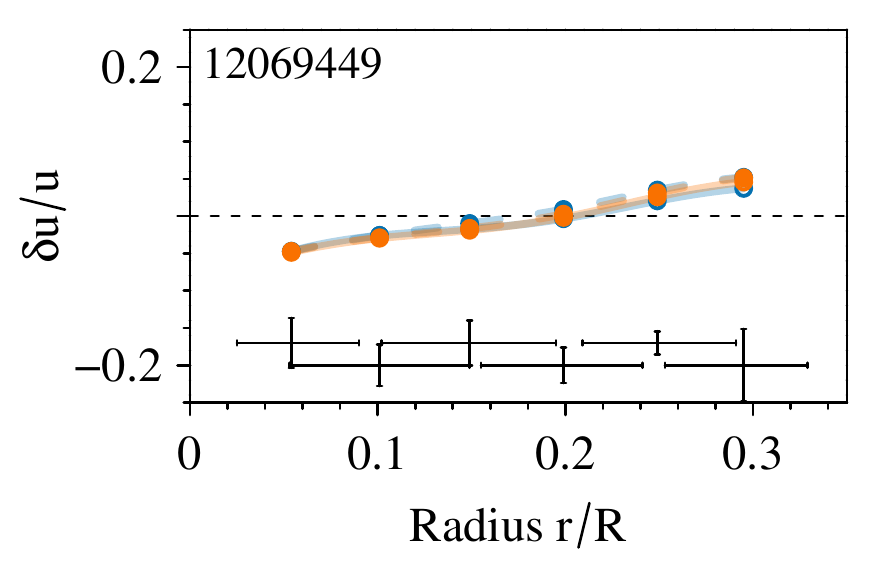}%
    }%
    \caption[]{(Caption on other page.)
    \label{fig:phy2}}
\end{figure}
\begin{figure}
    \contcaption{Core sound-speed profiles of LEGACY stars compared against stellar models constructed with different physics inputs: with/without diffusion (orange/blue, respectively) and with/without overshooting (filled/open points, respectively). 
    The quantity $\delta u/u$ is the relative difference in the isothermal speed of sound between the model and the star at that location in the stellar interior. 
    The uncertainties of the inversion results and the widths of the corresponding averaging kernels are shown as error bars in the bottom of each panel, and are vertically offset from one another for visibility. }
\end{figure}
    
    This work is soon to be submitted to the Astrophysical Journal.

    \item[Evolution inversions of evolved stars.] 
    There have been at least an order of magnitude more detections of solar-like oscillations in evolved stars such as red giants than in main-sequence stars. 
    When combined with kinematic information, determining the ages and chemical compositions of a large number of red giant stars will allow us to reconstruct the history of the Galaxy's development.
    
    In Chapter~\ref{chap:intro} I showed the future evolution of the Sun up through to core helium exhaustion. 
    Current ongoing work is the application of the techniques developed in Chapters~\ref{chap:ML} and \ref{chap:statistical} to these later stages of evolution.

    \item[Structure inversions of evolved stars.]
    In this thesis, I analyzed main-sequence solar-like oscillators. 
    After stars leave the main sequence, the $p$-modes in their envelopes mix with the $g$-modes in their deep interiors to give rise to mixed modes of oscillation. 
    Figure~\ref{fig:kernel-evol} shows the evolution of the kernel function for an ${\ell=1}$ mixed mode throughout the sub-giant phase of evolution. 
    After obtaining suitable reference models, for example using the technique mentioned in the previous point, I will invert mixed mode frequencies to determine the core structures of sub-giant and eventually red-giant stars. 
    This presents the exciting prospect for potentially learning more about the deep core structure of another star than we know about our own Sun. 

\begin{figure}
    \centering
    \hspace*{-1.35cm}\includegraphics[width=1.2\linewidth]{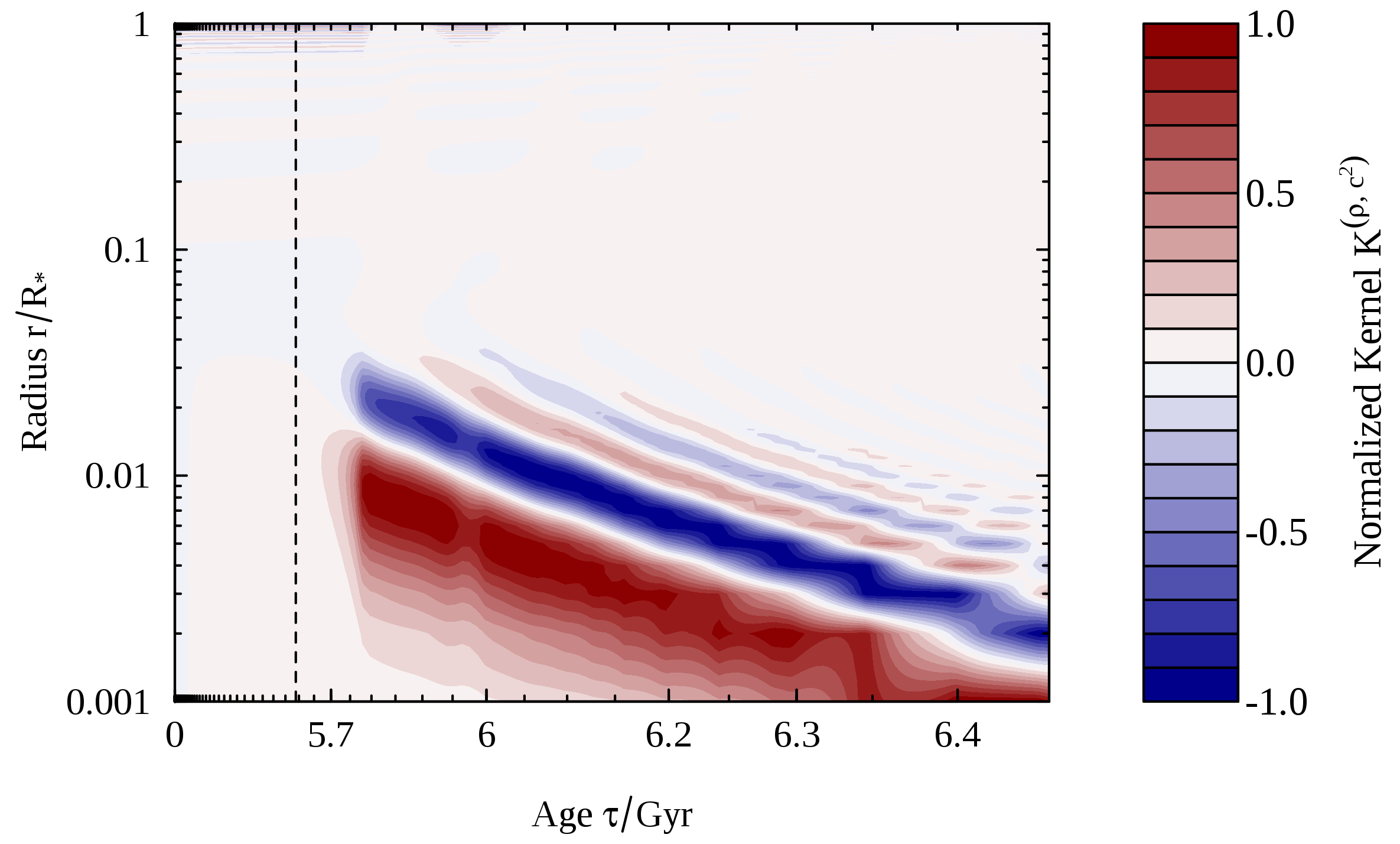}
    \caption[Kernel function evolution]{Evolution of the (${\rho, c^2}$) kernel function for the (${\ell=1}, {n=11}$) mode of a ${1.11\; M/M_\odot}$ star. 
        The vertical dashed line shows the end of the main sequence (TAMS). 
        As the mode mixes with a $g$-mode, it develops extreme sensitivity to the deep core structure of the star. 
        \label{fig:kernel-evol}}
\end{figure}

    \item[Evolution inversions for fundamental constants.]
    A problem of cosmological significance is the measurement of physical constants, and the determination of whether or not they really are constant. 
    The idea of using the Sun to constrain the cosmic variation of the gravitational constant $G$ goes back at least to the time of \citet{Dirac199}. 
    So far, this approach has not been undertaken using other stars. 
    I intend to use the tools discussed in this thesis to measure $G$ as well as other fundamental quantities that impact on stellar evolution and pulsation, such as the fine structure constant \citep[e.g.,][]{2008JCAP...08..010A, 2010AIPC.1269...21C}. 
    Though the Sun is the star with the best data, observations of a large number stars may be able to be combined into a more sensitive tool for these measurements. 
    Furthermore, the Sun's evolution only covers one third of the history of the Universe, and is therefore insensitive to any earlier variations to these quantities. 
\end{description}

In the longer term, there are other prospects that are quite exciting. 
\citet{2014ApJ...782....2L} predicted that $\ell=4$ modes would be observable in 16~Cyg~A and B from \emph{Kepler} data. 
With such data, it would be possible to resolve the sound speed profiles of the observed stars to even shallower layers, which would provide further constraints on theories of the stellar interior. 
However, recent data releases seem not to have produced any such detections. 
It does seem feasible within the coming decades that such observations could become available, perhaps through a combination of \emph{Kepler} data with SONG observations \citep{2014RMxAC..45...83A, 2017ApJ...836..142G} and possibly utilizing the forthcoming TESS and PLATO missions. 

In this thesis, I used artificial intelligence to assist in solving problems in stellar astrophysics. 
This is a form of so-called \emph{weak} AI. 
These tools will only get more powerful with the coming decades. 
Eventually, we may have \emph{strong} AI, which will be capable of fully driving scientific research. 
One day, it may be that AI will be able to determine on its own the set of astrophysical laws that are most harmonious with enormous quantities of empirical data. 



\chapter*{Bibliography\markboth{Bibliography}{Bibliography}}
\addcontentsline{toc}{chapter}{Bibliography}
\bibliographystyle{thesis.bst}
\setlength{\bibsep}{0.15cm}
\renewcommand{\bibsection}{}

\chapter*{Publications\markboth{Publications}{Publications}}
\addcontentsline{toc}{chapter}{Publications}
\textbf{\large Refereed publications}\\
\begin{enumerate}
    \item \textbf{Bellinger, E.~P.}, Basu, S., Hekker, S., \& Ball, W.: 2017, 
    ``\href{http://adsabs.harvard.edu/abs/2017ApJ...851...80B}{Model-independent Measurement of Internal Stellar Structure in 16~Cygni~A and B}'', 
    \\\emph{The Astrophysical Journal}, 851 (2), 80 
    
    \item \textbf{Bellinger, E.~P.}, Angelou, G.~C., Hekker, S., Basu, S., Ball, W., \& Guggenberger, E.: 2016, 
    ``\href{http://adsabs.harvard.edu/abs/2016ApJ...830...31B}{Fundamental Parameters of Main-Sequence Stars in an Instant with Machine Learning}'', 
    \emph{The Astrophysical Journal}, 830 (1), 20  
    
    \item Angelou, G.~C., \textbf{Bellinger, E.~P.}, Hekker, S., \& Basu, S.: 2017,
    ``\href{http://adsabs.harvard.edu/abs/2017ApJ...839..116A}{On the Statistical Properties of the Lower Main Sequence}'', 
    \emph{The Astrophysical Journal}, 839 (2) 116 (co-first author) 
    
    \item Guggenberger, E., Hekker, S., Basu, S., Angelou, G.~C., \& \textbf{Bellinger, E.~P.}: 2017, 
    ``\href{http://adsabs.harvard.edu/abs/2017MNRAS.470.2069G}{Mitigating the mass dependence in the $\Delta\nu$ scaling relation of red-giant stars}'',
    \emph{Monthly Notices of the Royal Astronomical Society}, 470 (2)
    
    \item Guggenberger, E., Hekker, S., Basu, S., \& \textbf{Bellinger, E.~P.}: 2016
    ``\href{http://adsabs.harvard.edu/abs/2016MNRAS.460.4277G}{Significantly improving stellar mass and radius estimates: A new reference function for the $\Delta\nu$ scaling relation}'', 
    \emph{Monthly Notices of the Royal Astronomical Society}, 461 (2)
    
    \item Glover, M., \textbf{Bellinger, E.~P.}, Radivojac, P., \& Clemmer, D.: 2015, 
    ``\href{https://www.ncbi.nlm.nih.gov/pubmed/26192015}{Penultimate Proline in Neuropeptides}'',
    \emph{Analytical Chemistry}, 87 (16), 8466-8472 
    
    \item Ji, C., Li, Y., \textbf{Bellinger, E.~P.}, Li, S., Arnold, R., Radivojac, P., \& Tang, H.: 2015, 
    ``\href{https://dl.acm.org/citation.cfm?id=2808750}{A maximum-likelihood approach to absolute protein quantification in mass spectrometry}'', 
    In refereed proceedings of the \emph{6th ACM Conference on Bioinformatics, Computational Biology, and Health Informatics} (pp.~296-305)
    
    \item Ngeow, C.~C., Kanbur, S.~M., \textbf{Bellinger, E.~P.}, Marconi, M., Musella, I., Cignoni, M., \& Lin, Y.~H.: 2012,
    ``\href{http://adsabs.harvard.edu/abs/2012Ap\%26SS.341..105N}{Period-luminosity relations for Cepheid variables: from mid-infrared to multi-phase}'', 
    \emph{Astrophysics and Space Science}, 341 (1), 105-113
\end{enumerate}

\newpage
\noindent \textbf{\large Conference proceedings}\\
\begin{enumerate}
    \item \textbf{Bellinger, E.~P.}, Angelou, G., Hekker, S., Basu, S., Ball, W., \& Guggenberger, E.: 2017,
    ``\href{http://adsabs.harvard.edu/abs/2017EPJWC.16005003B}{Fundamental Parameters in an Instant with Machine Learning: Application to Kepler LEGACY Targets}'', 
    in \emph{Seismology of the Sun and the Distant Stars}, 
    Vol.~60 of \emph{European Physical Journal Web of Conferences}, p.~05003 
    
    \item \textbf{Bellinger, E.~P.}, Wysocki, D., \& Kanbur, S.~M.: 2015,
    ``\href{http://adsabs.harvard.edu/abs/2016CoKon.105..101B}{Measuring amplitudes of harmonics and combination frequencies in variable stars}'', 
    in \emph{Communications from the Konkoly Observatory of the Hungarian Academy of Sciences}, 105 
    
    \item \textbf{Bellinger, E.~P.}, Kanbur, S.~M., \& Ngeow, C.~C.: 2012, 
    ``\href{https://static-content.springer.com/esm/chp\%3A10.1007\%2F978-3-642-29630-7_53/MediaObjects/299986_1_En_53_MOESM13_ESM.pdf}{New insights into the Cepheid PL Relation through the use of multiphase relations}'', 
    in proceedings of the \emph{20th Stellar Pulsations Conference} 
    
    \item \textbf{Bellinger, E.~P.}: 2012,
    ``\href{http://www.ncurproceedings.org/ojs/index.php/NCUR2012/article/view/300}{Multiphase Relations of Magellanic Cloud Cepheids}'', 
    in proceedings of the \emph{2012 National Conference on Undergraduate Research}
    
    \item \textbf{Bellinger, E.~P.}, Kanbur, S.~M., \& Ngeow, C.~C.: 2011,
    ``\href{http://aspbooks.org/custom/publications/paper/451-0311.html}{Multiphase Comparison of Period-Luminosity Relations for Magellanic Cloud Cepheids}'', 
    in proceedings of the \emph{9th Pacific Rim Conference on Stellar Astrophysics}, 451 (311)
    
    \item Hekker, S., Elsworth, Y., Basu, S., \& \textbf{Bellinger, E.~P.}: 2017,
    ``\href{http://adsabs.harvard.edu/abs/2017EPJWC.16004006H}{Evolutionary states of red-giant stars from grid-based modelling}'', 
    in \emph{Seismology of the Sun and the Distant Stars}, 
    Vol.~160 of \emph{European Physical Journal Web of Conferences}, p.~05003 
    
    \item Reyner, S., \textbf{Bellinger, E.~P.}, \& Kanbur, S.~M.: 2012,
    ``\href{https://static-content.springer.com/esm/chp\%3A10.1007\%2F978-3-642-29630-7_53/MediaObjects/299986_1_En_53_MOESM44_ESM.pdf}{The approximation of RR Lyrae and eclipsing binary light curves using cubic polynomials}'', 
    in proceedings of the \emph{20th Stellar Pulsations Conference}
\end{enumerate}

\vspace{2\baselineskip}

\noindent \textbf{\large Technical reports}\\
\begin{enumerate}
    \item \textbf{Bellinger, E.~P.}, Conner, D., Mittman, D., Magee, K., \& Heventhal, B.: 2012,
    ``\href{https://trs.jpl.nasa.gov/handle/2014/43122}{CASSIUS: the Cassini Uplink Scheduler}'', 
    \emph{JPL: NASA}, hdl:2014/43122
\end{enumerate}

\chapter*{Acknowledgements\markboth{Acknowledgements}{Acknowledgements}}
\addcontentsline{toc}{chapter}{Acknowledgements}
This thesis represents the culmination of my, by now, nearly ten-year-long fascination with variable stars, which began way back in my first year of university. 
It would have been all but impossible to chase this dream without the support of many individuals. 
I would like now to give thanks to all those who have supported me on this journey. 

I would first like to thank my doctoral advisors, Dr.\ ir.\ Saskia Hekker and Prof.\ Dr.\ Sarbani Basu, for their advice, guidance, and good ideas over the past three years. 
I appreciate the amount they pushed me to make this thesis what it is, and I look back with amazement at all the things I have been given the opportunity to learn about. 
I am proud of the hard work that they encouraged from me, and I look forward to continued collaboration in the future. 

During my studies, I have had the great fortune of being able to lean on the expertise of two post-docs, Dr.\ George Angelou and Dr.\ Warrick Ball. 
Without their help, I would have surely been stuck in the dark for far longer than I was. 
I want to especially thank George for teaching me about stellar evolution, and to thank Warrick for teaching me about kernels. 
I hope we will continue to collaborate long into the future! 

Next I want to thank the SAGE Group at the Max Planck Institute for Solar System Research and the Department of Astronomy at Yale University for hosting me over these three years. 
I have greatly enjoyed my stays, the exchange of ideas, and the numerous friendships that I've made in these places. 
I want to specifically thank Dr.\ Andr\'es Garc\'ia Saravia Ortiz de Montellano and Dr.\ Timo Reinhold for their valued help with this thesis. 
I also thank the IMPRS scientific coordinator, Dr.\ Sonja Schuh, and the staff at both Yale University and the MPS for all their assistance. 
I especially want to thank the IMPRS Student Group, which makes it easy for anyone from anywhere to fit in and make friends. 

Special thanks go to the Director of the Max Planck Institute for Solar System Research, Prof.\ Dr.\ Laurent Gizon; the Director of the GWDG, Prof.\ Dr.\ Ramin Yahyapour; and the Dean of Computer Science, Prof.\ Dr.\ Jens Grabowski for helping me to enroll into the 
G\"ottingen Ph.D.\ Programme in Computer Science. 
Additionally, I thank the remaining members of the examination board, Prof.\ Dr.\ Carsten Damm, Jun.\ Prof.\ Dr.\ Ing.\ Marcus Baum, and Prof.\ Dr.\ Yvonne Elsworth, FRS for agreeing to examine this thesis. 

I thank the National Physical Science Consortium for their very generous support in the form of a graduate fellowship over five years of my graduate studies. 
I also thank Dr.\ Judith E.\ Devaney Terrill for selecting me for the NPSC Fellowship, for hosting me at NIST for two summers, and especially for always encouraging a strong scientific mindset. 

I have had the privilege and honor of working with and (co-)supervising several wonderful students over the course of my graduate studies. 
I want to acknowledge: Felix Ahlborn (now a Ph.D.\ student at the Max Planck Institute for Astrophysics), Kenny Roffo (now employed at the NASA Jet Propulsion Laboratory and pursuing graduate studies at Johns Hopkins University), Marc Hon (finishing up his Ph.D.\ at the University of New South Wales in Sydney, Australia), and Alejandra Perea Rojas (in the midsts of applying to prestigious universities). 
I'm proud of you all - keep up the great work! 

At the Max Planck Institute for Solar System Research, we started a band called MegaGau{\ss} that practices every Monday evening and provides a much needed reprieve from the sometimes rollarcoaster-like nature of academia. 
I want to thank everyone who has played and participated over the last three years and over the many gigs we had; this list includes over twenty people! 
With no guarantee of completeness, the band included 
Abbey Ingram, 
Alessandro Cilla, 
Bastian Proxauf, 
Carla Wiles, 
ChiJu Wu, 
Daniel Maase,
David Marshall, 
Fatima Kahil, 
Felix Mackebrandt, 
Hans Huybrighs, 
Holly Waller, 
Katja Karmrodt, 
Kenny Roffo, 
Nils Gottschling, 
Robin Thor, 
Sudharshan Saranathan, 
Dr.\ Ankit Barik, 
Dr.\ David Martin Belda, 
Dr.\ Emanuele Papini, 
Dr.\ James Kuszlewicz, 
Dr.\ Keaton Bell, 
Dr.\ Theodosis Chatzistergos, and
Dr.\ Vera Dobos. 
Special thanks go out to my ``other half'' of the rhythm section, Helge Mi{\ss}bach, without whom there would have been no band! 

I want to take this opportunity to thank some of the teachers who have encouraged and inspired me over the years. 
This list includes my high school English, history, and physics teachers: Mr.\ Nelson, Mr.\ Kaufman, Mr.\ Battisti; and several of my college computer science professors: Prof.\ Vampola, Prof.\ Graci, and Prof.\ Dr.\ Early. 

To my `cohort' in the IMPRS school, Alessandro Cilla and Fatima Kahil, and to my other graduate student friends as well: best of luck with finishing your studies! 
To my friend K.\ Casey Shea, thank you for making this amazing thesis cover design for me! 
To all of my dear friends whom I have made over these years of study, thank you for making this journey more enjoyable than it certainly could have been. 
Special thanks go to Carla Wiles, for many things, including her support and her valued opinions on all the aesthetic aspects of this thesis. 

I want to thank my family for their unwavering support in my choice to study something as academic as the distant stars. 
I thank my mother Patricia, my father Paul, my sister Bobbie Lee, her partner Johnny, my brother Sean, my sister-in-law Valentina, my niece Nia, my nephews Rashay and Darius, and my step-parents Ron and Nina. 

Last, and certainly not least, I dedicate this thesis to my mentor, Prof.\ Dr.\ Shashi M.\ Kanbur, who has continuously and actively encouraged me over the past decade to pursue my ``academic dreams.'' 
Thank you, Shashi, for always being there for me, and for showing me the light of variable stars.

\chapter*{Curriculum vitae\markboth{Curriculum vitae}{Curriculum vitae}}
\addcontentsline{toc}{chapter}{Curriculum vitae}
{\Large\textbf{Earl Patrick Bellinger}}

  \section*{\sc\underline{Education}}
  \vspace*{-2mm}

  \noindent\textbf{Ph.D.~Candidate}, Institute of Computer Science, University of G\"ottingen\\ 
  International Max Planck Research School for Solar System Science\\
  Fellow of the National Physical Science Consortium\vspace{2mm}\\
  \noindent\textbf{M.Sc.~Computer Science}, Indiana University Bloomington, USA \hfill 2014\\
  Fellow of the National Physical Science Consortium \\
  GPA: 3.95/4.0\vspace{2mm}\\
  \textbf{B.Sc.~Applied Mathematics}, SUNY Oswego, NY, USA \hfill 2012\\
  \textbf{B.Sc.~Computer Science}, \emph{ibid.} \hfill 2012\\
  Presidential Scholar\\
  Honors Thesis: \emph{Multiphase Relations of Magellanic Cloud Cepheids} \\
  GPA: 3.81/4.0 (\emph{summa cum laude}, ranked \#1 in Computer Science)

  \section*{\sc\underline{Research Positions}}
  \vspace*{-2mm}
  
  \noindent\textbf{Max Planck Institute for Solar System Research} \emph{(Germany)} \hfill 2015 -- 2018\\
  Doctoral Candidate, Stellar Ages \& Galactic Evolution Group \vspace{1mm}\\
  \textbf{Yale University} \emph{(USA)} \hfill 2016 -- 2017\\
  Visiting Assistant in Research, Department of Astronomy\vspace{1mm}\\
  \textbf{Indiana University} \emph{(USA)} \hfill 2013 -- 2015\\
  Research Assistant, School of Informatics \& Computing \vspace{1mm}\\
  \textbf{NIST Information Technology Laboratory} \emph{(USA)} \hfill 2013 -- 2014\\
  Guest Researcher, Scientific Applications and Visualization Group \vspace{1mm}\\
  \textbf{National Center of Sciences} \emph{(Japan)} \hfill 2013\\
  Research Student, National Institute of Informatics \vspace{1mm}\\
  \textbf{NASA Jet Propulsion Laboratory} \emph{(USA)} \hfill  2012\\
  SURF Fellow, Cassini Mission to Saturn \vspace{1mm}\\
  \textbf{Federal University of Alagoas} \emph{(Brazil)} \hfill 2011\\
  REU Student, Institute of Physics\vspace{1mm}\\
  \textbf{Federal University of Santa Catarina} \emph{(Brazil)} \hfill 2010\\
  REU Student, Department of Physics
  
  \section*{\sc\underline{Teaching Positions}}
  \vspace*{-2mm}
  \textbf{Yale University} \hfill Spring 2017\\
  Teaching Assistant, Department of Astronomy\vspace{1mm}\\
  \textbf{University of G\"ottingen} \hfill Summer 2016\\
  Assistant, Institute for Astrophysics \vspace{1mm}\\
  \textbf{Indiana University} \hfill Fall 2012\\
  Associate Instructor, School of Informatics \& Computing\vspace{1mm}\\
  \textbf{SUNY Oswego} \hfill Fall 2010\\
  Seminar Leader, Honors Department
  
  \vspace*{-4.2mm}
  \section*{{\sc\underline{Selected Talks}} \hfill {\normalsize\normalfont $^{\scriptscriptstyle\bigstar}$\emph{invited}}}
  \vspace*{-2mm}
  \noindent $^{\scriptscriptstyle\bigstar}$\textbf{Stellar Astrophysics Centre Seminar} (Aarhus, Denmark) \hfill{2018} \\
  \hphantom{$^{\scriptscriptstyle\bigstar}$}``\emph{Determining stellar structure with asteroseismology using novel techniques}''
  
  \vspace{1mm}
  \noindent \hphantom{$^{\scriptscriptstyle\bigstar}$}\textbf{TESS/Kepler Asteroseismic Science Consortium} (Aarhus, Denmark) \hfill{2018} \\
  \hphantom{$^{\scriptscriptstyle\bigstar}$}``\emph{Testing stellar physics with asteroseismic inversions of solar-type stars}''

  \vspace{1mm}
  \noindent $^{\scriptscriptstyle\bigstar}$\textbf{Madison Seminar} (University of Wisconsin--Madison, USA) \hfill{2017} \\
  \hphantom{$^{\scriptscriptstyle\bigstar}$}``\emph{From Starlight to Stellar Ages with Asteroseismology}''
  
  \vspace{1mm}
  \noindent \hphantom{$^{\scriptscriptstyle\bigstar}$}\textbf{Rocks \& Stars II} (Max Planck Institute, G\"ottingen, Germany) \hfill{2017} \\
  \hphantom{$^{\scriptscriptstyle\bigstar}$}``\emph{The Seismic Structures of Solar-Type Stars}''
  
  \vspace{1mm}
  \noindent \hphantom{$^{\scriptscriptstyle\bigstar}$}\textbf{ERES-III} (Yale University, New Haven, CT, USA) \hfill{2017} \\
  \hphantom{$^{\scriptscriptstyle\bigstar}$}``\emph{Fundamental Parameters of Exoplanet Host Stars with Asteroseismology}''
  
  \vspace{1mm}
  \noindent $^{\scriptscriptstyle\bigstar}$\textbf{Science Today} (Public talk at SUNY Oswego, NY, USA) \hfill{2017} \\
  \hphantom{$^{\scriptscriptstyle\bigstar}$}``\emph{A Look Inside the Private Lives of Stars}''
  
  \vspace{1mm}
  \noindent $^{\scriptscriptstyle\bigstar}$\textbf{Red Giant Modeling Workshop} (G\"ottingen, Germany) \hfill{2016} \\
  \hphantom{$^{\scriptscriptstyle\bigstar}$}``\emph{Fundamental Stellar Parameters in an Instant with Machine Learning}''
  
  \vspace{1mm}
  \noindent \hphantom{$^{\scriptscriptstyle\bigstar}$}\textbf{RR Lyrae} (Visegr\'ad, Hungary) \hfill{2015} \\
  \hphantom{$^{\scriptscriptstyle\bigstar}$}``\emph{Resolving Combination Frequency Amplitudes of Multimode Pulsators}''
  
  \vspace{1mm}
  \noindent \hphantom{$^{\scriptscriptstyle\bigstar}$}\textbf{American Astronomical Society} (Seattle, WA, USA) \hfill{2015} \\
  \hphantom{$^{\scriptscriptstyle\bigstar}$}``\emph{Optimal Model Discovery of Periodic Variable Stars}''
  
  \vspace{1mm}
  \noindent $^{\scriptscriptstyle\bigstar}$\textbf{Delhi Workshop on Variable Stars} (Delhi, India) \hfill{2015} \\
  \hphantom{$^{\scriptscriptstyle\bigstar}$}``\emph{Calibrating the Cepheid Distances to the Magellanic Clouds}''
  
  \vspace{1mm}
  \noindent $^{\scriptscriptstyle\bigstar}$\textbf{Kerala Workshop on Stellar Astrophysics} (Kerala, India) \hfill{2014} \\
  \hphantom{$^{\scriptscriptstyle\bigstar}$}``\emph{Automated Supervised Classification of Variable Stars}''

  \vspace*{-4.2mm}
  \section*{\sc\underline{Honors \& Awards}}
  \vspace*{-2mm}
  Stellar Astrophysics Centre Postdoctoral Fellowship \hfill 2018 -- 2021\vspace{1mm}\\
  National Physical Science Consortium Graduate Fellowship    \hfill 2012 -- 2017\vspace{1mm}\\
  SUNY Oswego Presidential Scholarship                        \hfill 2008 -- 2012\vspace{1mm}\\
  Oebele Van Dyk Outstanding Computer Science Senior Award    \hfill 2012\vspace{1mm}\\
  SUNY Chancellor's Award 
  \hfill 2012\vspace{1mm}\\
  SUNY Oswego Student/Faculty Collaborative Challenge Grant   \hfill 2011

\end{document}